\documentclass[twoside,11pt]{book}

\usepackage[svgnames]{xcolor}
\usepackage{fancyhdr,indentfirst}
\usepackage[applemac]{inputenc}
\usepackage[T1]{fontenc}
\usepackage{palatino,textcomp,ulem}
\usepackage[Lenny]{fncychap}
\usepackage{varioref}
\usepackage{graphicx}
\usepackage{afterpage}
\usepackage{pdfpages}
\includepdfset{pages=-,link,pagecommand={\thispagestyle{fancy}}}
\usepackage{wrapfig,sidecap}
\usepackage{lscape,rotating}
\usepackage{longtable,dcolumn,colortbl,multicol,multirow}
\usepackage[labelfont=bf]{caption}
\usepackage{amsmath, amssymb}
\usepackage{natbib}
\bibpunct{(}{)}{;}{a}{}{,}
\usepackage[pdftex]{hyperref}   % doit toujours être chargé en dernier (sauf geometry): redéfinit certaines commandes
\hypersetup{bookmarks=true,colorlinks=true,linkcolor=black,citecolor=black,urlcolor=black}
\usepackage[a4paper,includeheadfoot,scale={0.82,0.88},heightrounded,centering,bindingoffset=.8cm]{geometry}
\graphicspath{{.}{fig/chap1/}{fig/chap2/}{fig/chap3/}{fig/chap4/}{fig/chap5/}{fig/appendA/}{fig/papers/}{fig/proc/}}

\def\aap{A\&A}%
\newcommand{\eprint}[2][]{{\tt\if!#1!#2\else#1:#2\fi}}

\begin{document}

\defcitealias{mm8}{Paper VIII}

\defcitealias{abn02}{ABN02}
\defcitealias{bcl02}{BCL02}

\defcitealias{chieflim04}{CL04}
\defcitealias{Hi07}{Hi07}
\defcitealias{tun07}{TUN07}

\defcitealias{Reimers75}{R75}
\defcitealias{dJ88}{dJ88}
\defcitealias{kupu00}{KP00}
\defcitealias{vink00}{VdKL00}
\defcitealias{vink01}{VdKL01}
\defcitealias{Kudr02}{K02}
\defcitealias{eldrvink06}{EV06}
\defcitealias{grafham08}{GH08}

\defcitealias{Salpeter55}{S55}
\defcitealias{milsca79}{MS79}
\defcitealias{nakum01}{NU01}

\defcitealias{cf88}{CF88}
\defcitealias{adelb98}{Adelb98}
\defcitealias{cfhz85}{CFHZ85}

% \citealt{Vink00} \citepalias[hereafter][]{Vink00}

\thispagestyle{empty}
{\centering\enlargethispage*{3cm}\bfseries\large
%\vspace*{-1.5cm}
\hspace*{-1cm}\textsc{Université de Genève}\hfill \textsc{Faculté des Sciences}\hspace*{-1cm}\\
\hspace*{-1cm}Département d'astronomie\hfill Professeur Georges Meynet\hspace*{-1cm}\\
\hspace*{-1cm}\hrulefill\hspace*{-1cm}\\[2cm]
%{\bfseries\Huge Nucleosynthesis\\[.25cm] in the Early Massive Stars\\[.25cm] and Abundances of Metal-Poor stars}\\
%{\bfseries\Huge The early stellar generations:\\[.25cm] massive stars \\[.25cm] and Carbon-Enhanced Metal-Poor stars}\\%\\[.25cm] and Abundances of Metal-Poor stars}\\
%{\bfseries\Huge Evolution and nucleosynthesis\\[.25cm] of the early generation of stars}\\%\\[.25cm] and Abundances of Metal-Poor stars}\\
%{\bfseries\Huge Evolution and nucleosynthesis\\[.25cm] of the early generation of stars}\\%\\[.25cm] and Abundances of Metal-Poor stars}\\
{\bfseries\huge The Early Generations of Rotating Massive Stars\\[.25cm] and the Origin of \\[.45cm] Carbon-Enhanced Metal-Poor Stars}\\%\\[.25cm] and Abundances of Metal-Poor stars}\\
\vspace{3cm}
{\Large \textsc{Thèse}}\\[3em]

présentée à la Faculté des sciences de l'Université de Genève\\
pour obtenir le grade de Docteur ès sciences,\\
mention Astronomie et Astrophysique\\[3em]

par\\[1em]
{\Large Arthur \textsc{Choplin}}\\[1em]
de\\[1em]
Gex (France)\\[2cm]

{\large Thèse N$^\mathrm{o}$ 5269}\\[3cm]
\large \textsc{Genève}\\
Observatoire de Genève\\
2018\\
}

%-------------------------------------------------------------------------------

\newpage
\thispagestyle{empty}
\centerline{\resizebox{!}{1.4\textwidth}{\includegraphics{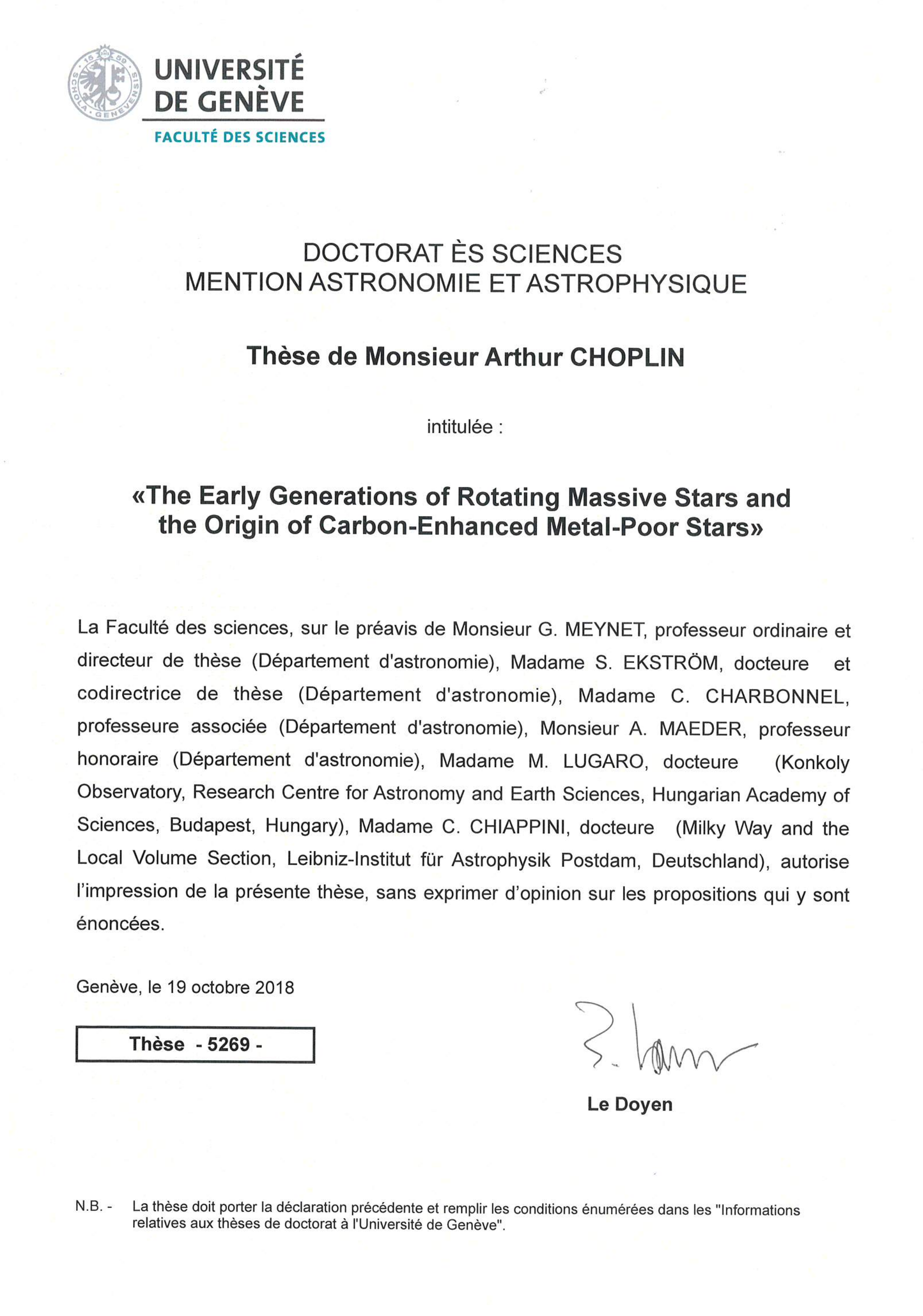}}}

\vspace{0.2cm}

%\textcolor{red}{ letter thesis}

%\begin{quote}
%Ce travail de th\`ese a donn\'e lieu \`a des publications dont la liste se trouve \`a la page \pageref{cpublilist}.
%\end{quote}

{
\frontmatter

\chapter*{Abstract}
\addcontentsline{toc}{chapter}{Abstract}
\markboth{ABSTRACT}{Abstract}
%\chapter*{Thesis Outline}
%\addcontentsline{toc}{chapter}{Thesis Outline}
%\markboth{THESIS OUTLINE}{Thesis Outline}

%Since the first massive stars released the first metals in the Universe, 
%The onset of the chemical of the universe was triggered by  generation of massive stars 
%The first stars in the universe released the first metals and set the chemical condition for the next generation.
The study of the long-dead early generations of massive stars is crucial in order to obtain a complete picture of the chemical evolution of the Universe, hence the origin of the elements. %that impacted in the chemical evolution of the Universe.
%The nature of these stars can be inferred indirectly thanks to the chemical composition of low-mass metal-poor stars observed in our galaxy, some of which are almost as old as the universe. 
The nature of these stars can be inferred indirectly by investigating the origin of low-mass metal-poor stars observed in our Galaxy, some of which are almost as old as the Universe. 
%Some of these metal-poor stars probably formed with the material ejected by only one or a few previous massive stars. %, hence providing a remarkable window on the early universe. 
The peculiar extremely iron-poor Carbon-Enhanced Metal-Poor (CEMP) stars, whose precise origin is still debated, are thought to have formed with the material ejected by only one or very few previous massive stars, hereafter called the source stars. 
%Rosetta stone
In this context, the source stars belong to the very first generations of massive~stars.

The main aim of this thesis is to explore the physics and the nucleosynthesis of the early generations of massive stars. %in the Universe.
It is achieved by combining stellar evolution modeling including rotation and full nucleosynthesis with observations of CEMP stars. By fitting the abundances of CEMP stars with the ejecta of new massive source star models, characteristics of the massive source stars can be deduced. %It allows

%As a first step, the observational steps to get abundances from CEMP stars as well as the scenarios for the origin of CEMP stars are reviewed. 
%The theoretical and modeling aspects of massive stellar evolution that are relevant for this work are then presented. 
%The origin of CEMP stars not significantly enriched in s- and r-elements (CEMP-no) is explored in chapter~\ref{cempno} with new very low or zero metallicity pre-supernova models of $20-60$ $M_{\odot}$ rotating source stars.
%The origin of CEMP stars enriched in s-elements (CEMP-s) is examined in chapter~\ref{cemps} in light of new massive source star models including full nucleosynthesis so as to follow the s-process. 
%The summary of these investigations, constraining the nature of the early massive stars, is given in chapter~\ref{Cconclu}. 

%I found that 
The efficient rotational mixing operating in rotating source stars triggers exchanges of material between the helium burning core and the hydrogen burning shell, and leads to a very rich and varied nucleosynthesis. This interplay between rotation and nucleosynthesis in massive source stars is able to cover the abundance scatter of light elements (carbon to silicon) of the CEMP stars with [Fe/H] $<-3$.
Some remaining discrepancies between models and observations can be alleviated if a late mixing process is included in the source star. It operates about 200 years before the end of the source star evolution, between the hydrogen burning shell and the helium burning shell. A detailed investigation of the abundances of light elements of 69 CEMP stars with [Fe/H] $<-3$ suggests that the best source stars are preferentially fast rotating 20 $M_{\odot}$ stars which experienced the late mixing process and an ejection of only their outer layers. 

The rotation-induced mixing also boosts the weak s-process in source stars, provided the initial metallicity is not strictly zero. The s-elements up to barium are strongly overproduced, especially if the initial rotation rate is high. 
The s-process boost induced by rotation was found to be significant for stars with initial masses below 60 $M_{\odot}$. Above this threshold, the interplay between rotation and nucleosynthesis is weaker so that rotating and non-rotating models give similar results. %the effect of rotation on the production of s-elements remains modest.  so that %were found to overproduced (mainly light) s-elements. 
%At lower [Fe/H] ratios, although less efficient, the s-process still operates. %, provided the initial rotation is not zero.
The signature of the rotation-induced s-process boost in $20-25$~$M_{\odot}$ source stars was found to be consistent with the abundance pattern of several CEMP stars enriched in s-elements.

The examination of the origin of CEMP stars in light of new massive source star models suggests that rotation was a dominant effect in the early generations of massive stars. Strong mixing events between burning shells are also expected to be a common feature in these early massive stars. Finally, early massive stars should have lost only their envelope and experienced rather strong fallback.

\chapter*{Résumé}
%\addcontentsline{toc}{chapter}{Résum}
\addcontentsline{toc}{chapter}{Résumé en fran\c cais}

\markboth{R\'ESUM\'E}{Résumé}

\begin{center}
\LARGE{Les premi\`eres g\'en\'erations d'\'etoiles massives en rotation \\ et l'origine des \'etoiles CEMP}\\
%\huge{Titre}\\
\end{center}

\vspace{1.cm}

\section*{Contexte}

Les premi\`eres g\'en\'erations d'\'etoiles massives\footnote{Au moins 8 $M_{\odot}$, i.e. 8 fois la masse du Soleil.} sont des acteurs essentiels dans l'histoire de l'Univers. 
\`A la fin de leur \'evolution, ces \'etoiles \`a courte dur\'ee de vie ont explos\'e en supernovae $-$ un \'ev\`enement dont la luminosit\'e peut atteindre 10 milliards de fois celle du Soleil $-$ et ont, pour la premi\`ere fois, enrichi l'Univers en m\'etaux\footnote{\'El\'ements plus lourds que l'h\'elium (par exemple l'oxyg\`ene).}, dont nous-m\^emes et notre Terre sommes constitu\'es.
La compr\'ehension, encore m\'econnue aujourd'hui, de l'\'evolution de ces \'etoiles, est primordiale afin d'acqu\'erir une connaissance globale de l'\'evolution de l'Univers, en particulier de son \'evolution chimique.

Bien qu'autour de nous, ces \'etoiles ne peuvent plus \^etre observ\'ees, leur nature peut \^etre \'etudi\'ee indirectement gr\^ace \`a l'observation, dans notre Galaxie, d'\'etoiles de faible masse et tr\`es pauvres en m\'etaux. Certaines de ces \'etoiles $-$ \textit{Carbon-Enhanced Metal-Poor} ou \textit{CEMP} $-$ se sont probablement form\'ees tr\`es t\^ot dans l'Univers, avec la mati\`ere \'eject\'ee d'une ou de quelques-unes des premi\`eres \'etoiles massives, appel\'ees ci-apr\`es \textit{\'etoiles sources}.
La composition chimique particuli\`ere des \'etoiles CEMP, d\'etermin\'ee par spectroscopie, donne de pr\'ecieuses informations sur la nucl\'eosynth\`ese, et donc sur la nature de leur(s) \'etoile(s) source(s). Les \'etoiles CEMP pourraient \^etre compar\'ees \`a des \textquotedblleft Pierre de Rosette\textquotedblright: leur d\'echiffrement permet d'\'etudier une civilisation ancienne, aujourd'hui disparue.

\section*{Objectifs et m\'ethodes}

Le but principal de ce travail de th\`ese est d'obtenir de nouveaux indices sur la physique des premi\`eres g\'en\'erations d'\'etoiles massives en utilisant les contraintes li\'ees \`a l'observation des \'etoiles CEMP.
Une telle \'etude est possible en combinant des mod\`eles num\'eriques d'\'etoiles sources avec les observations d'\'etoiles CEMP disponibles dans la litt\'erature.
Le code d'\'evolution stellaire de Gen\`eve, principal outil utilis\'e durant cette th\`ese, permet de mod\'eliser l'\'evolution des \'etoiles sources. 
%Comme les quelques codes de ce type existant dans le monde, 
%Outre les ingredients physiques standard (convection, ) 
Une des sp\'ecificit\'es de ce code est d'inclure une mod\'elisation sophistiqu\'ee de la rotation diff\'erentielle. 
%\oe{}uvrant dans les \'etoiles. 
En outre, ce code permet le calcul de la nucl\'eosynth\`ese compl\`ete, y compris le \textit{processus s} (capture lente de neutron par des noyaux atomiques, permettant la production d'\'el\'ements lourds). Ce processus, intervenant dans les \'etoiles de masse interm\'ediaire $-$ entre 1 et 8 $M_{\odot}$ $-$ et massives, est suppos\'e \^etre \`a l'origine d'environ 50~\% des \'el\'ements plus lourds que le fer, tels que le strontium, baryum, tungst\`ene ou plomb par exemple.

\section*{R\'esultats}

Le m\'elange induit par la rotation dans l'\'etoile source d\'eclenche des \'echanges successifs de mati\`ere entre 
%son c\oe{}ur 
la zone de combustion d'h\'elium et celle d'hydrog\`ene. Il en r\'esulte une nucl\'eosynth\`ese riche et vari\'ee, permettant la cr\'eation, en grande quantit\'e, de carbone, azote, oxyg\`ene, fluor, n\'eon, sodium, magn\'esium et aluminium. La composition chimique de cette mati\`ere, enrichie en \'el\'ements divers, et finalement \'eject\'ee par les vents stellaires et/ou lors de la supernova des mod\`eles d'\'etoiles sources, est en accord global avec la composition chimique des \'etoiles CEMP les plus pauvres en fer. 

Une comparaison plus d\'etaill\'ee entre ces \'etoiles CEMP et les mod\`eles d'\'etoiles sources a mis en \'evidence certains \'ecarts entre les mod\`eles et les observations. Un m\'elange additionnel, survenant dans une zone sp\'ecifique de l'\'etoile source, quelques centaines d'ann\'ees avant son explosion, permet de pallier ces d\'esaccords. Une proc\'edure automatique, d\'eterminant, parmi environ 35000 possibilit\'es, le meilleur mod\`ele d'\'etoile source pour une \'etoile CEMP donn\'ee, a \'et\'e appliqu\'ee \`a un \'echantillon de 69 \'etoiles CEMP. Les r\'esultats sugg\`erent que les \'etoiles sources sont pr\'ef\'erentiellement des \'etoiles en rotation rapide, d'environ 20 $M_{\odot}$, qui ont subi le m\'elange additionnel et qui n'ont \'eject\'e que leurs couches externes.

Le m\'elange rotationnel affecte \'egalement la synth\`ese d'\'el\'ements \textit{s} dans l'\'etoile source, et ce, d'autant plus que la vitesse de rotation initiale est \'elev\'ee. La mod\'elisation d'une nouvelle grille d'\'etoiles massives $-$ $10 < M_{\rm ini} < 150$~$M_{\odot}$ $-$, incluant rotation et \textit{processus s} montre que l'effet de la rotation sur le \textit{processus s} intervient de mani\`ere significative uniquement si la masse initiale de l'\'etoile est inf\'erieure \`a 60 $M_{\odot}$. Au-del\`a de ce seuil, l'interaction entre le m\'elange rotationnel et la nucl\'eosynth\`ese est moindre. Par cons\'equent, les mod\`eles d'\'etoiles sources en rotation donnent des r\'esultats similaires aux mod\`eles sans rotation. 
La signature chimique induite par la rotation sur la production d'\'el\'ements \textit{s} dans des \'etoiles sources de $20-25$ $M_{\odot}$ a \'et\'e identifi\'ee dans plusieurs \'etoiles CEMP enrichies en \'el\'ements \textit{s}.

\section*{Conclusion}

L'\'etude de l'origine des \'etoiles CEMP \`a la lumi\`ere de nouveaux mod\`eles d'\'etoiles sources vient corroborer l'id\'ee selon laquelle la rotation
%, en alt\'erant la nucl\'eosynth\`ese des premi\`eres g\'en\'erations d'\'etoiles massives, 
a jou\'e un r\^ole pr\'epond\'erant dans l'Univers jeune. 
%en alt\'erant la nucl\'eosynth\`ese des premi\`eres g\'en\'erations d'\'etoiles massives et donc l'enrichissement chimique . 
%les premi\`eres g\'en\'erations d'\'etoiles massives. 
Par ailleurs, ce travail sugg\`ere que (1) les premi\`eres g\'en\'erations d'\'etoiles massives pourraient, pour la plupart, avoir une masse d'environ 20 $M_{\odot}$ (2) de puissantes interactions entre diff\'erentes coquilles de combustion \'etaient \`a l'\oe{}uvre dans ces \'etoiles et (3) ces \'etoiles ont probablement \'eject\'e uniquement leur enveloppe, tandis que les r\'egions plus internes sont tomb\'ees sur l'objet compact central.

\chapter*{Remerciements}
\addcontentsline{toc}{chapter}{Remerciements}

\markboth{REMERCIEMENTS}{Remerciements}

%immense merci. Belle experience
Tout d'abord, un immense merci \`a Georges. 
C'est une grande chance d'avoir eu un directeur si enthousiaste et comp\'etent, dot\'e d'une curiosit\'e sans bornes, captivant en cours, conf\'erence ou lors d'une fondue bien arros\'ee. Merci de m'avoir laiss\'e si libre et autonome, tout en restant toujours imm\'ediatement disponible pour discuter, pour me conseiller, me rassurer ou me remettre sur les rails quand je m'\'egarais un peu. %C'est ‡ partir de ce socle fort robuste que j'ai pu mener 
%C'est un vrai r\'egal de c\^otoyer quelqu'un de si curieux, enthousiate 
%J'ai eu la chance de profiter pendant 4 ans de son incroyable (et communicatif) enthousiasme, de sa curiosit\'e \`a toute \'epreuve et de son experience. 
%En restant \`a Gen\`eve et ses alentours, j'adresse 
Un grand merci \`a Sylvia, en particulier pour son aide pr\'ecieuse dans les premiers mois. Il en va de m\^eme pour Andr\'e, ses conseils avis\'es et encouragements furent tr\`es appr\'eci\'es.
Merci \`a Cyril, toujours disponible et pr\^et \`a m'aider quand je l'ai sollicit\'e. Merci \`a Corinne pour m'avoir chaleureusement accueilli et pour sa bienveillance.
%Merci \`a toute la super \'equipe d'\'evolution stellaire. 
Je tiens \`a souligner que ce fut un vrai r\'egal de travailler au sein d'une telle \'equipe, accueillante, dynamique, dr\^ole. Bref, l'ambiance est bonne. Un merci tout sp\'ecial \`a Patrick, pour sa bonne humeur et son ind\'efectible humour. 
%egaye une discussion de couloir, pause caf\'e

En dehors de Gen\`eve, je remercie Raphael, avec qui c'est un vrai plaisir de collaborer. Merci pour m'avoir initi\'e au \textit{processus s}, au golf et pour m'avoir gracieusement permis de stocker quelques t\'eraoctets sur tes disques... Je remercie Cristina pour ses pr\'ecieux conseils scientifiques et encouragements. Merci \`a Alain Coc, Keith A. Olive, Jean-Philippe Uzan et Elisabeth Vangioni, sans lesquels une partie de cette th\`ese n'aurait pas pu exister. Merci \`a Alison Laird pour sa contribution et son inter\^et pour certains des travaux de cette th\`ese. Merci \`a Maria Lugaro pour ses commentaires bienveillants et constructifs sur mon travail.
Merci \`a Tim Beers, aux organisateurs de \textit{Gala of GALAH} et \`a Jinmi Yoon pour l'inter\^et qu'ils ont port\'e \`a mon travail et pour la confiance qu'ils m'ont accord\'ee.
Une pens\'ee aussi \`a mes anciens ma\^itres de stage de Grenoble, J\'er\^ome, V\'eronique et Jean-Charles avec qui j'ai fait mes premiers pas dans le monde de la recherche.

De retour \`a Gen\`eve, je dis grazie \`a Giovanni pour m'avoir \'epaul\'e au d\'emarrage, et aussi pour ses merveilleuses lasagnes. Merci \`a Mads pour nos premi\`eres soir\'ees sans limites \`a Gex, Sauverny ou Genthod. Merci \`a Lionel pour son introduction muscl\'ee au marxisme-l\'eninisme. Merci \`a Thibaut, un compagnon fort sympathique et enjou\'e pour cette fin de th\`ese. 
Merci \`a ceux qui excellent dans l'art d'\'egayer les pauses caf\'es, les parties de p\'etanque ou de ping-pong: Patrick bien s\^ur mais aussi St\'ephane et Thierry.
Merci \`a Song, qu'il me suffisait de croiser dans le couloir pour passer une bonne journ\'ee, tant sa joie et son affection sont grandes et communicatives. 
%dont la joie et l'affection r\'eveillerait le plus sinistre
Merci \`a Koh et Yu, mes coll\`egues \'eph\'em\`eres, curieux de tout, au travail comme en dehors. Merci \`a Carlo pour sa visite \`a Gen\`eve et nos discussions sur ces beaux objets que sont les CEMP.
Merci \`a Jose et Aline pour leur gentillesse et leur accueil bien amical au sein du groupe. Ce fut un vrai plaisir de les retrouver \`a Dublin pour un petit projet scientifique. Ce s\'ejour irlandais n'aurait pu \^etre aussi fructueux sans l'attention et la patience de Jose. En passant, merci aux coll\`egues dublinois, en particulier Laura, Ioana et Eoin. Assez rapidement, on se sent bien parmi eux, au bureau comme au pub.
Au risque d'en faire un peu trop, j'ai envie de remercier l'Observatoire en g\'en\'eral, car l'atmosph\`ere y est bien agr\'eable... je pense \`a l'administration, la cafet', la m\'ecanique (Robin pour la r\'eparation du piano...), les footeux du vendredi, les (not so) lazy runners et les grimpeurs. 
Un grand merci au groupe de musique de l'Obs. C'\'etait une superbe parenth\`ese hebdomadaire que de jouer avec vous dans ce bunker antiatomique. %emmen\'es par Thibaut, Maroussia et Francois, 
Je pense aussi \`a la belle f\^ete de No\"el 2016 de l'Obs organis\'ee avec vous, inoubliable! %: un peu dur au d\'emarrage mais finalement vraiment excellente.
Toujours dans la musique, merci \`a Lorenzo et Iris pour m'avoir fait connaitre puis permis de jouer le jeudi soir dans ce chouette endroit qu'est \textit{c'est ici l'endroit o\`u}. Musique encore: merci \`a la team du Cully Jazz Festival, sp\'ecialement Barbara et l'incroyable Job (je te suis reconnaissant d'avoir gard\'e ton sang froid \`a la fourri\`ere).
Enfin, merci \`a Uriel pour m'avoir fait d\'ecouvrir les joies du studio. %(un jour j'en aurai un comme toi).

%Bien que pas forc\'ement recommendable sur le papier, le pays de Gex peut finalement s'av\'erer extr\^emement doux \`a vivre quand on cotoie des personnes comme Louis, Florian, Tim, Olja, William, Remy, Aline, Baptiste... Merci \`a vous!

%En dehors de Gen\`eve, je remercie Raphael, avec qui c'est un vrai plaisir de collaborer. Merci pour m'avoir initi\'e au processus s, au golf et pour m'avoir gracieusement permis de stocker quelques t\'eraoctets sur ses disques... Je remercie Cristina pour ses pr\'ecieux conseils scientifiques et encouragements. Merci \`a Alain Coc, Keith A. Olive, Jean-Philippe Uzan et Elisabeth Vangioni, sans lesquels une partie de cette th\`ese n'aurait pas pu exister. Merci \`a Alison Laird et Camilla Hansen pour leur contribution et/ou inter\^et pour certains des travaux de cette th\`ese.
%Une pens\'ee aussi \`a mes anciens maitres de stages de Grenoble, J\'erome, V\'eronique et Jean-Charles avec qui j'ai fait mes premiers pas dans le monde de la recherche.

J'ai aussi des pens\'ees affectueuses et non sans \'emotion pour tous ceux que j'ai rencontr\'e voire rerencontr\'e lors de mes escapades, parfois dans le cadre professionnel, parfois en dehors.
Je pense d\'ej\`a \`a Taygun, Panos, Matt, Sanjana, Jacqueline, Brent et tout ce chouette petit monde crois\'e \`a Niigata. Je pense \`a Taygun encore, Melisa, Mehmet et son chat \`a Istanbul. 
Je pense aussi \`a Gen, Tilman, Rick, Tim, Camilla et Marc-Antoine \`a Melbourne, \`a Rana \`a Notre-Dame. %Une pens\'ee sp\'eciale pour Gen, sa gentillesse et son humour 
%Gr\^ace \`a Maxime, notamment, j'ai pu aller observer au Chili, c'\'etait une magnifique experience . 
Je pense \`a ceux du Chili, o\`u j'ai v\'ecu une magnifique experience d'observation sur Euler. Soyez salu\'es sur l'heure Justyn, Simona, Chiara (membres du LSVB ou La Silla VolleyBall), James, Vincent, Sophie, Am\'elie, Max, Mikaela et Lila!
Je pense aussi aux chamoniards et au refuge.
Je pense \`a Sol\`ene, Klimt et Bill Viola.
Je pense \`a Nicolas, Guillaume, Marco et autres tr\`es joyeux drilles de l'EJC 2016. Je pense \`a Simon, Damien et aux Genevois \`a Saas-Fee.
Je pense enfin \`a Michael, Eric et les amis de l'Observatoire Fran\c cois-Xavier Bagnoud pour avoir partag\'e avec moi cette conf\'erence, raclette et belle soir\'ee \`a Saint-Luc.

Pendant cette th\`ese, si j'ai pu garder (je crois) les pieds sur Terre, c'est d\'ej\`a en partie gr\^ace aux volleyeurs de Pr\'evessin, avec qui j'allais me d\'efouler, boire des canettes et manger des 'zia' chez da Ettore. C'est surement aussi gr\^ace \`a la sophrologie (quelle belle d\'ecouverte...) chez Philippe et Dominique. 
%J'en profite pour remercier Suzanne pour m'avoir aid\'e \`a sortir ma ch\`ere C3 bleue d'un bien mauvais pas.
C'est certainement aussi gr\^ace \`a mes tr\`es chers amis Gessiens, Chenevessiens et Divonnais. Je vous salue chaleureusement Louis, Florian, Tim, Olja, William, Nico, Remy, Aline, Baptiste...! Allez... quelques doux mots-cl\'es, au hasard de mes pens\'ees: lingot, chess championship \`a l'huile de foie de morue, bi\`ere-pizza (Charly's), Sal\`eve, parcours de sant\'e (10 tours), lactaire d\'elicieux (1), bobsleigh, Krazy House... Une franche accolade \`a Louis, pr\'esent (mais disparaissant parfois apr\`es quelques pintes) aussi bien dans les moments un peu difficiles que moins difficiles. %Plus
Si j'ai gard\'e la raison, c'est aussi gr\^ace \`a Fanny, qui a compt\'e, et \`a tous les chouettes moments pass\'es avec la bande, \`a Grenoble, Gex, Paris ou au fin fond de l'Ard\`eche.
C'est sans nul doute aussi gr\^ace \`a mes tous proches et leur bienveillance, mes parents et grand-parents, ma s\oe{}urette, Nathalie, Uli, Julia. Je remercie au passage les relecteurs pour vos corrections et suggestions en g\'en\'eral fort pertinentes (mention sp\'eciale \`a Laurent et Jacqueline).
%au risque de m'egarer un peu,
Avec certitude, si je n'ai pas d\'ebloqu\'e, c'est aussi gr\^ace \`a ma ch\`ere petite Perru***, arriv\'ee sur le tard mais bien arriv\'ee. Que dire? Comment dire? Merci pour ta p\'etillance, bonne humeur permanente, l\'eg\`eret\'e, humour, joie de vivre si communicative, pour avoir eu des paroles r\'econfortantes quand mon moral d\'egringolait. Merci surtout de m'avoir initi\'e aux plantes vertes, \`a Dalida, aux Colocs, \`a la brasse coul\'ee et j'en passe. En passant, justement, merci au reste de la famille Lap' pour votre accueil simple et chaleureux.
Pour terminer, si je n'ai pas trop ondul\'e de la toiture pendant ces ann\'ees, c'est peut \^etre aussi gr\^ace \`a mes amis de longue et tr\`es longue date. Je pense \`a Nathan, notre r\'ecent chavirement et notre future r\'egate, \`a R\'emi, au frelon disparu et aux micro trous noirs, \`a Bidou et au th\`eme ent\^etant de Taxi Driver, \`a Dri, mutique ou chantant John Legend, \`a Verrier ainsi qu'\`a son lit \`a m\'emoire de forme, \`a Dupin, aux m\'enisques bien faits et aux GBF bien r\'egl\'es, \`a Robin et Jeff Thirion, \`a Jouty et \`a nos frasques \`a la Machine du Moulin Rouge, \`a Fercot et au Rachais, \`a Antoine motocross et au robot de table, \`a F\'elix et au fameux jeu, \`a Manu et \`a l'inondation du 3\`eme \'etage, \`a Gorjup enfin, avec sa clochette et son poil soyeux.

%Maria, Antoine, Pierre

%saas fee
%Florian, Tim, Olja, William...

%Tim, gala of galah, Jinmi, check st evol team

%Saint luc conf

%Antoine G
%Georges, Sylvia, Raphael, Andr\'e, Cyril, Cristina, Corinne, Jose, 

%Tassos? Camilla, 

%Carlo, 

%Alain Coc, Elisabeth, etc AXIONS!!!

%Florian, Mads, Lionel, Patrick, St\'ephane, Thierry, William, Thibaut D., Louis, Remy, Olga, Tim, Yin, Aline, Maroussia, Nico B, Nico ?, Uriel, Baptiste, Giovanni, Song, Manu S.

%Bidou, Nathan, Remi, Manu, Felix, Robin, Fercot

%Dupin, Jouty 

%et les chamoniards, 

%solene
%Parents, G parents, Anna, Uli, Nat

%Japon: Taygun, Matt, Sanjana, Jacqueline

%Istanbul: Melisa, Mehmet 

%Musique: (Francois)

%CAP, cafet, admin (Claire? Myriam?), 

%Genthod, 

%meca (Robin, piano)

%Koh, Gen, Yamamoto, Eoin, Aline, 

%Chili: James, Simona, Chiara, Vincent, Justyn, Lila

%Jerome? JC Augereau? V\'eronique Vuitton?

%Australie: Tim B., 

%USA: Rana, 

%Fanny

%Marion, parents Marion, Camille

%Barbara festi Job

%Dri Verrier

%M2R

%Sophro

%Dublin Laura Ioana

%J'esp\`ere vraiment poursuivre ces collaborations.

%J'ai des pens\'ees amicales \`a mes 

%Merci aux 
%J'ai des pens\'ees toutes particuli\`eres pour mes coll\`egues mais surtout amis Louis, William, Tim, Olja, Remy, 

%Ping pong 

%Course \`a pied

%Prof anglais

%J'ai \'enorm\'ement appris \`a ses cot\'es. 

\tableofcontents
\listoffigures
\addcontentsline{toc}{chapter}{List of Figures}
\listoftables
\addcontentsline{toc}{chapter}{List of Tables}
}

{
\mainmatter
\pagestyle{fancy}

\chapter{Introduction}
\label{Cintro}

Stars are cosmic engines catching interstellar material, processing it and giving it back during their lives or when they die. 
Almost all the content in metal\footnote{Metals are elements heavier than helium.} in the Universe was synthesized in stars.
Especially, the first metals in the Universe were synthesized by the first stars, that were predominantly massive\footnote{Stars are considered as massive when they go through all hydrostatic burning stages, ultimately leading to an iron-core. The lower mass limit is around 10~$M_{\odot}$.} and that are long-dead now.
The comprehension of the entire chemical evolution of the Universe requires to understand its chemical evolution at all the different epochs, from the first stars era to the present day Universe. The onset of the chemical evolution, triggered by the early generation of massive stars, is a crucial and still largely missing piece of the puzzle. %, that set the chemical condition .

%For instance, i
%It is funny to think that our body is likely made with more than 50~\% of massive star ejecta: the human body contains more than 50~\% of oxygen, element which is almost only produced by massive stars. 
%Overall, the study of stars is tightly related to the understanding of our origins. 

%Because of the initial mass function scales with initial as $M$

%URS:
%In conclusion, massive stars are important for the formation and structure of the observed universe as well as for its chemical enrichment. They are therefore also fundamental physical constituents of our solar system and of life on earth.

%\textcolor{red}{First chapter: geography of the field. End of 1st chapter, good idea of the field, what is done in the thesis...}

%The early generation of massive stars is an essential piece of the puzzle.
%Why important?

%---------------------------------------------------------------------------------------------------------------------------------------
\section{From the Big Bang to the first stars}\label{before}

%\textcolor{red}{More reviews in this sec? General intro (small here)} Karlsson13?

\subsection{Big Bang nucleosynthesis}\label{bbn}

During the first minutes following the Big Bang, as the temperature decreased from $\sim 10^{32}$ K to about $10^8$ K, the first nucleosynthetic event in the Universe took place. The Big Bang nucleosynthesis (BBN) finished about 15 minutes after the Big Bang, when the density and temperature became too low for further nucleosynthesis to occur. At this time, for 100 g of Universe, about 24.8 g was in the form of $^{4}$He and almost all the rest in the form of protons \citep{cyburt03}. As shown in the left panel of Fig.~\ref{bbn2}, a bit of $^{2}$H, $^{3}$H, $^{3}$He and other light elements is also found. Metals up to oxygen could also be synthesized but in negligible amounts \citep[generally below $10^{-15}$ in mass fraction, see Fig.~\ref{bbn2}, right panel, also][]{iocco07, coc12, coc14}. 

380000 years after the BBN, the Universe became cool enough (about 3100 K), allowing protons and electrons to recombine and form neutral hydrogen atoms. 
%%%\textcolor{red}{Here maybe develop more. Maybe more info about the phase between recombination and first stars (formation?)}
From this point, the Universe entered the dark ages. It is transparent but no light-producing structures such as stars exist yet. %conditions were set for the first stars or Population III (Pop~III) stars. 

%As a consequence, the global amount of metal in the universe increased as a function of time.

%   \begin{figure*}
 %  \centering
  %    \includegraphics[scale=0.58, trim = 0cm 0cm 0cm 0cm]{bbn.png}
 %  \caption[Big bang nucleosynthesis]{Nucleosynthesis during the first minutes of the Universe \citep[from ][]{burles99, tytler00}.}
%\label{bbn}
  %  \end{figure*}

   \begin{figure*}[t]
   \centering
      \includegraphics[scale=0.58, trim = 0cm 0cm 0cm 0cm]{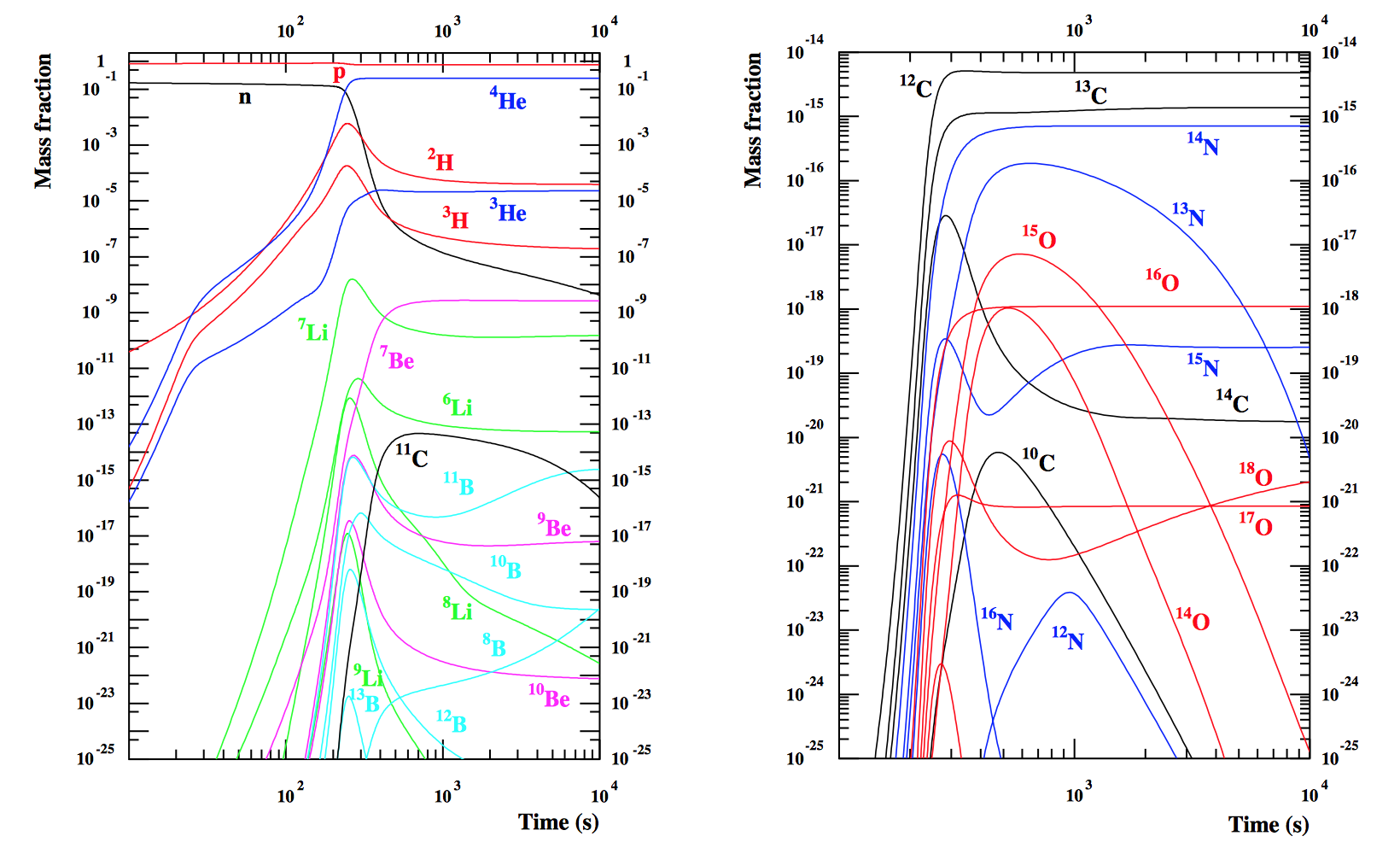}
   \caption[Big Bang nucleosynthesis]{Big Bang nucleosynthesis (mass fraction as a function of time in seconds). Note the different vertical scales between the two panels \citep[figure from][]{coc12}.}
\label{bbn2}
    \end{figure*}

   \begin{figure*}[t]
   \centering
      \includegraphics[scale=0.38, trim = 0cm 2cm 0cm 1cm]{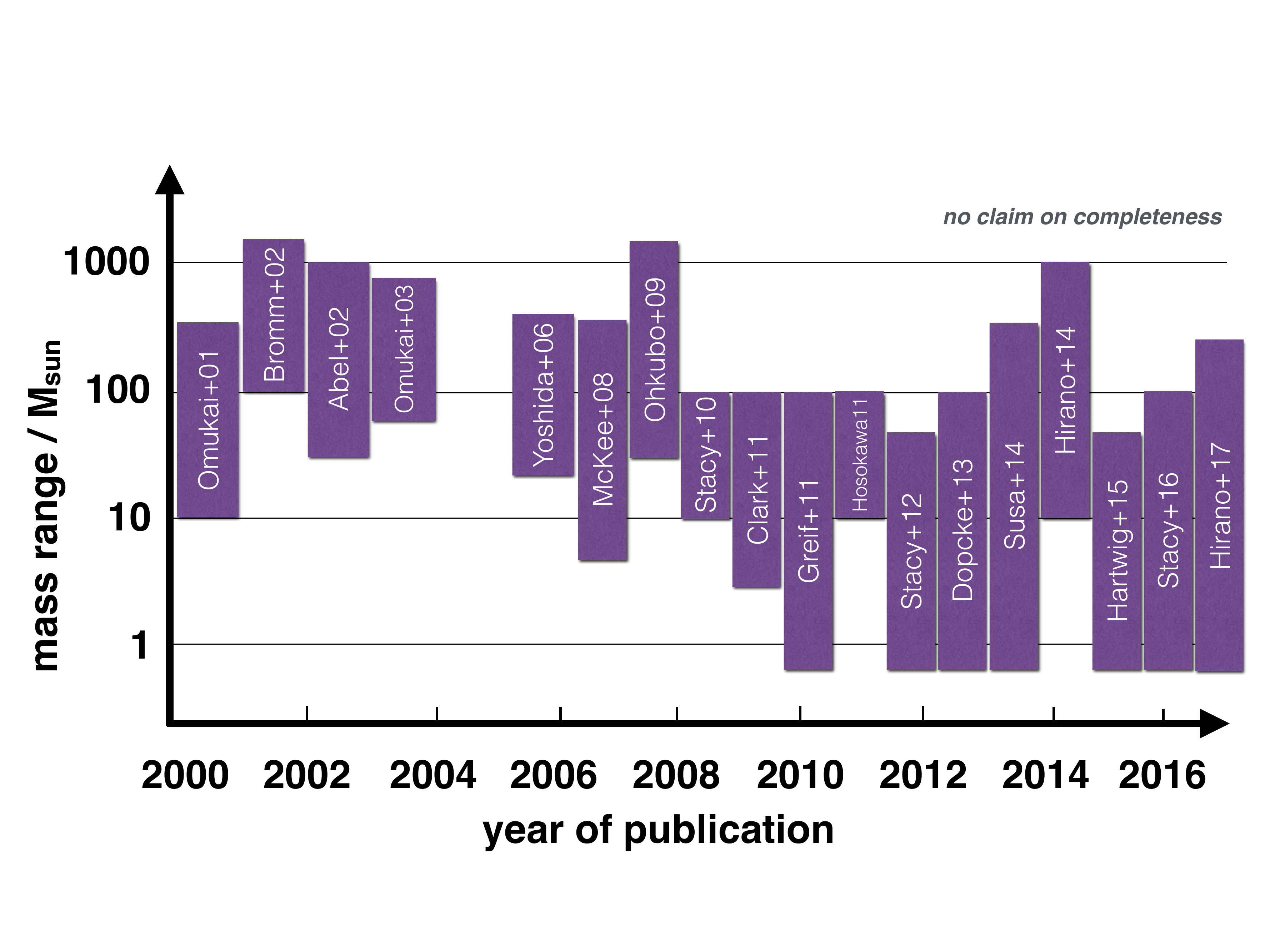}
   \caption[Mass range of the first stars]{Predicted mass range of the first stars as a function of the year of publication. Courtesy of T. Hartwig (adapted from a plot by N. Yoshida). No distinction is made between Pop~III.1 and Pop~III.2 stars (see text for details).}
\label{massr}
    \end{figure*}

%---------------------------------------------------------------------------------------------------------------------------------------
%\subsection{The formation of the first massive stars}
\subsection{The formation of the first stars}

The first stars (Population III or Pop~III stars) formed $\sim 200$ Myr later, with almost exactly 25~\% of Helium and 75~\% of Hydrogen in mass.
%The first stars are predicted to have formed in $\sim 10^6$ \msol~dark matter halos, at redshift $z \simeq 20-30$ \citep[e.g.][]{yoshida03a}.  %tegmark97
%For fragmentation to occur and a star to form, a gas cloud has to cool until its Jeans mass\footnote{the Jeans mass is the minimum mass of a gas cloud required in order the gravitational force to overcome the gas pressure.} 
For a star to form, a gas cloud has to cool until its Jeans mass\footnote{The Jeans mass is the minimum mass of a gas cloud required in order the gravitational force to overcome the gas pressure. When $M_J$ is reached, the cloud becomes unstable and collapses.} 
$M_J \propto T^{3/2} \rho^{-1/2}$ reaches the mass scale of individual stars. %$M_J$ in the local interst%When $M_J$ is reached, the cloud becomes unstable and collapses. 
%The Jeans mass decreases because of the cooling of the gas. 
In metal-rich environments, the cooling of the molecular cloud is mainly controlled by heavy elements such as carbon or oxygen. On the other side, it was found by \cite{saslaw67} that the cooling in a low-temperature and metal-free environment is achieved by molecular hydrogen H$_{2}$. The cooling efficiency of H$_{2}$ being low, the primordial gas is able to cool down to about $200$ K only \citep{abel02, bromm02}, against $\sim 10-20$ K at solar metallicity\footnote{The metallicity is the sum of the mass fraction of elements heavier than helium. The solar metallicity is $Z_{\odot} = 0.0134$ \citep{asplund09}.}. The higher temperature in primordial gas clouds implies a higher $M_J$ and makes the collapse and fragmentation of these clouds more difficult. %in metal-free environments. 
The fragments are consequently larger, suggesting that the first stars were more massive.

Various multidimensional cosmological simulations have shown that the first stars were predominantly massive \citep[generally $\gtrsim 30$~$M_{\odot}$, e.g.][]{abel02, bromm02}.
Some authors have refined this picture by suggesting the existence of two distinct star formation modes for Pop~III, leading to two classes of Pop~III stars: Pop~III.1 and Pop~III.2 \citep[e.g.][]{johnson06}. Pop~III.1 stars formed from metal-free $\sim 10^6$~$M_{\odot}$ dark matter minihalos. Pop~III.2 stars formed from a metal-free gas cloud but which was significantly ionized by photons from already existing Pop~III.1 stars.
%, prior to the onset of gravitational collapse. 
Such an ionization favors the creation of HD molecules in the star forming cloud, which increases the cooling efficiency and leads to lower initial stellar masses. The mass distribution of Pop~III.2 stars, possibly peaking at $\sim 10$~$M_{\odot}$ \citep{hirano14}, is shifted towards lower values compared to Pop~III.1 stars. However, if important turbulent motions were present in primordial gas clouds, the picture is reversed: Pop~III.2 stars are on average more massive than Pop~III.1 stars \citep{clark11}. Indeed, \cite{clark11} have shown that turbulence may be more vigorous in Pop~III.1 clouds than in Pop~III.2 clouds, hence facilitating the fragmentation of Pop~III.1 clouds, and ultimately leading to less massive stars on average. %In this case, the first stars would have form in group rather than in isolation and span a wide range of mass. 
%Recent high-resolution simulations have shown that the formation of $\sim 1$~$M_{\odot}$ Pop~III stars is possible because of an efficient fragmentation of the proto-stellar cloud \citep{stacy16}. 
The Pop~III initial mass range predicted by the most recent simulations still varies a lot from a study to another (cf. Fig.~\ref{massr}). For instance, \cite{stacy16} predicts masses between 0.05 and 20~$M_{\odot}$ while \cite{hosokawa16} between 15 and 600~$M_{\odot}$. This could be due to differences in the adopted resolution, number of minihaloes followed during the simulation, time of the simulation or types of feedback implemented in the simulation \citep[see Table 1 of][for an overview of the characteristics of some simulations of first star formation]{stacy16}.

%\cite{wade14}: among a sample of 90 Galactic O-type stars, $7 \pm 3~\%$ host magnetic fields (fossil fields?)

%\textcolor{red}{transition to Pop~II : Zcrit}

%   \begin{figure*}
%   \centering
 %     \includegraphics[scale=5.00]{eso1620a.jpg}
%   \caption[Evolution of the universe]{Schematic view of the evolution of the universe (credits: eso.org).}
%\label{exp}
 %   \end{figure*}

%\subsection{Chemical evolution of the Universe}

\section{Origin of the elements}

  \begin{figure*}[t]
   \centering
      \includegraphics[scale=0.39, trim = 0cm 0cm 1cm 0cm]{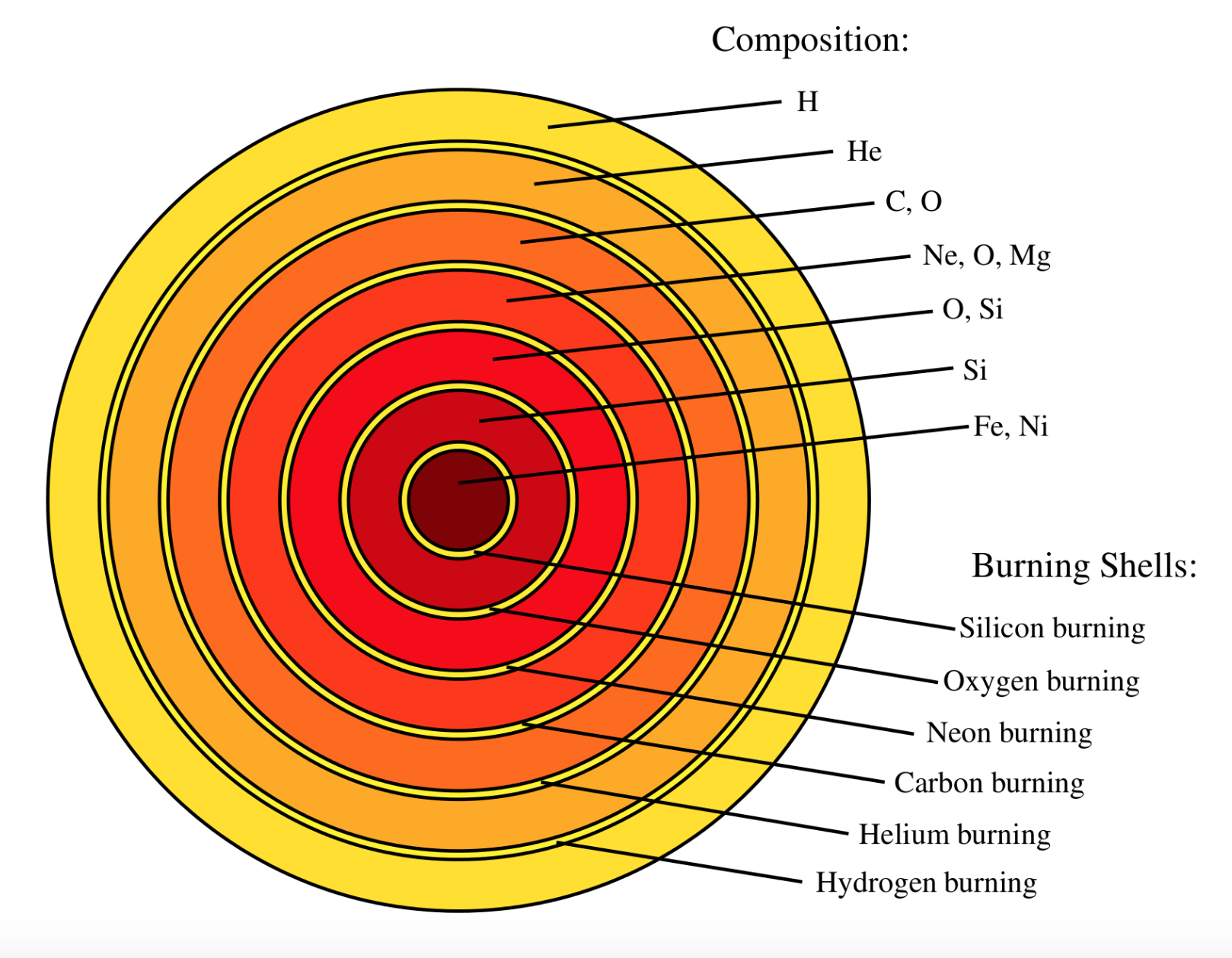}
   \caption[Schematic view of the pre supernovae structure of a massive star]{Schematic view of the stellar structure of a massive star before its death. The most abundant species in the different layers are indicated, as well as the different burning shells (figure from Frischknecht (2012) PhD thesis, originally from C. Winteler).}
\label{schemastar}
    \end{figure*}

%The first metals were produced and released by the first stars. 
%With time, new generations of stars formed with the material ejected by the previous generations and released new metals. 

%merrill52
%\paragraph{Fusion}
If gravity were the only force in stars, they would collapse in less than one hour (the free-fall timescale is about 30 minutes for the Sun). Stars have to produce energy to counteracts their own gravity.
%If the nuclear reactions were stopped now in the core of the Sun, we would not see any difference for a long time. 
The Kelvin-Helmholtz timescale $\tau_{\rm KH}$ gives an estimate of how long a star would shine with its current luminosity if the only power source were the conversion of gravitational potential to heat. For the Sun, $\tau_{\rm KH} = GM_{\odot}^2 / (R_{\odot} L_{\odot}) \simeq 30$ Myr. A lifetime of 30 Myr for the Sun is however too short in regards to geological and biological evidences that the Earth is billions of years old. Nuclear fusion provides another source of energy that sustains stars for a much longer time. Stars gain energy by fusion only up to iron because it is the most bound nucleus. Beyond iron, no fusion can occur.
Massive stars with initial masses $M_{\rm ini}>8~M_{\odot}$ undergo successive shorter and shorter core burning stages, lasting from millions of years to fractions of days, and ultimately leading to the onion-like structure shown in Fig.~\ref{schemastar}.
At the end of each core burning phase, the burning continues in a shell. As evolution proceeds, more and more burning regions are present in the massive star (Fig.~\ref{schemastar}).
Nucleosynthesis in massive stars is discussed in more details in Sect.~\ref{nucprocmass}.

%\paragraph{Neutron captures}
As mentioned, stellar fusion can synthesize elements up to iron. The main processes building elements beyond iron are neutron capture processes \citep{burbidge57}. Neutrons are captured by nuclei that can experience a $\beta-$decay reaction (a neutron is transformed into a proton) if they are unstable. Neutron capture processes are generally separated into two categories: slow and rapid (s- and r-processes), that are both expected to contribute to synthesize about 50~\% of the nuclei beyond iron. 
For the r-process, neutron captures operate on much smaller timescales than for the s-process, giving the possibility to reach very neutron rich (unstable) isotopes that finally decay to the valley of stability (i.e. to stable isotopes, see Fig.~\ref{sproc1}).
For the s-process, 
%\citep[e.g.][for a review]{kappeler11}, 
the neutron capture timescales $\tau_n$ are larger than the $\beta-$decay timescales $\tau_{\beta}$ of unstable isotopes. 
%($\lesssim 1$ day). 
It means that unstable isotopes $\beta-$decay before another neutron is captured (with the exception of the branching points where $\tau_{n} \sim \tau_{\beta}$). 
Starting from Fe, the s-process runs mainly along the s-process path which stays close to the valley of stability (black path in Fig.~\ref{sproc1}). The small neutron cross sections of the magic nuclei around A=88, 140, and 208 are bottlenecks for the flow and create distinct peaks in the typical s-process abundance pattern. These peaks can be seen in the abundance pattern of the Sun (top left panel of Fig.~\ref{sproc1}). The s-process ends at $^{208}$Pb, the last stable nucleus.

Some isotopes are called s-only (marked as \textit{s} in Fig.~\ref{sproc1}) because they cannot be synthesized with the r-process. An example is $^{82}$Kr. The very neutron-rich isotopes that $\beta-$decays to the valley of stability (following a diagonal in the (N, Z) plane) after r-process nucleosynthesis cannot reach $^{82}$Kr. This isotope is shielded by the stable isotope $^{82}$Se. 
\cite{clayton61} have shown that abundances of s-only isotopes in the Sun cannot be reproduced by a single neutron irradiation of an iron seed. 
In fact, the abundances of elements between $90 <$ A $< 204$ can be reproduced by a \textit{main} s-process component while for the elements below $A=90$, a \textit{weak} s-process component is required \citep{kappeler89}. 
%%%%There is also the so-called \textit{strong} component, that aims at reproducing the high Pb abundance in the solar system.
The main 
%%%%and strong 
component is associated to Thermally Pulsing (TP) Asymptotic Giant Branch (AGB) stars with $1 \lesssim M_{\rm ini} \lesssim 8$~$M_{\odot}$. The weak component is associated to massive stars with $M_{\rm ini} \gtrsim$ 8~$M_{\odot}$.
These two s-process components are discussed in Sect.~\ref{nucprocmass} and~\ref{sprocagb}.

   \begin{figure*}[t]
   \centering
      \includegraphics[scale=0.27]{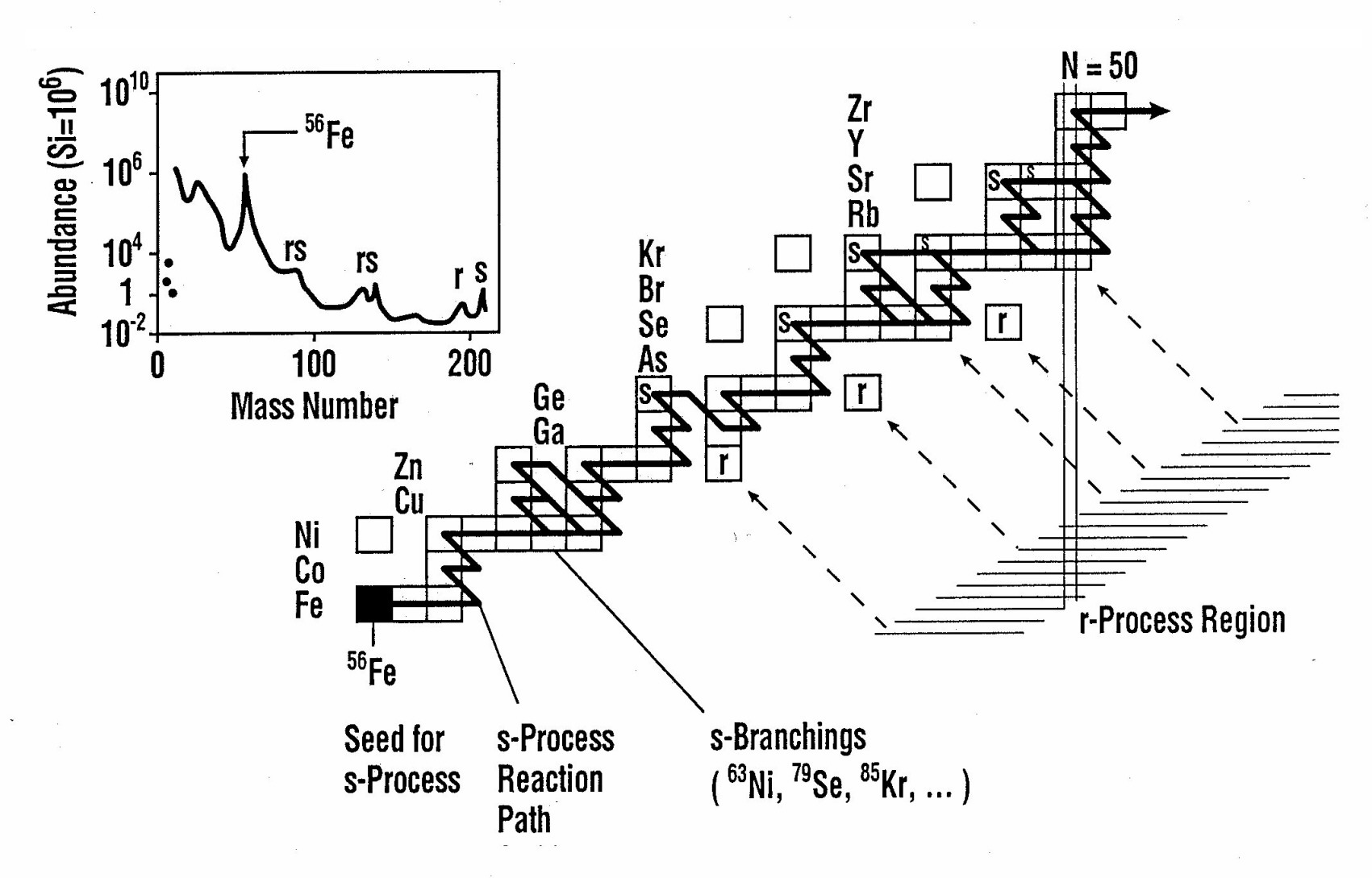}
   \caption[Neutron capture processes]{Illustration of the neutron capture processes responsible for the formation of the nuclei beyond iron. Only the stable nuclei between Fe and Zr are shown. The top left panel shows the solar system abundances with the abundance of silicon normalized to $10^6$ \citep[figure from][]{kappeler11}.}
\label{sproc1}
    \end{figure*}

\section{Stellar rotation}\label{sterot}

   \begin{figure*}[t]
   \centering
      \includegraphics[scale=0.55, trim = 0cm 0cm 0cm 0cm]{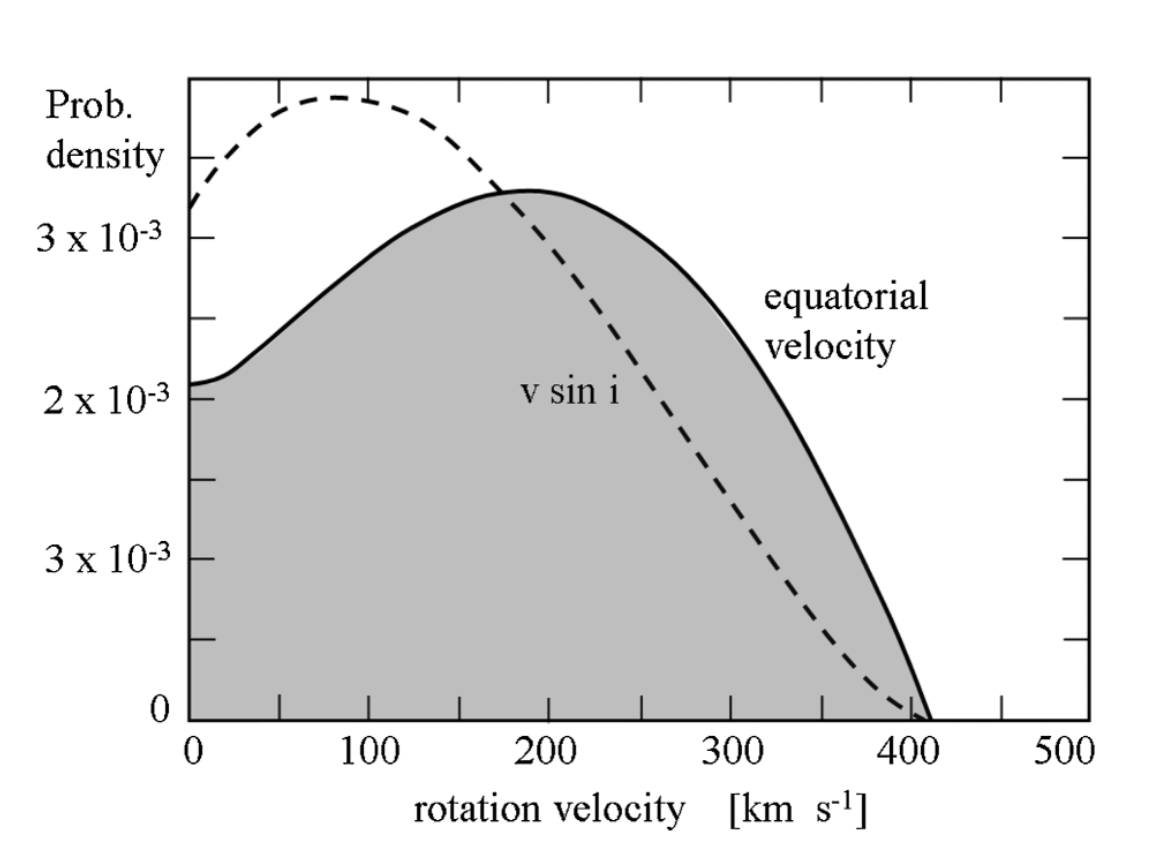}
   \caption[Probability density of the rotation velocities for 496 observed OB stars]{Probability density of the rotational velocities of 496 observed OB stars (masses between about 3 and 20~$M_{\odot}$). Figure from \cite{maeder12}, originally from \cite{huang06}.}
\label{distribv}
    \end{figure*}

%During the formation of a star, the initial molecular cloud 
%The initial molecular cloud 

%From the initial molecular cloud to the star

During the star formation process, the initial molecular cloud shrinks by $\gtrsim 5$ orders of magnitudes. The local conservation of angular momentum ($\Omega r^2 =$ constant) tells us that even if the initial gas cloud rotates very slowly, the newly formed star will rotate fast. 
%This simple consideration shows why stars must rotate.
A simple conservation of the angular momentum during star formation will however produce only very fast rotator. Since all stars are not very fast rotators, some processes able to dissipate angular momentum during the star formation process have to be invoked.

%Observationally, we know since several centuries that celestial bodies rotate. 
There are many observational evidences that stars rotate.
\cite{huang06} have reported the equatorial velocity distribution of 496 OB stars, whose velocity distribution peaks at 200 km~s$^{-1}$ (Fig.~\ref{distribv}). 
% . On average, these stars rotate fast, with the distribution peaking at 200 km~s$^{-1}$ (Fig.~\ref{distribv}). 
Because of the centrifugal force exerted by rotation, fast rotating stars have an ellipsoid-like shape with a bigger equatorial than polar radius. Thanks to interferometric observations, \cite{domiciano03} have reconstructed the shape of the fast rotating massive star Achernar (about 10~$M_{\odot}$). They reported an equatorial radius 1.56 $\pm 0.05$ larger than the polar radius.

Rotation is expected to affect the stellar luminosity, stellar lifetimes, mass loss or the chemical composition of the stellar surface for instance \citep[e.g.][]{meynet00b, heger00b}. Especially, massive stellar models predict surface nitrogen enhancements because of the combined effect of the CNO cycle operating in the stellar interior and rotational mixing. \cite{martins15} reported the CNO surface abundances of 74 observed O-type stars with masses between about 20 and 50~$M_{\odot}$. 80~\% of their sample can be explained by single stellar models including rotation.

\subsubsection{Rotation at low and zero metallicity}

%\textcolor{red}{Mettre dans chap 3?}

Low metallicity stars have less metals so that they are less opaques and more compact. As a consequence, for the same angular momentum content, low-metallicity stars rotate faster than solar metallicity stars.
%At low metallicity, stars are expected to rotate more rapidly \textcolor{red}{(cf. Sect.~XXX)}: 
%A first qualitative reasonable argument is that 
%if the initial angular momentum content is the same for solar metallicity and low metallicity stars, then low metallicity stars will rotate faster because they are more compact (they have less metals so that they are less opaques and more compact). 
Low metallicity stars generally lose little mass through winds meaning that they keep more angular momentum than their solar counterpart. This also favors a faster rotation.
Faster rotation at low metallicity is supported by different observations: \cite{hunter08b} have shown that stars in the Small Magellanic Cloud rotate on average faster than Galactic stars (the latter having higher metallicities in general). \cite{maeder99} and \cite{martayan07} have shown that the fraction of Be-type stars, whose existence is possibly linked to fast rotation, increases with decreasing metallicity. 
%%%%%The efficiency rotational mixing operating inside the star increases with initial rotation but is also increases with decreasing metallicity 
%The efficiency of the rotational mixing operating inside the star is then expected to increase with decreasing metallicity 
%%%%%because of a less efficient redistribution of the angular momentum inside low metallicity stars 
%(cf. Sect.~\ref{transportrot}). 
%Yields of low metallicity massive stellar models are strongly impacted by rotation, especially nitrogen \citep{meynet02b} or s-elements \citep{frischknecht12, frischknecht16}. It is important to note that the same physics of rotation is used at solar and low metallicity. The strong production of some elements like nitrogen is not artificial but appears naturally in low metallicity rotating massive stars. 

For Pop~III stars, \cite{stacy11} estimated their rotation speed thanks to smoothed particle hydrodynamic simulations. 
%of within a minihalo at $z \sim 20$. 
They have recorded the angular momentum of the sink particles falling into the growing protostar. The large amount of angular momentum suggests initial velocities of $1000$ km s$^{-1}$ or higher for stars with $M_{\rm ini} \geq 30$~$M_{\odot}$. %, indicating that Pop~III stars may have experienced strong rotational mixing. 
%, impacting their structure, nucleosynthesis but also final fate, likely. 
A caveat in this study is that they estimated the total angular momentum accreted within a radius of 50 UA from the center of the star, meaning that they do not resolve stellar scales. This prevents to see whether some angular momentum removal processes (e.g. stellar winds, disk-locking or magnetic torques) occur in the inner region.

Recently, by studying the angular momentum transfer in primordial discs including magnetic fields, \cite{hirano18} suggested that the final rotational state of Pop~III protostars should exhibit a net bimodality: either the protostar do not rotate at all, or it is a fast rotator, close to breakup speed. As mentioned in their paper, this study does not properly include MHD effects, which are likely required to self-consistently assess this possible bimodality.

Rotation and its effect in low metallicity massive stars is discussed in more details in the next chapters (especially in Sect.~\ref{cempnoscenar} and \ref{massivemodels}).

%\cite{clark08} found that vigorous fragmentation occurs at

%The UV photons emitted by the first stars broke the fragile H$_{2}$ molecule, suppressing the  molecule is fragile 

%\textcolor{red}{Maybe not put this Sect? Say a few words about formation of the stars in previous sect.}

%It was found by \cite{saslaw67} that cooling in the low-temperature primordial gas had to rely on molecular hydrogen H$_{2}$.

%---------------------------------------------------------------------------------------------------------------------------------------
\section{Observable signatures of the early generations of massive stars}\label{obssign}

%\textcolor{red}{REVIEW OF \cite{karlsson13} !!} 

%The first stars were different than the actual stars (\textcolor{red}{Mayber not here. Say why maybe before? Plot of Tilman?}). 
%\textcolor{red}{Citation book \cite{loeb10}}

We have many images of the Universe when it was older than a billion year.
With the Cosmic Microwave Background, we also have a picture of the Universe when it was only 380000 years old. One big challenge today is to obtain pictures in between these two periods, when the primordial Universe started to evolve into the incredibly rich zoo of objects we see today. To cite \cite{loeb10}: \textquotedblleft~the situation is similar to having a photo album of a person that begins with the first ultrasound image of him or her as an unborn baby and then skip to some additional photos of his or her years as teenager and adult~\textquotedblright. %\textcolor{red}{cite UMP stars?}
%To try to complete the album, one should look for the signature of the first stars and galaxies. 
%Even if images of the infant universe are still missing,
%There are several ways to obtain informations on the young universe. 

The first generations of massive stars in the Universe have specific signatures that one can look for to reveal their nature. 
%There are different approaches to infer the nature of the first stars:
%Although the present work focuses on one of these signatures (the metal-poor stars) 
%we list below all the existing observables for completeness.
The present work focuses on one of these signatures (the metal-poor stars) but the list below includes also other observables, for completeness.

%The first quasars likely formed later, at $z > 10$ \citep{haiman01, bromm03a}

%Stars live for $\sim 10$ $(M/M_{\odot})^{-3}$ Gyr

%:   fig frayons_fr
%\begin{figure}[h]
%\centering
%\resizebox{\textwidth}{!}{\includegraphics{}
%\captionsetup{listof=false}
%\caption{\textsl{Gauche:} variation du rayon en fonction de la masse initiale sur la ZAMS dans le cas sans rotation pour les quatre mÈtallicitÈs considÈrÈes. \textsl{Centre:} variation de la forme de l'Ètoile en fonction du taux de rotation (lÈgende sur la figure). \textsl{Droite:} variation de la tempÈrature effective en fonction de la colatitude avec le taux de rotation.}
%\label{frayons_fr}
%\end{figure}

\subsection*{Integrated light of high redshift galaxies}

Thousands of very distant galaxies with redshift $z > 5$ have been observed \citep[e.g.][]{bouwens07,bouwens09,oesch10,mclure10}. The most distant object observed today is a galaxy, GN-z11, with $z=11.1_{+0.08}^{-0.12}$, possibly formed $\sim 400$ Myr after the Big Bang \citep[][]{oesch16}. 
In the future, the James Webb Space Telescope (JWST) might be able to detect up to a thousand star-bursting galaxies with $z > 10$ \citep{pawlik11}. Their colors and spectra could be compared with predictions of the integrated light coming from such objects in order to interpret these observations in terms of stellar populations \citep[e.g.][]{schaerer02, Salvaterra11}. 
%. The light coming from such objects might be dominated by the light of the first stars. 
%Colors and spectral predictions of the integrated light coming from these objects are needed in order to interpret these observations in terms of stellar populations \citep[e.g.][]{schaerer02, Salvaterra11}. 

Abundance determination of high-z objects is challenging. Past and current observations managed to evaluate the abundances of a few elements (generally C, N, O, Si) of $z \lesssim 4$ objects. These objects are mostly damped Lyman $\alpha$ (DLA) systems with $2<z<3$ \citep[e.g.][]{lehner16}.
In general, the metallicity of such objects is estimated from considerations on the nebular emission lines formed in ionized gas at the sites of star formation \cite[e.g.][]{pettini04, pettini08iau}. To date, the most metal-poor DLA system is at $z=3.076$ and has [C/H] $=-3.43 \pm 0.06$, [O/H] $=- 3.05\pm 0.05$, [Si/H] $=-3.21\pm0.05$ and [Fe/H] $\leq -2.81$ \citep{cooke17}. 

One difficulty regarding high-$z$ objects is that most of the metal lines are redshifted to the infrared range (by a factor of $1+z$), a regime where the sky background is high. In space, although the current telescopes cannot reach the far infrared region (the Hubble Space Telescope, HST, goes up to $\lambda = 2.5$ $\mu$m), JWST should reach $\lambda = 28$~$\mu$m, which may allow to detect the metal lines of very high-$z$ objects.  

%The observed sample of galaxies with $z>10$ is however very small, since extremely challenging to detect. %Abundances of high-z galaxies are particularly hard to evaluate and interpret. %A few abundances (generally C,N,O, Si) of damped Lyman $\alpha$ (DLA) systems with $2.3<z<3.3$ were measured \citep{lehner16}. 

%The spectral properties of metal-free stellar populations have been studied \cite{schaerer02}

\subsection*{Reionization}

%From the recombination (cf. Sect.~\ref{bbn}) to the  
%By producing light, the first stars re-ionized the neutral hydrogen. It is called reionization and not ionization since the hydrogen was already ionized before the recombination (Sect.~\ref{bbn}). 
As the first stars formed and radiated energy, the Universe reverted from being neutral, to being ionized once again.
At the epoch of reionization and before significant expansion had occurred, the free electron density in the Universe was high enough for Cosmic Microwave Background (CMB) photons to undergo significant Thomson scattering. This let a detectable imprint on the CMB anisotropy map. 
Inspection of the tiny fluctuations in the CMB polarization by the Wilkinson Microwave Anisotropy Probe (WMAP) first suggested that reionization occured between $11 < z < 30$ \citep{kogut03}. Nine years of observations by WMAP have revised this result and suggested a reionization at $z \sim 10.4$ \citep{hinshaw13}. More recently, \cite{planck16a} found that the average redshift at which reionization occured lies between $z = 7.8$ and $8.8$. 
%Also, they derived that for $z > 9.4 \pm 1.2$, the universe was less than 10\% ionized.

Photons from reionization also altered the excitation state of the 21-cm hyperfine line of neutral hydrogen \citep[e.g. the reviews of][]{morales10, pritchard12}. 
Models predict that the 21 cm cosmic hydrogen signal will show an absorption feature at $z \simeq 20$, coming from the Lyman-$\alpha$ radiation of the earliest stars. %(REF SEE BARKANA 18 abstract).
The 21-cm transition is forbidden but since hydrogen amounts to $\sim 75$~\% of the gas mass present in the intergalactic medium, the line intensity is enough to be detected. The line frequency is $\nu = \nu_0 / (1 + z)$ MHz with $\nu_0$ the rest-frame frequency of 1420 MHz. For redshifts $6<z<50$ the corresponding frequencies are $30<\nu<200$ MHz. It corresponds to the radio frequency domain, making this line is a prime target for present and future radio interferometers like the Murchison Widefield Array, the Low Frequency Array or the Experiment to Detect the Global Epoch of Reionization Signature (EDGES). 
Recently, \cite{bowman18} reported the detection (with EDGES) of such an absorption feature peaking at 78 MHz consistent with expectations for the 21-cm signal induced by stars having existed by 180 million years ($z \simeq 20$) after the Big Bang. Some discrepancies between this observation and models further suggest that an unknown interaction between dark matter and baryons occurred at early times \citep{barkana18}. It may provide new clues on the nature of dark matter.

%\textcolor{red}{\cite{barkana06b}: reionization in High z galaxies}

%\textcolor{red}{Also : say a few words on 21-cm line, other way to study reionization bowman18 !! Reionization Signature. Hydrogen Epoch of Reionization Array (HERA). Morales10 (rev), Pritchard12 (rev), Yajima18}

%\textcolor{red}{Also : reionization imprints a thermal record on the IGM detectable in the $z \sim 5-6$ Ly alpha forest. Onorbe+17}

\subsection*{Supernovae and Gamma-Ray Bursts in distant galaxies}

Until now, the Swift Gamma-Ray Burst\footnote{Gamma-Ray Bursts are sudden and powerful gamma-ray flashes observed in the Universe, lasting from 0.001 seconds to about 15 minutes.} mission \citep{gehrels04} has detected 8 Gamma-Ray Bursts (GRB) at $z>6$ \citep[reported in Table 1 of ][]{salvaterra15}. The [S/H] ratio of the host system of GRB 050904 \citep[$z = 6.3$, ][]{kawai06} has been inferred to be [S/H] $= -1.6 \pm 0.3$ \citep{thone13}, suggesting a sub-solar metallicity for this object. It remains nevertheless still far from the zero or extremely low metallicity Universe. The fact that the metallicity is likely already rather high at $z=6$ somewhat supports the idea of a rapid enrichment in metals.
The next generation of space and ground telescopes, such as JWST, the European Extremely-Large Telescope (E-ELT) or the Giant Magellan Telescope (GMT) may not be able to directly see individual first stars \citep{pawlik11}. Even if such stars are gravitationally lensed, their detection will remain extremely challenging \citep{Rydberg13}. The explosion of the first massive stars, on the other hand, should be observable with the next generation of telescopes \citep{tanaka13, whalen13c,whalen13a,whalen13b}.

\subsection*{Gravitational waves}

The first stars were likely top-heavy and retained a significant part of their mass because of weaker stellar winds. As a consequence, they could have produced massive black holes (BH) that released gravitational waves (GW) if two of them merged.
Recent results suggested that the GW background coming from Pop~III binary black holes mergers might be detectable by Advanced LIGO/VIRGO detectors \citep{kinugawa14, inayoshi16}.

GW150914 was the first direct detection of gravitational waves. The signal came from the coalescence of a binary BH merger whose components had masses of $36^{+5}_{-4}$ and $29^{+4}_{-4}$~$M_{\odot}$ \citep{abbott16}. Cosmological simulations of \cite{hartwig16} show that there is a probability of $\gtrsim 1$~\% that GW150914 is of primordial origin.
These simulations also suggest that Advanced LIGO could detect $\sim 1$ primordial BH-BH merger per year. %for the final design sensitivity
The GW background from the explosion of the first massive stars might also be detectable but with the next-generation space GW detectors, like the Decihertz Interferometer Gravitational wave Observatory (DECIGO) and the Big Bang Observer \citep[BBO, ][]{sandick06, suwa07, marassi09}. 

\subsection*{Pockets of primordial material}

Pockets of primordial gas might be observable in the very distant Universe, but possibly also in the relatively local Universe if such pockets escaped metal pollution. \cite{fumagalli11} have observed two gas clouds at $z \sim 3$ with no discernible elements heavier than hydrogen. The derived upper limit for the metallicity is $Z<10^{-4}$ $Z_{\odot}$. \cite{simcoe12} have reported the discovery of another similar cloud at $z=7$. Such gas clouds might be used as laboratories to probe the formation of metal-free stars. However, in such clouds, the stellar formation process may be different from the formation process of true Pop~III stars, formed at the beginning of the Universe: in the early Universe, it is the mass of the cloud that triggers the collapse and the star formation process. In the more mature Universe, a cloud is believed to collapse because of one of several events (e.g. nearby supernovae, collision of molecular clouds...) that compress the cloud and initiate its gravitational collapse. The way the star formation process is triggered may change the initial stellar characteristics (e.g. initial mass).

%REPRENDRE ICI

%   \begin{figure*}[t]
%   \centering
%      \includegraphics[scale=0.41, trim = 0cm 0cm 0cm 0cm]{chemevol.png}
%   \caption[Schematic view of the chemical evolution of the Universe]{Schematic view of the chemical evolution of the Universe. Massive Pop~III stars formed with the products of big bang nucleosynthesis and released the first metals. Pop~II stars formed with the ejecta of the first generation of massive stars. Later, when the global metallicity in the Universe is a bit higher, other astrophysical sources like AGB stars contribute to the chemical enrichment \citep[figure from][]{jacobson14}.}
%\label{chemevol}
%    \end{figure*}

   \begin{figure*}[t]
   \centering
      \includegraphics[scale=0.26, trim = 5cm 10cm 5cm 0cm]{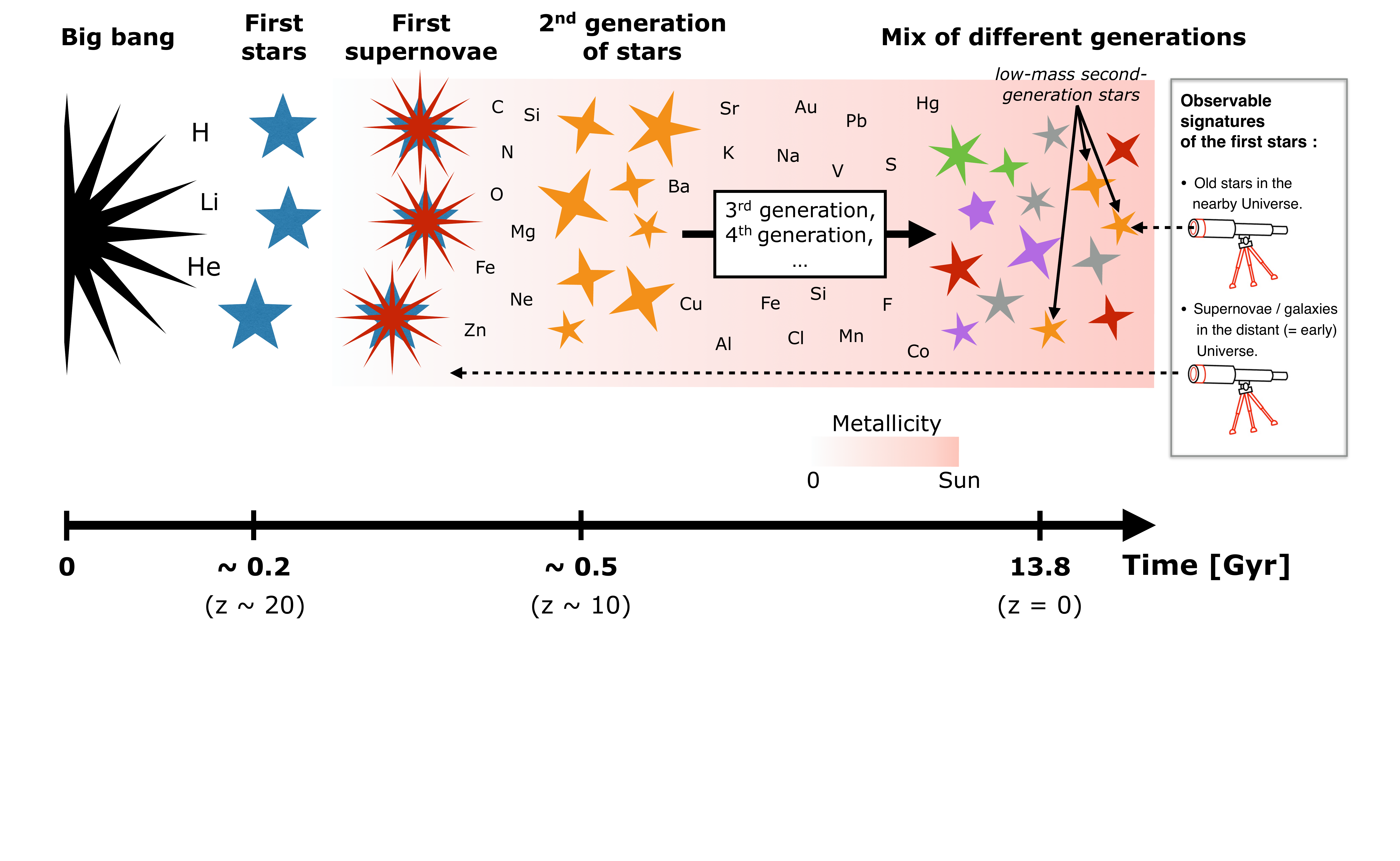}
   \caption[Schematic view of the chemical evolution of the Universe]{Schematic view of the chemical evolution of the Universe. Massive first stars (Pop~III) formed with the products of Big Bang nucleosynthesis and released the first metals. Then, successive stellar generations formed with the ejecta of the previous generations and released new metals. The nature of the early massive stars can be inferred by observing old low-mass stars in the nearby Universe or by observing the extremely high redshift Universe.}
\label{chemevol}
    \end{figure*}

%Then, successive stellar generations formed with the ejecta of the previous generations. Around us, nowadays, a mix of different stellar generations is observed. The nature of the first massive stars can be inferred by observing the rare 2nd generations of stars that are still alive today in our Galaxy. Another possibility is to observe the extremely distant Universe and try to catch the supernovae from the first generations of massive stars.

%\subsection*{Nucleosynthetic imprints in the most Metal-Poor stars of the Galaxy}
\subsection*{Nucleosynthetic imprints in low-mass metal-poor stars}

%\textcolor{red}{why low mass? Maybe no}

%The iron content observed at the surface of a star is often used as a proxy for its age. 
%Very metal deficient stars likely formed early in the Universe, with the material ejected by one or a few previous stellar generations. On the opposite, the sun for instance, which borned about 9 Gyr after the Big Bang, formed with a more metal rich material, processed by different previous generations of stars.
%Because of their lack in metals, metal-poor stars likely formed with the material ejected 

%During the past 3 decades, large surveys have revealed the existence of a population of low-mass metal-poor stars in the halo of our Milky Way. The small content in metal suggests that such stars formed early in the universe. The most iron-deficient stars might belong to the second generation of stars in the universe, formed with the ejecta of the first massive stars. Observation of metal-poor stars opened an unique window to the first massive stars, first supernovae and onset of the Galactic chemical evolution. The abundance of such stars give valuable clues on the first nucleosynthetic events (after the s) that took place in the Universe.

%This, after I think: The chemical composition of a star can be used as a proxy for its age.

As discussed in Sect.~\ref{bbn}, Big Bang nucleosynthesis formed almost only H and He. The first metals were created and released by the first stars. As time proceeded, successive generations of stars formed, evolved and exploded as supernovae, releasing more and more metals in the Universe (see Fig.~\ref{chemevol} for a schematic view). 
A consequence is that the global metallicity in the Universe increases with time. It also implies that the metallicity of a star can be used as a proxy for its age.
From this consideration follows the assumption that the most metal-poor stars are the oldest stars and then the best candidates to probe the beginning of the stellar era\footnote{This assumption is however challenged by recent works suggesting that more metal-rich stars (in the Galactic bulge) are better candidates (see the next point).}.
While massive metal-poor stars have short lifetimes and are then long dead, low-mass metal-poor stars that formed very early may still be alive around us, and hence observable. 
%The first stars formed with only Hydrogen and Helium. As time proceeds, metals were released by successive generations of stars so that the metal content of a star can be used as a proxy for its age. 
These low-mass and very metal-poor stars likely formed with the material ejected by the first short-lived massive stars. On the opposite, the Sun, which was born about 9~Gyr after the Big Bang, formed with a more metal rich material, processed by different generations of stars.

The quest for the most metal-deficient 
%low-mass\footnote{A 0.8~$M_{\odot}$ star has a lifetime comparable with the age of the universe.} 
stars in our neighborhood, called stellar archaeology, has started with the HK-survey \citep{beers85, beers92} and the Hambourg/ESO survey \citep{wisotzki96}. These surveys discovered a population of metal-poor stars in the Galactic halo. 
A significant fraction of these stars, especially the most iron-poor, are enriched in carbon \citep[they are called Carbon-Enhanced Metal-Poor or CEMP stars,][see also Chapter~\ref{Cintro}]{beers05}. 
%(with [Fe/H] $<-3$), 
%The small content in metal suggests that such stars formed early in the universe. The most iron-deficient stars might belong to the second generation of stars in the universe, formed with the ejecta of the first massive stars. Observation of metal-poor stars opened an unique window to the first massive stars, first supernovae and onset of the Galactic chemical evolution.
%that likely formed in the early universe because of their little metal content. 
%SMSS J031300.36-670839.3, is the observed star with the lowest Fe/H ratio \citep[this ratio is at least $10^7$ smaller than in the Sun]{keller14}.
Nowadays, several thousands of metal-poor stars are known, some of them possibly having an age close to the age of the Universe 
%\citep[e.g.][that have given an estimation of the age of such stars by means of the radioactive abundances]{sneden96,sneden08}.
\citep[e.g.][]{sneden96,sneden08}.
%Nowadays, several hundreds of stars with [Fe/H] $<-3$ are known
% \citep[SAGA database\footnote{http://sagadatabase.jp}, ][]{suda08}. 
%In total, eight stars with [Fe/H] $<4.5$ were observed \citep[listed in the Table 1 of ][]{frebel15}.
This opened a new window on the first stars: the chemical composition of the most metal-poor stars delivers valuable clues on the first nucleosynthetic events that took place in the Universe, hence on the nature of early massive stars.
Chapter~\ref{Cintro} discusses metal-poor stars in more detail.
%\textcolor{red}{Sentence on dwarf galax. Or Sect. Plot of Tilman? Proba of finding a PopIII in different systems?}

%\textcolor{red}{dire que ici avantage c'est qu'on a PLEIN de contraintes. Chaque element chimique est une contrainte. Bien sur ca marche pour les etoiles qui ont un peu de metaux, pas ultra mp. Bien pour contraindre early times but maybe not very very early}

\subsection*{Nucleosynthetic imprints in the metal-poor stars of the Galactic bulge}

%The stars belonging to the bulge of the Milky Way show a low-metallicity end at ?
The star formation rate is likely more intense in the bulge than in the other parts of the Milky Way. It implies that the metal content increased more rapidly so that there is a different metallicity-age relation in the bulge compared to the Galactic halo. A given metallicity likely corresponds to an earlier age in the bulge than in the halo.
Using a chemodynamical model that mimics the formation of the Milky Way, \cite{tumlinson10} suggested that of all the stars with very little metals (less than 1/1000 of the solar metallicity), those found in the central regions of the Galaxy were more likely to have formed earlier. It is however very challenging to find metal-poor stars in the bulge since this is the place in the Galaxy where the average metallicity is the highest (the most metal-rich stars known are in the bulge). Another difficulty is that the bulge is a crowded region with high extinction due to dust, making observations more difficult. Finally, the large distance to the bulge (around 8.5 kpc) precludes the observation of too faint stars, unless they are gravitationally microlensed \citep{bensby13}.
%\cite{casey15} reported the discovery of 3 bulge stars with [Fe/H] $\lesssim -2.7$ \textcolor{red}{define bracket or not put} whose abundance patterns is rather similar to metal-poor halo stars. 
%Theoretical models of \cite{tumlinson10} suggest that there is a 70\% chance that at least one of these 3 stars formed at $10\lesssim z \lesssim 15$.

The globular cluster (GC) NGC 6522 \citep[first observed by][]{baade46}, is an interesting metal-poor object located in the bulge. The low metal content derived from the spectra of 8 stars in this GC \citep[about 1 tenth of the solar metallicity,][]{barbuy09} suggests that this GC is old. Images from HST were obtained for this cluster and compared to isochrones \citep{kerber18}. The derived age is $\sim 12.5 \pm 0.5$ Gyr. High abundances of s-element (Y and Sr) were also detected, possibly pointing towards a signature of massive rotating stars, that enriched the original gas from which the GC formed \citep{chiappini11}.\\

%Nowadays, among these observable signatures, the best window on the first stars is probably the observation of metal-poor stars. \textcolor{red}{maybe remove. Or a bit more. Transition}

%\textcolor{red}{Best window : nucleo...}

\section{Overview of this thesis}

The aim of the present work is to get clues on the nature of the early generations of massive stars. 
Informations on these short-lived hence long-dead massive stars are obtained by combining predictions from stellar evolution models including rotation with the observation of metal-poor low-mass stars observed in the halo of the Milky Way. 

% is used to obtain clues on the previous generation of massive stars.

Chapter~\ref{Cintro} reviews the observational steps to get abundances from metal-poor stars as well as the scenarios for the origin of the peculiar Carbon-Enhanced Metal-Poor (CEMP) stars. %Internal mixing processes in CEMP stars are discussed 
The mixing \& fallback and spinstar scenarios are discussed in detail.

Chapter~\ref{stevol} discusses the theoretical and modeling aspects of massive stellar evolution that are relevant for this work. The specificities of low metallicity models are stressed. 
The physical ingredients used and/or implemented in the stellar evolution code are presented.
A study on the effect of axions on Pop~III stars is also included.

Chapter~\ref{cempno} and \ref{cemps} are the main results of this thesis. The origin of the CEMP (mainly CEMP-no and CEMP-s) stars is investigated using predictions from new massive stellar models I have computed.
Chapter~\ref{cempno} investigates the origin of the most iron-poor CEMP stars, generally not significantly enriched in s- and/or r-elements. 
A simplified nucleosynthesis model is first used and compared to observations. Then, massive source star models are computed, their physics (especially nucleosynthesis) is discussed before comparing the chemical composition of their ejecta to observations.
Chapter~\ref{cemps} extends the investigation by following the s-process in massive source star models. New confrontations between source star models including extended nucleosynthesis and observations of CEMP stars enriched in s-elements are presented.

Chapter~\ref{Cconclu} summarizes the main results of this work, goes to the conclusions and considers the perspectives.\\

\chapter{CEMP stars: observations and origins}
\label{Cintro}

This chapter reviews observational aspects and existing scenarios for the origin of CEMP stars. Metal-poor stars are first discussed before going into the special case of CEMP stars. What are the observational and modeling steps to obtain their abundances? What are the specificities of CEMP stars? Can these chemically peculiar stars be explained by internal mixing processes? If no, what kind of external sources should be invoked?

%---------------------------------------------------------------------------------------------------------------------------------------
\section{Generalities on metal-poor stars}\label{generalmp}

%\textcolor{red}{Intro maybe}

%\textcolor{red}{Where are found these stars? Structure of MW...}

%\textcolor{red}{Normal halo stars and then CEMP and CEMP-s and no}

%\textcolor{red}{Observation review... uncertainties on abundances, stars in binary, abundances etc... what we know about CEMP}

%\textcolor{red}{Nucleosynthetic imprints of the first stars in the most metal-poor halo stars}

%\subsection{The structure (and kinematics?) of the Milky Way}

%\textcolor{red}{Maybe a bit more detailed discussion here. Section about the structure and kinematics of MW? (Kinematics of stars in different components)}

%\begin{itemize}
%\item A thick disk,
%\item A thin disk,
%\item A bulge,
%\item A spheroidal halo
%\end{itemize}

%If the most metal-poor stars around us haves
%Assuming that the most metal-poor stars around us have ages comparable to the age of the Universe, they give us a precious window in the early Universe.
%Before looking for very old stars around us, 
%If we could observe stars that have almost the age of the universe is 
Observing the oldest stars is a crucial task to better understand the early Universe. 
As a first step, it seems natural to wonder whether some stars born very early could live for long enough to be still alive today and around us.
%In stellar archaeology, a fundamental question that one should answer first of all would be: \textit{can the early generation of stars in our galaxy live for long enough to be still alive today?}.
At first order, the lifetime of a star $\tau_{\star}$ is equal to the mass of fuel available divided by the stellar luminosity. 
%, assuming that the luminosity is constant. 
It yields
%The main sequence, corresponding the burning of Hydrogen is by far the longest stage.

\begin{equation}
%\tau \simeq 10 \left[\frac{M}{M_{\odot}}\right] . \left[\frac{L_{\odot}}{L}\right] \simeq 10 \left[\frac{M}{M_{\odot}}\right] \left[\frac{M}{M_{\odot}}\right]^{a} \simeq 10 \left[\frac{M}{M_{\odot}}\right]^{1-a} \rm{Gyr}
\tau_{\star} \simeq 10 \left[\frac{M}{M_{\odot}}\right] . \left[\frac{L_{\odot}}{L}\right] \simeq 10 \left[\frac{M}{M_{\odot}}\right]^{1-a} \rm{Gyr}
\end{equation}
where the mass-luminosity (ML) relationship $L \sim M^{a}$ is used. 
%Since the first suggestion of the ML relationship by \cite{halm1911}, different works have determined empirically the parameter $a$ with an increased accuracy. 
Setting a standard value of $a=3.5$ leads to lifetimes of 17.5 and 13 Gyr for 0.8 and 0.9~$M_{\odot}$ stars, respectively. It means that nowadays, stars with mass $\lesssim 0.9$~$M_{\odot}$ that formed early in the Universe are still alive, hence possibly observable. 
%On the opposite, early massive stars with $M \gtrsim 1$~$M_{\odot}$ are already dead.

%This is what makes the viability of the field called \textit{stellar archaeology}, aiming at  
%This relation shows the possibility of observing, today and around us, low mass stars that formed very early in the universe.

%\textcolor{red}{parler de l'age des etoiles? astero et paragraphe sur radioactive species? (later in the intro)}

%\textcolor{red}{next par. mettre qu'une fois. Ici ou plus tard}

%\textcolor{red}{attention: repetition avec petit par. sur les mp stars je pense. Mettre ce par. avant? la ou on parle des mp pour la 1ere fois?}
As discussed in Sect.~\ref{obssign}, the metallicity of a star can be used as a proxy for its age. As a consequence, the oldest stars may be searched among the most metal-poor stars. 
However, determining the metallicity Z at the surface of a star would require to determine the abundance of all elements heavier than helium, which is not possible. Iron is often taken as an indicator of the metallicity of a star since it is one of the most abundant and easily measurable element in a stellar spectrum. It can be obtained from every observed star. 
%It allows an easy classification of stars. 
%We note that the most iron poor stars are not necessarily the stars with the lowest metallicity Z since Z contains all metals. 
The iron content at the surface of a star is given with the [Fe/H] ratio. For two elements X and Y one has
\begin{equation}
[\textrm{X/Y}] =\log_{10}(N_{\rm X} / N_{\rm Y})_{\star} - \log_{10}(N_{\rm X} / N_{\rm Y})_{\odot}
\label{bracketratio}
\end{equation}
with $N_{\rm X}$ and $N_{\rm Y}$ the number density\footnote{In the Sun, $N_{\rm H}/N_{\rm Fe} \simeq 32000$ \citep{asplund09} meaning that for each iron atom there are 32000 hydrogen atoms.} of elements X and Y. For instance, [Fe/H] $= -4.0$ corresponds to an iron abundance of 1/10000 that of the Sun. This notation requires the abundances of the observed star and the ones of the Sun, taken as a reference. %\textcolor{red}{Ref for sun ? Asplund?}
\cite{chamberlain51} first reported a deficiency in iron for 2 stars: HD19445 and HD140283. They derived values of [Fe/H]~$=-0.8$ and $-1.0$. Nowadays, several hundreds of stars with [Fe/H] $<-3$ are known. Eight stars with [Fe/H] $<-4.5$ were observed \citep[listed in the Table 1 of ][]{frebel15}. 
%The impact of metals in the emergent stellar spectrum is shown in Fig.~\ref{spectrum}: 
The spectra of the Sun and 3 stars with decreasing [Fe/H] ratios are shown in Fig.~\ref{spectrum}. 
%(without brackets, X/Y is like Eq.~\ref{bracketratio} but without $\log$). 
We clearly see the decreasing line strength with decreasing metallicity. We also see that at very low metallicity, the iron line is the (or among the) strongest lines.

\cite{beers05} first established a classification of metal-poor stars based on the [Fe/H] ratio (Table \ref{MPtable}). According to this terminology, iron-poor stars are classified as metal-poor stars.
%, even if we do not know their total metallicity. In other words, iron is taken as a proxy for the total content in metal.
Rigorously, this association is correct as long as all the abundances (compared to the Sun) of a given iron-poor star scale down like iron, i.e. if a star having [Fe/H]~$=-4$ has also [X/H]~$=-4$ (where X are all metals). For instance, an iron-poor star with [Fe/H]~$=-4$ and [X/H]~$=1$ (where X are all metals except iron) is iron-poor but not metal-poor. It will have a super-solar metallicity. However, observations of iron-poor stars show that, in general, other metals roughly scale down in proportion to iron. Notable exceptions, which are the object of this thesis, nevertheless exist: some iron-poor stars are strongly enhanced in other metals and then their chemical pattern is highly non-solar (cf. Sect.~\ref{pecEMP}). But even in such extreme cases, the star very likely stays metal-poor. As an illustrative example, a star with [X/H] $=-4$ (X are all metals) 
%(equivalent to [X/Fe] $=0$ where X refer to all metals) 
has a total metallicity $Z = Z_{\odot} / 10^4 \simeq 1.4~10^{-6}$. Taking the same star but with [C/H] $=0$ (equivalent to [C/Fe] $=+4$) gives $Z=2.2~10^{-3}$, which is $\sim 6$ times lower than $Z_{\odot}$. % It shows that the most iron-poor stars are not necessarily those with the lowest metallicity. }
%To conclude, 
It shows that even if some metals are strongly enhanced in iron-poor stars, one can securely guess that the total metal content of an iron-poor star stays low (except in very rare cases\footnote{Taking the same star as before but this time with [C/H] $=1$ (equivalent to [C/Fe] $=+5$) will have $Z>Z_{\odot}$. Such kind of stars with very little iron but with $Z>Z_{\odot}$ are extremely rare. HE~1045-1434 \citep{beers07} and SDSS~J1245-0738 \citep{bonifacio15}, with respectively [Fe/H] $=2.55$ and [C/Fe] $=3.2$, and [Fe/H] $=-3.21$ and [C/Fe] $= 3.45$, may belong to this rare class of stars.}). This said, it is important to keep in mind that the most iron-poor stars are not necessarily the most metal-poor stars.

   \begin{figure*}[t]
   \centering
      \includegraphics[scale=0.42, trim = 12cm 0.5cm 11cm 1cm]{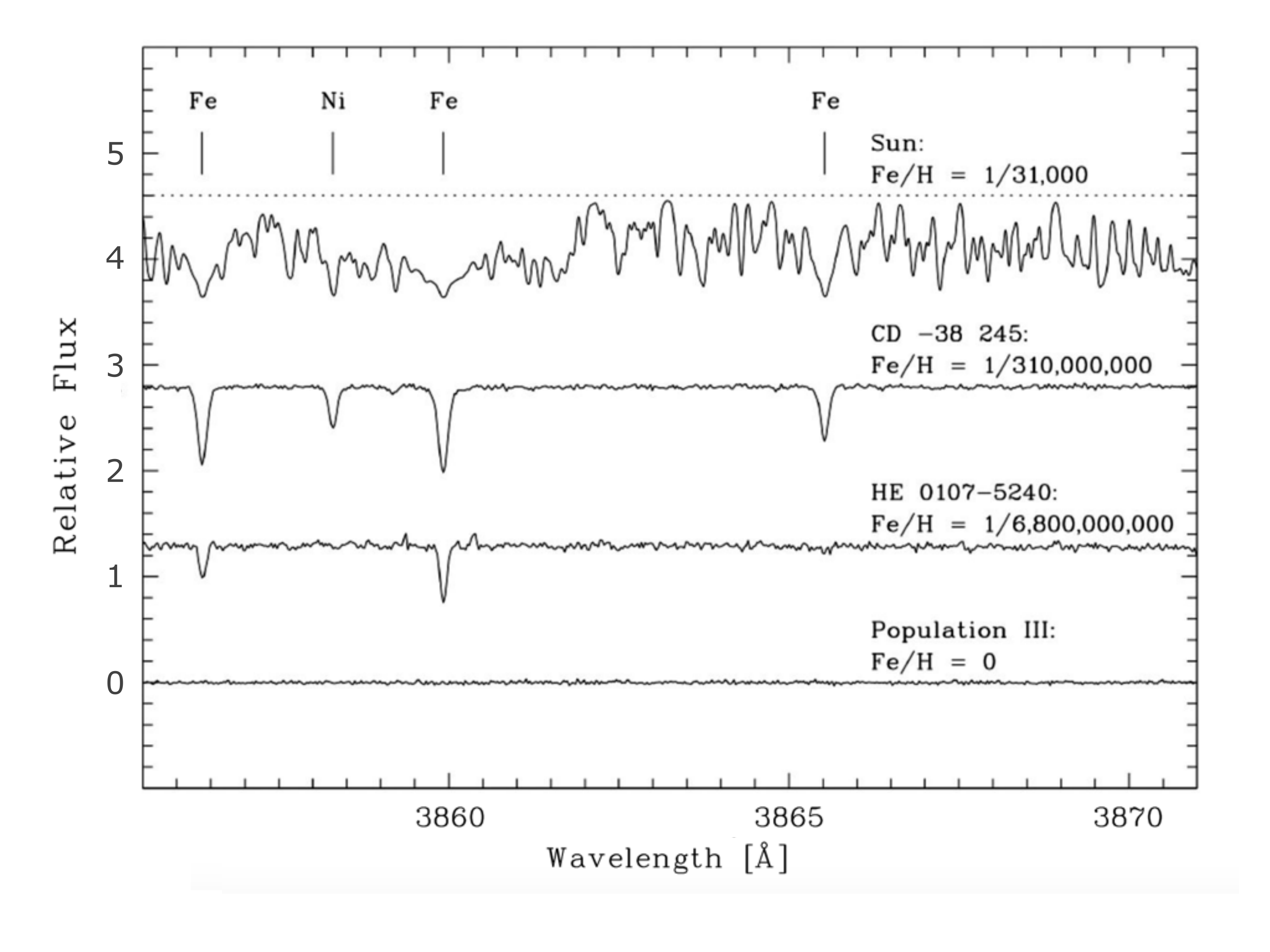}
   \caption[Spectrum of stars with different metallicities]{Spectrum of stars with different metallicities. The spectra are on the same scale and have been offset arbitrarily in the vertical direction (source: www.eso.org).}
\label{spectrum}
    \end{figure*}

%If observing stars randomly, the probability of finding a [Fe/H] $< -3.5$ is very small.
%In the next section we discuss the observational steps needed to get the surface chemical composition of metal-poor stars. 
%It starts with the observational techniques developed to quickly spot the good metal-poor candidates among the field stars.

%In the next section we discuss the observational techniques developed to efficiently spot the good metal-poor candidates among the field stars.

%maybe next section?

%Nowadays, several hundreds of stars with [Fe/H] $<-3$ are known
% \citep[SAGA database\footnote{http://sagadatabase.jp}, ][]{suda08}. 
%In total, eight stars with [Fe/H] $<4.5$ were observed \citep[listed in the Table 1 of ][]{frebel15}.

%:   tab tvelocp3
\begin{table}[t]
\caption[Nomenclature of iron-deficient stars.]{Nomenclature for stars of different metallicites, as defined in \cite{beers05}. The last column is the number of Milky Way stars observed below the given [Fe/H] ratio (numbers are from the SAGA database). For the MMP class, the recently discovered star J0023+0307, with [Fe/H] $<-6.6$ \citep{aguado18}, was added since not in SAGA yet.}
\label{MPtable}
\centering
\begin{tabular}{r r r r }
\hline\hline
    [Fe/H] & Term & Acronym & N   \\
\hline
        $< -1$ & Metal-poor & MP & $>500$\\
        $< -2$ & Very Metal-poor & VMP & $>500$\\
        $< -3$ & Extremely Metal-poor & EMP & 499\\
        $< -4$ & Ultra Metal-poor & UMP & 29\\
        $< -5$ & Hyper Metal-poor & HMP & 5\\
        $< -6$ & Mega Metal-poor & MMP & 2\\
\hline
\end{tabular}
\end{table}

As a final remark, let us mention that the direct association of the lack of metals with the stellar age is not straightforward.
%This simple view has however its limits: 
For instance, some metal-poor stars could have formed later in isolated pockets of interstellar material that remained metal-poor. Also, some more metal-rich stars could have form early if the star formation rate (hence steepness of age-metallicity relation) was very high.
Nevertheless, it is worth noting that the age of some very metal-poor stars was determined owing to the detection of radioactive elements like Th and U at their surface. $^{238}$U and $^{232}$Th have half lives of about 4.5 and 14 Gyr respectively. Assuming that Th and U were synthesized before the birth of the metal-poor star 
%(e.g by a supernova or a neutron star merger) 
and supposing that the initial abundances of Th and U are known, one can compare the initial abundances of Th and U to the observed abundances to estimate the age of the star. Ages of about $13-14$ Gyr were derived for several metal-poor stars with [Fe/H] $\sim -3$ \citep[][]{hill02, sneden03, frebel07b}. It gives some support to the fundamental hypothesis of stellar archaeology stating that very metal-poor stars are very old stars. The uncertainties related to this dating method are however large (several Gyr). It is mainly because our understanding of the nuclear processes responsible for the formation of Th is still incomplete \citep[especially the uncertainties in the nuclear data and unknown thermodynamic conditions in which the nuclear processes, able to form Th, take place,][]{goriely99}.
\section{Observation and abundances of metal-poor stars}\label{obsmp}

%\textcolor{red}{work of Hartwig? Tumilson?}

Metal-poor stars are very rare objects: in the solar neighborhood, only $\sim 1/10^{5}$ field star would have [Fe/H] $< -3.5$ \citep[][see also Table~\ref{MPtable}]{frebel15}. %However, as we will see in Sect.~\re{wasurvey}, some techniques 
How and where to find them in the Milky Way which contains billion of stars? 
%How and where the most metal-poor stars can be found in our Galaxy?
As schematically shown in Fig.~\ref{arch}, our Galaxy contains four main components: a thick disk, a thin disk, a bulge and a spheroidal halo. The disk is about 30 kpc.
\cite{carollo07} have shown that the halo can be divided into two broadly overlapping structural components: the inner (radius $\lesssim 15$ kpc) and outer halo (radius $\gtrsim 15$ kpc). These two components have different spatial density profiles, stellar orbits and stellar metallicities. The metallicity of the inner and outer halo peaks at [Fe/H] $\simeq -1.6$ and $-2.2$, respectively. 
Surveys carried out over the past decades have shown that the most metal-poor stars are found in the halo of the Milky Way. 
It is reviewed below the observational steps required to efficiently spot the good metal-poor candidates among the field stars.

%Halo stars have also more eccentric orbits \citep{carollo10}.

\subsection{The observational steps}

   \begin{figure*}[t]
   \centering
      \includegraphics[scale=0.78, trim = 0cm 0cm 0cm 0cm]{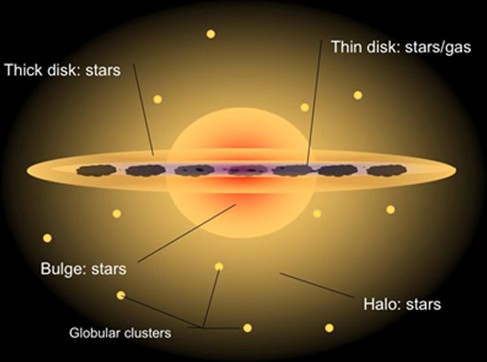}
   \caption[Schematic view of the architecture of the Galaxy]{Schematic view of the architecture of the Galaxy (source : http://www.cefns.nau.edu).}
\label{arch}
    \end{figure*}

   \begin{figure*}[h!]
   \centering
      \includegraphics[scale=0.84, trim = 12cm 0.5cm 11cm 1cm]{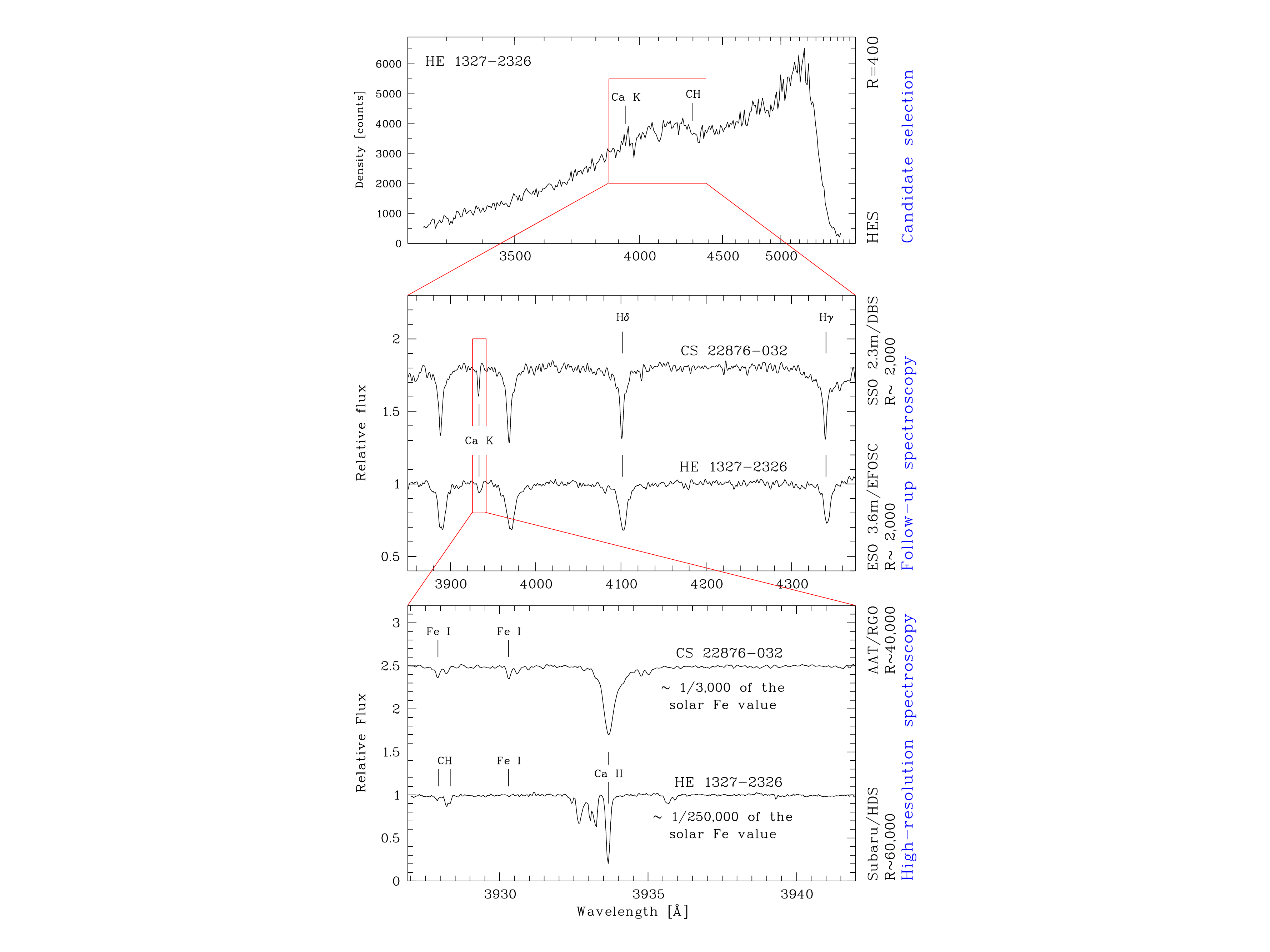}
   \caption[Spectrum of a metal-poor star: from low to high resolution]{Spectrum of the metal-poor star HE~1327-2326 with [Fe/H] $=-5.6$, from low to high resolution. It shows the three steps required to obtain accurate abundances. CS~22876-032, with [Fe/H]~$= -3.7$ \citep{norris00} is shown for comparison \citep[figure from][]{frebel05}.}
\label{obs1}
    \end{figure*}

%maybe later?
The process of getting the surface chemical composition of metal-poor stars is long.
As discussed in the review of \cite{beers05}, there are 3 main observational steps to obtain accurate abundances of metal-poor stars:

\begin{itemize}
\item Step 1 : a wide-angle survey in order to select the metal-poor candidates.
\item Step 2 : a moderate-resolution spectroscopic follow-up of the candidates to validate their lack of metals.
\item Step 3 : a high-resolution spectroscopy of the most interesting candidates to accurately determine the abundances.
\end{itemize}

Typical spectra obtained from each of these three steps are shown in Fig.~\ref{obs1}. The star is HE~1327-2326 \citep{frebel05,frebel05nat,frebel06b,frebel06,frebel08,aoki06}, having [Fe/H] $= -5.6$. From step 1 to 3, the resolution $R = \lambda/\Delta \lambda$ increases from 400 to 60000. The high resolution spectroscopy (step 3) was performed with the High Dispersion Spectrograph \citep[][]{noguchi02} on the 8.2-meter optical-infrared SUBARU telescope in Hawaii. Below I give some details on these three observational steps.

\subsubsection*{Step 1 : wide-angle surveys}\label{wasurvey}	%maybe no subsections?

The first step, finding metal-poor candidates, can be achieved in different ways.

%List?
\paragraph{Proper-motion surveys}
%\paragraph{Stellar kinematics}

A significant fraction of halo stars has high proper motion\footnote{The proper-motion of a star corresponds to its apparent angular motion across the sky with respect to more distant stars.} compared to Galactic disk stars. It arises because halo stars have elliptical orbits that are not in the Galactic plane (i.e. very different orbits than disk stars like the Sun) and because halo stars have typically larger space velocities than disk stars. This last point implies that a halo star will likely have a higher proper motion than a disc star at the same distance. %\textcolor{red}{REF}}
%As a consequence, the use of proper-motion catalogues is a possibility in identifying metal-poor stars.
%Proper-motion\footnote{The proper-motion of a star corresponds to the observed changes in its apparent place in the sky.} measurements of stars in the Galaxy is a possibility. 
\cite{schwarzschild50} and \cite{roman50} first suggested a possible correlation between the space velocity of stars and the weakness of their lines (attributed to the lack of metals in their atmosphere). These pioneering works motivated further studies that confirmed the connection between the kinematics of stars and their abundances: metal-poor stars are preferentially found among the stars with a high space velocity. For instance, the first star having [Fe/H] $< -3.0$ (G64-12) was discovered by \cite{carney81} as the result of its high space velocity. 
\cite{ryan91} carried out a large spectroscopic follow-up of a sample of stars with high proper-motion. About 10~\% of the sample was found to have [Fe/H] $< -2.0$.

Among the first surveys measuring proper motion, which were based on photographic plates, one finds the Lick Northern Proper Motion survey \citep{klemola87}, the Southern Proper Motion survey \citep{girard98} or the SuperCOSMOS Sky Survey \citep{hambly01}. More recent surveys are the Sloan Digital Sky Survey \citep[SDSS,][both photometry and spectroscopy]{york00} or the Gaia mission \citep[e.g][]{perryman01}. 
%\textcolor{red}{Talk more about 2 last surveys? Later..?}

%Proper-motion surveys is a way to select metal-poor candidates. As an example,  %\textcolor{red}{Say more}.
%maybe next sect
%Maybe say that allow to see distrib of Z of the halo? peak at -1.6, -2? inner outer halo (Zuo+17)

%Reprendre review Frebel (et Beers?) Sec 3.2.2

%Beers05 : fewer than 10% of the stars in these catalogs have [Fe/H] < ?2.0. This is a direct result of the fact that the metallicity distribution function (MDF) of halo stars peaks around [Fe/H] = ?1.6 and falls rapidly with declining metal abundance.

\paragraph{Colorimetric surveys}

%Photomotry low-resolution method for discovering metal-poor candidates

The metals (particularly Fe) in the atmospheres of stars absorb preferentially the blue light so that metal-rich stars appear redder while metal-poor stars appear bluer. The color of a star and more specifically the ultraviolet excess $\delta (U-B)$ can be used as a metallicity indicator \citep[e.g. the early works of][]{wallerstein60, wildey62}.

The SkyMapper Southern Sky Survey \citep[SMSS,][]{keller07} is a photometric program of the Southern Hemisphere started in 2008 and giving accurate estimates of atmospheric parameters of metal-poor stars (effective temperature, surface gravity and stellar metallicity). This survey led to the discovery of the most metal-poor star currently known: SM 0313-6708 with [Fe/H] $< -7.3$ \citep{keller14}.

The Pristine survey \citep{starkenburg17} is a photometric survey using a narrow filter with a width of $\sim 100$ \r{A} that covers the wavelengths of the Ca doublet lines (at 3968.5 and 3933.7 \r{A}). A spectroscopic follow-up of promising metal-poor candidates selected by the Pristine survey has led to the discovery of numerous metal-poor stars with [Fe/H] $<-3$ \citep{youakim17}.

%\textcolor{red}{Completer}

\paragraph{Objective-prism surveys}

%Until now, wide-angle spectroscopic surveys with low-resolution ($R \simeq 400$) have been the most efficient way to find metal-poor stars. 
Such surveys are wide-angle spectroscopic surveys with low-resolution ($R \simeq$~400). 
%They have been the most efficient way to find metal-poor stars. 
%A large number of low-resolution spectra are obtain. \textcolor{red}{Either the color of the star,} or the strength of the Ca\textsc{II} K line is examined, leading to a first estimate of the content in metal of the star. 
The idea is generally to examine the strength of the Ca\textsc{II} K line so as to obtain a first estimate of the metal content of the star. 
\cite{bond70} and \cite{bidelman73} first carried out such surveys, aiming at identifying metal-poor stars based on spectroscopy only. 
The HK survey \citep{beers85, beers92} and the Hamburg/ESO (HES) survey \citep{wisotzki96} are other surveys of this type, that were prolific sources of metal-poor stars. 

%The HK survey delivers a large number of low-resolution spectra for stars with magnitude $m$ between $11.0$ and $15.5$. 
After a spectroscopic follow-up of metal-poor candidates selected from the HK survey, it appeared that 11~\% of the candidates had [Fe/H] $< -2.0$ \citep{beers05}.
A drawback regarding the HK survey is that metal-deficient cool stars would likely be missed. Indeed, such stars have enhanced strengths of their Ca \textsc{II} K lines because of the lower temperature %, and that even if the metallicity is low. 
and therefore are considered as metal-rich stars.  
%At the same time, hot stars with an intermediate metal content are included in the metal poor candidates because of the weakness of the Ca\textsc{II} K line. 
%The HK-II survey \citep[][and subsequent unpublished work by J. Rhee and collaborators]{rhee01} aimed at identifying the metal-poor red giant cool stars that may have been missed in the original analysis due to the temperature bias against cool stars. 
The HK-II survey \citep{rhee01} aimed at identifying the metal-poor red giant cool stars that may have been missed in the original analysis due to the temperature bias against cool stars. 
%Colors in the JHK bands can be estimated with the help of the Two Micron All Sky Survey \citep[2MASS][]{skrutskie06}.
%Including the $B-V$ selection in the HK survey (cf. previous discussion), raises the efficiency of the selection from 11 to 32~\%.
It allowed to raise the efficiency of the HK survey to detect metal-poor stars from 11 to 32~\%.

The HES survey can observe fainter objects than the HK survey (about 2 magnitude deeper). Because of the broad wavelength range covered ($3200<\lambda<5200$ against $3875<\lambda<4025$ for the HK survey), the $U-V$ color of stars can be directly obtained from the 
%prism 
spectra and combined to the spectroscopy to select metal-poor candidates with more security. A spectrum obtained from the HES survey is shown on the upper panel of Fig.~\ref{obs1}.
Spectroscopic follow-up of metal-poor candidates selected from the HES survey have shown that $50-60$~\% of the candidates have [Fe/H]~$< -2.0$. %\citep{beers05}. 

%The works of \cite{roman55} and \cite{eggen62} reported a correlation between the proper motion of stars and their ultraviolet excess. This excess is attributed to the lack of metals in the stellar atmosphere. Indeed, the density of absorption lines due to metals is high in the blue region of the spectra, hence the UV excess if metals are not abundant. These 2 pioneering studies motivated 

\subsubsection*{Step 2 : moderate-resolution spectroscopy}\label{followup}	%maybe no subsections?

After step 1, a moderate-resolution spectroscopic follow-up of the candidates is required to validate the metal-poor stars among the sample. It also exists surveys targeting hundredth thousands of stars with a moderate-resolution and thus allowing to skip the step 1.%directly determine the metal-poor nature of a star, without the need for follow-up confirmation.

The SDSS survey do both photometry and medium resolution spectroscopy ($R \sim 2000$). SDSS has gone through 3 stages: SDSS-I (from 2000 to 2005), SDSS-II (from 2005 to 2008) and SDSS-III \citep[from 2008 to 2014,][]{eisenstein11}. 
%SDSS-II includes three individual surveys, among it the Sloan Extension for Galactic Understanding and Exploration 1 (SEGUE-1) survey \citep{yanny09}. 
SDSS includes several individual surveys, among them the Sloan Extension for Galactic Understanding and Exploration (SEGUE) survey \citep{yanny09}.
%The SEGUE-2 survey took place from 2008 to 2009. 
The two phases of SEGUE, SEGUE-1 and -2, obtained spectra for 240000 and 118151 stars, respectively, with typical errors in $T_{\rm eff}$, [Fe/H] and $\log g$ of 117 K, 0.22 and 0.26 dex, respectively. 
%\footnote{source: http://www.sdss3.org/surveys/segue2.php}. 
SDSS/SEGUE surveys were prolific sources of metal-poor stars \citep[e.g.][]{fukugita96, adelman-McCarthy08, abazajian09} and motivated numerous high-resolution observations \citep[e.g.][where 137 metal-poor SDSS candidates were observed with the high dispersion spectrograph of the Subaru telescope.]{aoki13}. 

The Large Sky Area Multi-Object Fiber Spectroscopic Telescope (LAMOST) survey \citep{deng12} started in 2008. 
%and is a Galactic structure survey. 
It should provide spectra for 2.5 million stars. The resolution is about $R=$~1800. This survey also triggered high-resolution observations. \cite{li15} reported $R=$~36000 spectra for two stars with [Fe/H] $\sim -4$, selected in the LAMOST database.

From 2003 to 2013, the RAdial Velocity Experiment \citep[RAVE,][]{steinmetz03} targeted bright stars ($8 < m < 12$, it represents stellar distances up to $\sim$ 3 kpc from the Sun). This survey delivered spectra for about 480000 stars with a resolution 
%\footnote{source: https://www.rave-survey.org/project/} 
$R=7000$. In a RAVE database of approximately 200000 stars belonging to the Galactic disk or halo, \cite{fulbright10} identified 631 stars with [Fe/H] $<-2$ ($\sim$ 0.3~\%). In the Galactic bulge, the ARGOS survey (Abundances and Radial velocity Galactic Origins Survey) revealed 16 stars with [Fe/H] $<-2$ within a sample of 14150 stars \citep[$\sim$~0.1~\%, ][]{ness13}. In the frame of the EMBLA (Extremely Metal-poor BuLge stars with AAOmega) survey, \cite{howes14, howes15, howes16} reported the discovery of about 30 other bulge stars with [Fe/H] $<-2$ (among them a star with [Fe/H] $=-3.94 \pm 0.16$). No star with [Fe/H]~$<-4$ was found in the bulge. %These differences between the bulge and the halo are can be expected when knowing that the average metallicity is higher in the bulge.

\subsubsection*{Step 3 : high-resolution spectroscopy}\label{highres}	%maybe no subsections?

Once the promising candidates are identified, the next step consists in obtaining high-resolution (R  $\geq$ 30000), high S/N (signal-to-noise) spectra. %so as to accurately determine the chemical abundances. 
%, isotopic ratios (of very few elements), and sometime stellar ages.

Numerous studies conducted high-resolution spectroscopic observations on metal-poor star candidates, originally selected from the HE, HK, SDSS or SEGUE surveys \citep[e.g.][]{mcwilliam95,carretta02, cohen04, cohen08, cohen13,barklem05,aoki07,aoki13,hollek11, caffau11, bonifacio12, norris13,yong13, roederer14a, hansen14}.
There are at least three public database, grouping the abundances of metal-poor stars derived from high-resolution spectra: 
\begin{itemize}
\item The SAGA database\footnote{\href{http://sagadatabase.jp/}{http://sagadatabase.jp}.} \citep{suda08}.
\item The \cite{frebel10} database\footnote{\href{http://www.metalpoorstars.com}{http://www.metalpoorstars.com}.}. 
\item The JINAbase\footnote{\href{http://jinabase.pythonanywhere.com}{http://jinabase.pythonanywhere.com}.} \citep{abohalima17}.
%\item the Hypatia catalog database \footnote{https://www.hypatiacatalog.com/} \citep{hinkel14,hinkel17}
\end{itemize}

%The APO Galactic Evolution Experiment \citep[APOGEE][]{allende08} was one of the four experiments in SDSS-III.

Among the future high resolution surveys one finds 4MOST \citep[e.g.][]{feltzing17}, WEAVE, a spectrograph that will offer two possible resolutions: 5000 and 20000 or GALAH \citep[Galactic Archaeology with the HERMES spectrograph][]{martell16proc} which aims at collecting R $\sim$ 28000 spectra for one million stars in the Milky Way. 

%WFMOS \citep{freeman08}

%WEAVE Gal archaeo: Hill) (first light expected in 2019)

%PFS \citep{takada14}

%\cite{hansen14} : mid and high res spectra. Candidates from HES survey

%\subsubsection{a}

%Several types of surveys

%Maybe describe method to get MP and then all survey? Or both at the same time maybe

\subsection{Abundances and uncertainties}\label{secabuncer}

%When the spectra is obtained, an atmosphere model is required to derive the stellar abundances (see Sect.~\ref{secabuncer}). 

Once a high-resolution spectrum is acquired from a star, the abundances in term of numbers have to be determined (e.g. [X/Fe] ratios). This is a highly nontrivial task that required a realistic stellar atmosphere model. 
Stellar abundances are therefore not \textit{observed} but deduced by \textit{modeling} the spectral lines.
The stellar abundances are generally determined by inspection of the equivalent width of spectral absorption lines. The observational uncertainty varies as $\sqrt{\rm FWHM}$ / (S/N) where FWHM is the full width at half maximum of the line. %\citep{cayrel04}. 
A spectrograph with a high resolving power and a high S/N minimizes the uncertainties, allowing the detection of weaker features, which is primordial when observing stars with little metals, hence with weak lines.
%While the spectrum is obtained through observations, the determination of the abundances require a model.
%A realistic stellar atmosphere model from which the light originates is needed. 

A perfect atmosphere model would be a 3D model taking into account departures from local thermodynamic equilibrium (LTE) and where all the atomic and molecular physics (determining the absorption lines) is included. Such a model does not exist. Often, abundances are derived using a 1D LTE model. 
%\cite{asplund05b} 
%For instance, for the star HE1327-2326 ([Fe/H] $=-5.7$) \cite{frebel05nat} used a 1D model and found [C/Fe] $=4.1 \pm 2$. A new analysis in 

\cite{frebel08}, have estimated by how much the abundances are affected when deriving the abundances with either a 1D or a 3D model. They investigated the star HE~1327-2326 ([Fe/H]~$=-5.7$). They found a 3D-1D correction for C, N and O of about $-0.7$ dex. 

\cite{ezzeddine17} studied anew the stellar parameters of 20 stars with [Fe/H] $<-4$ using a 1D NLTE (non LTE) atmosphere model instead of a 1D LTE model. They derived [Fe/H] corrections up to 1 dex compared to the 1D LTE case. These corrections are larger at lower [Fe/H]: at [Fe/H] $=-4$ and $-7$, the corrections are about 0.5 and 1 dex respectively. \cite{lind12} reported a similar trend: they derived NTLE corrections of $\lesssim 0.1$ and $\lesssim 0.5$ dex at solar and low metallicity, respectively. Also, these corrections apply only for the derivation of Fe abundances from neutral lines (Fe~\textsc{\small{I}}). The corrections are mostly insignificant if using Fe~\textsc{\small{II}} lines. Determining the Fe abundance with accuracy is important since it is generally a prerequisite for the determination of the abundance of other elements.

%The atomic and molecular data included in the atmosphere model is crucial in order to identify all the absorption lines. Building the most complete linelist is a way to reduce uncertainties when deriving abundances from a spectrum. For instance, \cite{masseron14} combined the analysis of molecules in metal-poor spectra with stellar-radiative transfer codes to built a new extensive CH linelist, that can be used to improve the molecular data in atmosphere codes.

   \begin{figure}[t]
   \centering
      \includegraphics[scale=0.54, trim = 2.5cm 0cm 0cm 0cm]{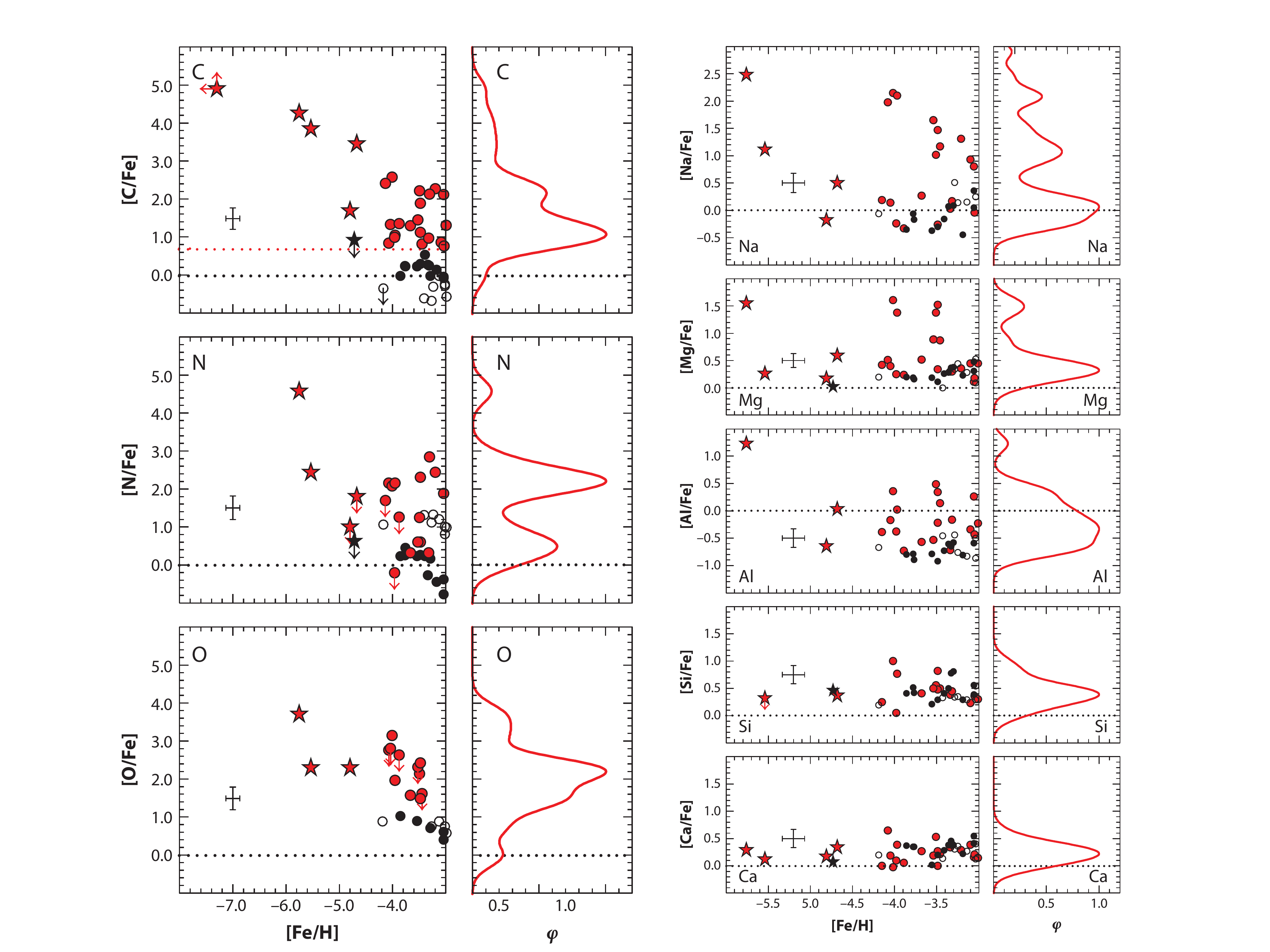}
   \caption[ X/Fe ratios of the most iron-poor stars]{[X/Fe] ratios of the most iron-poor stars. Red symbols show C-rich stars (CEMP-s, CEMP-r, and CEMP-r/s stars are excluded). Black symbols show C-normal stars. Stars show objects with [Fe/H]~$<-4.5$. Histograms of the abundances are shown on the right. \citep[figure adapted from][]{frebel15}.}
\label{abundlight}
    \end{figure}

\section{CEMP stars as peculiar metal-poor stars}\label{pecEMP}

This section aims at describing the characteristics of CEMP stars. Their possible origin is discussed in the next section.

%As spectroscopy progressively revealed the chemical composition of metal-poor stars, it appeared that their abundances 

A natural guess could be that in metal-poor stars, all metals are scaled down compared to the Sun by roughly the same factor. 
%It would mean that for a star with [Fe/H] $-3$, [C/H] $=-3$ (or equivalently, [C/Fe] $=0$). 
However, observations progressively revealed that metal-poor stars generally have non-solar-like abundance patterns. In particular, the carbon to iron ratio is super-solar in many metal-poor stars. 
The first carbon stars were observed about 150 years ago by Angelo Secchi \citep{secchi1868}. At that time, he only reported a peculiar banding in the stellar spectra. These stars were later identified by \cite{rufus16} as stars enriched in carbon.
Since, many carbon-rich stars were discovered, particularly among the most iron-poor stars. Stars enriched in carbon and depleted in iron were called Carbon-Enhanced Metal-Poor \citep[CEMP,][]{beers05}.
This name might seem somewhat contradictory since carbon is a metal. However, CEMP stars likely stay metal-poor compared to the Sun, even if they show overabundances in carbon or other metals (cf. discussion in Sect.~\ref{generalmp}). 
%\textcolor{red}{REPETITION For instance, a star with [Fe/H] $=-4$ and [X/Fe] $=0$ (where X refer to all elements) has a total metallicity $Z = Z_{\odot} / 10^4 \simeq 1.4~10^{-6}$. The same star with [C/Fe] $=+3$ has $Z=2.2~10^{-4}$. It shows that the most iron-poor stars are not necessarily those with the lowest metallicity. 
%%%%As an example, HE1327-2326 is a HMP star with [Fe/H] $=-5.7$ and [C/Fe] $=3.8$, meaning that [C/H] $=-1.9$, i.e. the 'absolute' carbon abundance is still much lower than in the Sun. 
%However, even if the total metallicity of the star cannot be assessed (since it would require to determine the abundance of all elements), one can securely guess that the total content of an iron-poor star is low. }
%\cite{rossi05} developed a method 
\cite{beers05} proposed two criteria defining a CEMP star: [Fe/H] $< -1.0$ and [C/Fe] $> 1.0$. The criterium of [C/Fe] $> 0.7$ is also often used in the literature \citep[first proposed by][]{aoki07}. 
For what follows, depending on the criterium used by the considered author, I use the names
\begin{itemize}
\item CEMP$_{1.0}$ if the condition is [C/Fe] $> 1.0$.
\item CEMP$_{0.7}$ if the condition is [C/Fe] $> 0.7$.
\end{itemize}

% ([C/Fe] $> 1.0$ or [C/Fe] $> 0.7$).

%\subsubsection*{Are CEMP stars really old?}

%\textcolor{red}{ALREADY PUT BEFORE. PUT ONCE. here or much before maybe?? Be careful, some are not CEMP. Some metal-poor stars were found to be enriched in heavy radioactive elements like Th and U which are interesting elements for age estimation. $^{238}$U and $^{232}$Th have a half life of about 4.5 and 14 Gyr respectively. Assuming that Th and U were synthesized before the birth of the metal-poor star (e.g by a supernova or a neutron star merger) and supposing that we know the initial abundances of Th and U, one can compare the initial abundances of Th and U to the observed abundances to estimate the age of the star. Ages of about $13-14$ Gyr were derived for several stars with [Fe/H] $\sim -3$ \cite{hill02, sneden03, frebel07b}, giving some support to the fundamental hypothesis of stellar archaeology assuming that very metal-poor stars have ages comparable to the age of the Universe. The uncertainties related to this dating method are however large (several Gyr), mainly because our understanding of the r-process nucleosynthesis is still incomplete \citep[especially uncertainties in the nuclear data, unknown thermodynamic conditions in which the r-process takes place][]{goriely99}.}

\subsubsection*{Are CEMP stars enriched in other elements?}

%\textcolor{red}{a bit more on that? Heart of the phd...}

%other elements
CEMP stars have various lithium abundances, from\footnote{A(X) = $\log \epsilon (\rm X)$ = $\log (N_{\rm X}/N_{\rm H}) + 12$, where X represents a given element.} A(Li) $\simeq 2$ \citep[i.e. close to the Spite plateau of $2.05\pm 0.16$,][]{spite82} to A(Li) $< 0.62$ for HE~1327-2326 with [Fe/H] $=-5.7$ \citep{frebel08}. Some observational works have suggested a melt-down of the Spite plateau below [Fe/H] $\simeq -3$ 
%\citep[][see also the top left panel of the Fig.~\ref{12ab} of the present work]{aoki09, sbordone10, bonifacio12}. 
\citep[][]{aoki09, sbordone10, bonifacio12}. 
%As mentioned in \cite{frebel15}, 
The melt-down has nevertheless to be further confirmed since some iron-poor stars could have experienced severe surface Li depletion episodes (especially those who are giants). 
\cite{korn09} have computed models of the star HE~1327-2326 including the effects of atomic diffusion. They predict a maximal Li depletion of 1.2 dex from the birth of the star to the present day. It gives a maximal initial Li abundance of about 1.8, which is much closer to the Spite plateau. Other scenarios in the same work predict a milder Li depletion, of the order of 0.2 dex, which would give an initial Li well below the Spite plateau. In any case, these initial Li abundances are not compatible with the WMAP-based primordial Li abundance of 2.63 predicted from primordial nucleosynthesis \citep{spergel07}. 
%The most iron-poor star \citep[SM 0313-6708,][]{keller14}, is a CEMP RGB star with A(Li) $= 0.7$. 
%\cite{korn07} predicted the Li depletion due to atomic diffusion in stars belonging to the globular cluster NGC 6397 with [Fe/H] $= -1.9$. 
%Applying their predictions for the most iron-poor star SM 0313-6708 (their predictions are used as rough guide since the metallicity differ significantly) would give a Li depletion factor of more than 1 dex. It means that the Li abundance of SM 0313-6708 was much closer to the Spite plateau before going giant. 

Many CEMP stars have high N/Fe, O/Fe, Na/Fe or Mg/Fe ratios. For stars with [Fe/H]~$<-$~3, the scatter of the [X/Fe] ratios globally decreases from X $=$ C to Ca \citep{frebel15}. For instance, the [C/Fe], [O/Fe], [Na/Fe], [Mg/Fe] and [Ca/Fe] ratios span about 5, 4, 2.5, 1.5 and 0.5 dex, respectively (see Fig.~\ref{abundlight}).

A significant amount of CEMP stars are also enriched in elements heavier than iron. These elements are thought to be mostly synthesized through the slow and rapid neutron capture processes (also the intermediate process or i-process, see Sect.~\ref{origcemp}). 
%(s- and r-processes, see Sect.~\ref{sproctheo} for more details). %for a discussion about the possible astrophysical sites where these processes take place). 
\cite{aoki00} and \cite{vaneck01} discovered the first VMP stars enriched in Pb. Many other CEMP stars enriched in s- and/or r-elements were discovered, down to [Fe/H] $\sim -3$ \citep{burris00,simmerer04,sivarani04,lai07,placco13}. Below this threshold, enhancements in s- and r- elements are generally very modest (see Fig.~\ref{12ab}). %\cite{sneden03} reported the discovery of CS 22892-052, with [Fe/H] $=-2.9$ and enriched in r-elements (e.g. Eu). 
%discuss agb?

%CEMP-r/s, -i

%   \begin{figure*}[t]
 %  \centering
  %    \includegraphics[scale=0.6, trim = 0cm 0cm 0cm 0cm]{cfe.png}
  % \caption[ Carbon over iron for the ]{[C/Fe] as a function of [Fe/H] for a sample of VMP stars. Different symbol size for a same color denotes different sources for the observations. \citep[Figure from][]{norris13}. \textcolor{red}{more stars or say not complete?}}
%\label{cfe}
   % \end{figure*}

\cite{beers05} established a classification of CEMP stars based on their heavy element abundances (see Table \ref{subclass}). One finds CEMP stars enriched in s-elements (CEMP-s), r-elements (CEMP-r), both r- and s-elements (CEMP-r/s) and without significant enhancement in s-/r-elements (CEMP-no). As they note, this classification should be viewed as a first approximation and used as a guideline for the future. Indeed, the location of the class boundaries are somewhat arbitrary and in some cases, it may exist a continuity between the classes rather that very distinct groups. 
Eu and Ba were chosen to define such classes because (1) they are generally readily measurable in a stellar spectrum and (2) they are expected to be produced in a different amount by the s- and r-process (Eu: mainly r-process, Ba: mainly s-process). 
It has been shown \citep[especially][]{yoon16} that CEMP-no stars are generally found on the \textit{low carbon band} with A(C) $\leq$ 7.1 and CEMP-s/rs on the \textit{high carbon band} with A(C) $>$ 7.1. It might be linked to the different formation channel for CEMP-no and CEMP-s/rs stars (Sect.~\ref{origcemp}). 
Interestingly, this separation into 2 bands, allows (with some level of confidence) to classify CEMP stars only based on their C abundance rather on e.g. Ba, whose abundance determination requires much higher resolution spectra. 
Fig.~\ref{12ab} shows that CEMP-s stars mainly lie at [Fe/H] $>-3$ (green symbols). CEMP-no stars (blue symbols) are generally found at lower [Fe/H]. Some CEMP stars (magenta symbols) are still unclassified, according to the criteria of Table~\ref{subclass}.

%:   tab tvelocp3
\begin{table}[t]
\caption[Classification of CEMP stars.]{Classification of CEMP stars, as defined in \cite{beers05}.}
\label{subclass}
\centering
\begin{tabular}{l l }
\hline\hline
    Term & Conditions    \\
\hline
        CEMP & [C/Fe] $> 1.0$\\
        CEMP-r & [C/Fe] $> 1.0$	\hspace{0.7cm}	[Eu/Fe] $> 1.0$\\
        CEMP-s & [C/Fe] $> 1.0$	\hspace{0.7cm}	[Ba/Fe] $> 1.0$	\hspace{0.7cm}		[Ba/Eu] $> 0.5$\\
        CEMP-r/s & [C/Fe] $> 1.0$\hspace{0.7cm}	$0.0 <$ [Ba/Eu] $< 0.5$\\
        CEMP-no & [C/Fe] $> 1.0$\hspace{0.7cm}	[Ba/Fe] $< 0$\\
\hline
\end{tabular}
\end{table}

%Whether CEMP-r/s show a superimposed s- and r-process signature 
%%%%(from one source or 2 different sources) 
%or a single intermediate pattern is a debated question \citep{jonsell06,lugaro12}. \cite{hampel16} proposed the existence of an 'intermediate neutron-capture process' (i-process) and to rename CEMP-r/s as 'CEMP-i'. \textcolor{red}{Gull18: r+s star in prep (see Frebel slide Melbourne)}.
%%%%%%%%%This classification links some classes (CEMP-s, -r and r/s) to nucleosynthetic processes (s- and r-processes). 

\afterpage{
   \begin{figure}[h!]
   \centering
      \includegraphics[scale=0.95, trim = 0cm 0cm 0cm 0cm]{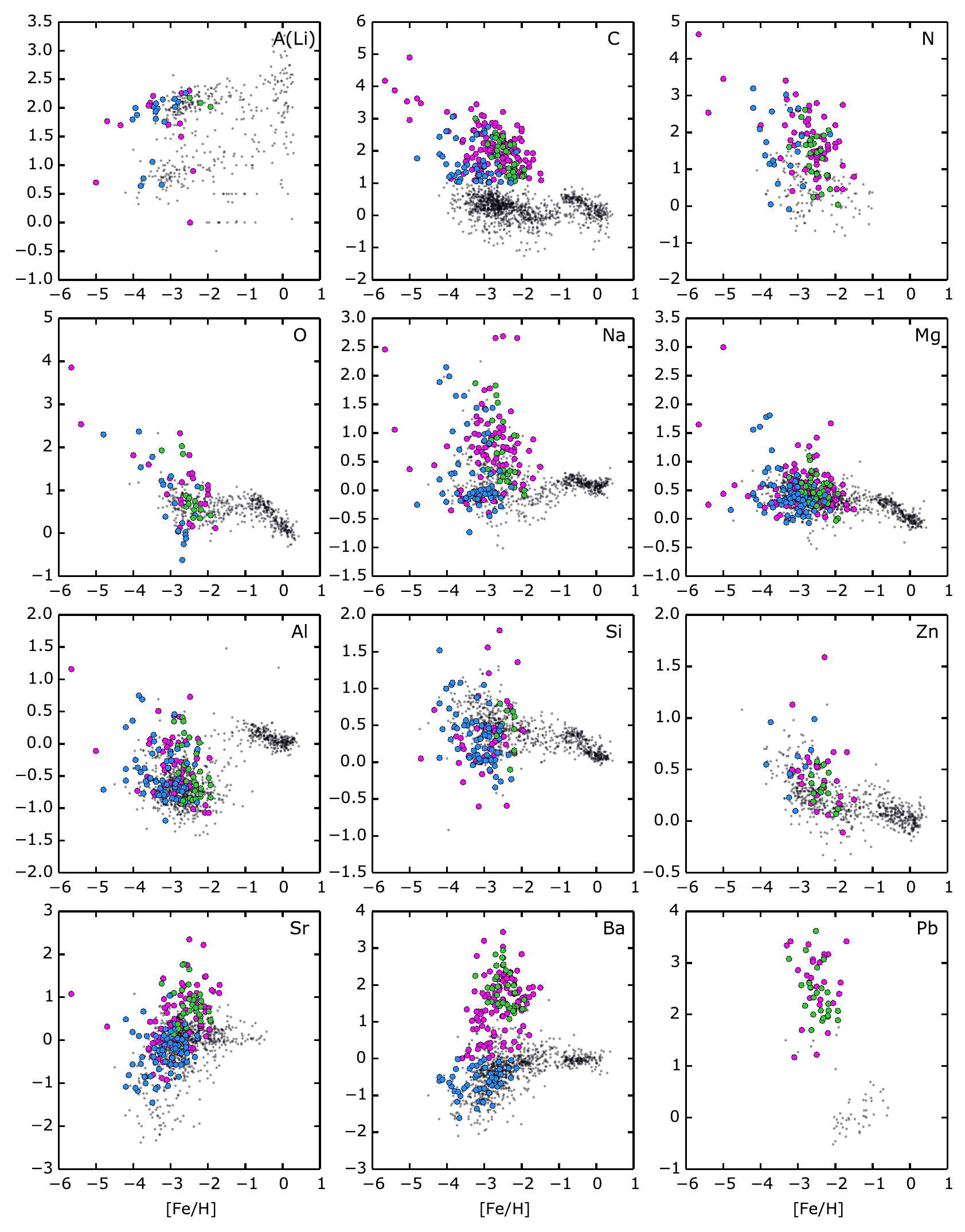}
   \caption[Abundances of iron-poor stars]{A(Li) (upper left) and [X/Fe] as a function of [Fe/H]. Green symbols are CEMP-s, blue are CEMP-no, according to criteria of Table \ref{subclass}. Magenta symbols are unclassified CEMP because of missing abundances (Ba, Eu). Light grey dots show carbon normal stars. Abundances with upper limits are not plotted. The typical uncertainty is $\pm 0.3$ dex. When abundances from different authors are available for one star, the most recent one is selected. The abundance data is taken from the SAGA database \citep[][last update on Sept. 2017]{suda08}. }
\label{12ab}
    \end{figure}
\clearpage
}

\subsubsection*{Are CEMP stars frequent?}
\cite{carbon87} and \cite{norris97} first emitted the possibility of a higher frequency of C-rich stars with decreasing [Fe/H]. \cite{marsteller05} reported a possibly high fraction of about 50~\% of CEMP$_{1.0}$ stars\footnote{I remind here that CEMP$_{1.0}$ means that the author used the condition [C/Fe] $>1$ for a star to be CEMP. CEMP$_{0.7}$ means that the author used the condition [C/Fe] $>0.7$.} among [Fe/H] $<-2$ stars selected in the HES survey. However, \cite{cohen05} showed that the [Fe/H] ratio of some stars in the HES survey was overestimated by $\sim 0.5$ dex. Consequently, they derived a CEMP$_{1.0}$ fraction of $14 \pm$ 4~\% (instead of 50~\%) among the stars with [Fe/H] $<-2$. In 240 stars with [Fe/H] $\leq -2$, \cite{lucatello06} found a CEMP$_{1.0}$ fraction of 21 $\pm 2$~\%. 
%Inspecting 121 giant stars with [Fe/H] $<-2$, \cite{frebel06} found that $\sim 9 \pm 2$~\% are CEMP$_{1.0}$. They note that the low frequency might be due to the underestimation of the [C/Fe] ratio by about 0.5 dex.
\cite{lee13} considered a sample from SDSS/SEGUE of about 247000 stars and found that 12~\% of [Fe/H] $<-2$ stars are CEMP$_{0.7}$. 
%Rossi05 method

One of the possible bias to these frequencies is that in giant stars, the first dredge-up has occurred. Amongst other, it changes the surface carbon abundance (cf. Sect.~\ref{selfenr}). 
For a sample of 505 stars with [Fe/H] $<-2$, 
%and with no observed overabundances of neutron-capture elements, 
\cite{placco14c} have estimated the effect of the first dredge-up on the surface carbon abundance. The recognized CEMP stars enriched in s- and/or r-elements were excluded from their analysis\footnote{Note that this does not mean that all stars they consider are CEMP-no. Some CEMP stars in the sample without a determined Ba abundance might appear to be CEMP-s, -r/s or -r in the future.}. The stellar models they used predict that the correction on the [C/Fe] ratio induced by the first dredge-up is generally about 0.5 dex for the evolved stars. They recovered the initial [C/Fe] ratio of the 505 stars of their sample (the initial ratio is higher in case the star has experienced the first dredge-up). They finally found that 20, 43 and 81~\% of stars with [Fe/H] $<-2$, $<-3$ and $-4$ are CEMP$_{0.7}$, respectively. The first dredge-up effect corrected, it likely gives the fraction of stars which were born as CEMP$_{0.7}$. 
%(note that the recognized CEMP stars enriched in s- and/or r-elements were excluded from their analysis).
%(note that the recognized CEMP-s, -r/s and -r stars were excluded from their analysis). }

%\textcolor{red}{fig of placco14?}

%The CEMP frequency also rises from the inner to the outer halo 
There is also a spatial variation of the CEMP frequency: \cite{carollo12} have shown that the CEMP$_{0.7}$ frequency is increasing with the distance to the Galactic plane. Almost all CEMP$_{0.7}$ stars belong to the halo and the fraction is higher in the outer than in the inner halo. Considering the most distant stars (more than 9 kpc from the Galactic plane, i.e. mainly in the outer halo), they derived a CEMP$_{0.7}$ frequency of 20~\%.
%CEMP-no stars are more frequent at low [Fe/H] (see Fig.~\ref{cfe}), making them very interesting to study the most early stages of the Universe.
\cite{carollo14} reported a fraction of CEMP-no stars of 43 and 70~\% in the inner and outer halo respectively. For the CEMP-s stars, they found 57 and 30~\% in the inner and outer halo respectively. It suggests that the dominant source of CEMP stars in the two halo components were different. 
%In the Sculptor dwarf galaxy, for a time, only one CEMP-no \citep{Skuladottir15} and three CEMP-s \citep{lardo16, salgado16} stars were detected in a sample of hundreds of stars. 
A recent study suggested that the CEMP fraction (among the [Fe/H] $<-3$ stars and after excluding the likely CEMP-s and -r/s stars) in the Sculptor dwarf galaxy is $\sim 36 \pm 8~\%$ \citep{chiti18}, i.e. similar to the $\sim 43~\%$ of the Galactic halo \citep{placco14c}.

%the carbon enhancement appears to be somewhat less pronounced than in the galaxy: the fraction of CEMP-no to C-normal stars in the sample of Sculptor stars with [Fe/H] $<-2$ is 4.5$^{+10.5}_{-3.8}$~\% \citep{Skuladottir15}. Despite the limited sample (11 stars), it can be compared to the 20~\% fraction of \cite{placco14c} found in the galaxy.

\subsubsection*{Are CEMP stars in binary systems?}\label{arebinaries}

\cite{lucatello05} and \cite{starkenburg14} showed that the whole sample of CEMP-s stars is consistent with the hypothesis of them all existing in binary systems. A careful monitoring of the radial velocity of 22 CEMP-s stars over several years has revealed a clear orbital motion for 18 stars ($\sim 82~\%$)
% giving sup- port to the AGB scenario. 
while 4 stars appeared to be single \citep{hansen16a}. The probability of finding one face-on system in their sample is about $0.01~\%$, meaning that it is extremely unlikely that all the four apparently single CEMP-s stars are in fact binary systems that were seen face-on. Apparently single stars might nevertheless have a companion with a long orbital period (about $10^3 - 10^4$ days at minimum), which would prevent to detect radial motions of the CEMP-s stars.
A similar study was carried out for CEMP-r and CEMP-no stars. The binary frequency was found to be $18 \pm 6~\%$ for CEMP-r \citep{hansen15a} and $17 \pm 9~\%$ for CEMP-no stars \citep{hansen16b}, i.e. much lower than for CEMP-s stars. It suggests that the CEMP-r and CEMP-no stars are likely disconnected from a binary origin.
% \cite{jonsell06}: see other classification table.

%\cite{behara10} : CEMP r+s

%---------------------------------------------------------------------------------------------------------------------------------------
%\section{Origin of the CEMP stars}
\section{Internal mixing processes in CEMP stars}\label{selfenr}
%\section{self-enrichment}\label{selfenr}

%\textcolor{red}{origin of CEMP is a bit like origin of the elements?}

%\textcolor{red}{The origin of metal-poor stars is a fascinating question that, if answered, would greatly enhanced our knowledge about the first generations of stars and of nucleosynthetic processes like the s- and process.}

%\textcolor{red}{put this before? in prev sect? when starting to talk about CEMP}

CEMP stars formed with the material ejected by previous stars (at least some of it). 
%, possibly among the very first stellar generations. 
One may see the nucleosynthetic signature from these previous stellar generations in the CEMP star surface chemical composition. However, the CEMP star surface composition can be altered during the life of the CEMP star itself. If this alteration is important enough, it could erase the chemical imprint let by the previous star(s). 
%Is so, the peculiar abundances of CEMP stars would be the effect of these processes. %the  explained by \textit{in}
%From this consideration, we can divide the scenarios investigating the origin of CEMP stars in two broad categories. The first one assumes that the CEMP stars are explained thanks to an event (or multiple events) that took place during the life of the CEMP star. The second one proposes that CEMP stars are explained thanks to an event that took place before the CEMP star life.
Two different categories of processes can affect the surface abundances of CEMP stars: external and internal processes. 
\begin{itemize}
\item External processes can be accretion of interstellar material or accretion of material from a binary companion. While accretion of interstellar material was found to not have a significant effect on metal-poor stars \citep{frebel09,johnson11}, the accretion of material ejected by a companion is certainly an important process, especially for CEMP-s stars (cf. Sect.~\ref{cempsext}).
\item Internal processes happen in the CEMP star itself. There are for instance the first dredge-up or thermohaline mixing. 
\end{itemize}

%Different internal processes can affect the surface abundances of stars with initial masses about $0.8$~$M_{\odot}$.
Below it is discussed the main (known) internal mixing processes that can happen in CEMP stars and possibly alter their surface composition. %These processes are important to consider in order to gauge whether CEMP stars can be 
A sample of CEMP stars is considered and compared with low-mass stellar models. The sample comprises the CEMP stars with [C/Fe] $>1$ and [Fe/H] $<-3$. The CEMP stars enriched in s- and/or r-elements are excluded\footnote{This likely excludes most of the CEMP stars whose surface abundances were modified because of the accretion of material from a companion (especially the CEMP-s stars, see Sect.~\ref{cempsext}).}. 
%\footnote{Such stars, (especially CEMP-s) may have been polluted by a companion (see Sect.~\ref{cempsext})}.

%We gather these processes into a general scenario called 'self-enrichment'. 

%This scenario raises the following question: 'do the abu CEMP stars can ' that allow explain CEMP stars thanks to internal processes.

%\textcolor{red}{\cite{frebel09} and \cite{johnson11} found that interstellar accretion is not a significant effect for metal-poor stars. Thus, the abundances measured in metal-poor stars likely reflect the chemical conditions at their formation.}

%\subsection{Self-enrichment in CEMP stars}\label{selfenr}
%\subsection{Self-enrichment}\label{selfenr}

%\textcolor{red}{fujimoto, campbell}

   \begin{figure*}[t]
   \centering
   \begin{minipage}[c]{.49\linewidth}
       \includegraphics[scale=0.43]{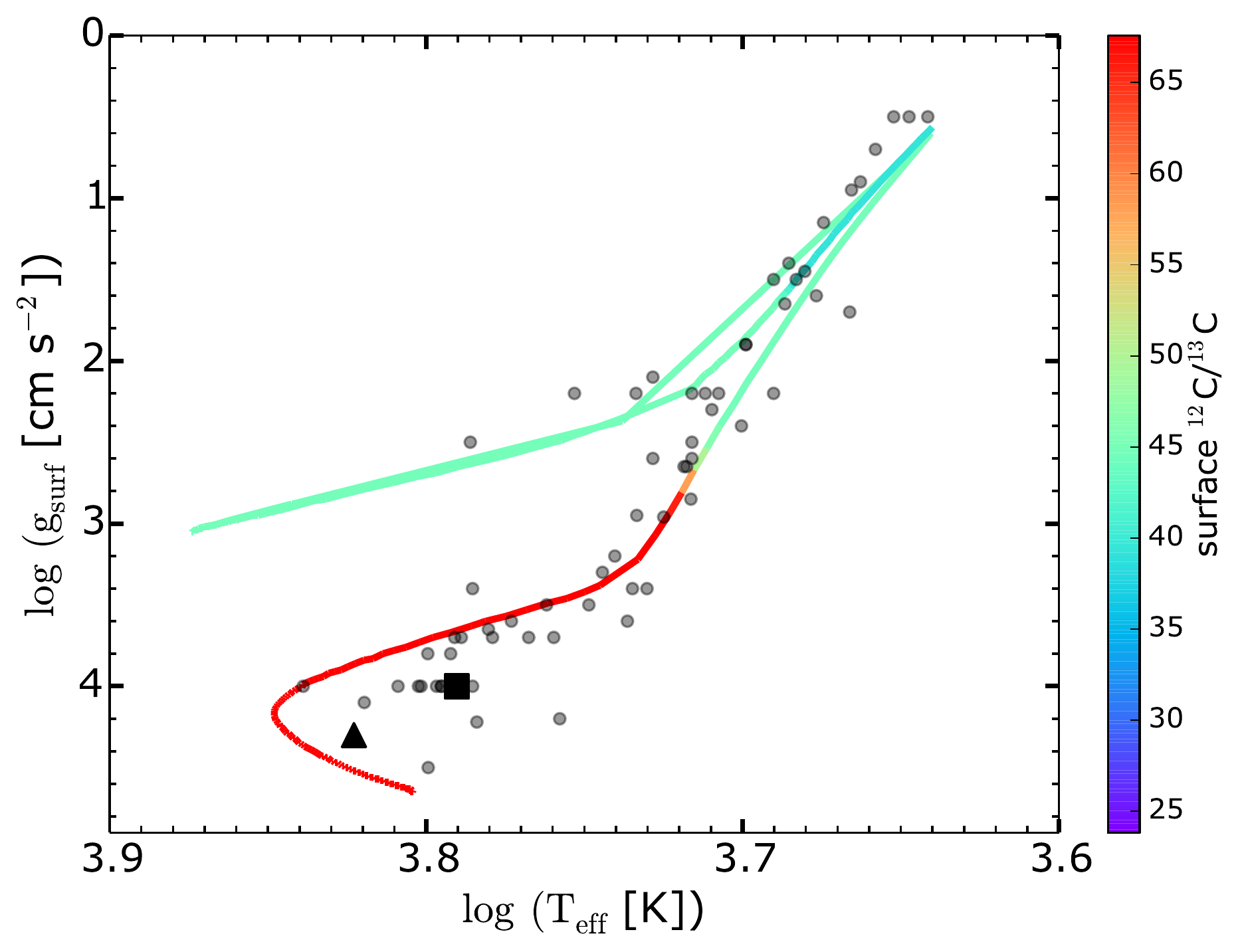}
   \end{minipage}
   \begin{minipage}[c]{.49\linewidth}
       \includegraphics[scale=0.43]{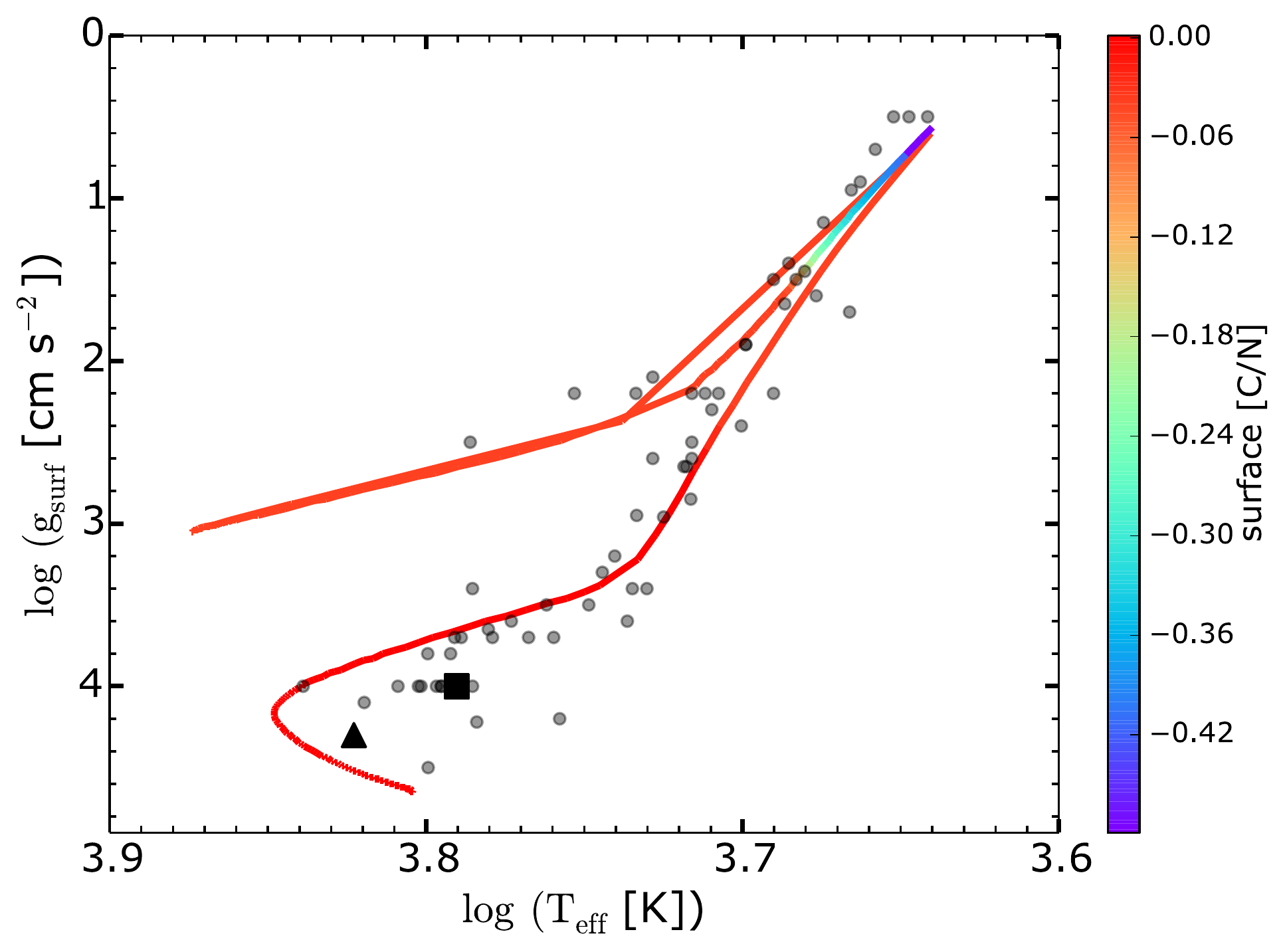}
   \end{minipage}
   \begin{minipage}[c]{.49\linewidth}
       \includegraphics[scale=0.43]{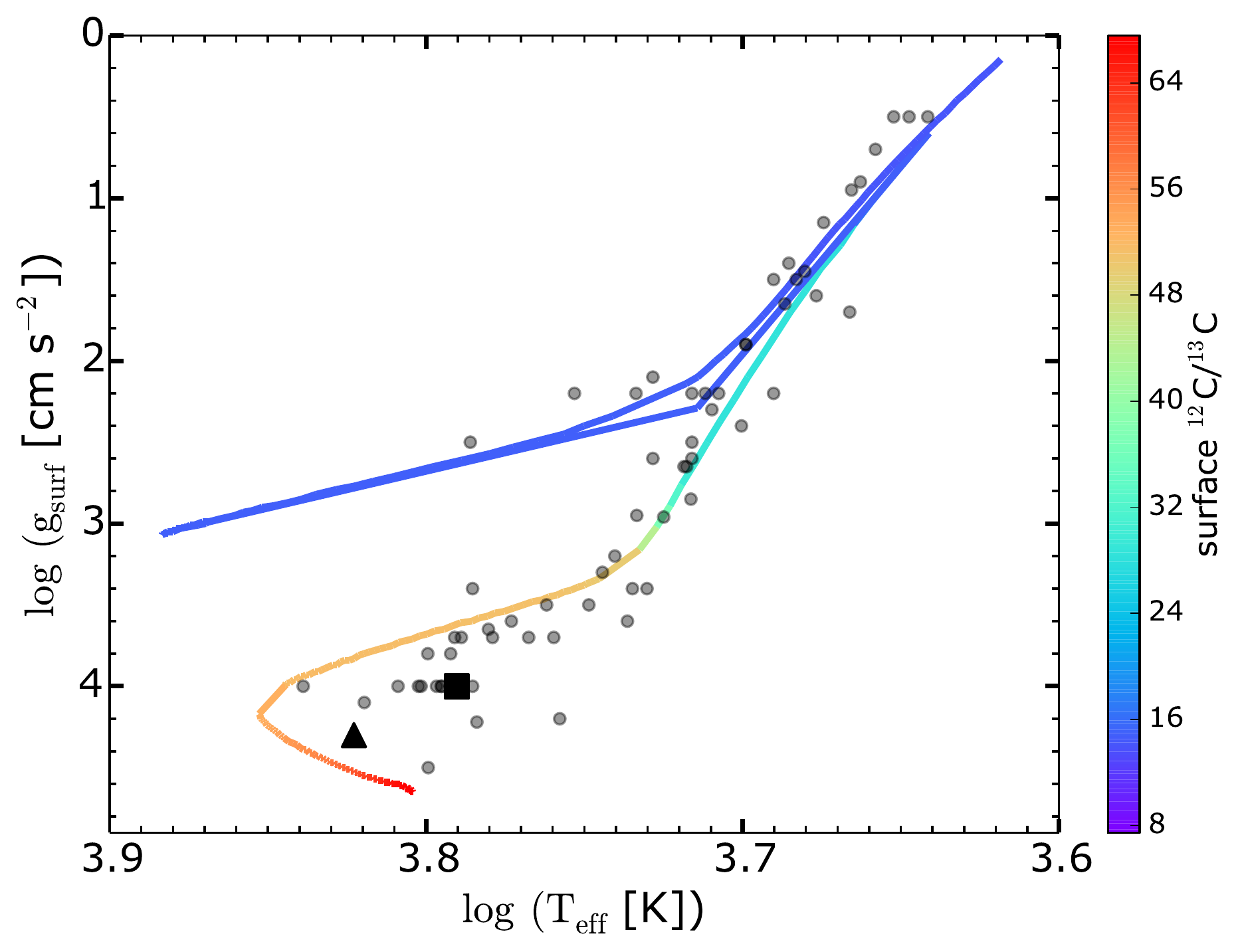}
   \end{minipage}
   \begin{minipage}[c]{.49\linewidth}
       \includegraphics[scale=0.43]{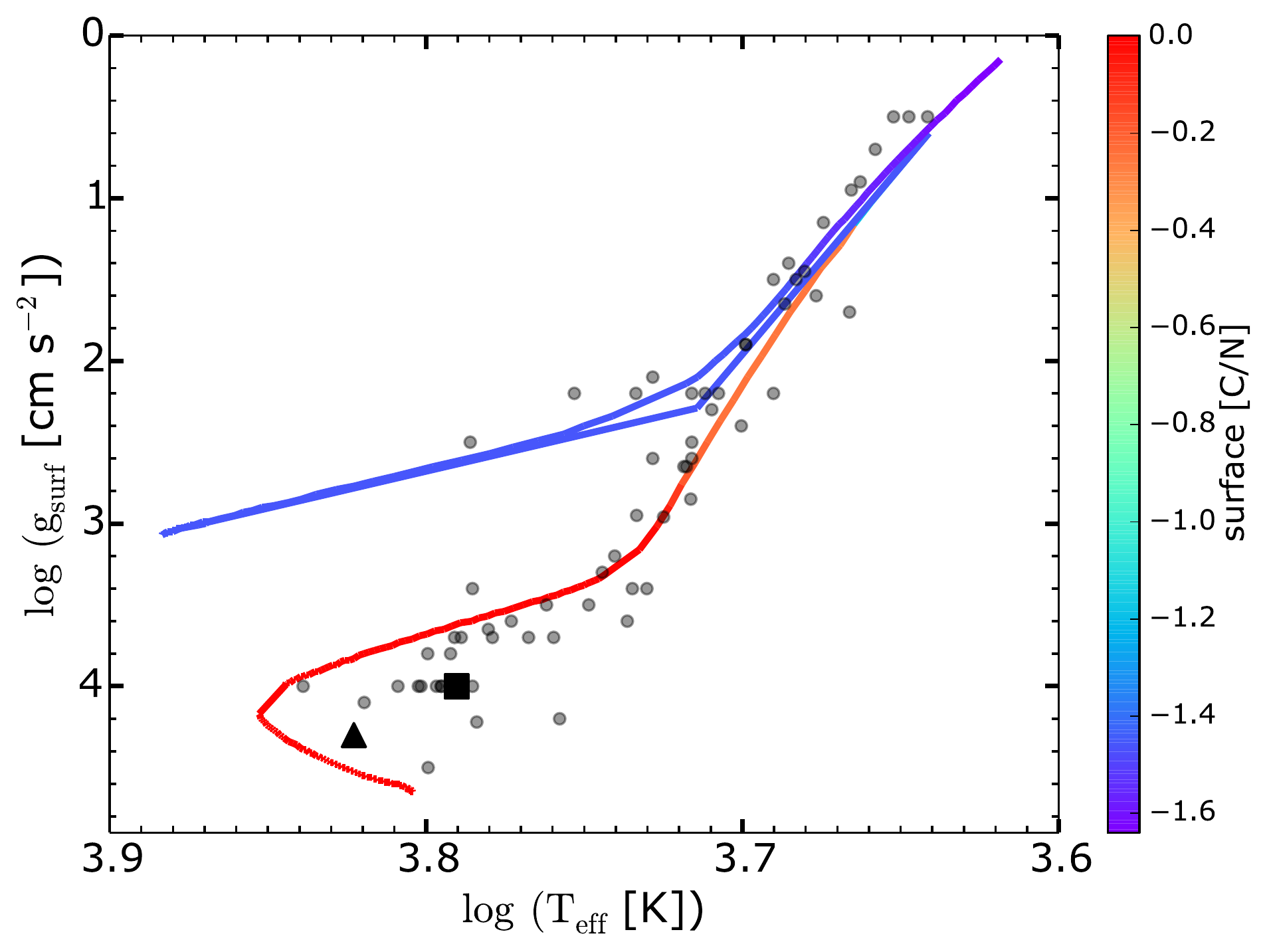}
   \end{minipage}
   \caption[]{Evolutionary tracks of 0.85~$M_{\odot}$ models at $Z=0.0001$ without thermohaline and without rotation (top panels), with thermohaline and with rotation \citep[bottom panels, models from][]{lagarde12}. The color shows the surface $^{12}$C/$^{13}$C (left panels) and [C/N] ratios (right panels). Symbols show CEMP stars with [C/Fe] $>1$ and [Fe/H] $<-3$ (recognized CEMP-s, -r and -r/s are not plotted). The triangle shows HE~1029-0546 with $^{12}$C/$^{13}$C $= 9$ and [C/N] $=-0.26$. The square shows CS~29528-041 with [C/N] $=-1.47$ ($^{12}$C/$^{13}$C is unknown).}
\label{thermohaline}
    \end{figure*}

%    \textcolor{red}{Charbonnel94,95...}
%internal

\paragraph{The first dredge-up.} Stellar models predict that as a low-mass star runs out of hydrogen, its envelope expands, the surface temperature decreases, making the star evolving to the red giant branch (RGB). During that stage, the outer convective envelope expands inward and penetrates hotter regions, where the CN-cycle is active. Some CN-processed material is consequently brought to the surface and alters the star's surface light element abundances. This mixing episode is called the first dredge-up \citep{iben64}. Since the CN-cycle mainly transforms $^{12}$C into $^{14}$N and to a lesser extent into $^{13}$C, the effect of the first dredge-up is to decrease the surface C/N and $^{12}$C/$^{13}$C ratios. 
\cite{charbonnel94} has shown that if starting with $^{12}$C/$^{13}$C $\simeq 65$ and $^{12}$C/$^{14}$N $\simeq 3.7$ (equivalent to [C/N $=0$]) in a 1~$M_{\odot}$ model at $Z=10^{-3}$, the first dredge-up decreases $^{12}$C/$^{13}$C and $^{12}$C/$^{14}$N to about 25 and 2 (equivalent to [C/N] $\sim -0.3$) respectively.
%Observation of stars with $-2<$ [Fe/H] $<-1$ that just experienced the first dredge-up have abundances of light elements in agreement with predictions from classical evolutionary models \citep{gratton00}. 
%%%%\cite{placco14c} studied the effect of 1st dredge-up in the carbon surface abundance of 505 CEMP. They found that for evolved stars, the 1st dredge-up decreases the [C/Fe] by about 0.5 dex. In particular, they found a maximum decrease of 0.8 dex.
In the top panels of Fig.~\ref{thermohaline} the surface $^{12}$C/$^{13}$C and [C/N] ratios along the evolution of a standard 
%(no rotation, no thermohaline mixing, see next point) 
low metallicity 0.85~$M_{\odot}$ model are shown. The first dredge-up occurs at $\log (T_{\rm eff}) \sim 3.7$ and $\log g \sim 3$ and decreases the surface $^{12}$C/$^{13}$C from about 70 to 45. The [C/N] ratio is barely affected by the first dredge-up (top right panel). 
Dredge-up events generally cannot explain CEMP stars since (1) they tend to deplete the carbon while we look for the opposite and (2) some CEMP stars are still unevolved (Fig.~\ref{thermohaline}) meaning that they did not experienced any dredge-up episode.
Quantifying this process for each CEMP star is nevertheless important in order to correct the surface abundances of evolved stars and recover their initial surface abundances, that reflect more directly the abundances in their natal cloud. 
%For the peculiar case of the CEMP stars, as mentioned in Sect.~\ref{pecEMP}, 
This was done in \cite{placco14c} for the carbon abundance. Their CEMP stellar models suggest that the first dredge-up correction on the surface C abundance is about 0.5 dex (0.8 dex in the most extreme case). 

%\textcolor{red}{Copied from obs sect. Remove or include in previous par. A possible bias to these frequencies is that when stars become giants, the surface cools and becomes convective. Some material from inner layers is therefore brought up to the surface. This is the first dredge-up. The inner material brought to the surface could have experienced nuclear burning through the CN cycle that transforms $^{12}$C into $^{14}$N. Consequently, in such a material, N is enhanced and C depleted. After the 1st dredge-up, the surface [C/Fe] ratio of a star is likely reduced. ref to another section?}

\paragraph{Thermohaline mixing and rotation.} Thermohaline mixing occurs after the first dredge-up so that unevolved CEMP stars are not affected by this process (provided they do not accrete heavy material from a companion, cf. Sect.~\ref{cempsext}). 
%Thermohaline mixing can be triggered because of the reaction $^{3}$He($^{3}$He,2p)$^{4}$He that reduces the molecular weight in the H-burning shell, just below the convective envelope. 
Thermohaline mixing can occur in giant stars because of an inversion of the molecular weight around the top of the H-burning shell, just below the convective envelope. 
The negative $\nabla_\mu$ ($\mu$, the mean molecular weight, is growing outward) results in a mixing event, following the first dredge-up, and decreasing again the C/N and $^{12}$C/$^{13}$C ratios \citep[e.g.][]{charbonnel07, eggleton08}. 
Rotation transports the H-burning products outwards. It likely adds another mechanism to decrease the surface C/N and $^{12}$C/$^{13}$C ratios. 
The bottom panels of Fig.~\ref{thermohaline} show the surface $^{12}$C/$^{13}$C and [C/N] ratios during the evolution of a low metallicity 0.85~$M_{\odot}$ model including thermohaline mixing and rotation. 
In this case, the surface ratios are significantly reduced, particularly near the end of the evolution, where $^{12}$C/$^{13}$C $\sim 8$ and [C/N] $\sim -1.6$. According to these models, some of the evolved CEMP stars shown in Fig.~\ref{thermohaline} may have experienced an important modification of their surface C and N abundances \citep[cf. also the work of][]{stancliffe09}. 
%An important fraction of CEMP stars is still relatively unevolved and likely no mixing yet or a modest mixing. 
Among the unevolved CEMP stars, some have low $^{12}$C/$^{13}$C ratios, like HE~1029-0546 with $^{12}$C/$^{13}$C $=9$ \citep[triangle in Fig.~\ref{thermohaline},][]{hansen15}. Some other unevolved stars have low [C/N] ratios like CS~29528-041 with [C/N] $=-1.47$ \citep[square in Fig.~\ref{thermohaline},][]{sivarani06}. These low ratios likely cannot be explained by internal processes. Such ratios probably reflect some processes at work in external sources (e.g. previous massive stars).

%Mixing processes (e.g. thermohaline) should be taken into account for the evolved CEMP stars. 
%%%%%%%\cite{lagarde10} have shown that a [C/N] of a solar metallicity star goes from 0 to $-0.21$ (after the 1\textsuperscript{st}) to 
%Ideally, one would also need to consider this process when investigating, for instance, the frequency of CEMP among MP stars. %and provide extra energy (exothermic)

\paragraph{Atomic diffusion.} Such a process has the effect of separating the elements. It groups together the effects of gravitational settling, thermal diffusion and radiative acceleration \citep{michaud15}. 
\cite{richard02} predicted that atomic diffusion can alter the surface composition of metal-poor stars by $\sim 0.1 - 1$ dex. It depends on the chemical species considered, effective temperature and the evolutionary stage. For C, O, Na, Mg, Al and Si, the alteration does not exceed 0.5 dex, except in some hot models ($T_{\rm eff} \gtrsim$ 6300 K). \cite{richard02} have also shown that the impact of atomic diffusion becomes very small (about $\sim 0.1$ dex) if including an additional turbulence effect that is required to account for the chemical anomalies of some stars \citep[the AmFm stars,][]{richer00}.\\

Overall, although internal processes in CEMP stars can modify their surface abundances, these processes likely cannot account for the unevolved or relatively unevolved CEMP stars. For the most evolved CEMP stars, internal mixing processes might have erased the initial surface chemical composition (hence the signature of previous stellar generations) for some elements like C or N.
%These mixing processes make more difficult the link between the surface composition of evolved CEMP stars and the material ejected by their progenitors. (also dilution,...)
An important work would be to compute a grid of CEMP stellar models including various mixing processes in order to quantify the degree of mixing experienced by each observed evolved CEMP star. The aim would be to recover the initial surface abundances of all elements. %for the CEMP stars. 
%\citep[as in][that corrected the effect of the first dredge-up on the surface carbon abundance]{placco14c}. 
Such a correction would give the chemical composition of the CEMP natal cloud. It will establish a more direct link between the evolved CEMP stars and the ejecta of the previous generation of stars.

%Overall, when investigating the origin of CEMP stars, internal processes are important to consider since they can modify their surface abundances. 

%The question is then: were the nucleosynthetic signatures of the CEMP source stars erased by these internal processes? We have seen that, at least for the relatively unevolved CEMP stars, it should not be the case.
 
%To date however, theses processes likely cannot explain the CEMP star sample. especially those who are unevolved.

%\subsection{CEMP stars and external sources: AGB, massive stars, supernovae, neutron stars}\label{origcemp}
\section{The origin of CEMP stars: external sources}\label{origcemp}

The origin of the CEMP stars is closely linked to the origin of the elements. When investigating the origin of CEMP stars, the aim is generally to find what is the astrophysical source able to provide the carbon, nitrogen, s-, r-elements...

It really seems that external sources are needed to explain the peculiarities of CEMP stars. The external sources can be objects that lived before and enriched the cloud in which the CEMP star formed and/or companions stars that transferred material to the (future) CEMP star. 
%As discussed below, the first category is generally associated to CEMP-no stars and the second to CEMP-s stars. %Almost every scenario for the origin of CEMP stars can be categorized in the 
%Almost every scenario for the origin of CEMP stars can be categorized in th
%In this global frame, many different scenarios have emerged to explain CEMP stars. 
In this global frame, the existing scenarios for the origin of the different classes of CEMP stars are discussed below. %Almost every scenario for the origin of CEMP stars can be categorized in the 

%  depending on their abundances.  Depending on the class of the CEMP

%Here we mostly focus on CEMP stars that are intriguing objects because of their generally highly non-solar-like patterns. \textcolor{red}{The fact that analogs of CEMP stars are not found at higher metallicity suggests that (some of) the nucleosynthesis events in the early Universe were different. CEMP stars are then interesting if one wants to know more about the peculiarities of the early Universe.}

%Two kinds: event during the CEMP life (AGB), or before. AGB is link to previous section because it is a pollution event (but from an external source)

%\textcolor{red}{general idea first: CEMP formed with source star}

%\textcolor{red}{origin of CEMP-s, no, r... discuss quickly each process ? especially in agb. refer to last sec for weak s-process}

%\textcolor{red}{The origin of CEMP stars is closely related to the origin of elements in the universe. These objects are unique observables to test prediction for the origin of} 

\subsection{CEMP-s stars}\label{cempsext}

%\textcolor{red}{schema formation CEMP-s?}

   \begin{figure*}[t]
   \centering
      \includegraphics[scale=0.35, trim = 3cm 0cm 0cm 0cm]{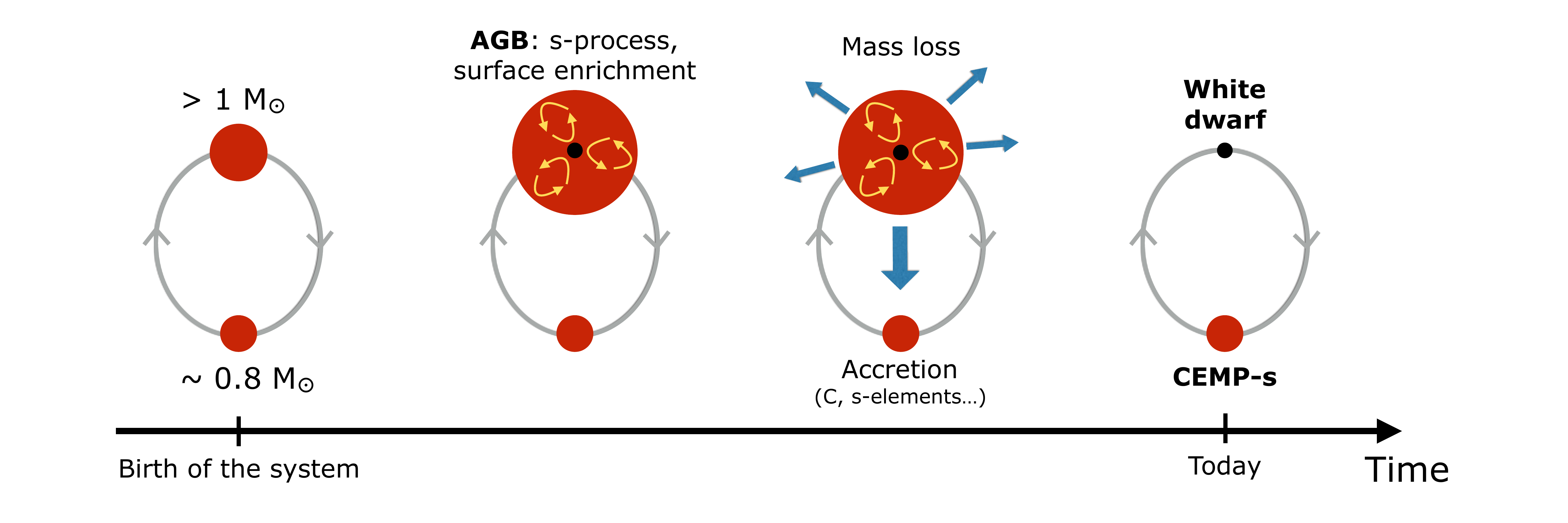}
   \caption[Schematic view of the binary scenario for CEMP-s stars]{Schematic view of the binary scenario to explain the peculiar abundance pattern of CEMP-s stars. In a binary system composed of a $0.8$~$M_{\odot}$ and a $>1$~$M_{\odot}$ stars, the more massive star goes through the Asymptotic Giant Branch (AGB) phase and loses mass. This lost material is enriched, among other, in carbon and s-elements. Some of this material is accreted by the secondary, that becomes a CEMP-s star. The primary ends its life as a white dwarf (see text for more details).}
\label{cempsform}
    \end{figure*}

%Maybe in another place.. BEFORE OR AFTER

%In the solar system, about half of the nuclei between Fe and Bi are produced by the slow neutron capture process \citep[s-process,][]{burbidge57,cameron57}. Most of the remainder are produced by the rapid neutron capture process \citep[r-process][]{burbidge57,cameron57,becker59,seeger65}. There is also a class of $\sim 35$ neutron deficient stable isotopes that cannot be produced neither by s-process nor by r-process. They are believed to be produced by the proton process \citep[p-process][]{burbidge57,ito61,audouze75, arnould76,woosley78}, in explosive events .

%The detection of the unstable element Tc ($T_{1/2} \sim$ Myr) at the surface of an AGB star (lifetime $\gg$ Myr) was the proof of the production of heavy element during the life of stars \citep{merrill52}. The AGB phase (occurring for stars with $M_{\rm ini} \lesssim 8$~$M_{\odot}$) was identified to produce a material enriched in s-elements \cite{iben82}.\\

Near the end of their lives, stars with initial masses between about 1 and 8~$M_{\odot}$ evolve to the Asymptotic Giant Branch (AGB). AGB stars are known to be at the origin of the \textit{main} s-process component (cf. Sect.~\ref{sprocagb}) and also to produce light elements such as carbon. Are CEMP-s AGB stars?
Actually no: many CEMP-s stars are still rather unevolved and therefore do not have gone through the AGB phase, where s-elements can be produced. 
%%%%The s-elements in CEMP-s stars likely come from an external source. 
%However, if the (future) CEMP-s star formed in a binary system where its companion was slightly more massive (e.g. 1.5~$M_{\odot}$), this companion may experience the AGB phase, produce and finally transfer (among others) the s-elements to the primary, that becomes a CEMP-s star.
The main scenario (see Fig.~\ref{cempsform} for a schematic view) considers that CEMP-s were born as $\sim 0.8$~$M_{\odot}$ normal stars (i.e. not C-rich, not s-rich) in a binary system with a more massive companion (e.g. 1.5~$M_{\odot}$). During its AGB stage, the companion produces and ejects through winds a material enriched in light and s-elements. 
%The wind material overfills the Roche-lobe\footnote{The Roche-lobe is a distinctively shaped region that surrounds a star in a binary system. In this region, the material is gravitationally bound to the star. The material from the primary that exceeds the Roche-lobe radius is no longer gravitationally bound to the primary and can flow toward the secondary.} and some of this material is accreted by the companion, that becomes a CEMP-s star. The AGB star ends its life as a white dwarf.
Some of the wind material is accreted by the companion, that becomes a CEMP-s star. The AGB star ends its life as a white dwarf.

In the scenario described above, the mechanism for the transfer of mass is wind mass transfer. %\citep[or wind Roche-lobe overflow, WRLOF,][]{mohamed07}. 
The transfer of mass may indeed preferentially occur through wind accretion rather than through Roche-lobe\footnote{The Roche-lobe is a distinctively shaped region that surrounds a star in a binary system. In this region, the material is gravitationally bound to the star. If the primary exceeds the Roche-lobe radius, some material can flow towards the secondary.} overflow because when the donor is an AGB star, Roche-lobe overflow may be unstable \citep{paczynski65} and brings the system to the common envelope phase. In this phase, 3D hydrodynamic simulations suggest that accretion is very inefficient \citep{ricker08}, meaning that the formation of CEMP-s stars might not happen.
%The fact that the transfer of material should occur through wind accretion and not through a standard mass transfer episode (where the stellar surface itself fills the Roche-lobe) is motivated by the fact that when the donor is an AGB star, standard mass transfer may be unstable \citep{paczynski65} and bring the system to the common envelope phase. In this phase, 3D hydrodynamic simulations suggest that accretion is very inefficient \citep{ricker08}. 
%%%The wind mass transfer \citep[or wind Roche-lobe overflow, WRLOF,][]{mohamed07} scenario for CEMP-s stars was considered in e.g. \cite{abate13,abate15b,abate15a}.
By investigating the origin of 15 CEMP-s stars with known orbital periods, \cite{abate15a} have shown that the wind mass transfer model generally agrees well with observations but in most of the cases, the wind accretion rate should be enhanced by a factor of $5-10$ to account for the observations.

Atomic diffusion and thermohaline mixing on CEMP-s stars are important ingredients that may induce important surface composition changes \citep[e.g.][]{stancliffe07}. The accreting CEMP-s star undergoes thermohaline mixing because the material coming from the AGB companion has a larger mean molecular weight than the material at the surface of the accreting future CEMP-s star\footnote{In this case, thermohaline mixing can occur during the main sequence.}. After the accretion, thermohaline can reduce the surface carbon abundance in the CEMP-s by $\sim$1 dex \citep{stancliffe07}. A too strong thermohaline 
%or atomic diffusion 
would make the s-elements sink and disappear from the stellar surface. 
However, if CEMP-s stars rotate, the horizontal turbulence induced by rotation may reduce or even kill the thermohaline mixing \citep{denissenkov08, maeder13}. 
%Interestingly, it was suggested that horizontal turbulence operating , the thermohaline mixing may be reduced or even killed\cite{maeder13}, \cite{maeder13} have investigated the interactions of various instabilities in rotating stars: if the horizontal turbulence is strong, the thermohaline mixing may be reduced or even killed (their Eq.~14).
Models of rotating CEMP-s stars computed by \cite{matrozis17b} supports this view by showing that even a mild rotational mixing severely inhibits thermohaline mixing (also atomic diffusion is inhibited).

%A possible difficulty however is that 
Worth to mention is also that if a substantial amount of material (hence angular momentum) is accreted from the AGB companion, the CEMP-s star should spin up significantly. \cite{matrozis17a} have shown that $\sim 0.05$~$M_{\odot}$ can be accreted before the CEMP-s reaches critical rotation and therefore stops accreting more material. However, more than $0.1$~$M_{\odot}$ is often needed to reproduce the chemical composition of observed CEMP-s stars \citep[e.g.][]{abate15b,abate15a}. 
%This may suggest that an additional process allowing significant angular momentum loss is required.
This said, a generally good agreement is found between the yields of AGB stars and chemical composition of CEMP-s stars \citep[with nevertheless several persisting discrepancies, especially for C, N, F, Na, and $^{12}$C/$^{13}$C,][]{lau09,stancliffe10,bisterzo11,bisterzo12,lugaro12}. 

The AGB binary scenario is also well supported by the fact that a large fraction of CEMP-s stars is in a binary system (cf. Sect.~\ref{arebinaries}).
The apparently single CEMP-s stars 
%%%%\citep[4 out of 22 in the sample of][]{hansen16a} 
nevertheless challenge this scenario, even if one can imagine that the single CEMP-s stars have lost their AGB companion or that they are in a binary system with a very long period, explaining the non-detection of radial velocity variation. %In case of very wide orbits, one may wonder whether enough mass from the AGB donor could have been accreted by the CEMP-s star.
%On the other hand, it is also not excluded that some CEMP-s stars in binary systems do not contain the nucleosynthetic signature of their AGB companion since AGB stars may fail the third dredge-up, as it was observed recently \citep{kamath17}. In this case, the abundance pattern of the CEMP-s star should come from the previous generation(s) of stars. 
%%%%On the other hand, it is possible that single CEMP-s stars were formed through the AGB channel since (1) single CEMP-s stars might have lost their companion or (2) they might be in a binary system with very long period, explaining the non-detection of radial velocity variation. 

In general, even if CEMP-s stars are explained with the AGB scenario, it does not mean that the previous generations of massive stars did not contribute. Likely, it means that the nucleosynthetic signature let by the AGB companion on the CEMP-s star is the dominant one and has blurred the nucleosynthetic signature let by the previous generation of stars. In Chapter~\ref{cemps}, I investigate whether the nucleosynthetic imprint of previous massive stars can be identified in CEMP stars enriched in s-elements.
%Massive stars ($M_{\rm ini} \gtrsim 10$~$M_{\odot}$) also produce s-elements (cf. Sect.~\ref{sprocmassive})

%   \begin{figure*}[t]
%   \centering
%      \includegraphics[scale=0.27, trim = 0cm 0cm 0cm 0cm]{wrlof.pdf}
%%% %  \caption[Schematic view of the wind mass transfer mechanism for CEMP-s stars.]{Schematic view of the wind mass transfer mechanism for CEMP-s stars. $R_{\star}$ is the AGB star radius. $R_{\rm w}$, defines the front of the wind (or wind radius). $R_{\rm L1}$ is the Roche-lobe radius. The wind flowing through the lagrangian point $L_1$ can be accreted by the (future) CEMP-s star.}
   %\caption[Schematic view of the wind mass transfer mechanism for CEMP-s stars.]{Schematic view of the wind mass transfer mechanism for CEMP-s stars. The dashed lines represents the front of the AGB star wind. The wind flowing through the lagrangian point $L_1$ can be accreted by the (future) CEMP-s star (adapted from C. Abate, PhD thesis, 2014).}
%\label{wrlof}
 %  \end{figure*}

%\cite{kamath17}: AGB fail dUP. Single CEMP-s... Hansen

%AGB models were compared to observed CEMP-s stars by different authors \citep[e.g.][]{stancliffe08, lau09,bisterzo12,lugaro12,abate13,abate15b,abate15a,hollek15}. A generally good agreement is found between yields of AGB and chemical composition of CEMP-s stars, 

%proof of s-process in AGB (especially Tc) : \cite{merrill52}, \cite{smith90}

\subsection{CEMP-r, -r+s, -r/s, -i stars}

The origin of CEMP-r stars is tightly linked to the astrophysical site for the r-process, which is still unknown. %The r-process operates
The two main possible r-process sites are magnetorotationally driven supernovae \citep[MRSNe, e.g.][]{thielemann11, winteler12, nishimura15} and neutron star (NS) merger \citep[e.g.][]{symbalisty82, freiburghaus99, thielemann11, thielemann17, wanajo14}. Recently, LIGO detected GW170817, the first observed binary neutron star inspiral \citep{abbott17}. The light curve of the electromagnetic counterpart was found to be compatible with a NS merging event producing r-elements \citep{smartt17,pian17}. 

NS mergers alone may nevertheless face difficulties to explain the most iron-poor CEMP-r stars \citep[e.g. SM 0248$-$6843 with [Fe/H\text{]} $=-3.71$ and [Eu/Fe\text{]}$=1$,][]{jacobson15}, as shown by inhomogeneous chemical evolution models \citep{argast04,wehmeyer15}. As summarized in the review of \cite{thielemann17}, one important reason is that the merging of two NS requires two prior supernova events, that can already lead to a substantial floor of Fe. This floor may be too high compared to the CEMP-r stars with the lowest [Fe/H] ratios. 
This issue may nevertheless be solved if the two NS have undergone kicks at the time of the supernova (SN) explosion: in this case, they were removed from the SN debris and moved into a medium with low Fe pollution.
%an and (2) some time before merging and ejecting r-elements. 
%Both the Fe ejected by the two prior SN and the possible additional Fe coming from other sources during the inspiral time could significantly raise [Fe/H] and therefore set unfavorable conditions for producing very iron-poor CEMP-r stars. 
With chemical evolution models, \cite{wehmeyer15} have also shown that if the NS-NS system takes too long to coalesce, further nucleosynthesis events can occur and enrich the interstellar medium (ISM) in Fe. When the NS-NS system finally merges, the surrounding ISM has a higher [Fe/H] than at the birth of the NS-NS system. In particular, they show that considering NS merger alone in their models cannot explain the [Eu/Fe] ratios at [Fe/H] $<-2.5$. %This implies an overall Eu production shift towards higher metallicities.
Considering also MRSNe as sources of r-elements can provide a solution to account for the enrichment in Eu at [Fe/H] $<-2.5$ \citep{cescutti14,wehmeyer15}.

%Other stars seems to show an intermediate neutron i-process, meaning that neither a s-process pattern, nor a r-process pattern and nor a combination of both can fit the abundances \citep{roederer16}. 
Some other stars have abundance patterns that neither a s-process pattern, nor a r-process pattern and nor a combination of both can fit the abundances. 
In such cases, an intermediate neutron capture process \citep[i-process, first named by][]{cowan77}, operating at neutron densities in between the s- and r-process may be the solution \citep[e.g.][]{lugaro12, roederer16, hampel16}. The astrophysical sites for the i-process are unknown. Such a process may operate in AGB or super-AGB stars \citep{herwig11,jones16b}, in accreting white dwarfs in close binary systems \citep{denissenkov17} or in the He burning shell of Pop~III massive stars \citep{clarkson18}. The s-process boost induced by rotation in the massive stars models of the present work (cf. Chapter~\ref{cemps}) does not meet the required conditions (especially neutron density) for the i-process to occur.

Finally, some other stars seems to show an r+s pattern \citep{jonsell06,gull18}. \cite{gull18} reported the discovery of a CEMP star with [Fe/H] $=-2.2$ whose best fit is achieved by a combination of s- and r-process pattern. In this case, the CEMP star would have formed from at least two sources, one experiencing the s-process, the other experiencing the r-process. 

%\cite{cowan77}, 
%\cite{herwig11}, \cite{clarkson18}, \cite{vanhoof17}, \cite{roederer16}, 
%Hampel,% \cite{banerjee17}

%\cite{lugaro12}, i-process

   \begin{figure*}[t]
   \centering
      \includegraphics[scale=0.31, trim = 0cm 0cm 0cm 0cm]{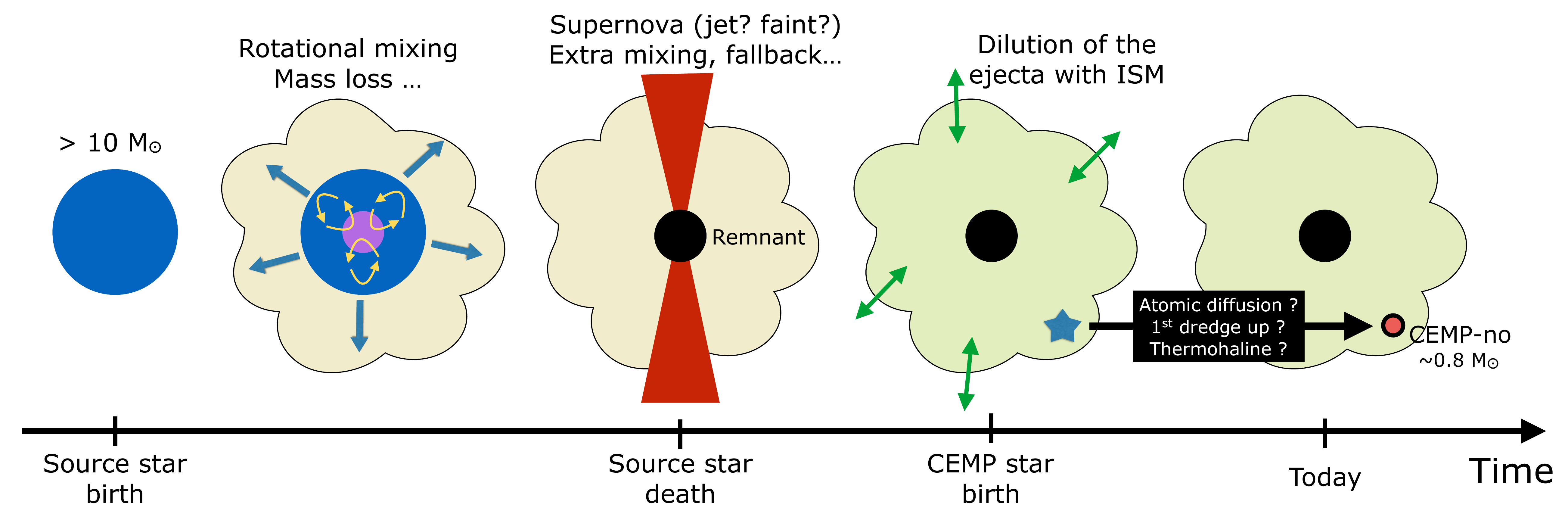}
   \caption[Schematic view of a possible formation channel for CEMP-no stars]{Schematic view of a possible formation channel for CEMP-no stars. What is shown are more some of the interesting ingredients to explain CEMP-no stars rather than an exact scenario. A previous massive source star (Pop~III or low metallicity) is at the origin of the material used to form the CEMP-no star (a combination of more than one source is also possible). The massive star can experience rotational mixing, mass loss and a supernova (that might be asymmetric because of the fast rotation of the core). The material ejected through winds and/or supernova can be diluted with the ISM and finally be used to form a CEMP-no star. From its birth to now, the CEMP-no star may have experienced internal mixing processes, blurring the nucleosynthetic signature of the previous massive star (cf. text for more details).}
\label{cempnoform}
    \end{figure*}

\subsection{CEMP-no stars}\label{cempnoscenar}

%\textcolor{red}{main diff. between CEMP(-no) and MP : reservoir (GCE) or 1 star (st. models)}

%\textcolor{red}{more in origin?: yoon beers, 3 groups? Gen Chiaki. Also separation of CEMP-no / s with A(C)}

At the present day, no CEMP-s, -r, -r/s stars were found below [Fe/H] $\lesssim-3.5$. %There are either CEMP-no or CEMP. 
Below this threshold, all the CEMP stars are either CEMP-no or just CEMP in case no Ba (and Eu) abundance is available to classify them. %as CEMP-s, -r/s or -no. 
%%%CEMP-no stars dominate at very low [Fe/H]. 
%A substantial fraction of the most iron-poor with [Fe/H] $<-4$ stars are however just CEMP 
%One might reasonably guess that only CEMP-no would be found at [Fe/H] $\lesssim -4$ since (1) at the present day, no CEMP-s, -r, -r/s stars were found below [Fe/H] $\lesssim-3.5$ and (2) some iron is needed to synthesized s- and r- elements. 
For convenience, I use the term CEMP-no also for the most iron-poor stars whose rigorous classification is just CEMP.

While interesting in the case of CEMP-s stars, the AGB binary scenario applied to CEMP-no stars \citep{suda04, suda10, campbell10} faces difficulties to account for the observables.
%While the mass transfer scenario (Fig.~\ref{cempsform}) seems very interesting in the case of the CEMP-s stars, it faces more difficulties for the CEMP-no stars because most CEMP-no stars appear to be single \citep{hansen16b} and because at very low [Fe/H], AGB may not had time to contribute to the chemical enrichment yet.
One reason is that most CEMP-no stars appear to be single stars \citep{hansen16b}. 
The self-enrichment scenario \citep{fujimoto00} seems unlikely, especially to explain the unevolved stars (cf. Sect.~\ref{selfenr}). 
\cite{venn08} proposed that the high C and low Fe may be explained by the separation of gas and dust beyond the CEMP star surface, followed by the accretion of dust-depleted gas. In this case, a dusty disk should exist around CEMP stars, whose presence would be betrayed by a mid infrared excess. Later, \cite{venn14} have shown that no mid infrared excess was found for six out of the seven stars they examined (the exception is for HE~0107-5240, that shows a small excess). Also, this dust-gas separation process should affect in a similar way the different isotopes of a given element. The carbon isotopic ratio for instance, will not be modified by this process. The very low $^{12}$C/$^{13}$C ratios on some CEMP-no stars need to be explained by another process.

In the end, this let us with the interesting possibility that CEMP-no stars are the rather direct daughters of a previous generation of stars. These previous stars should be massive since at the very low metallicity considered here, low or intermediate mass stars may not had time to contribute to the chemical enrichment yet. Also, these previous massive stars should have contributed little in iron.

%Contrary to MP-no stars, 
The fact that the CEMP-no class of star shows very large abundances scatter suggests that they formed with the ejecta of only one or a few previous stars. If they were formed from a well mixed reservoir of many previous stars, the scatter should be much smaller, as is it observed at higher metallicities or in normal metal-poor stars. %, for instance. 
%%%Possibly, 
%This is a possible major difference between CEMP-no and MP-no stars: while CEMP-no formed from $\sim 1$ source star, MP-no formed from $\gg 1$ source stars.
%The big question is to determine what characteristics the source star should have to explain the abundances of CEMP-no stars.
%%%Possibly, one major difference between CEMP-no and MP-no is that CEMP-no formed with only the ejecta of $\sim 1$ massive stars while MP-no stars formed in the reservoir that was already well enriched by $\gg 1$ previous stars. 
%%%Generally, all stars at [Fe/H] $\lesssim -4$ are believed to be CEMP-no stars, even if for a substantial fraction of them, no Ba abundance is available. 
%%%%The majorities of works investigating the origin of CEMP-no stars
Over the past $\sim 15$ years, multiple scenarios, aiming at characterizing the nature of the CEMP-no source stars, emerged: %Probably the two main ones are 'mixing and fallback + faint SN' and 'spinstars'. 

\begin{itemize}
\item Mixing and fallback + faint SN \citep{umeda02, umeda03, umeda05, iwamoto05, tominaga07b, heger10, tominaga14, ishigaki14}.
\item Very low metallicity stars experiencing fast rotation \citep[spinstars,][]{meynet05iau, meynet06, meynet10, hirschi10, maeder15a, maeder15b, choplin16, choplin17a}.
\item Normal SN from 15~$M_{\odot}$ + faint SN with strong fallback from 35~$M_{\odot}$ \citep{limongi03}.
\item Jet-induced SN from $25-40$~$M_{\odot}$ Pop~III stars \citep{maeda03, tominaga09}.
\item Rotation + faint SN \citep[no fallback,][]{takahashi14}.
\item $15-40$~$M_{\odot}$ rotating Pop~III stars with mixing and fallback \citep{joggerst10}.
%\item rotation + late mixing + faint SN \citep{choplin17a}.
\item H-ingestion event in the He burning shell of a 45~$M_{\odot}$ Pop~III star \citep{clarkson18}.
%\item mass transfer from a companion \citep{suda04, suda10, campbell10},
%\item Self-enrichment \citep{fujimoto00},
%\item Dust-gas around CEMP-no stars \citep{venn08}.
\end{itemize}

This is a tentative list. Some scenario of the list above may be grouped together. Also, for a given scenario, some works occasionally consider slightly different methods or physical ingredients, so that they may be considered as separated scenarios. 

%%%%%%%The variety of existing scenario suggests that the CEMP-no stars have multiple origins. It may reflect that a variety of events occurred in the early generations of massive stars. 
Some works suggest that more than one class of source star is required to explain the CEMP-no stars.
For instance, the A(C) $-$ [Fe/H] diagram (where A(C) = $\log (N_{\rm C}/N_{\rm H})$ + 12) suggests that the [Fe/H] $\lesssim -4$ stars separate in at least two groups \citep{yoon16}, so that at least two different classes of progenitor might be needed. Also, comparisons of the chemical composition of CEMP-no stars with ejecta of massive source stars often show that only one class of progenitor faces difficulty to reproduce large sample of CEMP-no stars. \cite{placco16} have used the yields of non-rotating Pop~III stars of \cite{heger10} and shown that 5 out of the 12 investigated CEMP-no stars can be reproduced by the predicted source star yields while 7 stars are let without an acceptable source star.

The scenarios listed above all consider that (1) the surface abundances of CEMP-no stars reflect rather directly the abundances in their natal clouds and (2) the natal cloud was polluted by one or very few zero or very low metallicity massive source star (schematic view in Fig.~\ref{cempnoform}). 
Each scenario has nevertheless its own specificity. %What differs is the behavior of the source star.
Among the quoted works in the list above, frequent ingredients considered in source star models are rotation, extra mixing events \citep[in the sense of][see details below]{umeda02} and strong fallback. Occasionally, both are considered. Occasionally, refinements or additional ingredients were proposed. 
Below are discussed two scenarios in more detail: mixing \& fallback and fast rotating massive stars.

%   \begin{figure*}[t]
 %  \centering
 %     \includegraphics[scale=0.54, trim = 0cm 0cm 0cm 0cm]{znfe.png}
  % \caption[a]{ \citep[Figure from][]{cayrel04} Explain mix fall in appendix?}
%\label{znfe}
 %   \end{figure*}

   \begin{figure*}[t]
   \centering
      \includegraphics[scale=0.56, trim = 0cm 0cm 0cm 0cm]{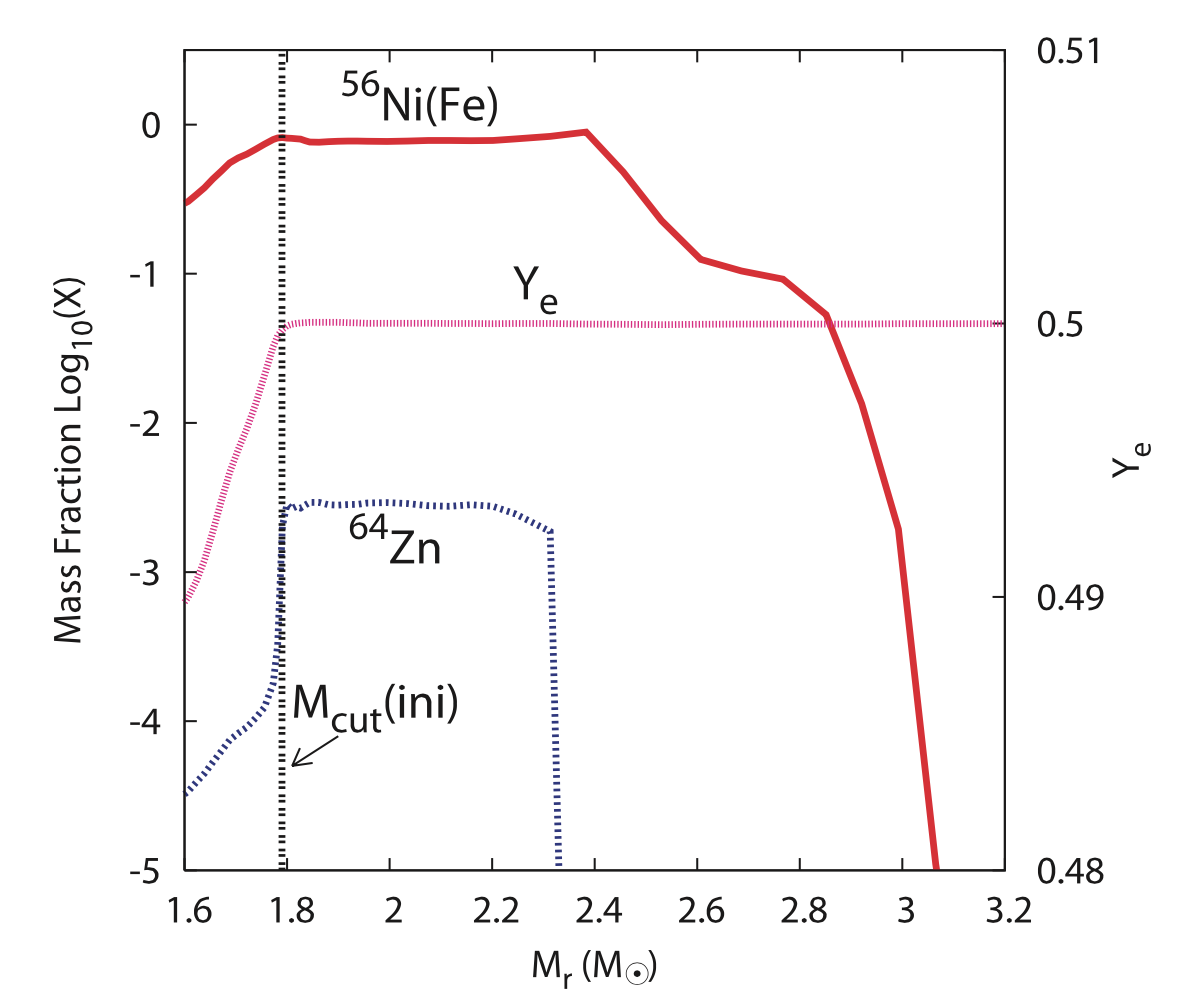}
   \caption[Inner abundances of a Pop~III 25~$M_{\odot}$ after explosive nucleosynthesis]{Inner abundances of a Pop~III 25~$M_{\odot}$ after explosive nucleosynthesis. $Y_{\rm e}$ is the electron fraction \citep[figure from][]{tominaga07b}}
\label{mixfall}
    \end{figure*}

\subsubsection{Mixing and fallback}

%The \textit{mixing \& fallback} scenario was originally developed by \cite{umeda02} to explain the peculiarities (especially the rising [Zn/Fe] at low [Fe/H]) of normal metal-poor stars. 
%%%%(\textcolor{red}{see Appendix XX for more details? or put here?}). 
%%%%%Originally the mixing and fallback scenario was developed by \cite{umeda02} to explain some peculiarities in the abundances of metal-poor stars. 
%From [Fe/H] $=-2$ to [Fe/H] $=-4$, 

\paragraph{Mixing.} The [Zn/Fe] ratio of normal metal-poor stars was found to rise with decreasing [Fe/H] \citep[][their figure 12]{cayrel04}. Models of massive stars including explosive nucleosynthesis show that at the end of the evolution, Zn is located in deeper regions than Fe (Fig.~\ref{mixfall}). Consequently, if some Zn is ejected from the massive star (through a supernova event), also a large amount of Fe is ejected and the trend of [Zn/Fe] observed at low [Fe/H] cannot be explained. In order to expel some Zn without expelling too much Fe, \cite{umeda02} have included an extra mixing event in the source star models. 
%the \textit{mixing \& fallback} scenario. 
The mixing event occurs at the time of the massive star explosion and brings material from inner layers closer to the surface. 
% (see the left panel of Fig.~\ref{mixfall2} and explanation in the legend). 
This allows to bring some Zn upward, without having too much Fe. When considering such a mixing, the supernova yields can reproduce the observed [Zn/Fe] ratios at low [Fe/H].

   \begin{figure*}[t]
   \centering
      \includegraphics[scale=0.46, trim = 3cm 13cm 3cm 0cm]{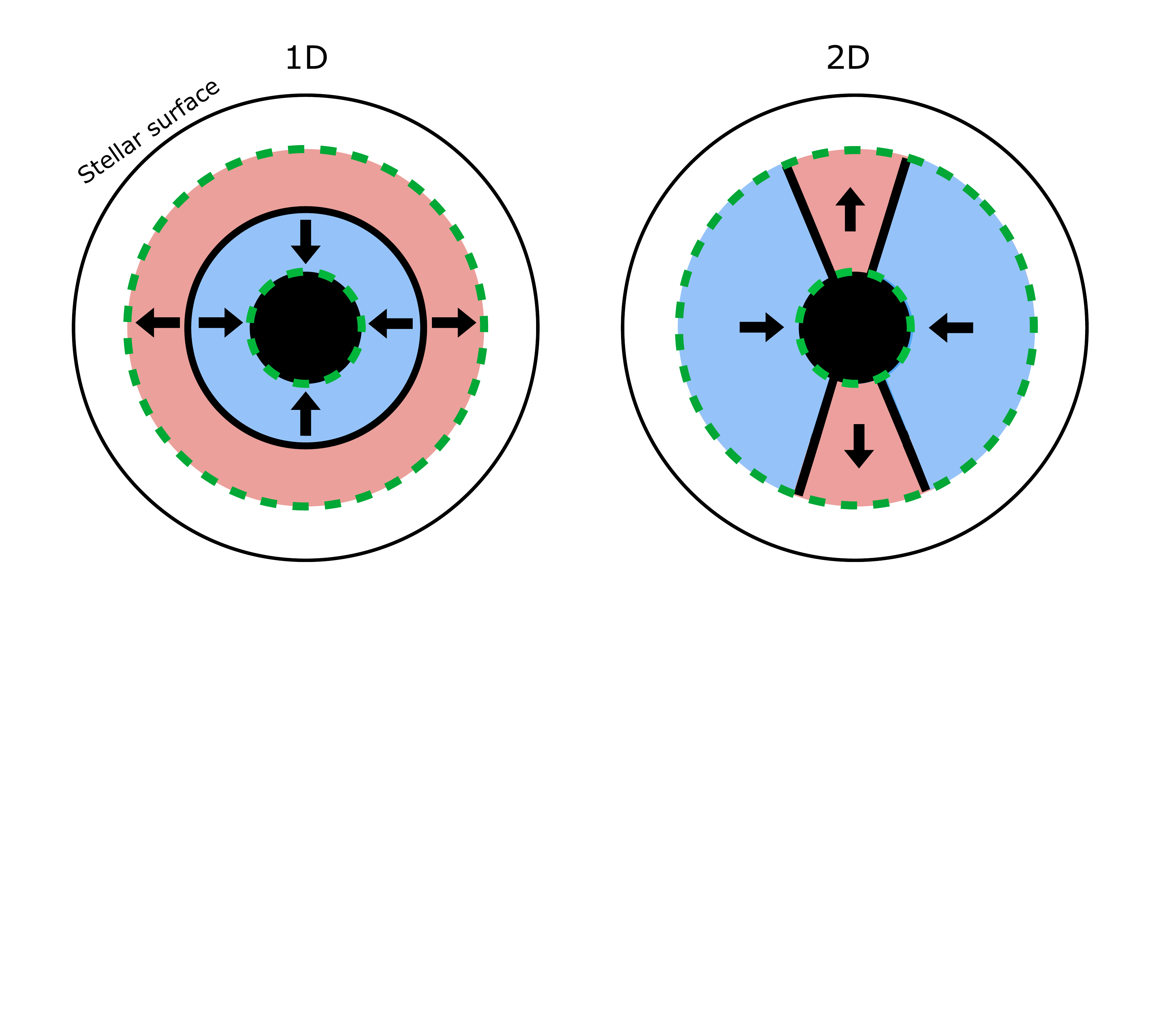}
   \caption[Schematic view of the mixing and fallback model]{Schematic view of the 1D mixing and fallback model (left) that mimics the 2D jet SN model (right). At the end of the massive star evolution, a region, delimited here by the 2 green dashed circles, is mixed. A part of this region is expelled (red), the other part (blue) falls onto the central remnant. In the 2D jet model, the material is ejected along the jet axis \citep[figure adapted from][]{tominaga07b}. }
\label{mixfall2}
    \end{figure*}

%Some minor differences between 1D and 2D models nevertheless exist: 2D models gives more Co and Zn for instance.
%Fig.~\ref{mixfall2} shows 

\paragraph{Fallback.} For stars having a large [C/Fe] and low [Fe/H] (CEMP stars), the concept of faint supernova with strong fallback was also introduced \citep{nomoto03, umeda03}. The idea of a strong fallback comes from the fact that in the massive source star, carbon is located closer to the surface than iron. Therefore, in order to have a large [C/Fe] ratio in the ejecta, not too deep layers should be expelled so as to eject mostly carbon and little iron. 
The supernova is faint because very little $^{56}$Ni (whose decay into $^{56}$Fe is mainly responsible for the luminosity of the supernova) is ejected. If too much $^{56}$Ni is ejected, one will ultimately get too much $^{56}$Fe compared to the low [Fe/H] ratios of the most iron poor stars.
The concept of strong fallback and faint supernova for the source star is somewhat consistent with the fact that low-metallicity stars are more compact and hence might explode less easily.\\

The mixing \& fallback scenario is a combination of both events. 
%considers the massive star at the pre-supernova stage. 
To fit the observations with a massive source star experiencing mixing and fallback, a massive source star model is evolved until the end of its evolution. Then a region inside the model, between two limiting shells (the two green dashed circles in Fig.~\ref{mixfall2}), is considered to be fully mixed at the time of the supernova explosion. A part of the mixed region is ejected (red part in Fig.~\ref{mixfall2}), a part is kept into the remnant (blue part). The extension of the mixed region and the fraction of this region which is expelled are chosen so as to reproduce the abundances of observed metal-poor stars.
%have shown that 2D jet-induced explosion of a 40~$M_{\odot}$ Pop~III star leads to a material having a chemical composition close to the material  

Although the source star models with mixing and fallback are generally 1D models, \cite{tominaga09} has shown that for a Pop~III stellar model of 40~$M_{\odot}$, the 1D mixing and fallback model gives similar chemical yields than a 2D jet-induced explosion with fallback (Fig.~\ref{mixfall3}). 
It can be roughly understood by the fact that the mixed region is ejected differently in 1D (spherically) and 2D (along the jet axis) but has a similar chemical composition. 
Nevertheless, some chemical species are overproduced in the 2D jet SN model (e.g. Sc, see Fig.~\ref{mixfall3}). This is due to the higher energy concentration in the jet-induced explosion, that leads to a bit different explosive nucleosynthesis. 
Overall, it suggests that the spherical mixing and fallback model is rather well mimicking a 2D jet-induced explosion.

The mixing and fallback scenario for CEMP-no stars has the advantage to well explain the iron-peak elements like Zn or Fe. One caveat is that it faces difficulties in reproducing the high N abundance of some metal-poor stars \citep{tominaga07b} except if additional mixing is considered between the He-convective shell and the H-rich envelope just before the supernova \citep{iwamoto05,tominaga14}.

   \begin{figure*}[h]
   \centering
      \includegraphics[scale=0.6, trim = 8cm 18cm 8cm 0cm]{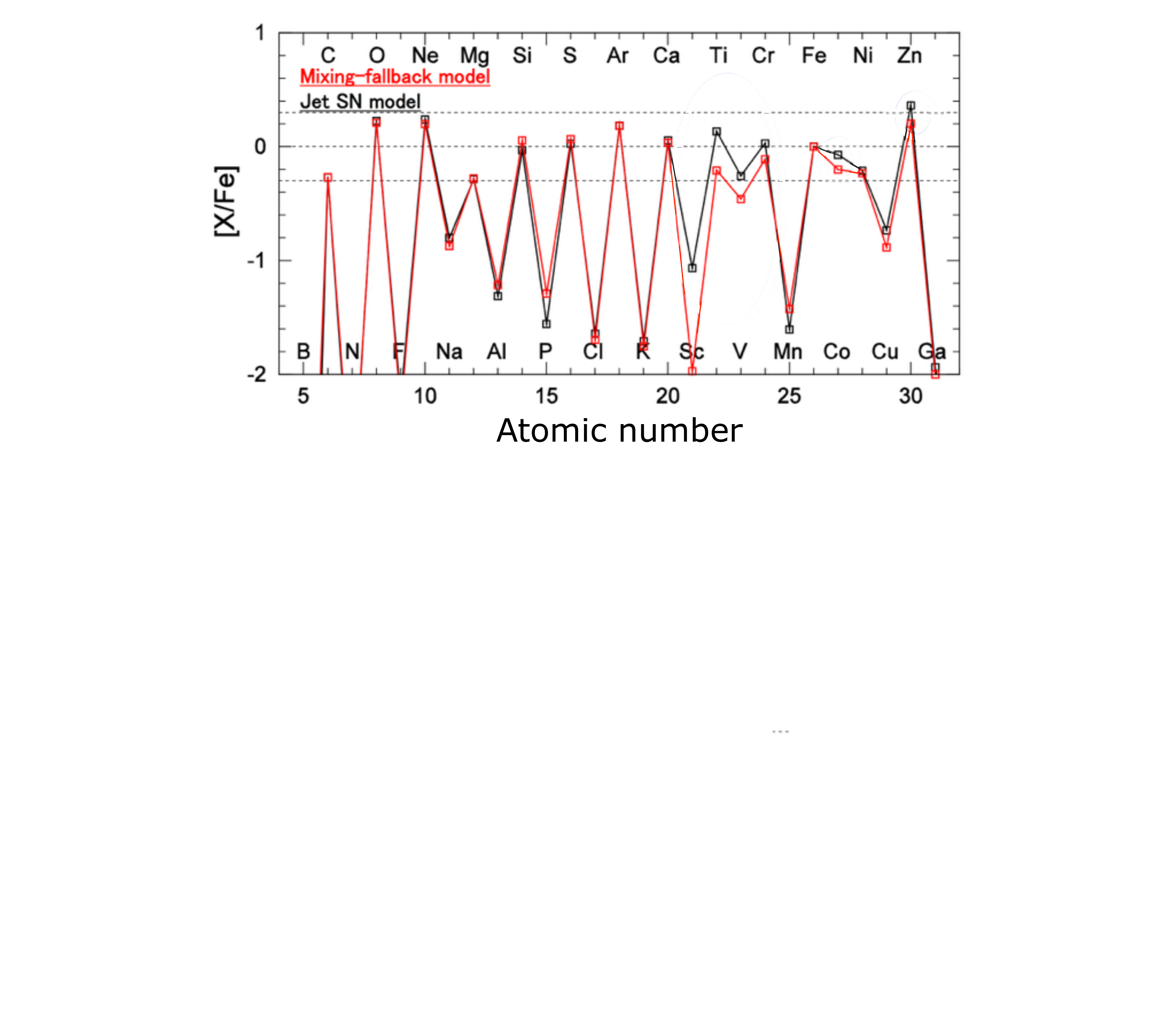}
   \caption[Post-supernova yields of a non rotating Pop~III 40~$M_{\odot}$ model]{Post-supernova yields of a non rotating Pop~III 40~$M_{\odot}$ model. The red pattern stands for the 1D mixing and fallback model, the black pattern results from a 2D jet-SN model. Figure adapted from a slide of K. Nomoto (\href{http://www.astro.rug.nl/~archeocosmic/PDF\_TALKS/Nomoto.pdf}{http://www.astro.rug.nl/~archeocosmic/PDF\_TALKS/Nomoto.pdf}). See also the Fig.~6 of \cite{tominaga09}. }
\label{mixfall3}
    \end{figure*}

\subsubsection{Fast rotating massive stars}

%Mettre ca dans CHAP 1 ou 3

%At low metallicity, stars are expected to rotate more rapidly and the effects of rotation are likely stronger (cf. Sect.~\ref{sterot} and \ref{massivemodels}).

In a broad context, the fast rotating massive star scenario or spinstar\footnote{The present work focuses on massive stars but, originally, the term spinstar also designates intermediate mass stars.} scenario proposes that fast rotation at low metallicity is a dominant process that largely impacts the outputs of massive stars (especially the stellar yields).
It is motivated by several aspects (see also Sect.~\ref{sterot} and \ref{massivemodels}). One of the aspect, further discussed in Sect.~\ref{transportrot}, is that the efficiency of the rotational mixing operating inside the star is expected to increase with decreasing metallicity. This effect arises naturally, as a result of the physics of rotation in stars (discussed in Sect.~\ref{massivemodels}).
As a consequence of this efficient rotational mixing, the yields of some elements like C, N, O \citep[e.g.][]{meynet02b} or s-elements \citep[e.g.][]{pignatari08, frischknecht16} are impacted a lot by rotation in low metallicity massive stellar models.  %(Sect.~\ref{secback} for more details). 
%can be impacted a lot by rotation , particularly nitrogen \citep[e.g.][]{meynet02b} or s-elements \citep[e.g.][]{frischknecht16} .
%%%It is important to note that the same physics of rotation is used at solar and low metallicity. 
%The strong production of some elements like nitrogen is not artificial but appears naturally in low metallicity rotating massive stars. 
%The efficiency rotational mixing operating inside the star increases with initial rotation but is also increases with decreasing metallicity 
%The fast rotating massive star scenario or \textit{spinstar} scenario in the early universe appears naturally, as a consequence of the physics of rotation, whose effects are more important at low than at solar metallicity (see Sect.~\ref{sterot} and \ref{massivemodels}). 
The spinstar scenario tells us that the low-mass stars that formed early in the Universe should contain mostly (but possibly among others) the nucleosynthetic signatures of previous fast rotating massive stars. % that lived before. %The spinstar scenario proposes that this nucleosynthetic signature is among the dominant ones in the early universe. %Other signature from supernovae, 
%, but not the only one: for instance, non-rotating or slow rotating massive stars should also have existed.

%The efficiency of the rotational mixing operating inside the star is then expected to increase with decreasing metallicity 
%because of a less efficient redistribution of the angular momentum inside low metallicity stars 
%(cf. Sect.~\ref{transportrot}). 

Thanks to Galactic chemical evolution models, it was shown  that the N/O, C/O, $^{12}$C/$^{13}$C and Sr/Ba ratios of very metal-poor (mostly normal, i.e. not carbon-enhanced) halo stars are better reproduced if including the yields of fast rotating massive stars \citep{chiappini06,chiappini08,cescutti13}. The spinstar scenario has also been proposed to explain the abundance anomalies (particularly the Mg-Al anticorrelation) in Globular Clusters \citep{decressin07}.

%the low-mass stars formed from massive star ejecta should contain th since the early generation of massive stars were predominantly (not only) rot
%In the frame of this scenario, the abundances of the CEMP stars should contain the nucleosynthetic signature of these fast rotating massive stars. %This signature is probably mixed with the signature of non-rotating or slow rotating massive stars since we generally expect a distribution of velocity and not only fast rotators. What the spinstar scenario propose is that the signature of fas be Probably, not all early massive stars were fast rotators so CEMP stars should contain various 
When applied to CEMP stars, the spinstar scenario proposes that most CEMP stars have formed from the material ejected through winds and/or supernova by a previous fast rotating massive source star. 
\cite{meynet06} have shown that the [C/Fe], [N/Fe] and [O/Fe] ratios in the ejecta (wind + supernova) of a very low-metallicity fast rotating 60~$M_{\odot}$ are similar to the surface ratios of the unevolved CEMP stars G77-61 and HE~1327-2326 (with [Fe/H] $<-4$). In the same line, \cite{hirschi07} has shown that the CNO abundances of HE~1327-2326 can be explained by the wind of a low metallicity fast rotating 40~$M_{\odot}$ model, whose ejecta was diluted 600 times with ISM (i.e. 600 times more ISM material than stellar ejecta). \cite{meynet10} have predicted that CEMP stars formed from a fast rotating massive source star should be helium rich, provided the ejecta was little diluted with pristine ISM. \cite{maeder15a} and \cite{maeder15b} proposed that the back-and-forth, partial mixing at work between the He- and H-regions of the rotating source star may account for the wide variety of CEMP-no stars abundances (more details in Sect.~\ref{secback}).\\

%investigated in details the origin of 3 CEMP stars with [Fe/H] $<-4.5$, that were the 3 most iron-poor stars at that time.

%\citep[e.g.][]{meynet06, hirschi07, meynet10}.

%\textcolor{red}{A bit more details. Why my work is new?} 

%In the spinstar scenario for CEMP stars, the massive CEMP source star is considered to be a fast rotator that experienced a progressive mixing resulting from shear instabilities and meridional circulation. The CEMP star forms with this material, enriched in various elements, that was ejected (through winds and/or supernova) by the previous rotating source star \citep{meynet06, hirschi07, meynet10}.

It is important to make the difference between the two different kinds of mixing at work in the spinstar and mixing \& fallback scenarios: 
\begin{itemize}
\item Spinstar: the mixing is induced by rotation and occurs progressively during the evolution of the star.
\item Mixing \& fallback: the mixing arises only at the time of the supernova.
\end{itemize}
Contrary to the mixing \& fallback process, the process of rotation, which is at the heart of the spinstar scenario, was not introduced in stellar models so as to explain the abundances of MP or CEMP stars. As mentioned, spinstars are expected to exist at very low metallicity, as a result of the physics of rotation (more details in Sect.~\ref{massivemodels}).
One strength of the spinstar scenario is that it provides a natural physical process to explain the abundances of most elements in CEMP stars. 
Another strength is that spinstars can account for other observables like the abundances of normal halo stars (cf. previous discussion). Rotation may also be a key process to explain the variation of surface abundances observed in solar metallicity, nearby massive stars \citep[e.g.][]{martins15}. 
Rotation is linked to different observables, one of which could be CEMP stars.
%In the end, it is worth keeping in mind that originally, the process of rotation, at the heart of the spinstar scenario, was not introduced in stellar models so as to explain the abundances of MP or CEMP stars. }

The spinstar scenario alone does not offer a solution to explain the little enrichment in iron-peak elements of CEMP stars (e.g. Fe, Zn). 
For instance, in the material ejected by the stellar winds of a rotating Pop~III star, no iron is ejected so that [Fe/H] $=0$. Iron, zinc, ... can be explained by the mixing process in the sense of \cite{umeda02} or they can come from another source. % (in this case the CEMP source star is not a Pop. III star but a low metallicity star).
%The spinstar scenario is compatible the mixing \& fallback scenario.
Overall, the spinstar scenario proposes that rotation in massive stars played a major role in the chemical enrichment of the early Universe. It does not exclude other ingredients like mixing \& fallback.
As noted in \cite{maeder15a}, spinstar and mixing \& fallback scenarios are more complementary than contradictory: one can imagine a rotating star experiencing an additional mixing event at the end of its evolution.

\section{Summary}

Carbon-Enhanced Metal-Poor (CEMP) stars are peculiar MP stars observed in the halo of the Galaxy and showing overabundances in light and sometimes heavy s- and/or r- elements. CEMP stars are more frequent at low [Fe/H]: about 80~\% of stars with [Fe/H] $<-4$ are CEMP. The most iron-poor stars are generally weakly enriched in s- and/or s-elements (CEMP-no). They can be used to constrain the very early chemical evolution of the Universe.

Evolved CEMP stars probably underwent internal mixing processes (especially the first dredge-up and thermohaline mixing), altering their surface abundances. Although important, these processes are likely not able to explain the peculiar abundances of this class of stars (especially the unevolved stars). The abundances of CEMP-s stars may be explained by an accretion episode from the material (enriched in carbon and s-elements) ejected from an AGB companion. The fact that many CEMP-s stars are in a binary system supports this scenario. CEMP-no stars are generally single stars so that another scenario should be invoked. CEMP-no stars (and generally the most iron-poor stars) are thought to have kept the nucleosynthetic imprint of only one or very few previous massive source stars. In some cases however, this imprint may have been blurred by internal mixing processes in the CEMP-no star itself. 
Rotation, pre supernova mixing and strong fallback are among the interesting characteristics the source stars should have to account for the abundances of observed CEMP-no stars.

\chapter{Massive source stars: nucleosynthesis and models with rotation}
%\chapter{Massive stars evolution and nucleosynthesis}
\label{stevol}

%\vspace{1.0cm}

This chapter first discusses the nucleosynthesis in massive stars. A section is dedicated to the s-process in AGB stars for completeness. 
%In a second time, the interplay between rotation and nucleosynthesis in massive stars is presented. 
%The stellar evolution theoretical and modeling aspects relevant for this work are then presented. Emphasis is placed on the specificities of low metallicity models.
The theoretical and modeling aspects of massive stellar evolution that are relevant for this work are then presented. Emphasis is placed on the specificities of stellar evolution at low metallicity.
%The stellar evolution code used important ingredients needed to compute massive stellar models

%== Z=0 UNIVERSE ==============================================================
%\section{Zero-metallicity Universe}
\section{Nucleosynthesis processes in massive stars}\label{nucprocmass}

%\textcolor{red}{next par already put in sect 1}

%\textcolor{red}{If gravity were the only force in a star, it would collapse in less than an hour (the freefall timescale is about 30 minutes for the Sun). Stars have to produce energy to counteracts their own gravity.
%%%%%%If the nuclear reactions were stopped now in the core of the Sun, we would not see any difference for a long time. 
%The Kelvin-Helmholtz timescale $\tau_{\rm KH}$ gives an estimate of how long a star would shine with its current luminosity if the only power source were the conversion of gravitational potential to heat. For the Sun, $\tau_{\rm KH} = GM_{\odot}^2 / (R_{\odot} L_{\odot}) \simeq 30$ Myr. A lifetime of 30 Myr for the Sun is however too short in regards to geological and biological evidence that the Earth is billions of years old. Nuclear fusion provides another source of energy that sustain stars for a much longer time. Stars gain energy by fusion only up to iron because it is the most bound nucleus, meaning that no fusion can occur beyond iron.
%Massive stars with $M_{\rm ini}>8~M_{\odot}$ undergo successive shorter and shorter burning stages, lasting from million of years to a fraction of day, and ultimately leading the \textit{onion-like} structure shown in Fig.~\ref{schemastar}.
%}

\subsection{Stellar fusion}

\paragraph{Core hydrogen burning.}

The first burning stage, lasting for $\sim 1$ Myr to several tens of Myr, is the fusion of hydrogen to helium. The pp-chains and the CNO-cycle both contribute to transform H into He. Above 17 MK (corresponding the H-burning temperatures of a $\sim 1.2$~$M_{\odot}$ star), the CNO cycle becomes more efficient than the pp-chain. Provided a little bit of CNO elements are available, this is the CNO-cycle that dominates the production of energy during the H-burning stage of massive stars. The main CNO loop (CNOI) is shown by the colored loop in Fig.~\ref{cnocycle}. One loop transforms four protons into one $^{4}$He.
%\begin{equation}
%^{12}C(p,\gamma)^{13}N(e^+\nu_e)^{13}C(p,\gamma)^{14}N(p,\gamma)^{15}O(e^+\nu_e)^{15}N(p,\alpha)^{12}C
%\end{equation}
As shown by the colors of the arrows, $^{14}$N($p,\gamma$) is the slowest reaction, followed by $^{12}$C($p,\gamma$) and $^{13}$C($p,\gamma$). After a timescale determined by $^{14}$N($p,\gamma$) (the slowest reaction) the CNOI cycle reaches an equilibrium where $^{12}$C/$^{14}$N~$= 0.025$ and $^{12}$C/$^{13}$C~$=$~$3.3$. Because $^{14}$N($p,\gamma$) is the bottleneck reaction, the main effect of CNOI is to transform the initial C nuclei into $^{14}$N. The branching point at $^{15}$N is the starting point of other CNO loops that take place at higher temperatures and that allow the synthesis of other isotopes, particularly $^{16}$O. %The loop $^{15}N(p,\gamma)^{16}O(p,\gamma)^{17}F(e^+\nu_e)^{17}O(p,\alpha)^{14}N$ contributes to synthesize $^{16}$O. 

During this first stage, the Ne-Na and Mg-Al chains (Fig.~\ref{nenamgal}) can also be activated for temperatures higher than $\sim 35$ MK for Ne-Na and $\sim 50$ MK for Mg-Al. The rates of the reactions at work in these chains are more uncertain than the reactions rates of the CNO-cycle. At the branching point $^{23}$Na, the $^{23}$Na($p,\alpha$)/$^{23}$Na($p,\gamma$) ratio equals 159 and 7 at 40 and 70 MK respectively according to \cite{iliadis10b}. For the rates of \cite{cyburt10}, these ratios are 0.3 and 0.6 (still at 40 and 70 MK). 
We see that depending on the literature source used and of the temperature, the Ne-Na and Mg-Al chains can either be considered as closed loops or not. They are closed loops if the $^{23}$Na($p,\alpha$)/$^{23}$Na($p,\gamma$) ratio is high. 

Generally, the production of Ne, Na, Mg and Al in massive stars can be significantly affected if considering nuclear reaction rates from different sources. 
\cite{decressin07} have shown that multiplying the rate of $^{24}$Mg($p,\gamma$) by $10^3$ around 50 MK changes the production of Mg-Al isotopes during the main sequence of low metallicity massive stars by $0.5-1$ dex. %Such an increase of this reaction rate provides a solution to reproduce the amplitude of the observed Mg-Al anticorrelation observed in Globular Clusters.
%The main effect of the NeNa chain is to produce $^{23}$Na. 

  \begin{figure*}[t]
   \centering
      \includegraphics[scale=0.41, trim = 0cm 0cm 1cm 0cm]{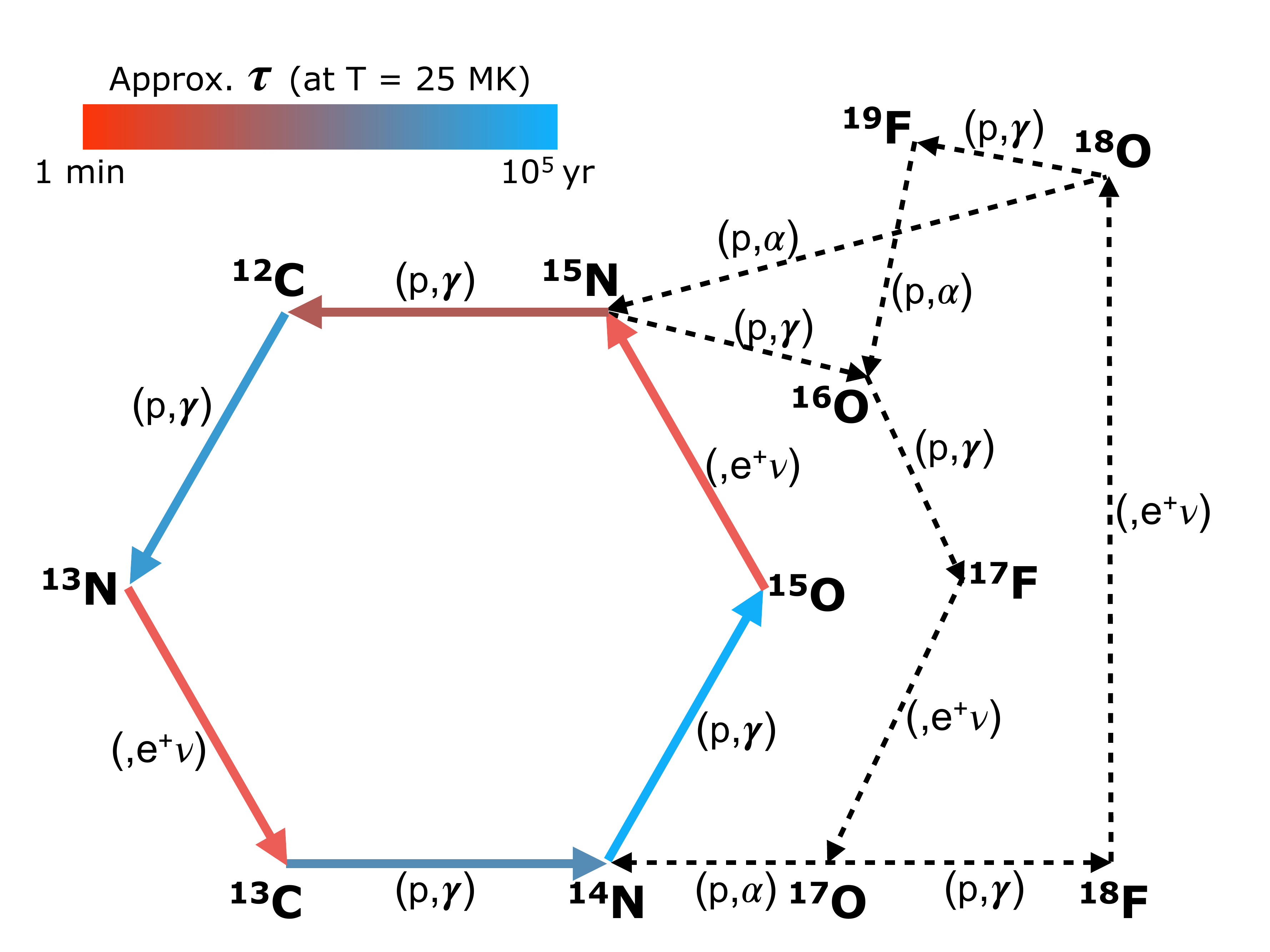}
   \caption[The CNO cycle]{Illustration of the CNO cycle. The colored hexagon shows the main loop (CNOI). The other loops are shown by dashed lines. The colored arrows show the approximate timescales for the associated reactions \citep[adapted from][]{maeder09}.}
\label{cnocycle}
    \end{figure*}

  \begin{figure*}[h!]
   \centering
      \includegraphics[scale=0.41, trim = 0cm 3cm 0cm 1cm]{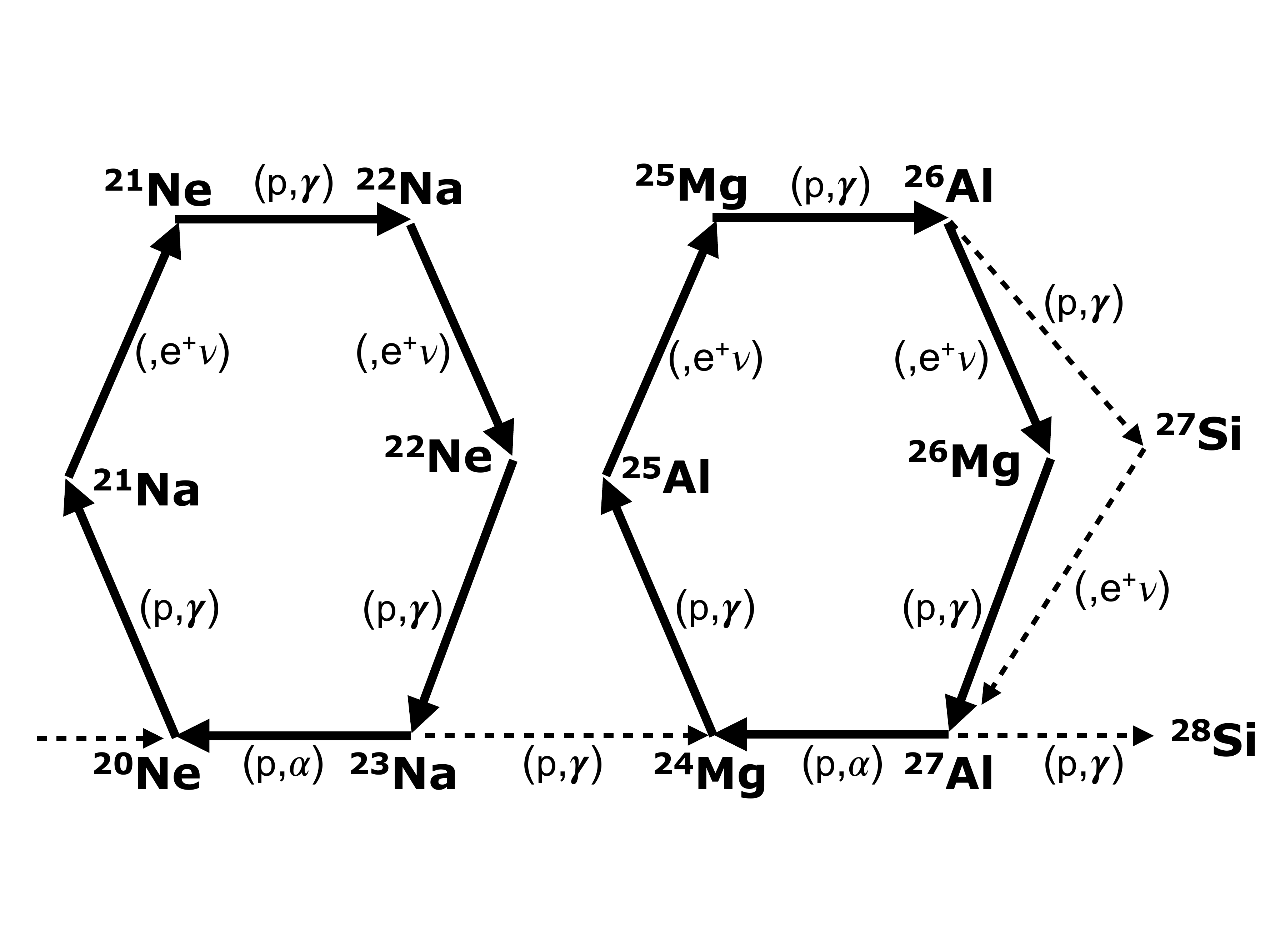}
   \caption[The Ne-Na and Mg-Al chains]{Ne-Na and Mg-Al chains (or cycles) \citep[adapted from][]{maeder09}.}
\label{nenamgal}
    \end{figure*}

\paragraph{Core helium burning.}

When running out of hydrogen, the core contracts until the central temperature is hot enough (about 100 MK) for the $3\alpha$ process to start. The second burning stage is the core helium burning phase. It lasts for $\sim 0.1-1$ Myr, which is about 1/10 of the core hydrogen burning phase. 
At the beginning of core He-burning, the $^{14}$N synthesized during core H-burning through the CNO cycle (cf. previous discussion) is quickly transformed into $^{22}$Ne via the chain $^{14}$N($\alpha,\gamma$)$^{18}$F($e^+ \nu_e$)$^{18}$O($\alpha,\gamma$)$^{22}$Ne. The $^{22}$Ne($\alpha,n$)$^{25}$Mg reaction, activated at $T \sim 220$ MK releases free neutrons that can be captured by seeds like iron and heavier elements. This is the slow neutron capture process (s-process) which is discussed below in Sect.~\ref{sproctheo}.
%Finally, during this phase, the chain $^{14}$N($\alpha,\gamma$)$^{18}$F($e^+ \nu_e$)$^{18}$O($\alpha,\gamma$)$^{22}$Ne($\alpha,n$)$^{25}$Mg releases free neutrons that can be captured by seeds like iron and heavier elements. This is the slow neutron capture process (s-process, see Sect.~\ref{sproctheo}). 

During core He-burning, a hydrogen shell burns above the helium core. The H-shell burns at a slightly higher temperature than the H-core. It induces a somewhat different nucleosynthesis in the H-shell. For instance, the Ne-Na and Mg-Al cycles are more active in the H-shell.
Rotational mixing can transport elements from the He-burning core to the H-burning shell (and vice-versa) and trigger a rich and varied nucleosynthesis (Sect.~\ref{secback}).

%Of importance is the reaction $^{12}$C($\alpha,\gamma$)$^{16}$O, that determines the C/O ratio in the core at the end of core He-b. This ratio plays an important role in determining the stellar structure during the last evolution stages \citep[e.g.][]{tur07}. It can ultimately affects the final fate of the star.

\paragraph{Advanced burning stages.}
At the end of the core He-burning stage, the most abundant species in the core are $^{12}$C and $^{16}$O. Since the $^{12}$C + $^{12}$C reaction has the lowest coulomb barrier, the next burning stage is core carbon burning.
After carbon burning comes neon photodisintegration, oxygen and then silicon burning. The advanced stages last for $10-1000$ yr (C), $0.1-1$ yr (Ne), $0.1-1$ yr (O) and $0.1 - 10$ days \citep[Si, e.g.][]{heger00,hirschi04}. During these stages, most of the energy goes out from the star in the form of neutrinos. To compensate for this strong energy loss, the rates of nuclear reactions increase. This leads to a quick burning of chemical species, explaining the short duration of the advanced stages. During these stages, the core is decoupled from the rest of the star. It implies that the star does not move anymore in the Hertzsprung-Russell (HR) diagram.
The main products of C-burning are $^{20}$Ne, $^{23}$Na and $^{24}$Mg. Neon photodisintegration produces mainly $^{16}$O, $^{24}$Mg and $^{28}$Si, oxygen burning mostly $^{28}$Si and $^{32}$S and silicon burning $^{56}$Ni \citep[cf.][especially their Table~19]{woosley95}.
At the end of each core burning phase, the burning continues in a shell. As evolution proceeds, more and more different burning regions are present in the massive star. % (Fig.~\ref{schemastar}).

%\textcolor{red}{mixing timescale long compared to evolution timescale?}

%\textcolor{red}{Say at which T CNO Ne Na and Mg Al are activated roughly}

\subsection{The weak s-process}\label{sproctheo}

Massive stars are generally associated to the \textit{weak} s-process, mainly responsible for the elements with $A<90$.
\cite{cameron60} identified that at the beginning of the core He-burning phase of massive stars, the secondary\footnote{An isotope is synthesized through the \textit{secondary} channel if it comes from the initial metal content of the star. It is formed through the \textit{primary} channel if produced from the initial hydrogen and helium content of the star.} $^{14}$N (synthesized during the main sequence thanks to the CNO-cycle) is converted into $^{22}$Ne by successive $\alpha-$captures and can provide a source of neutrons with the $^{22}$Ne($\alpha,n$) reaction. This reaction is efficiently activated at $T > 220$ MK. 
It was later recognized that the s-process in massive stars occurs principally in the He-burning core of massive stars \citep{peters68, couch74, lamb77, langer89, prantzos90, raiteri91a}. 
In solar metallicity stars with $M_{\rm ini} \gtrsim 30$~$M_{\odot}$, some $^{22}$Ne is left at the end of core He-burning phase so that s-process can occur during later stages \citep{couch74}.
The carbon shell burning is the second efficient s-process production site inside massive stars \citep[it contributes to $\sim 20$~\% at solar metallicity,][]{raiteri91b, the07}. Neutrons are released by $^{22}$Ne($\alpha,n$) with the $\alpha$ particles provided by the $^{12}$C($^{12}$C,$\alpha$)$^{20}$Ne reaction. The neutron density is typically $\sim 10^{11}$ cm$^{-3}$, which is $\sim 4$ dex higher than in the He-burning core. 
He-burning shell and C-burning core do not contribute significantly in producing s-elements \citep{arcoragi91}.
At low metallicity, the $^{13}$C($\alpha, n$)$^{16}$O reaction provides a small burst of neutrons at the very beginning of the core He-burning phase that can synthesize some light s-elements \citep[$A\lesssim85$,][]{baraffe92}.

The synthesis of s-elements (Sr, Ba...) is due to neutron capture on heavy seeds (e.g. Fe). 
However, neutrons are also captured by light elements. 
$^{12}$C and $^{16}$O are abundant in the He-core and are important poisons for the s-process. $^{16}$O captures neutrons via $^{16}$O($n,\gamma$). Then, either the neutron is recycled with $^{17}$O($\alpha,n$) or it is definitely lost with $^{17}$O($\alpha,\gamma$). The ratio of these two reaction rates plays an important role on the efficiency of the s-process. At low metallicity, the poisoning effect of $^{16}$O is stronger because the ratio of $^{16}$O to s-process seeds (e.g. iron) is higher \citep{prantzos90, rayet00}. 
During the shell carbon burning phase, important additional neutron poisons ($^{20}$Ne and $^{24}$Mg) limit the efficiency of the s-process. 

\cite{meynet06} and \cite{hirschi07} have shown that extra $^{22}$Ne can be synthesized during the core He-burning phase of massive rotating stars (the interplay between rotation and nucleosynthesis in low-metallicity massive stars is discussed in Sect.~\ref{secback}). Compared to non-rotating models, rotation gives more neutrons (source) with the same amount of seed. A higher source over seed ratio shifts the production of s-elements towards higher atomic masses \citep{gallino98}. \cite{pignatari08} have shown that rotational mixing would allow the production of s-elements up to  $A \simeq 140$. \cite{frischknecht12,frischknecht16} started to explore the s-process in rotating massive stars, by computing models (about 30 models in total) of different masses ($15<M_{\rm ini}<40$~$M_{\odot}$) and metallicities ($10^{-7}<Z<Z_{\odot}$) while following the complete s-process during the evolution. The Chapter~\ref{cemps} of this thesis aims at extending this work by computing a new grid of models and compare the yields to low mass metal-poor stars enriched in s-elements.

\subsection{Explosive nucleosynthesis}\label{explonuc}

The chemical composition of the massive star can be further modified by explosive nucleosynthesis \citep[e.g.][]{thielemann96}. It is activated by a shock wave that heats the material, processes it and finally ejects the envelope of the star while the central part will form a neutron star or a black hole. The innermost layers, from the center to the C-shell can burn explosively. The outer stellar layers are mostly unprocessed by explosive nucleosynthesis. \cite{thielemann90} reported that above the mass coordinate $\sim 2$~$M_{\odot}$ of a 20~$M_{\odot}$ model (corresponding roughly to the location of the O-burning shell), the material is not significantly altered by explosive nucleosynthesis. 
In the region affected by explosive nucleosynthesis, the abundances of elements lighter than neon and heavier than zinc are marginally modified. By contrast, the iron-peak elements are the most affected \citep{woosley95, thielemann96, woosley02, chieffi03, limongi03, nomoto06}. 
The s-elements produced during the evolution (cf. Sect.~\ref{sproctheo}) are not expected to be strongly affected by the explosion \citep{rauscher02, tur09}. The shock of the explosion may nevertheless affect the isotopic distribution of heavy nuclei in the He-shell of the progenitor \citep{meyer00, pignatari13}. In some cases, massive stars may also experience magneto-rotational supernovae, which is a possible site for the r-process \citep{winteler12}.

In this work, the explosive nucleosynthesis is not calculated. Pre-supernova models provide nevertheless a good estimation for the many species that are not much affected by the explosion. Likely, pre-supernova models also provide a good estimate of the nucleosynthesis of all elements above the carbon or oxygen burning shell, where explosive burning may not operate significantly.

%%%%%\cite{thielemann90}: explo nucleo on a 20 Mo. 'Material beyond 2 Mo is essentially unaltered by the propagating shock front.' (globally the pos. of O-shell)

%\cite{prantzos90b}: p-process

%\cite{woosley02}: review. Read beg of part on s proc. Well explained. 'Between roughly 2 and 3 $10^9$ K the s-process nuclei produced in helium and carbon burning, as well as those incorporated  into  the  original  star,  experience a  partial meltdown to the iron group. '

%%%%%At the end of its evolution, a massive star with can experience 

   \begin{figure}[t]
   \centering
      \includegraphics[scale=0.55, trim=0cm 0cm 0cm 0cm,clip]{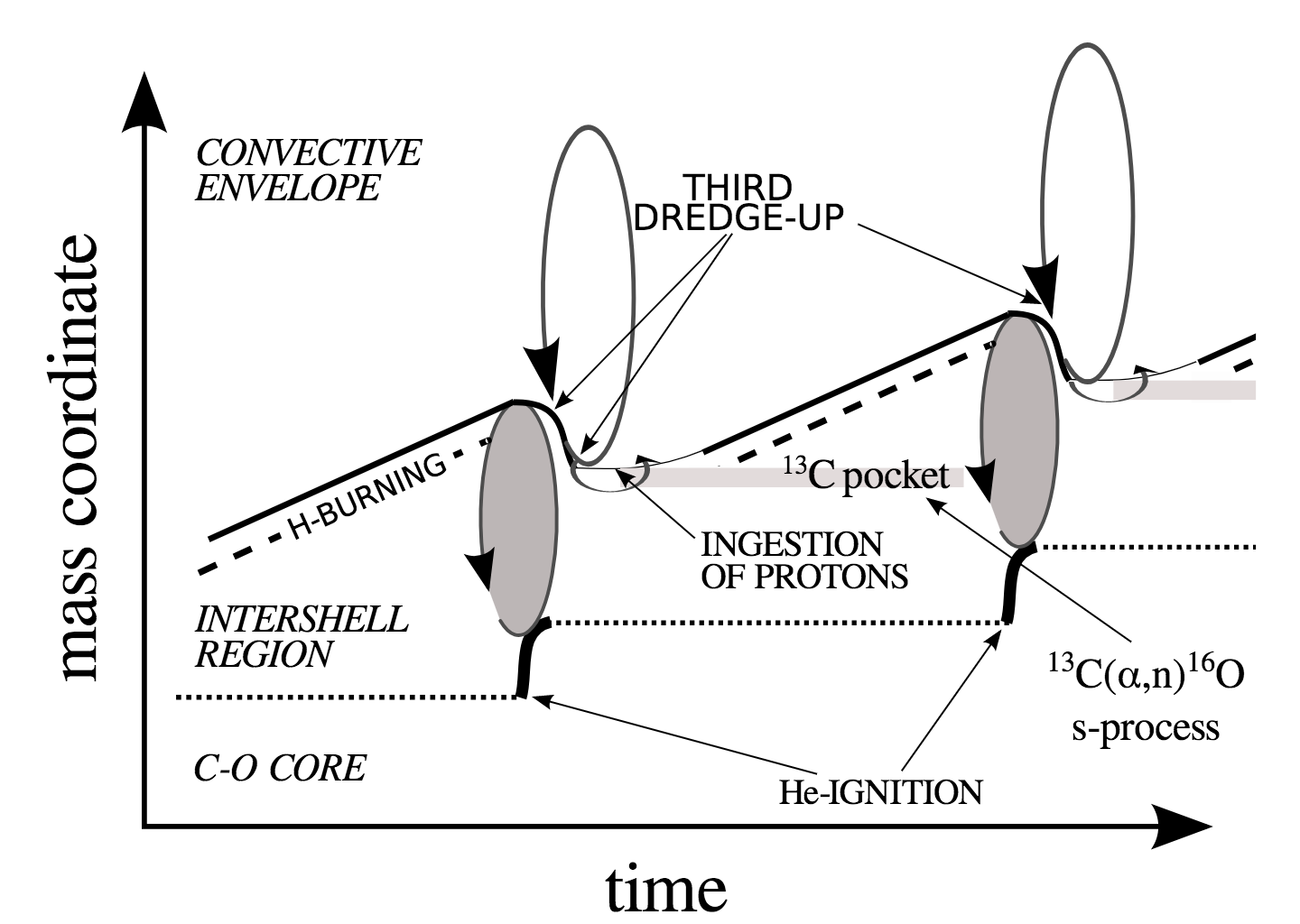}

\caption[Schematic view of the interior of a star during the AGB phase]{Schematic view of the interior of a star during the AGB phase. The horizontal grey bars represents the zone where protons are assumed to be ingested by $^{12}$C to make $^{13}$C \citep[see text for explanation, from C. Abate, PhD thesis, adapted from][]{busso99}.}
            
  \label{agbschema}
    \end{figure}

\section{The s-process in AGB stars}\label{sprocagb}

%\textcolor{red}{a work on the MAIN sprocess. Can synthesize elements up to Pb. Neutron fluxes... Why diff than massive stars?}

The present work focuses on massive stars but the s-process in AGB stars is also discussed for completeness. Thermally pulsing AGB stars are associated to the \textit{main} s-process component, responsible for the elements with $90<$ A $<209$.
The AGB corresponds to the last stage of the life of stars having initial masses between about 1 and 8~$M_{\odot}$ \citep[e.g. the reviews of][]{iben83, busso99, herwig05,karakas14}. At this phase, the star becomes larger and more luminous and strong mass loss episodes may occur (up to $10^{-4}$~$M_{\odot}$~yr$^{-1}$). AGB stars have a degenerate carbon-oxygen core. The core is surrounded by the helium and hydrogen shells, separated by the intershell region. Above is a convective envelope. %, the hydrogen burning shell and finally the convective envelope. 
The energy production is dominated by the H-burning shell. The He shell stays quiet most of the time. 
As H is burning in its shell, helium is synthesized, making the intershell region growing in mass (Fig.~\ref{agbschema}). When a critical value is reached, He-burning is suddenly ignited and a thermal pulse occurs. The energy produced drives convection so that the products of He-burning are brought upward, to the intershell region. The energy produced also expands the outer layers of the star. It has the effect of moving the H-burning shell outward and therefore stopping the burning in the H shell. It also makes the inner boundary of the convective envelope moving inward and eventually reaching the intershell region. Then, products of internal nucleosynthesis are brought up to the surface (this episode is called third dredge-up or TDU, Fig.~\ref{agbschema}).
The process described above occurs cyclically ($\sim 50000$ yrs between 2 thermal pulses) until the convective envelope is entirely removed through winds. % have removed  times before being stopped because of too much mass lost from the star \citep{herwig05}.

During the TDU, protons are mixed down to the intershell region and they are directly captured by the abundant $^{12}$C to form $^{13}$C through the chain $^{12}$C($p,\gamma$)$^{13}$N($\beta^{+}$)$^{13}$C. It forms a $^{13}$C pocket. During the next thermal pulse, the reaction $^{13}$C($\alpha, n$)$^{16}$O occurs, providing neutrons and thus triggering the s-process. If too many protons are mixed in the intershell region, $^{14}$N is also significantly formed from $^{12}$C (thanks to the CNO cycle) and suppresses the s-process by capturing the neutrons via $^{14}$N($n,p$)$^{14}$C. 
%The formation of the $^{13}$C pocket is a critical ingredient in AGB models and is not fully understood yet \citep[e.g.][]{lattanzio05}. 
% if any seed (e.g. iron) is present. 
%As shown in Fig.~\ref{agbschema}, 

The $^{22}$Ne($\alpha,n$)$^{25}$Mg reaction is not expected to contribute much in producing the main s-process component in AGB stars. One reason is that this reaction is activated in AGB stars with initial masses $M_{\rm ini} \gtrsim 3$~$M_{\odot}$ \citep{truran77} while the solar s-only abundances are best reproduced by $1.5-3$~$M_{\odot}$ AGB models \citep[e.g.][]{gallino88,hollowell88}. This said, a contribution of $^{22}$Ne($\alpha,n$) is nevertheless expected to occur in low mass AGB stars \citep[e.g.][]{straniero95}, and particularly in the low metallicity regime, for initial masses above about 2~$M_{\odot}$ \citep{lugaro12}.

With time, more and more sophisticated AGB stellar models with different initial masses, metallicities, reaction networks, etc... were computed \citep[e.g.][]{renzini81, gallino98,herwig04b,karakas10,cristallo11,lugaro12,doherty14,cristallo15}. Several important uncertainties remain like for instance the formation mechanism of the $^{13}$C pocket\footnote{In AGB models, a partial mixing zone is added artificially in a post-processing algorithm in order to mix protons with $^{12}$C into the intershell region \citep{lugaro04}. It forms the $^{13}$C-pocket which is needed to produce the neutrons required for the s-process.} which is a critical ingredient and is not fully understood yet \citep[e.g.][]{lattanzio05}, the third dredge-up (more generally the convection) or the mass loss \citep[cf. Sect.~4 of][for an extensive discussion about uncertainties in AGB models]{karakas14}.

%straniero97

%Two other reasons are that the extent in mass of the flash-driven convective zone 

Despite these uncertainties, the detection of CNO and s-elements at the surface of observed AGB stars provides evidences of the ongoing nucleosynthesis in such stars \citep[e.g.][]{smith90}. The best evidence is probably the detection of the unstable element Tc at the surface of AGB stars \citep{merrill52, smith83}: the longest-lived Tc isotope ($^{99}$Tc) has a mean life of about 1/10 of the duration of the AGB phase, meaning that it has to be synthesized by the AGB star itself. %and not by a previous generation of stars. 
%It is likely a proof of the s-process happening in AGB stars.

%\textcolor{red}{nucleo and then rotation I think}

\section{Massive stellar models including rotation}\label{massivemodels}

%Also say that 1D and multi D complementary: 1D explore large number of parameters. Multi D: can be used to constrained 1D models (prescriptions). E.g. convective boundaries.
To study the nucleosynthetic processes mentioned in the previous section, and more generally the evolution of stars, stellar models are required. 
Although stars are 3D objects, following their entire life and structure using multi dimensional codes is computationally too expensive. At the present day, only specific stages/zones of a star can be followed using multi dimensional codes \citep[e.g.][]{herwig14, cristini17}. %It is amongst the future big challenges in stellar evolution. 
In stellar evolution models, a star is often considered as a spherically symmetric object. This greatly simplifies the modeling since reducing the problem to one dimension. Compared to multi dimensional codes, 1D codes have the advantage to be able to model the entire life of stars and to explore a large parameter space. They have the disadvantadge to use a parametrized physics to describe the multi dimensional processes (e.g. turbulence). 

Interestingly, rotating stars can also be modeled with 1D codes, following the theory of shellular rotation \citep{zahn92}. The fundamental assumption of this theory is that in rotating stars, the horizontal turbulence (on an isobaric surface, i.e. surface of same pressure) is much stronger than the vertical one. This is motivated by the stable density stratification in the vertical direction, opposing a strong force to vertical fluid motions. 
In this theoretical frame, internal rotation depends essentially on the distance to the stellar center and little on latitude. 
%In particular, chemical homogeneity can be assumed in a thin shell of the star.
%Shellular rotation considers that the angular velocity is constant on isobaric shells (which are not spherical but oblate) and depends on the radius $r$ and on the colatitude $\theta$. This 2D problem can be simplified to 1D if associating an average radius $r_P$ to an isobar, defined by $V_P = (4 \pi /3)r_P^3$ where $V_P$ is the volume enclosed by the isobar \citep{meynet97}. For any quantity $X$ which is not constant over an isobar (i.e. that depends on $\theta$), a mean value $\langle X \rangle$ is derived by integrating this quantity over the isobar and dividing by the surface $S_P = 4\pi r_P^2$ of the isobar
%\begin{equation}
%\langle X \rangle = \frac{1}{S_P} \oint_{\rm isobar} X(r,\theta) \rm{d} \sigma
%\end{equation}
%where d$\sigma$ is a small surface element of the isobar.
%\begin{equation}
%\Omega (r,\theta) = \bar{\Omega (r)} + \Omega_2 (r) P_2 (\cos \theta)
%\end{equation}
%where $P_2$ is the second Legendre polynomial.
%Multi-D codes are ideal to simulate rapid phenomena, such as convection (Dearborn et al. ApJ 639,405 2006), but cannot be used to evolve a star over its lifetime.
The generalized 1D equations of hydrostatic equilibrium, mass conservation, energy conservation and energy transport, describing a star in shellular rotation are \citep{meynet97}:

\begin{flalign}
\frac{{d} P}{{d} M_P} = - \frac{G M_P}{4\pi r_P^4} f_P &&
\end{flalign}

\begin{flalign}
\frac{{d} r_P}{{d} M_P} = \frac{1}{4\pi r_P^2 \bar{\rho}} &&
\end{flalign}

\begin{flalign}
\frac{{d} L_P}{{d} M_P} = \epsilon_{\rm nucl} - \epsilon_{\nu} + \epsilon_{\rm grav} &&
\end{flalign}

\begin{flalign}
\frac{{d} \ln T}{{d} M_P} = - \frac{G M_P}{4\pi r_P^4} f_P~\rm{min} \left[ \nabla_{\rm ad}, \nabla_{\rm rad} \frac{f_T}{f_P} \right] &&
\end{flalign}

\noindent with\\

\noindent $f_P = \frac{4\pi r_P^4}{G M_P S_P} \frac{1}{\langle g_{\rm eff}^{-1} \rangle}$ and $f_T = \left(\frac{4\pi r_P^2}{S_P}\right)^2 \frac{1}{\langle g_{\rm eff} \rangle \langle g_{\rm eff}^{-1} \rangle}$\\

\noindent and $M_P$ the mass enclosed by the sphere of radius $r_P$, $g_{\rm eff}$ the effective gravity (resulting from both gravitational and centrifugal acceleration), $\nabla_{\rm rad}$ and $\nabla_{\rm ad}$ the radiative and adiabatic temperature gradients, $\bar{\rho}$ the average density between two isobars, $\epsilon_{\rm nucl}$,  $\epsilon_{\nu}$,  $\epsilon_{\rm grav}$ the rates of nuclear energy production, neutrino losses, and gravitational energy (energy released/absorbed by contraction/expansion of the star).
The quantities in brackets are averaged over an isobar according~to

\begin{equation}
\langle X \rangle = \frac{1}{S_P} \oint_{\rm isobar} X(r,\theta) \rm{d} \sigma
\end{equation}
where $S_P$ is the surface of the isobar, defined by $S_P = 4 \pi r_P^2$ with $r_P$ the radius of the equivalent sphere, having the same volume than the oblate isobar. d$\sigma$ is a small surface element on the isobar.

In this work, the Geneva stellar evolution code \textsc{genec} \citep{eggenberger08} is used. It was mainly developed and used for the computation of massive stellar models \citep[e.g.][]{maeder87, maeder92} with a sophisticated treatment of rotation \citep{meynet97}. 
The physical and modeling ingredients relevant for this work are discussed in the following sections. %Important specificities of low metallicity massive stars are highlighted. 

%Since an isobaric surface is oblate and not spherical,   Shellular rotation considers that the angular velocity is constant on isobaric shells (which are not spherical but oblate) and depends on the radius $r$ and on the colatitude $\theta$. This 2D problem can be simplified to 1D if associating an average radius $r_P$ to an isobar, defined by $V_P = (4 \pi /3)r_P^3$ where $V_P$ is the volume enclosed by the isobar. For any quantity $X$ which is not constant over an isobar (i.e. that depends on $\theta$), a mean value $\langle X \rangle$ is derived by integrating this quantity over the isobar and dividing by the surface $S_P = 4\pi r_P^2$ of the isobar

%The structure of stars is described by 4 equations that give the luminosity $L$, the temperature $T$, the mass $M$ and the pressure $P$ as a function of the radius $r$. These equations are supplemented with a set of equations describing the evolution of the chemical elements with time. These equations, with the effect Including the effects of shellular rotation

%It is based on the Kippenhahn code (Kippenhahn et al. 1967), and methods and improvements were described e.g. in Meynet & Maeder (1997) and Eggenberger et al. (2008).
 
% \textcolor{red}{Low Z?}

%== Z=0 UNIVERSE ==============================================================
%\section{Zero-metallicity Universe}

%\section{Meridional circulation. Chand title?}

\subsection{Transport induced by rotation}\label{transportrot}

%\textcolor{red}{a bit more about specificity at low Z I think}

In a differentially rotating star, the equipotentials are differently spaced as a function of the colatitude $\theta$ because the star is distorted by the centrifugal effect. The centrifugal force makes the equatorial radius larger than the polar radius so that the equipotentials are closer to each other at the poles and more spaced at the equator. Consequently, the effective gravity is larger at the poles. The radiative flux is proportional to the local effective gravity \citep{zeipel24} meaning that there is an excess of flux along the polar axis and a deficiency close to the equatorial plane. This imbalance triggers global circulation motions called meridional circulation. Such currents, which are advective, contribute to the transport of angular momentum and chemical elements in rotating stars.
In differentially rotating stars, because the stellar layers have different angular velocities, shear instabilities arise and also contribute (in a diffusive way) to the transport of angular momentum and chemical elements.

\subsubsection{Transport of angular momentum}

In the radial direction, the transport of angular momentum obeys the advective-diffusive equation \citep{chaboyer92}
\begin{equation}
\rho \frac{d }{d t} (r^2 \bar{\Omega})_{M_r} = \underbrace{\frac{1}{5r^2}\frac{\partial}{\partial r}(\rho r^2 U(r) \bar{\Omega})}_\textrm{advection} + \underbrace{\frac{1}{r^2}\frac{\partial}{\partial r}\left( \rho r^4 D \frac{\partial \bar{\Omega}}{\partial r} \right)}_\textrm{diffusion}
\label{angmomeq}
\end{equation}
where $\bar{\Omega}$ is the angular velocity of a shell, $\rho$ the density, $D$ the total diffusion coefficient, taking into account the various instabilities transporting angular momentum (especially convection and shear turbulence) and $U(r)$ the amplitude of the radial component of the meridional velocity \citep{maeder98}.
%The meridional circulation is mainly described by its velocity, that can be derived starting from the equation of the energy conservation. This velocity expands into the horizontal $V(r)$ and radial component $U(r)$. 
A large\footnote{$U(r)$ can be positive or negative. Here the absolute value is considered.} $\textbar U(r) \textbar$ implies an efficient transport of angular momentum in the radial direction. $ U(r) $ scales with $ E_{\Omega} $, a term of particular importance in the expression of $U(r)$ and that can be approximated by %\citep{zahn92,maeder98, maeder12}

\begin{equation}
E_{\Omega} \simeq \frac{8}{3} \left[ 1 - \frac{\Omega^2}{2\pi G \bar{\rho}} \right] \left( \frac{\Omega^2 r^3}{GM_r} \right).
\label{opik}
\end{equation}
%is of particular importance. $\textbar U(r) \textbar$ increases with $\textbar E_{\Omega} \textbar$. 
%%%%meaning that the chemical elements and angular momentum are more efficiently transported if $\textbar E_{\Omega} \textbar$ is large. 
%%%%$E_{\Omega}$ is proportional to the ratio of the centrifugal to the gravitational force (term in parenthesis). 
%In Eq.~\ref{opik}, 
The second term in brackets is the Gratton-\"{O}pik term. 
In the outer layers of stars, the density $\bar{\rho}$ is small so that $E_{\Omega}$, hence $U(r)$, can become largely negative (see solid line in Fig.~\ref{merid}). It contributes to redistribute efficiently the angular momentum and therefore smooth the $\Omega$-profile.
%It triggers a circulation cell that goes down along the polar axis from the surface inwards, and then rises in the equatorial plane outwards to the surface.
%\textcolor{red}{Mayber later? Or talk about metallicity, compactness... before?} 

Low metallicity stars contain less metals so they are less opaque and therefore more compact. Consequently, $\bar{\rho}$ is larger so that the Gratton-\"{O}pik term is generally smaller. This reduces $U(r)$ (see Fig.~\ref{merid}), meaning that the angular momentum is less redistributed inside the star. It leads to steeper internal $\Omega$-profiles. This tends to produce larger $\Omega$-gradients which lead to stronger shear instabilities and then to a more efficient shear mixing. %(Fig.~\ref{lowzeffect}).
%It generally becomes negative in the outer layers of stars, where $\bar{\rho}$ becomes very small. It leads to U(r)<0 and ... 

%Say : merid IMPLIES grad of omega (advection). This implies shear instabilities and then mixing!

   \begin{figure*}[t]
   \centering
      \includegraphics[scale=0.55, trim = 2cm 0cm 0cm 0cm]{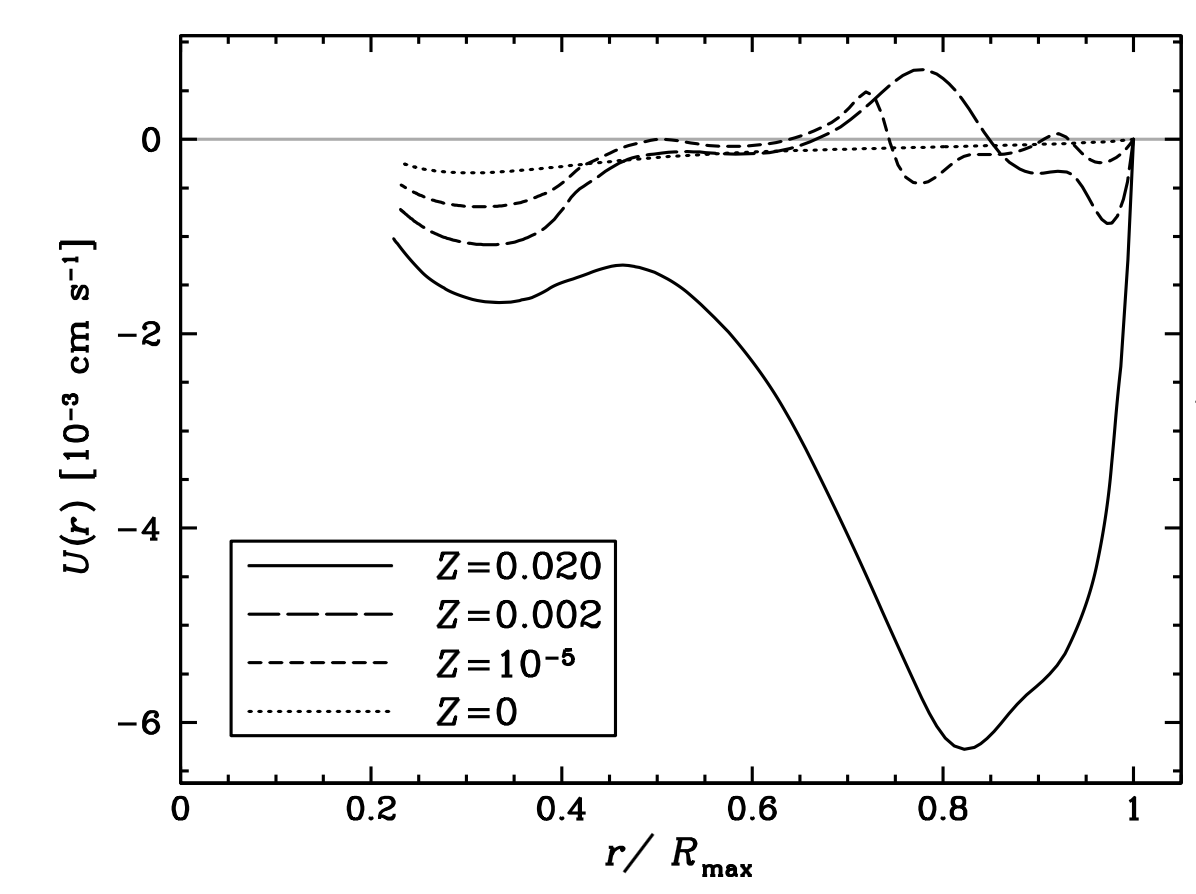}
   \caption[Internal $U(r)$ profile in rotating 20~$M_{\odot}$ models with various metallicities]{Internal profile of the amplitude of the radial component of the meridional velocity $U(r)$ in rotating 20~$M_{\odot}$ models with various metallicities. The radius is normalized to the total radius $R_{\rm max}$. For all the models, the central H mass fraction is about 0.40 \citep[figure from][]{ekstrom08}.}
\label{merid}
    \end{figure*}

%== Z=0 UNIVERSE ==============================================================
\subsubsection{Transport of chemical elements}\label{transportchem}

\cite{chaboyer92} have shown that for chemical elements, the combination of meridional circulation and horizontal turbulence can be described by a pure diffusive process. The associated diffusion coefficient is
\begin{equation}
D_{\rm eff} = \frac{1}{30}\frac{\textbar r U(r) \textbar ^2}{D_{\rm h}}
\end{equation}
with $D_{\rm h}$ the horizontal (i.e. on an isobaric surface) shear diffusion coefficient from \cite{zahn92}. It is written as $D_{\rm h} = c_{\rm h}^{-1} r \textbar 2V - \alpha U(r) \textbar$ with $\alpha = \frac{1}{2}\frac{\text{d}\ln(r^2\bar{\Omega})}{\text{d}\ln r}$ and $c_{\rm h} = 1$.
As a consequence, in contrast with the angular momentum transport equation (Eq.~\ref{angmomeq}), the equation for the change of abundance of a given chemical element $i$ in a given shell at coordinate $r$ is a pure diffusive equation:

\begin{equation}
\rho \frac{d X_i}{d t} = \frac{1}{r^2}\frac{\partial}{\partial r}\left[ \rho r^2 (D + D_{\rm eff}) \frac{\partial X_i}{\partial r} \right] + \left(\frac{dX_i}{dt}\right)_{\rm nucl}.
\label{changespec}
\end{equation}
The last term accounts for the changes due to the nuclear reactions (cf. Sect.~\ref{nucnetw}). 
%In \textsc{genec}, $D_{\rm eff}$ is calculated only during the main sequence.  
In radiative zones, $D = D_{\rm shear}$ with $D_{\rm shear}$ the diffusion coefficient accounting for the shear instabilities. %induced by stellar layers having different angular velocities.
Several prescriptions exist in the literature \citep{maeder97, talon97, maeder13}. The one used in this work is from \cite{talon97}. It can be expressed as

%Differential rotation generates turbulent motions in radiative zones. %In convective zones, the turbulent motion due to convection overwhelm the ones induced by rotation. 
%Turbulence is stronger in the horizontal than in the vertical direction because the stable thermal gradient in the vertical direction opposes a strong force to the fluid motions. In 1D stellar evolution codes, the horizontal and vertical turbulence are modeled through the $D_{\rm h}$ and $D_{\rm shear}$ diffusion coefficients respectively.

%In a differentially rotating star, a shear exists between layers having different angular velocities. Such a shear induces turbulence that leads to the mixing of chemical species between the different layers. In 1D stellar evolution codes, this mixing is included through a diffusion coefficient $D_{\rm shear}$. Two prescriptions exist in the literature. The first one, from \cite{talon97} can be expressed as

\begin{equation}
%D_{\rm shear}^{\rm TZ97} = 2\ Ri_{\rm c} \frac{H_p}{g\delta} \frac{ K +D_h}{(\nabla_{ad} - \nabla) + \frac{\varphi}{\delta}\nabla_\mu(\frac{K}{D_h}+1) } \left(\frac{9 \pi}{32} \Omega \frac{{\rm d} \ln \Omega}{{\rm d} \ln r}\right)^2.
D_{\rm shear}^{\rm TZ97} = f_{\rm energ} \frac{H_p}{g\delta} \frac{ K +D_h}{(\nabla_{ad} - \nabla) + \frac{\varphi}{\delta}\nabla_\mu(\frac{K}{D_h}+1) } \left(\frac{9 \pi}{32} \Omega \frac{{\rm d} \ln \Omega}{{\rm d} \ln r}\right)^2
\label{dshtz97}
\end{equation}
where $K$ is the thermal diffusivity, %$D_{\rm h}$ the horizontal diffusion coefficient describing the horizontal turbulent transport (at a given radius).
and $f_{\rm energ} = 2 Ri_{\rm c}$ with $Ri_{\rm c}$ the critical Richardson number below which the medium becomes dynamically unstable and turbulent. 
Although the canonical value for $Ri_{\rm c}$ is 1/4 some works have shown that turbulence may happen for larger Richardson number. \cite{canuto02} has shown that turbulence may arise for $Ri$ $\leq$  $Ri_{\rm c} \sim 1$. \cite{bruggen01} have done numerical simulations suggesting that shear mixing occurs already at $Ri \sim 1.5$. The physical range of values may be around $0.25<Ri_{\rm c}<2$, corresponding to $0.5<f_{\rm energ}<4$.

%and $f_{\rm energ}$ a parameter used to calibrate the efficiency of the shear mixing. It is adjusted so as to reproduced some observational properties, generally the surface N/H of OB stars. 
I computed two 15~$M_{\odot}$ models at solar metallicity with $\upsilon_{\rm ini} = 300$ km~s$^{-1}$ and with $f_{\rm energ} = 1$ and 4. At core H depletion, the surface N/H ratios are enhanced by a factor of 2 ($f_{\rm energ}=1$) and 3 ($f_{\rm energ}=4$) compared to the initial surface N/H ratio. Such surface enrichments qualitatively agree with observation of $10 - 20$~$M_{\odot}$ rotating stars \citep[e.g.][]{gies92, villamariz05, hunter09}. 
In this work, except stated explicitly, $f_{\rm energ} = 4$.
The effect of $f_{\rm energ}$ on the internal mixing of low metallicity massive stars is investigated in Sect.~\ref{secvarpar}.

   \begin{figure*}[t]
   \centering
      \includegraphics[scale=0.68, trim = 0cm 0cm 0cm 0cm]{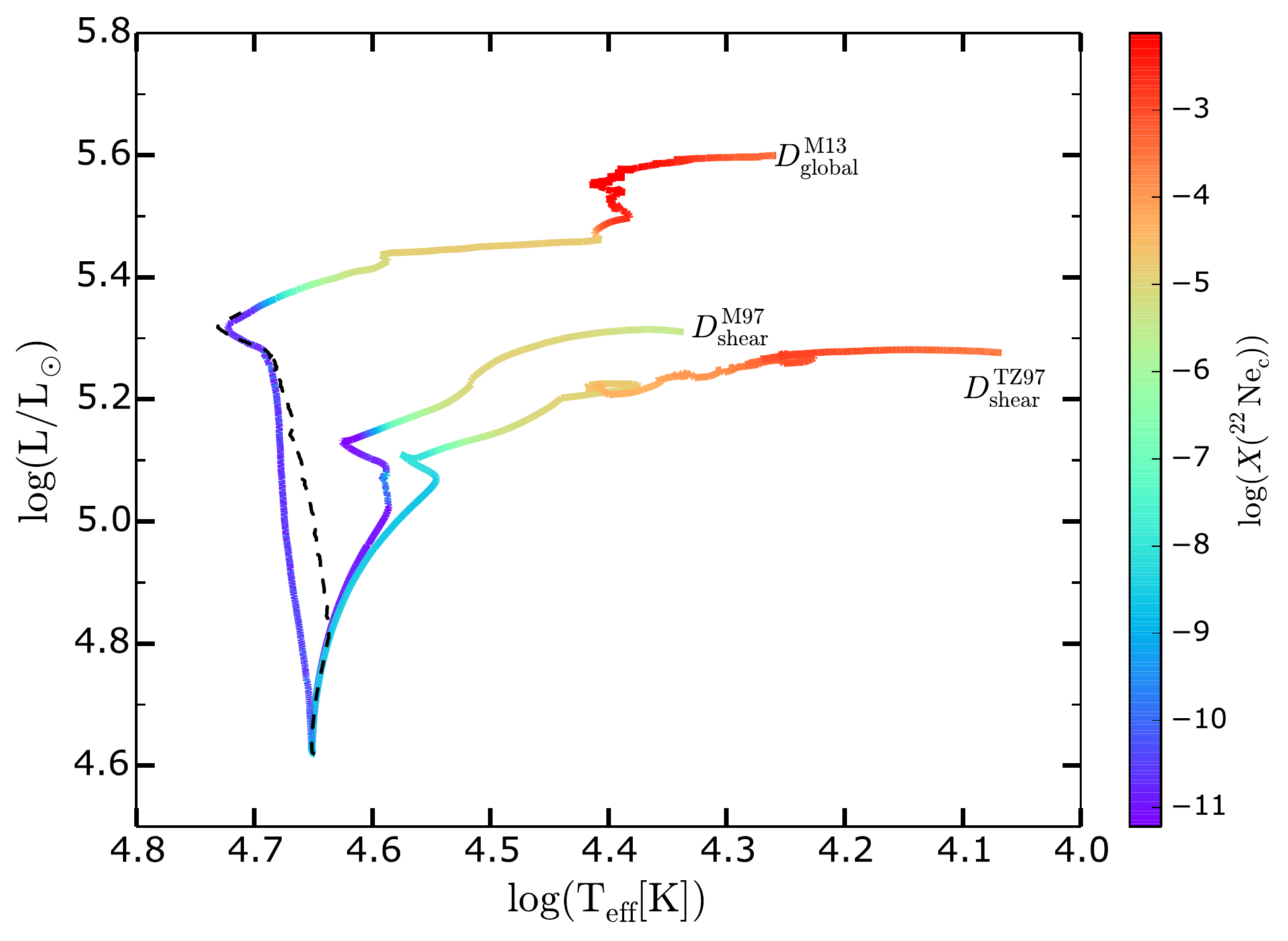}
   \caption[HR diagram of rotating 20~$M_{\odot}$ computed with different $D_{\rm shear}$ coefficients]{Hertzsprung-Russell (HR) diagram of a 20~$M_{\odot}$ model with $Z=10^{-5}$ and $v_{\rm ini}/v_{\rm crit} = 0.4$ with different $D_{\rm shear}$ coefficients. The color shows the central mass fraction of $^{22}$Ne. The dashed black track is computed with the same physics as the $D_{\rm shear}^{\rm TZ97}$ model but with the $D_{\rm shear}$ coefficient multiplied by 10. The 3 colored models are computed until the end of the core helium burning phase. The dashed model is computed until the end of the main sequence.}
\label{hrdico}
    \end{figure*}

\subsubsection{A comparison of 3 different prescriptions for the shear mixing}

Although not used in the present work, I have included in \textsc{genec} the global diffusion coefficient proposed by \cite{maeder13}, taking into account different instabilities together and their possible interaction (see Appendix~\ref{dshear} for calculation details).

%provides more details as well as a comparison between the different prescriptions for shear mixing.
Fig.~\ref{hrdico} shows the tracks of 20~$M_{\odot}$ models with $\upsilon_{\rm ini}/\upsilon_{\rm crit} = 0.4$, $Z=10^{-5}$ and computed with various $D_{\rm shear}$ coefficients: from \cite{maeder97}, from \cite{talon97} and the global coefficient from \cite{maeder13}. The new coefficient $D_{\rm global}^{\rm M13}$ combining the different instabilities leads to a bluer and more luminous evolution, characteristic of stars experiencing a high degree of mixing. The surface $^{4}$He mass fraction at the end of the main sequence is the highest in the $D_{\rm global}^{\rm M13}$ model (0.56 against 0.25 and 0.31 for the $D_{\rm shear}^{\rm TZ97}$ and $D_{\rm shear}^{\rm M97}$ models, respectively), which is also a signature of an efficient rotational mixing. An estimation of the mixing efficiency in the deep stellar interior can be obtained with the amount of central $^{22}$Ne after the main sequence (more details in Sect.~\ref{secback}). $^{22}$Ne in the center reaches a much higher mass fraction in the $D_{\rm shear}^{\rm TZ97}$ and $D_{\rm global}^{\rm M13}$ models. It suggests that, after the main sequence, a more efficient mixing in the deep interior is at work in these two models compared to the $D_{\rm shear}^{\rm M97}$ model. The efficiency of the s-process in massive stars is impacted by the amount of $^{22}$Ne (main neutron source, cf. Sect.~\ref{sproctheo}). From Fig.~\ref{hrdico}, we see that the s-process should operate similarly in the $D_{\rm shear}^{\rm TZ97}$ and $D_{\rm global}^{\rm M13}$ because of the similar $^{22}$Ne content. On the opposite, the $D_{\rm shear}^{\rm M97}$ model predicts a less efficient operation of the s-process.

The dashed track in Fig.~\ref{hrdico} shows the main sequence of the $D_{\rm shear}^{\rm TZ97}$ model but with $f_{\rm energ} =$~40 instead of 4. In this case, the track is rather close to the track of the model with the global diffusion coefficient. A higher $f_{\rm energ}$ value for the $D_{\rm shear}^{\rm TZ97}$ model may mimic the more complex physics contained in the $D_{\rm global}^{\rm M13}$ coefficient. It shows that finally, even if $f_{\rm energ} = 4$ was proposed as a possible upper limit for the $D_{\rm shear}$ of \cite{talon97}, a higher $f_{\rm energ}$ (i.e. higher mixing efficiency) probably cannot be excluded.

%Generally, $\alpha$ is used to calibrate the efficiency of mixing. It is adjusted so as to reproduced some observational properties. 

%In the prescription of \cite{talon97}, we see that $f_{energ} = 2 Ri_c$ can likely take values higher than 1. The physical range of values might be around $0.25<Ri_c<2$ so $0.5<f_{energ}<4$.\\

%where $\nabla = \nabla_{rad}$ in stellar evolution models and where $9 \pi /32$ is sometimes written as its numerical value $0.8836$. $2 Ri_c$ is also often written as $f_{energ}$ \cite[see][for instance]{meynet13}.

%Then Dshear, etc... prescription can lead to more or less Ne22 .. etc. Also smoothing, in appendix. It can affect s process nucleo

\subsection{Convection}

In a star, a heat excess with respect to what radiation can transfer drives turbulent convective motions. It provides an additional and efficient way of transporting the energy. %Convection is another way of transporting energy together with radiative transfer. 
In this work, the Schwarzschild criterion is used to determine the stability of a given stellar layer. 

%The diffusion coefficient in convective zones is defined according to the mixing-length theory (MLT): $D_{\rm conv} = 1/3 \alpha_{\rm MLT} H_{\rm P} \upsilon_{\rm conv}$

%with an overshoot parameter dover/HP = 0.10 from and above

%The mixing-length theory (MLT) is used. The dimensionless mixing length parameter $\alpha= l / H_{\rm P}$ with $l$ the mixing length

In the interior of the Sun, the turnover timescale $t_{\rm turn}$ (timescale for an element of material to complete a loop in a convective cell) can be estimated using the mixing-length theory. It is about 0.5~yr \citep{maeder09}, suggesting that chemical homogeneity is quickly reached in a convective zone. 
Since the lifetime of a star is much longer than the turnover timescale, the convective mixing can be considered as instantaneous during most of the stellar life. The last stages of massive stars (after central carbon burning) are however very short \citep[less than 1 yr, e.g.][]{hirschi04} so that the approximation of instantaneous convective mixing is no more valid. In this case, the convection is treated as a diffusive process. 
%In any case, the turbulent motion due to convection overwhelm the turbulence induced by rotation, meaning that in convective zones, the $D$ coefficient in Eq.~\ref{angmomeq} and \ref{changespec} is equal to the convective diffusion coefficient.

During the H- and He-burning phases of the models computed in the present work, overshoot is considered: the size of the convective core is extended by $d_{\rm over} = \alpha H_{\rm P}$ where $H_{\rm P}$ is the pressure scale height and $\alpha = 0.1$ for initial masses above $1.7$~$M_{\odot}$. $\alpha$ was calibrated so as to reproduce the observed main-sequence width of stars with $1.35<M<9$~$M_{\odot}$ \citep{ekstrom12}.

   \begin{figure}[t]
   \centering
      \includegraphics[scale=0.46, trim = 0cm 1.5cm 0cm 0cm]{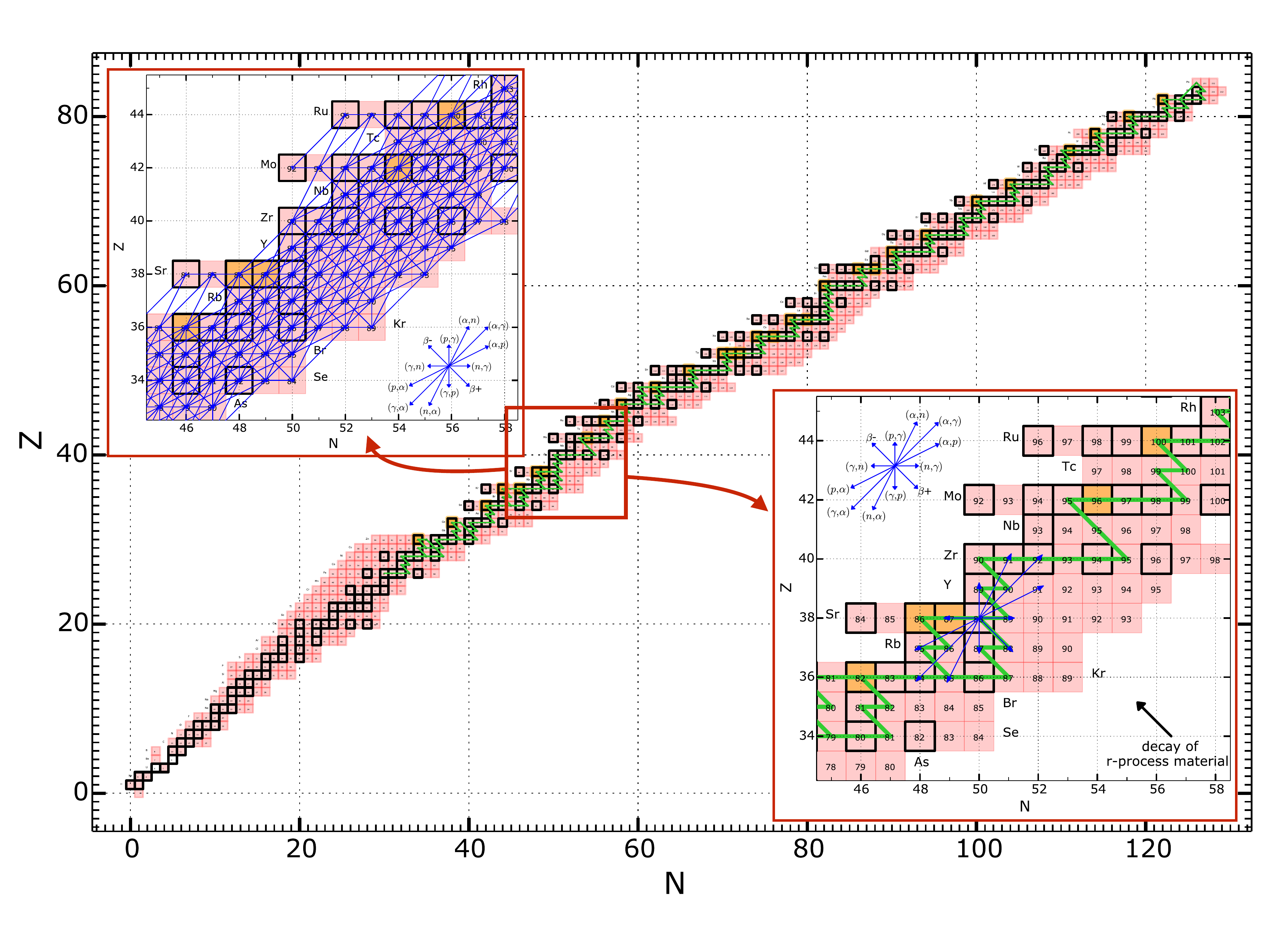}
   %\caption[a]{\textcolor{red}{maybe more: illustration of the nuclear network used (in chapXX)}. Number of neutrons as a function of number of protons. The black squares show the stable isotopes and the red shaded ones show the 737 isotopes considered in the extended network used in this work (Chapter~/ref{cemps}). The green path shows the principal s-process path. \textit{Lower right panel}: zoom on a specific region. Blue arrows show the nuclear reactions starting from $^{88}$Sr. Orange squares denote stable isotopes that can be synthesized by the s-process (also p-process eventually) but not by the r-process. \textit{Upper left panel}: all reactions included in the extended network, from $^{78}$As to $^{103}$Rh.} %\textcolor{red}{s-only isotopes and r-only? stable iso thicker in subplots?}}
   \caption[Visualization of the large network used in the present work]{The red shaded squares show the 737 isotopes considered in the extended network used in this work (Chapter~\ref{cemps}). The black squares show the stable isotopes. The green path shows the principal s-process path. \textit{Lower right panel}: zoom on a specific region. Blue arrows show the nuclear reactions starting from $^{88}$Sr. Orange squares denote stable isotopes that can be synthesized by the s-process (also by p-process, possibly) but not by the r-process. \textit{Upper left panel}: blue segments show all reactions included in the extended network, from $^{78}$As to $^{103}$Rh.} %\textcolor{red}{s-only isotopes and r-only? stable iso thicker in subplots?}}
\label{networkfig}
    \end{figure}

%== Z=0 UNIVERSE ==============================================================
%\section{Zero-metallicity Universe}
\subsection{Nuclear network}\label{nucnetw}

%\textcolor{red}{say 2 sections, one without sproc one with}{}

%The variation Eq.~\ref{changespec}
In a stellar evolution code, to follow the abundance changes due to nuclear reactions, one has to express the nuclear term in Eq.~\ref{changespec} (last term). It is written as
%in change of the mass fraction $X_i$ of an isotope $i$ over time then becomes

\begin{equation}
%\frac{\partial Y_i}{\partial t} = \underbrace{\sum_{j} N_{j}^{i} \lambda_j Y_j}_\textrm{1 body reactions} + \underbrace{\sum_{j,k} N^{i}_{j,k}\rho N_A \langle \sigma v \rangle_{j;k} f_{j;k} Y_j Y_k}_\textrm{2 body reactions} + \underbrace{\sum_{j,k,l} N^{i}_{j,k,l}\rho^2 N_A^2 \langle \sigma v \rangle_{j;k;l} f_{j;k;l} Y_j Y_k Y_l}_\textrm{3 body reactions}
%\frac{\partial Y_i}{\partial t} = \underbrace{\sum_{j} N_{i} \lambda_j Y_j}_\textrm{1 body reactions} + \underbrace{\sum_{j,k} \frac{N_{i}}{1+\delta_{jk}} \rho N_A \langle \sigma v \rangle_{j;k} f_{j;k} Y_j Y_k}_\textrm{2 body reactions} + \underbrace{\sum_{j,k,l} \frac{N_{i}}{1+\Delta_{ijk}} \rho^2 N_A^2 \langle \sigma v \rangle_{j;k;l} f_{j;k;l} Y_j Y_k Y_l}_\textrm{3 body reactions}
\begin{split} 
%\left(\frac{d X_i}{d t} \right)_{\rm nucl} = & \underbrace{A_i \sum_{j} N^{i}_{j} \lambda_j Y_j}_\textrm{1-body reactions} + \underbrace{A_i \sum_{j,k}  \frac{1}{1+\delta_{jk}} N^{i}_{j,k} \rho N_A \langle \sigma v \rangle_{j;k} f_{j;k} Y_j Y_k}_\textrm{2-body reactions}\\ 
%&+ \underbrace{A_i  \sum_{j,k,l}  \frac{1}{1+\Delta_{ijk}} N^{i}_{j,k,l} \rho^2 N_A^2 \langle \sigma v \rangle_{j;k;l} f_{j;k;l} Y_j Y_k Y_l}_\textrm{3-body reactions}
\left(\frac{d X_i}{d t} \right)_{\rm nucl} = & \underbrace{A_i \sum_{j} N^{i}_{j} \lambda_j \frac{X_j}{A_j}}_\textrm{1-body reactions} + \underbrace{A_i \sum_{j,k}  \frac{1}{1+\delta_{jk}} N^{i}_{j,k} \rho N_A \langle \sigma v \rangle_{j;k} f_{j;k} \frac{X_j X_k}{A_j A_k}}_\textrm{2-body reactions}\\ 
&+ \underbrace{A_i  \sum_{j,k,l}  \frac{1}{1+\Delta_{ijk}} N^{i}_{j,k,l} \rho^2 N_A^2 \langle \sigma v \rangle_{j;k;l} f_{j;k;l} \frac{X_j X_k X_l}{A_j A_k A_l}}_\textrm{3-body reactions}
\end{split}
\label{network}
\end{equation}
where $A_i$, $A_j$, $A_k$ and $A_l$ are the mass number of isotopes $i$, $j$, $k$ and $l$, $N^{i}_{j}$, $N^{i}_{j,k}$ and $N^{i}_{j,k,l}$ are the number of particles of isotope $i$ destroyed $<0$ or produced $>0$ in a particular reaction, $f_{j;k}$ and $f_{j;k;l}$ the screening factors, $\Delta_{ijk} = \delta_{jk} + \delta_{jl} + \delta_{kl} + 2\delta_{jkl}$ and $\delta$'s are factors preventing double counting. 1-body reactions refer to photo-disintegrations, $\beta$-decays, $\nu$-, $e^-$- and $e^+$ captures. $\lambda_j$, $\langle \sigma v \rangle_{j;k}$ and $\langle \sigma v \rangle_{j;k;l}$ are the nuclear reaction rates. These rates are estimated experimentally \citep[e.g.][]{caughlan88} or theoretically \citep[e.g.][]{rauscher00} and tabulated over a certain range of temperature. %\textcolor{red}{Some reaction 1 body do not depend on T}

The set of differential equation \ref{network} is called \textit{stiff} because of the strong variety of reaction timescales, ranging from fraction of seconds to Gyr. An explicit method is not possible to solve these equations because it would require extremely small timesteps in order to follow the fastest reactions. In the new version of \textsc{genec}, these equations are solved in a implicit manner by the backward Euler method\footnote{Details can be found in the Sect.~3.5 of Frischknecht, PhD thesis (2012, \href{https://edoc.unibas.ch/21287/}{https://edoc.unibas.ch/21287/}).}. %\textcolor{red}{explain quickly}...

In Chapter~\ref{cempno} of this work, I rely on models computed with a small nuclear network, that follows 31 species in total: $n$, $^{1}$H, $^{3,4}$He, $^{12,13,14}$C, $^{14,15}$N, $^{16,17,18}$O, $^{18,19}$F, $^{20,21,22}$Ne, $^{23}$Na, $^{24,25,26}$Mg, $^{26,27}$Al, $^{28}$Si, $^{32}$S, $^{36}$Ar, $^{40}$Ca, $^{44}$Ti, $^{48}$Cr, $^{52}$Fe and $^{56}$Ni. It allows to track the reactions that contribute significantly to generate nuclear energy. This network is also enough to follow the nucleosynthesis of the main species beyond Fe until the ultimate stages of evolution. Finally, this network is well suited for investigating the origin of CEMP-no stars since such stars do not show strong overabundances in heavy elements.
In Chapter~\ref{cemps} instead, a network of 737 isotopes is considered, from $^{1}$H to $^{212}$Po (Fig.~\ref{networkfig}), that allows to follow the full s-process nucleosynthesis. Some of the reactions linking the isotopes are shown by the blue segment in Fig.~\ref{networkfig}. 
The nucleosynthesis is followed thanks to the BasNet code \citep[Basel Network, ][]{arnett85, thielemann85} which is coupled to \textsc{genec}\footnote{Details can be found in the Sect.~3.7 of Frischknecht, PhD thesis (2012, \href{https://edoc.unibas.ch/21287/}{https://edoc.unibas.ch/21287/}).}. It allows to follow consistently all the nucleosynthesis during the entire evolution.

%code is coupled to the evolution of stellar models and used throughout all of it. In this case, the nucleosynthesis is followed thanks to the code BasNet \citep[Basel network, ][]{arnett85, thielemann85} which is coupled to \textsc{genec}\footnote{Details can be found in the Sect.~3.7 of Frischknecht, PhD thesis, 2012 (\href{https://edoc.unibas.ch/21287/}{https://edoc.unibas.ch/21287/})}.

%For stellar modelers, the interesting output of a rate measurement in a laboratory is a table that gives values of the rate for about 50 temperature values of the . 
%In the original version of the \textsc{GENEVA} code, following about 20 isotopes, nuclear rates are in tabular form (one column for the temperature, one for the rate).
%To fasten the computation of large nuclear reaction networks, a specific format is used for the nuclear reactions. The rates are written in analytical form, as 

%\textcolor{red}{REF on fig with nuc net}

 % \begin{figure*}[h!]
  % \centering
  %    \includegraphics[scale=0.47, trim = 0cm 0cm 0cm 0cm]{vinichem.png}
  % \caption[a]{Maybe later? Put also same but for He burning zone (H+He or just He?) and also effect of mass cut for 1 model. And dilution effect?}
%\label{vinichem}
   % \end{figure*}

\subsection{Initial composition and opacity tables}\label{inicompo}

In this work, the initial composition of the low metallicity massive source stars is taken $\alpha$-enhanced. Details are given below.
In \textsc{genec}, the initial abundances are in mass fraction. The solar mass fraction $X_{\rm i,\odot}$ of an isotope $i$, associated to an element $ e$ can be written as
\begin{equation}
X_{\rm i,\odot} = 10^{\log \epsilon_{\rm e,\odot}-12}~f_{\rm i}~A_{\rm i}~X_{\rm H, \odot}
\label{solfrac}
\end{equation}
where $\log \epsilon_{\rm e,\odot}$ is the solar abundance of element $e$ taken from \cite{asplund05}, except for Ne, which is from \cite{cunha06}. $f_{\rm i}$ is the isotopic fraction in the solar system \citep{lodders03}. $A_{\rm i}$ is the mass number of isotope $i$ %and $X_{\rm H, \odot}$ is the solar hydrogen mass fraction. 
and $X_{\rm H, \odot} = 0.7399$ is the solar hydrogen mass fraction, at the present day and corresponding to the Asplund+Cunha mixture. 
As a remark, since the Sun is 4.57 Gyr old, it has likely experienced atomic diffusion since its birth. Atomic diffusion progressively decreases the surface helium and increases the surface hydrogen. Then, the initial solar abundances are different than the present-day abundances. \cite{ekstrom12} have calibrated a 1~$M_{\odot}$ model including atomic diffusion to reproduce the solar radius, luminosity and surface chemical composition at the age of the Sun. It gives an initial hydrogen and helium mass fraction of 0.720 and 0.266 respectively. The associated initial solar metallicity is $Z_{ \odot} = 0.014$.

In case the initial metallicity $Z$ of the model is not solar, $X_{\rm He}$ is calculated according to the relation $X_{\rm He}=X_{\rm He,p} + \Delta X_{\rm He} / \Delta Z \times Z$ where $X_{\rm He,p}$ is the primordial helium abundance and $\Delta X_{\rm He} / \Delta Z = (X_{\rm He,\odot}-X_{\rm He,p})/Z_{\odot}$ the average slope of the helium-to-metal enrichment law. I set $X_{\rm He,p}=0.2484$ \citep{cyburt03}.
%derived from \cite{asplund05}. 
Once the helium mass fraction is calculated, the initial hydrogen mass fraction is deduced from $X_{\rm H} = 1-X_{\rm He}-Z$. 
%In the case of our models with $Z=10^{-5}$, $X \simeq 0.752$ and $Y \simeq 0.248$.

When $Z$ is not solar, if the mixture of metal remains solar-like, the mass fractions of metals are just multiplied by $Z/Z_{\odot}$ i.e. $X_{\rm i} = X_{\rm i,\odot}~Z/Z_{\odot}$ with $X_{\rm i,\odot}$ from Eq.~\ref{solfrac}. This leads to a solar-scaled mixture, meaning that [X/Fe] $=0$.
However, at [Fe/H] $<0$, it appears that [$\alpha$/Fe] $>0$ where $\alpha$ represent the $\alpha$-elements \citep[e.g.][]{reddy06}. 
In \textsc{genec}, for $-1<$ [Fe/H] $<0$, [$\alpha$/Fe] $=-B_{\rm \alpha}$[Fe/H] 
with $B_{\rm \alpha}$ = 0.562, 0.886, 0.500, 0.411, 0.307, 0.435, 0.300, 0.222 and 0.251 for the $\alpha$-elements $^{12}$C, $^{16}$O, $^{20}$Ne, $^{24}$Mg, $^{28}$Si, $^{32}$S, $^{36}$Ar, $^{40}$Ca and $^{48}$Ti respectively.
These numbers were derived by fitting the abundance trends [$\alpha$/Fe] versus [Fe/H] from halo and thick disc F- and G-dwarfs stars with $-1 \lesssim$ [Fe/H] $\lesssim 0$ 
%for $-1 \lesssim$ [Fe/H] $\lesssim 0$ 
\citep{reddy06}. 
%(i.e. linear increase of [$\alpha$/Fe] with decreasing [Fe/H]) and 
%At low metallicity, I used 
At [Fe/H] $<-1$, [$\alpha$/Fe] $=B_{\rm \alpha}$ is set (i.e. no dependance on metallicity). 
The mass fractions of the $\alpha$-isotopes are calculated according to
\begin{equation}
X_{\rm i,\alpha} = X_{\rm i,\odot}~\frac{X_{\rm Fe}}{X_{\rm Fe,\odot}}~10^{B_{\rm \alpha}}~f_{\rm i}~A_{\rm i}
\end{equation}
where $X_{\rm Fe}$ is the sum of the mass fractions of all Fe isotopes. 
%It is calculated thanks to the secant method.
The mass fraction of the non $\alpha$-isotopes are calculated with the same equation but with $B_{\rm \alpha} = 0$. 
Accordingly to this $\alpha$-enhanced mixture, the opacity tables were computed with the OPAL tool\footnote{\href{http://opalopacity.llnl.gov}{http://opalopacity.llnl.gov}.}. They are complemented at low temperatures by the opacities from \cite{ferguson05}.

In general, at low metallicity, the initial mixture is poorly known and other initial mixtures cannot be excluded. An $\alpha$-enhanced mixture may be the most natural choice. However, it is worth noting that for most of the elements and low-metallicity models considered in this work, the abundances in the ejecta of the massive star models are very different from the initial ones. It implies that the dependance of the chemical yields on the initial composition is weak.

Also worth to mention is that with Galactic chemical evolution models, \cite{chiappini08} have shown that if massive fast rotators were common in the early Universe, $^{12}$C/$^{13}$C would be between about 30 and 300 in the (almost) primordial ISM. Without fast rotators, $^{12}$C/$^{13}$C would be between about 4500 and 31000.
Here I take $^{12}$C/$^{13}$C $=300$. Lower or higher values cannot be excluded. This is discussed in the next chapters.

%Faux? : $\alpha$-enhanced mixtures are commonly used when computing low metallicity models \citep[e.g.][]{meynet06,hirschi07,frischknecht16}.\\

%== Z=0 UNIVERSE ==============================================================
%\section{Zero-metallicity Universe}
%\subsection{Mass loss and opacity}
\subsection{Mass loss}\label{secmassloss}

AGB and massive stars show spectroscopic evidences of stellar winds \citep[][for a review]{kudritzki00}. 
Stellar winds are mainly driven by the absorption of UV-photons by metal lines. Winds are consequently stronger in hot metal-rich stars. The radiation pressure expels stellar material. It can make a 120~$M_{\odot}$ solar metallicity star losing more than 70~\% of its initial mass \citep{ekstrom12}. % A 120~$M_{\odot}$ at solar metallicity ends its life with about 31~$M_{\odot}$ \citep{ekstrom12}.
%The strong radiation pressure of luminous stars expels the material and implies mass loss. 
%The strong radiation pressure expels the material and implies mass loss. 
%Mass loss is an important ingredient that can alter the evolution of the Massive stars lose mass
At low metallicity, radiative winds are weaker because of the smaller content in metals. Radiative mass loss rates follow a scaling relation with metallicity $Z$:
\begin{equation}
\dot{M} (Z) = \left( \frac{Z}{Z_{\odot}} \right)^{a} \dot{M}(Z_{\odot})
\label{mdotz}
\end{equation}
where $\dot{M}(Z_{\odot})$ is the solar metallicity mass loss rate and $0.5 < a < 0.85$ \citep{vink01}. %when the prescription of \cite{vink01} is used and it is 0.5 in all the other cases. %%%%mass loss of \cite{dejaeger88} is used and it is equal to 0.85 when \cite{vink01} is used.
In this work, radiative mass-loss rates are from \cite{vink01} when $\log (T_{\rm eff}) \geq 3.95$ and from \cite{jager88} when $\log (T_{\rm eff}) < 3.95$.
For the models of Sect.~\ref{implem} however, the prescription of \cite{kudritzki00} is used instead of \citet[][]{vink01}. Radiative winds at low metallicity are generally small so that no big impact is expected. Also, for $\log (T_{\rm eff}) < 3.95$, \cite{jager88} is used in any case.

The stellar surface can also reach the Eddington luminosity $L_{\rm EDD}$ (luminosity when the radiative force balances the gravitational force) which triggers additional mass loss. In \textsc{genec}, the mass-loss rate is increased by a factor of 3 when the luminosity of any layer in the stellar envelope becomes higher than five times the Eddington luminosity \citep{ekstrom12}.

Another recent mass loss mechanism was proposed for metal-free stars with $M_{\rm ini} \geq 150$~$M_{\odot}$ \citep{moriya15}: if $\log (T_{\rm eff}) < 3.7$ the envelope of such stars may become pulsationally unstable near the end of their evolution and undergo extreme mass-loss events, of the order of $10^{-4} - 10^{-2}$~$M_{\odot}$~yr$^{-1}$. Although not used in the models of this work, I have added this prescription in \textsc{genec} (details in Appendix~\ref{pulsmassloss}).

%\cite{moriya15} have shown that the envelope of massive Pop. III stars can become pulsationally unstable near the end of their evolution and then undergo extreme mass-loss events.

%\textcolor{red}{See paper sproc. Different mass losses with rotation. Enumerate}

\subsubsection{Effect of rotation on mass loss}

%\textcolor{red}{un mot on mass loss equatorial and polar}

Rotation is expected to affect the mass loss of stars. 
First, it makes the stellar winds anisotropic by favoring an ejection of matter through both the poles and the equator \citep{maeder99b}. The reason is that the polar regions are hotter than the equatorial regions (cf. Sect.~\ref{transportrot}) so that the radiation pressure, hence the mass loss rate is stronger at the poles. 
Also, for stars with equatorial temperatures $T \lesssim 23000$ K (rotation tends to reduce the temperature at the equator), the high opacity around the equator triggers enhanced radiative winds that push the matter outward and produce an equatorial ring. 

When the stellar surface reaches the critical velocity, the outer layers are no longer bound and some mass is removed mechanically. 
%(and angular momentum) is removed, making the surface slowing down.
In rotating stars having an Eddington factor $\Gamma = L/L_{\rm EDD} > 0.639$, because of the interplay between radiation and rotation, the critical velocity is reduced \citep{maeder00}. The consequence is that mechanical mass loss occurs more easily (i.e. at a smaller rotation rate). 
Mechanical mass losses on the rotating massive Pop~III models of \cite{ekstrom08phd} removes $< 10$~\% of the total stellar mass. 

More generally, if considering 2 stars of same mass $M$, one rotating at $\Omega$ and one non-rotating, at about the same location in the HR diagram, the mass loss ratio $\dot{M} (\Omega)$/$\dot{M} (0)$ is greater than one and depends, among other, on $\Omega$ and $\Gamma$ \citep{maeder00}. Models including rotation are corrected by this factor, that becomes very high for large $\Gamma$ factors.
%The interplay between radiation and rotation also enhances the radiative mass loss in rotating stars \citep{maeder00}.
%: in a rotating star, the polar regions are hotter so that the radiation pressure and consequently the mass loss is stronger.

Also, by changing the distribution of the chemical species in the stellar interior, rotation modifies the tracks in the HR diagram. Since radiative mass loss rates depend on $L$ and $T_{\rm eff}$, the mass loss experienced by the star is changed. For instance, rotation tends to produce bigger cores, increasing the luminosity $L$ of the star hence increasing the mass loss rate $\dot{M}$, that varies as $L^{2.2}$ \citep{vink01}.

Fast rotation may also induce extreme radiative winds during the core He-burning stage of very low massive metallicity stars. It occurs if a large amount of metals from the stellar interior is brought up to the surface. In this case, the surface metallicity increases dramatically so that strong radiative winds are triggered (Eq.~\ref{mdotz}). It can make a very low metallicity ($Z=10^{-8}$) massive star to lose about 70~\% of its mass \citep{hirschi07}. %have shown that a 85~$M_{\odot}$ model at $Z=10^{-8}$ can lose about 65~$M_{\odot}$
Such strong mass loss episodes are however not happening in all low/zero metallicity rotating models \citep{hirschi07,ekstrom08} and are sensitive to different parameters (see Sect.~\ref{secother} for additional discussions).

\subsection{Chemical composition of the ejecta}

For a stellar model, the ejected mass $m^{\rm ej}_{\rm i}$ in $M_{\odot}$ under the form of an isotope $i$ is given by
\begin{equation}
m^{\rm ej}_{\rm i} =  \underbrace{\int_{0}^{\tau} \dot{M}(t) X_{\rm i,s}(t) \text{d}t}_\textrm{wind contribution} + \underbrace{\int_{M_{\rm cut}}^{M_{\rm fin}} X_{\rm i}(M_{\rm r}) \text{d}M_r}_\textrm{supernova contribution},
\label{yie}
\end{equation}
where the bounds $\tau$, $M_{\rm{fin}}$, and $M_{\rm{cut}}$ are the total lifetime, mass at the end of the evolution and mass cut\footnote{At the end of the evolution, the mass cut is the mass coordinate that delimitates the part of the star which is expelled from the part which is kept into the remnant.} of the model, respectively. %(both given in \textcolor{red}{Table~XX}). 
$X_{\rm i,s}(t)$ and $\dot{M}(t)$ are the surface mass fraction and the mass loss rate at time $t$ respectively. $X_{\rm i}(M_{\rm r})$ is the mass fraction of the isotope $i$ at coordinate $M_{\rm r}$, at the end of the evolution. 
From Eq.~\ref{yie} and knowing the solar composition, the [X/Y] ratios in the ejected material can be computed\footnote{In the bracket notation, X and Y are in number so that the mass ejected $m_{\rm i}^{\rm ej}$ of an isotope $i$ has to be divided by its atomic mass $A_{\rm i}$. For instance, [C/H] $= \log \left( \frac{m_{^{12}\textrm{C}}^{\rm ej}/12 + m_{^{13}\textrm{C}}^{\rm ej}/13}{m_{^{1}\textrm{H}}^{\rm ej}} \right) - \log (\frac{\textrm{C}}{\textrm{H}})_{\odot}$. }. 

The stellar yield of an isotope $i$ is expressed as
\begin{equation}
%m_{\rm i} =  \underbrace{\int_{0}^{\tau} \dot{M}(t) X_{\rm i,s}(t) \text{d}t}_\textrm{wind contribution} + \underbrace{\int_{M_{\rm cut}}^{M_{\rm fin}} X_{\rm i}(M_{\rm r}) \text{d}M_r}_\textrm{supernova contribution},
m_{\rm i} = \underbrace{\int_{0}^{\tau} \dot{M}(t) (X_{\rm i,s}(t) - X_{\rm i,0}) \text{d}t}_\textrm{wind contribution}   +   \underbrace{\int_{M_{\rm cut}}^{M_{\rm fin}} (X_{\rm i}(M_{\rm r}) - X_{\rm i,0}) \text{d}M_r}_\textrm{supernova contribution},
\label{yie2}
\end{equation}
with $X_{\rm i,0}$ the initial mass fraction of isotope $i$. The only difference with Eq.~\ref{yie} is that the mass under the form of the isotope $i$ that was initially present in the star is deduced. Hence, $m_{\rm i}$ can be either positive or negative.

Finally, for an isotope $i$, the production factor $f_{\rm i}$ is defined as
\begin{equation}
f_i = \frac{M_{\rm ej}}{M_{\rm ini}}\frac{X_{\rm i, ej}}{X_{\rm i, 0}},
%f_{\rm i} = \frac{\int_{0}^{\tau} \dot{M}(t) (X_{\rm i,s}(t) - X_{\rm i,0}) \text{d}t  +   \int_{M_{\rm cut}}^{M_{\rm fin}} (X_{\rm i}(M_{\rm r}) - X_{\rm i,0}) \text{d}M_r}   {},
\label{fact}
\end{equation}
with $M_{\rm ej}$ the total mass ejected by the star, $M_{\rm ini}$ the initial mass and $X_{\rm i,ej}$ the mass fraction of isotope $i$ in the ejecta. It expresses, for a given isotope, the ratio of what is given back by the star divided by what was present initially in the whole star. % It is greater than zero and is a convenient way to see by how much a model produces a given isotope.

   \begin{figure*}[t]
   \centering
      \includegraphics[scale=0.48, trim = 0cm 0cm 0cm 0cm]{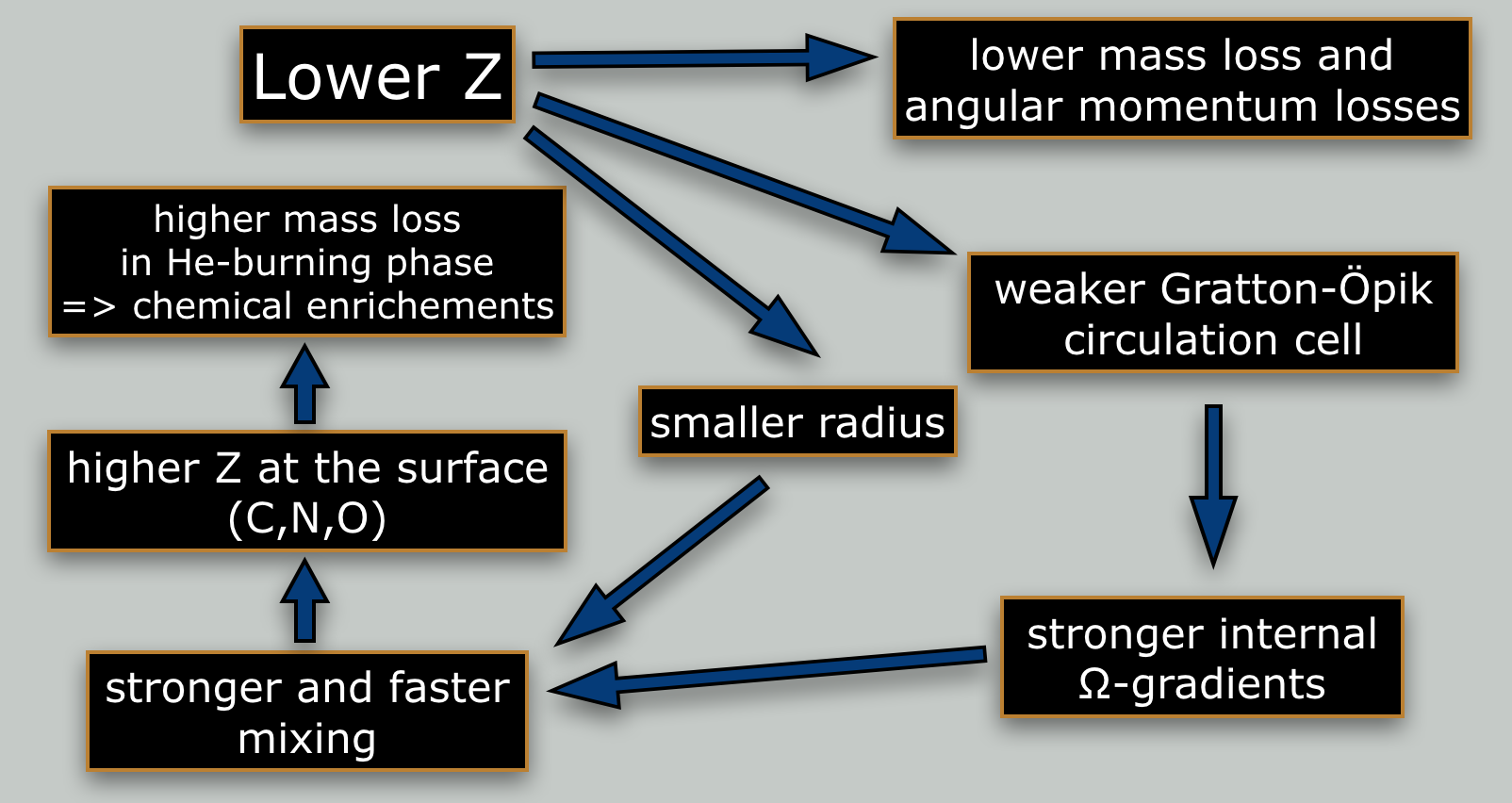}
   \caption[Some important effects of rotation at low metallicity]{Some important effects of rotation at low metallicity \citep[adapted from][]{maeder09}.}
\label{lowzeffect}
    \end{figure*}

\section{Axions and their impact on Pop~III stars}\label{secaxion}

This section presents a side project of the thesis, related to the evolution of metal-free massive stars including axion cooling.
%\section{Effects of axions on Population III stellar models} %????
Axions are hypothetical elementary particles that are of interest as possible components of cold dark matter \citep{preskill83, abbott83, dine83}.
They interact weakly with matter. In stars, they can be considered as an energy sink, taking away the energy, like neutrinos. An axion loss rate as a function of the stellar temperature and density 
%(that depends mostly on temperature and density) 
can be derived and introduced in stellar evolution codes in order to see how such losses affect the life of stars. \cite{friedland13} found that axion losses can shorten and eliminate the blue loop phase of solar metallicity $8-12$~$M_{\odot}$ stars. 
%The idea is to see whether axions may let distinct observable signatures on stars, that might then allow to verify their presence or constrain their physics.

In collaboration with 
%\href{mailto:taoso@iap.fr}
{Alain Coc} (Centre de Sciences Nucl\'{e}aires et de Sciences de la Mati\`{e}re (CSNSM), France), Keith A. Olive (William I. Fine Theoretical Physics Institute, School of Physics and Astronomy, University of Minnesota, USA), Jean-Philippe Uzan and Elisabeth Vangioni (Institut d'Astrophysique de Paris (IAP), France), I have studied the effects of axions on Pop~III stars in the article (the full paper can be found page \pageref{axions} of this thesis):\\
%\hyperlink{pdfwimps.1}{page \pageref*{pwimps}}

\noindent \href{http://adsabs.harvard.edu/abs/2017A\%26A...605A.106C}{
\textbf{Effects of axions on Population III stars}\\
A. Choplin, A. Coc, G. Meynet, K. A. Olive, J.-P. Uzan \& E. Vangioni, 
2017 A\&A \textbf{605}, 106}\\

First, I introduced the axion losses $\epsilon_{\rm ax}$ in \textsc{genec}. The derivatives of $\epsilon_{\rm ax}$ as a function of the pressure $P$ and temperature $T$ were also introduced in \textsc{genec} so as to help for convergence when solving the structure equations (see Appendix~\ref{snumchem} for more details and calculations).
%In a second time, 8, 10 and 12~$M_{\odot}$ models at solar metallicity were computed with and without axion losses. We confirm the disappearance
Then, I computed a grid of non-rotating Pop~III massive stars with initial masses of 20, 25, 32, 40, 60, 85, 120 and 150~$M_{\odot}$, with and without axion losses, until the end of the core C-burning phase. 

During the main sequence, the energy is mainly transported by radiation and convection. Axion losses have a modest effect: the main sequence lifetime is shorten by less than 3~\% in models including axions. Although still modest, axion losses are stronger during the core He-burning stage, because of higher temperature. Consequently, the core He-burning lifetime is reduced by $7 - 10$~\%. 
%The phase during which axions induce the most significant changes in the evolution is 
The phase from core He depletion to core C-burning ignition lasts for $10^3 - 10^4$ yr. During this phase, the carbon-oxygen core contracts so that the central temperature increases and axion losses become important. Neutrino losses are not strong yet. This is during this short axion-dominated phase that axions are likely inducing the most significant changes in the evolution. From core C-burning ignition to the end of the evolution, axion losses become negligible compared to neutrino losses so that stellar evolution is no more altered. 

We found that globally, the effects of axions on stellar evolution are very modest and will be very hardly observable. 
The strongest effect of axions arises in the 85 and 120~$M_{\odot}$ models. Differences in central temperature, density and chemical composition appear after core He depletion, for a short period of time, as a transitory reaction of the model to the increased loss of energy in the central regions (cf. previous discussion). Such differences might prevent the 85 and 120~$M_{\odot}$ models with axion losses to experience a pulsational pair-instability supernova (PPISN), as it may be expected for stars with initial masses between 70 and 140~$M_{\odot}$ \citep{woosley17}. If so, such stars with axions could produce black holes in a mass range where standard stellar models do not predict them. It might induce a potential signature in gravitational waves. 
As discussed in the paper, however, the uncertainties of different stellar model parameters produce effects at least as strong as axions on the structure of the star, particularly during the late evolution stages. 
We are therefore still far from a situation where Pop~III stellar models can be used as a physics laboratory to verify the presence of axions or constrain their properties.

\section{Summary}
A quick summary of nucleosynthesis, modeling and theoretical aspects important for what follows is given.
In the H-burning region of massive stars, the operation of the CNO cycle, Ne-Na and Mg-Al-Si chains can synthesize elements from C to Si. %The nuclear reaction rates involved in the Ne-Na and Mg-Al-Si chains are uncertain, 
The $^{14}$N transformed in $^{22}$Ne at the core He-burning ignition will later (when the central temperature $T > 220$ MK) provides free neutrons. It leads to the principal component of the \textit{weak} s-process in massive stars (the other and smaller component comes from the C-burning shell), able to synthesize elements with A~$<$~90. Rotation-induced mixing provides additional $^{22}$Ne, which can shifts the s-process pattern towards heavier elements. %(provided some heavy seed like Fe is present). 
Among others light elements, $^{16}$O, which is abundant in the He-burning core, can poison the s-process by capturing neutrons. The \textit{main} s-process, responsible for elements with $90<$ A $<209$ is associated with thermally pulsing AGB stars, having initial masses between about 1 and 8~$M_{\odot}$.

%At low metallicity, 
%The theory of shellular rotation 
Some important physical ingredients were presented in Sect.~\ref{massivemodels} for the modeling of massive stars including rotation.
At low metallicity, a natural effect of the physics of rotation in stars is that the rotation-induced mixing of chemical elements is more efficient. One reason is due to the less efficient redistribution of the angular momentum by meridional currents. Rotation enhances the mass loss in different ways, particularly because of the interplay of radiation and rotation. The mass loss is generally reduced at low metallicity but extreme mass loss events can nevertheless happen in fast rotating massive models (Fig.~\ref{lowzeffect} for a schematic view). The nuclear network considered in this work comprises either 31 isotopes (Chapter~\ref{cempno}) or 737, from $^{1}$H to $^{212}$Po (Chapter~\ref{cemps}). While the small network is suitable for the study of CEMP-no stars, the second is required to study CEMP stars enriched in heavier elements like strontium or barium.

\chapter{Mixing in CEMP-no source stars}
\label{cempno}

%\textcolor{red}{Maybe : redo analysis of individual CEMP with X/Fe and not X/H (since X/H = X/Fe + Fe/H is not exact) and also take only xi $< 1.5$ maybe (0.5 dex by abunadnce in average at worst case)}

%\textcolor{red}{Or just take X/H tables from saga. Also say a work about Fe and iron peak elements. Mixing fallback}

%\textcolor{red}{Add Clarkson+18 somewhere, Herwig works about p ingesiton}

%\textcolor{red}{interplay not before core He not after. Because... slides}

%In this chapter, the origin of the CEMP stars with [Fe/H] $<-3$ and not significantly enriched in s- and/or r-elements is investigated. \textbf{These stars are referred to as CEMP-no stars, even if sometime, not 
In this chapter, the origin of the CEMP stars with [Fe/H] $<-3$ is investigated. The recognized CEMP-s, -r/s and -r stars are not considered. For convenience, the stars considered here are referred to as CEMP-no stars, even if strictly speaking, some of them are just CEMP (because of missing abundance data to classify them).
Sections~\ref{secbox} and \ref{seclate} include the results published in the first and second articles mentioned below, respectively (the full papers can be found page~\pageref{pcempno} and \pageref{pbox} of this thesis).\\ %Other sections discuss additional, unpublished results.\\
%Together with additional results and discussions, this chapter includes the results published in the articles mentioned below. Sect.~\ref{secback} and \ref{seclate} (\hyperlink{pbox}{page \pageref*{pbox}} of this thesis):\\

\noindent \href{http://adsabs.harvard.edu/abs/2016A\%26A...593A..36C}{
\textbf{Constraints on CEMP-no progenitors from nuclear astrophysics}\\
A. Choplin, A. Maeder, G. Meynet, \& C. Chiappini, 
2016 \aap \ \textbf{593}, 36}\\

\noindent \href{http://adsabs.harvard.edu/abs/2017A\%26A...605A..63C}{
\textbf{Pre-supernova mixing in CEMP-no source stars}\\
A. Choplin, S. Ekstr\"{o}m, G. Meynet, A. Maeder, C. Georgy, \& R. Hirschi, 
2017 \aap \ \textbf{605}, 63}\\

%Section~\ref{secback} discusses mainly the first article while Sec.~\ref{seclate} and \ref{secorig} the second article. Additional results and discussions are presented in other sections.

%\noindent Electronic tables of the models (ZAMS and evolutionary sequence files) are available at\\ \url{http://obswww.unige.ch/Recherche/evol/Critical-limit-and-Be-stars}.

%== SN, Mixing? Lithium? ici ou a la fin de la sect?=====================================================================
%\section{Linking the CEMP-no and its source \label{srotzamsphysinput}}

%== INTRO ======================================================================
%\section{back-and-forth mixing in the source star\label{backforth}}

%== INTRO ======================================================================
\section{The back-and-forth mixing process \label{secback}}
%\section{Interplay between rotation and nucleosynthesis in massive stars\label{secback}}

%\textcolor{red}{speak (plot?) about mixing timescale vs burning phase timescale. After He burning, mixing not important since tau mixing $>$ tau burning}

%\textcolor{red}{word on primary and secondary?}

%The previous sections presented some aspects of nucleosynthesis and rotation in massive stars. 

%Here I discuss how rotation impacts the nucleosynthesis in rotating massive stars during the core He-burning phase.

In the previous chapter, some aspects regarding the nucleosynthesis and the physics of rotation were discussed. The interplay between rotation and nucleosynthesis in rotating massive stars is now discussed.

During the core H-burning and He-burning phase, the mixing induced by rotation changes the distribution of the chemical elements inside the star. In advanced stages (C-burning and after), the burning timescale becomes small compared to the rotational mixing timescale so that rotation barely affects the distribution of chemical elements. %(only the inner layers are affected by further nucleosynhtesis). 
During the core He-burning phase, two different burning regions exist in the star (He-burning core and H-burning shell).
The rotational mixing triggers exchanges of material between the convective He-burning core and the convective H-burning shell: He-burning products are transported to the H-burning shell, processed by H-burning, transported back to the He-burning core, etc... %It can lead to a rich and varied nucleosynthesis. Deja mis, repetition
The main steps of this mixing process are (see Fig.~\ref{schemadiff} for a schematic view):

\begin{enumerate}
\item In the He-burning core, the triple alpha process synthesizes $^{12}$C. $^{16}$O is formed by $^{12}$C($\alpha,\gamma$)$^{16}$O.
\item $^{12}$C and $^{16}$O are mixed into the H-burning shell. It boosts the CNO cycle and creates primary CNO elements, especially $^{14}$N (and $^{13}$C to a smaller extent).
\item The products of the H-burning shell (among them primary $^{13}$C and $^{14}$N) are mixed back into the He-core. From the primary $^{14}$N, the reaction chain $^{14}$N($\alpha,\gamma$)$^{18}$F($e^+ \nu_e$)$^{18}$O($\alpha,\gamma$)$^{22}$Ne allows the synthesis of primary $^{22}$Ne. The reactions $^{22}$Ne($\alpha,n$) and $^{22}$Ne($\alpha,\gamma$) make $^{25}$Mg and $^{26}$Mg respectively. 
The neutrons released by the $^{22}$Ne($\alpha,n$) reaction produce $^{19}$F, $^{23}$Na, $^{24}$Mg and $^{27}$Al by $^{14}$N($n,\gamma$)$^{15}$N($\alpha,\gamma$)$^{19}$F, $^{22}$Ne($n,\gamma$)$^{23}$Ne($e^- \bar{\nu}_e$)$^{23}$Na, $^{23}$Na($n,\gamma$)$^{24}$Na($e^- \bar{\nu}_e$)$^{24}$Mg and $^{26}$Mg($n,\gamma$)$^{27}$Mg($e^- \bar{\nu}_e$)$^{27}$Al, respectively.
Free neutrons can also be captured by heavier seeds like $^{56}$Fe and boost the s-process (cf. Sect.~\ref{sproctheo}). This point is investigated in details in Chapter~\ref{cemps}.
%\item when $T \gtrsim 220$ MK (generally corresponding to the middle-end of core He-burning), $^{22}$Ne($\alpha,n$) and $^{22}$Ne($\alpha,\gamma$) make $^{25}$Mg, $^{26}$Mg and s-elements.
\item The newly formed elements in the He-burning core can be mixed again into the H-burning shell. It boosts the Ne-Na and Mg-Al chains: additional Na and Al are produced.
\end{enumerate}

  \begin{figure*}[t]
   \centering
      \includegraphics[scale=0.165, trim = 0cm 0cm 15cm 0cm]{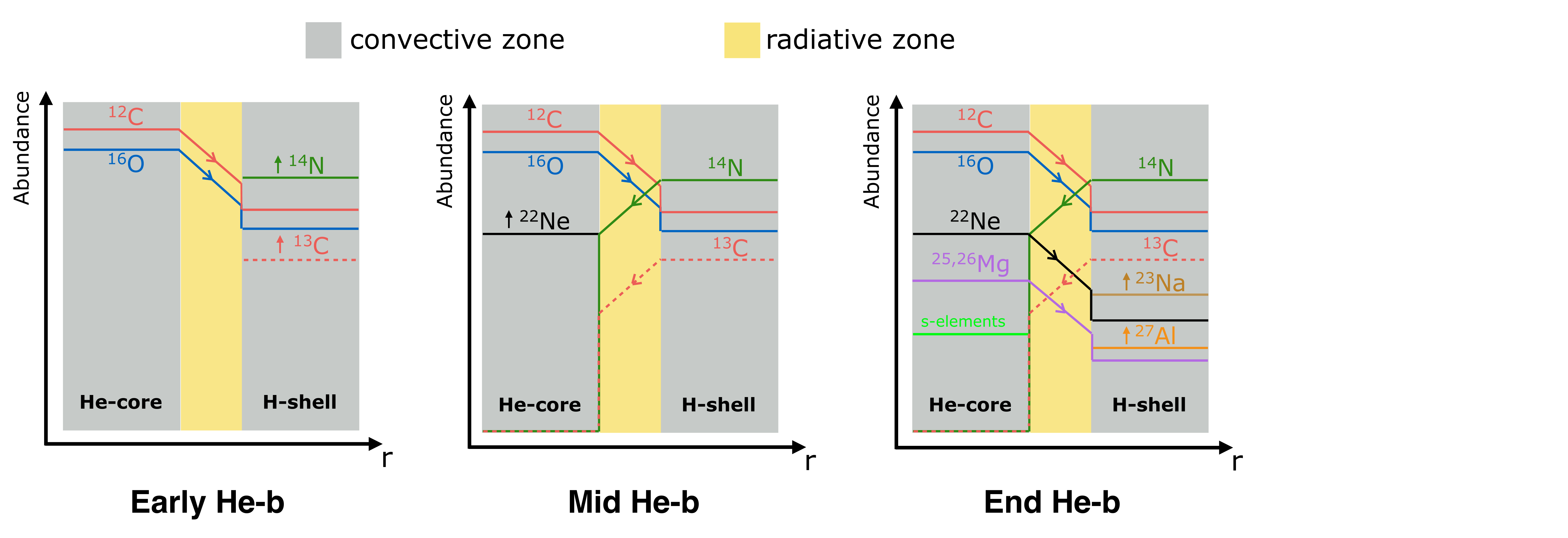}
   \caption[Schematic view of the back-and-forth mixing process]{Schematic view of the back-and-forth mixing process at work in a fast rotating massive star (for clarity, only some chemical species are represented). The small vertical arrows denote some of the elements whose abundance is increased by the arrival of the products of the other burning zone.
   %It occurs during the core helium burning phase and it is an exchange of chemical species between the helium burning core and the hydrogen burning shell.
   }
\label{schemadiff}
    \end{figure*}

  \begin{figure*}[h!]
   \centering
      \includegraphics[scale=0.47, trim = 0cm 0cm 0cm 0cm]{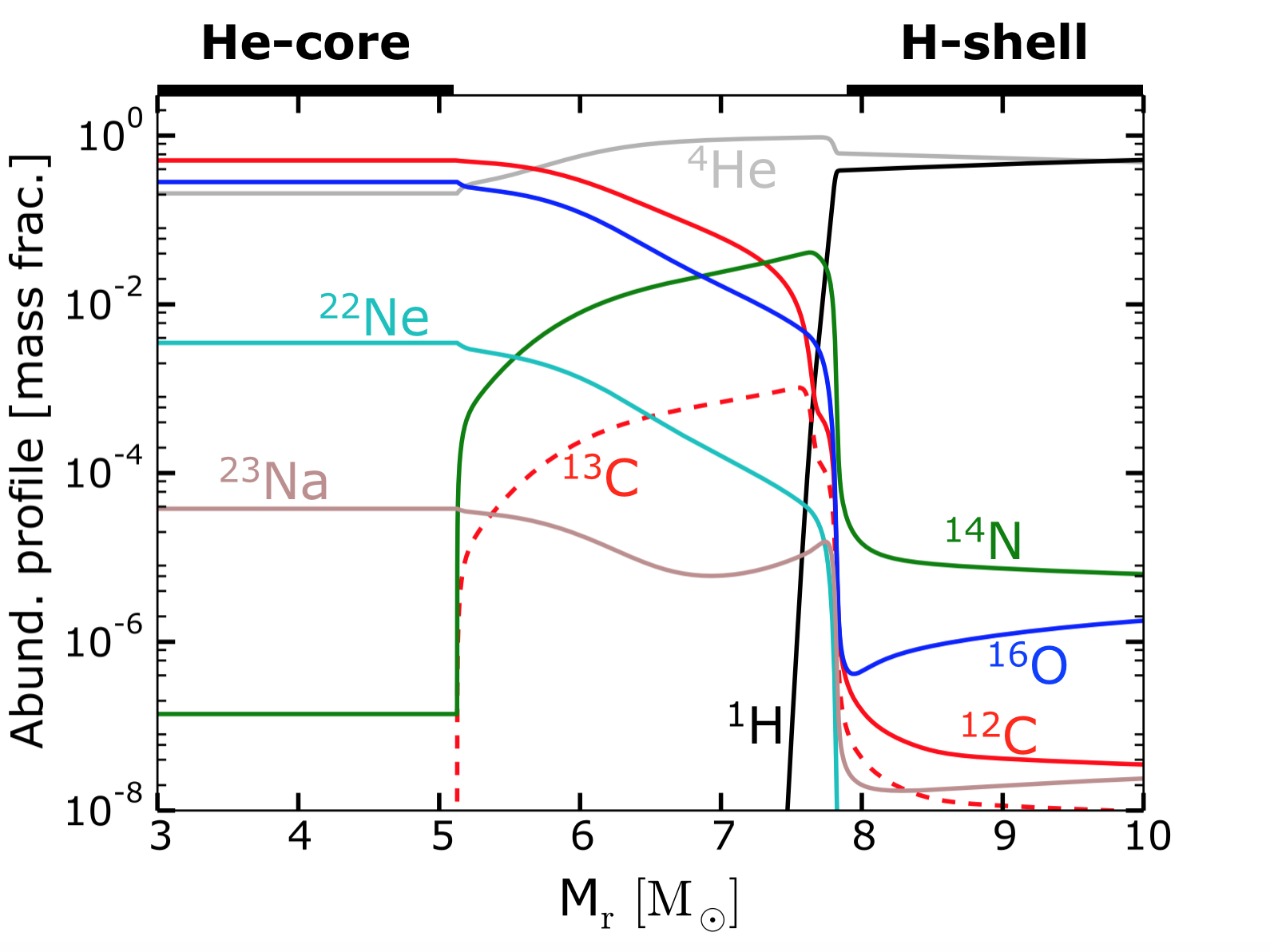}
   \caption[Abundances profile of a 20~$M_{\odot}$ with $\upsilon_{\rm ini}/\upsilon_{\rm crit} = 0.7$ at $Z=10^{-5}$]{Abundance profile of a fast rotating 20~$M_{\odot}$ model with $\upsilon_{\rm ini}/\upsilon_{\rm crit} = 0.7$ at
   %$\upsilon_{\rm ini}/\upsilon_{\rm crit} = 0.7$ and 
   $Z=10^{-5}$. The model is near the end of the core He-burning stage. The thick black lines on the top labelled He-core and H-shell show the location of the convective He-burning core and H-burning shell respectively.}
\label{mixmodel}
    \end{figure*}

%\textcolor{red}{say that it is a simplified view In reality all steps occur at the same time more or less. Also rotation BOOSTS the reactions but still some F, Ne, Na... can be formed in a secondary channel}

%\textcolor{red}{Put nucleo in previous chapter ? including a section on interplay between nucleo and rot... Maybe...}

A very fast rotator will go through all the steps while a slow rotator only through the first one. %Maybe not true. All rotation go through all stages but in a different degree}
An important effect (not shown in Fig.~\ref{schemadiff}) is the growth of the convective He-burning core that helps reaching layers that had been previously enriched in H-burning products (e.g. $^{13}$C, $^{14}$N). 
%There is a contribution of both the growing of the convective core and the backward diffusion of chemical elements.
Both the growing of the convective He core and the backward diffusion of chemical elements impact the nucleosynthesis in the He-burning core of rotating models.
 
%\textcolor{red}{maybe remove this:}

Fig.~\ref{mixmodel} shows the results of this mixing process in a 20~$M_{\odot}$ stellar model, when the central $^{4}$He mass fraction is about 0.2. The $^{14}$N and $^{13}$C peaks can be seen and to a smaller extent, the $^{23}$Na peak (at $M_{\rm r} \sim 8$~$M_{\odot}$). In the He-core, $^{22}$Ne has been enhanced because of the ingestion of the extra $^{14}$N. 
Complete stellar models are discussed in more details in Sect.~\ref{sec20}.
 
%\textcolor{red}{ALSO: growing of He core}

\section{Nucleosynthesis in a box \label{secbox}}

\cite{maeder15a} suggested that the wide range of 
%C, N, O, Na, Mg, Al and Sr
abundances covered by CEMP-no stars could come from a material ejected by massive source stars having experienced various degree of rotational mixing. 
In this scenario, the variety of CEMP-no star abundances are mainly explained by the interplay between rotation and nucleosynthesis at work during the core He-burning phase of the massive source star (described in Sect.~\ref{secback}).
%In this scenario, the variety of abundances are mainly made during the core He-burning stage of the source stars. 
\cite{maeder15b} built a new classification scheme for CEMP-no stars by considering the successive steps in the back-and-forth mixing process. Five classes were proposed, the first one showing a complete absence of mixing and the fifth one a high degree of mixing (see Fig.~\ref{schemaclass}). For instance, HE~1327-2326, with [Fe/H] $=-5.7$ \citep{frebel08} and showing strong overabundances in light elements (e.g. CNO) as well as in strontium, is belonging to class 4. In \cite{maeder15b}, they considered 46 stars, 4 appeared to be of class 2, 17 of class 3, 9 of class 4 and 16 unclassified because of missing abundance data. No star belonging to class 0 or 1 were found.

  \begin{figure*}[t]
   \centering
      \includegraphics[scale=0.24, trim = 0cm 0cm 3cm 0cm]{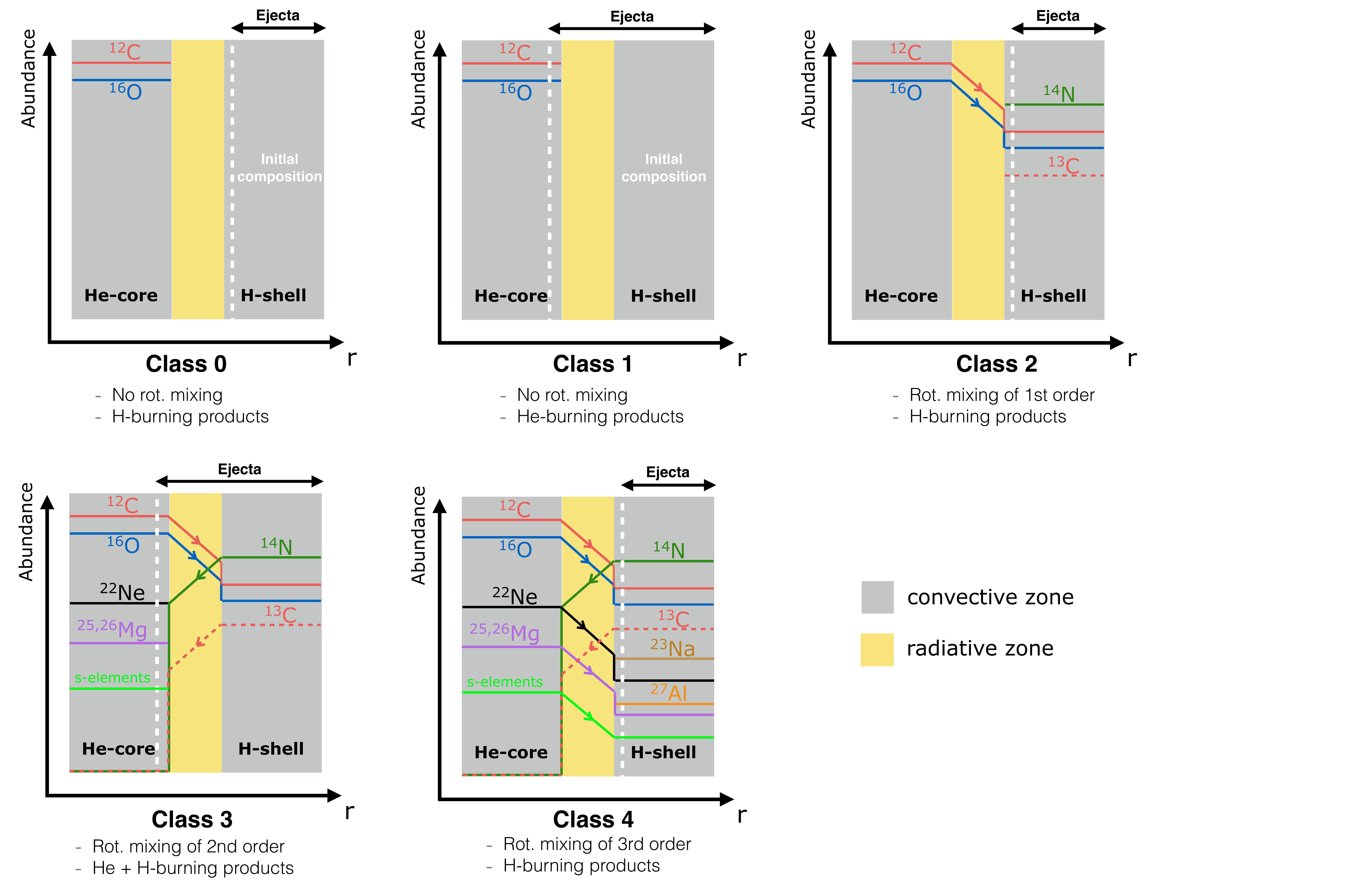}
   \caption[A possible classification of CEMP-no stars]{Five different possible chemical compositions of the CEMP-no source star at the end of the core He-burning phase. Each diagram corresponds to a different degree of mixing (top left: no mixing, bottom right: highest degree of mixing). The corresponding classes of CEMP-no stars are indicated, following the classification (simplified) of \cite{maeder15b}. \textit{Ejecta} denotes the material that should be used to form the CEMP-no star. %It occurs during the core helium burning phase and it is an exchange of chemical species between the helium burning core and the hydrogen burning shell.
   }
\label{schemaclass}
    \end{figure*}

  \begin{figure*}[h!]
   \centering
      \includegraphics[scale=0.9, trim = 0cm 0cm 0cm 0cm]{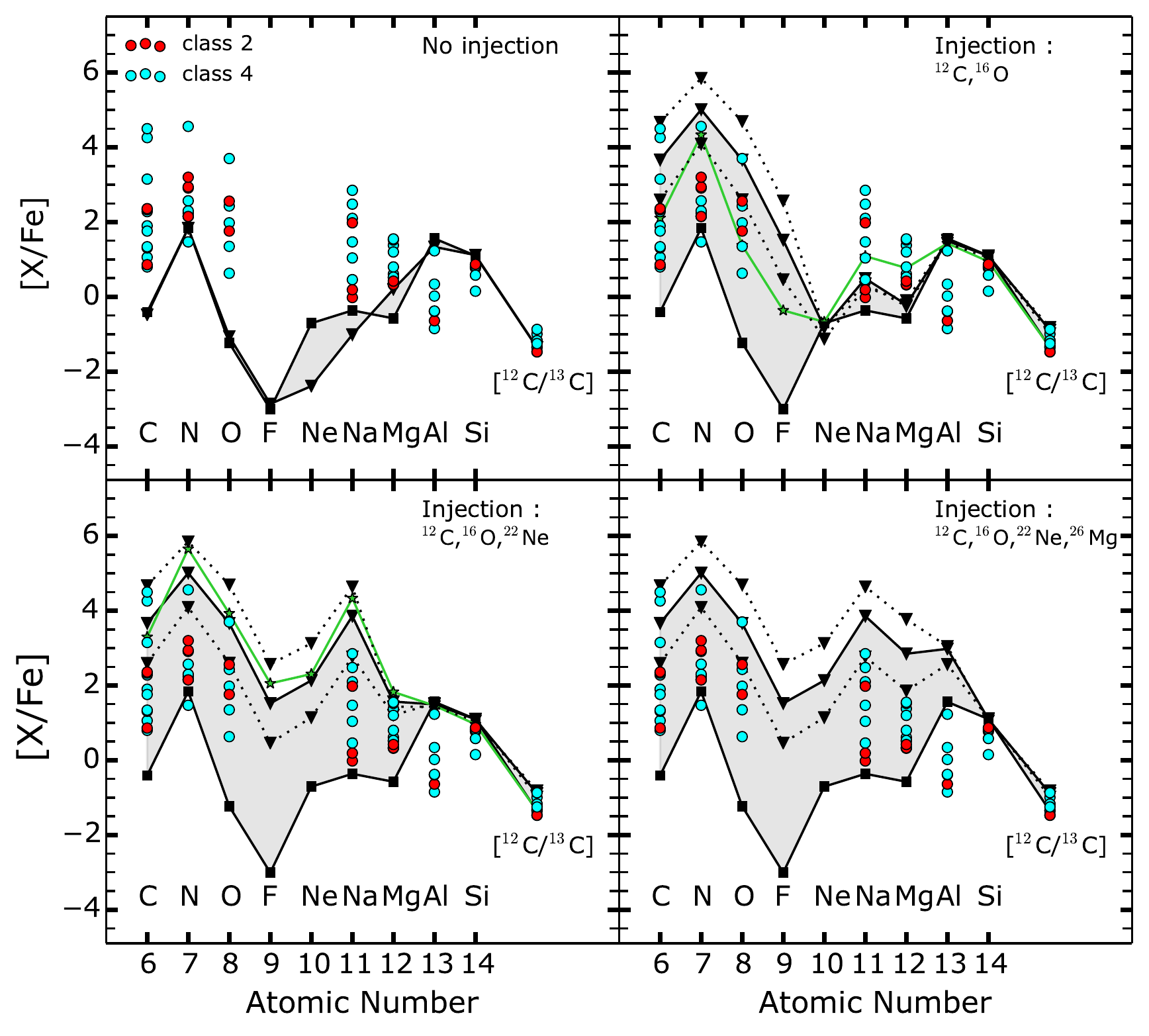}
   \caption[Nucleosynthesis in a one-zone model while injecting $^{12}$C, $^{16}$O, $^{22}$Ne and $^{26}$Mg]{Composition of the box at H-exhaustion (or when the time $t$ exceeds 10 Myr) for four injection cases. The density and temperature are $\rho = 1$ g cm$^{-3}$ and $T_6= 50$ MK. The initial (final) composition is represented by black lines with squares (triangles). The lower (upper) black dotted lines represent the final composition when the rates of injection are divided (multiplied) by 100.
%The shaded area shows the range of ratios reachable if considering dilution with initial ISM. 
The green pattern on the top right panel shows the composition in the H-burning shell at the end of the core He-burning phase of a 20 $ M_{\odot}$ stellar model with $\upsilon_{\rm ini}/\upsilon_{\rm crit} = $ 30~\%. The green pattern on the bottom left panel shows the same stellar model but with $\upsilon_{\rm ini}/\upsilon_{\rm crit} = $ 70~\%. The circles show observed CEMP-no stars of classes 2 and 4 (upper limits are excluded and the typical uncertainty is $\pm 0.3$ dex).% with $-4.1<$ [Fe/H] $<-3.5$ (i.e. around the metallicity [Fe/H] $=-3.8$ of the model). 
%HE 1310-0536, that was missing in the paper, was added here. It does not impact the results and discussions.
   }
\label{injbox}
    \end{figure*}

%\subsection{Nucleosynthesis in a box \label{secbox}}

In \cite{choplin16}, we aimed at investigating quantitatively the effect of the back-and-forth mixing process using a one-zone (or box) nucleosynthesis code I developed. This code allows the injection of chemical species in the box while nucleosynthesis is calculated. It mimics the effect of rotational mixing at work in complete rotating stellar models. In the paper, I used the one-zone code to mimic the hydrogen burning shell of a massive rotating star in which $^{12}$C, $^{16}$O, $^{22}$Ne, and $^{26}$Mg are injected (these species are supposed to come from the He-burning core of the rotating massive star). We studied the nucleosynthesis of the CNO cycle and the Ne-Na Mg-Al chains at different temperatures, densities, and with different nuclear reaction rates while injecting chemical species.
%We report here the main results and discuss additional points. Small improvements were done (especially Fig...)

\subsubsection{Initial setup}

For the initial composition of the box, a rotating 60 $\rm M_{\odot}$ model is used, with $Z = 10^{-5}$ (corresponding to [Fe/H] $=-3.8$) and computed with an $\alpha-$enhanced mixture. 
%\textcolor{red}{talk about initial mixture here or before. Very uncertain but most plausible assumption...?}. 
The initial abundances in the box are taken from the H-burning shell of this stellar model, when it starts the core He-burning stage (central $^{4}$He mass fraction equals to 0.98). The corresponding initial [X/Fe] ratios in the box are shown in Fig.~\ref{injbox} by the patterns with squares. The initial composition of the box is different than the initial composition of the stellar model since some nuclear burning has already operated in the stellar model during the H-burning phase. In particular, CNO burning has operated (that mainly transforms C and O into N) explaining the high initial [N/Fe] and lower [C/Fe] and [O/Fe] in the box.
The nuclear reaction rates of the CNO cycle are from \cite{angulo99} except for $^{14}$N($p,\gamma$)$^{15}$O which is from \cite{mukhamedzhanov03} if $T \leq 0.1$ GK and \cite{angulo99} otherwise. The rates related to the Ne-Na and Mg-Al chains are from \cite{iliadis01}. Only $^{20}$Ne($p,\gamma$)$^{21}$Na and $^{22}$Ne($p,\gamma$)$^{23}$Na are taken from \cite{angulo99} and \cite{hale02}, respectively.
For $20-60$~$M_{\odot}$ stellar models at such metallicity, $T \sim 50$ MK and $\rho \sim 1$ g cm$^{-3}$ in the H-burning shell. There are the default temperature and density taken in the box. 
The box simulations are stopped either at H-exhaustion ($^{1}$H mass fraction is below $10^{-8}$) or when the time exceeds 10 Myr.

Four separate injection experiments were carried out: (1) no injection, (2) injection of $^{12}$C and $^{16}$O, (3) injection of $^{12}$C, $^{16}$O and $^{22}$Ne and (4) injection of $^{12}$C, $^{16}$O, $^{22}$Ne and $^{26}$Mg. The species are injected at a constant rate. %given in~$M_{\odot}$ yr$^{-1}$. 
During a time $\Delta t$, a mass $\Delta m_{\rm i} = R_{\rm i} \Delta t$ under the form of the isotope $i$ is injected in the burning box. $R_{\rm i}$ is the injection rate of the isotope $i$ expressed in $M_{\odot}$~yr$^{-1}$.
$R_{\rm i}$ was calibrated using results from complete stellar models. 
Details on this calibration are now given \citep[see also the appendix of][page~\pageref{pbox} of this thesis]{choplin16}.

\subsubsection{Injection of chemical species}

The rate $R_{\rm ^{12}C}$ at which $^{12}$C is injected in the box was calibrated by quantifying $M_{\rm ^{14}N}^{\rm prim}$ which is the mass of primary $^{14}$N formed during the core helium burning phase of a rotating massive star. 
Primary $^{14}$N during the core helium burning phase of rotating stars is formed because of the arrival of $^{12}$C and $^{16}$O from the helium core (cf. Sect.~\ref{secback}).
$M_{\rm ^{14}N}^{\rm prim}$ can be expressed as the total amount of $^{14}$N in the star at core helium exhaustion minus the amount of $^{14}$N that can be formed with the initial CNO content (the secondary $^{14}$N):
\begin{equation}
M_{\rm ^{14}N}^{\rm prim} =  \left( \int_{0}^{M} X_{\rm ^{14}N}(M_{\rm r}) \, \mathrm{d}M_{\rm r} \right)_{\rm Y_{\rm c} = 0} - M~(X_{\rm C,ini} + X_{\rm N,ini} + X_{\rm O,ini})
\end{equation}
where $Y_{\rm c}$ is the central $^{4}$He mass fraction, $X_{\rm ^{14}N}(M_{\rm r})$ the mass fraction of $^{14}$N at coordinate $M_{\rm r}$, $X_{\rm C,ini}$, $X_{\rm N,ini}$, $X_{\rm O,ini}$ the initial mass fractions of the CNO elements in the stellar model, and $M$ the total mass of the star at the end of the core helium burning phase.
Let us suppose that all the $^{12}$C and $^{16}$O diffusing from the helium core to the hydrogen shell are transformed into $^{14}$N. 
In this case, an average injection rate $R_{\rm ^{12}C + ^{16}O}$ of $^{12}$C + $^{16}$O can be defined
%to get a mass $M_{\rm ^{14}N}^{\rm prim}$ of primary nitrogen in the star at the end of core helium burning, an average injection rate of ($^{12}$C + $^{16}$O) is needed in the hydrogen shell of
\begin{equation}
R_{\rm ^{12}C + ^{16}O} = \frac{M_{\rm ^{14}N}^{\rm prim}}{\tau_{\rm HeB}  }
\end{equation}
where $\tau_{\rm HeB}$ is the duration of the core helium burning phase. 
It gives the average rate at which $^{12}$C + $^{16}$O should be injected in the H-burning shell so as to obtain $M_{\rm ^{14}N}^{\rm prim}$ in the model at the end of core helium burning. 
For a 60 $\rm M_{\odot}$ model at $Z=10^{-5}$ and with\footnote{The critical velocity $\upsilon_{\rm crit}$ is reached when the gravitational acceleration is counterbalanced by the centrifugal force. It is expressed as $\upsilon_{\rm crit} = \sqrt{\frac{2}{3}\frac{GM}{R_{\rm p,c}}}$ with $R_{\rm p,c}$ the polar radius at the critical limit.} $\upsilon_{\rm ini} / \upsilon_{\rm crit} = 0.7$, $R_{\rm ^{12}C + ^{16}O}$ is equal to $5~10^{-8}$~$M_{\odot}$~yr$^{-1}$.

Of course, the amount of primary nitrogen synthesized, hence the value of $R_{\rm ^{12}C + ^{16}O}$, can change significantly depending on the rotation, the mass of the stellar model, or the prescription for the rotational mixing for instance.
The $R_{\rm ^{12}C + ^{16}O}$ rate estimated above gives an idea of possible values but other injection rates around this value are possible.
Here I set $R_{\rm ^{12}C} = R_{\rm ^{16}O} = 10^{-8}$~$M_{\odot}$~yr$^{-1}$. The two rates are taken equal since there is roughly as much $^{12}$C as $^{16}$O in the helium core so that roughly as much $^{12}$C as $^{16}$O will enter in the H-burning shell. 
Finally, $R_{\rm ^{22}Ne} = R_{\rm ^{26}Mg} = R_{\rm ^{12}C} / 100 = 10^{-10}$~$M_{\odot}$~yr$^{-1}$. The factor 100 between the two rates accounts for the fact that $^{22}$Ne and $^{26}$Mg are about 100 times less abundant than $^{12}$C and $^{16}$O in the helium burning core so that about 100 times less $^{22}$Ne and $^{26}$Mg will enter by rotational mixing into the H-burning shell.
%In the paper (page \pageref{pbox}), 
Injection rates divided and multiplied by 100 were also investigated: $10^{-10}$ and $10^{-6}$~$M_{\odot}$~yr$^{-1}$ for $R_{\rm ^{12}C}$ and $R_{\rm ^{16}O}$ and therefore $10^{-12}$ and $10^{-8}$~$M_{\odot}$~yr$^{-1}$ for $R_{\rm ^{22}Ne}$ and $R_{\rm ^{26}Mg}$.

\subsubsection{Results}

As shown in Fig.~\ref{injbox}, C, N, O and F are boosted by $\sim 4$ dex when injecting $^{12}$C, $^{16}$O because of the injection and the effect of CNO cycle. Injection of $^{22}$Ne gives Ne, Na and Mg with Na being the main product. With the nuclear rates considered, the $^{23}$Na($p,\alpha$)$^{20}$Ne reaction is just 1.6 stronger than the $^{23}$Na($p,\gamma$)$^{24}$Mg reaction. It means that $^{23}$Na goes almost 50$-$50 to $^{20}$Ne and $^{24}$Mg. It allows a significant production of $^{24}$Mg. No Al is produced because the $^{24}$Mg($p,\alpha$)$^{25}$Al reaction, activating the Mg-Al cycle, is too slow at the considered temperature. Injection of $^{26}$Mg boosts Mg and Al. The [$^{12}$C/$^{13}$C] ratio\footnote{[$^{12}$C/$^{13}$C] = $\log$($^{12}$C/$^{13}$C)$_{\star}$ $-$ $\log$($^{12}$C/$^{13}$C)$_{\odot}$. In the Sun, ($^{12}$C/$^{13}$C$)_{\odot}$ = 89, i.e. $\log$($^{12}$C/$^{13}$C)$_{\odot}$ = 1.95. } stays always around $-1.5$ (equivalent to $^{12}$C/$^{13}$C $\sim 3$), which is the CNO-equilibrium value. At $T = 50$ MK, the timescale for the $^{12}$C/$^{13}$C ratio to reach CNO-equilibrium is about $10$ yr, i.e. reached almost instantaneously.

This simple one-zone model gives a good estimation of what happens in complete stellar models: green patterns in Fig.~\ref{injbox} are from complete 20~$M_{\odot}$ stellar models with $\upsilon_{\rm ini}/\upsilon_{\rm crit} =$ 30~\% and 70~\%, at the end of the core He-burning phase. The abundances are taken in the stellar shell where the energy released by hydrogen burning is the highest. These patterns are close enough to the results obtained with the box model.

\subsubsection{Comparison with CEMP-no stars of classes 2 and 4}

From the classification done in \cite{maeder15b}, the classes 2 and 4 CEMP-no stars were selected (13 stars). These classes of stars are made of a material that was processed by H-burning, with various degree of enrichment (Fig.~\ref{schemaclass}). I considered only stars having a [Fe/H] ratio close to the [Fe/H] $=-3.8$ of the box. 
HE~0057-5959 was classified as a class 2+Na star since it shows the characteristics of a class 2 star, except for Na, which is overabundant ([Na/Fe] = 1.98). 

The best match between models and observations is when $^{12}$C and $^{16}$O are injected for the class 2 and when $^{12}$C, $^{16}$O and $^{22}$Ne are injected for the class 4. Injecting only $^{12}$C and $^{16}$O does not allow to reproduce the stars of class 4, which are generally Na- and Mg-rich. Injecting additional $^{26}$Mg raises the [Al/Fe] ratio and leads to a larger discrepancy between models and observations. In any case, the models overestimate the [Al/Fe] ratios. Below are discussed the uncertainties on the observed and predicted [Al/Fe] ratios.

\paragraph{Uncertainties on the aluminum abundance of CEMP-no stars.}
The first point is that the CEMP sample is not homogeneous\footnote{A sample is homogeneous if the same procedure is used to obtain all observations and abundances.}. The abundance data considered comes from several sources. Ideally, an homogeneous sample is needed because a mixture of abundances from various authors sometimes adds an undesirable scatter that makes harder the use of the sample to constrain the models. However very metal-poor stars are very rare and no large homogeneous sample exists yet.
The second point is that in the sample considered, most of the abundances are derived based on 1D LTE atmosphere models.
For turnoff stars at [Fe/H] $\simeq -3$, \cite{andrievsky08} derived a correction $\Delta$[Al/Fe] $=$ [Al/Fe]$_{\rm NLTE}$ $-$ [Al/Fe]$_{\rm LTE} \simeq 0.6$ dex. In addition to NLTE, 3D effects can add another correction factor. \cite{nordlander17} have shown that the difference between 3D NLTE and 1D LTE for Al in metal-poor stars can amount $+0.5$ dex.
In the stars plotted in Fig.~\ref{injbox}, there is for instance CS~22949-037 with [Al/Fe] $=0.02$. This ratio was derived based on 1D LTE model-atmosphere analyses \citep{norris13}. According to the previous discussion, NLTE/3D effects could rise [Al/Fe] up to $\sim 0.6$ dex. This is still $\sim 1$ dex below the values predicted by the models.

\paragraph{Uncertainties on the aluminum abundance predicted by the models.}
As mentioned, the initial [Al/Fe] $\simeq 1.5$ in the box comes from the H-burning shell of a complete stellar model at core He-ignition. The initial [Al/Fe] of this stellar model is 0 (i.e. solar). The nucleosynthesis during the main sequence and very beginning of core He-burning has synthesized some Al (especially through the Mg-Al chain), up to [Al/Fe] $\simeq 1.5$. I have investigated how the Al production is affected (considering both the stellar model and the box model) when varying some specific nuclear reaction rates.
Several literature sources provide nuclear reaction rates for the three major reactions involving $^{27}$Al in H-burning zones: $^{26}$Mg($p,\gamma$)$^{27}$Al, $^{27}$Al($p,\gamma$)$^{28}$Si and $^{27}$Al($p,\alpha$)$^{24}$Mg. For instance, there are six sources for $^{27}$Al($p,\gamma$)$^{28}$Si spanning $\sim 4$ order of magnitude at $T \sim 50$ MK (according to the JINA reaclib\footnote{\href{http://jinaweb.org/reaclib/db/}{http://jinaweb.org/reaclib/db/}.}). 
%For the reaction $^{27}$Al($p,\gamma$)$^{28}$Si there is a factor of about $10^4$ (at 50 MK) and about  
Using the rates in the literature that favor the destruction (or smallest production) of $^{27}$Al leads to a final [Al/Fe] in the box of about 0, i.e. $\sim 1.5$ dex below the [Al/Fe] ratio shown in Fig.~\ref{injbox}. It would match better the observations. Finally, because of important uncertainties existing on some nuclear rates, caution is required when interpreting the nucleosynthetic predictions of some elements like Al. 
%\textcolor{red}{However, the scatter between nuclear rates of different source generally decreases with increasing temperature. The H-burning shell of a massive star spans a range of temperatures between about 30 and 80 MK. The nucleosynthesis at 80 MK is somewhat less uncertain than the nucleosynthesis at lower temperatures. The effect of changing the nuclear rates on a stellar model is complex and cannot be inferred just by using the one-zone model.}

\paragraph{The $^{12}$C/$^{13}$C ratio.}
The CEMP-no stars in Fig.~\ref{injbox} have low $^{12}$C/$^{13}$C ratios compared to the Sun: $-1.5 \lesssim$ [$^{12}$C/$^{13}$C] $\lesssim -1$, which is equivalent to $3 \lesssim$ $^{12}$C/$^{13}$C $\lesssim 9$ \citep[in the solar system, the carbon isotopic ratio is about 90,][]{lodders03}. Some CEMP-no stars are giants and may therefore have experienced mixing events like the first dredge-up. In this case, the observed $^{12}$C/$^{13}$C may have changed since the formation of the CEMP-no star. If so, the link between the CEMP-no star and its source star is more difficult to establish (further discussions in Sect.~\ref{secdup}, cf. also Sect.~\ref{selfenr}). 
By contrast, the surface composition of unevolved CEMP-no stars reflects more directly the composition of the cloud in which they formed, hence the composition of the material ejected by the previous source star.
Some CEMP-no stars in the considered sample are still rather unevolved: CS~22958-042 for instance, has $T_{\rm eff} = 6250$ K, $\log~g=3.5$ and $^{12}$C/$^{13}$C $=9$. Such a $^{12}$C/$^{13}$C ratio is consistent with a material processed only by H-burning in the previous massive source star. In He-burning regions of massive stars, $^{13}$C is destroyed so that $^{12}$C/$^{13}$C $\sim \infty$. Consequently, it is unlikely that this CEMP-no star formed from a source star material processed by He-burning. This will be discussed in more details while considering yields from complete source star models (Sect.~\ref{sec20} and~\ref{compacemp}).

%NLTE effects is another source of uncertainties that can affect the [Al/Fe] ratio up to about 0.5 dex \citep{baumueller97, andrievsky08}
\subsubsection{Conclusion}

The conclusion from this box experiment is that the range of abundances of class 2 and 4 CEMP-no stars is overall well reproduced by a material processed by H-burning at a temperature and density characteristic of $20-60$~$M_{\odot}$ source stars. During its burning, the material was enriched in He-burning products: $^{12}$C, $^{16}$O for class 2, $^{12}$C, $^{16}$O and $^{22}$Ne for class 4. This is consistent with the scenario of \cite{maeder15b} proposing that CEMP-no stars are made of a material ejected from a massive star having experienced various degree of rotational mixing.

%=====================================================================================================================================================================
%\section{A grid of massive source stars with various initial velocity \label{sec20}}
\section{Massive source stars with rotation\label{sec20}}

As a next step, complete stellar models with various initial rotation rates are considered. It allows to give more realistic abundance predictions than with a one-zone model. Other source star parameters such as mass and metallicity are also investigated.

\subsection{Physical ingredients}

  \begin{figure*}[t]
   \centering
      \includegraphics[scale=0.68, trim = 0cm 0cm 0cm 0cm]{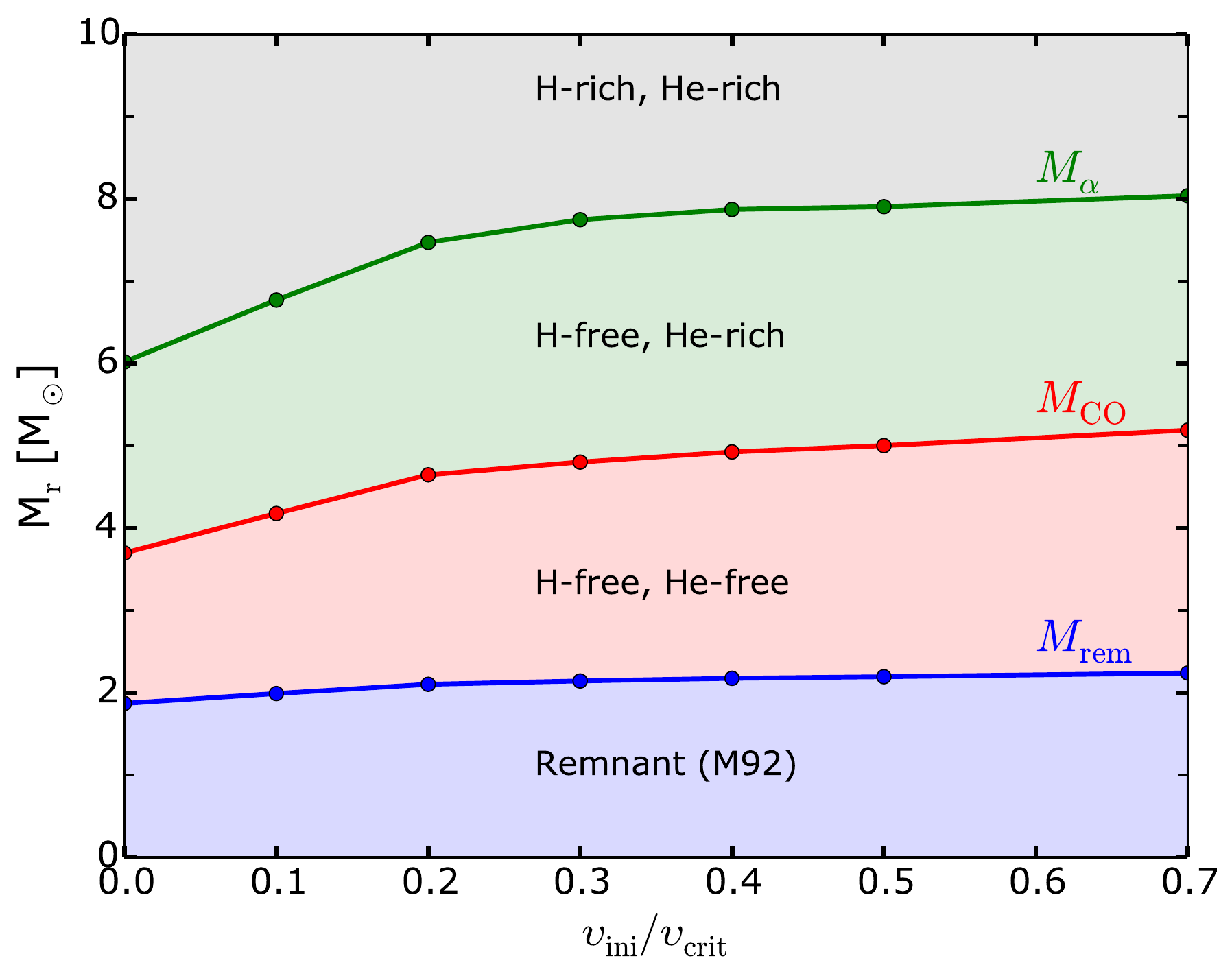}
   \caption[Masses of the cores as a function of initial velocity for 20~$M_{\odot}$ model with $Z=10^{-5}$]{Masses of the various cores as a function of initial velocity for 20~$M_{\odot}$ models with $Z=10^{-5}$, at the end of the core carbon burning phase. $M_{\alpha}$ denotes the $\alpha$-core (defined where the mass fraction of $^{1}$H drops below $10^{-3}$), $M_{\rm CO}$ the CO-core (defined where the mass fraction of $^{4}$He drops below $10^{-3}$) and $M_{\rm rem}$ shows the remnant mass \citep[from the relation of][that links the mass of the CO-core with the mass of the remnant]{maeder92}.}
\label{cores20}
    \end{figure*}

   \begin{figure*}
   \centering
   \begin{minipage}[c]{.49\linewidth}
       \includegraphics[scale=0.44]{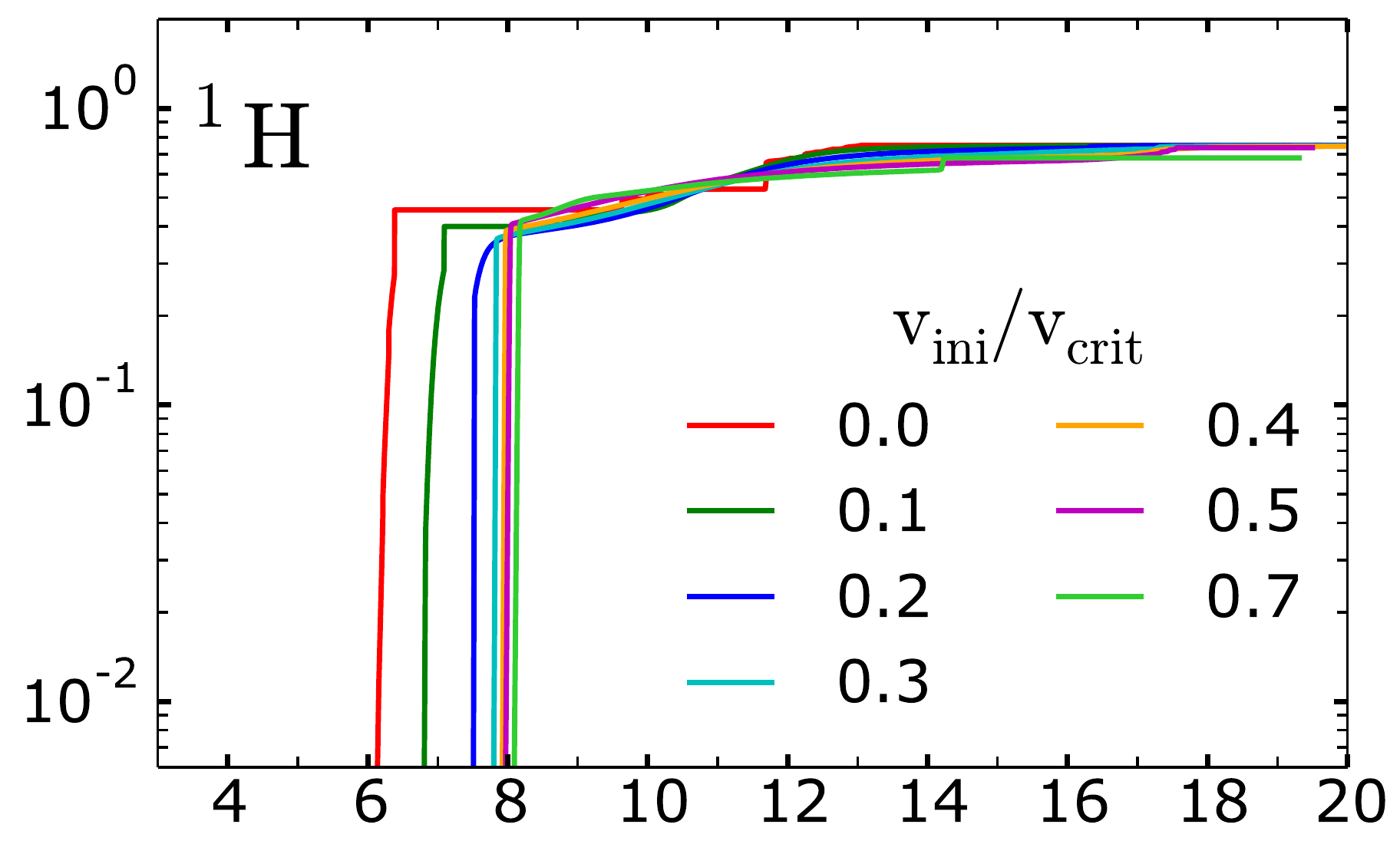}
   \end{minipage}
   \begin{minipage}[c]{.49\linewidth}
       \includegraphics[scale=0.44]{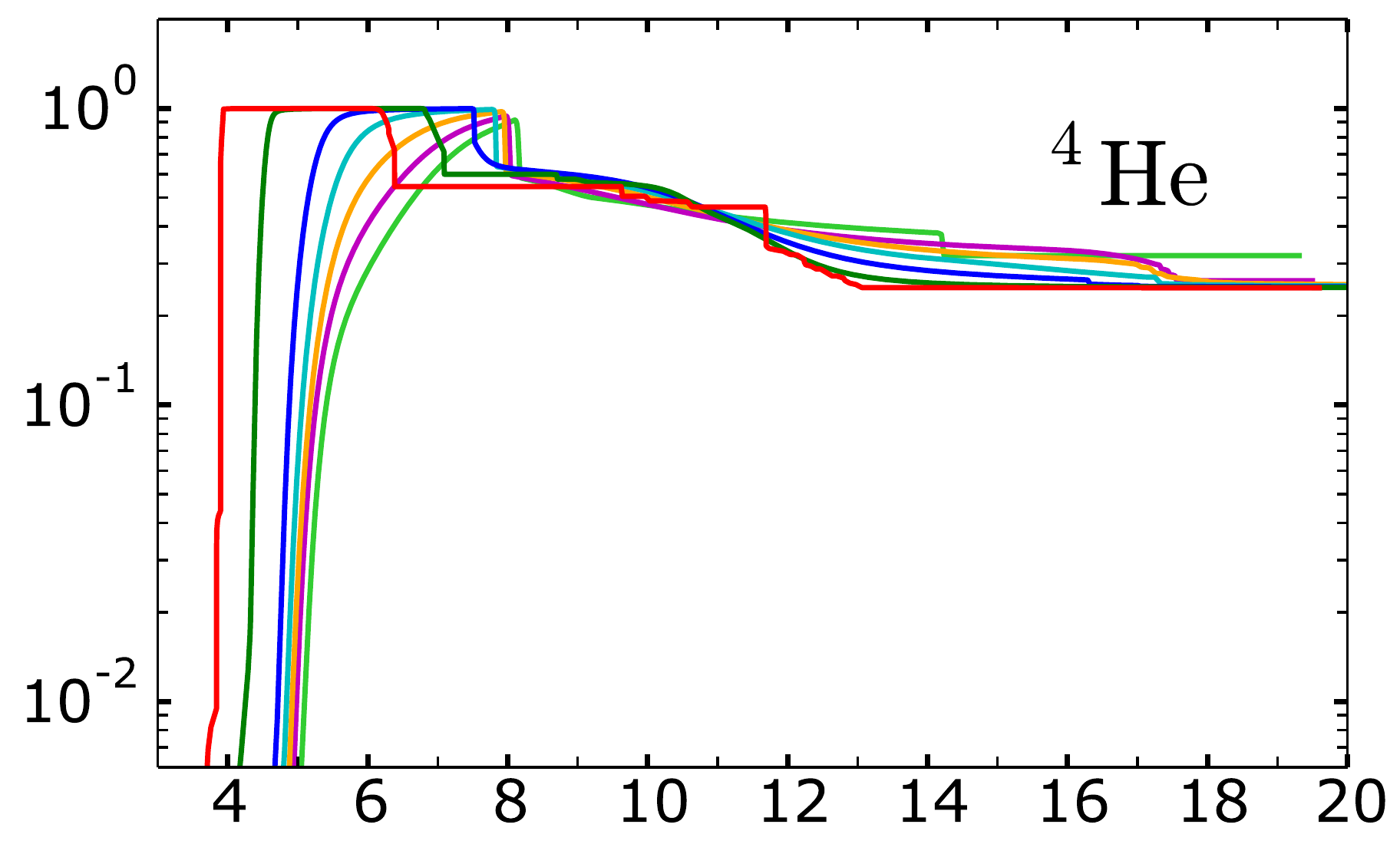}
   \end{minipage}
   \begin{minipage}[c]{.49\linewidth}
       \includegraphics[scale=0.44]{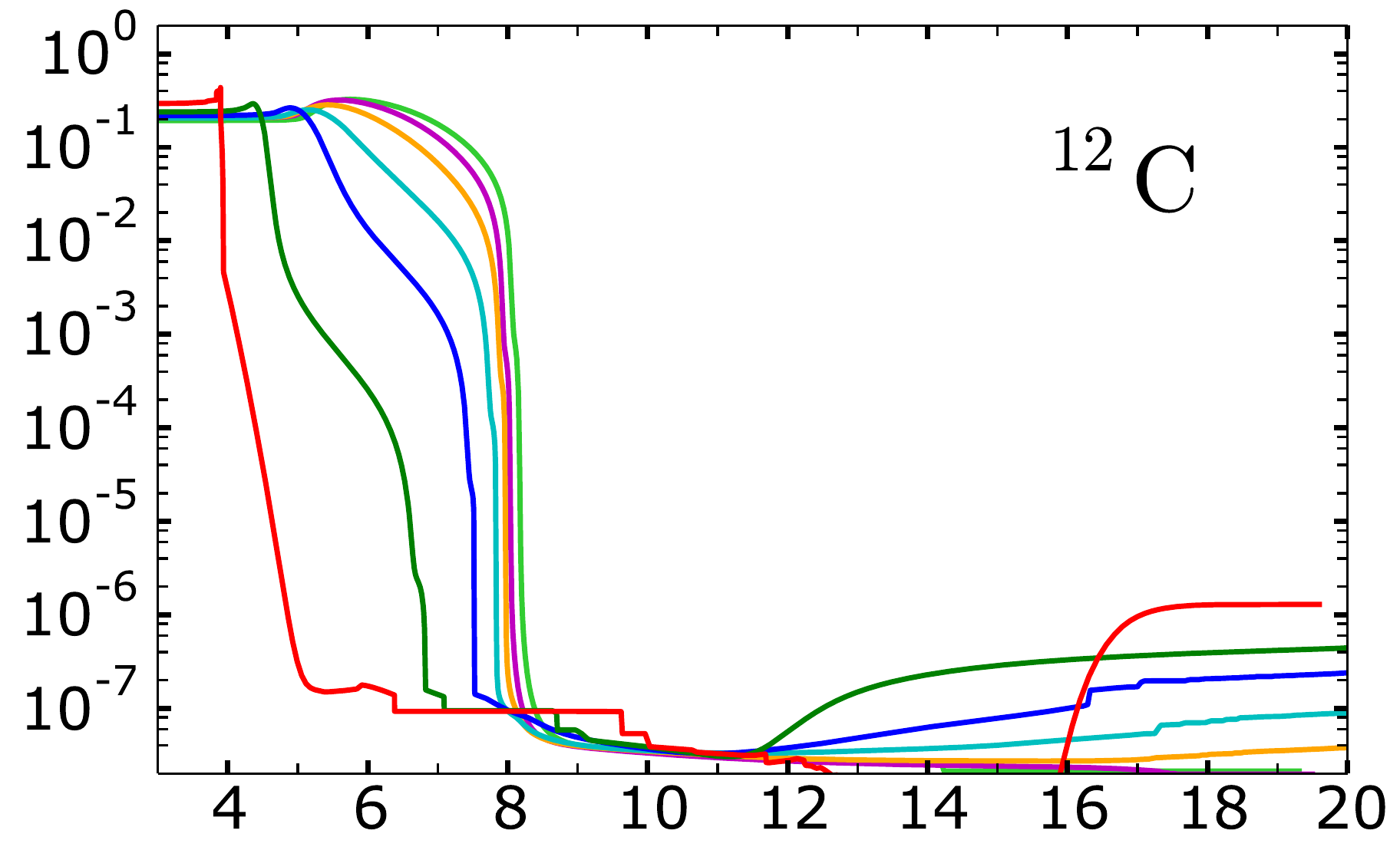}
   \end{minipage}
   \begin{minipage}[c]{.49\linewidth}
       \includegraphics[scale=0.44]{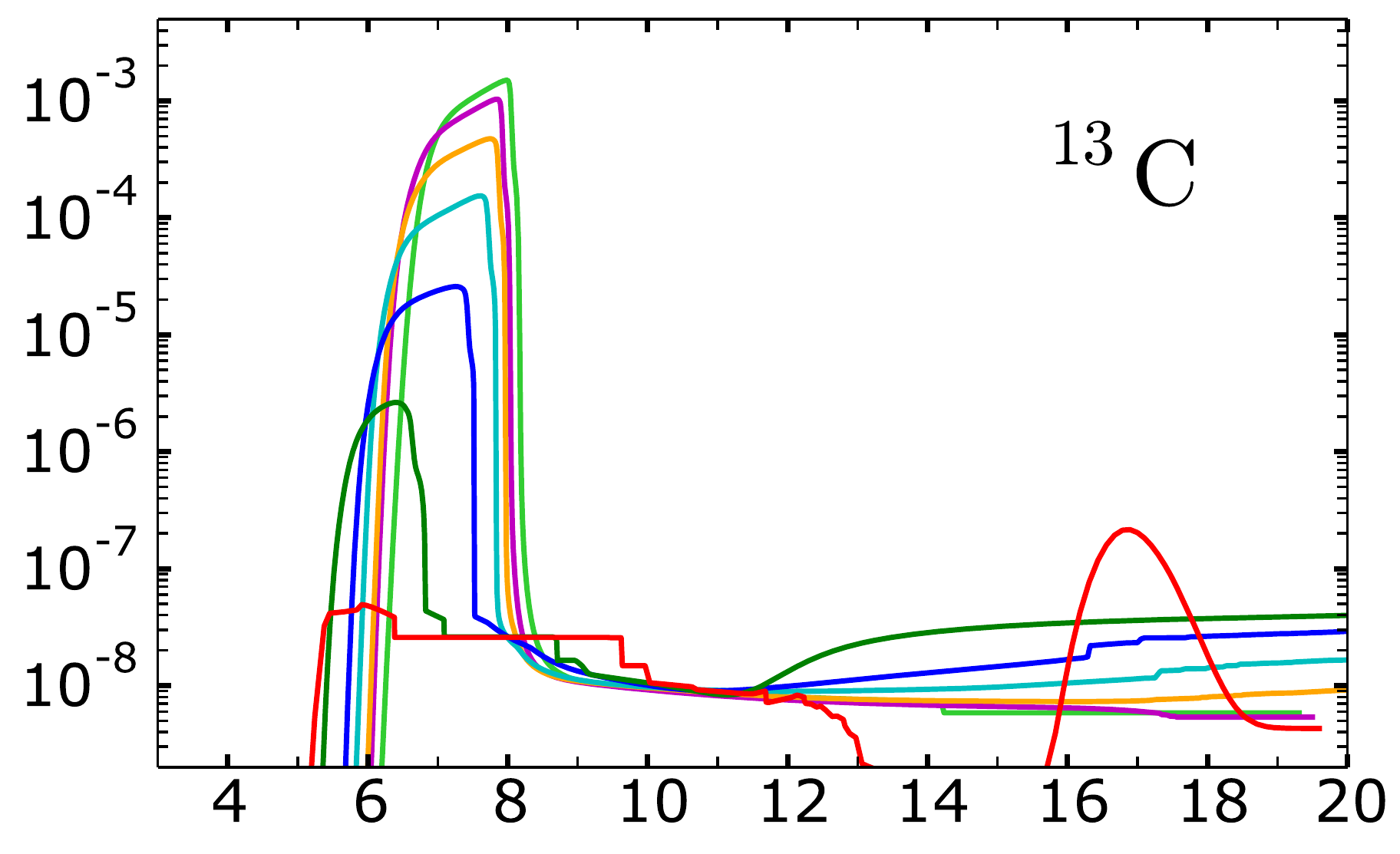}
   \end{minipage}
   \begin{minipage}[c]{.49\linewidth}
       \includegraphics[scale=0.44]{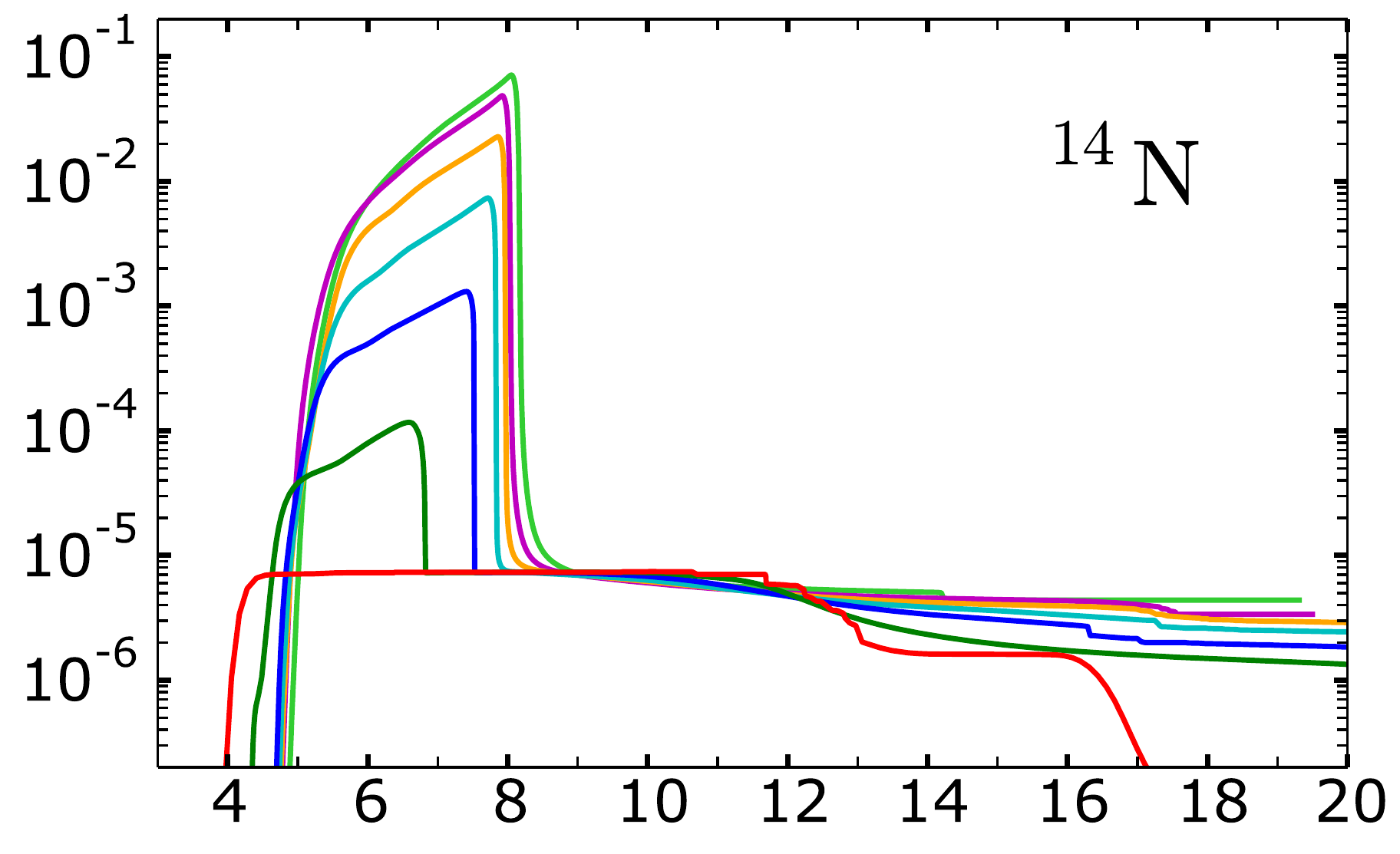}
   \end{minipage}
   \begin{minipage}[c]{.49\linewidth}
       \includegraphics[scale=0.44]{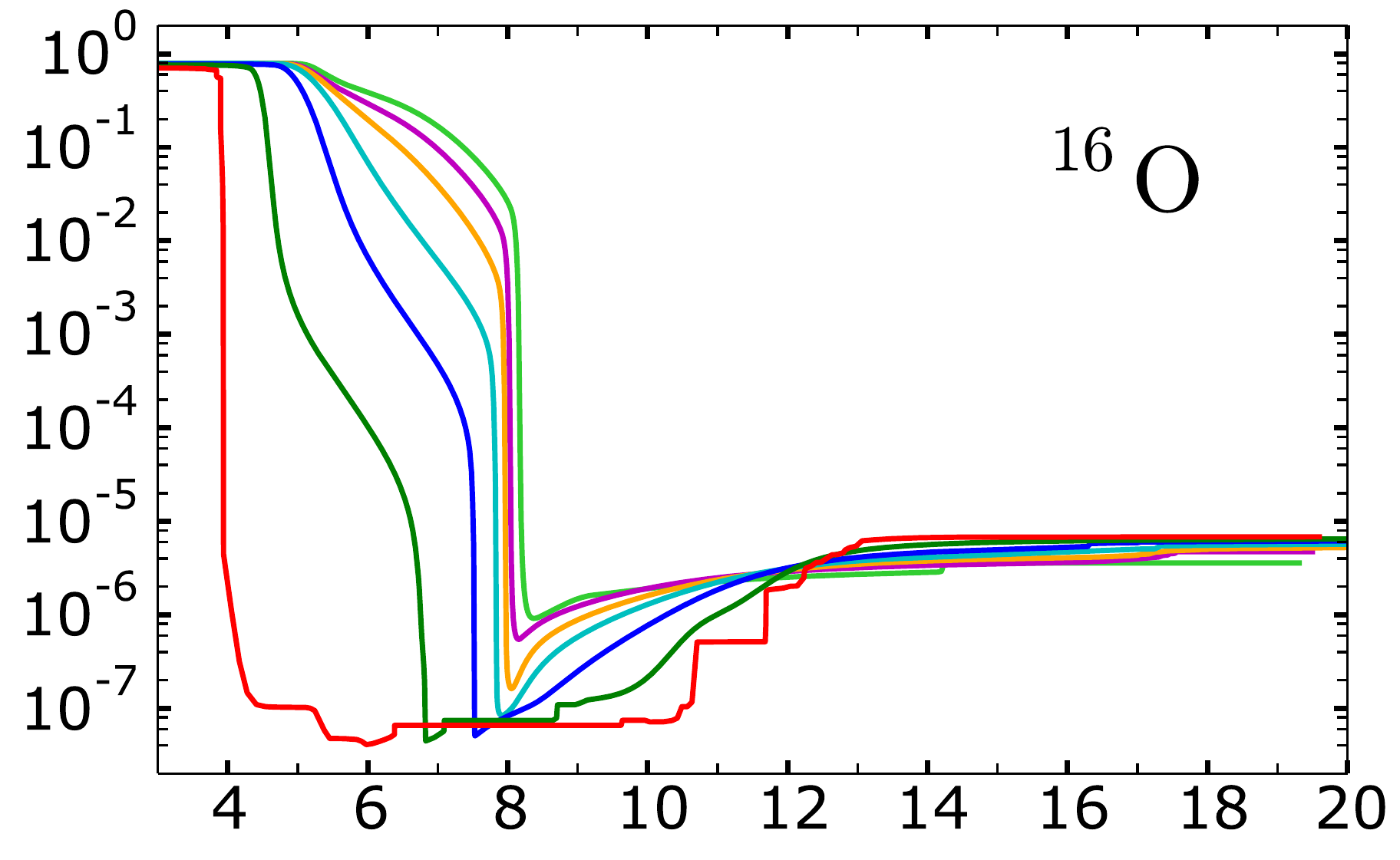}
   \end{minipage}
   \begin{minipage}[c]{.49\linewidth}
       \includegraphics[scale=0.44]{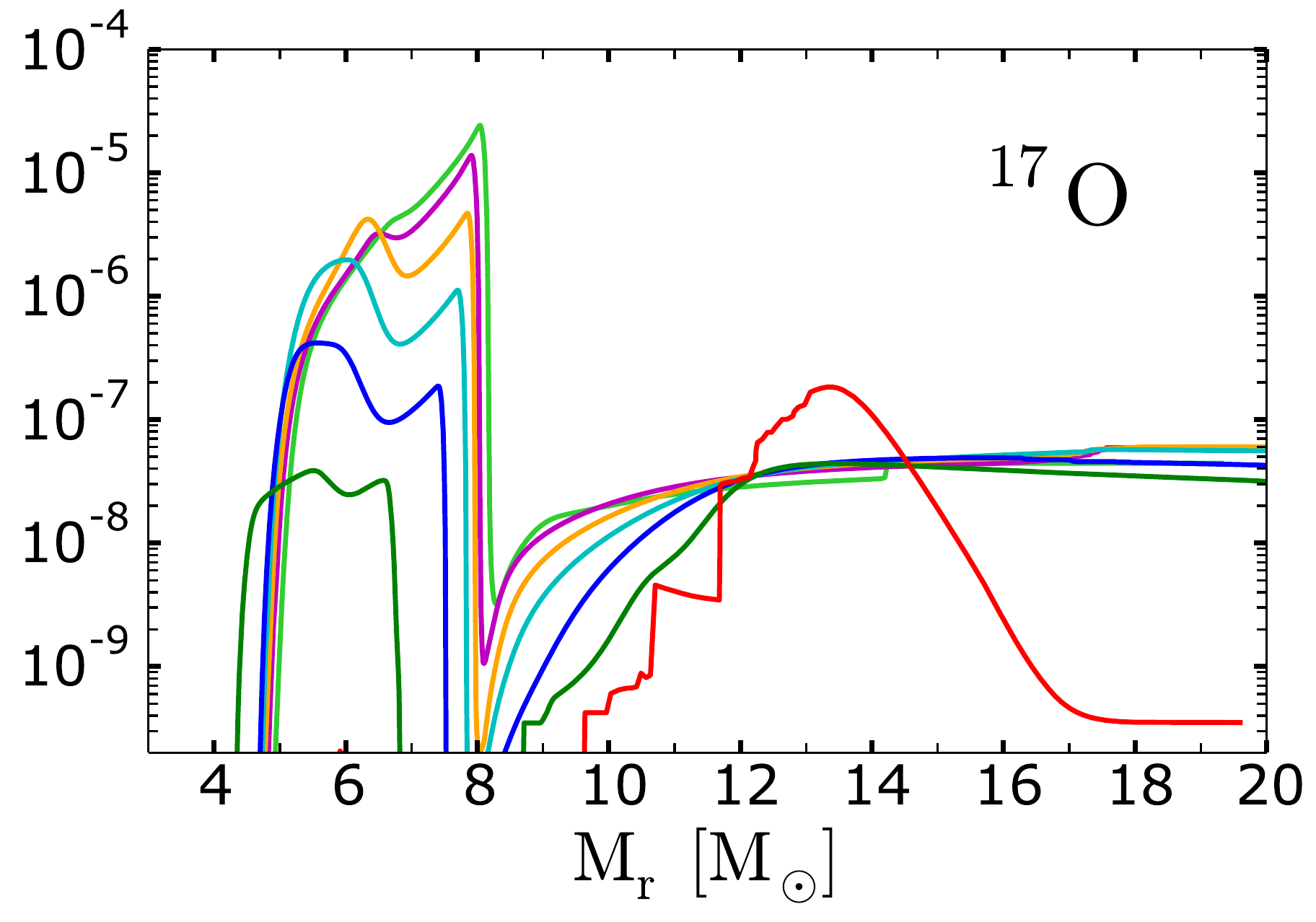}
   \end{minipage}
   \begin{minipage}[c]{.49\linewidth}
       \includegraphics[scale=0.44]{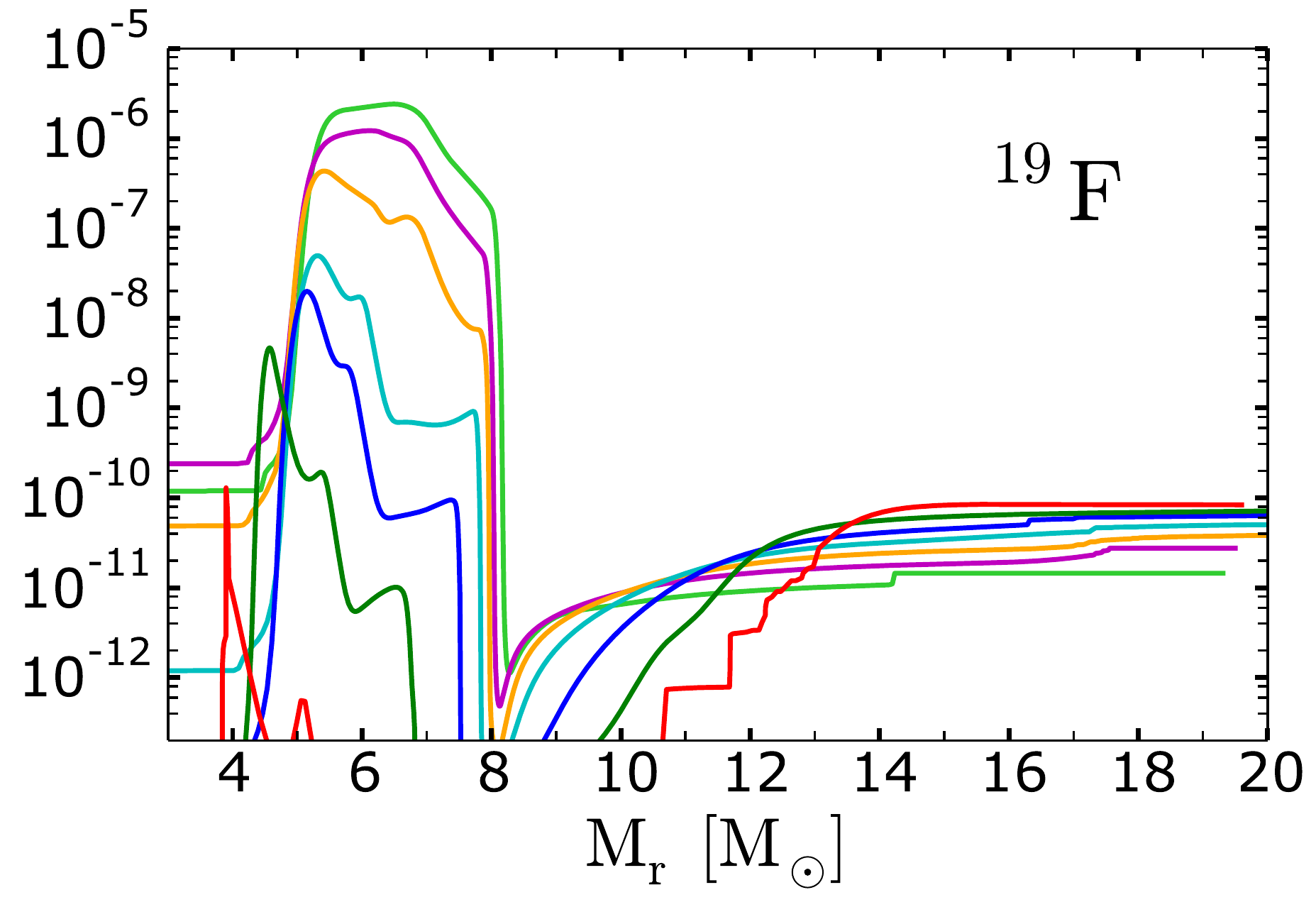}
   \end{minipage}
   \caption[Internal abundance profiles of 20~$M_{\odot}$ models at $Z=10^{-5}$ with various initial rotations]{Internal abundance profiles in mass fraction of 20~$M_{\odot}$ models at $Z=10^{-5}$ with various initial rotation rates, at the end of the core helium burning phase, between $M_{\rm r} = 3$ and 20~$M_{\odot}$. %$M_{\alpha}$ and $M_{\rm CO}$ are shown in the top panels for the non rotating model.
   }
\label{abrotmod}
    \end{figure*}

   \begin{figure*}
   \centering
   \begin{minipage}[c]{.49\linewidth}
       \includegraphics[scale=0.44]{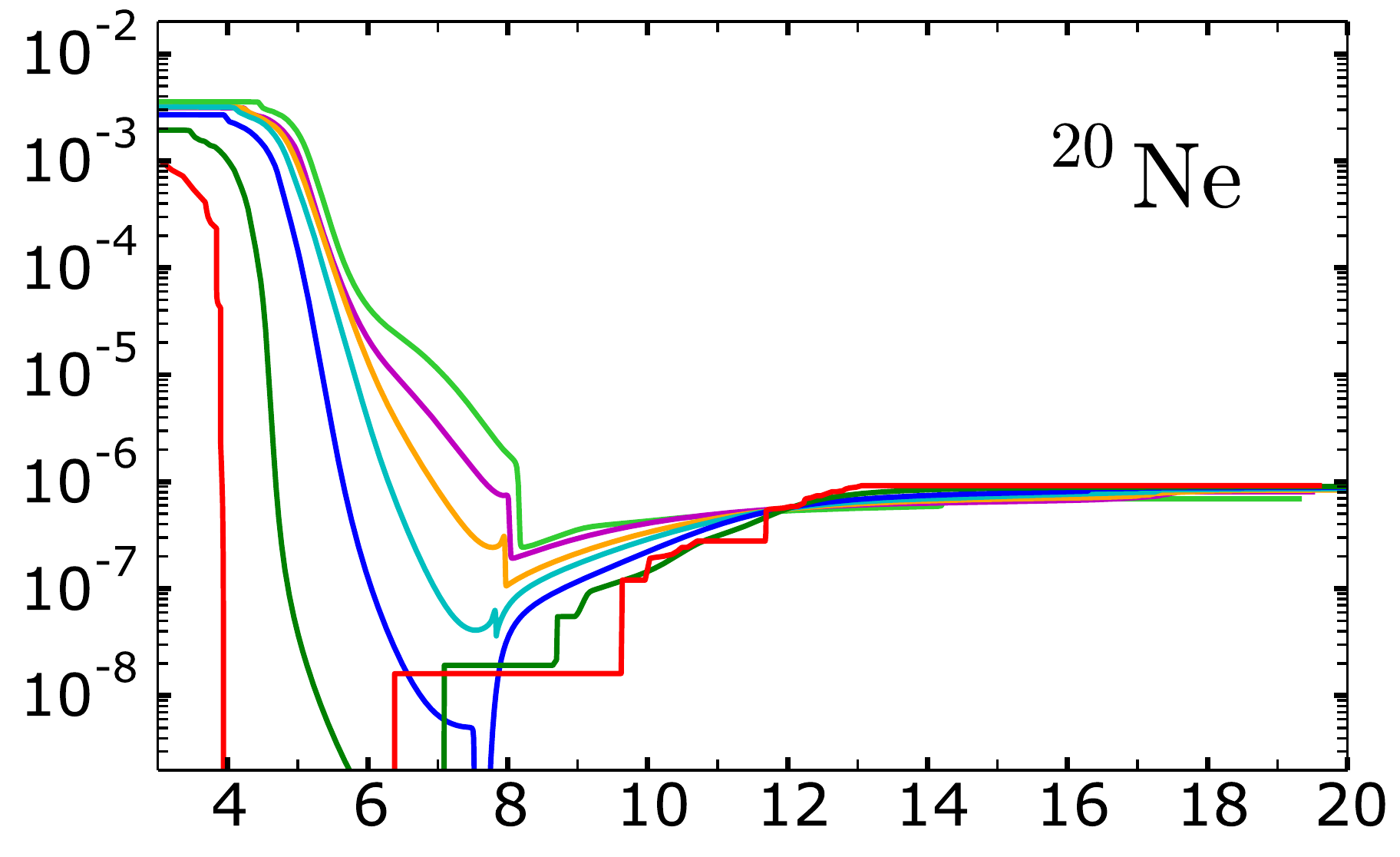}
   \end{minipage}
   \begin{minipage}[c]{.49\linewidth}
       \includegraphics[scale=0.44]{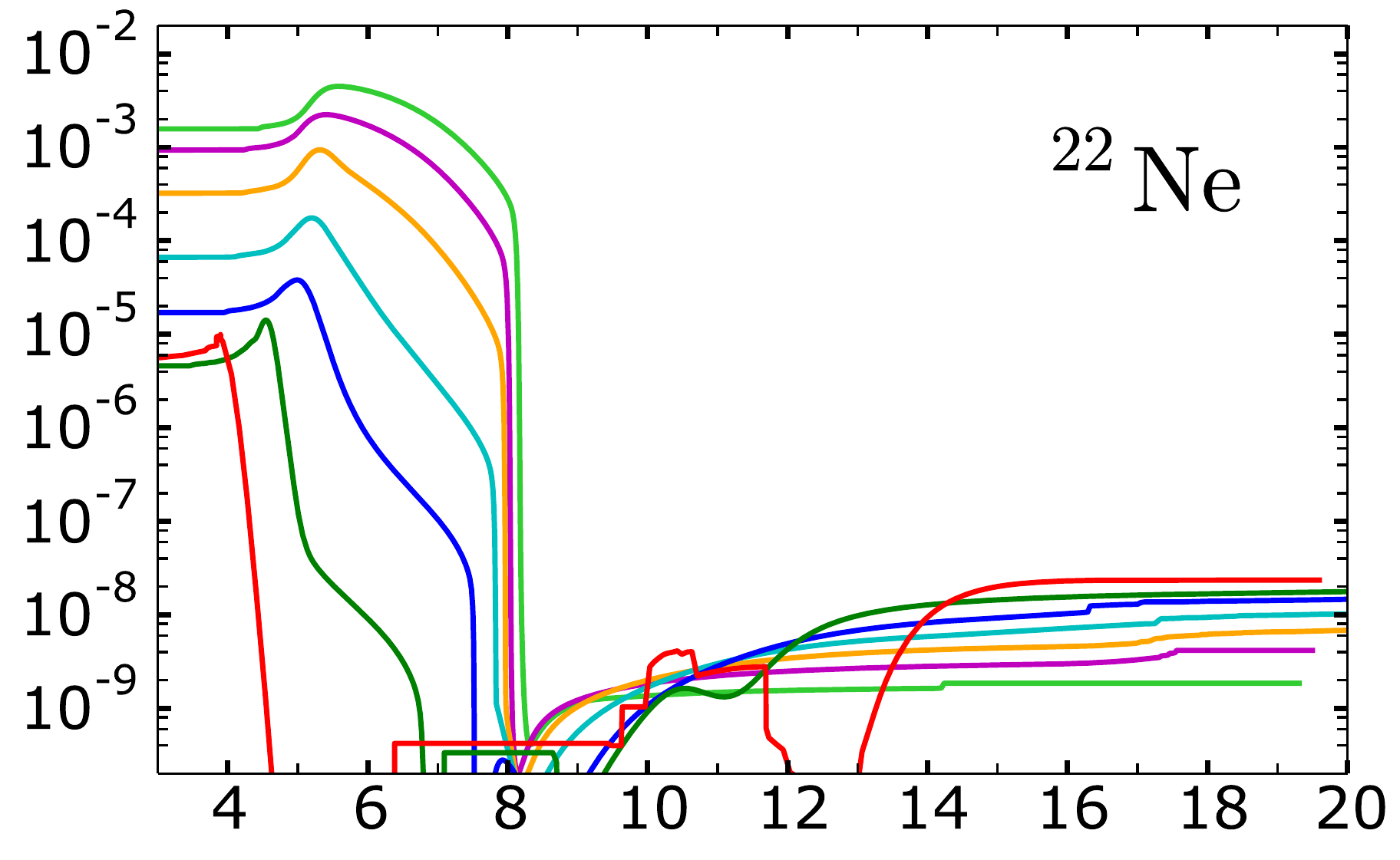}
   \end{minipage}
   \begin{minipage}[c]{.49\linewidth}
       \includegraphics[scale=0.44]{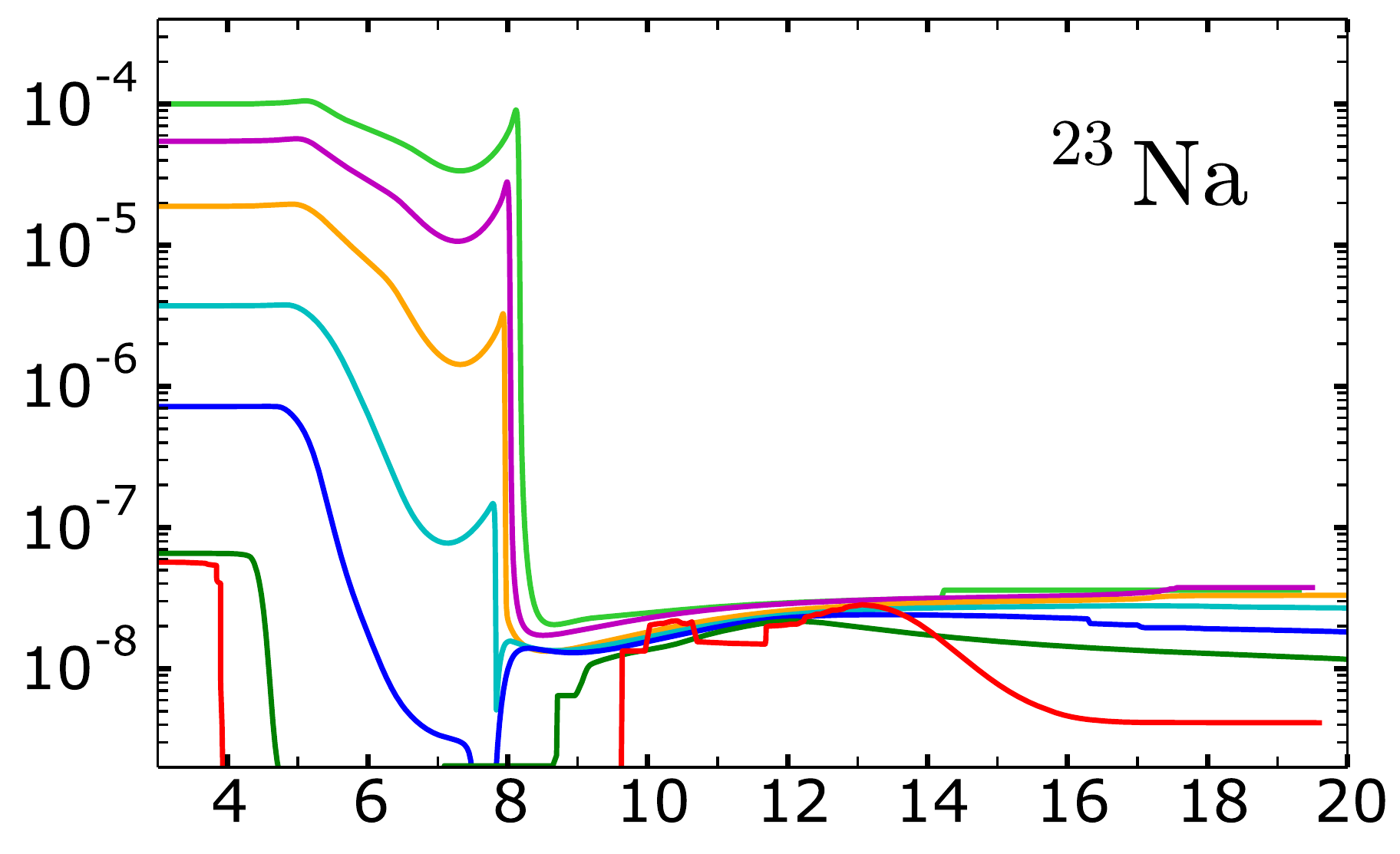}
   \end{minipage}
   \begin{minipage}[c]{.49\linewidth}
       \includegraphics[scale=0.44]{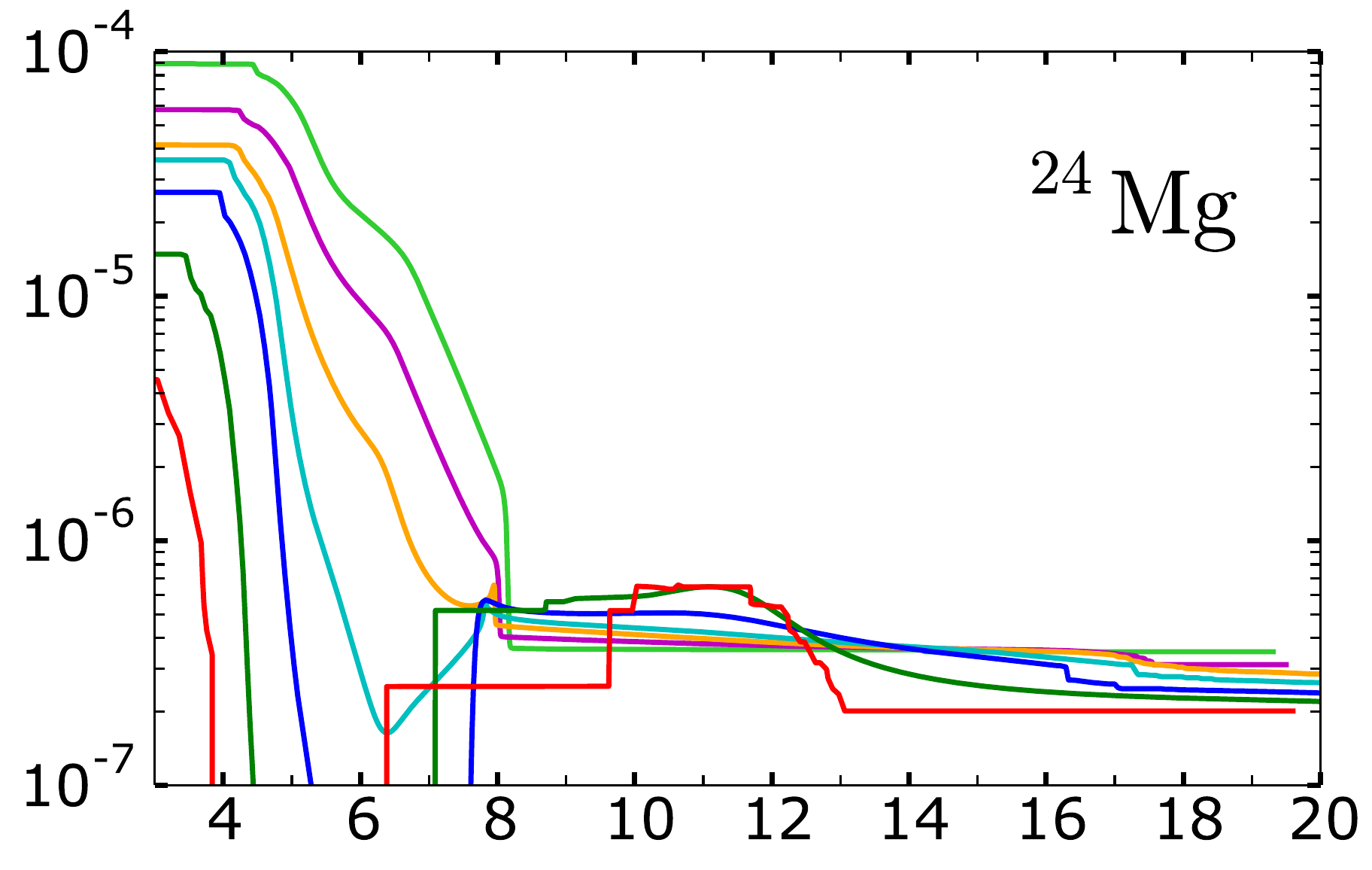}
   \end{minipage}
   \begin{minipage}[c]{.49\linewidth}
       \includegraphics[scale=0.44]{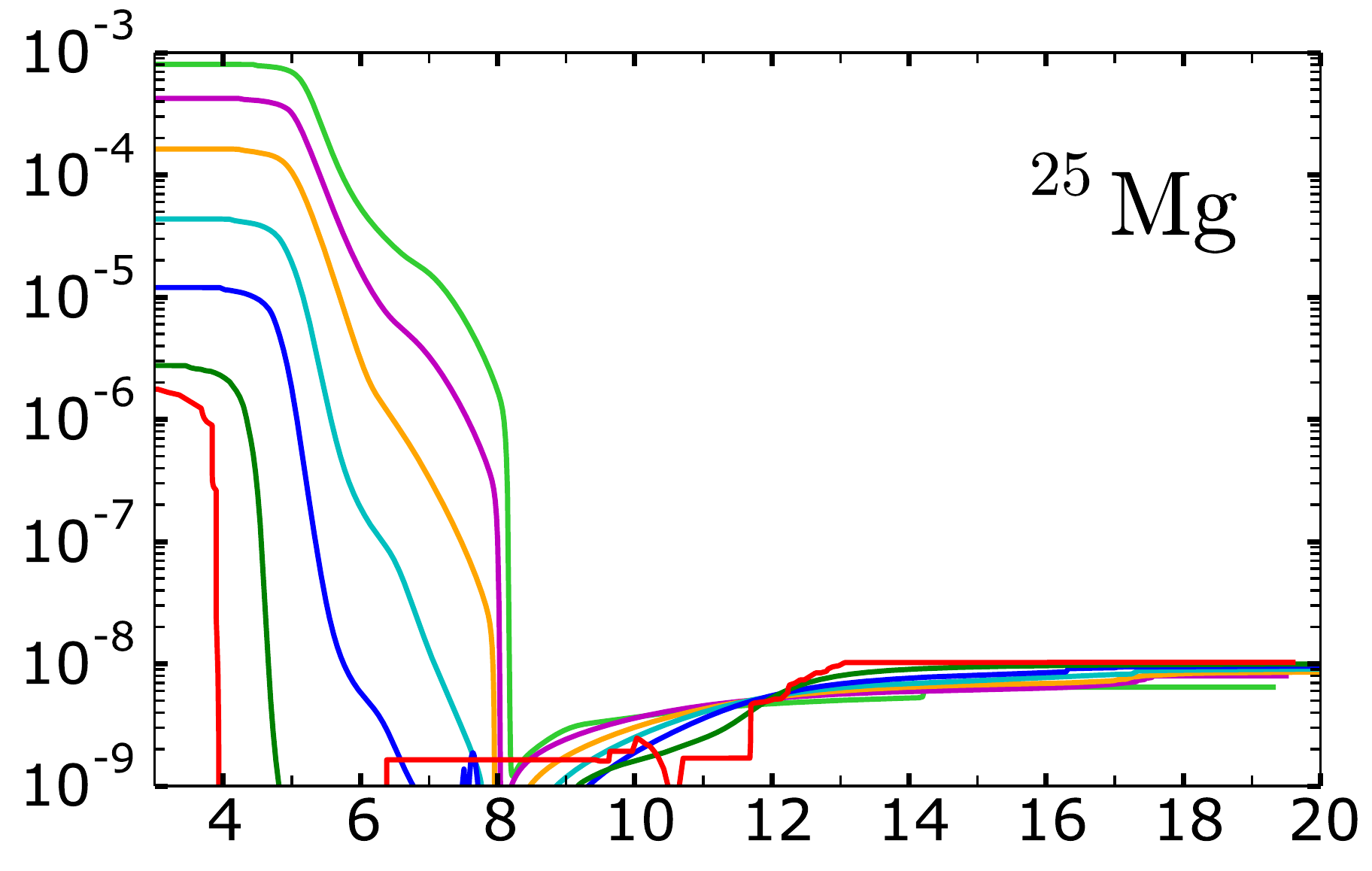}
   \end{minipage}
   \begin{minipage}[c]{.49\linewidth}
       \includegraphics[scale=0.44]{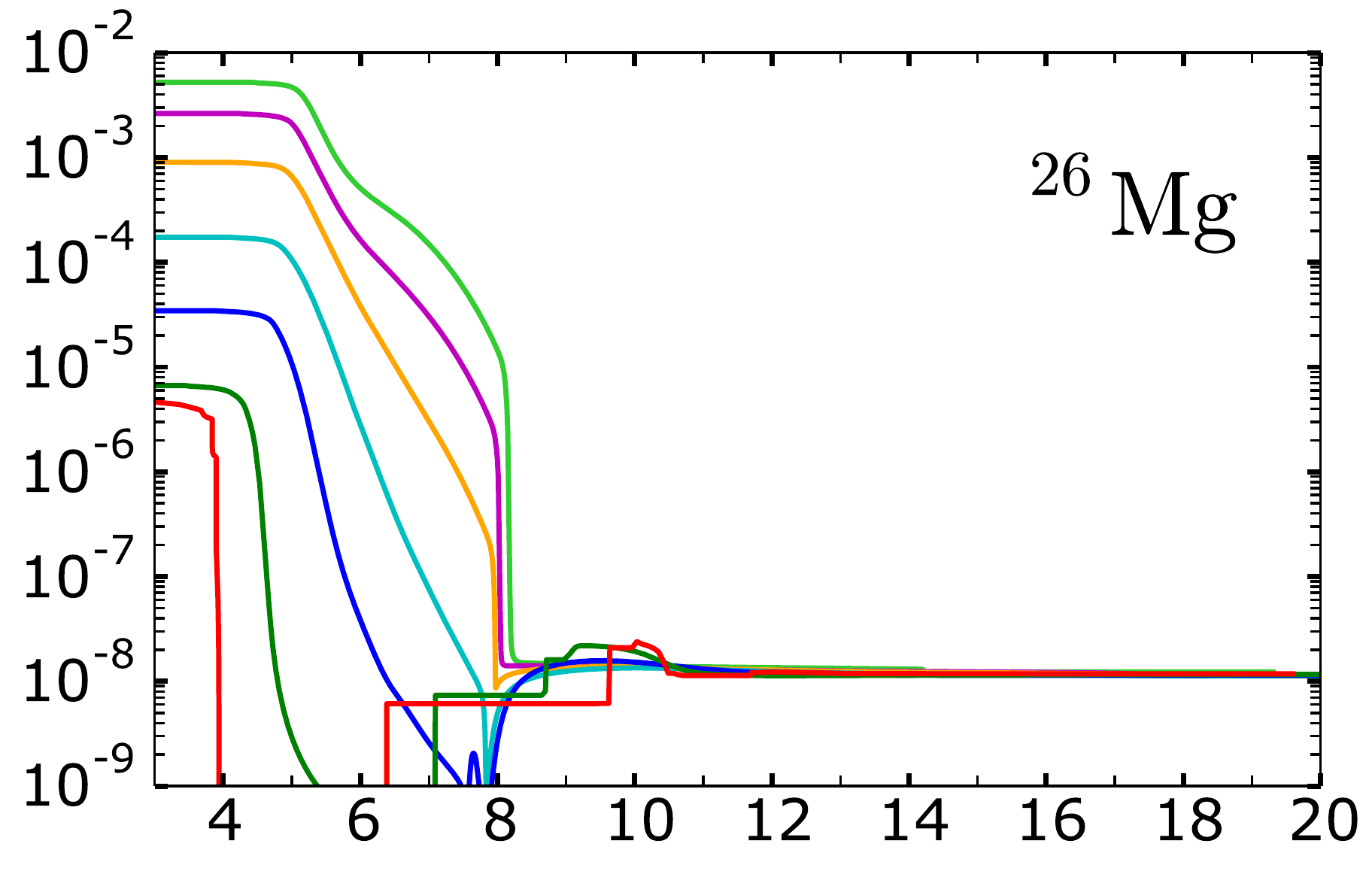}
   \end{minipage}
   \begin{minipage}[c]{.49\linewidth}
       \includegraphics[scale=0.44]{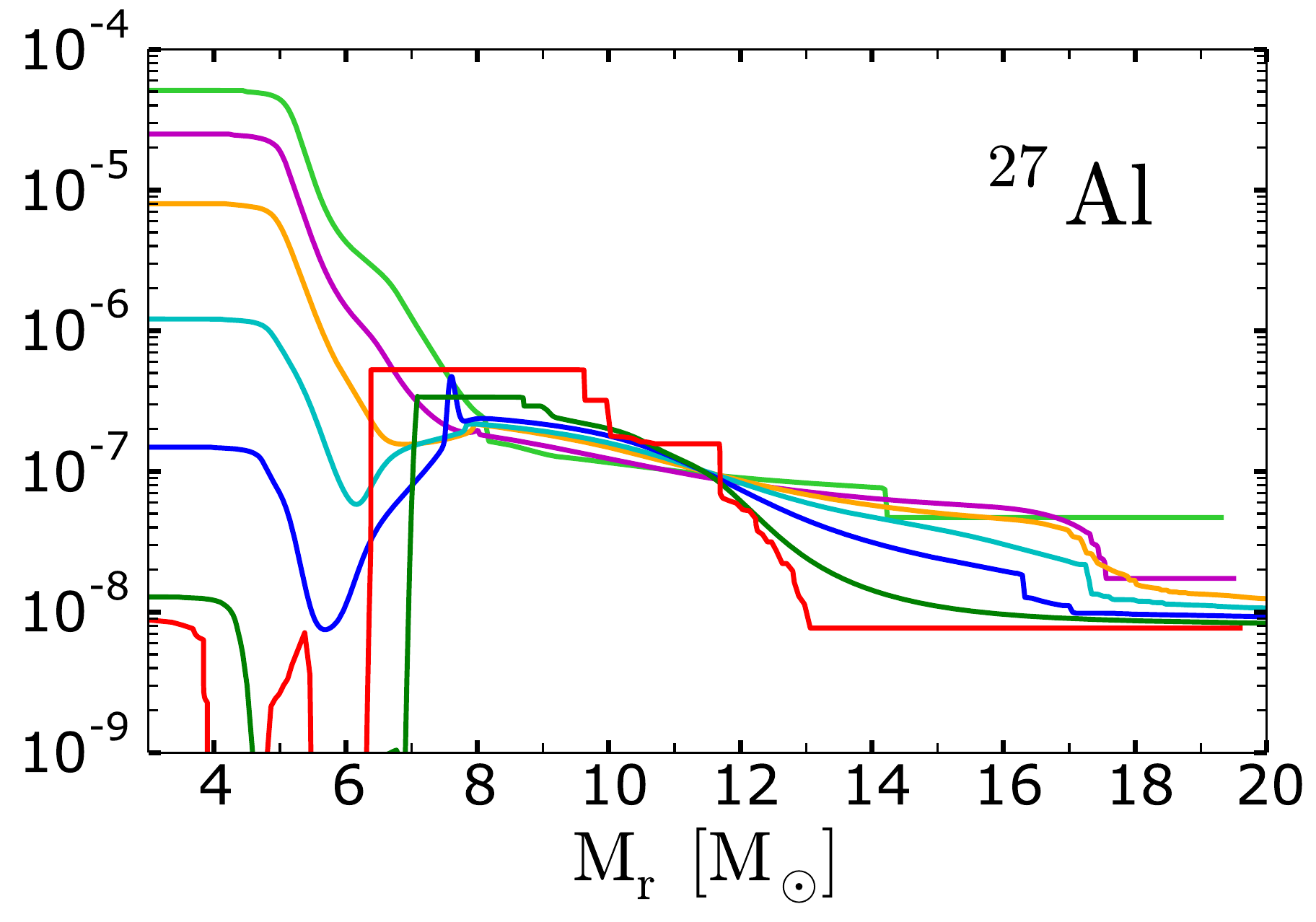}
   \end{minipage}
   \begin{minipage}[c]{.49\linewidth}
       \includegraphics[scale=0.44]{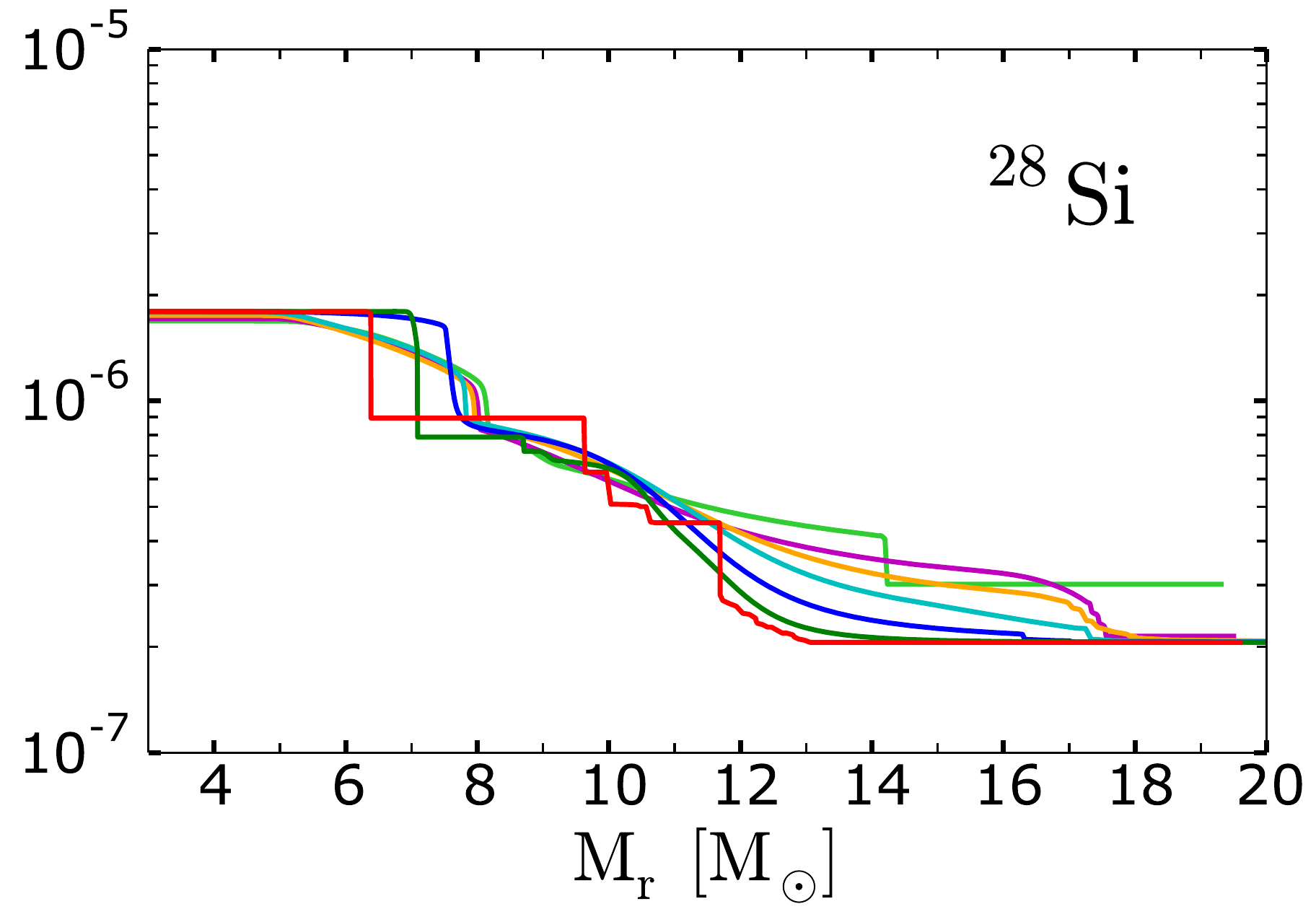}
   \end{minipage}
   \caption[Same as Fig.~\ref{abrotmod} for other isotopes]{Same as Fig.~\ref{abrotmod} for other isotopes.}
\label{abrotmod2}
    \end{figure*}

%\textcolor{red}{main species. About 40 isotopes. 38 maybe. Depend on burning stage, zone of the star }

I computed 20~$M_{\odot}$ models at a metallicity $Z=10^{-5}$ 
%([Fe/H] $=-3.8$) 
with\footnote{The critical velocity $\upsilon_{\rm crit}$ is reached when the gravitational acceleration is counterbalanced by the centrifugal force. It is expressed as $\upsilon_{\rm crit} = \sqrt{\frac{2}{3}\frac{GM}{R_{\rm p,c}}}$ with $R_{\rm p,c}$ the polar radius at the critical limit.} $\upsilon_{\rm ini}/ \upsilon_{\rm crit} =$ 0, 0.1, 0.2, 0.3, 0.4, 0.5 and 0.7.
I also computed a 40 and a 60~$M_{\odot}$ model at $Z=10^{-5}$ and with $\upsilon_{\rm ini}/ \upsilon_{\rm crit} =$ 0.4 to explore different initial masses. Finally, a 20~$M_{\odot}$ at $Z=0$ and with $\upsilon_{\rm ini}/ \upsilon_{\rm crit} =$ 0.4 was computed to evaluate the effect of metallicity.
%The mass-loss rates are from \cite{vink01} when $\log T_{\rm eff} \geq 3.95$ and from \cite{jager88} when $\log T_{\rm eff} < 3.95$. $D_{\rm shear}$ is from \cite{talon97} and $D_{\rm h}$ from \cite{zahn92}. 
The nuclear reaction rates are the same than for the box model.
The evolution is stopped at the end of core carbon burning. The very last burning stages (also explosive nucleosynthesis, cf. Sect.~\ref{explonuc}) significantly modify only the abundances of the most inner layers. 
%Also, explosive nucleosynthesis (not considered here) mostly affect the iron-group elements in the innermost layers of the star \citep{woosley95, thielemann96, limongi03, nomoto06}. 
The yields of the models presented here provide good predictions, 
% for the light elements, 
provided the most inner layers of the star are not ejected and considered in the yields. %(which is the case, as discussed after).
More details on the other physical ingredients (shear mixing, mass loss, initial composition...) can be found in Sect.~\ref{massivemodels}.

%Models de base: nuc rates standard (liste table in O2013 is STD, no ths8, etc...)

%Models al7gths8 : idem MAIS juste al7g from ths8

%Models dans papier 2 : al7a\_ths8, al7g\_ths8 and mg6g\_nacr

%\textcolor{red}{Initial composition here? (now it is in chap 2 but maybe better here)}

%\textcolor{red}{Say somewhere winds are low. Or table}

%\subsection{Interplay between rotation and nucleosynthesis\label{secinter}}
\subsection{Evolution with rotation \label{secinter}}

This section discusses the effects of rotation by focusing on the 20~$M_{\odot}$ models with various initial rotation rates.

%\textcolor{red}{Maybe repetition here. Talk about all nucleo in back forth mixing only once. Here or before.}
%It boosts the burning (of H, He, it depends on the burning stage), raises the temperature and tends to make the convective core bigger. 
%On the other hand, 
The effective gravity of the stellar core is decreased by rotation because of the effect of the centrifugal force. It decreases the temperature in the convective core and tends to make it smaller. 
On the other hand, rotation tends to produce bigger cores because of the rotational mixing that provides additional fuel to the core. %It tends to make it bigger. 
Overall, in the models presented here, rotation produces bigger cores at the end of the evolution (Fig.~\ref{cores20}). For instance the size of the carbon-oxygen core (CO-core, defined where the $^{4}$He mass fraction drops below $10^{-3}$) increases by about 1.5~$M_{\odot}$ from the non rotating model to the fast rotating one. 
A consequence is that the base of the H-rich envelope and the He-rich region are shifted towards higher mass coordinates when initial rotation increases (Fig.~\ref{abrotmod}, top panels). %It therefore shifts the abundance profiles toward higher mass coordinates. 

As the initial rotation increases, the He-burning products (e.g. $^{12}$C, $^{16}$O, $^{22}$Ne) transit quicker and in greater amount to the H-burning shell so that more $^{13}$C, $^{14}$N, $^{19}$F, $^{23}$Na are synthesized (Fig.~\ref{abrotmod} and \ref{abrotmod2}). Around the mass coordinate 7~$M_{\odot}$, $^{13}$C and $^{14}$N are boosted by $\sim 4$ dex from $\upsilon_{\rm ini}/\upsilon_{\rm crit} = 0$ to 0.7. 
%$^{17}$O and $^{19}$F are also largely boosted. 
Rotation increases the production of $^{19}$F, $^{22}$Ne, $^{23}$Na, $^{24,25,26}$Mg and $^{27}$Al in the He-burning core (cf. Sect.~\ref{secback}). 
%$^{22}$Ne is built by $^{14}$N($\alpha,\gamma$)$^{18}$F($e^+ \nu_e$)$^{18}$O($\alpha,\gamma$)$^{22}$Ne, $^{25,26}$Mg by $^{22}$Ne($\alpha,\gamma$)$^{26}$Mg and $^{22}$Ne($\alpha,n$)$^{25}$Mg. The neutrons released by $^{22}$Ne($\alpha,n$) produce $^{23}$Na, $^{24}$Mg and $^{27}$Al by $^{22}$Ne($n,\gamma$)$^{23}$Ne($e^- \bar{\nu}_e$)$^{23}$Na, $^{23}$Na($n,\gamma$)$^{24}$Na($e^- \bar{\nu}_e$)$^{24}$Mg and $^{26}$Mg($n,\gamma$)$^{27}$Mg($e^- \bar{\nu}_e$)$^{27}$Al respectively. To a smaller extent, $^{19}$F is also boosted in the He-core by $^{14}$N($n,\gamma$)$^{15}$N($\alpha,\gamma$)$^{19}$F. 
%Close to the end of the core He-burning phase, the convective He-burning core runs out of fuel so that it contracts and the temperature increases. In the central convective region, $^{19}$F($\alpha,\gamma$)$^{22}$Ne becomes efficient and destroys $^{19}$F. 
Near the end of the core He-burning phase, $^{19}$F is also destroyed in the central regions (below 5~$M_{\odot}$) by $^{19}$F($\alpha,p$)$^{22}$Ne. 
%%%%Rotational mixing transports back these new elements to the H-burning shell. 
%The peak of $^{23}$Na in the H-burning shell at $\sim 8$~$M_{\odot}$ (Fig.~\ref{abrotmod2}) is due to proton captures on the $^{22}$Ne that has diffused from the He-core to the H-shell. 
%%%%In the end, it mainly enhances the production of $^{23}$Na: the abundance of $^{23}$Na at $\sim 8$~$M_{\odot}$ increases with initial rotation mainly because of proton captures on the $^{22}$Ne coming from the He-core (Fig.~\ref{abrotmod2}).
While the Ne-Na chain is boosted in the H-burning shell of rotating models (see the peak of $^{23}$Na at $\sim 8$~$M_{\odot}$), the Mg-Al chain is not efficiently activated (no similar peak in the H-burning shell). 
%because of the too low temperature H-burning shell ($T \lesssim 50$ MK). 
This is mainly because the temperature in the H-burning shell is too low ($T \lesssim 45$ MK). 
Another reason is that the synthesis of extra Al in the H-burning shell needs extra Mg, which is only built in the He-core when $T \gtrsim 220$ MK (through an $\alpha$-capture on $^{22}$Ne). In a 20~$M_{\odot}$ model, this temperature corresponds to the end of core He-burning phase. The extra Mg created in the core has then little time to be transported to the H-burning shell and boosts the Mg-Al chain. 
$^{28}$Si is the only isotope that is barely affected by rotation (bottom right panel of Fig.~\ref{abrotmod2}).

%much since Mg, Al and Si are not overproduced in the H-burning shell because of the too low temperature ($T \lesssim 50$ MK) which is insufficient to activate efficiently the Mg-Al-Si chain. 
%\textcolor{red}{say somewhere T at which CNO, Ne-Na, Mg-Al activated (see Decressin+07 par ex)}

%The effect of rotation on surface abundances is rather modest. 
As initial rotation increases, the surface is enriched in H-burning products. This is because of the rotational mixing that transports these elements from H-burning layers up to the surface. We indeed see that (1) the surface $^{4}$He is increased in the faster rotating model, (2) the effect of CNO burning appears at the surface (less $^{12}$C and $^{16}$O, more $^{14}$N) and (3) the products of the Ne-Na and Mg-Al cycles are more abundant (mostly $^{23}$Na, $^{24}$Mg and $^{27}$Al). The surface abundances change mostly during the main sequence phase because of its longer duration (the chemical species have more time to be transported). When reaching the stellar surface, chemical species are potentially expelled through winds. In these models however, the mass loss metallicity relation ($\dot{M} \propto Z^{0.85}$, cf. Sect.~\ref{secmassloss}) plays a major role and thus prevents significant radiative mass loss episodes. Also the mechanical mass loss (cf. Sect.~\ref{secmassloss}) stays small for these models. The 20~$M_{\odot}$ model with $\upsilon_{\rm ini}/\upsilon_{\rm crit} = 0.7$ loses about 0.12~$M_{\odot}$ through mechanic mass loss and 0.6~$M_{\odot}$ through radiative mass loss. The total mass lost through winds is < 1~$M_{\odot}$ for all the 20~$M_{\odot}$ models. We note that the mass loss may be strongly enhanced for higher mass stars.
% the mass lost through winds remains modest (< 1~$M_{\odot}$):  weakness of the winds due to the low metallicity of the models (cf. Sect.~\ref{secmassloss}) is the dominant effect here and it prevent significant mass loss. % since none of the effects induced by rotation mentioned in Sect.~\ref{}.
%The surface abundance of heavier elements is affected by less than 1 dex (Fig.~\ref{abrotmod2}). The surface $^{23}$Na, $^{24}$Mg and $^{27}$Al are the most affected. It is due to the operation of the Ne-Na and Mg-Al chains inside hot H-burning layers.

%After core helium burning, the burning timescale becomes short compared to the rotational mixing timescale. Consequently, rotational mixing barely operates after helium is exhausted in the core and hence does not alter much the abundance profiles shown in Fig.~\ref{abrotmod} and \ref{abrotmod2}. In general, from the end of core helium burning, only the abundances of the inner regions ($\lesssim 5$~$M_{\odot}$) change, because of further nucleosynthesis (core carbon burning, ...). 

%\textcolor{red}{A word about winds however...}
%At the end, the total mass lost through winds is $\lesssim 1$~$M_{\odot}$ for all models.

%\subsection{Effect of various parameters on the yields\label{secvarpar}}
%\subsection{Yields\label{secvarpar}}
\subsection{Composition of the ejecta\label{secvarpar}}

The effect of varying the initial rotation, mass cut, dilution, efficiency of rotational mixing, mass and metallicity on the chemical composition of the source star ejecta is considered.

%To compute the chemical composition in the ejecta, one needs 

  \begin{figure*}[t]
   \centering
      \includegraphics[scale=0.74, trim = 0cm 0cm 0cm 0cm]{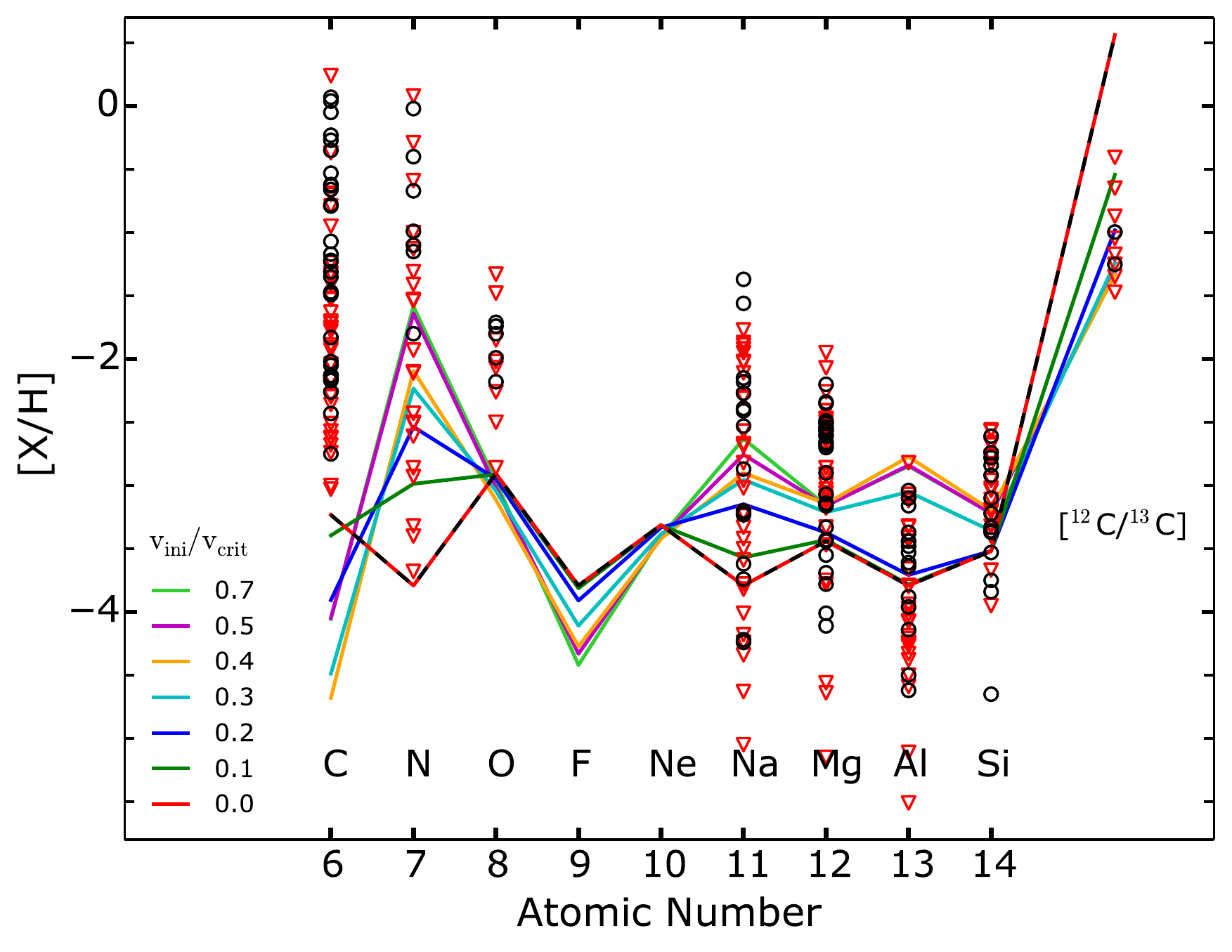}
   \caption[Wind composition of 20~$M_{\odot}$ models at $Z=10^{-5}$ with various initial rotations]{Composition of the wind of 20~$M_{\odot}$ models at $Z=10^{-5}$ with various initial rotation rates. The dashed black pattern (superimposed with the red pattern) shows the composition of the ISM. Black circles and red triangles show main-sequence and giant CEMP stars respectively, with [Fe/H$]~<-3$ (Table~\ref{tabcemp}). Recognized CEMP-r, -s and -r/s are excluded. Abundances with upper/lower limits are not plotted. The typical uncertainty is $\pm 0.3$ dex. %Abundance data is from the SAGA database \citep[][last update on Sept. 2017]{suda08}. 
  % \textcolor{red}{Faire un plot avec effet Z?? Pop~III !!! Write on the top: wind, M alpha, Mco , Mrem}
  }
\label{vinichem_w}
    \end{figure*}

  \begin{figure*}[t]
   \centering
      \includegraphics[scale=0.65, trim = 0cm 0cm 0cm 0cm]{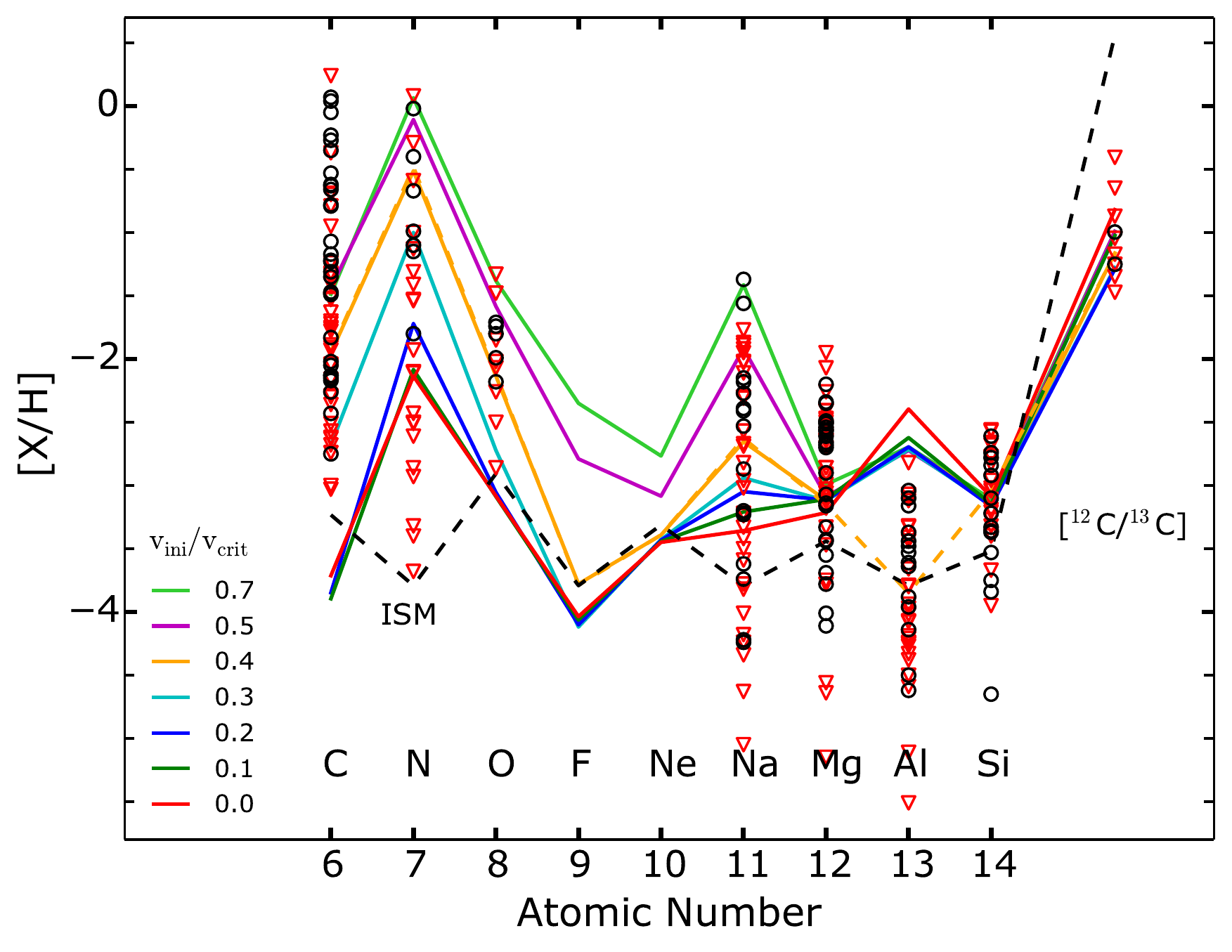}
   \caption[Same as Fig.~\ref{vinichem_w} but for the composition of all the H-rich material]{Same as Fig.~\ref{vinichem_w} but for the composition of all the H-rich material. This material includes the wind material plus the material from a supernova with a mass cut set at $M_{\alpha}$. $M_{\alpha}$ corresponds to the bottom of the H-envelope (the bottom of the H-envelope is defined where the $^{1}$H mass fraction drops below $10^{-3}$). The dashed orange line shows the $\upsilon_{\rm ini}/\upsilon_{\rm crit} = 0.4$ model with the rate of the $^{27}$Al($p,\gamma$)$^{28}$Si reaction from \cite{cyburt10} instead of \cite{iliadis01}.}
\label{vinichem_h}
    \end{figure*}
%\vspace{0.6cm}

  \begin{figure*}[h!]
   \centering
      \includegraphics[scale=0.65, trim = 0cm 0cm 0cm 0cm]{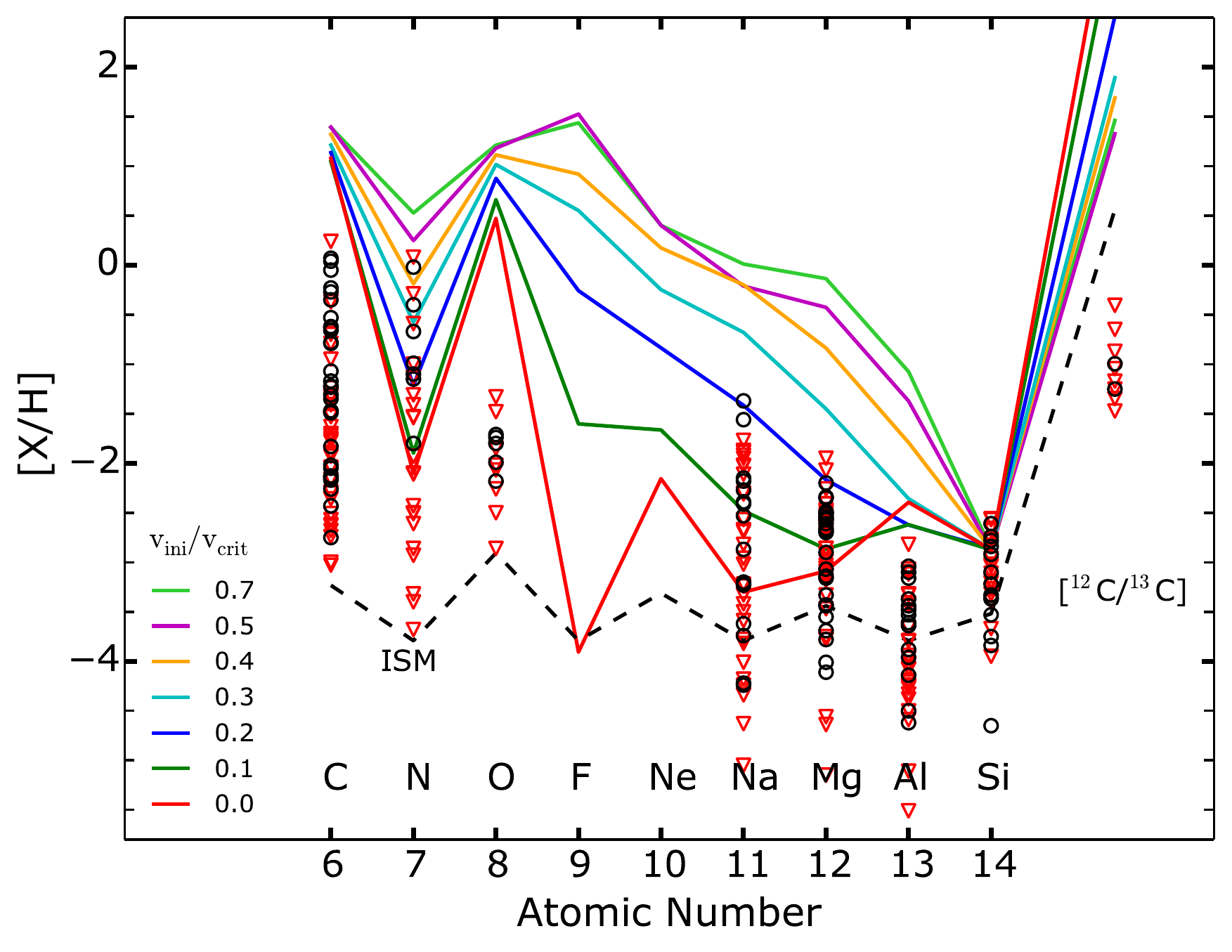}
   \caption[Same as Fig.~\ref{vinichem_h} but for the composition of all the H-rich + He-rich material]{Same as Fig.~\ref{vinichem_h} but for the composition of all the H-rich + He-rich material. 
 This material includes the wind material plus the material from a supernova with a mass cut set at $M_{\rm CO}$. $M_{\rm CO}$ corresponds to the bottom of the He-shell (the bottom of the He-shell is defined where the $^{4}$He mass fraction drops below $10^{-3}$). $M_{\rm CO}$ also corresponds to the top of the CO-core.  
   %(wind + a supernova defined with the mass cut set at $M_{\rm CO}$ the bottom of the He-shell (defined where the $^{4}$He mass fraction drops below $10^{-3}$). It corresponds to the top of the CO-core. 
   %\textcolor{red}{put al7g ths8 model? Link He rich region and CO core. Take Y 0.01 or 0.001 every where to define co core maybe.? XH with Mcut Maeder92?}
   }
\label{vinichem_hhe}
    \end{figure*}

\subsubsection{Initial rotation}

\paragraph{Composition of the wind.}
Fig.~\ref{vinichem_w} shows the [X/H] ratios in the wind of the models. For all models, the wind contribution is less than 1~$M_{\odot}$ and comes from the H-rich region. The non-rotating model (red) is superimposed with the ISM pattern since the ejected material has kept exactly the same composition as the ISM. As rotation increases, the wind is overall more and more enriched in a material processed by the CNO cycle, Ne-Na and Mg-Al chains (cf. Sect.~\ref{secinter}). For the faster rotators, the $^{12}$C/$^{13}$C ratio is equal to the CNO equilibrium value of $\sim 4$ (equivalent to [$^{12}$C/$^{13}$C] $=-1.35$).
%Na/H and Al/H (Mg/H and Si/H to a smaller extent) are boosted with rotation because of the same reasons as before but this time for Ne-Na and Mg-Al(-Si) chains products. Is the Al synthesized because of back-and-forth??? Maybe more built during MS... because Fig 3.6: Al does not depend on rot. explain

  \begin{figure*}[t]
   \centering
      \includegraphics[scale=0.38, trim = 0cm 0cm 0cm 0cm]{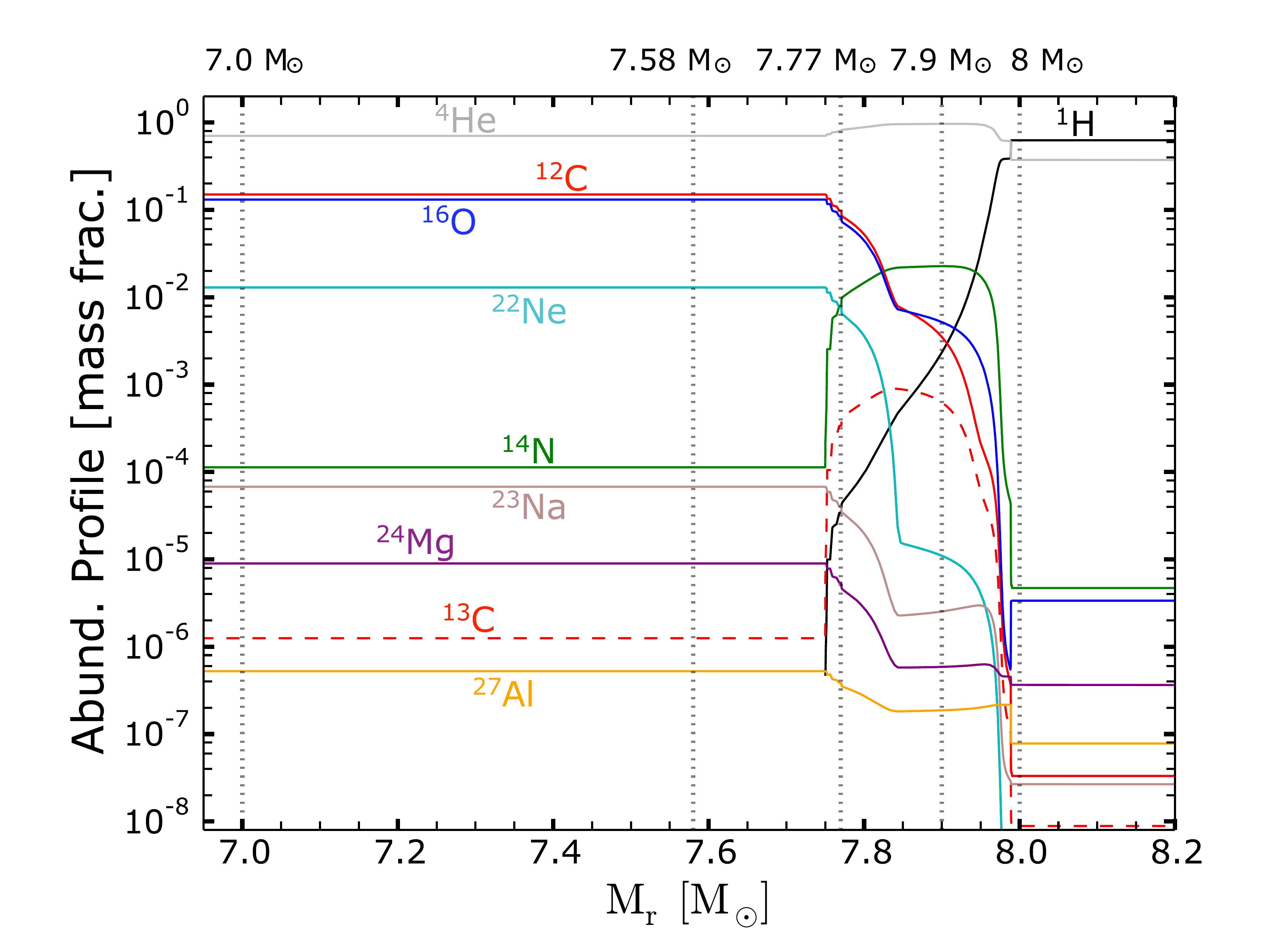}
   \caption[Abundances of the 20~$M_{\odot}$ at $Z=10^{-5}$ and $\upsilon_{\rm ini}/\upsilon_{\rm crit} = 0.4$ (pre-SN)]{Abundance profile of the 20~$M_{\odot}$ model with $Z=10^{-5}$ and $\upsilon_{\rm ini}/\upsilon_{\rm crit} = 0.4$ at the pre-SN stage. Dashed vertical lines corresponds to the mass cuts of Fig.~\ref{vinichem_mcut}.}
\label{ab_mcut}
    \end{figure*}
%\vspace{1cm}
  \begin{figure*}[h!]
   \centering
      \includegraphics[scale=0.68, trim = 0cm 0cm 0cm 0cm]{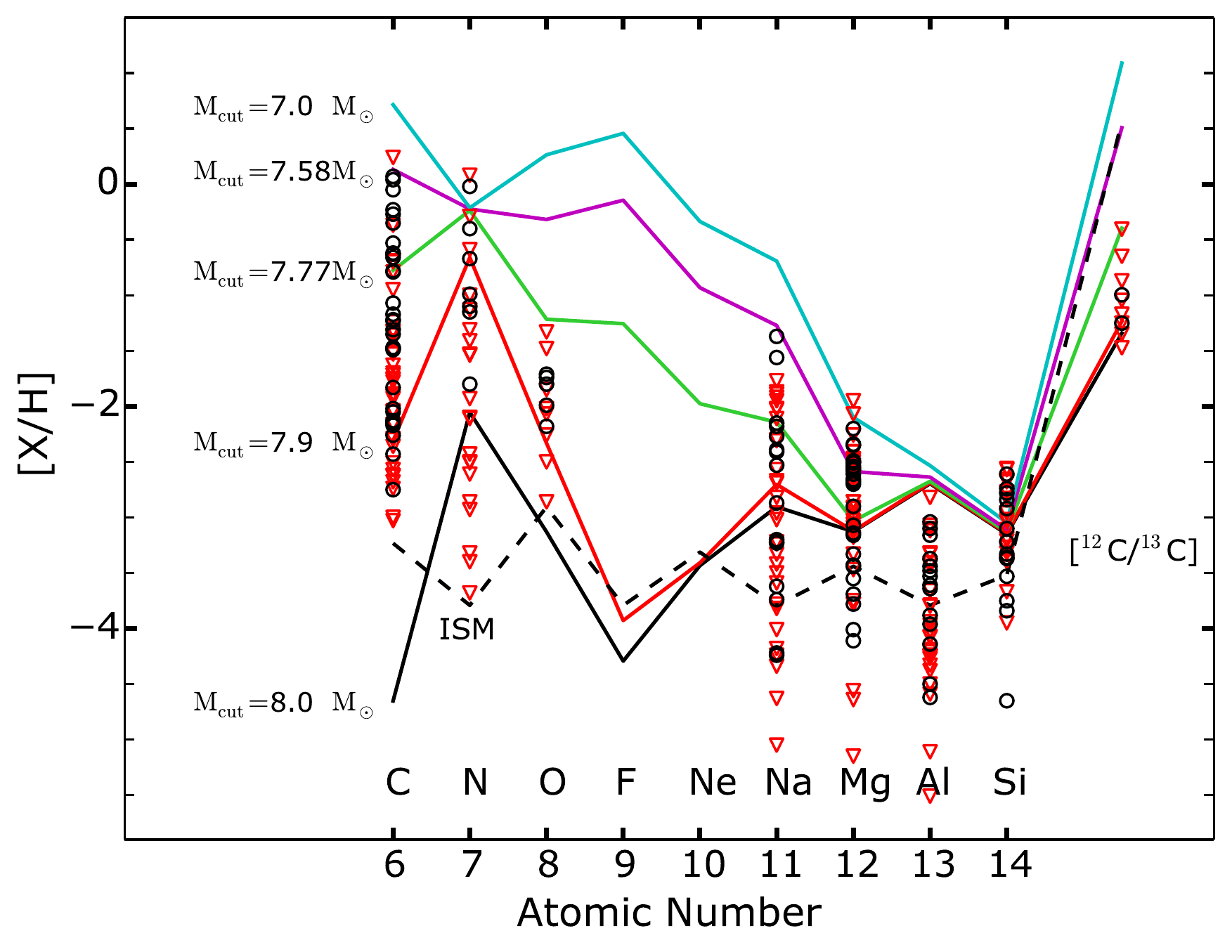}
   \caption[Ejecta composition of a 20~$M_{\odot}$ model with $Z=10^{-5}$ and $\upsilon_{\rm ini}/\upsilon_{\rm crit} = 0.4$ for various mass cuts]{Composition of the ejecta of a 20~$M_{\odot}$ model with $Z=10^{-5}$ and $\upsilon_{\rm ini}/\upsilon_{\rm crit} = 0.4$ for various mass cuts taken in between the H-shell and He-shell. These mass cuts are also represented in Fig.~\ref{ab_mcut}.}
\label{vinichem_mcut}
    \end{figure*}

\paragraph{Composition of the H-rich material.}
Fig.~\ref{vinichem_h} shows the chemical composition of the ejecta when considering the wind plus all the material above the bottom of the H-rich region. In this case, $M_{\rm cut} = M_{\alpha}$, where $M_{\alpha}$ is the mass of the helium core set where the mass fraction of $^{1}$H drops below 10$^{-3}$. From the top left panel of Fig.~\ref{abrotmod} we see that it corresponds to a total ejected mass of $\sim 12-14$~$M_{\odot}$, depending on the model. The typical CNO pattern appears for all the models (more N, less C and O) but the sum of CNO elements increases with rotation, as a result of $^{12}$C and $^{16}$O having diffused to the H-burning shell. [Na/H] spans $\sim 2$ dex while [Mg/H], [Al/H] and [Si/H] do not vary more than 0.5 dex. Mg and Al are overproduced with rotation (Fig.~\ref{abrotmod2}) but only in the He-rich region, which is not consider in the yields of Fig.~\ref{vinichem_h}. $^{12}$C/$^{13}$C is very close to the CNO equilibrium value.

\paragraph{Composition of the H-rich + He-rich material.}
Fig.~\ref{vinichem_hhe} is similar as Fig.~\ref{vinichem_h} but it considers the wind plus all the material above the bottom of the He-rich region. In this case, $M_{\rm cut} = M_{\rm CO}$.
%, where $M_{\rm CO}$ is the mass of the CO-core and is set where the mass fraction of $^{4}$He drops below 10$^{-3}$. 
The total ejected mass is about $15-16$~$M_{\odot}$, depending on the model (Fig.~\ref{cores20}). Compared to the previous case, the additional $\sim 2$~$M_{\odot}$ ejected are H-free, so that is raises a bit the [X/H] ratios (by < 0.5 dex). The CNO pattern is reversed compared to Fig.~\ref{vinichem_h} because in He-burning regions, $^{14}$N is depleted while $^{12}$C and $^{16}$O are abundant (Fig.~\ref{abrotmod}). No or very little little additional $^{14}$N is added to the ejecta compared to Fig.~\ref{vinichem_h}. The ratios containing He-burning products ([Ne/H], [Na/H], [Mg/H] and [Al/H]) are largely boosted compared to Fig.~\ref{vinichem_h}. They also greatly increase with initial rotation as a result of the back-and-forth mixing process. In He-burning regions, $^{13}$C is destroyed by $^{13}$C($\alpha,n$)$^{16}$O so that the $^{12}$C/$^{13}$C ratio is largely enhanced compared to the case where only H-rich ejecta is considered. 
%FAUX je pense:
%$^{12}$C/$^{13}$C decreases with initial rotation as a result of the reaction $^{12}$C($n,\gamma$)$^{13}$C, which is more active in rotating models because of the larger neutron density induced by the reaction $^{22}$Ne($\alpha,n$)$^{25}$Mg.
%cest plutot du au primary C13 dans H rich region...
Considering deeper mass cuts raise again C/H, O/H, Ne/H, Mg/H, and $^{12}$C/$^{13}$C but, overall, it does not change significantly the trends of Fig.~\ref{vinichem_hhe}.

%\paragraph{Composition of the ejecta with $M_{\rm cut} = M_{\rm rem} $.} $M_{\rm rem}$ is defined according to the relation of \cite{maeder92} that links the mass of the CO-core to the mass of the remnant. Remnant masses are about 2~$M_{\odot}$ for the 20~$M_{\odot}$ models (Fig.~\ref{cores20}). \textcolor{red}{Check behavior of S0 and S1 models and explain quickly.}

\subsubsection{Mass cut in the intershell region}

Fig.~\ref{vinichem_mcut} shows the chemical composition of the wind plus an ejecta defined with various mass cuts, for the $\upsilon_{\rm ini}/\upsilon_{\rm crit} = 0.4$ model. The 5 considered mass cuts are represented in Fig.~\ref{ab_mcut} which shows the abundance profile of the 20~$M_{\odot}$ model with $\upsilon_{\rm ini}/\upsilon_{\rm crit} = 0.4$ at the pre-SN stage. The zone between $7.8 \lesssim M_{\rm r} \lesssim 8$~$M_{\odot}$ corresponds to the intershell region, in between the H- and He-burning shells. As $M_{\rm cut}$ decreases, deeper layers are ejected and we move progressively from the H-rich only ejecta to a mixed H+He ejecta. Varying the mass cut from 7 to 8~$M_{\odot}$ leads to a quick increase of the [X/H] ratios, up to $\sim 5$ dex for [C/H]. It illustrates the high sensitivity of the yields on the mass cut when varying it around the intershell region.

 % \begin{figure*}%[h!]
  % \centering
  %    \includegraphics[scale=0.72, trim = 0cm 0cm 0cm 0cm]{dilpat_H.pdf}
  % \caption[a]{Composition of the H-rich ejecta of a 20~$M_{\odot}$ model with $Z=10^{-5}$ and $\upsilon_{\rm ini}/\upsilon_{\rm crit} = 0.7$ for various dilution factor $D$. The $D=0$ pattern corresponds to the green pattern in Fig.~\ref{vinichem_h}.}
%\label{vinichem_dilh}
   % \end{figure*}

 % \begin{figure*}%[h!]
   %\centering
   %   \includegraphics[scale=0.72, trim = 0cm 0cm 0cm 0cm]{dilpat_HHe.pdf}
  % \caption[a]{Same as Fig.~\ref{vinichem_dilh} but for H- and He-rich ejecta. The $D=0$ pattern corresponds to the green pattern in Fig.~\ref{vinichem_hhe}}
%\label{vinichem_dilhhe}
   % \end{figure*}

   \begin{figure*}
   \centering
   \begin{minipage}[c]{.49\linewidth}
      \includegraphics[scale=0.44, trim = 0cm 0cm 0cm 0cm]{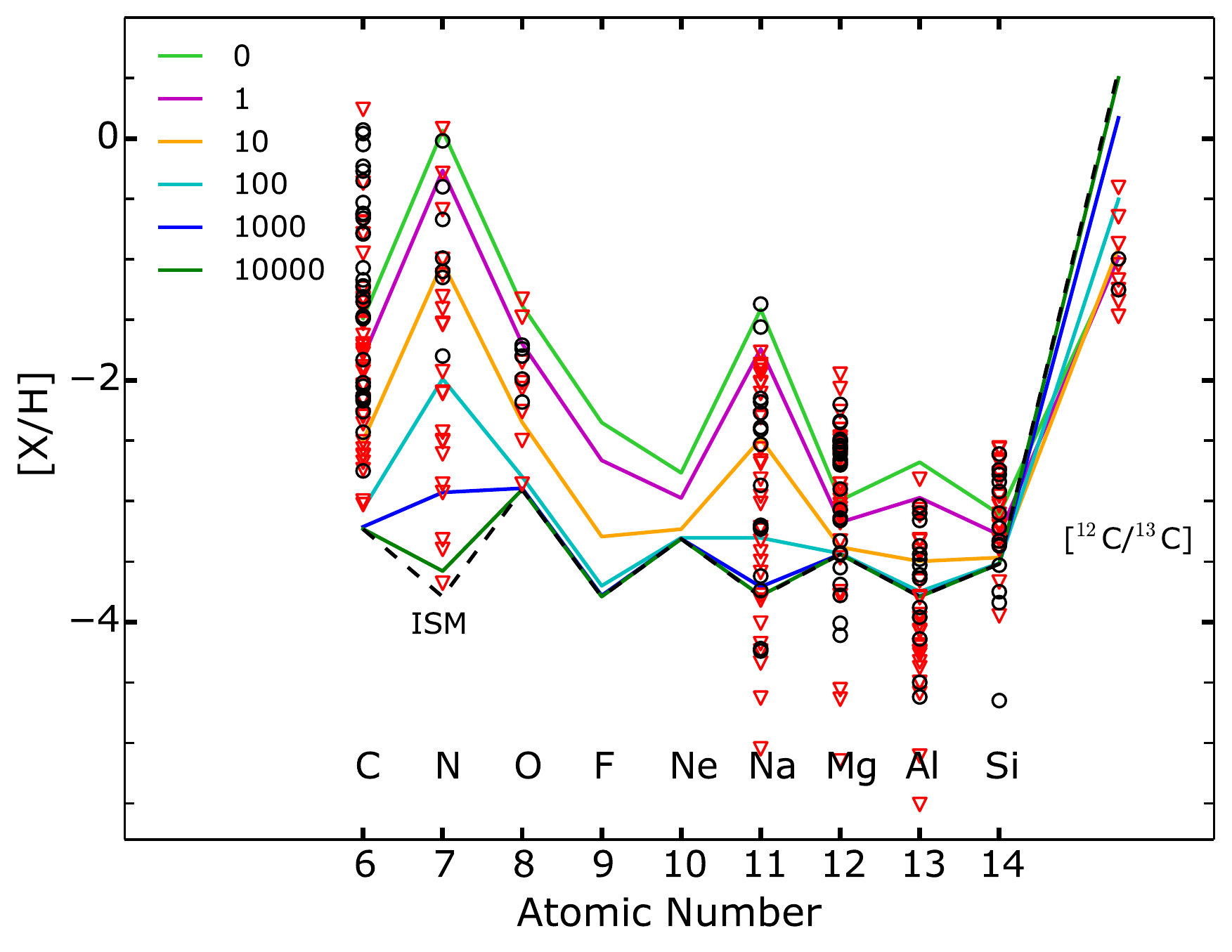}
   \end{minipage}
   \begin{minipage}[c]{.49\linewidth}
      \includegraphics[scale=0.44, trim = 0cm 0cm 0cm 0cm]{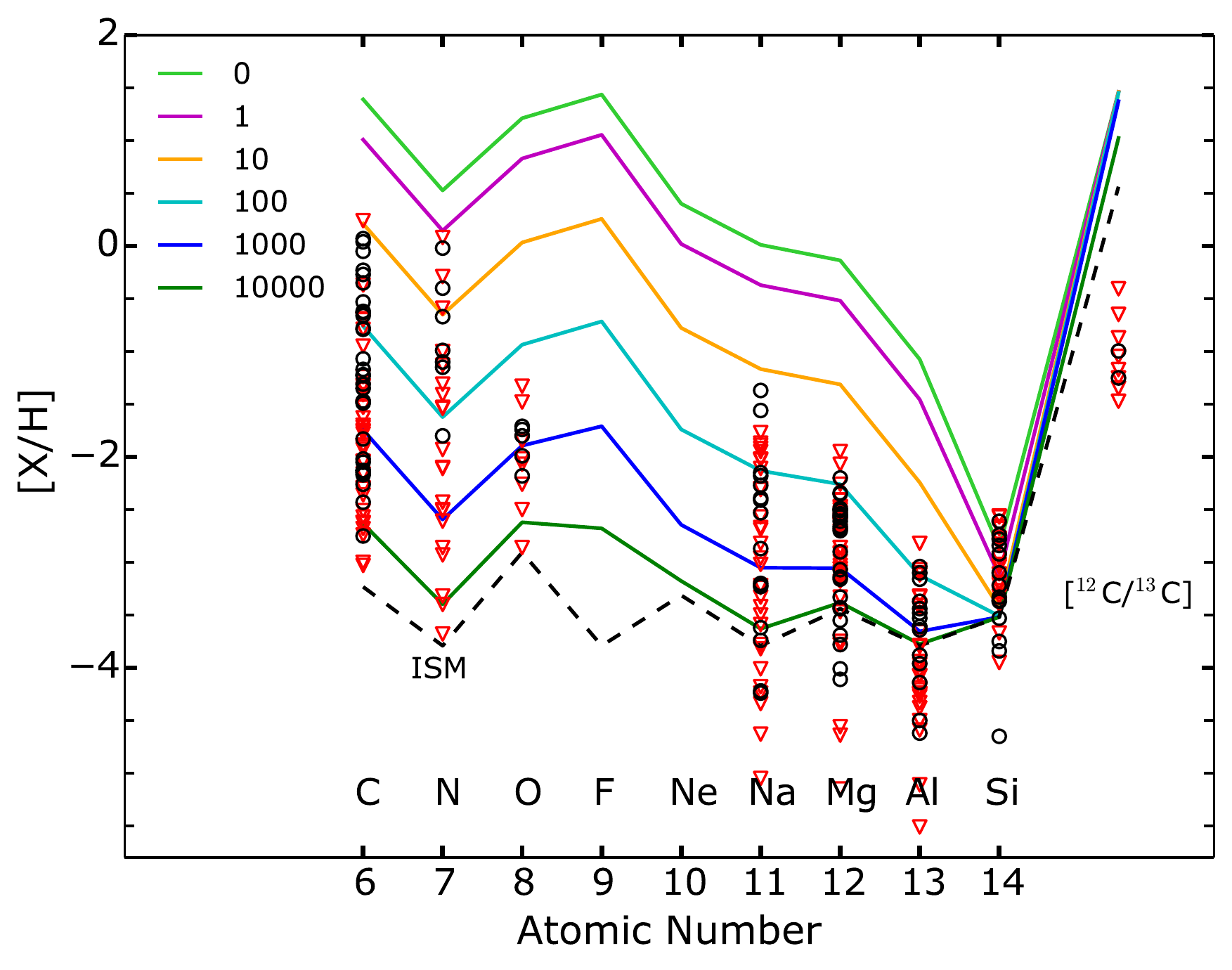}
   \end{minipage}
   \caption[Impact of the dilution in the ejecta composition for a fast rotating 20~$M_{\odot}$]{\textit{Left panel}: composition of the H-rich ejecta of the 20~$M_{\odot}$ model with $\upsilon_{\rm ini}/\upsilon_{\rm crit} = 0.7$ for various dilution factors $D$. The $D=0$ pattern corresponds to the green pattern in Fig.~\ref{vinichem_h}. \textit{Right panel}: same as left panel but for H-rich + He-rich ejecta. The $D=0$ pattern corresponds to the green pattern in Fig.~\ref{vinichem_hhe}.}
\label{vinichem_dil}
    \end{figure*}

\subsubsection{Dilution of the source star ejecta with ISM}

The material ejected from the star can be mixed with the material in the ISM. The effect of the dilution depends on the composition of the ISM, which is not known at low metallicity (a choice has nevertheless to be made when computing low metallicity models, as done in this work). 
Generally, in the case of the scenario proposing that CEMP-no stars were formed from one or very few source stars, we can probably just guess that the ISM is either metal-free or very metal-poor. Table~\ref{ismtable} reports how are affected the different abundance ratios in the source star ejecta while considering either a dilution with a metal-poor ISM or a metal-free ISM. A dilution with a metal-free ISM adds some H so that the [X/H] ratios are affected (where X refers to metals). On the other hand, such an ISM does not modify the [X/Fe] (where X refers to metals) or the isotopic ratios. In almost all the cases, when the [X/H] or [X/Fe] ratios are affected by dilution with ISM, they are reduced (cf. Fig.~\ref{vinichem_dil} and next paragraph). This is because in the source star ejecta, the H and Fe abundances are similar to the H and Fe abundances in the ISM while light metals (C to Al) are generally strongly overproduced by the source star compared to what is present in the ISM.

%:   tab tvelocp3
\begin{table}[h]
\caption[Effect of dilution on the source star ejecta composition]{Effect of diluting the source star ejecta with two different kinds of ISM: metal-poor and metal-free. Different kinds of abundance ratios are considered. X refers to metals.}
\label{ismtable}
\centering
\begin{tabular}{l | l l l}
\hline\hline
	 				& [X/H] 			& [X/Fe] 			& isotopic ratios    \\
\hline
	Metal-poor ISM 	& towards ISM value	& towards ISM value	& towards ISM value\\
	Metal-free ISM		& towards ISM value	& unchanged		& unchanged\\
\hline
\end{tabular}
\end{table}
%The material ejected from the star can be mixed with the material in the ISM. We can define the dilution factor $D$ as
%\begin{equation}
%D = \frac{M_{\rm ISM}}{M_{\rm ej}}
%\end{equation}
%with $M_{\rm ej}$ the total mass ejected by the source star and $M_{\rm ISM}$ the mass of initial ISM added to the massive star ejecta.
%Fig.~\ref{vinichem_dilh} shows the chemical composition of the wind plus all the material above the bottom of the H-rich region for the $\upsilon_{\rm ini}/\upsilon_{\rm crit} = 0.7$ model. Different dilution factors are considered. $D=100$ means that there is 100~$M_{\odot}$ of ISM for 1~$M_{\odot}$ of ejecta. As $D$ increases, the composition is shifted toward the ISM composition. Fig.~\ref{vinichem_dilhhe} is similar but for the material ejected are the H-rich + He-rich regions (like Fig.~\ref{vinichem_hhe}). In this case, the quantity of carbon, nitrogen, etc... are more important so that higher $D$ values are needed to reach the ISM pattern.
%\textcolor{red}{Dilution. Put this here or after when talking about obs? Talk once about dil?}
%We compare the chemical abundances of the CEMP sample with the ejecta of the source star models. 

In this work, when diluting the source star ejecta with ISM, I consider that the ISM material is the same than the ISM material used to form the source star ($\alpha$-enhanced ISM, cf. Sect.~\ref{inicompo} for details on the initial composition).
To be more quantitative, a dilution factor $D$ can be defined as
\begin{equation}
D = \frac{M_{\rm ISM}}{M_{\rm ej}}
\end{equation}
with $M_{\rm ej}$ the total mass ejected by the source star and $M_{\rm ISM}$ the mass of initial ISM added to the massive star ejecta. 
The left panel of Fig.~\ref{vinichem_dil} shows the chemical composition of the wind plus the H-rich ejecta for the $\upsilon_{\rm ini}/\upsilon_{\rm crit} = 0.7$ model. Different dilution factors are considered. $D=100$ means that there is 100~$M_{\odot}$ of ISM for 1~$M_{\odot}$ of source star ejecta. As $D$ increases, the composition is shifted towards the ISM composition. 
The right panel of Fig.~\ref{vinichem_dil} is similar but for the H-rich + He-rich ejecta (like in Fig.~\ref{vinichem_hhe}). In this case, the amount of light elements is more important in the source star ejecta so that higher $D$ values are needed to reach back the ISM pattern.
The ISM considered here is $\alpha$-enhanced, which explains why the [C/H], [O/H]... ratios are enhanced compared to the [N/H], [F/H]... ratios. Also, in this ISM, [$^{12}$C/$^{13}$C] $\simeq 0.5$, which corresponds to $^{12}$C/$^{13}$C $=300$ (cf. Sect.~\ref{inicompo}). 
%For comparison, in the Sun, $^{12}$C/$^{13}$C $\simeq 90$ \citep{lodders03} so that [$^{12}$C/$^{13}$C] $=0$. Such a lower $^{12}$C/$^{13}$C in the ISM will not change much the appearance of the patterns in Fig.~\ref{vinichem_dil}. 
A more extended discussion about the $^{12}$C/$^{13}$C ratio can be found in Sect.~\ref{compacemp}.

%in the present work, the composition  acts differently on the abundance ratios considered (e.g. [X/H], [X/Fe], isotopic ratios). Also  th ratios on the kind of ISM considered (e.g. metal-free or low metallicity) on if the The table below 

At the present day, the dilution factor cannot be strongly constrained. 
%The stellar wind, which escapes at \textcolor{red}{XXX} km s$^{-1}$ might be not diluted so much while the supernova, escaping at \textcolor{red}{XXX} km s$^{-1}$ might be more diluted. 
The point explosion or Sedov-Taylor explosion can be used as an approximation to estimate how fast will a shock wave travel and what would be left behind it. 
%SEE : http://www.ita.uni-heidelberg.de/~dullemond/lectures/num_fluid_2011/Chapter_10.pdf
For a standard explosion energy of $E_{51} = 1$ ($E_{51}$ is the SN energy in units of $10^{51}$ ergs) and typical ISM densities, the ejecta will be mixed with about $M_{\rm ISM} = 10^{4}$~$M_{\odot}$ of ISM \citep{cioffi88, ryan96, wehmeyer15}. For the models of Fig.~\ref{vinichem_dil}, the total mass ejected is between $\sim 12$ and $\sim 16$~$M_{\odot}$. Taking $10^{4}$~$M_{\odot}$ of ISM leads to a dilution factor of $10^{4}/16 < D < 10^{4}/12$, which gives $600 \lesssim D \lesssim 800$ (in between the cyan and blue patterns in Fig.~\ref{vinichem_dil}). This would be for standard explosions energies. In the case of a low-energetic supernova with $E_{51} = 0.1$ for instance, $60 \lesssim D \lesssim 80$ \citep[i.e. $D$ is divided by 10, this is because $D \sim M_{\rm ISM} \sim E_{51}^{0.95}$,][their Sect.~5.6.1]{ryan96}. %This gives $60 \lesssim D \lesssim 80$.
%the explosion energy can be E $\ll 10^{51}$ erg so that $D$ might be much smaller than 600.

Another way to constrain the dilution factor may be the surface Li abundance of the CEMP-no stars. We can assume that Li is completely destroyed in massive stars so that X(Li)$_{\rm ej}$, the Li mass fraction in the source star ejecta is 0. The Li mass fraction $X(\textrm{Li})_{\rm CEMP}$ in the CEMP-no star natal cloud (possibly different than the value observed today) is then
\begin{equation}
X(\textrm{Li})_{\rm CEMP} =  \frac{M_{\rm ej}X(\textrm{Li})_{\rm ej} + M_{\rm ISM} X(\textrm{Li})_{\rm ISM}}{M_{\rm ej} + M_{\rm ISM}} = \frac{M_{\rm ISM} X(\textrm{Li})_{\rm ISM}}{M_{\rm ej} + M_{\rm ISM}}
\end{equation}
where $M_{\rm ej}$ is the mass ejected by the source star, and $X(\textrm{Li})_{\rm ISM}$ the Li abundance in the ISM. The dilution factor can then be expressed as
\begin{equation}
D = \frac{M_{\rm ISM}}{M_{\rm ej}} =  \frac{X(\textrm{Li})_{\rm CEMP}}{X(\textrm{Li})_{\rm ISM} - X(\textrm{Li})_{\rm CEMP}}.
\end{equation}
We can suppose that the Li abundance in the initial ISM is equal to the WMAP value of $A(\textrm{Li})_{\rm ISM}=2.72$ \citep{cyburt08}. If, as a first guess, we consider that in situ processes changing the Li abundance did not occur during the CEMP-no star life, we can take $X(\textrm{Li})_{\rm CEMP}$ equal to the observed Li abundance. %In this case $D$ can be calculated.
If the CEMP-no star has no Li at its surface, $D=0$ and the CEMP-no star is made of pure source star ejecta. If the Li abundance is high, more ISM is needed to form the CEMP-no star.
The maximal detected $A(\textrm{Li})$ for CEMP-no stars is about 2.1 \citep[i.e. around the Spite Plateau value][]{spite82}. It leads to a maximal dilution factor of $D_{\rm max} \sim 0.3$, i.e. all CEMP-no stars would be made of almost pure source star ejecta, even the most Li-rich.

This estimation of $D$ stays correct as long as (1) in situ processes changing the Li abundance did not occur in CEMP stars and (2) the Li ISM abundance is indeed $A(\textrm{Li})_{\rm ISM}=2.72$ (value from WMAP). If instead, $A(\textrm{Li})_{\rm ISM}=2.2$ for instance, $D_{\rm max} \sim 4$, i.e. the source star ejecta may still largely dominate compared to the ISM (see Fig.~\ref{vinichem_dil}). 
The main uncertainty in this method are the possible internal processes changing the surface Li abundance of the CEMP-no stars. \cite{korn09} have used models including atomic diffusion to estimate by how much the surface Li abundance of the dwarf CEMP-no star HE~1327-2326 was depleted since its birth. At maximum, Li was depleted by 1.2 dex. It would correspond to an initial Li of $A(\textrm{Li})_{\rm ini} < 1.82$ for HE~1327-2326 and lead to a small $D$ value. Overall, this method of guessing $D$ is uncertain so precaution is required. It nevertheless suggests very small dilution factors. This is inconsistent with the high $D$ values guessed from the Sedov-Taylor theory assuming standard explosions energies. It might nevertheless be more consistent with low energetic SNe ($E_{51} < 1$), that give smaller $D$ values (for $E_{51} = 0.1$ however, there is still 1 order of magnitude of difference at least). % that should give much smaller $D$ values.

   \begin{figure*}[t]
   \centering
   \begin{minipage}[c]{.49\linewidth}
       \includegraphics[scale=0.44]{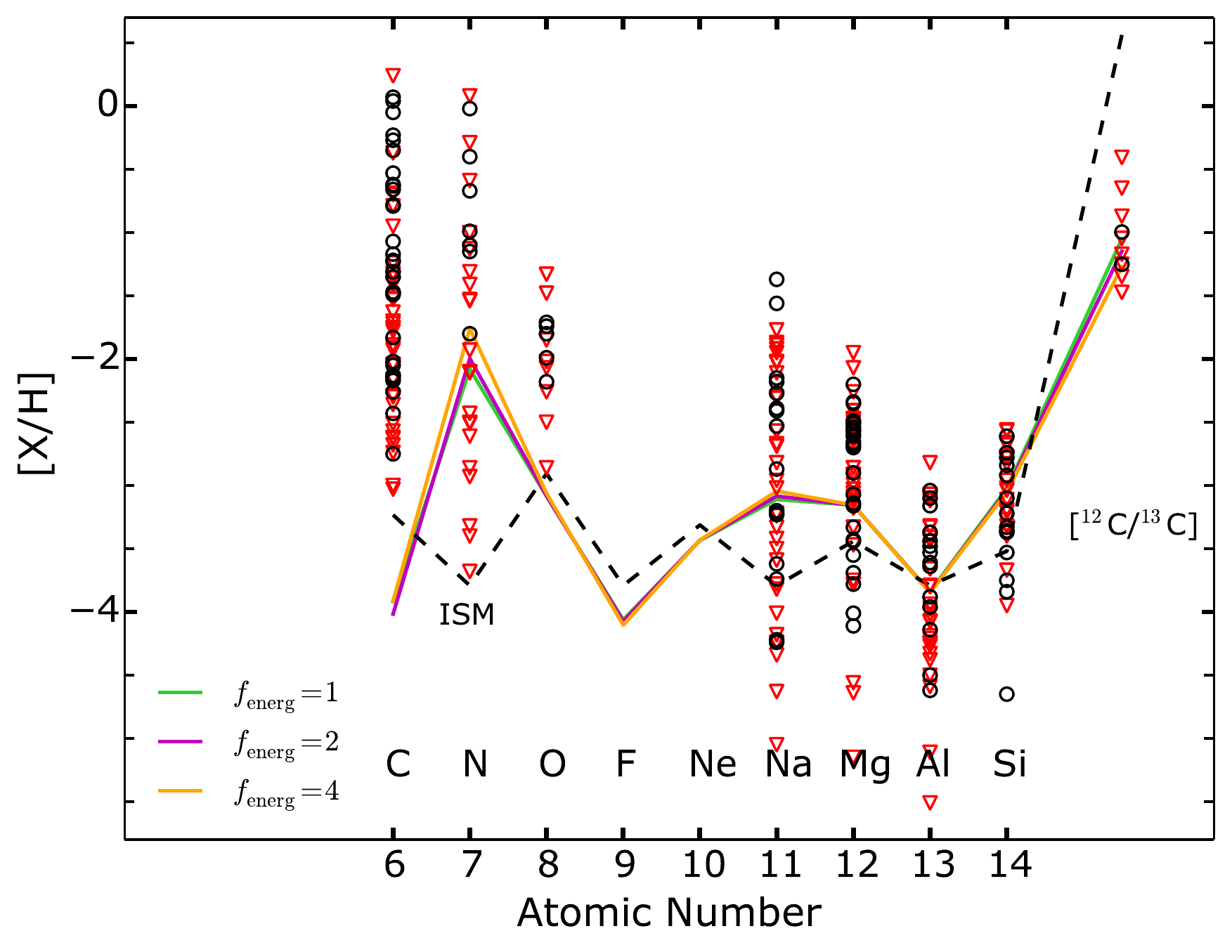}
   \end{minipage}
   \begin{minipage}[c]{.49\linewidth}
       \includegraphics[scale=0.44]{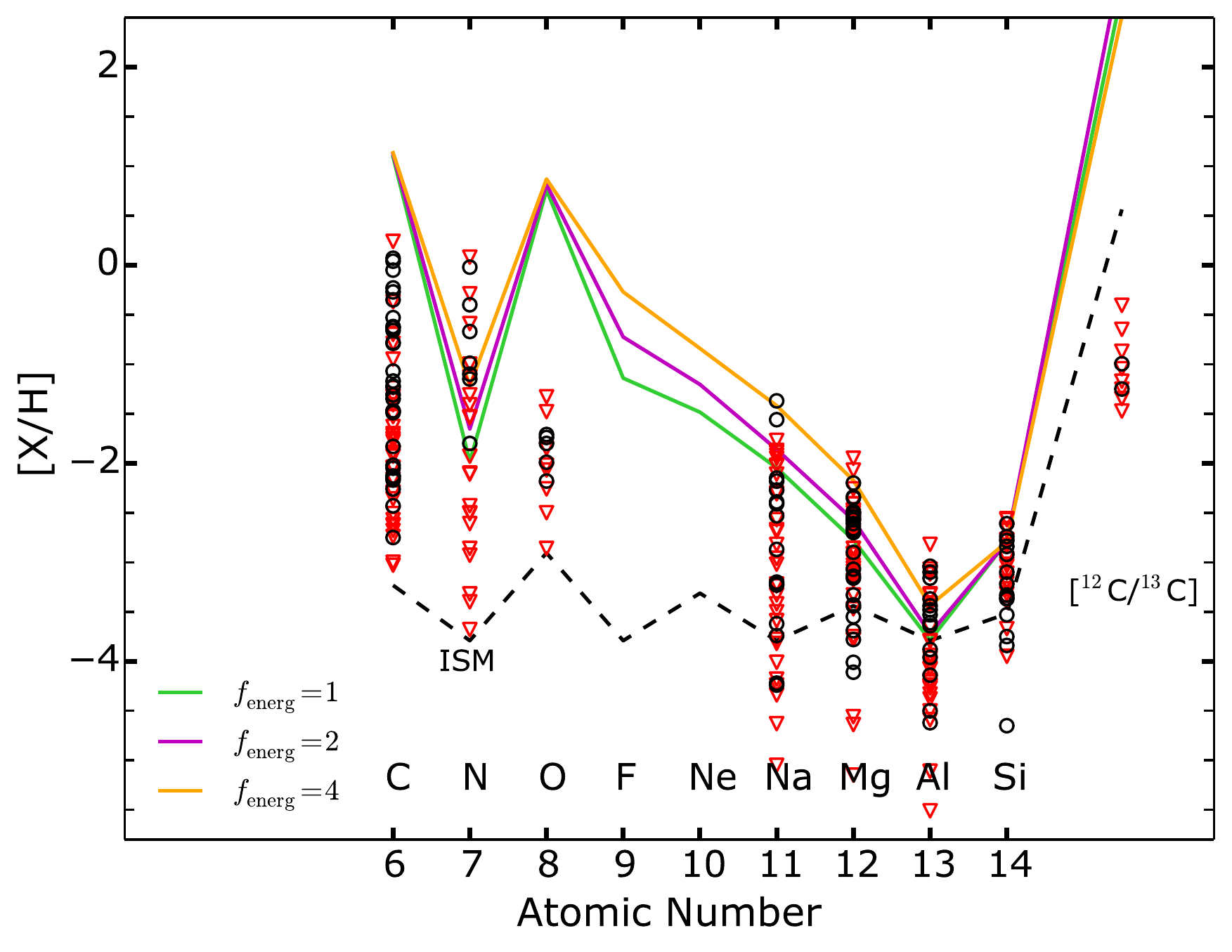}
   \end{minipage}
   \begin{minipage}[c]{.49\linewidth}
       \includegraphics[scale=0.44]{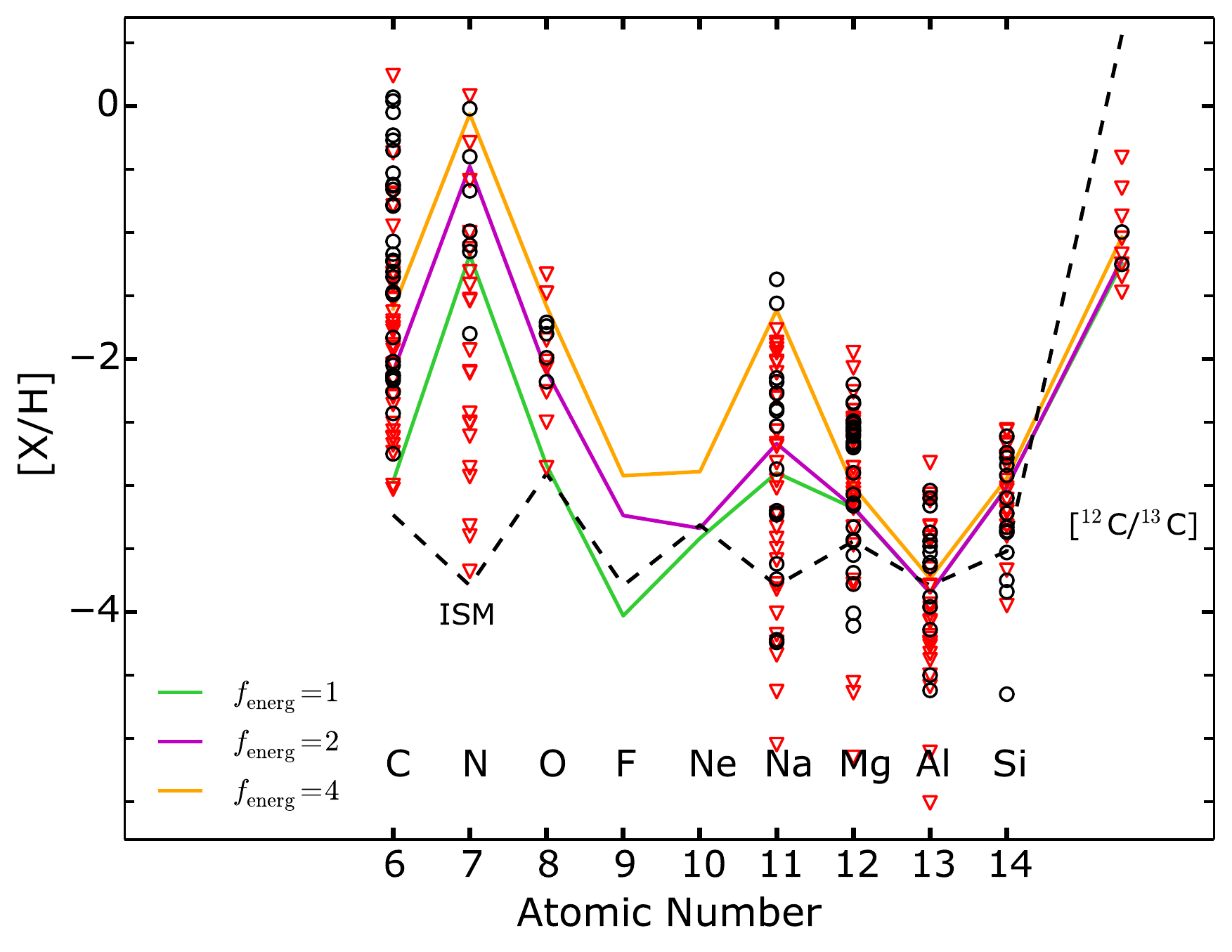}
   \end{minipage}
   \begin{minipage}[c]{.49\linewidth}
       \includegraphics[scale=0.44]{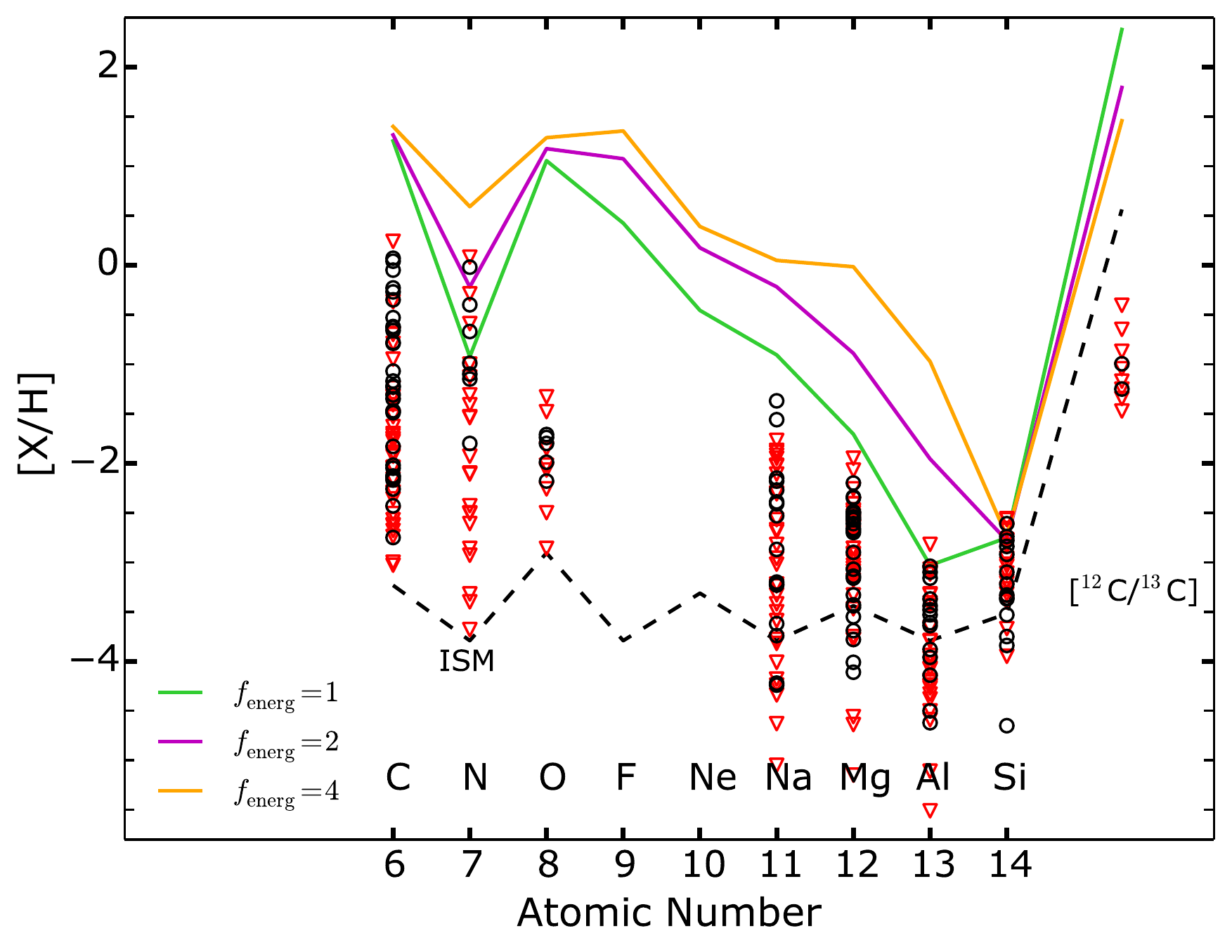}
   \end{minipage}
   \caption[Effect of varying the efficiency of rotational mixing ($f_{\rm energ}$) on the yields]{Effect of varying the efficiency of rotational mixing ($f_{\rm energ}$) on the yields. The top panels show the 20~$M_{\odot}$ models with $\upsilon_{\rm ini}/\upsilon_{\rm crit} = 0.2$. The bottom panels show the 20~$M_{\odot}$ models with $\upsilon_{\rm ini}/\upsilon_{\rm crit} = 0.7$. The left panels show the H-rich ejecta. The right panels show the H-rich + He-rich ejecta.}
\label{fenergeff}
    \end{figure*}

\subsubsection{Efficiency of the rotational mixing}

%\textcolor{red}{say a word about calibration of 10-15 Mo to reproduce N enrichment}
%Also, as discussed in Sect.~\ref{transportrot}, both $f_{\rm energ}=1$ and 4 are consistent with the surface N/H enrichments of observed $10 - 20$~$M_{\odot}$ rotating stars.

The $f_{\rm energ}$ parameter in the $D_{\rm shear}$ expression (Eq.~\ref{dshtz97}) was taken equal to 4 in these models. 
It is consistent with the surface N/H enrichments of observed $10 - 20$~$M_{\odot}$ rotating stars (cf. Sect.~\ref{transportrot}). 
$f_{\rm energ}=4$ corresponds to a critical Richardson number\footnote{Physical values of $Ri_{\rm c}$ are likely between 0.25 and 2, cf. Sect~\ref{transportrot}.} $Ri_{\rm c} = 2$. 
$f_{\rm energ}=1$ ($Ri_{\rm c} = 0.5$) is also a reasonable choice, likely consistent with the observations (cf. Sect.~\ref{transportrot}).

Fig.~\ref{fenergeff} shows the effect of changing $f_{\rm energ}$ from 1 to 4. The 2 top panels show the H-rich (left) and H-rich + He-rich ejecta (right) of the 20~$M_{\odot}$ models with $\upsilon_{\rm ini}/\upsilon_{\rm crit} = 0.2$ and with various $f_{\rm energ}$. The yields are barely affected when changing $f_{\rm energ}$. The same plots for the 20~$M_{\odot}$ model with $\upsilon_{\rm ini}/\upsilon_{\rm crit} = 0.7$ are shown below. In this case, the effect is much larger and leads to differences of $1 - 1.5$ dex at maximum. The $f_{\rm energ} = 1$ model with $\upsilon_{\rm ini}/\upsilon_{\rm crit} = 0.7$ has similar yields than the $f_{\rm energ} = 4$ model with $\upsilon_{\rm ini}/\upsilon_{\rm crit} = 0.3$. 
Changing $f_{\rm energ}$ does not change the general trends but shifts the yields upward or downward, for a given initial rotation. To obtain similar yields as the $\upsilon_{\rm ini}/\upsilon_{\rm crit} = 0.7$, $f_{\rm energ}=4$ model while setting $f_{\rm energ}=1$, it is probably needed to raise $\upsilon_{\rm ini}/\upsilon_{\rm crit}$ up to about 1.
The $f_{\rm energ} \sim$ $Ri_{\rm c}$ parameter is important, especially for fast rotators, but its accurate value is not known. % but is likely between $0.5 < f_{\rm energ} < 4$ ($0.25<$ Ri$_{\rm c}<2$).

   \begin{figure*}[t]
   \centering
   \begin{minipage}[c]{.49\linewidth}
       \includegraphics[scale=0.44]{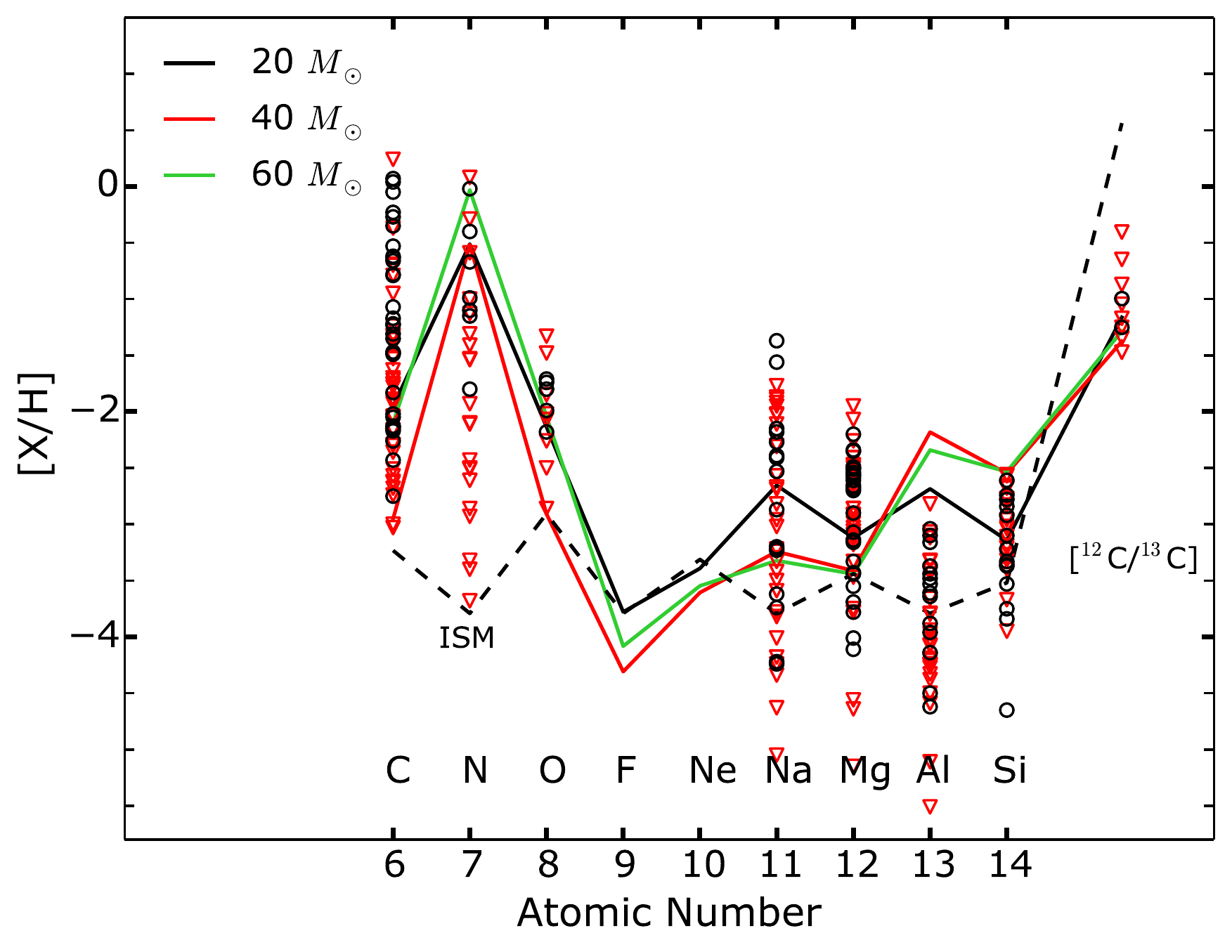}
   \end{minipage}
   \begin{minipage}[c]{.49\linewidth}
       \includegraphics[scale=0.44]{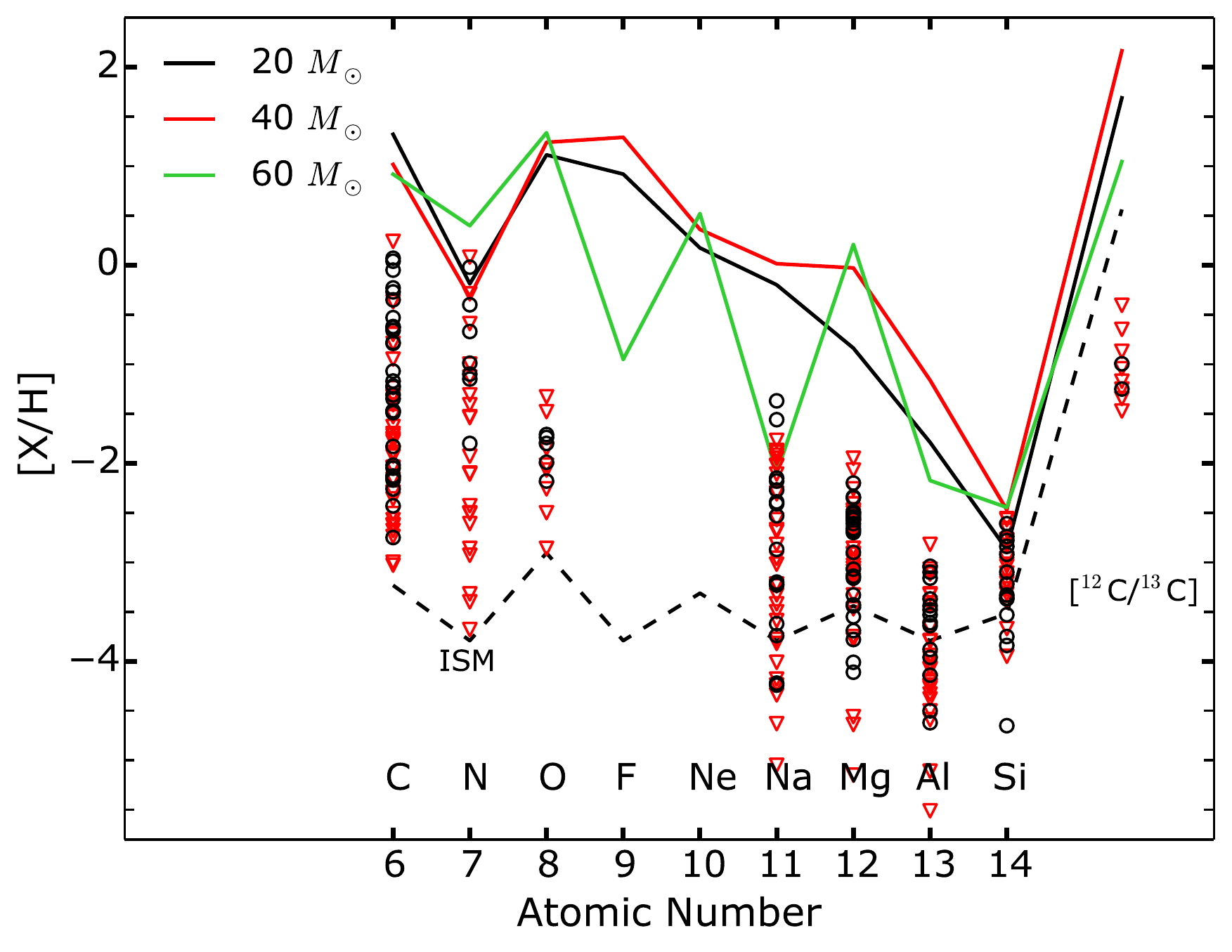}
   \end{minipage}
   \caption[Ejecta composition of $Z=10^{-5}$, $\upsilon_{\rm ini}/\upsilon_{\rm crit} = 0.4$ models with different initial masses]{Composition of the H-rich (\textit{left panel}, same as Fig.~\ref{vinichem_h}) and H-rich + He-rich (\textit{right panel}, same as Fig.~\ref{vinichem_hhe}) ejecta of $Z=10^{-5}$, $\upsilon_{\rm ini}/\upsilon_{\rm crit} = 0.4$ models with different initial masses.}
\label{masseff}
    \end{figure*}

   \begin{figure*}[h!]
   \centering
   \begin{minipage}[c]{.49\linewidth}
       \includegraphics[scale=0.44]{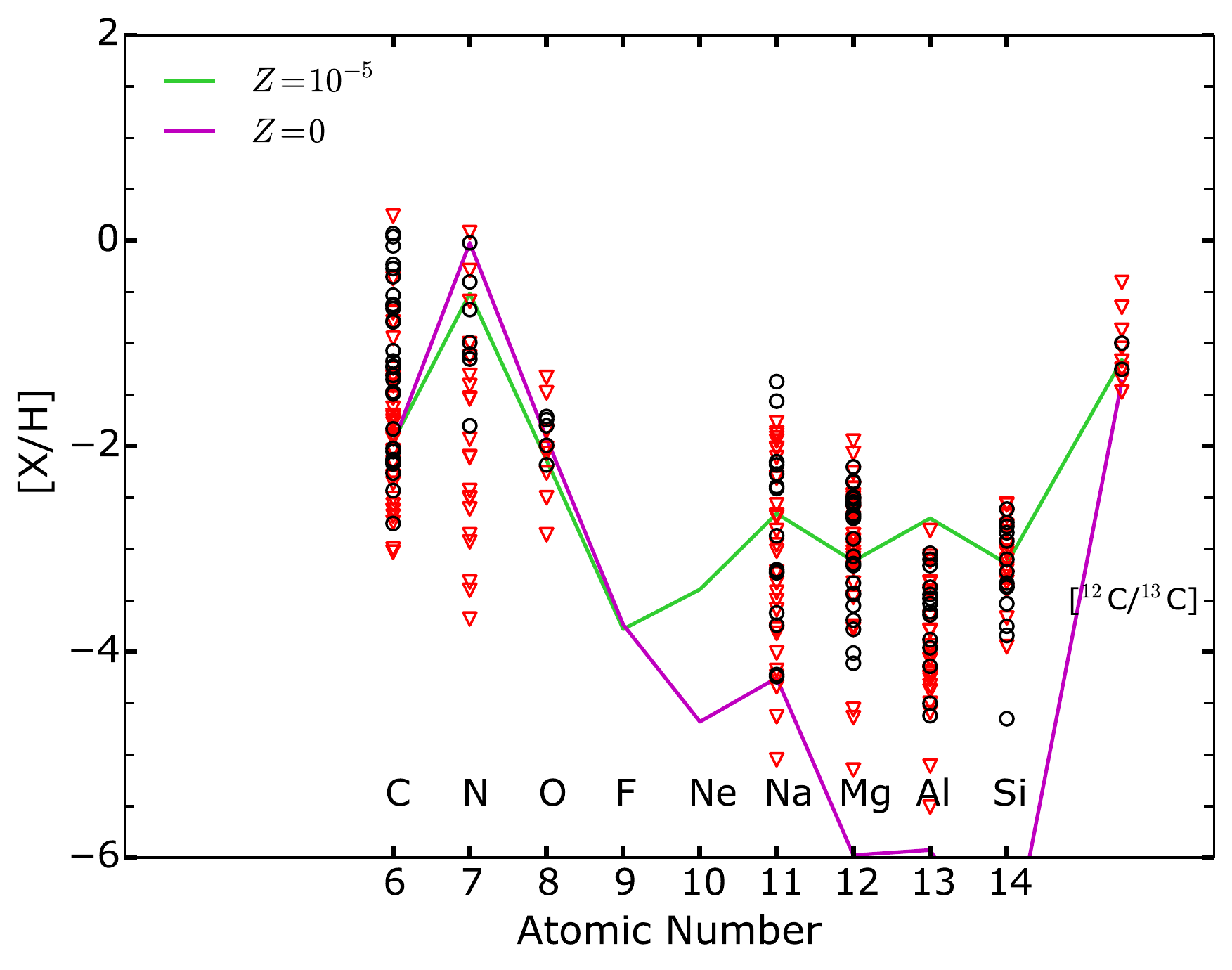}
   \end{minipage}
   \begin{minipage}[c]{.49\linewidth}
       \includegraphics[scale=0.44]{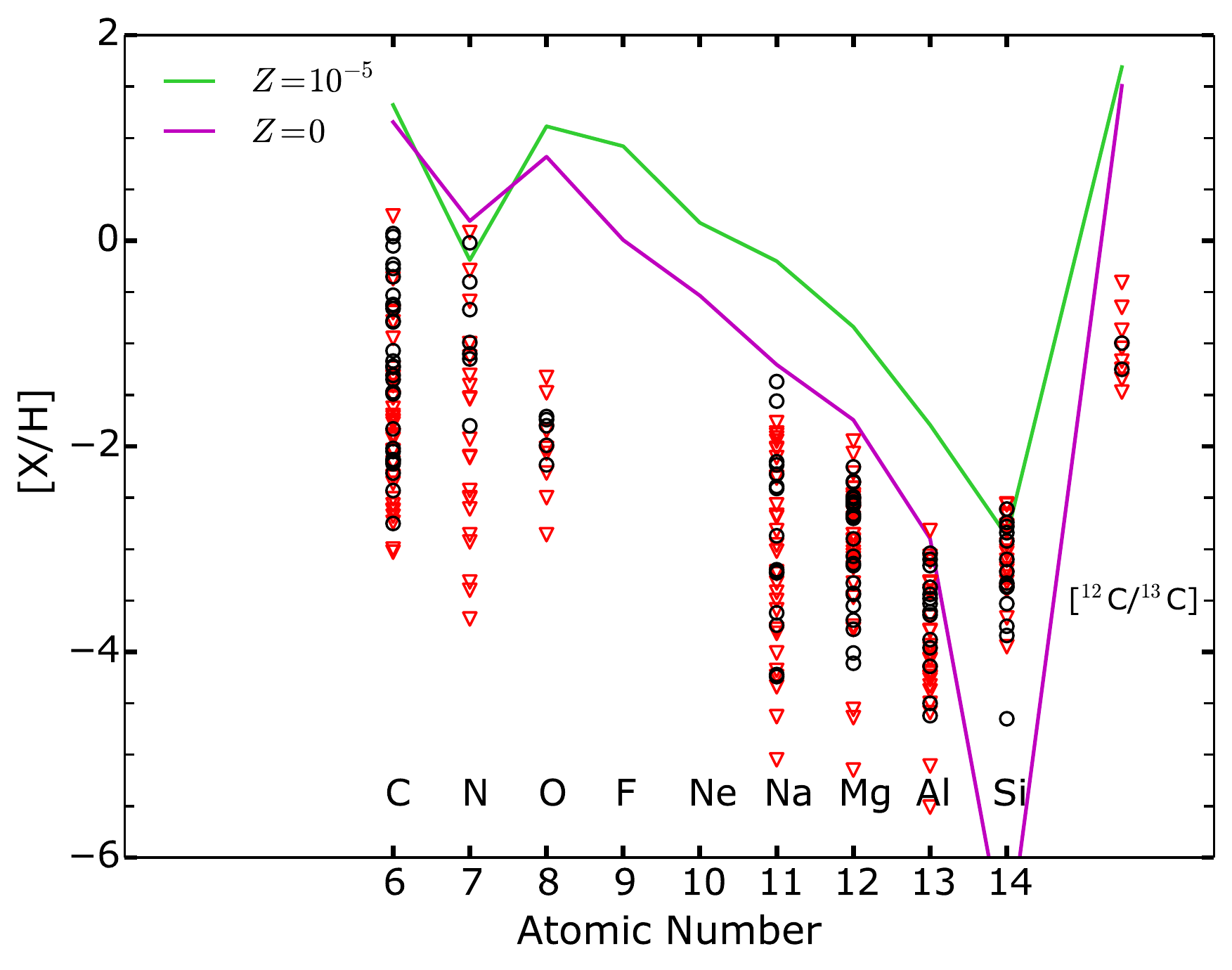}
   \end{minipage}
   \caption[Ejecta composition of 20~$M_{\odot}$, $\upsilon_{\rm ini}/\upsilon_{\rm crit} = 0.4$ models with different metallicities]{Same as Fig.~\ref{masseff} but for 20~$M_{\odot}$ models with $\upsilon_{\rm ini}/\upsilon_{\rm crit} = 0.4$ and with two different initial metallicities.}
\label{meteff}
    \end{figure*}

\subsubsection{Mass}

Figure~\ref{masseff} shows the yields of 20, 40 and 60~$M_{\odot}$ models with $\upsilon_{\rm ini}/\upsilon_{\rm crit} = 0.4$ at $Z=10^{-5}$. The final yields are rather similar. The 60~$M_{\odot}$ produces more N because its H-shell is more active than in lower mass models. In the H-rich layers, more Na is produced in the 20~$M_{\odot}$ models because the back-and-forth mixing process (Fig.~\ref{schemadiff}) is more efficient in lower masses. It is mainly because the process has more time to operate due to the longer duration of the core helium burning stage. Also, Al and Si are overproduced in the H-rich layers of the 40 and 60~$M_{\odot}$ because of the higher H-burning temperature, increasing the efficiency of the Mg-Al-Si chain. The yields of the 60~$M_{\odot}$ in the H-rich + He-rich layers show significant differences compared to the other models (right panel). It is explained by the fact that the He-burning shell becomes convective in the 20 and 40~$M_{\odot}$ models while it does not for the 60~$M_{\odot}$ model. In the 60~$M_{\odot}$ model, the H-burning shell is very active and therefore limits the activation of the He-burning shell. In the 20 and 40~$M_{\odot}$ models, while becoming convective, the He-shell engulfs some $^{14}$N from the H-rich region. It boosts the production of F and Na through the chains $^{14}$N($n,\gamma$)$^{15}$N($\alpha,\gamma$)$^{19}$F and $^{14}$N($\alpha,\gamma$)$^{18}$F($e^+ \nu_e$)$^{18}$O($\alpha,\gamma$)$^{22}$Ne($n,\gamma$)$^{23}$Ne($e^- \bar{\nu}_e$)$^{23}$Na. The neutrons required to activate this chain come from $^{22}$Ne($\alpha,n$).

\subsubsection{Metallicity}

Figure~\ref{meteff} shows the yields (H-rich: left panel, H-rich + He-rich: right panel) of the 20~$M_{\odot}$ models with $\upsilon_{\rm ini}/\upsilon_{\rm crit} = 0.4$ at $Z=10^{-5}$ (green pattern) and $Z=0$ (magenta). The two models are computed with the same physics.
The [C/H], [N/H] and [O/H] ratios do not change by more than 0.5 dex from the $Z=10^{-5}$ the $Z=0$ model. 
The reason is that the transport by rotation of $^{12}$C and $^{16}$O from the convective He-burning core to the convective H-burning shell operates at a similar efficiency in the $Z=0$ model. 
%In particular, rotational mixing leads to a similar production of primary N in both models.
In the H-rich ejecta (left panel), the elements from Ne to Si are much less abundant in the $Z=0$ model compared to the $Z=10^{-5}$ model. 
Considering also the He-rich ejecta raises considerably the abundances of these elements in both models. The exception is for the [Si/H] ratio in the $Z=0$ model, which stays several dex below the [Si/H] ratio of the $Z=10^{-5}$ model.
The differences are mainly due to the fact that the H-burning shell of the $Z=0$ model is more active, which limits the growth of the convective He-burning core (at the middle of the core He-burning phase, the size of the convective He-cores are 3.9~$M_{\odot}$ and 2.7~$M_{\odot}$ in the $Z=10^{-5}$ and $Z=0$ models, respectively). 
The consequence is that less H-burning products (e.g. $^{14}$N) let in between the He-core and the H-shell are engulfed by the convective He-core of the $Z=0$ model. 
Globally, it has the effect of reducing the efficiency of the exchanges of material between the He-core and the H-shell. Finally, the nucleosynthesis is less rich and varied in the $Z=0$ model.
Also, Si is barely affected in the models considered (see also Fig.~\ref{abrotmod2}) so that the [Si/H] ratio in the ejecta mostly reflects the initial [Si/H] ratio. In the $Z=0$ model, the initial Si abundance is zero so that [Si/H] $\sim$ $\log$(Si/H) $\sim -\infty$ in the ejecta.

%As a result, the efficiency of the back-and-forth mixing process is reduced at the first stage (cf. Fig.~\ref{schemadiff})
%As a result, $^{22}$Ne, $^{25}$Mg, ... are not strongly overproduced in the core. It further implies that very little $^{22}$Ne diffuses again to the H-burning shell, 

%Considering the He-rich ejecta raises considerably the abundances of these elements except for Si which is still several dex below the $Z=10^{-5}$ model.
%In the $Z=0$, the back-and-forth mixing process is less efficient than in the $Z=10^{-5}$ model. 
%The main reason is that 

%When considering the H-rich ejecta (left panel), the $Z=0$ model produce .
%When decreasing the metallicity to $Z=0$, the back-and-forth mixing process becomes less efficient. The main reason is  

%\textcolor{red}{Here I would like to say a few words about effect of Z. I am trying to compute two additional 20 Mo models with $\upsilon_{\rm ini}/\upsilon_{\rm crit} = 0.4$ and with $Z=0$ and $Z=10^{-8}$}

\section{Comparison with CEMP-no stars}\label{compacemp}

%\textcolor{red}{Maybe remove known CEMP-s? or consider only $<-3.5$ stars?}
It is now discussed whether the abundances of observed metal-poor stars can be reproduced by the massive stellar models discussed previously. Particular attention is paid to the constraints provided by the $^{12}$C/$^{13}$C ratio.

%\subsubsection{The CEMP sample}
For the observations, I consider CEMP stars with [Fe/H] $<-3$ and [C/Fe] $>1.0$. Recognized CEMP-r, -s and -r/s stars are excluded (according to the criteria of Table~\ref{subclass}). 
I took into account the $\Delta$[C/Fe] of \cite{placco14c} which allows to correct the effect of the first dredge-up of evolved stars and recover the initial C abundance. For the considered stars, $0<$ $\Delta$[C/Fe] $<0.74$. With this correction, 12 stars become CEMP and are then added to the sample.
It gives the 69 stars of Table~\ref{tabcemp}. 
Stars with $T_{\rm eff} \geq 5500$ K and $\log g \geq 3.25$ are classified as main-sequence stars and the other as giants stars \footnote{With the exception of G77-61, with $T_{\rm eff} = 4000$ K and $\log g = 5.05$ that is considered as a main-sequence star, following \cite{plez05} and \cite{beers07}.}.
%We compare the chemical abundances of the CEMP sample with the ejecta of the source star models.
%The comparison with models is made through the [X/H] ratios and [$^{12}$C/$^{13}$C] ratios calculated as [X/H] $=$ [X/Fe] $+$ [Fe/H] and [$^{12}$C/$^{13}$C] = $\log$($^{12}$C/$^{13}$C)$_{\star}$ -  $\log$($^{12}$C/$^{13}$C)$_{\odot}$ with $\log$($^{12}$C/$^{13}$C)$_{\odot} = 1.95$ \textcolor{red}{REF}.

%\subsection{$^{12}$C/$^{13}$C to constrain the source star mass cut}

%The [X/H] ratios from CEMP stars span between $\sim 2$ and $\sim 4$ dex depending on the considered specie (e.g. Fig.~\ref{vinichem_dilhhe}). 

%\textcolor{red}{talk a bit about CNONa.. etc and say that star 2 star analysis is done after?}

\subsection{A global comparison}\label{globalcomp}

%Pop~III
Let us first mention that $Z=0$ source star models may face more difficulties in reproducing the abundance patterns of CEMP-no stars than very low metallicity source star models: the 20~$M_{\odot}$ Pop~III model in Fig.~\ref{meteff} either produces enough C, N, O but not enough Na, Mg, Al (left panel) or it produces enough Na, Mg, Al but then C, O and $^{12}$C/$^{13}$C are overestimated (right panel). Also [Si/H] is underestimated by several dex.

%ALL OTHER MODELS
Considering all the other source star models discussed in Sect.~\ref{secvarpar}, we see that a global match can be found with the abundances of CEMP-no stars. Source star ejecta with various initial rotation rates, mass cuts and dilution factors allow to reproduce all or nearly all the range of CEMP-no star abundances. 
However, a global match does not mean that individual CEMP-no stars can be well reproduced. The next step is to compare source star models with individual CEMP-no stars. As it will be discussed thereafter (especially in Sect.~\ref{secpuzzle}), a more in-depth comparison reveals some difficulties for models to reproduce some of the abundances of individual CEMP-no stars.

Interestingly, we also note that the CNO pattern of the material ejected by the source star has generally either a $\wedge$-shape (H-rich ejecta, Fig.~\ref{vinichem_h}) or a $\vee$-shape (H-rich + He-rich ejecta, Fig.~\ref{vinichem_hhe}). It is nevertheless not always the case (e.g. Fig.~\ref{vinichem_mcut}). In Table~\ref{tabcemp}, very few stars have C, N and O abundances available together, without upper limits. CS~29498-043 has C,N and O abundances available and it has a clear $\vee$-shape. By contrast, HE~1327-2326 has a clear $\wedge$-shape. Another example is the CNO pattern of HE~0107-5240, which has a different shape (it as much N as O). When considering the entire sample of Table~\ref{tabcemp}, about 70~\% is compatible with a $\wedge$-shape for the CNO pattern. Similarly, about 70~\% is compatible with a $\vee$-shape. About the same fraction (70~\%) of the sample is compatible with an almost flat CNO pattern. Detailed comparisons between source star models and CEMP stars are likely required to obtain more informations on the CEMP source stars (Sect.~\ref{seclate} and \ref{secstat}).

%\textcolor{red}{The range covered is similar for MS (black circle) than for RGB stars (red triangles), meaning that the effect of first dredge-up (and possibly other mixing processes occurring during the giant phase) is probably small compared to the dispersion of initial abundances.}

%\pagestyle{empty}
\begin{landscape}
%\begin{table}[htbp]
%\begin{center}

\begin{footnotesize}

\begin{longtable}{lllllllllllllll}
\caption[Stars with \text{[}Fe/H\text{]} $<-3$ and \text{[}C/Fe\text{]} $>1$ (recognized CEMP-s, -r/s and -r excluded)]{Stars with [Fe/H] $<-3$ and [C/Fe] $>1$. [C/Fe] is corrected from $\Delta$[C/Fe] (also reported in the table), that accounts for the effect of the first dredge-up \citep{placco14c}. Recognized CEMP-s, -r/s and -r stars are excluded. Abundance data is from the SAGA database \citep[][]{suda08}. When multiple abundances exist for a star, the abundance from the most recent literature source is selected. \label{tabcemp}}\\ %\textcolor{red}{Put class of CEMP? Trier etoiles par nom maybe} }\\

\hline
%\textbf{First entry} & \textbf{Second entry} & \textbf{Third entry} & \textbf{Fourth entry} \\
 %Star                 & $T_{\rm eff}$   & $\log~g$ & [Fe/H] & A(Li)  & [C/H] & [N/H]  & [O/H]  & [Na/H] & [Mg/H] & [Al/H] & [Si/H] &$^{12}$C/$^{13}$C & $\Delta_{\rm C}$ & Ref \\
 Star	& $T_{\rm eff}$   & $\log~g$ & [Fe/H] & A(Li)  & [C/Fe] & [N/Fe]  & [O/Fe]  & [Na/Fe] & [Mg/Fe] & [Al/Fe] & [Si/Fe] &$^{12}$C/$^{13}$C & $\Delta$[C/Fe] & Ref \\
\hline
\endfirsthead
\multicolumn{14}{c}%
{\tablename\ \thetable\ -- continued} \\
\hline
%\textbf{First entry} & \textbf{Second entry} & \textbf{Third entry} & \textbf{Fourth entry} \\
 %Star                 & $T_{\rm eff}$   & $\log~g$ & [Fe/H] & A(Li)  & [C/H] & [N/H]  & [O/H]  & [Na/H] & [Mg/H] & [Al/H] & [Si/H] &$^{12}$C/$^{13}$C & $\Delta_{\rm C}$ & Ref \\
 Star	& $T_{\rm eff}$   & $\log~g$ & [Fe/H] & A(Li)  & [C/Fe] & [N/Fe]  & [O/Fe]  & [Na/Fe] & [Mg/Fe] & [Al/Fe] & [Si/Fe] &$^{12}$C/$^{13}$C & $\Delta$[C/Fe] & Ref \\
\hline
\endhead
\hline %\multicolumn{4}{r}{\textit{Continued on next page}} \\
\endfoot
\hline
\endlastfoot

 BD+44\_493            & 5430 & 3.4  & -3.8   & 0.64   & 1.23   & \ensuremath{<}2.2   & 1.54   & 0.3     & 0.89    & -0.44   & 0.48    & \ensuremath{>}15.0    & 0.0  & 1,2,3            \\
 CS22877-001          & 4790 & 1.45 & -3.24  & 0.66   & 1.54   & -0.08  & 1.22   & -       & 0.38    & -0.73   & -       & 35.0     & 0.44 & 2                \\
 CS22891-200          & 4490 & 0.5  & -3.88  & \ensuremath{<}0.13  & 1.26   & 1.38   & \ensuremath{<}1.67  & 0.29    & 0.82    & -0.37   & 1.05    & \ensuremath{>}6.0     & 0.73 & 2,4              \\
 CS22897-008          & 4550 & 0.7  & -3.73  & 0.77   & 1.05   & 0.05   & \ensuremath{<}1.86  & -0.05   & 0.6     & -0.77   & 0.46    & \ensuremath{>}20.0    & 0.45 & 5,6,7            \\
 CS22949-037          & 4630 & 0.95 & -4.2   & \ensuremath{<}0.13  & 1.9    & 2.67   & \ensuremath{<}2.39  & 1.89    & 1.56    & 0.26    & 1.52    & \ensuremath{>}4.0     & 0.74 & 5,8,2            \\
 CS22950-046          & 4380 & 0.5  & -3.64  & \ensuremath{<}-0.05 & 1.13   & 1.13   & \ensuremath{<}2.39  & -0.18   & 0.58    & -0.63   & 1.08    & -        & 0.52 & 5,4              \\
 CS22953-037          & 6150 & 3.7  & -3.05  & 1.97   & 1.0    & \ensuremath{<}2.18  & \ensuremath{<}0.5   & -       & 0.55    & -0.56   & 0.21    & -        & 0.0  & 5                \\
 CS22957-027          & 5220 & 2.65 & -3.0   & \ensuremath{<}0.86  & 2.63   & 1.69   & \ensuremath{<}1.2   & 0.98    & 0.08    & -0.12   & -       & 6.0      & 0.02 & 2                \\
 CS22960-053          & 4860 & 1.65 & -3.33  & \ensuremath{<}0.55  & 1.42   & 3.41   & -      & -       & 0.77    & 0.51    & -       & -        & 0.02 & 5                \\
 CS29498-043          & 4440 & 0.5  & -3.85  & \ensuremath{<}-0.05 & 3.06   & 1.74   & 2.37   & 1.03    & 1.78    & 0.75    & 1.08    & 8.0      & 0.31 & 2                \\
 CS29502-092          & 4820 & 1.5  & -3.2   & \ensuremath{<}0.45  & 1.45   & 1.1    & 1.13   & -0.02   & 0.5     & -0.6    & -       & 12.0     & 0.39 & 2                \\
 CS29527-015          & 6240 & 4.0  & -3.55  & 2.091  & 1.29   & -      & -      & -0.19   & 0.41    & 0.07    & 0.18    & -        & 0.0  & 9,10,11,12,13,14 \\
 CS29528-041          & 6170 & 4.0  & -3.06  & 1.71   & 1.57   & 3.04   & \ensuremath{<}1.38  & 0.67    & 0.38    & -0.38   & -0.16   & -        & 0.0  & 12,15            \\
 G77-61               & 4000 & 5.1  & -4.0   & \ensuremath{<}3.31  & 3.21   & 2.2    & 1.82   & 0.77    & 0.56    & -       & -       & 5.0      & 0.0  & 16,17            \\
 HE0015+0048          & 4600 & 0.9  & -3.07  & \ensuremath{<}0.92  & 1.29   & -      & -      & -       & 0.66    & -0.37   & 0.5     & -        & 0.67 & 18               \\
 HE0017-4346          & 6198 & 3.8  & -3.07  & -      & 3.11   & -      & -      & 1.51    & 0.87    & 0.03    & -       & -        & 0.0  & 8                \\
 HE0102-0633          & 6012 & 3.7  & -3.1   & -      & 1.08   & -      & -      & \ensuremath{<}1.06   & 0.44    & -0.27   & 0.32    & -        & 0.0  & 8                \\
 HE0107-5240          & 5100 & 2.2  & -5.4   & \ensuremath{<}1.12  & 3.88   & 2.54   & 2.54   & 1.06    & 0.25    & \ensuremath{<}-0.08  & \ensuremath{<}0.48   & \ensuremath{>}50.0    & 0.06 & 19,20            \\
 HE0134-1519          & 5500 & 3.2  & -4.0   & 1.27   & 1.0    & \ensuremath{<}1.0   & \ensuremath{<}2.9   & -0.24   & 0.25    & -0.38   & 0.05    & \ensuremath{>}4.0     & 0.0  & 21,22            \\
 HE0146-1548          & 4636 & 1.7  & -3.46  & -      & 1.57   & -      & -      & 1.16    & 0.87    & 0.14    & 0.5     & -        & 0.73 & 23,24            \\
 HE0233-0343          & 6100 & 3.4  & -4.7   & 1.77   & 3.48   & \ensuremath{<}2.8   & \ensuremath{<}4.0   & \ensuremath{<}0.5    & 0.59    & \ensuremath{<}0.03   & 0.05    & \ensuremath{>}5.0     & 0.0  & 21               \\
 HE0251-3216          & 5750 & 3.7  & -3.15  & -      & 2.53   & -      & -      & 0.97    & 0.61    & 0.05    & -0.6    & -        & 0.0  & 8                \\
 HE0450-4902          & 6300 & 4.5  & -3.1   & \ensuremath{<}1.98  & 2.03   & 2.0    & \ensuremath{<}3.5   & 0.23    & 0.53    & -0.78   & 0.0     & -        & -    & 21               \\
 HE0557-4840          & 4900 & 2.2  & -4.8   & \ensuremath{<}0.7   & 1.77   & \ensuremath{<}0.96  & 2.3    & -0.25   & 0.16    & -0.71   & -       & -        & 0.01 & 25,26            \\
 HE1005-1439          & 5000 & 1.9  & -3.2   & -      & 2.51   & 1.79   & -      & 1.17    & 0.59    & -       & -       & -        & 0.02 & 27               \\
 HE1012-1540          & 5230 & 2.65 & -3.76  & \ensuremath{<}0.75  & 2.4    & 1.15   & \ensuremath{<}2.18  & 1.65    & 1.81    & 0.69    & 0.65    & \ensuremath{>}30.0    & 0.0  & 2,8              \\
 HE1029-0546          & 6650 & 4.3  & -3.3   & \ensuremath{<}2.0   & 2.64   & 2.9    & \ensuremath{<}3.7   & -       & -0.03   & \ensuremath{<}-0.42  & -0.03   & 9.0      & -    & 21               \\
 HE1150-0428          & 5200 & 2.5  & -3.21  & -      & 2.51   & 2.62   & -      & 1.44    & 0.35    & -       & -       & -        & 0.02 & 8                \\
 HE1201-1512          & 5725 & 4.2  & -3.89  & -      & 1.14   & \ensuremath{<}1.23  & -      & -0.35   & 0.2     & -0.73   & -       & -        & 0.0  & 23,24            \\
 HE1249-3121          & 5373 & 3.4  & -3.23  & -      & 1.91   & -      & -      & -       & 0.24    & -0.79   & -       & -        & 0.0  & 28,29            \\
 HE1300+0157          & 5550 & 3.3  & -3.49  & 1.06   & 1.34   & \ensuremath{<}0.8   & 1.78   & -0.13   & 0.33    & -0.15   & 0.57    & -        & 0.0  & 8,30             \\
 HE1300-0641          & 5308 & 2.96 & -3.14  & -      & 1.25   & -      & -      & -       & 0.02    & -1.19   & -       & -        & 0.01 & 28,29            \\
 HE1305-0331          & 6081 & 4.22 & -3.26  & -      & 1.09   & -      & -      & -       & -       & -0.7    & -       & -        & 0.0  & 31,28            \\
 HE1310-0536          & 5000 & 1.9  & -4.2   & \ensuremath{<}0.8   & 2.44   & 3.2    & \ensuremath{<}2.8   & 0.19    & 0.42    & -0.39   & 0.8     & 3.0      & 0.08 & 21               \\
 HE1327-2326          & 6180 & 3.7  & -5.66  & \ensuremath{<}0.7   & 4.18   & 4.67   & 3.86   & 2.46    & 1.65    & 1.16    & -       & \ensuremath{>}5.0     & 0.0  & 32,33            \\
 HE1338-0052          & 5856 & 3.7  & -3.0   & -      & 1.53   & -      & -      & -       & 0.44    & -0.16   & 0.39    & -        & 0.0  & 8                \\
 HE1351-1049          & 5204 & 2.85 & -3.46  & -      & 1.74   & -      & -      & -       & 0.28    & -0.75   & -       & -        & 0.01 & 28,29            \\
 HE1410-0004          & 5605 & 3.5  & -3.02  & \ensuremath{<}1.32  & 2.24   & -      & 1.28   & 0.61    & 0.53    & -       & -       & -        & 0.04 & 8,34             \\
 HE1413-1954          & 6302 & 3.8  & -3.5   & 2.035  & 1.67   & -      & -      & -       & -       & -       & -       & -        & 0.0  & 10,29            \\
 HE1456+0230          & 5664 & 2.2  & -3.32  & -      & 2.37   & 3.03   & -      & 0.3     & 0.29    & -0.09   & 0.54    & -        & 0.02 & 8                \\
 HE1506-0113          & 5016 & 2.4  & -3.54  & -      & 1.49   & 0.61   & -      & 1.65    & 0.89    & -0.53   & 0.5     & -        & 0.02 & 23,24            \\
 HE2123-0329          & 4725 & 1.15 & -3.22  & \ensuremath{<}0.58  & 1.06   & -      & -      & -       & 0.58    & -0.57   & 0.56    & -        & 0.66 & 18               \\
 HE2139-5432          & 5416 & 2.2  & -4.02  & -      & 2.61   & 2.09   & -      & 2.15    & 1.61    & 0.36    & 1.0     & -        & 0.01 & 23,24            \\
 HE2318-1621          & 4846 & 1.4  & -3.67  & -      & 1.04   & 1.24   & -      & 0.71    & 0.2     & -0.58   & -       & -        & 0.5  & 35               \\
 HE2331-7155          & 4900 & 1.5  & -3.7   & \ensuremath{<}0.37  & 1.34   & 2.57   & \ensuremath{<}1.7   & 0.46    & 1.2     & -0.38   & \ensuremath{<}0.25   & 5.0      & -    & 21               \\
 LAMOSTJ125346.09 & 6030 & 3.65 & -4.02  & 1.8    & 1.59   & -      & -      & -0.2    & 0.24    & -       & -       & -        & -    & 36               \\
 LAMOSTJ131331.18 & 4750 & 1.6  & -4.12  & \ensuremath{<}0.68  & 1.83   & -      & -      & -0.06   & 0.34    & -       & 0.45    & -        & -    & 36               \\
 LAMOSTJ1626+1721     & 5930 & 3.6  & -3.2   & -      & 1.07   & -      & -      & -0.02   & 0.5     & -0.33   & 0.46    & -        & -    & 37               \\
 LAMOSTJ1709+1616     & 5780 & 3.5  & -3.71  & -      & 1.58   & -      & -      & -       & 0.28    & -       & 0.18    & -        & -    & 37               \\
 SDSSJ0002+2928       & 6150 & 4.0  & -3.26  & -      & 2.63   & -      & -      & 0.99    & 0.36    & -       & -       & -        & -    & 38               \\
 SDSSJ0126+0607       & 6900 & 4.0  & -3.01  & \ensuremath{<}2.2   & 3.08   & -      & -      & 0.86    & 0.66    & -       & -       & -        & -    & 38,39            \\
 SDSSJ0212+0137       & 6333 & 4.0  & -3.59  & 2.04   & 2.28   & \ensuremath{<}2.66  & 1.6    & -       & 0.52    & -0.55   & 0.23    & -        & -    & 40               \\
 SDSSJ0351+1026       & 5450 & 3.6  & -3.18  & -      & 1.55   & -      & -      & -0.15   & 0.71    & -       & -       & -        & -    & 38               \\
 SDSSJ0723+3637       & 5150 & 2.2  & -3.32  & -      & 1.79   & -      & -      & -0.1    & 0.23    & -       & -       & -        & -    & 38               \\
 SDSSJ1035+0641       & 6262 & 4.0  & \ensuremath{<}-5.07  & \ensuremath{<}1.1   & 3.54   & -      & -      & -       & \ensuremath{<}-0.06  & -       & -       & -        & -    & 40               \\
 SDSSJ1114+1828       & 6200 & 4.0  & -3.35  & -      & 3.3    & 2.2    & -      & -       & -       & -       & -       & \ensuremath{>}60.0    & -    & 41               \\
 SDSSJ1143+2020       & 6240 & 4.0  & -3.15  & -      & 2.8    & 2.48   & -      & -       & -       & -       & -       & 20.0     & -    & 41               \\
 SDSSJ1245-0738       & 6110 & 2.5  & -3.21  & -      & 3.45   & -      & -      & 1.29    & 0.68    & -0.12   & 0.02    & -        & -    & 40               \\
 SDSSJ131326.89 & 5200 & 2.6  & -5.0   & \ensuremath{<}0.8   & 2.96   & 3.46   & -      & 0.37    & 0.44    & -0.11   & \ensuremath{<}0.21   & -        & -    & 42               \\
 SDSSJ1349-0229       & 6200 & 4.0  & -3.24  & -      & 3.01   & -      & -      & 1.87    & 0.73    & -       & -       & \ensuremath{>}30.0    & -    & 38,43            \\
 SDSSJ1422+0031       & 5200 & 2.2  & -3.03  & -      & 1.7    & -      & -      & 0.36    & 0.77    & -       & -       & -        & -    & 38               \\
 SDSSJ1613+5309       & 5350 & 2.1  & -3.33  & -      & 2.09   & -      & -      & 0.76    & 0.92    & -       & -       & -        & -    & 38               \\
 SDSSJ161956+170539   & 6191 & 4.0  & -3.57  & -      & 2.34   & -      & -      & -       & 0.02    & -       & -0.27   & -        & -    & 44               \\
 SDSSJ1646+2824       & 6100 & 4.0  & -3.05  & -      & 2.52   & -      & -      & -       & 0.71    & -       & -       & -        & -    & 38               \\
 SDSSJ1742+2531       & 6345 & 4.0  & -4.8   & -      & 3.63   & -      & \ensuremath{<}3.03  & \ensuremath{<}0.7    & \ensuremath{<}0.27   & -       & \ensuremath{<}0.34   & -        & -    & 40               \\
 SDSSJ1746+2455       & 5350 & 2.6  & -3.17  & -      & 1.24   & -      & -      & 0.48    & 0.69    & -       & -       & -        & -    & 38               \\
 SDSSJ2209-0028       & 6440 & 4.0  & -3.96  & -      & 2.61   & -      & -      & -       & -       & -       & -       & -        & -    & 41               \\
 SMSSJ005953.98 & 5413 & 2.95 & -3.94  & 2.0    & 1.2    & -      & -      & 1.99    & 0.61    & -0.25   & 0.73    & -        & -    & 45               \\
 SMSSJ031300.36 & 5125 & 2.3  & \ensuremath{<}-7.3   & 0.7    & 4.9    & \ensuremath{<}3.8   & \ensuremath{<}5.0   & \ensuremath{<}1.8    & 3.0     & \ensuremath{<}1.1    & \ensuremath{<}3.0    & -        & -    & 46               \\

\end{longtable}

\textbf{References}.  1 - \cite{placco14b}; 2 - \cite{roederer14c}; 3 - \cite{ito13}; 4 - \cite{venn04}; 5 - \cite{roederer14a}; 6 - \cite{cayrel04}; 7 - \cite{spite06}; 8 - \cite{cohen13}; 9 - \cite{spite12}; 10 - \cite{sbordone10}; 11 - \cite{bonifacio09}; 12 - \cite{andrievsky07}; 13 - \cite{andrievsky10}; 14 - \cite{andrievsky08}; 15 - \cite{sivarani06}; 16 - \cite{beers07}; 17 - \cite{plez05}; 18 - \cite{hollek11}; 19 - \cite{bessell04}; 20 - \cite{christlieb04}; 21 - \cite{hansen15}; 22 - \cite{hansen14}; 23 - \cite{norris13}; 24 - \cite{yong13}; 25 - \cite{norris12}; 26 - \cite{norris07}; 27 - \cite{aoki07}; 28 - \cite{barklem05}; 29 - \cite{zhang11}; 30 - \cite{frebel07}; 31 - \cite{ren12}; 32 - \cite{frebel08}; 33 - \cite{aoki06}; 34 - \cite{cohen06}; 35 - \cite{placco14a}; 36 - \cite{li15}; 37 - \cite{li15}; 38 - \cite{aoki13}; 39 - \cite{aoki08}; 40 - \cite{bonifacio15}; 41 - \cite{spite13}; 42 - \cite{frebel15}; 43 - \cite{behara10}; 44 - \cite{caffau13}; 45 - \cite{jacobson15}; 46 - \cite{keller14}

\end{footnotesize}

%\end{center}
%\end{table} 
\end{landscape}
%\pagestyle{plain}

%=====================================================================================================================================================================

\subsubsection{Constraints from the $^{12}$C/$^{13}$C ratio}
%C12C13
%The $^{12}$C/$^{13}$C ratios are generally very low. 
%It has to be noted that some CEMP, that are not plotted, have a lower limit on the $^{12}$C/$^{13}$C ratio. 
%In the $\alpha$-enhanced mixture considered in the source star models, $^{12}$C/$^{13}$C $=300$ (corresponding to [$^{12}$C/$^{13}$C] $\simeq 0.5$).
The CEMP-no stars with a measured $^{12}$C/$^{13}$C ratio have $3<$ $^{12}$C/$^{13}$C $<35$. Two stars are still dwarfs, with $^{12}$C/$^{13}$C $\sim 6$. 
As mentioned previously (Sect.~\ref{secbox}), the $^{12}$C/$^{13}$C ratio at the surface of an unevolved CEMP-no star is probably similar to the $^{12}$C/$^{13}$C ratio in the cloud in which the star formed, hence similar to the $^{12}$C/$^{13}$C ratio in the source star ejecta.
%For comparison, in the Sun, $^{12}$C/$^{13}$C $\simeq 90$ \citep[][]{lodders03}. 
%As a consequence, the ratio may also be similar to the ratio in the source star ejecta.

The $^{12}$C/$^{13}$C ratios of the CEMP-no star sample are well reproduced by the H-rich ejecta of the source star models (Fig.~\ref{vinichem_h}). By contrast, they are largely overestimated by the H-rich + He-rich ejecta \citep[Fig.~\ref{vinichem_hhe}, cf. also][Sect.~5.1, page~\pageref{pbox} of this thesis]{choplin16}. 
Of course one could imagine to dilute the H-rich + He-rich ejecta, having a high $^{12}$C/$^{13}$C, with an ISM having a very low $^{12}$C/$^{13}$C (as discussed in Sect.~\ref{inicompo}, here I take $^{12}$C/$^{13}$C = 300 in the ISM but a lower value may also be chosen). The final mixture may present a low $^{12}$C/$^{13}$C ratio, consistent with observations. 
In this case however, it just pushes back the problem: the need for a source producing a large amount of $^{13}$C in the early Universe remains and has to be explained by some mechanism.

It is worth mentioning here that Galactic evolution models predict that a standard population of very low metallicity massive source stars will lead to a $^{12}$C/$^{13}$C ratio of $4500 - 31000$ in the (almost) primordial ISM \citep{chiappini08}. If instead, this population is dominated by massive fast rotators, Galactic evolution models predict $30 <$ $^{12}$C/$^{13}$C $<300$.
Rotation in massive stars is indeed a way to produce large amounts of $^{13}$C (e.g. Fig.~\ref{abrotmod}). 
The fact that it exists CEMP-no stars (especially unevolved stars) with a $^{12}$C/$^{13}$C ratio even lower that 30 suggests that a very special material is required to form them, maybe coming from the relatively external layers of one specific rotating massive source star.
%However a $^{12}$C/$^{13}$C ratio of 30 is still high compared to the $^{12}$C/$^{13}$C ratio of most CEMP-no stars. 

%Also, This ratio is typical of a CNO-processed material. 
As a remark, let us mention that the fact that there is no CEMP star with a $^{12}$C/$^{13}$C ratio below the CNO-equilibrium value (which is about~4) suggests that the dominant source of $^{13}$C in the early Universe comes from CNO burning. If there was another important source of $^{13}$C, we should observe CEMP-no stars with lower $^{12}$C/$^{13}$C ratios. Moreover if such stars exist, they should be observable since low $^{12}$C/$^{13}$C ratios are easier to detect: if the ratio is low, $^{13}$C is abundant and then $^{13}$C lines are stronger.
This last point also shows that it can exist a bias towards low $^{12}$C/$^{13}$C. % If $^{12}$C/$^{13}$C is high, $^{13}$C lines are weak and hard to detect. 
Some CEMP-no stars have a lower limit for $^{12}$C/$^{13}$C \citep[e.g. HE~1201-1512 and HE~1327-2326 with $^{12}$C/$^{13}$C $>20$ and $>5$ respectively,][]{aoki06,norris13}. Future observations may reveal a population of stars with higher $^{12}$C/$^{13}$C ratios. %Future observations might reveal CEMP with higher $^{12}$C/$^{13}$C ratios.

\subsubsection{Can the H-rich material of massive source stars explain the abundances of CEMP-no stars?}

In the previous discussion and in Sect.~\ref{secbox}, it was proposed that CEMP-no stars could have formed with only the H-rich material of the source star. %(see also Fig.~\ref{vinichem_h}).} %Is there s -process elements on these stars?  Check maybe
%This could mean that only the H-rich envelope of the source star was expelled. 
%%%%%This may be achieved for low energetic supernova, so that deep layers are not unbound and expelled.
However, if considering only the H-rich source star ejecta, several issues arise (see Fig.~\ref{vinichem_h}): 
\begin{enumerate}
\item The ranges of Mg/H, Al/H and Si/H ratios are not covered by source star models (also N and Na to a smaller extent).
\item The predicted [Al/H] ratios are too high (also true if considering the H- + He-rich ejecta, Fig.~\ref{vinichem_hhe}).
\item Such an ejecta shows a clear CNO processed signature with a very characteristic CNO pattern ($\wedge$-shape), while some CEMP-no stars are not compatible with this pattern.
\item The most C-rich CEMP-no stars, with [C/H] $\gtrsim -1.5$ cannot be explained.
\end{enumerate}

First, some scatter in the CEMP star abundances is probably induced by the fact that the abundance data is not homogeneous, together with the possible 3D/NLTE corrections on abundances (cf. Sect.~\ref{secabuncer} and \ref{secbox}). This could alleviate the issue 1. 

Second, some nuclear reaction rates of the Ne-Na and Mg-Al chains are uncertain (cf. Sect.~\ref{secbox}). Changing these rates will change the predicted Na/H, Mg/H, Al/H ratios but in a similar way for all the models, i.e. the predicted scatter will not change. I computed again the $\upsilon_{\rm ini}/\upsilon_{\rm crit} = 0.4$ model with the nuclear rate of $^{27}$Al($p,\gamma$)$^{28}$Si from \cite{cyburt10} instead of \cite{iliadis01}. This is an extreme case, favoring Al destruction, since the rate of \cite{cyburt10} is the highest one below 100 MK. At $T=50$ MK, the \cite{cyburt10} rate is $\sim 100$ times larger than the \cite{iliadis01} rate. The orange dashed line in Fig.~\ref{vinichem_h} shows that the [Al/H] ratio in the ejecta of the model with the \cite{cyburt10} decreases by $\sim 1$ dex compared to the standard case. We note that the [Si/H] is barely modified. Nuclear rate uncertainties can lead to significant differences in the predicted yields and make the [Al/H] consistent with the bulk of observed abundances. This helps with the issue 2.

%\textcolor{red}{here talk about 20 Mo models, plot these models in CNCC, also 60 and maybe others models. Then other models? I think so. Heger, Limongi... Models that are interesting have a common feature. Then new sec. We explore the late mixing process. New grid with a bit improved physical ingredient. Or maybe other models (Hirschi... ) after, at the end?}

The issues 3 and 4 can be solved if considering the H-rich + He-rich ejecta (Fig.~\ref{vinichem_hhe}), which gives much more carbon and reverses the CNO pattern. Some dilution with the ISM might provide abundance patterns able to reproduce the observations (right panel of Fig.~\ref{vinichem_dil}). However, in this case, the predicted $^{12}$C/$^{13}$C is too high. The difficulty here is to get a high enough C/N ratio together with a low $^{12}$C/$^{13}$C ratio, as it is observed on many CEMP-no stars (Fig.~\ref{cnccrotfirst}). This issue is discussed in the next section.

\subsection{The C/N $-$ $^{12}$C/$^{13}$C puzzle\label{secpuzzle}}

\subsubsection{Standard source star models}

In source star models, either both C/N and $^{12}$C/$^{13}$C ratios are low (H-rich ejecta), or both ratios are high (H-rich + He-rich ejecta).
The crosses in Fig.~\ref{cnccrotfirst} show the ratios in the total wind of some source star models discussed in the previous section. The tracks show the variation of the ratios in the ejecta with $M_{\rm cut}$ let as a free parameter. A specific $M_{\rm cut}$ value gives one point in such a track. Here, for a given source star model, all possible mass cuts are considered: in Eq.~\ref{yie}, $M_{\rm cut}$ is varied between 0 and $M_{\rm fin}$. It gives a collection of points, which can be seen as a line.
%The crosses in Fig.~\ref{cnccrot} show the ratios in the total wind of 20, 32 and 60~$M_{\odot}$ source stars models \textcolor{red}{Table XXX}. There are other models than the models presented in this section but the physical ingredients are rather similar. We discuss these new models in details in the next section. The tracks show the variation of the ratios in the ejecta with $M_{\rm cut}$ let as a free parameter. A specific $M_{\rm cut}$ value will give one point in this diagram. Here we consider all possible mass cuts. In Eq.~\ref{yie}, $M_{\rm cut}$ is varied between 0 and $M_{\rm fin}$. 
%%%The yields (hence the C/N and $^{12}$C/$^{13}$C ratios) are calculated according to Eq.~\ref{yie}
%%%The lines represent the ratios in the ejecta when varying $M_{\rm cut}$ between $M_{\rm fin}$ and 0, according to Eq.~\ref{yie}.

We see that whatever the source star ejecta considered, none can match the bulk of observed stars (the tracks are nevertheless consistent with some lower limit $^{12}$C/$^{13}$C ratios). All models show a similar behavior. CNO processed material gives $\log$($^{12}$C/$^{13}$C) $\sim 0.6$ and [C/N] $\sim 2$ and He-processed material $\log$($^{12}$C/$^{13}$C) $\gg 1$ and [C/N] $\gg 1$. A mix of CNO and He-processed material gives something in between, but not in the region where most of the CEMP-no stars lie.

  \begin{figure*}[t]
   \centering
      \includegraphics[scale=0.39, trim = 0cm 0cm 0cm 0cm]{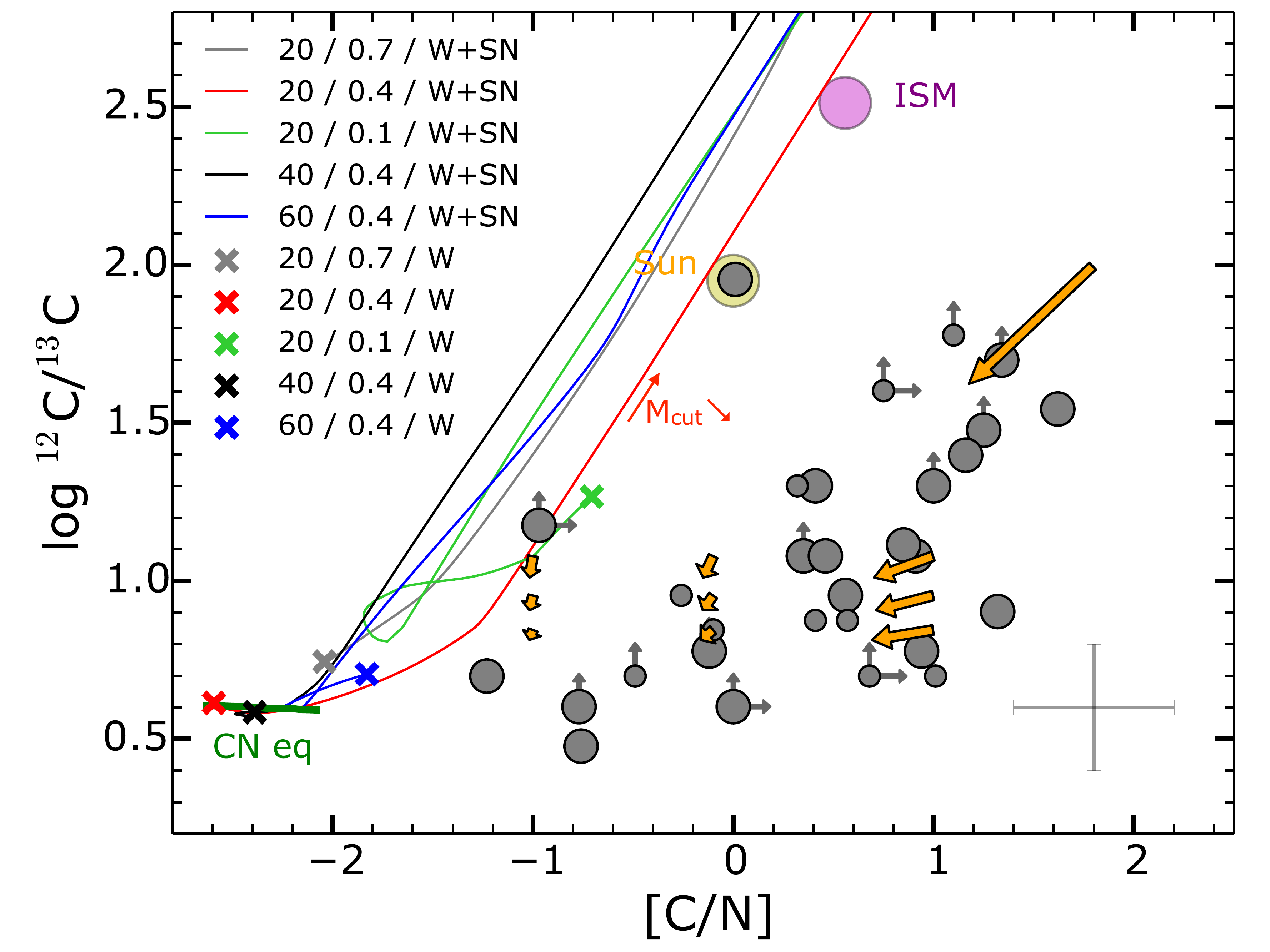}
   \caption[C/N vs. $^{12}$C/$^{13}$C: CEMP-no stars and source star models]{Grey circles are observed main sequence (small circles) and giants (big circles) CEMP stars with [Fe/H] $<-2.5$ (a few stars were added compared to Table~\ref{tabcemp}). Recognized CEMP-r, -s and -r/s are excluded. The grey arrows indicate that only limits are deduced from spectroscopy. The typical uncertainty is shown. The orange arrows show the effect of the first dredge-up on the surface composition of low metallicity 0.8~$M_{\odot}$ models with various initial C/N and $^{12}$C/$^{13}$C ratios. The yellow and purple circles show the solar ratios and the ratio in the $\alpha$-enhanced ISM considered here, respectively. The crosses show the ratios in the wind (W) at the end of core carbon burning. The tracks represent the integrated ratios as more and more layers of the final structure are ejected and added to the wind (W+SN). %, effect of the mass cut) 
The red arrow shows in which direction the tracks are going when decreasing the mass cut (i.e. ejecting deeper layers). In the legend, the first and second numbers refer to initial masses and rotation rates, respectively. The thick green line labelled \textit{CN eq} represents the ratios obtained in a one-zone model at CN-equilibrium for 30 $<T<$ 80 MK.}
              \label{cnccrotfirst} 
    \end{figure*}

\subsubsection{Effect of the first dredge-up in CEMP-no stars \label{secdup}}

%\textcolor{red}{refer to thermohaline mixing in prev sect, or put it here? Charbonnel94, 95}

Some CEMP-no stars in Fig.~\ref{cnccrotfirst} are giants (big grey circles) and have likely experienced the first dredge-up. This process can change the location of the stars in the C/N $-$ $^{12}$C/$^{13}$C diagram.
Below is investigated by how much and in which direction the first dredge-up can move a CEMP-no star in this diagram. % and maybe alleviate the problem for giant CEMP stars.

First, I started from a reference 0.8~$M_{\odot}$ model having the composition of the star CS~22958-042 at the Zero-Age Main-Sequence (ZAMS). CS~22958-042 is a subgiant with [Fe/H] $=-2.99$, [C/Fe] $=2.15$, [N/Fe] $=2.25$ and $^{12}$C/$^{13}$C $= 7$ \citep{roederer14c}. Then, 9 other 0.8~$M_{\odot}$ models were computed with different initial [C/N] and $^{12}$C/$^{13}$C ratios.
Table~\ref{tabledup} gives the initial ratios of the 10 models. The standard model, having the composition of CS~22958-042, initially has $\log$($^{12}$C/$^{13}$C) $=0.85$ and [C/N]~$=$~$-0.1$.

\begin{table*}[h!]
%\scriptsize{
\caption[Initial and final surface $\log$($^{12}$C/$^{13}$C) and \text{[}C/N\text{]} ratios for 0.8~$M_{\odot}$ models]{Initial and final (after the first dredge-up) surface $\log$ ($^{12}$C/$^{13}$C) and [C/N] ratios for the ten 0.8~$M_{\odot}$ models computed. \label{tabledup}}
\begin{center}
\resizebox{16.5cm}{!} {
\begin{tabular}{l | cccccccccc} 
\hline % inserts double horizontal lines
\hline % inserts single horizontal line
 %& Model 1 & Model 2 & Model 3 & Model 4 & Model 5 & Model 6 & Model 7 & Model 8 & Model 9 & Model 10  \\ % table heading
% & [$M_\odot$]&   & [km/s]  & [Myr] & [Myr]  &[Myr] &[yr] &[day] &[day] &[day]  & [$M_\odot$] & [$M_\odot$]  \\
%\hline % inserts single horizontal line
%& \multicolumn{10}{c}{Initial ratio}\\
$\log$ ($^{12}$C/$^{13}$C)$_{\rm ini}$		&$0.85$	&$0.85$	&$0.85$	&$0.95$	&$0.95$	&$0.95$	&$1.08$	&$1.08$	&$1.08$	&    2			\\
$\log$ ($^{12}$C/$^{13}$C)$_{\rm 1DUP}$	&$0.81$	&$0.81$	&$0.81$	&$0.91$	&$0.91$	&$0.91$	&$1.01$	&$1.01$	&$1.01$	&   $1.17$		\\
 \hline
%& \multicolumn{10}{c}{Final surface ratio}\\
$[$C/N]$_{\rm ini}$						&	$-1$	&  $-0.1$	&	1	&	$-1$	&  $-0.1$	&	1	&	$-1$	&  $-0.1$	&	1	&   $1.8$		\\
$[$C/N]$_{\rm 1DUP}$					&$-1.02$	&$-0.16$	&$-0.69$	&$-1.02$	&$-0.15$	&$0.71$	&$-1.02$	&$-0.15$	&$0.70$	&   $1.62$		\\
\hline
\end{tabular}
}
\end{center}
%}
\end{table*}

The evolution is stopped after the first dredge-up. The thermohaline mixing is not considered.
The surface $\log$($^{12}$C/$^{13}$C) and [C/N] ratios after the first dredge-up are reported in Table~\ref{tabledup}. The effect of the first dredge-up on the surface [C/N] and $\log$($^{12}$C/$^{13}$C) ratios of these models is also shown in Fig.~\ref{cnccrotfirst} by the orange arrows. 
In all the cases, the surface abundance ratios are decreased. The decrease ranges from 0 to 0.63 dex. % in the most extreme case.
The lower the initial $^{12}$C/$^{13}$C and C/N ratios, the smaller the effect of the dredge-up. This is because if starting with $^{12}$C/$^{13}$C and C/N ratios already close to CN equilibrium in the stellar envelope, then, adding some material at CN equilibrium barely affects the ratios. On the opposite, if the initial $^{12}$C/$^{13}$C and C/N ratios are far from equilibrium, adding a bit of CN-processed material in the envelope has a stronger impact on the ratios. 
%\textbf{This suggests that only the evolved CEMP stars with high $^{12}$C/$^{13}$C and/or high C/N may have experienced a significant modification of their surface abundances by the 1$^{\rm st}$ dredge-up. It is shown by the increasing size of orange arrows in Fig.~\ref{cnccrotfirst} at higher  $^{12}$C/$^{13}$C and C/N.}
These results suggest a rather modest effect of the first dredge-up, except for the CEMP stars that were formed from a material in which the C/N and $^{12}$C/$^{13}$C ratios were very far from CN equilibrium (see the biggest orange arrow in Fig.~\ref{cnccrotfirst}). 
%Even a CEMP star that initially has a surC/N 

%REPRENDRE ICI

\subsubsection{Other mixing processes in CEMP-no stars}

In Fig.~\ref{cnccrotfirst}, some observed evolved CEMP-no stars may also have experienced other mixing processes like thermohaline (cf. Sect.~\ref{selfenr}). As discussed, however, a significant fraction of CEMP-no stars may not have experienced too much mixing, except the most evolved stars (cf. Fig.~\ref{thermohaline}). 
Also, it has to be noted that unevolved CEMP-no stars lie globally at similar positions than evolved CEMP-no stars in Fig.~\ref{cnccrotfirst}. If the internal mixing processes had a strong impact on the surface C/N and $^{12}$C/$^{13}$C ratios, we should probably see distinct groups of CEMP-no stars: the bulk of evolved CEMP-no stars on the one hand, the bulk of unevolved CEMP-no stars on the other hand.
%the  suggests that only the evolved CEMP-no stars with high $^{12}$C/$^{13}$C and/or high C/N may have experienced a significant modification of their surface abundances by the 1$^{\rm st}$ dredge-up. It is shown by the increasing size of orange arrows in Fig.~\ref{cnccrotfirst} at higher  $^{12}$C/$^{13}$C and C/N.
The fact that it is not the case may finally indicate that the composition of the evolved CEMP-no stars shown in Fig.~\ref{cnccrotfirst} reflects quite well the composition of the cloud in which they formed. 

\section{The late mixing process in the source star\label{seclate}}

A solution to the C/N $-$ $^{12}$C/$^{13}$C puzzle is to introduce a \textit{late mixing process} in the source star, occurring between the H- and He-burning shell, about 200 yr before the end of the evolution. %This mixing process is the main object of \cite{choplin17a}. 
%Fig.~\ref{cnccrotmix} shows the same models as Fig.~\ref{cnccrot} but with the late mixing process. The 20~$M_{\odot}$ model can now provide a material able to reproduce the bulk of CEMP stars. 
Below I give more details on this process and discuss how it can naturally produce a material with a high C/N ratio and a low $^{12}$C/$^{13}$C ratio, able to improve the fit between models and observations. In Sect.~\ref{secother}, I investigate whether other models available in the literature can provide a solution.
%Other models possibly providing a solution are discussed in a second time.
%\textcolor{red}{maybe first explain rot models in CNCC plot and then mixing with late mixing and then explanation of the process in details (in this sec).}
This mixing process was investigated in \citet[][\hyperlink{pbox}{page \pageref*{pcempno}} of this thesis]{choplin17a}. 
%The process is schematically shown in Fig.~\ref{latemix}. 
The main points of the paper are summarized in this section.%, especially by discussing other possible models, able to account for .

  \begin{figure*}[h!]
   \centering
      \includegraphics[scale=0.68, trim = 0cm 0cm 0cm 0cm]{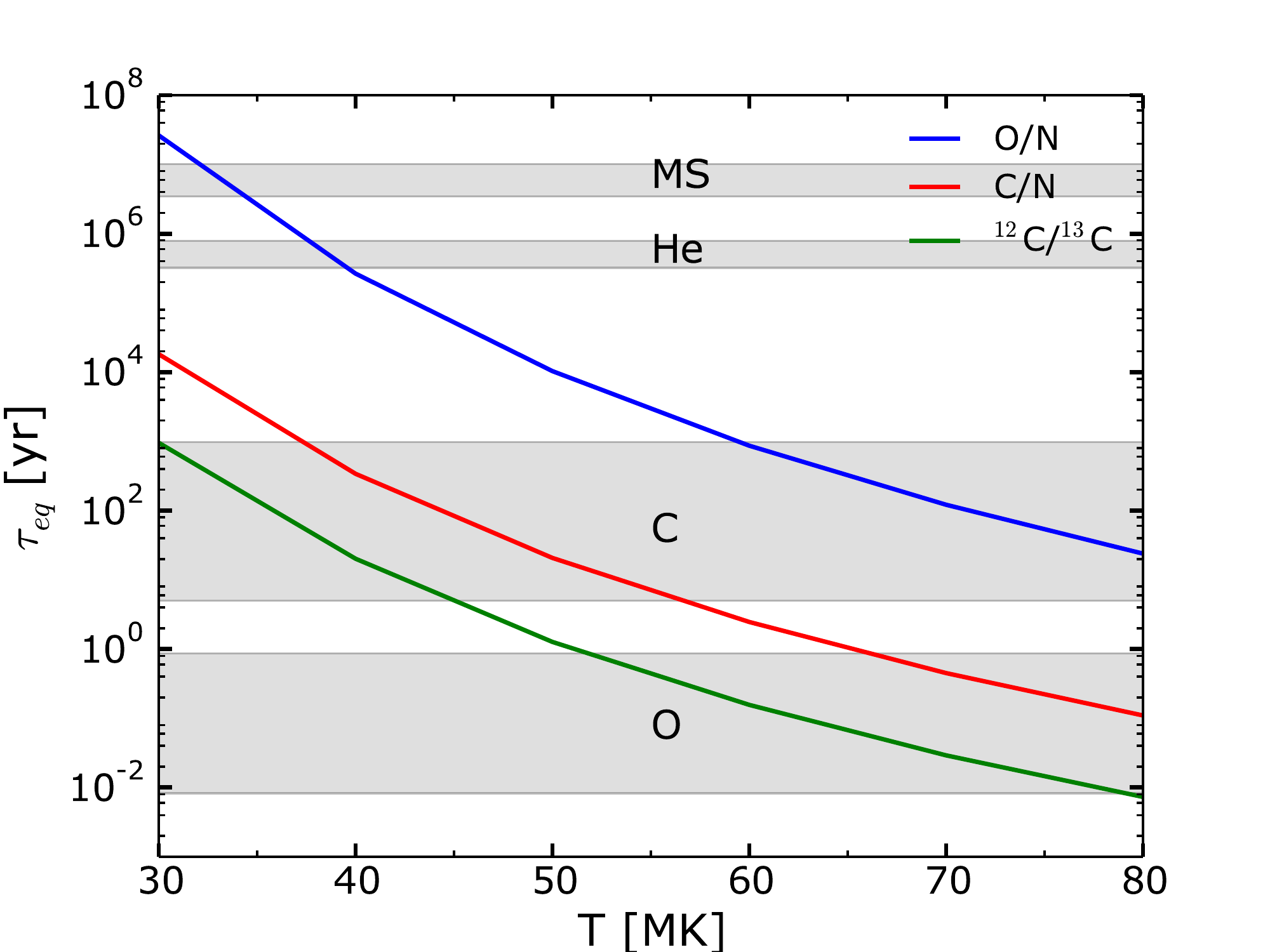}
   \caption[CNO equilibrium timescales as a function of the temperature]{CNO equilibrium timescales of the O/N, C/N, and $^{12}$C/$^{13}$C ratios as a function of the temperature. A one-zone model at density $\rho =$ 1 g cm$^{-3}$ is used. The shaded areas show the ranges of duration for the various burning stages (main sequence, He-, C-, and O-burning) of the models presented in Table \ref{modtab}.}
              \label{eqtimes} 
    \end{figure*}

   \begin{figure*}[t]
   \centering
      \includegraphics[scale=0.32, trim = 0cm 0cm 0cm 0cm]{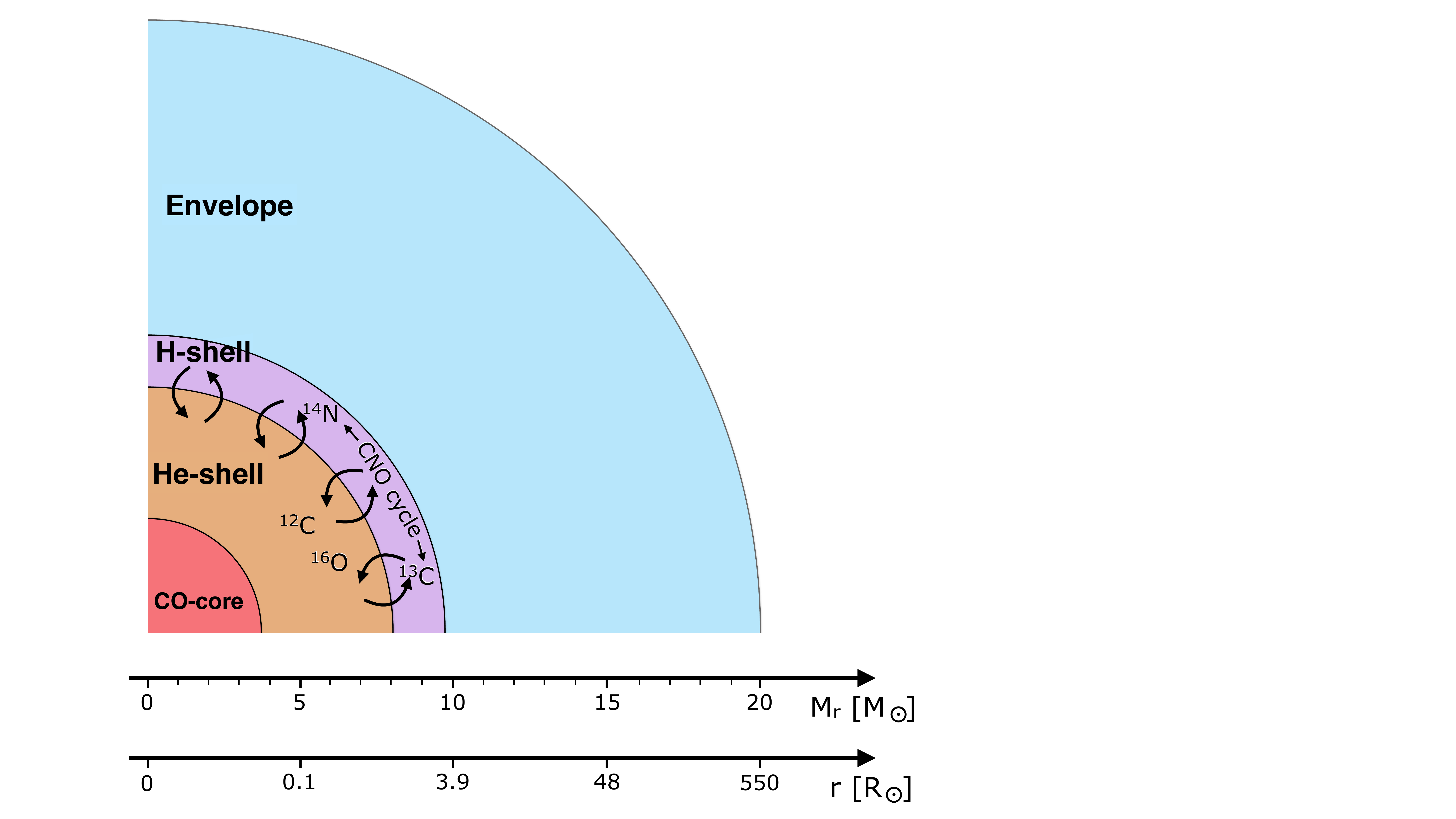}
   \caption[Schematic view of the late mixing process at work in the source star]{Schematic view of the late mixing process at work in the source star. The mixing starts during the core carbon burning phase. The curved black arrows show the region where it operates. The scheme is at scale in mass. Corresponding radii are indicated.}
\label{latemix}
    \end{figure*}

\subsection{General idea}

Figure~\ref{eqtimes} shows the time it takes for the C/N, O/N and $^{12}$C/$^{13}$C ratios to reach their CNO equilibrium value in a H-burning zone at $30<T<80$ MK. 
%This are typical temperatures in the H-burning shell of a low metallicity massive star. 
Results are obtained using the one-zone model described in Sect.~\ref{secbox}.
Equilibrium values are reached quicker when the temperature increases. 
%From 30 to 80 MK, the timescales decrease by $\sim 4-5$ orders of magnitude.
Whatever the temperature, $^{12}$C/$^{13}$C reaches equilibrium $\sim 10$ times faster than C/N, and C/N reaches equilibrium $100-1000$ times faster than O/N. 
In the H-burning shell of a complete stellar model, there is a gradient of temperature and the temperature $30 \lesssim T \lesssim 80$ MK. The global equilibrium timescale of the shell cannot be deduced directly from Fig.~\ref{eqtimes}. However, the relative difference between the different timescale stays the same.

Let us now assume that some $^{12}$C is injected in a single burning zone at $T=30$ MK which is at CNO equilibrium. The extra $^{12}$C disturbs the CNO equilibrium. Fig.~\ref{eqtimes} tells us that after 2000 yr, $^{12}$C/$^{13}$C will have reached back its equilibrium value while C/N and O/N will not. It means that if the H-burning shell of the massive star is burning at 30 MK and some $^{12}$C is injected 2000 yr before the end of the evolution, then, at the pre-SN stage, the H-shell will have C/N above equilibrium and $^{12}$C/$^{13}$C at equilibrium. The average temperature in the H-burning shell of a massive source star is about $40$ MK. Then, to obtain a partially processed CN material, with C/N above equilibrium and $^{12}$C/$^{13}$C at equilibrium, some $^{12}$C has to be injected in the H-burning shell $\sim 100 - 200$ yr before the end of the evolution. At this time, the source star burns carbon in its core. For $20-60$~$M_{\odot}$ models, the core carbon burning stage lasts for $10-1000$ yr (grey area in Fig.~\ref{eqtimes}).

\subsection{Implementation in source star models}\label{implem}

\begin{table*}
\scriptsize{
\caption[Properties of the source star models]{Properties of the source star models: model label (column 1) initial mass (column 2), $\upsilon_{\rm ini}/\upsilon_{\rm crit}$ (column 3), initial equatorial velocity (column 4), total lifetime (column 5), duration of the main sequence, helium, carbon, neon, oxygen, and silicon-burning phases (column $6 - 11$), mass of the model at the end of the evolution (column $12$), remnant mass according to the relation of \citet[][column $13$]{maeder92}. \label{modtab}}
\begin{center}
\resizebox{16.5cm}{!} {
\begin{tabular}{lcccccccccccc} 
\hline % inserts double horizontal lines
\hline % inserts single horizontal line
Model & $M_\text{ini}$ &  $\upsilon_\text{ini}/\upsilon_\text{crit}$  &  $\upsilon_\text{ini}$  & $\tau_\text{life}$ &  $\tau_\text{MS}$& $\tau_\text{He}$  & $\tau_\text{C}$ & $\tau_\text{Ne}$ & $\tau_\text{O}$ & $\tau_\text{Si}$ & $M_\text{final}$ & $M_\text{rem}$  \\ % table heading
 & [$M_\odot$]&   & [km/s]  & [Myr] & [Myr]  &[Myr] &[yr] &[day] &[day] &[day]  & [$M_\odot$] & [$M_\odot$]  \\
\hline % inserts single horizontal line
\multicolumn{13}{c}{No rotation}\\
\hline  
20s0    &  20           &               0                              &  0 &        8.93    &       8.02    &       0.79    &       978                     & 168 & 318 & 3.1           &       19.98       &    1.88         \\
32s0    & 32            &               0                              &  0 &        5.78    &       5.24    &       0.48    &       124                     & 22 & 47 & 0.6               &       31.94       &     2.98       \\
60s0    & 60            &               0                              &  0 &        3.81    &       3.44    &       0.33    &       15                      & 7 & 7 & 0.5           &       59.80            &   6.35     \\
 \hline
 \multicolumn{13}{c}{Rotation}\\
\hline
20s7    & 20            &               0.7                         &  610 &      11.0    &       10.1    &       0.76    &       400                     & 277 & 128 & 1.4    &       19.50       &    2.20        \\
32s7    & 32            &               0.7                         &  680 &      7.14    &       6.61    &       0.47    &       45                      & 7 & 15 & 0.6             &       30.71       &  3.69  \\
60s7    & 60            &               0.7                         &  770 &      4.69    &       4.33    &       0.32    &       5                       & 1 & 3 & 0.2 & 47.65        &    8.88     \\
\hline
\multicolumn{13}{c}{No rotation, late mix}\\
\hline  
20s0mix    &  20           &               0                            &  0 &        8.93    &       8.02    &       0.79    &       993                     & - & - & -           &       19.98        &   1.88         \\
32s0mix    & 32            &               0                            &  0 &        5.78    &       5.24    &       0.48    &       157                     & - & - & -               &       31.94    &   2.98            \\
60s0mix    & 60            &               0                            &  0 &        3.81    &       3.44    &       0.33    &       18                      & - & - & -           &       59.80         &    6.35       \\
\hline
 \multicolumn{13}{c}{Rotation, late mix}\\
\hline
20s7mix    & 20            &               0.7                        &  610 &      11.0    &       10.1    &       0.76    &       412                     & - & - & -    &       19.50           &   2.20     \\
32s7mix    & 32            &               0.7                        &  680 &      7.14    &       6.61    &       0.47    &       51                      & - & - & -            &       30.70    &   3.69    \\
60s7mix    & 60            &               0.7                        &  770 &      4.69    &       4.33    &       0.32    &       5                       & - & - & - & 47.64        &     8.73    \\
\hline
\end{tabular}
}
\end{center}
}
\end{table*}

This mixing process was investigated for a grid of six models whose characteristics are given in Table~\ref{modtab}. Small differences exist between the input parameters of these models and the models of the previous section. First, when $\log (T_{\rm eff}) \geq 3.95$, the mass-loss rates are from \cite{kudritzki00}, instead of \cite{vink01}. Radiative winds are generally small at low metallicity so that no big impact is expected. Second, following the study on Al discussed in Sect.~\ref{secbox}, the nuclear rates from the literature that minimize the production of Al are selected: \cite{angulo99} for $^{26}$Mg($p,\gamma$)$^{27}$Al, \cite{cyburt10} for $^{27}$Al($p,\gamma$)$^{28}$Si and $^{27}$Al($p,\alpha$)$^{24}$Mg. Third, $f_{\rm energ}$ in the expression of $D_{\rm shear}$ (Eq.~\ref{dshtz97}) was taken equal to 1 instead of 4.

First, the models were computed normally (without late mixing) until the end of the central silicon-burning phase. The computation is stepped when the mass fraction of $^{28}$Si in the core is less than $10^{-8}$. For rotating models, the effects of rotation were taken into account until the end of the carbon burning phase. Last stages were computed without rotation. It saves a lot of computational time and leads to only very small differences in the abundance profiles since the duration of the last stages is short ($\sim 1 - 300$ days, cf. Table~\ref{modtab}) compared to the rotational mixing timescale.
In a second step, I have computed again the end of the evolution for the six models while triggering the late mixing process $\sim$ 200 yr before the end of the evolution (Fig.~\ref{latemix} for a schematic view). For these models, the evolution was stopped at the end of core carbon burning. Last stages are very short and change only the composition of the most inner layers. It gives four categories of models: (1) no rotation, no late mixing, (2) no rotation, late mixing, (3) fast rotation, no late mixing and (4) fast rotation, late mixing. %Models with intermediate rotation are not considered. Fast rotating models can be taken as an upper limit.

%The end of this phase is defined such that X($^{12}$C)$_c$, the central mass fraction of $^{12}$C is equal to 10$^{-5}$.
%The duration of core C burning can be shorter than 200 yr. In such cases, the late mixing begins before the C starts to burn but still after the core He-burning phase. 
%Late mixing is only operating in radiative zones and is triggered around the bottom of the H-burning shell. 
To model the late mixing process in rotating models, the shear diffusion coefficient $D_{\rm shear}$ is multiplied by a factor of 100 in between the H- and He-shell. %$D_{\rm shear}$ exist only in radiative zones so that the late mixing process occurs only in radiative zones.
%Non-rotating models are also studied to see whether or not the effect of the late mixing alone might be sufficient to explain the abundances observed at the surface of CEMP-no stars. 
In non-rotating models, an artificial and constant diffusion coefficient $D_{\rm no~rot} = 10^{9}\,\text{cm\,s}^{-1}$ is set in the mixing zone. This is a typical value of the diffusion coefficient found in rotating models including late mixing. 
%These new models\footnotemark 
%\footnotetext{These new models do not appear in Table~\ref{table:1}. They have the same properties as the models without late mixing, shown in Table~\ref{table:1}. Only the duration of the C-burning phase changes slightly.}
%are identified with `mix' (20s0mix, 32s0mix...). 
Although modeled through the shear diffusion coefficient, it is not assumed that the physical origin of the late mixing process is linked to the shear. Its possible physical origin is discussed in Sect.~\ref{secorig}.
%Its possible physical origin is discussed in Sect. \ref{late}. Also, a parametric study of this late mixing is done in Sect. \ref{sec:5}.

   \begin{figure*}[t]
   \centering
   \begin{minipage}[c]{.49\linewidth}
       \includegraphics[scale=0.23]{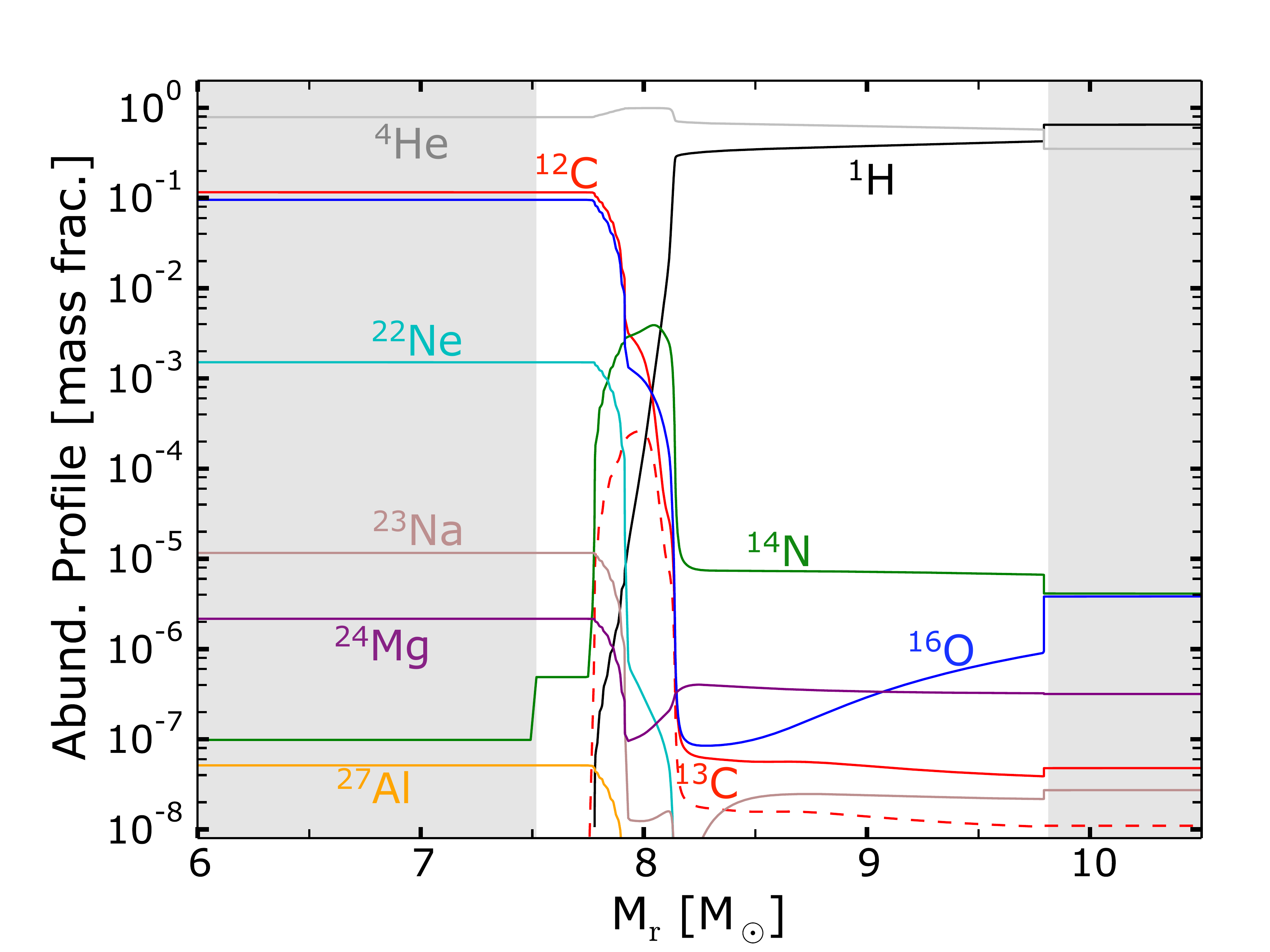}
   \end{minipage} \hfill
   \begin{minipage}[c]{.49\linewidth}
      \includegraphics[scale=0.23]{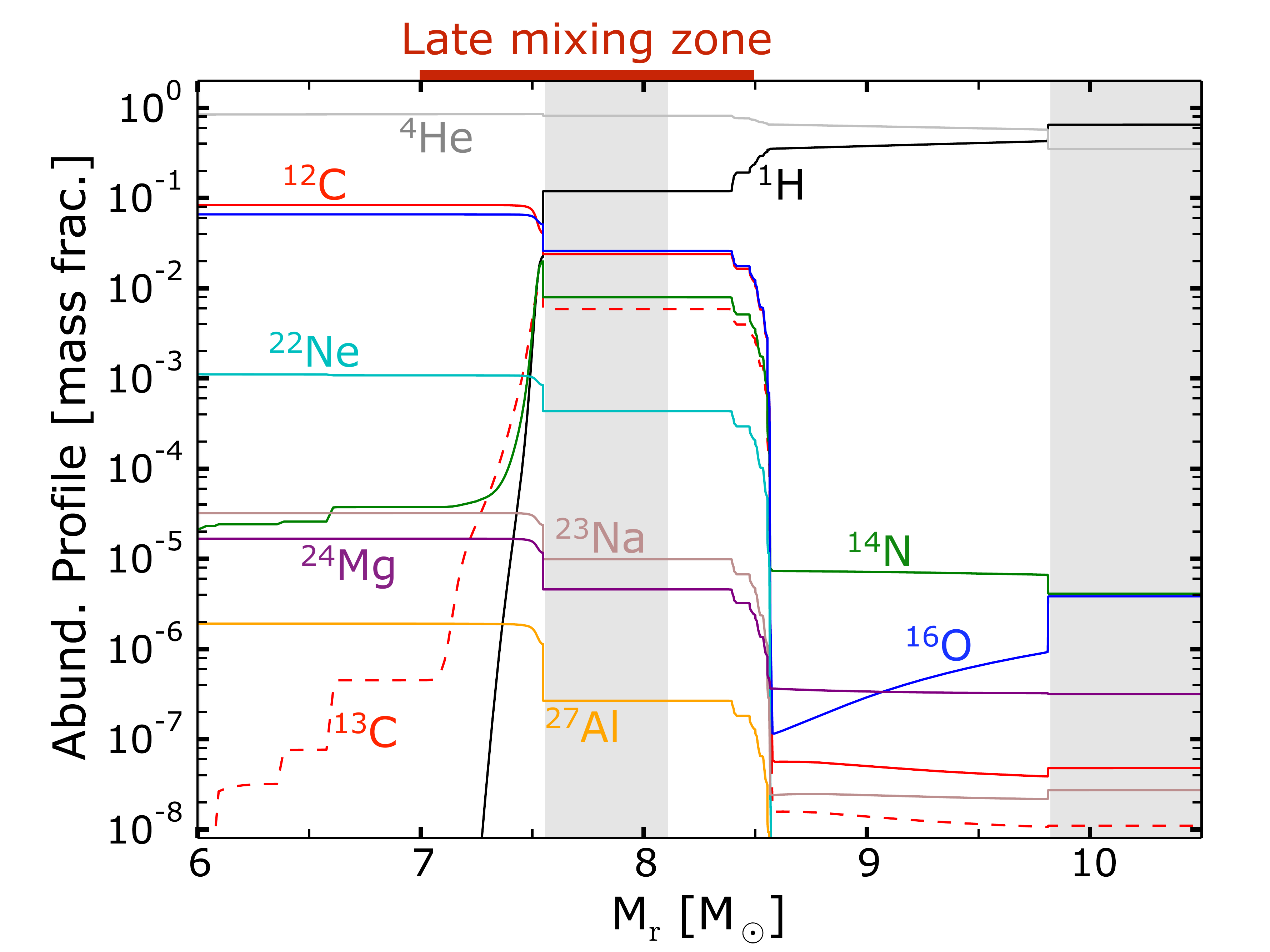}
   \end{minipage}
   \caption[Abundances of the 20~$M_{\odot}$ with $\upsilon_{\rm ini}/\upsilon_{\rm crit} = 0.7$ with/without late mixing]{Abundance profiles of the 20~$M_{\odot}$ models with $\upsilon_{\rm ini}/\upsilon_{\rm crit} = 0.7$ at the end of the core carbon-burning phase with no late mixing (left) and with late mixing (right). Shaded areas show the convective zones. The zone where the late mixing process occurs is indicated on the top.
}
\label{abundprofmix}
    \end{figure*}

  \begin{figure*}[h!]
   \centering
      \includegraphics[scale=0.45, trim = 0cm 0cm 0cm 0cm]{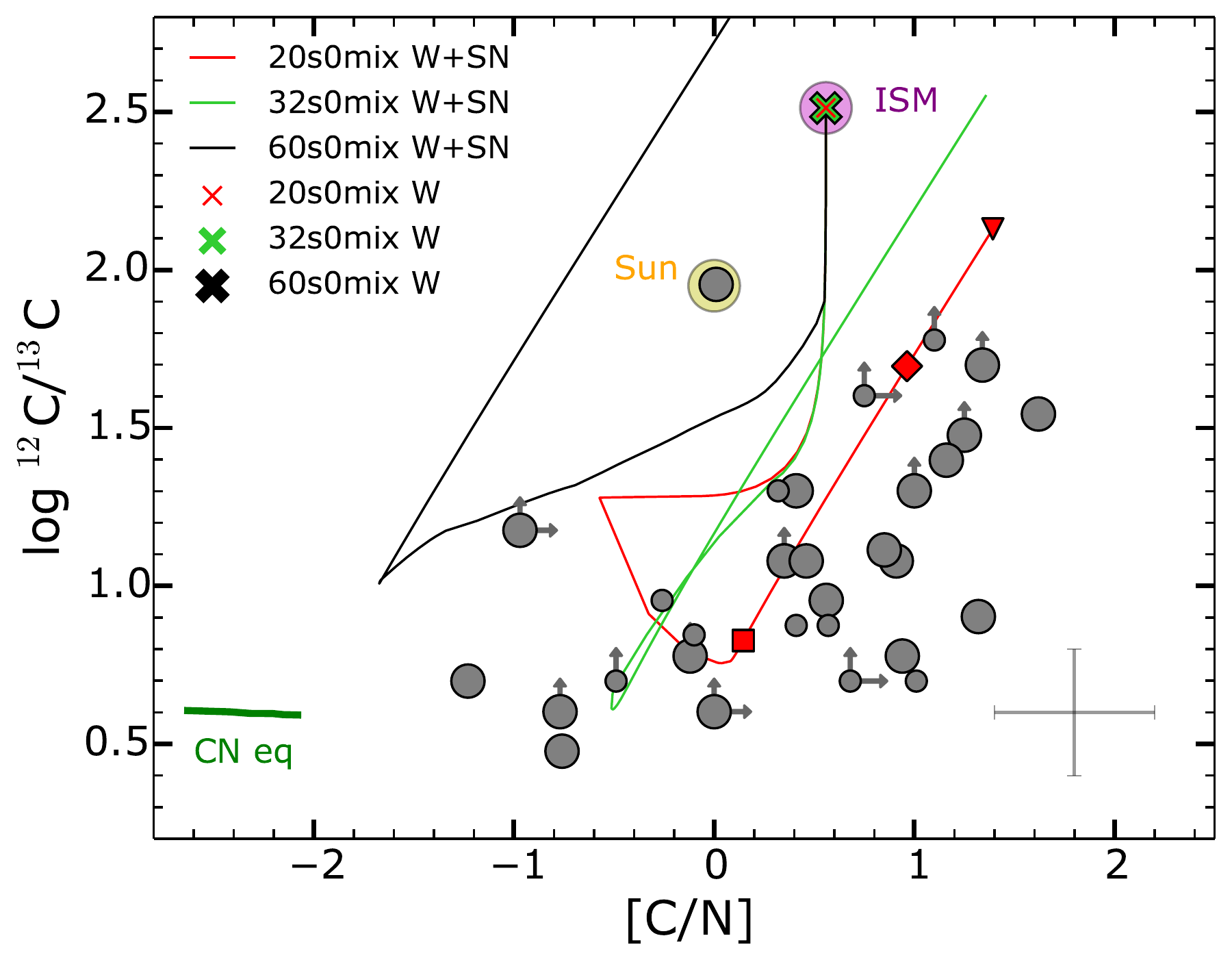}
      \includegraphics[scale=0.45, trim = 0cm 0cm 0cm 0cm]{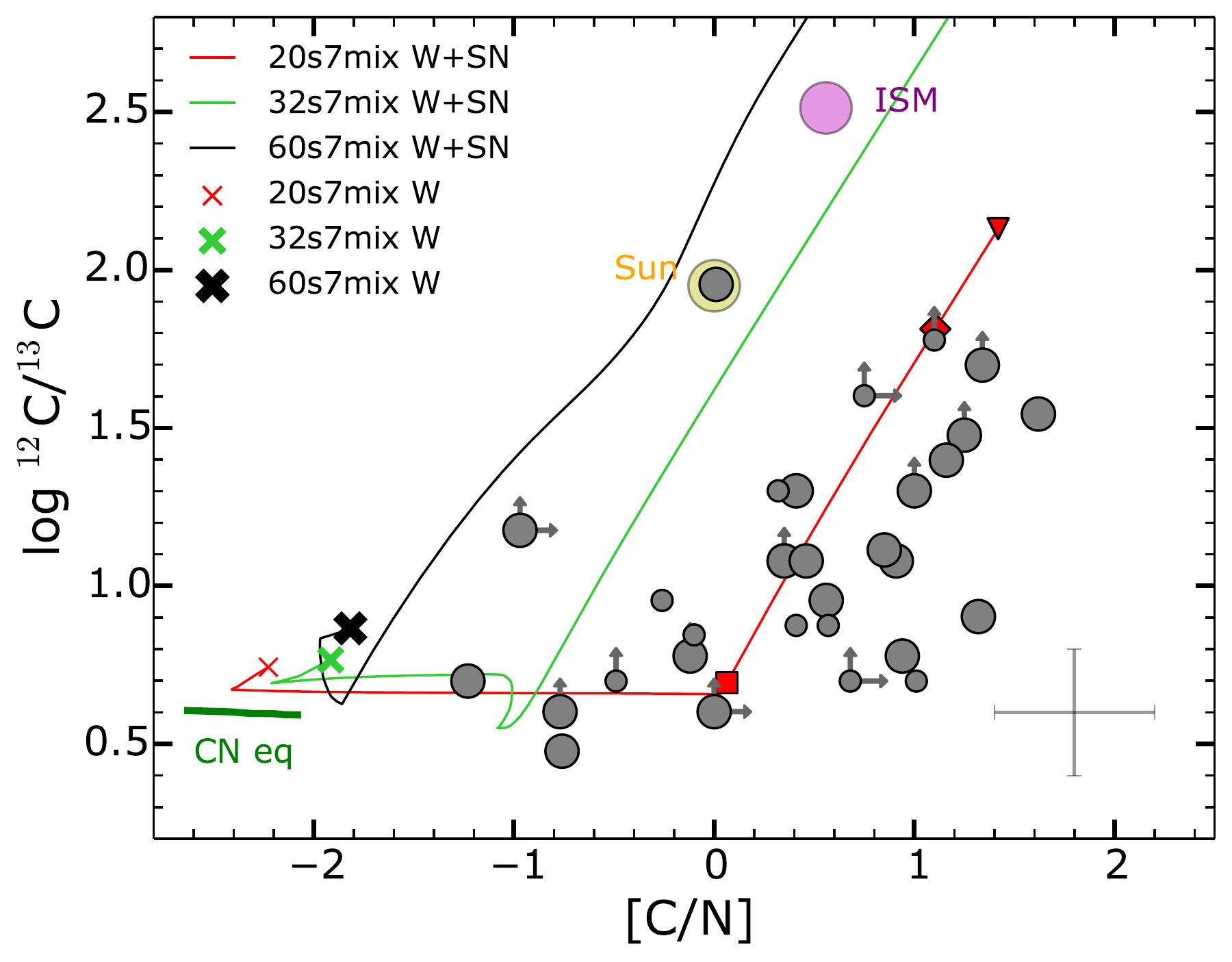}
  % \caption[a]{Same as Fig.~\ref{cnccrotfirst} but for the 20, 32 and 60~$M_{\odot}$ models with $\upsilon_{\rm ini}/\upsilon_{\rm crit} = 0.7$ and including the late mixing process. The red square, diamond and triangle show the composition of the ejecta of the 20~$M_{\odot}$ when the mass cut is equal to $M_{\rm \alpha}$, $M_{\rm CO}$, and $M_{\rm rem}$ respectively. \textcolor{red}{ALSO SHOW NO ROT TO SHOW THAT IT WORKS}}
   \caption[C/N vs. $^{12}$C/$^{13}$C: CEMP-no stars and source star models including late mixing]{Same as Fig.~\ref{cnccrotfirst} but for the 20, 32 and 60~$M_{\odot}$ models of Table~\ref{modtab} including the late mixing process and with $\upsilon_{\rm ini}/\upsilon_{\rm crit} = 0$ (left) and 0.7 (right). The red square, diamond and triangle show the composition of the ejecta of the 20~$M_{\odot}$ model when the mass cut is equal to $M_{\rm \alpha}$, $M_{\rm CO}$, and $M_{\rm rem}$, respectively.}
              \label{cnccrotmix} 
    \end{figure*}

\subsection{Comparison with CEMP-no stars}

When the late mixing process is included, additional $^{12}$C and $^{16}$O enter into the H-shell, boosting the CNO cycle and then releasing more energy. The H-shell becomes convective, extends in mass so that more He-burning products are engulfed. The fresh $^{12}$C starts to be transformed into $^{13}$C and $^{14}$N in the H-shell. However, the time remaining before the end of the evolution being short, the [C/N] equilibrium value of $\sim -2.3$ is not reached. The right panel of Fig.~\ref{abundprofmix} shows the abundance profile of the rotating 20~$M_{\odot}$ model with late mixing, at the end of core carbon burning. We see that the convective H-shell contains a lot of CNO elements, and has X(C)/X(N)~$>1$ while X($^{12}$C)/X($^{13}$C) is at equilibrium, around 4. %The right panel of Fig. \ref{kip} shows this intermediate zone where [C/N] is about 0 at the end of evolution. 
This process builds a zone which is, at the end of evolution, partially processed by the CN cycle in the source star, where C/N is high and $^{12}$C/$^{13}$C at equilibrium.

Fig.~\ref{cnccrotmix} shows the ejecta of the non-rotating (left panel) and rotating (right panel) 20, 32 and 60~$M_{\odot}$ models including the late mixing process. 
With increasing mass, the tracks in Fig.~\ref{cnccrotmix} are shifted to the left, away from observations. This is mainly due to the fact that higher-mass models have a higher temperature in the H-burning shell. This implies that the CN cycle operates faster. In this case, the injected $^{12}$C is transformed more rapidly into $^{14}$N. Then, the [C/N] ratio in the H-burning shell is closer to the equilibrium value ($\sim -2$) at the end of evolution. It finally implies that the ejecta of higher-mass models cannot reach high [C/N] ratios together with low $^{12}$C/$^{13}$C ratios.
%However, if the late mixing occurs sufficiently late (later than $\sim 200$ yr before the end of the evolution), the [C/N] ratio may not have time to reach back equilibrium, even in the hot H-burning shell of a $60\,M_\odot$ source star. %, where the CN cycle is faster.
There are several reasons that may make the late mixing process more likely to occur in $\sim 20$~$M_{\odot}$ source stars than in $\sim 60$~$M_{\odot}$ source stars: 
\begin{itemize}
\item In a $60$~$M_{\odot}$ model, the late mixing process should occur very late in the evolution, so as to end up with a high C/N (see above). If it occurs too early, C/N has enough time to reach back equilibrium. It means that the time window for the late mixing to operate is shorter in a 60~$M_{\odot}$ model than in a $20$~$M_{\odot}$ model.
\item In more massive stars, the mixing process should be extremely strong so as to compensate for the short time available.
\item In more massive stars, the distance between the H- and He-burning shells is greater, so that the connection between the two shells might be less likely.
\end{itemize}
Overall, $\sim 20\,M_\odot$ source stars might be better candidates for the late mixing process, hence for reproducing the observations.
%It shows that both the non-rotating and rotating 20~$M_{\odot}$ models best reproduce the bulk of observations. 
%Non-rotating models including late mixing (not shown on the Figure for clarity) provide a similar material. 

Fig.~\ref{cnccrotmix} shows that both the non-rotating and rotating 20~$M_{\odot}$ models can reproduce the bulk of observations. 
These two kind of models show however important differences regarding other elements.
%It is discussed later in this section (by considering other elements) how non-rotating models including late mixing differ from rotating models including late mixing.
In fact, the late mixing process changes the distribution of chemical species in the source star (Fig.~\ref{abundprofmix}) but in the models presented here, it implies further nucleosynthesis almost only for the elements\footnote{This process may also form heavy elements through neutron captures, see Sect.~\ref{secother}.} $^{12}$C, $^{13}$C and $^{14}$N. This is because the burning timescales for the other species considered in the network (e.g. O, Ne, Na, Mg, Al) are longer compared to the remaining time before the end of the evolution. For instance, there is little time for the Ne-Na cycle to operate and produce additional $^{23}$Na in models including late mixing. Instead, a progressive mixing, achieved by rotation during the core helium burning stage can form extra $^{23}$Na (Sect.~\ref{secinter} and Fig.~\ref{abrotmod2}). This is an important difference between non-rotating (little Na) and rotating models (high Na) including late mixing.
Many CEMP-no stars are enriched in Na. It may suggest that, at least for a part of the CEMP-no star sample, two kind of mixing are needed: rotational mixing (for Na) and the late mixing process (for a high C/N with a low $^{12}$C/$^{13}$C).

\subsection{Physical origin \label{secorig}}

   \begin{figure*}[t]
   \centering
      \includegraphics[scale=0.6, trim = 0cm 0cm 1cm 0cm]{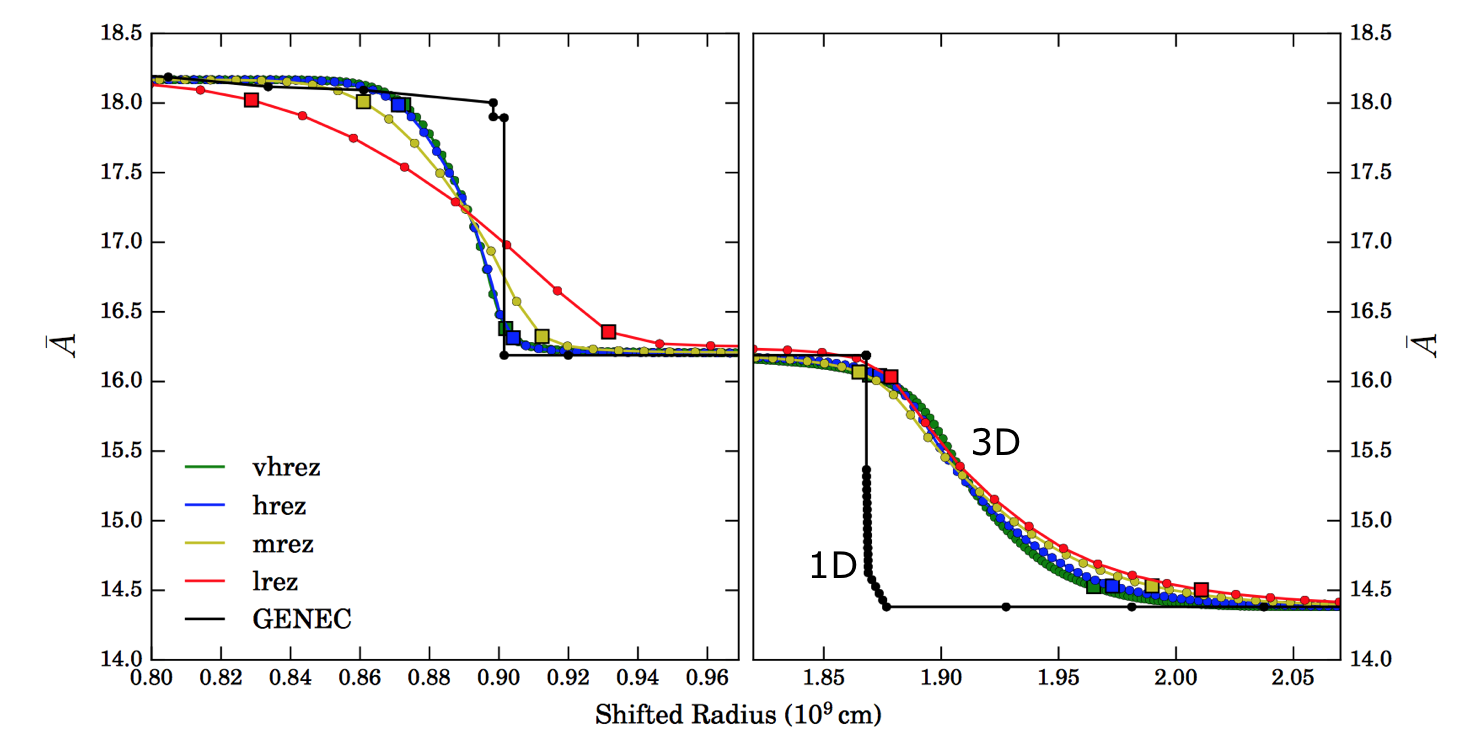}
   \caption[Convective boundary regions of the C-burning shell of a 15~$M_{\odot}$ model]{Radial profile of the averaged atomic weight $\bar{A}$ 
   %Radial compositional profiles 
   at the lower (left) and upper (right) convective boundary regions of the C-burning shell of a 15~$M_{\odot}$ model. The radius of each profile is shifted such that the boundary position coincides with the boundary position of the \texttt{vhrez} model.
The symbols denote the mesh points. The black profile is from \textsc{genec} while the other profiles are from 3D hydrodynamic simulations of various resolutions. The range of radius shown corresponds approximately to the range of mass coordinates $1.2 - 1.8$~$M_{\odot}$ \citep[figure from][]{cristini17}.}
\label{convbound}
    \end{figure*}

In the present work, the late mixing process is triggered artificially. 
It has to be noted that in some cases, similar events occur naturally. 
Indeed, various authors reported sudden ingestion events of H-burning material into the He-burning core or shell in low or zero metallicity massive stellar models \citep[with or without rotation,][see also Sect.~\ref{secother}]{hirschi07, ekstrom08, heger10, limongi12}. 
Although observed in some stellar models, the occurrence conditions and physical process(es) responsible for such events remain unknown. No specific behaviour is observed with any stellar parameter.

%The conclusions would be strengthened if a known physical process were able to explain this mixing. 

%Our main aim here was to see to what extent the CEMP-no sample could be reproduced by standard source- star models and if it could not, to try to find which ingredients are missing. 
The treatment of the convection in stellar evolution codes may impact the shell/shell interaction or the core/shell interaction and thus the occurrence of the late mixing event. 
%(also called H-ingestion event). 
In \textsc{genec}, the convective boundaries are determined using the Schwarzschild criterion. 
%, and the convection is assumed to be adiabatic11. 
The boundaries of the convective zones are sharp (step functions). During the main sequence and core He-burning phase, the convective core is extended using a penetrative overshoot. 
%the length of which is proportional to a fraction of the pressure scale height at the edge of the core. 
The overshoot is applied neither for the more advanced phases of stellar evolution, nor for the intermediate convective shells. %, and is thus not relevant to the above discussion.
This prescription probably does not capture the whole physics of convection \citep{arnett15}.
Multi-dimension hydrodynamics numerical simulations of convection in deep stellar interior show that the chemical composition of each side of the convective boundary makes a smooth transition and is not a step function. 
%The boundaries are not strict barriers for the chemical elements, and part of them can be mixed through the boundary 
\citep{herwig06, meakin07b, arnett11, cristini17}. Fig.~\ref{convbound} shows a comparison between 1D code (\textsc{genec}) and 3D simulations, for the carbon burning shell. We see that in 3D, the carbon shell extends further in mass compared to 1D models. 
Multi-D simulations of H/He burning-zones during earlier stages (e.g. core He-burning stage) do not exist since they are computationally too expansive, mainly because the burning timescales are long. % and the radiative effects are important.

%\cite{arnett16} noticed the following fact: during the core He-burning stage, He-burning produces C and O, which increase the opacity. Higher opacity will in turn increase convective mixing and core size, so that more C and O are produced and the opacity increases again. It may lead to a positive feedback and makes the interaction between the He-core and H-shell easier. This positive feedback might stand as well in more advanced stages for the H- and He-burning shell and consequently naturally trigger the late mixing process.

It might be that improving the way convection is treated in classical 1D codes to follow more closely the behavior observed in multi-dimensional simulations strengthens the exchanges between the H- and He-burning shells. This could naturally induce the creation of the late mixing zone. If so, the late mixing invoked in this work would result from an overly poor description of the convective boundaries in 1D stellar evolution models.
\section{A systematic study of the CEMP-no source star population \label{secstat}}

\begin{table}[h!]
\begin{small}
\caption[Parameters of the best source star models]{Parameters of the best source star models (column 2) for the considered CEMP-no stars (column 1). $f_{\rm H}$, $f_{\rm He}$ and $f_{\rm CO}$ correspond to the ejected fraction of the H-envelope (above $M_{\alpha}$), He-shell (between $M_{\alpha}$ and $M_{\rm CO}$) and CO-core (below $M_{\rm CO}$), respectively. $D$ is the dilution factor, $X$(He) is the mass fraction of helium predicted for the CEMP-no star, $\xi$ the error on the fit and $N$ the number of abundances used to make the fit.}
\label{bestable}
\begin{center}
\resizebox{16.5cm}{!} {
\begin{tabular}{llllllllll}
\hline
 Star	& Best model  & $M_{\rm cut}$ [$M_{\odot}$] & $f_{\rm H}$ & $f_{\rm He}$  & $f_{\rm CO}$ & $D$  & $X$(He) & $\xi$  & $N$ \\
% 	& 	 	      & [$M_{\odot}$] &		    &			     &			     & 		 & 	    & 	\\
\hline
 BD+44\_493                 & 60s7mix & 29.88 & 1    & 0.72 & 0    & 1000 & 0.25 & 0.46 & 4 \\
 CS22877-001               & 20s7mix &  5.07 & 1    & 0.97 & 0    & 1000 & 0.25 & 1.32 & 5 \\
 CS22891-200               & 20s7mix &  8.5  & 0.92 & 0    & 0    &   10 & 0.26 & 0.62 & 4 \\
 CS22897-008               & 60s0    & 22.7  & 1    & 0.47 & 0    & 1000 & 0.25 & 0.31 & 4 \\
 CS22949-037               & 20s7mix &  8.54 & 0.92 & 0    & 0    &    0 & 0.38 & 1.38 & 4 \\
 CS22950-046               & 20s7mix &  8.53 & 0.92 & 0    & 0    &   10 & 0.26 & 0.21 & 4 \\
 CS22957-027               & 20s7mix &  7.4  & 1    & 0.03 & 0    &   10 & 0.26 & 1.03 & 5 \\
 CS22960-053               & 20s7    &  7.97 & 1    & 0.02 & 0    &    0 & 0.39 & 1.11 & 3 \\
 CS29502-092               & 20s7mix &  7.32 & 1    & 0.07 & 0    &  100 & 0.25 & 1.19 & 6 \\
 CS29527-015               & 60s7mix & 28.04 & 1    & 1    & 0    & 1000 & 0.25 & 0.1  & 3 \\
 CS29528-041               & 32s7mix & 19.05 & 0.75 & 0    & 0    &    0 & 0.42 & 0.99 & 4 \\
 G77-61                    & 20s0mix & 12.78 & 0.5  & 0    & 0    &    0 & 0.27 & 1.28 & 6 \\
 HE0017-4346               & 20s7mix &  7.77 & 0.98 & 0    & 0    &    0 & 0.4  & 0.28 & 3 \\
 HE0134-1519               & 60s0mix & 24.21 & 1    & 0.1  & 0    &   10 & 0.27 & 0.71 & 3 \\
 HE0146-1548               & 20s7mix &  8.51 & 0.92 & 0    & 0    &    0 & 0.38 & 1.05 & 3 \\
 HE0251-3216               & 20s7mix &  7.09 & 1    & 0.16 & 0    &   10 & 0.26 & 0.34 & 3 \\
 HE0450-4902               & 20s7mix &  8.43 & 0.92 & 0    & 0    &    0 & 0.38 & 0.94 & 4 \\
 HE1005-1439               & 20s7mix &  6.86 & 1    & 0.25 & 0    &   10 & 0.27 & 0.66 & 4 \\
 HE1029-0546               & 20s0mix & 11.58 & 0.58 & 0    & 0    &    0 & 0.31 & 0.55 & 4 \\
 HE1150-0428               & 20s7mix &  8.31 & 0.93 & 0    & 0    &    0 & 0.38 & 0.45 & 4 \\
 HE1201-1512               & 60s0    & 24.29 & 1    & 0.13 & 0    &   10 & 0.27 & 0.63 & 3 \\
 HE1300+0157               & 60s7    & 28.22 & 1    & 1    & 0.01 & 1000 & 0.25 & 0.27 & 4 \\
 HE1310-0536               & 32s0mix & 14.99 & 0.79 & 0    & 0    &   10 & 0.27 & 0.89 & 5 \\
 HE1327-2326               & 60s7    & 33.51 & 1    & 0.21 & 0    &    0 & 0.62 & 1.08 & 5 \\
 HE1410-0004               & 20s7mix &  8.43 & 0.92 & 0    & 0    &    0 & 0.38 & 1.04 & 4 \\
 HE1456+0230               & 20s7    &  7.86 & 1    & 0.06 & 0    &    0 & 0.4  & 0.96 & 4 \\
 HE2318-1621               & 20s0mix & 13.63 & 0.44 & 0    & 0    &    0 & 0.25 & 0.73 & 4 \\
 HE2331-7155               & 20s7    &  7.99 & 1    & 0.02 & 0    &    0 & 0.39 & 1.14 & 5 \\
 LAMOSTJ125346.09 & 60s0mix & 24.22 & 1    & 0.1  & 0    &    0 & 0.53 & 0.5  & 3 \\
 LAMOSTJ131331.18 & 60s0    & 24.33 & 1    & 0.12 & 0    &    0 & 0.53 & 0.42 & 3 \\
 LAMOSTJ1626+1721          & 20s0    &  5.8  & 1    & 0.16 & 0    &    0 & 0.38 & 0.7  & 3 \\
 SDSSJ0002+2928            & 20s7mix &  7.18 & 1    & 0.12 & 0    &   10 & 0.26 & 0.16 & 3 \\
 SDSSJ0126+0607            & 32s7mix & 10.73 & 1    & 1    & 0.03 &   10 & 0.27 & 0.18 & 3 \\
 SDSSJ0212+0137            & 20s7    &  7.89 & 1    & 0.05 & 0    &    0 & 0.4  & 0.21 & 3 \\
 SDSSJ0351+1026            & 60s7mix &  8.84 & 1    & 1    & 0.68 & 1000 & 0.25 & 0.51 & 3 \\
 SDSSJ0723+3637            & 20s0    &  5.78 & 1    & 0.16 & 0    &    0 & 0.38 & 0.23 & 3 \\
 SDSSJ1143+2020            & 20s7mix &  6.48 & 1    & 0.41 & 0    &   10 & 0.27 & 0.51 & 3 \\
 SDSSJ1245-0738            & 20s7    &  3.72 & 1    & 1    & 0.26 &   10 & 0.26 & 0.23 & 3 \\
 SDSSJ1349-0229            & 20s7mix &  5.78 & 1    & 0.69 & 0    &   10 & 0.27 & 0.48 & 3 \\
 SDSSJ1422+0031            & 20s7mix &  3.03 & 1    & 1    & 0.4  &  100 & 0.25 & 0.67 & 3 \\
 SDSSJ1613+5309            & 20s7mix &  3.3  & 1    & 1    & 0.34 &  100 & 0.25 & 0.68 & 3 \\
 SDSSJ1746+2455            & 20s7mix &  8.51 & 0.92 & 0    & 0    &    0 & 0.38 & 0.79 & 3 \\
 SMSSJ005953.98   & 20s7mix &  8.55 & 0.92 & 0    & 0    &    0 & 0.38 & 1.13 & 3 \\
\hline
\end{tabular}
}
\end{center}

\end{small}

\end{table}

  \begin{figure*}[t]
   \centering
      \includegraphics[scale=0.44, trim = 0cm 0cm 0cm 0cm]{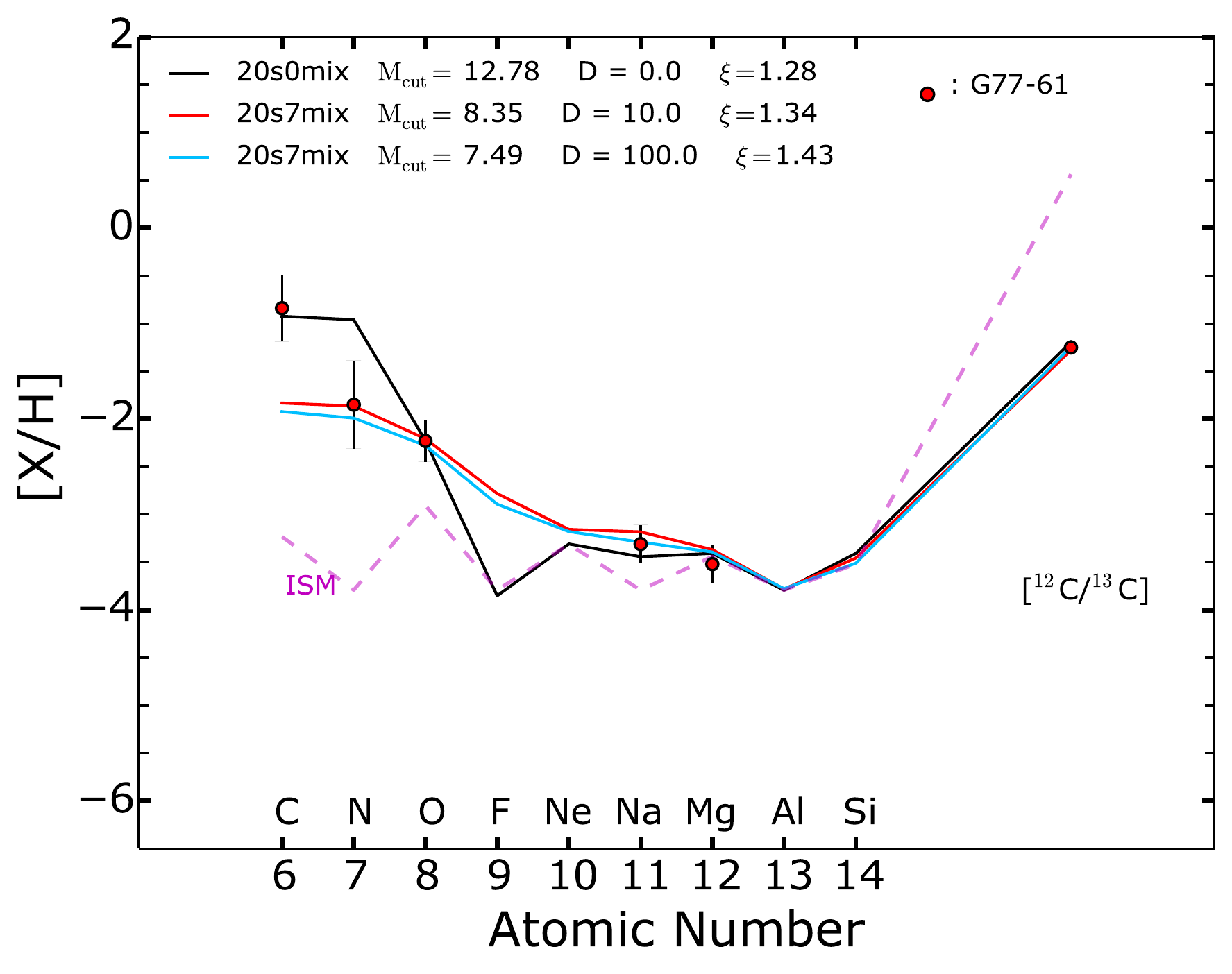}
      \includegraphics[scale=0.44, trim = 0cm 0cm 0cm 0cm]{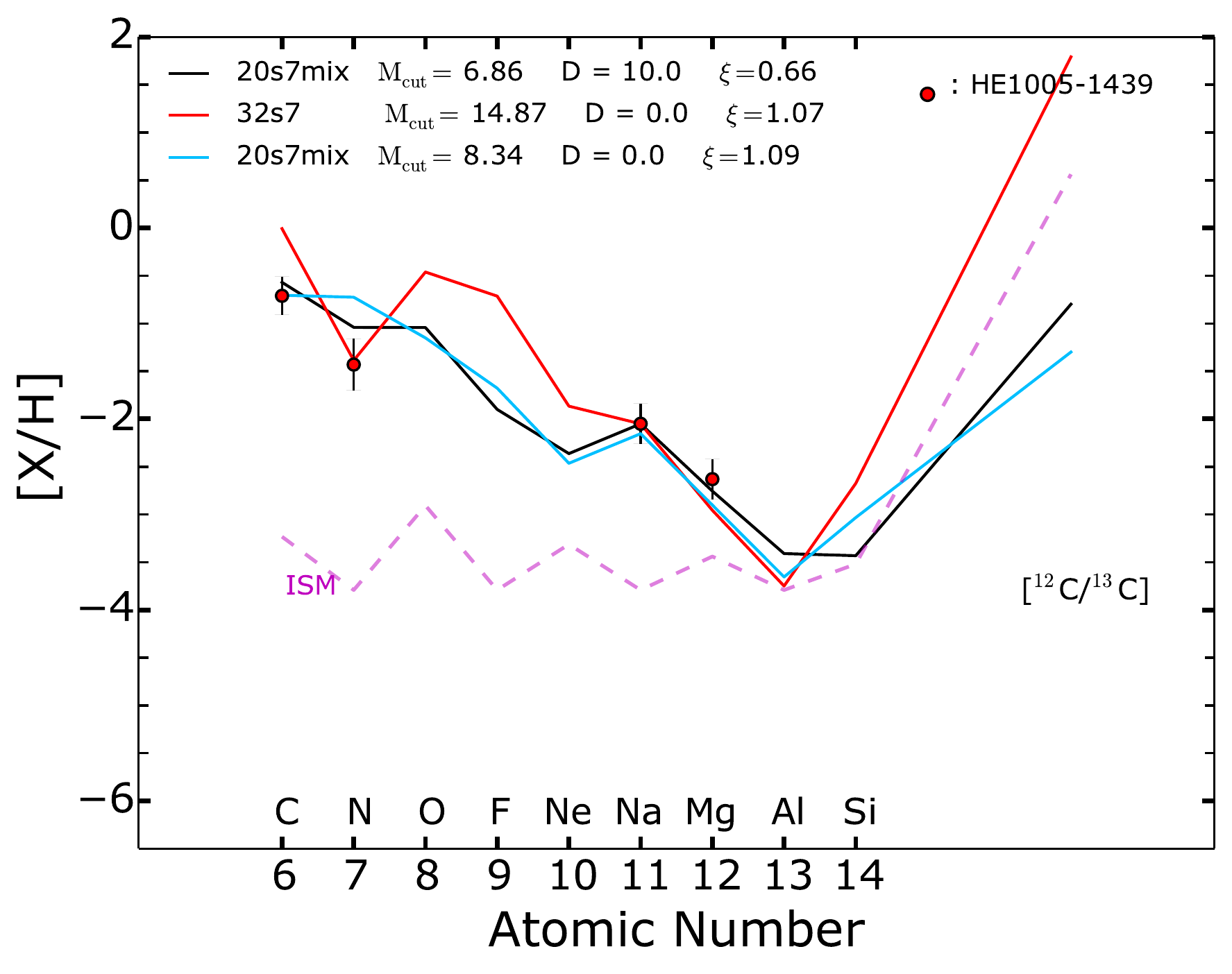}
   %\caption[a]{Fit of CS29502-092 (left) and HE0450-4902 (right) with the models of Table~\ref{modtab}. The best model, with the lowest $\xi$, is shown (black, red and green patterns sorted by increasing $\xi$). Uncertainties and limits on abundances are shown by vertical bars and arrows.}
   \caption[Best fits of G77-61 and HE~1005-1439 with the source star models of Table~\ref{modtab}]{Best abundance fits of G77-61 (left) and HE~1005-1439 (right) with the source star models of Table~\ref{modtab}, when considering the dilution factor and the mass cut as free parameters. The three best models, with the lowest $\xi$, are shown (black, red and green patterns sorted by increasing $\xi$). Uncertainties and limits on CEMP abundances are shown by vertical bars and arrows, respectively.}
              \label{XH3best} 
    \end{figure*}

  \begin{figure*}[t]
   \centering
      \includegraphics[scale=0.94, trim = 0cm 0cm 0cm 0cm]{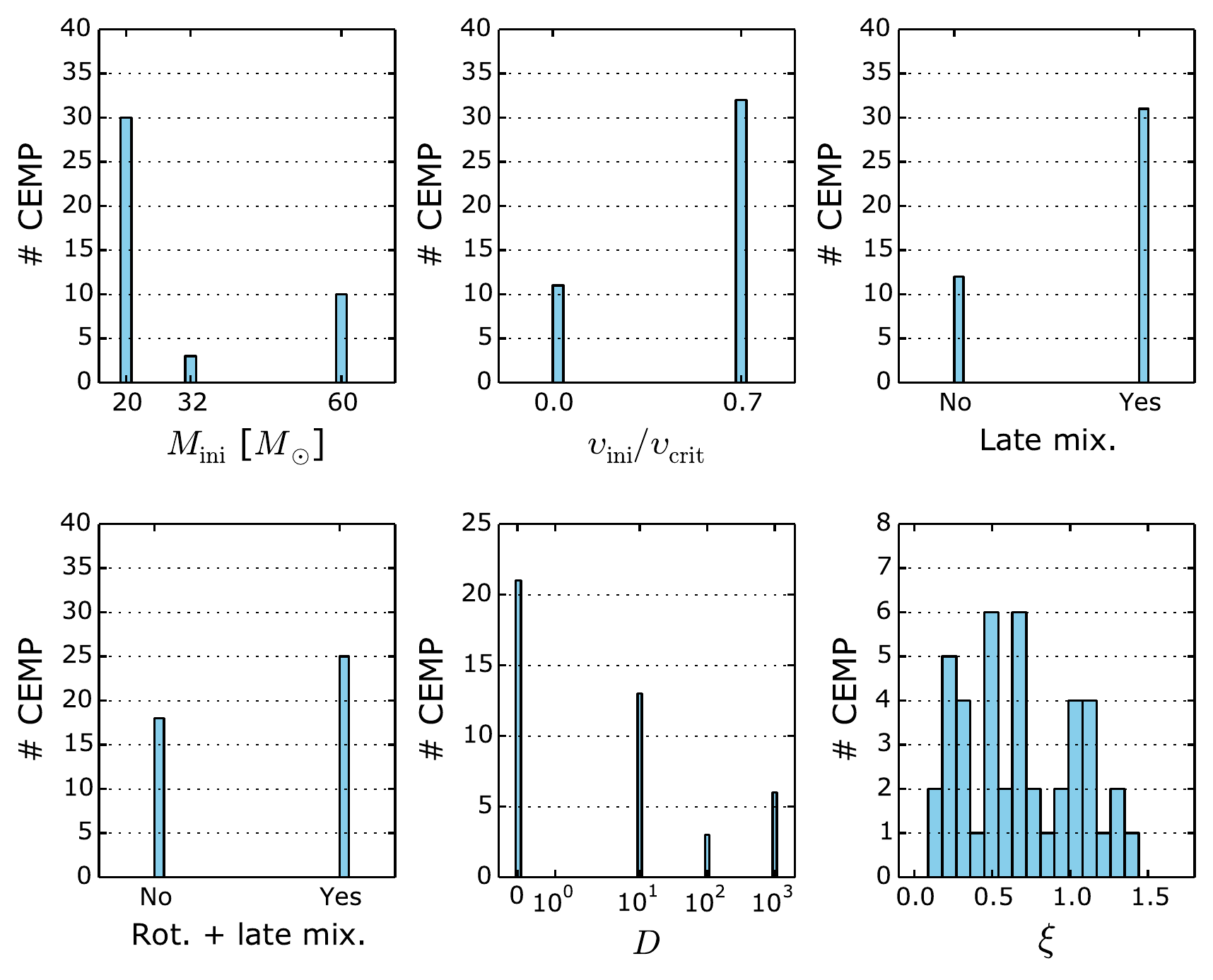}
   \caption[Characteristics of the best source star models]{Occurrence of various parameters for the best source star models (with the lowest $\xi$) of the 43 CEMP-no stars of Table~\ref{bestable}. Shown are the initial mass of the source star, initial velocity, if the model includes late mixing, if it includes late mixing and rotation, the dilution factor and the error on the fit $\xi$.}
              \label{histfin1} 
    \end{figure*}

  \begin{figure*}[t]
   \centering
      \includegraphics[scale=0.55, trim = 0cm 0cm 0cm 0cm]{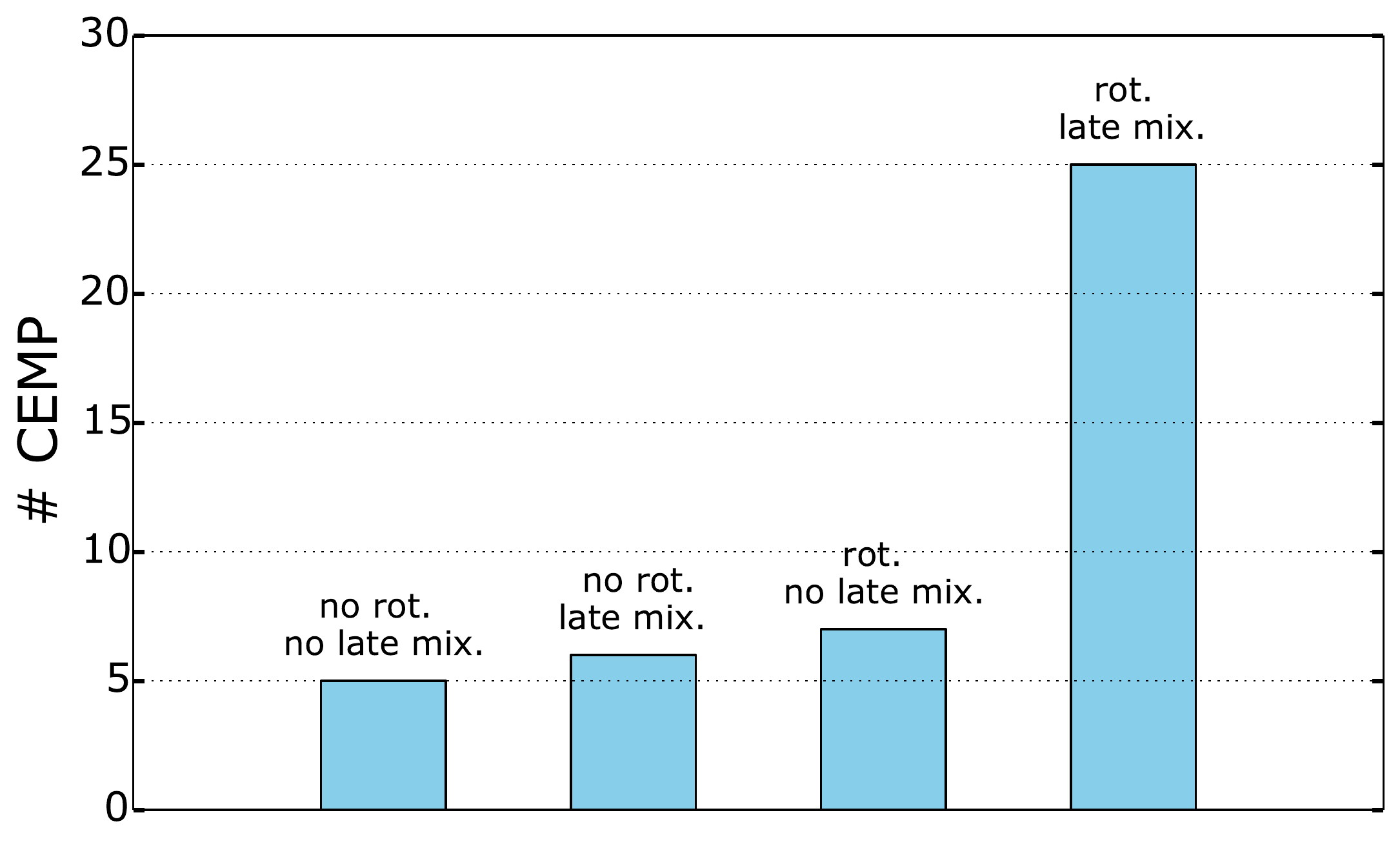}
   \caption[Occurrence of the four categories of the best source star models]{Occurrence of the four categories of the best source star models (with the lowest $\xi$) of the 43 CEMP-no stars of Table~\ref{bestable}.}
              \label{histfin2} 
    \end{figure*}

  \begin{figure*}[h!]
   \centering
      \includegraphics[scale=0.55, trim = 0cm 0cm 0cm 0cm]{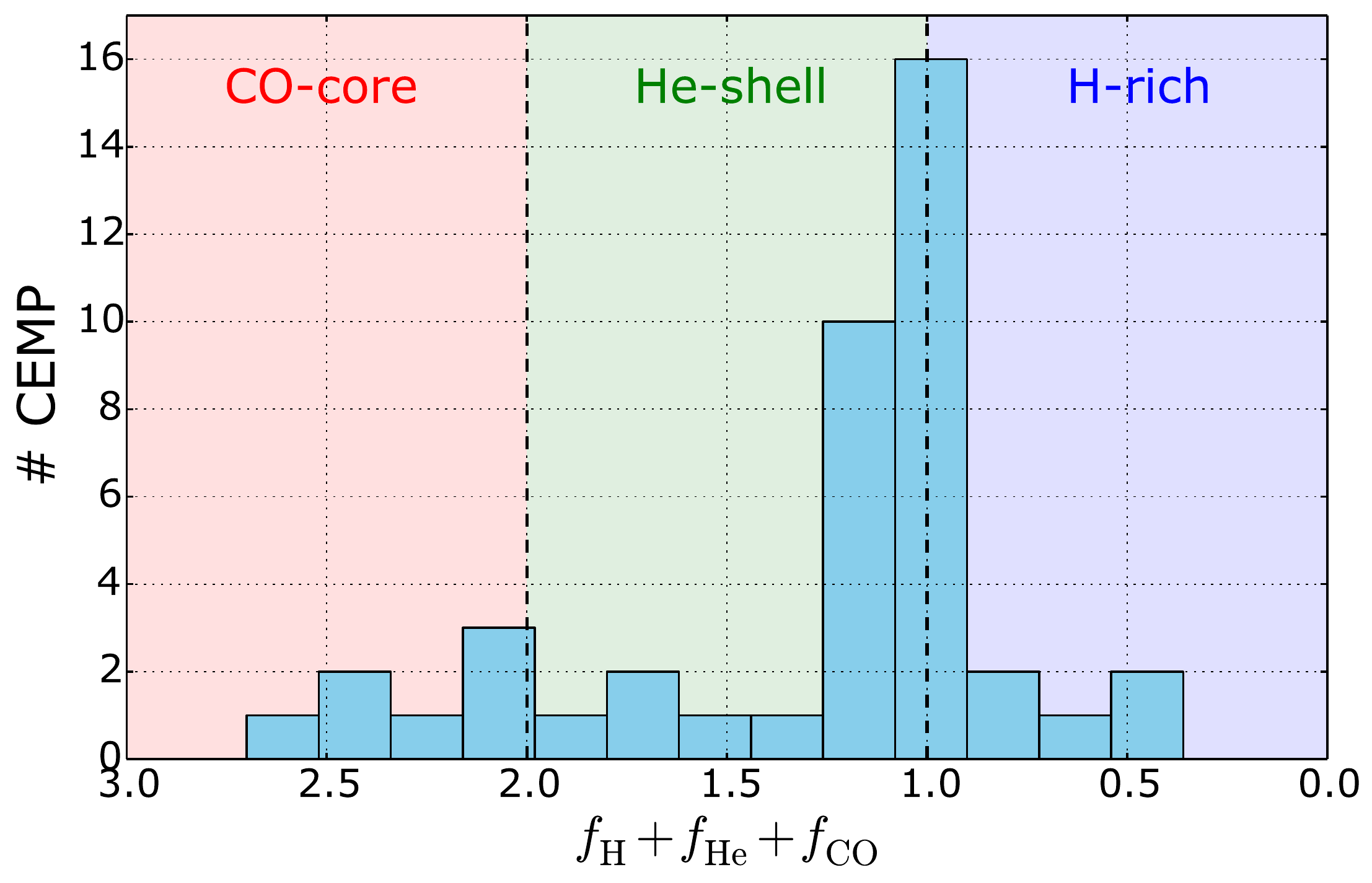}
%   \caption[Distribution of the location of the mass cut for the best source star models]{\textbf{Distribution of the location of the mass cut (arbitrary unit) for the best source star models of Table~\ref{bestable} (see text for details). }
%   %The mass cut is computed as the sum $f_{\rm H} + f_{\rm He} + f_{\rm CO}$ (i.e. ejected fraction of the H-rich envelope, He-shell and CO-core, cf. Table~\ref{bestable}). 
%   %Distribution of the location of the mass cut \textbf{(arbitrary unit)} for the best source star models (Table~\ref{bestable}). 
%   The purple, green and red shaded area represent the H-rich region, He-rich region and CO-core respectively (not at scale). The interface between the blue and green region corresponds to $M_{\rm r} = M_{\alpha}$ and interface %   between the green and red region corresponds to $M_{\rm r} = M_{\rm CO}$.}
   \caption[Distribution of $f_{\rm H} + f_{\rm He} + f_{\rm CO}$ for the best source star models]{Distribution of $f_{\rm H} + f_{\rm He} + f_{\rm CO}$ (cf. Table~\ref{bestable}) for the best source star models of Table~\ref{bestable}. This sum corresponds to the ejected fraction of the H-rich envelope, He-shell and CO-core (see text for details). 
   %The purple, green and red shaded area represent the H-rich envelope, He-shell and CO-core respectively (not at scale)
   %The mass cut is computed as the sum $f_{\rm H} + f_{\rm He} + f_{\rm CO}$ (i.e. ejected fraction of the H-rich envelope, He-shell and CO-core, cf. Table~\ref{bestable}). 
   %Distribution of the location of the mass cut \textbf{(arbitrary unit)} for the best source star models (Table~\ref{bestable}). 
   The purple, green and red shaded area delimitate these three different regions (the extension of each region is normalized to 1).}
   %The interface between the blue and green region corresponds to $M_{\rm r} = M_{\alpha}$ and interface between the green and red region corresponds to $M_{\rm r} = M_{\rm CO}$.}
              \label{histfin3} 
    \end{figure*}

   \begin{figure*}[t]
   \centering
       \includegraphics[scale=0.63]{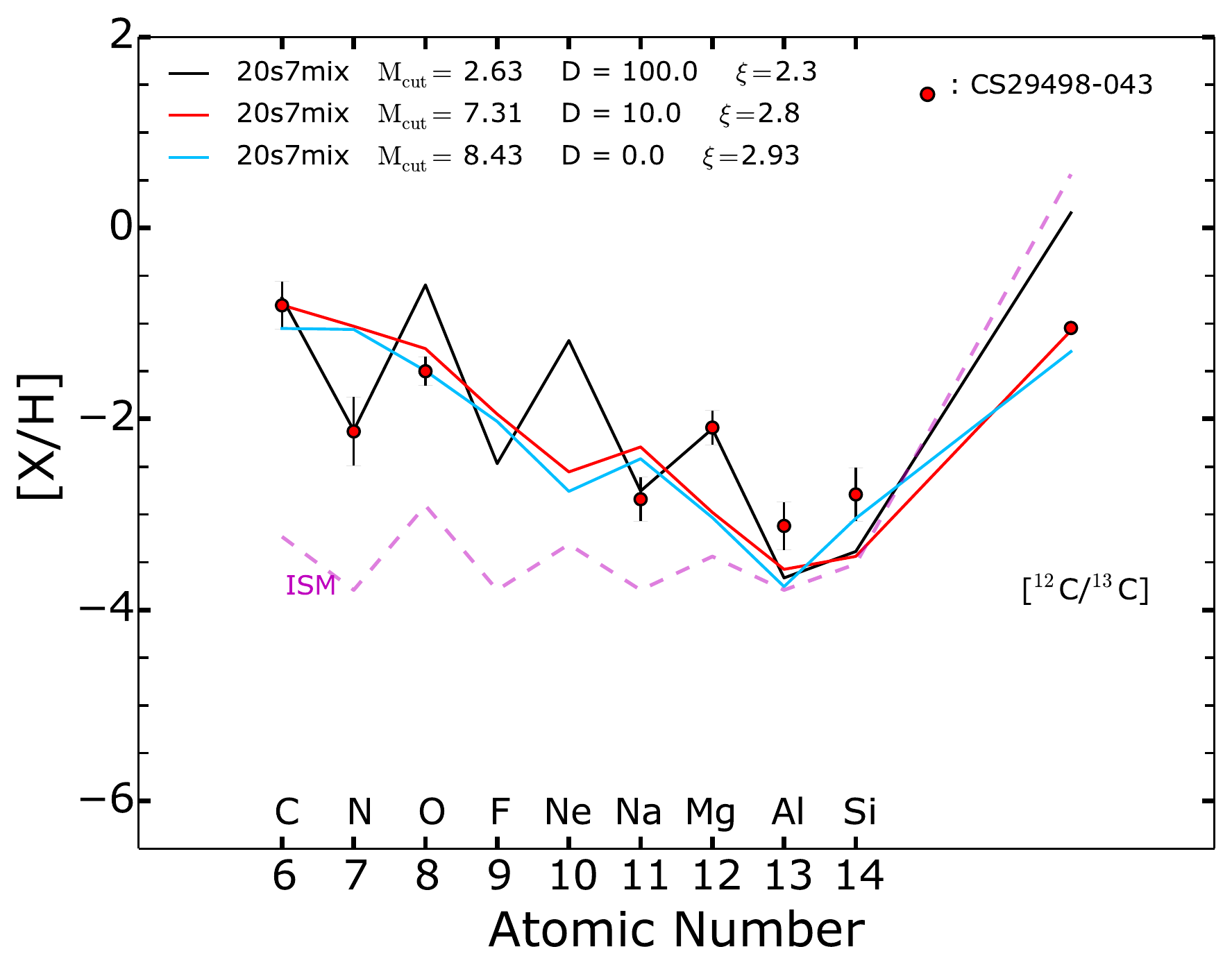}
   \caption[Same as Fig.~\ref{XH3best} but for CS~29498-043]{Same as Fig.~\ref{XH3best} but for CS~29498-043.}
\label{CS29498}
    \end{figure*}

%\textcolor{red}{How many models in total? Roughly}

%\textcolor{red}{Change title. Change plots of individual stars. Only the best model?}

I now investigate what is the best source star model for each individual CEMP-no star. All the 69 CEMP stars of Table~\ref{tabcemp} are considered. 
%We consider light elements, from C to Si and the $^{12}$C/$^{13}$C ratio. The effect of the 1$^{\rm st}$ dredge-up is corrected following \cite{placco14c}. 
To fit the abundance pattern of each CEMP star, I consider the 12 source star models of Table \ref{modtab} and vary the dilution factor and mass cut for each models. 
%As a first step, the 12 models of Table~\ref{modtab} are considered. 
The final structure of the source star is taken at the end of the core carbon burning stage for all the 12 models since ultimate stages were not computed for the 6 models including late mixing. 
%For each of the 12 models, the mass cut is varied between the final mass and $M_{\rm rem}$ (cf. Table~\ref{modtab}).
Four dilution factors are considered : $D=0$, 10, 100 and 1000. It gives 12 $\times$ 4 = 48 combinations. For each of these 48 cases, the mass cut of the given source star model is varied between its final mass (i.e. only winds) and $M_{\rm rem}$ (cf. Table~\ref{modtab}). The mass cut is chosen so as to minimize the sum of the residuals (in absolute value):
%For a given CEMP star, the mass cut of the source star is set so that the sum of the residuals (in absolute value)

\begin{equation}
\xi = \sum_{i} \mid \textrm{A}^{\rm obs}_{i} - \textrm{A}^{\rm mod}_{i} \mid .
\end{equation}
$\textrm{A}^{\rm obs}_{i}$ are the [C/H], [N/H], [O/H], [Na/H], [Mg/H] and $\log(^{12}$C/$^{13}$C) ratios of the CEMP-no star to fit\footnote{Table~\ref{tabcemp} reports the [X/Fe] ratios. For the present analysis, I used the [X/H] ratios (taken from the SAGA database).}. $\textrm{A}^{\rm mod}_{i}$ are the same ratios but in the source star ejecta, after the eventual dilution with ISM. $\xi$ is then evaluated using 1 to 6 abundance ratios. When no abundance or just an upper limit is available for a CEMP star, the abundance is not considered to estimate $\xi$. 
For rotating and non-rotating source star models, the number of possible mass cuts is about 1000 and 500 respectively (it depends on the resolution, which is higher in rotating models). 
In total, there are about 35000 possible ejecta compositions. 

%%%For a given CEMP star, the mass cut that minimizes $\xi$ is calculated for the 48 models (12 source stars models $\times$ 4 dilution factors). 
%For each of the 48 cases, the model with the minimal $\xi$ is saved.
For each of the 48 cases, the ejecta composition with the mass cut that minimizes $\xi$ is saved. Among these 48 ejecta, the 3 best ones (the 3 lowest $\xi$), are shown in Fig.~\ref{XH3best} for two CEMP stars. This procedure was applied for the 69 CEMP stars (all the plots are shown in Appendix \ref{fitstar}). In this sample, I selected the cases where $\xi$ was estimated with at least 3 abundances. Fits using only 1 or 2 abundances can lead to a high degree of degeneracy (many source star models provide a good solution). After this selection, it gives a subsample of 50 CEMP-no stars (19 stars excluded). In this new sample, the CEMP-no stars for which no satisfactory fit was found were excluded: fits with $\xi \geq 1.5$ are excluded. It finally gives 43 stars. The parameters of the best source star models for these 43 CEMP-no stars are reported in Table~\ref{bestable}.
%whose $\xi$ distribution is shown on Fig.~\ref{histfin1} (bottom right panel).

Figures \ref{histfin1},  \ref{histfin2} and  \ref{histfin3} show the results. 20~$M_{\odot}$ source stars are preferred in 70~\% of the cases, before the 60~$M_{\odot}$ (23~\%) and 32~$M_{\odot}$ (7~\%) source stars. Rotation gives the best fit for 74~\% of the sample. The late mixing process appears in 72~\% of the best models. $D \leq 10$ in 79~\% of the cases. 
Figure~\ref{histfin2} shows that rotation + late mixing gives the best fit in 58~\% of the cases. %No rotation and no mixing are preferred for only 2 stars.
The histogram of Fig.~\ref{histfin3} shows the distribution of the sum $f_{\rm H} + f_{\rm He} + f_{\rm CO}$ (i.e. the sum of the ejected fraction of the H-rich envelope, He-shell and CO-core, cf. Table~\ref{bestable}) for the best source star models. This sum varies between 0 (no material ejected from the source star) and 3 (all the material is ejected). A value of 1.5, for instance, means that 100~\% of the H-rich envelope plus 50~\% of the He-shell of the source star was ejected. The sum $f_{\rm H} + f_{\rm He} + f_{\rm CO}$ can also be seen as a mass cut value normalized to 3.
Figure \ref{histfin3} shows that in most of the cases (about 70~\%), $f_{\rm H} + f_{\rm He} + f_{\rm CO}$ is about 1 or just above~1. It means that the CEMP-no stars mostly formed with the H-rich envelope of the source star plus possibly a small part of the He-shell. In other other words, a mass cut located around the bottom of the H-rich region of the source star is preferred. It corresponds to $M_{\rm cut} = M_{\alpha}$. 
%This may be consistent with low-energetic supernovae expelling not too deep layers. 
In $15-20$~\% of the cases, $f_{\rm H} + f_{\rm He} + f_{\rm CO} \geq 2$. In this case, it suggests a deeper mass cut in the source star, located around the top or in the CO core. %, as in 'standard' supernovae explosions.

%disucssion of results

The results suggest that the CEMP-no source stars are preferentially rotating $20$~$M_{\odot}$ models experiencing the late mixing process. In most cases, only the outer layer of the source star should be expelled. Also, the results suggest a modest dilution with ISM.
%a low energetic supernova and a modest dilution with ISM. 
In \cite{choplin17a}, we investigated the individual abundances of only six CEMP-no stars in details but reached similar conclusions. 
%One of the main question that emerges from these results is: does fast rotation is compatible with a low-energetic supernova? Fast rotation could instead lead to energetic supernovae %\citep{woosley93}

%\textcolor{red}{Limits: degeneracy. Maybe weight to the best models... instead of selecting THE best}

%This gives some support to the idea that mixing (progressive ) has played an important role on the early and massive star generations.
%Considering different initial rotation rate might 

It is worth noting that the late mixing process, developed to explain the abundances of a small sample of CEMP-no stars in the C/N vs. $^{12}$C/$^{13}$C diagram (cf. Sect.~\ref{seclate}), is still favored when considering a larger sample of CEMP stars that often do not have a measured $^{12}$C/$^{13}$C ratio and/or the N abundance available. One reason is that standard models with or without rotation tend to give either much more N than C or much more C than N while numerous CEMP-no stars have roughly as much C as N. The late mixing process allows to get as much C as N in the ejecta. % (where e.g. $^{12}$C/$^{13}$C is not available).
Also to note is that, even if both non-rotating and rotating models including late mixing can provide a solution in the C/N vs. $^{12}$C/$^{13}$C diagram (cf. Fig.~\ref{cnccrotmix}), rotating source stars models are preferred. One reason is that rotation can provide the additional Na and Mg needed to reproduce some CEMP-no stars while the late mixing process cannot.

\subsubsection{Possible improvements}

This analysis can be improved in different ways. Only the best source star models were considered here. The second best models could also be considered with smaller weights than the best models. It can also be done for the third best models, etc... This would allow to also include the good (but not best) models. It may increase the statistics of the results. An other improvement could be to attribute weights to the different abundances used to perform the fit. For instance,  the abundances that are strongly affected by NLTE/3D effects should have a lower weight. The abundances that are the most affected by nuclear rate uncertainties in stellar models should also have a lower weight.

In the future, it would be interesting to do the same analysis while considering larger grids of source star models with more initial masses and rotation rates, and with different initial metallicities. This could allow to derive a initial mass or velocity function for the early generations of massive stars. However, the sample of CEMP-no stars may still be quite small. Larger (and homogeneous) samples are required so as to obtain meaningful statistic results. 

Finally, let us mention that among the 50 stars fitted (69 in total but 19 were excluded because of a too small abundance data) 7 stars could not be fitted correctly.
\begin{itemize}
\item 3 out of these 7 stars (HE~0107-5240, HE~0557-4840 and SDSS~J131326.89) have very low [Na/H] and [Mg/H] ratios, at least 1 dex below the ISM values considered here. No solution can give such low values. Zero metallicity source star models may provide a better solution. As discussed at the end of Sect.~\ref{secvarpar}, zero metallicity source stars may produce as much CNO as low metallicity source stars but less Na, Mg, Al and Si. This remains to be tested.
\item 3 other stars (HE~1506-0113, HE~1012-1540 and HE~2139-5432) have high [Na/H] together with low [N/H] ratios. This is difficult to explain with the present source star models, where N and Na are rather well correlated. In particular, both are produced in rotating models.
\item The last star (CS~29498-043) has a CNO pattern very typical of an He-burning material but a very low $^{12}$C/$^{13}$C ratio, typical of an H-burning material (see Fig.~\ref{CS29498}). Although the models with extra mixing can approach the solution closer than standard models, no satisfactory fit can be found. 
\end{itemize}
Overall, these seven stars may require different kinds of progenitors. In any case, these stars will require further attention. This is among the next steps of this analysis.

\section{Are CEMP-no stars helium-rich? \label{herich}}

   \begin{figure*}[t]
   \centering
      \includegraphics[scale=0.55]{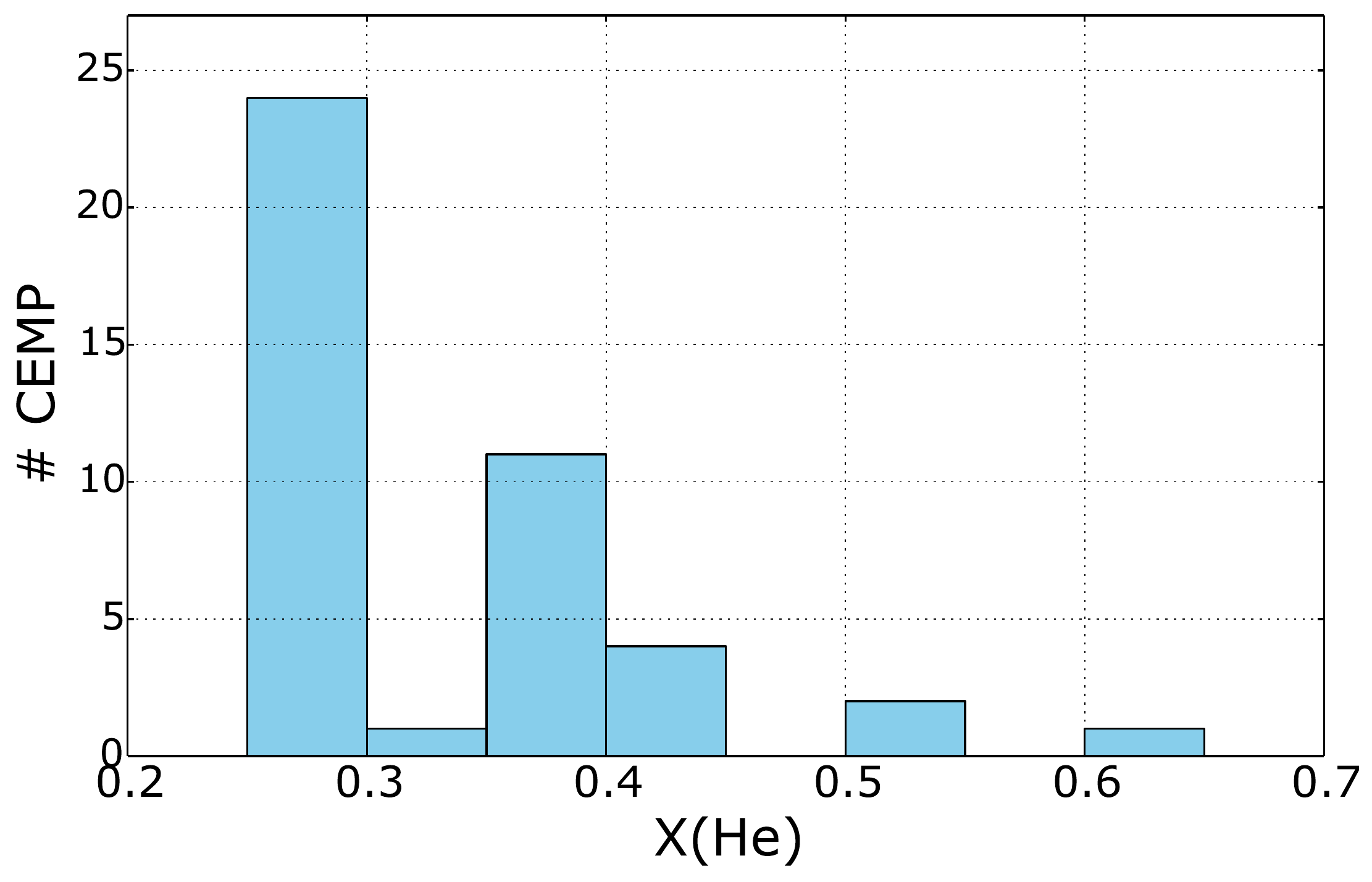}

\caption[Distribution of the helium mass fraction in the ejecta of the best source star models]{Distribution of the helium mass fraction in the ejecta of the best source star models (Table~\ref{bestable}). It corresponds to the helium mass fraction in the source star ejecta, mixed with some ISM material if $D \neq 0$. This distribution shows the helium mass fraction predicted for the 43 CEMP-no stars of Table~\ref{bestable}. }
            
  \label{hehisto}
    \end{figure*}

A signature of the operation of the CNO cycle is an enrichment in helium. Thus, the CEMP-no stars that formed with almost only the envelope of massive source stars (i.e. $D \simeq 0$) should be He-rich \citep{meynet10}. 
%\textcolor{red}{more details?} 
If instead the source star ejecta was significantly mixed with ISM, the mass fraction of helium $X$(He) in the CEMP-no star should be around the primordial value of 0.248 \citep{cyburt03}. 
%Detection of helium on CEMP stars will then provide valuable clues. 

Table~\ref{bestable} reports the predicted helium mass fraction in the ejecta (after dilution) of the best source star models. It shows for instance that the star CS~22960-053 should have formed with 0.39 of helium in mass fraction.
Fig.~\ref{hehisto} shows the distribution of helium mass fractions. While $\sim 60$~\% of the CEMP star sample is predicted to have formed with rather normal helium mass fractions ($X$(He) $\leq0.3$), about 40~\% should have formed with $X$(He) $ > 0.3$.
%CS~22949-037, with $X$(He) = 0.38, is a He-rich candidate. 
I discuss below what are the effects on the evolution of CEMP stars if changing the initial He. The specific case of the He-rich candidate CS~22949-037, with $X$(He) = 0.38, is then investigated.

\subsubsection{Low-mass metal-poor models enriched in helium}

\cite{chantereau15} have investigated the evolution of $0.3-1$~$M_{\odot}$ stars at [Fe/H] $=-1.75$ with an initial He mass fraction $Y_{\rm ini}$ between 0.248 and 0.8.
As $Y_{\rm ini}$ increases, the opacity in the stellar interior is reduced, making the star more compact. This shifts the evolutionary tracks towards higher $T_{\rm eff}$. The higher compactness of the star makes the central temperature higher, increasing the rate at which hydrogen is burnt. This, together with the fact that there is less hydrogen fuel in helium rich stars, lead to shorter main sequence lifetimes for the helium rich models.
Their 0.8~$M_{\odot}$ models with $Y_{\rm ini} = 0.248$ (standard), $0.4$ and $0.8$ spend 12.9, 4.62 and 0.16 Gyr on the main sequence, respectively.

To investigate the specific case of the CEMP-no star CS~22949-037, I computed non-rotating 0.5, 0.6, 0.7, 0.8 and 0.9~$M_{\odot}$ stellar models starting with the chemical composition of CS~22949-037. 
%\textcolor{red}{thermohaline...! see teff and logg... This star is a red giant so that its surface has likely experienced some mixing, its initial composition could be different than the observed one. However, this star has [C/N] $=-1.5$ and $^{12}$C/$^{13}$C $=4$ according to \cite{spite06} \citep[][have however only reported a lower limit with $^{12}$C/$^{13}$C $>4$]{roederer14c}. Because the observed [C/N] and $^{12}$C/$^{13}$C \citep[according to][]{spite06} are close to their equilibrium values, the initial [C/N] and $^{12}$C/$^{13}$C ratios should not be very different (cf. discussion on the effect of the first dredge-up, Sect.~\ref{secpuzzle}).}
For each mass, different initial helium mass fractions are tested: 0.25, 0.3, 0.4 and 0.5. What is added to the initial helium mass fraction is removed from the initial hydrogen mass fraction. The OPAL tool was used to compute new opacity tables accordingly to the particular composition of CS~22949-037. For the other input parameters, I followed \cite{ekstrom12}.

%As the initial helium mass fraction $\rm Y_{ini}$ increases, the opacity in the stellar interior is reduced, making the star more compact. This shifts the evolution tracks toward higher $T_{eff}$ (see Fig.~\ref{evol}). The higher compactness of the star makes the central temperature higher, increasing the rate at which hydrogen is burnt. This, together with the fact that there is less hydrogen in helium rich models, leads to shorter main sequence lifetimes for the helium rich models.

%%%% (compare for instance the 0.9~$M_{\odot}$ tracks with different helium content in Fig.~\ref{evol}).

   \begin{figure*}[t]
   \centering
      \includegraphics[scale=0.81]{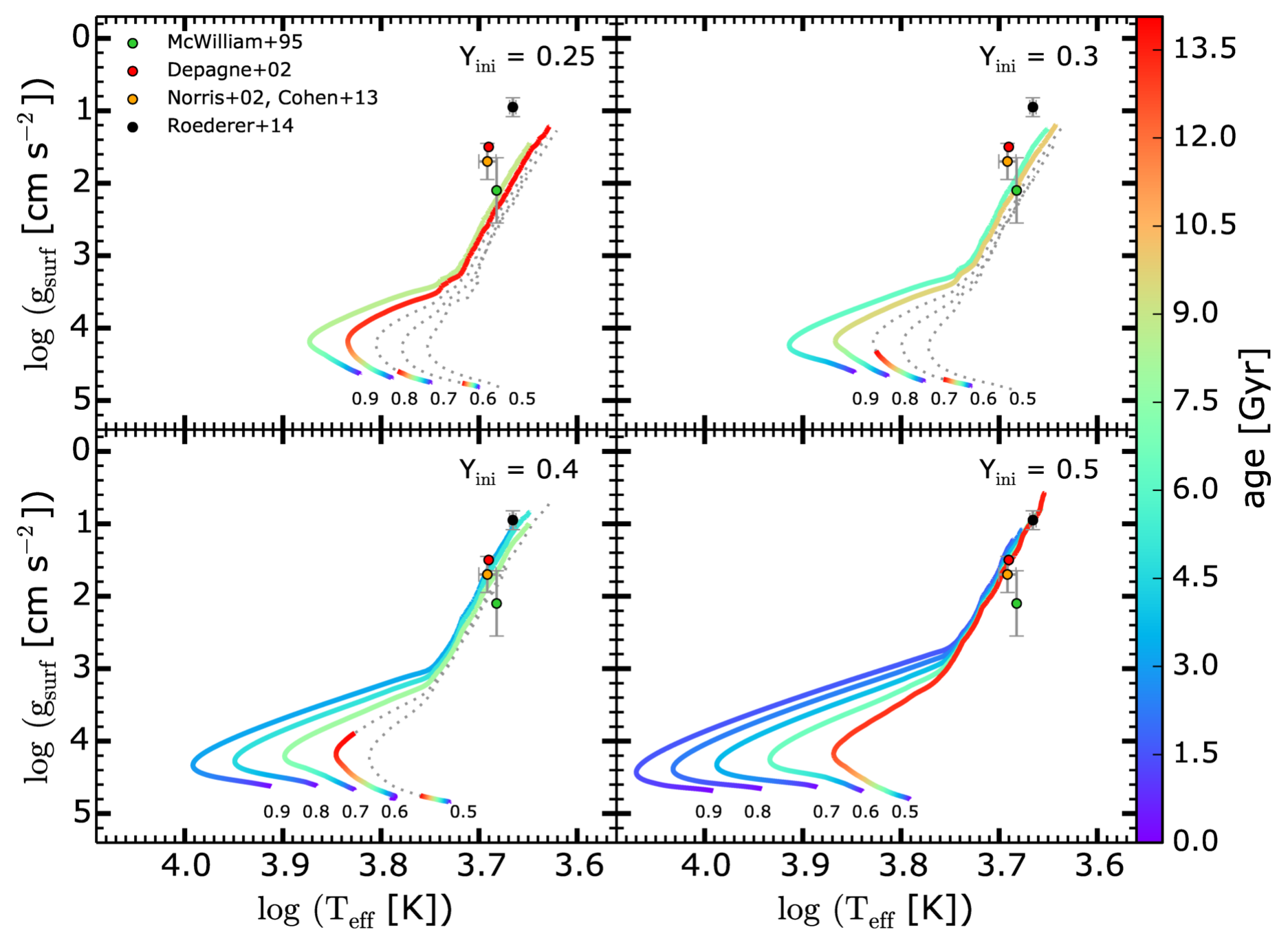}

\caption[$\log g$ vs. $\log (T_{\rm eff})$ for CEMP-no star models with various initial helium content]{Surface gravity as a function of effective temperature. Lines represent stellar models between 0.5 and 0.9 $\rm M_{\odot}$. The age of the models in Gyr is given by the color. Each panel corresponds to a different initial helium mass fraction. When the lifetime of the model exceeds 14 Gyr (age of the Universe), the rest of the track is represented by a dotted line. Squares are observations of CS~22949-037 \citep[from][uncertainties are indicated if available]{mcwilliam95, depagne02, norris02, cohen13, roederer14a}.}
            
  \label{evol}
    \end{figure*}

%   \begin{figure}[t]
 %  \centering
 %     \includegraphics[scale=0.6, trim = 1cm 0cm 0cm 0cm ]{mtau.pdf}
%%%%\caption[]{$\tau$ (time needed to the model to reach $\log$($g_{\rm surf}$) $= 2$, see text) as a function of the initial mass. Each line show a series of models with a different initial $^{4}$He mass fraction. The grey band shows the possible range of age for the star CS~22949-037. The upper limit is 13.8 Gyr, the age of the Universe. The lower limit is 11.4 ($\pm 0.7$) Gyr, the age of the local field halo stars derived by \cite{kalirai12}.}
%\caption[]{$\tau$ (time needed for the model to reach $\log$($g_{\rm surf}$) $= 2$, see text) as a function of the initial mass. Each line show a series of models with a different initial $^{4}$He mass fraction. The dashed lines shows the maximal age for CS~22949-037 (age of the Universe).}
            
 % \label{mtau}
  %  \end{figure}

Fig.~\ref{evol} shows the tracks of these models. 
The observations of CS~22949-037 are better reproduced by the tracks having $Y_{\rm ini}$ $=0.4$ and $0.5$. 
This is consistent with the analysis of Sect.~\ref{secstat} which predict $Y_{\rm ini} = 0.38$ for CS~22949-037. 
%\textcolor{red}{However, the red giant branch could be shifted depending on how the physics is treated (more details?)}. 
However, because of uncertainties on both the modeling and observational sides, it is difficult to make meaningful comparisons between the observations and the tracks of Fig.~\ref{evol}.

Considerations on the age of CS~22949-037 can give additional hints. The color of the track shows the age of the model.
To match the observables, the tracks of Fig.~\ref{evol} have to reach the $\log g$ and $T_{\rm eff}$ of CS~22949-037 at the age of CS~22949-037. For instance, while the 0.9~$M_{\odot}$ model with $Y_{\rm ini} = 0.4$ seems a good candidate, it reaches the $\log g$ and $T_{\rm eff}$ of CS~22949-037 after only $\sim 4$ Gyr. It is unlikely that CS~22949-037 is only 4 Gyr old. %Also, %Even if the age of CS~22949-037 is not known, we know that it is younger than 13.8 Gyr (age of the Universe). Also
On the other hand, the 0.5~$M_{\odot}$ model (still with $Y_{\rm ini} = 0.4$) is not a solution since it would still be unevolved today, with a much higher $\log g$ than CS~22949-037. 
We can see that if CS~22949-037 has $Y_{\rm ini} = 0.4$, it should be between 0.6 and 0.7~$M_{\odot}$.

%Fig.~\ref{mtau} shows the time needed for the CS~22949-037 models to reach $\log$($g_{\rm surf}$) $= 2$ as a function of the initial mass of the model. 
%%%In any case, CS~22949-037 is younger than 13.8 Gyr (age of the Universe, dashed line).
%The analysis of Sect.~\ref{secstat} predicts $Y_{\rm ini} = 0.38$ for CS~22949-037. 
%If $Y_{\rm ini} = 0.38$ (close to the cyan line), the mass of CS~22949-037 should be greater than 0.6~$M_{\odot}$: if below 0.6~$M_{\odot}$, CS~22949-037 would be older than the Universe (dashed line). On the other hand, if CS~22949-037 is 0.8~$M_{\odot}$, it means that it is only 5 Gyr old, which is unlikely. Assuming that CS~22949-037 is older than 10 Gyr gives a mass between 0.6 and 0.65~$M_{\odot}$.

%For a given $Y_{\rm ini}$, Fig.~\ref{mtau} allows to see what would be the approximate mass of CS~22949-037. $\tau$ defines the time needed for the model to reach $\log$($g_{\rm surf}$) $= 2$, which corresponds to the surface gravity of CS~22949-037. 
%The analysis of Sect.~\ref{secstat} predicts $Y_{\rm ini} = 0.38$ for this star. For $Y_{\rm ini} = 0.38$ (close to the cyan line), the mass of CS~22949-037 should be $\sim 0.60 - 0.65$~$M_{\odot}$ (i.e. the range of mass where the cyan line and the grey shaded area intersect). 

We see here that the scenario investigated in this work to explain CEMP-no stars predicts that some of the CEMP-no stars (the ones that are He-rich) will have rather small masses, due to their He-richness. Both the determination of the helium abundance and the mass of the CEMP-no stars may provide further clues on their origin.

\subsubsection{Measuring the helium abundance and the mass of CEMP-no stars}

Helium is challenging to detect and direct measurements are rare. One possibility is to focus on the HeI line at $1.08$ $\mu$m that forms in the upper chromosphere of cool stars. In doing so, \cite{pasquini11} managed to measure the difference in helium abundance for two giants stars ([Fe/H]~$=-$~1.22 and $-1.08$) in the globular cluster NGC 2808. With their chromospheric model, they found a difference of 0.17 in helium mass fraction, meaning that if one of the two star has a solar He abundance, the other would have $Y \sim 0.42$. Also, \cite{dupree13} reported an helium abundance of $Y<0.22$ and $0.39<Y<0.44$ in two other giant stars ([Fe/H] $=-1.86$ and $-1.79$) belonging to the globular cluster $\omega$ Centauri. 
In CEMP-no stars and generally in very iron-poor stars, no helium measurements has been performed yet.

%Other CEMP-no stars in Table~\ref{bestable} are interesting He-rich candidates. There is for instance CS~22960-053 with $X$(He) $=0.46$. It has [C/N] $-2$ which is characteristic of CN-processing. There is also HE2331-7155, with $X$(He) $=0.39$, $^{12}$C/$^{13}$C $=5$ and [C/N] $=-1.23$.

Let us mention that asteroseismology may provide the mass of giants CEMP-no stars. The asteroseismic information (hence the mass) of cool giant stars with $1.9 < \log~g <3.2$ may be obtained with the Kepler K2 mission \citep[e.g.][]{stello15}. Another way to have a direct estimation of the mass is if the CEMP-no star is in an eclipsing binary system: since the inclination of an eclipsing system is known, masses of the two stars can be determined from the Kepler's laws. Tracks of stellar models computed with the right masses and with various helium abundances may allow to get an estimation of the helium abundance of these observed stars.

\subsubsection{Proxies for the helium abundance }

Some abundances of other elements may give a proxy for the helium abundance. 
%It is natural to He-burning products 
%We may imagine that the He-rich CEMP stars should also present overabundances in He-burning products From the results of Table~\ref{bestable}, we can see that 
A high helium abundance requires a small dilution factor otherwise the helium abundance goes back towards the ISM value of $\sim 0.25$. A small dilution factor might be related to a low Li abundance (Sect.~\ref{secvarpar}, paragraph about dilution), meaning that He-rich CEMP stars could also be Li-poor (caution is required since Li is a fragile element that can be heavily affected by various processes in the CEMP star itself). Interestingly, the He-rich candidate CS~22949-037 has a very low Li abundance (A(Li) $<0.13$). Two other He-rich candidates, CS~22960-053 ($X$(He) $=0.39$) and HE~2331-7155 ($X$(He) $=0.39$), have A(Li) $<0.55$ and A(Li) $<0.37$, respectively. He-rich candidates are nevertheless not always Li-poor (e.g. CS~22960-053 with a predicted helium abundance of $X$(He) $=0.42$ and a Li abundance, deduced from observations, of A(Li) $=1.71$).

It is also worth noting that the He-rich CEMP stars should probably not be searched among the CEMP stars having a $\vee$-shape CNO pattern (cf. Sect.~\ref{globalcomp}). It may be surprising since it is a pattern characteristic of a material processed by He-burning, hence potentially He-rich. However, when a CEMP star is predicted to be formed with the He-processed material of the source star, a high dilution factor is generally also predicted (often $D =1000$, see Table~\ref{bestable}) so that the final helium abundance (the one of the CEMP) is close to the ISM value of 0.25. 
Instead, it is predicted here that many He-rich CEMP stars should have a $\wedge$-shape CNO pattern. This pattern is characteristic of an ejecta processed by H-burning. Such an ejecta is also enriched in helium (provided the dilution with ISM is small).
\section{Other CEMP-no source star models \label{secother}}

   \begin{figure*}[t]
   \centering
      \includegraphics[scale=0.24, trim = 0cm 0cm 0cm 0cm]{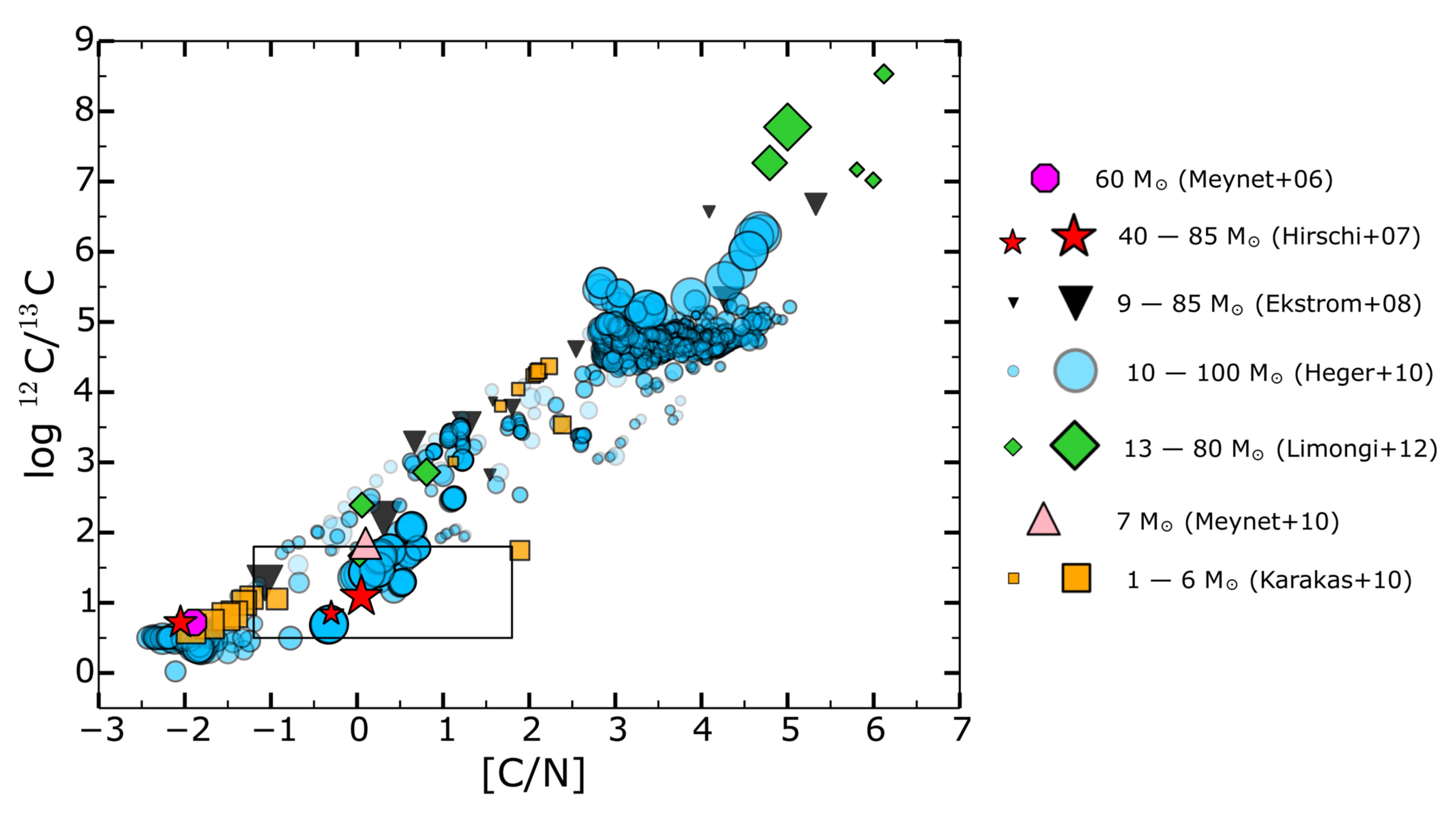}
   \caption[Ejecta composition of massive and AGB star models from various authors]{Composition of the ejecta of massive and AGB star models from various authors. The symbol size scales with the initial mass (see the legend on the right). The black frame shows where observed CEMP-no stars lie. Symbols show the composition of the ejecta of rotating 60~$M_{\odot}$, $Z=10^{-5}$ \citep{meynet06}; rotating 40, 60 and 85~$M_{\odot}$, $Z=10^{-8}$ \citep{hirschi07}; non-rotating and rotating $9-85$~$M_{\odot}$, $Z=0$ \citep{ekstrom08}; rotating 7~$M_{\odot}$ early AGB, $Z=10^{-5}$ \citep{meynet10}; non-rotating $1-6$~$M_{\odot}$ AGB, $Z=10^{-4}$ \citep{karakas10}; non-rotating $10-100$~$M_{\odot}$, $Z=0$ \citep{heger10}; and non-rotating $13-80$~$M_{\odot}$, $Z=0$ \citep{limongi12}.}
\label{cnccmods}
    \end{figure*}

   \begin{figure*}[t]
   \centering
      \includegraphics[scale=0.24, trim = 0cm 0cm 0cm 0cm]{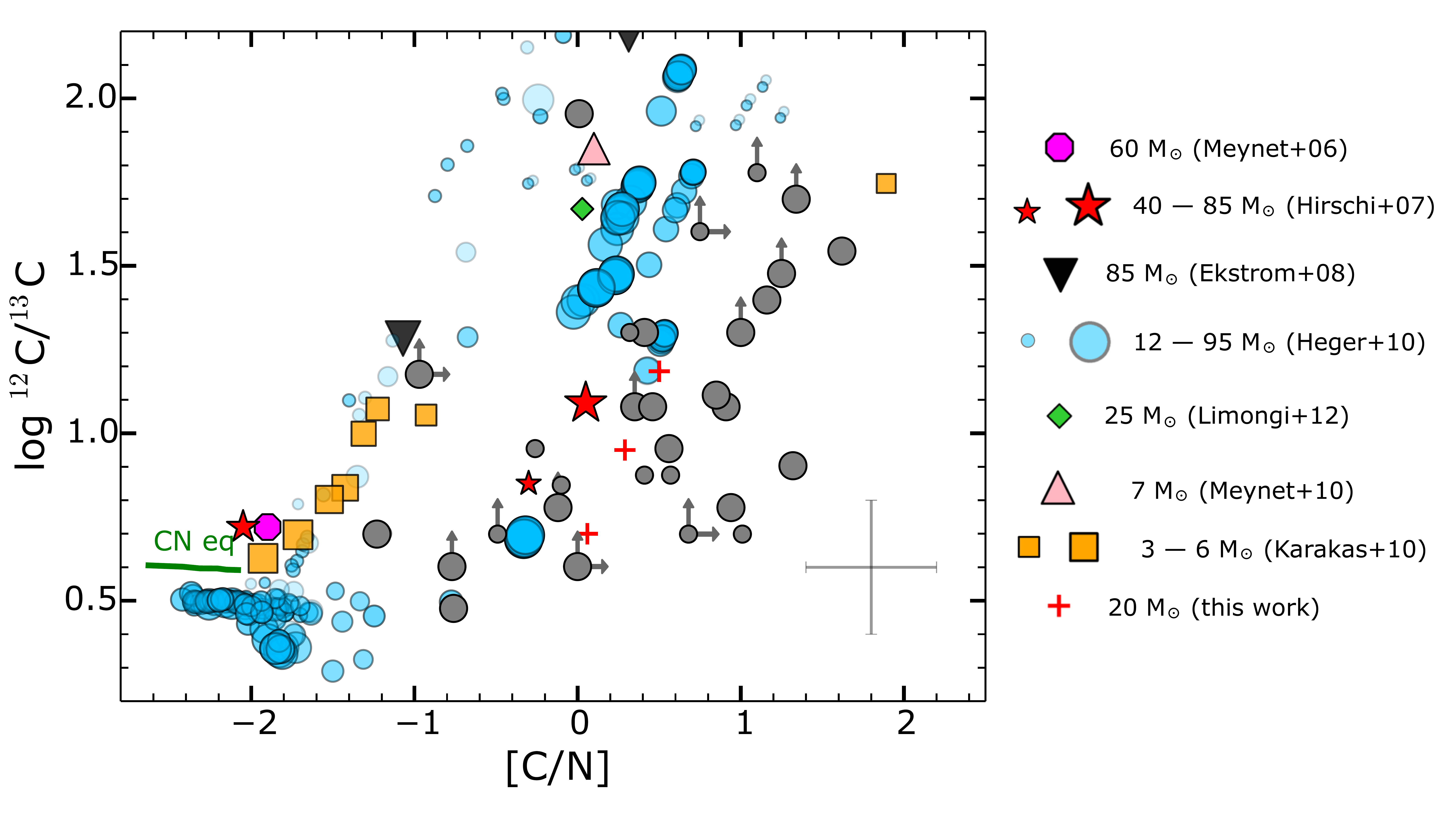}
   \caption[Same as Fig.~\ref{cnccmods} but in the region where observations lie]{Same as Fig.~\ref{cnccmods} but in the region where the observations lie (grey circles). Red crosses show models from the present work (20~$M_{\odot}$ with $\upsilon_{\rm ini}/\upsilon_{\rm crit} = 0.7$, $Z=10^{-5}$ and with the late mixing process). The 3 crosses correspond to 3 different mass cuts.  }
\label{cnccmodszoom}
    \end{figure*}

%This section discusses some other source star models from different authors, mostly in the light of the C/N $-$ $^{12}$C$^{13}$C diagram. 

%Models of mixing \& fallback from \cite{} are not discussed here.  I refer to Sect.~\ref{cempnoscenar}.

%   \begin{figure*}
  % \centering
 %     \includegraphics[scale=0.7, trim = 0cm 0cm 1cm 0cm]{CNCC_H07.pdf}
 %  \caption[a]{Chemical composition of the ejecta of the fast rotating 85~$M_{\odot}$ model with $Z=10^{-8}$ from \cite{hirschi07}. The thick red track represents the integrated ratios in the wind, as evolution proceeds (this model loses about $\sim 65$~$M_{\odot}$). The cross shows the ratios in the wind at the end of evolution. The thin red line shows the effect of the mass cut, as in Fig.~\ref{cnccrotfirst}. \textcolor{red}{REMOVE THIS FIG MAYBE}}
%\label{cncch07}
  %  \end{figure*}

%Here, other source star models from various authors, are discussed. 

\subsection{An energetic H-ingestion event in the source star}

Let us first mention the recent work of \cite{clarkson18}, that investigated the origin of the 3 most iron-poor stars with a non-rotating 45~$M_{\odot}$ source star model experiencing an energetic H-ingestion event during shell He burning. It is interesting in the frame of the present work since the ingestion process they consider is similar to the late mixing process described in Sect.~\ref{seclate}. In their work, nucleosynthesis was calculated in post-processing, using a one-zone model. They found that after the H-ingestion event, neutron densities of $10^{-13}$ cm$^{-3}$ are reached (typical of the i-process), leading to neutron captures on light seeds. These neutron captures allow the synthesis of small amounts of heavier elements like Ca (not much heavier elements than Ca are synthesized). 
Globally they found a reasonable agreement for 2 out of the 3 iron-poor stars considered. One difficulty is that neutron captures in the source star model do not produce enough Fe so that an additional enrichment process is required. Also, one of the star they reproduce (HE~1327-2326) has [Sr/Fe] $=1.08$. It is not mentioned if their models can account for it. The origin of Sr in HE~1327-2326 is investigated in the Sect.~\ref{origHE1327} of this work.

\subsection{Other source star models in the C/N $-$ $^{12}$C/$^{13}$C diagram}

It is now investigated whether various massive source star models from different authors can provide a solution in the C/N $-$ $^{12}$C/$^{13}$C diagram. 
Low or zero metallicity massive star models with available yields are considered. Some intermediate mass models are also included (see the legend of Fig.~\ref{cnccmods} for the complete list of models).
In the non-rotating $10-100$~$M_{\odot}$ Pop~III models of \cite{heger10}, explosive nucleosynthesis is calculated and various assumptions for the mixing \citep[in the sense of the mixing \& fallback process of][]{umeda02} and the explosion energy are considered. Explosive nucleosynthesis is also calculated in the non-rotating $13-80$ Pop~III stars of \cite{limongi12} . 

The composition of the ejecta of all the considered models is shown in Fig.~\ref{cnccmods}. 
The observations lie in a very small zone (black frame) compared to the region occupied by the source star models. Worth to remind is that some CEMP-no stars have an upper limit for N and/or a lower limit for $^{12}$C/$^{13}$C. It means that it might exist an unseen population of stars with higher $^{12}$C/$^{13}$C and C/N ratios. For what regards the observations without upper/lower limits, only a small fraction of the source star models could possibly account for them.

Fig.~\ref{cnccmodszoom} shows a zoom in the region where observations lie.
%Models from \cite{meynet10} and \cite{karakas10} are AGB while other are massive stars. 
Models of \cite{limongi12} consider standard core collapse supernovae, with $E_{51} \sim 1$ and deep mass cuts. As they say (see also Fig.~\ref{cnccmods} and \ref{cnccmodszoom}), this assumption likely cannot account for the majority of the most metal-poor stars.
It is similar for the models of \cite{ekstrom08} where a deep mass cut is considered, following the relation of \cite{maeder92}. %\textcolor{red}{put a plot of this relation?? discuss? put a X/H with this Mcut?}. 
%This material does not match the observation. 
Higher mass cut values (closer to the stellar surface) in the models of \cite{limongi12} and \cite{ekstrom08} may improve the fit between models and observations. %Higher mass cut values, closer to the surface, might give different ratios.
AGB models may account for some observations but there are generally away from the majority of CEMP-no stars. 
Models approaching the most the bulk of observed stars are (1) some non-rotating $50-100$~$M_{\odot}$ Pop~III models from \cite{heger10}, represented by big blue circles, and (2) fast-rotating 40 and 85~$M_{\odot}$ models with $Z=10^{-8}$ from \cite{hirschi07}.

\subsubsection{A fast rotating massive source star with extreme mass loss}

Let us first summarize the peculiar evolution of the fast rotating 85~$M_{\odot}$ model of \cite{hirschi07}, that leads to [C/N] $\simeq 0$ and $\log$($^{12}$C/$^{13}$C) $\simeq 1.1$ in the ejecta (big red star in Fig.~\ref{cnccmodszoom}). From the beginning of the core helium burning stage, $^{12}$C and $^{16}$O diffuse progressively to the radiative H-shell. At the middle of the core helium burning phase, the helium convective core encroaches the H-shell so that the CNO cycle is suddenly boosted in the H-shell and a copious amount of $^{14}$N and $^{13}$C is quickly formed. The energy released makes the H-shell convective. This CNO-rich convective zone quickly reaches the stellar surface so that the surface metallicity is enhanced and dramatic mass loss occurs, removing about 60~$M_{\odot}$ from the star. The material ejected is only partially processed by the CNO cycle because the convective zone extends very quickly in cold regions where the CNO cycle operates slowly or does not operate at all. In addition, the strong mass loss episode occurs also quickly after the encroaching of the H-shell, saving definitely the material from further processing. From the instant of the encroaching of the H-shell to the instant where strong mass loss stops, only $\sim 0.03$ Myr have past. 
%(about 9~\% of the duration of the core helium burning stage for this model.
Finally, the model (which is now about 25~$M_{\odot}$) finishes to burn helium in its core (and then subsequent burning stages occur normally).
The big red star in Fig.~\ref{cnccmodszoom} shows the chemical composition in the wind of this model, which is indeed partially CNO processed.

\subsubsection{H/He interaction in non-rotating massive source stars}

Some $50-100$~$M_{\odot}$ models of \citet[blue circles]{heger10} experience a similar encroaching (cf. previous paragraph) of the He-core with the H-shell, as they mention in their paper. Also, as they note, in some models, the encroaching occurs during the core carbon burning stage, between the He- and H-burning shells (like for the late mixing process, cf. Sect.~\ref{seclate}). In both cases, it produces a lot of $^{13}$C and $^{14}$N. Mass loss was taken equal to zero in their models so that no material is ejected through winds. However, some material is likely saved in cold regions from further CNO processing until the end of the source star evolution. It leads to a partially processed CNO material in the ejecta as shown by some of the blue points in Fig.~\ref{cnccmodszoom} that approach the bulk of observed CEMP-no stars.

\subsubsection{A strong H/He interaction: a characteristic shared by the best source star models?}

All the models able to fit the observations in the [C/N] $-$ $^{12}$C/$^{13}$C diagram (including the models of this work discussed in Sect.~\ref{seclate}) likely share a common feature: a strong interaction between the H- and He-region occurred in the source star, boosting the CNO cycle, but right after, the material was saved from further CNO processing (either by being ejected quickly, or by being saved in cold regions of the source star until a supernova expels these layers). 
The fact that this interaction occurs also in non-rotating models \citep[e.g.][]{heger10} may indicate that the origin of the phenomena is not entirely due to rotation. Rotation may facilitate such interactions. As discussed in Sect.~\ref{secorig}, the main ingredient might be convection. Convection constitutes a major uncertainty that should be better constrained in the future thanks to multi-D simulations. 
%However, what can do rotation is to help transporting the metals up to the surface and trigger strong winds. %This is strength of the model of \cite{hirschi07} is that 

This said, the present work suggests that this strong interaction between the H- and He-regions is required but not sufficient to explain the CEMP-no stars. The special nucleosynthesis induced by the progressive rotational mixing at work in the source star also appears to be needed to account for the abundances of CEMP-no stars (e.g. Sect.~\ref{secstat}).
%The source star models computed in this work show that the progressive mixing induced by rotation is also required.

\subsubsection{The mass of the source stars}

%One difference between the models of this work and the models of \cite{hirschi07} and \cite{heger10} is the initial mass of the possiblemodels. 
Models of \cite{hirschi07} and \cite{heger10} in Fig.~\ref{cnccmodszoom} suggest that the CEMP-no source stars could have initial masses of $50-100$~$M_{\odot}$ while this work suggest lower masses, about 20~$M_{\odot}$ (e.g. Fig.~\ref{histfin1}). \cite{heger10} investigated the origin of the CEMP stars HE~0107-5240 and HE~1327-2326 with their Pop~III massive models. In both cases, the best fits are given by their $10.5-25$~$M_{\odot}$ source star models. 
Similarly to \cite{umeda03}, an artificial mixing episode is included is their models at the time of the supernova in order to mix a little bit of iron in upper stellar layers. The iron is then expelled and the little iron content of HE~0107-5240 and HE~1327-2326 can be reproduced. 
\cite{placco16} have used the models of \cite{heger10} to investigate the origin of 12 stars with [Fe/H]~$<-4$. They found that the best fits are given by $12.6 - 41$~$M_{\odot}$ models (in both studies, the $^{12}$C/$^{13}$C ratio is not considered). 
Also, in these works, comparisons are made through the [X/H] ratios but no dilution with primordial ISM was considered. It likely means that the yields of the source stars models considered should be seen as upper limits since any dilution with primordial ISM will decrease the [X/H] ratios. Some caution is therefore required.

From Fig.~\ref{cnccmods} and \ref{cnccmodszoom}, we see that the $10-40$~$M_{\odot}$ of \cite{heger10} are not reproducing the bulk of observations. Fig.~\ref{cnccmodszoom} rather suggests that their models with $M>40$~$M_{\odot}$ may be better source stars. 
In the present section, since the comparison is made using only few abundances (C and N), it is not possible to conclude further. It suggests that tensions probably exist and that the preferred masses of the CEMP-no source stars may require further investigations.
%\textcolor{red}{In the end, it might indicate that $10-40$~$M_{\odot}$ are nevertheless the best CEMP source star candidates but with some additional ingredient needed to account for the C/N and $^{12}$C/$^{13}$C ratios}.

%It is not excluded that unknown internal processes inside CEMP stars are responsible for the low $^{12}$C/$^{13}$C. For unevolved stars, one has however to find a process occurring during the Main-Sequence. It might be interesting to 

%If such event did not occur, we would expect (at least some of) the observations to lie

%\textcolor{red}{Similar behavior was not observed in our models}

%\textcolor{red}{Talk about convection and uncertainties. All start from the encroaching in H07 model}

%Report CNO inj with figures! ??

\section{Summary}

This chapter has investigated the origin of the CEMP stars with [Fe/H] $<-3$ and not significantly enriched in s- and/or r-elements (CEMP-no stars). The main results are summarized below.

The high abundance scatter (especially for C, N, O, Na, Mg and Al) of these stars was suggested to originate from a material ejected by previous massive source stars that experienced various degree of rotational mixing. A part of the CEMP-no star sample is thought to have formed mainly with the H-burning shell of the massive source star.
%During the core He-burning phase of the source star, the rotational mixing induces exchanges of material between the He-burning core and the H-burning shell. 
%During the core He-burning phase of the source star, rotational mixing enriches the H-burning shell in He-burning products. 
During the core He-burning phase of rotating source stars, He-burning products are transported to the H-burning shell by the operation of the rotational mixing. Once into the H-shell, these products are further processed by H-burning and transported back to the He-burning core, etc. %enriches the H-burning shell in He-burning products. 
%The H-burning shell is enriched in He-burning products %In particular, He-burning products like $^{}$
%triggered rotation and nucleosynthesis leads to an overproduction of C, N, O, Na, Mg and Al. 
I developed a one-zone nucleosynthesis model that mimics the H-burning shell of a rotating massive star. The effect of rotation was modeled by progressively injecting He-burning products ($^{12}$C, $^{16}$O, $^{22}$Ne and $^{26}$Mg) in the burning zone. 
The ranges of abundances of the considered CEMP-no stars were found to be overall well reproduced by a material processed by H-burning at a temperature and density characteristic of $20-60$~$M_{\odot}$ source stars. To reproduce the observations, this material should be enriched, while burning, in $^{12}$C, $^{16}$O and occasionally $^{22}$Ne. This suggests that the CEMP-no source stars experienced mid to strong rotational mixing. 

%20~$M_{\odot}$ 
Source stars models with various initial masses, rotation rates and metallicities were computed. As the result of the rotation induced mixing during the core He-burning phase, the yields of elements from C to Al scale with initial rotation. In the fastest rotating model, the [X/H] ratios in the ejecta can be boosted by $\sim 2$ to $\sim 5$ dex compared to the non-rotating model. The mass cut and the dilution of the source star ejecta with ISM are important and weakly constrained parameters, that strongly affect the source star yields.
Overall, a global match between the abundances of CEMP-no stars and very low metallicity 20~$M_{\odot}$ source star models with various initial rotation rates can be found. The low $^{12}$C/$^{13}$C ratio (especially on unevolved CEMP-no stars) requires that no or little material processed by He-burning was expelled from the source star. In this case however, [C/N]~$\sim -2$, which does not account for the bulk of CEMP-no stars, that generally have higher [C/N] ratios.
The source star ejecta either have low $^{12}$C/$^{13}$C ratios together with low [C/N] ratios or high $^{12}$C/$^{13}$C ratios together with high [C/N] ratios. It cannot fit the bulk of CEMP-no stars.
 
A solution to improve the fit is to include a \textit{late mixing process} in the massive source star, operating in between the H-burning and the He-burning shell, about 200 yr before the end of the source star evolution. 
It may occur preferentially in $\sim 20$~$M_{\odot}$ source stars (both rotating and non-rotating).
This additional process was studied by computing new source star models with initial masses of 20, 32 and 60~$M_{\odot}$, without and with fast rotation, without and with the late mixing process (12 models). 
Using the yields of these 12 source star models, an automatic abundance fitting procedure was applied to the CEMP-no stars with [Fe/H] $<-3$. %The mass cut of the source star and dilution factor were let as free parameters.
I found that the best CEMP-no source stars are preferentially rotating 20~$M_{\odot}$ models that experienced the late mixing process, that ejected the layers above or just below the interface between the H-rich and He-rich region and whose ejecta underwent a modest dilution with the ISM.\\

Finally, to avoid confusion, I recall the three different kinds of mixing that may arise in the source star:

\begin{itemize}
\item The mixing induced by rotation that occurs progressively and everywhere, during the entire life of the source star.
\item The late mixing process (Sect.~\ref{seclate}) arising between the H-burning shell and He-burning shell, shortly before the end of the evolution\footnote{It may also occur earlier in the evolution, between the H-burning shell and He-burning core, cf. Sect.~\ref{secother}.}.
\item The mixing in the sense of \cite{umeda02} that occurs between two limiting shells, at the time of the supernova.
\end{itemize}

%\chapter{Massive rotators and CEMP-s stars}
\chapter{The s-process in CEMP source stars}
\label{cemps}

%\textcolor{red}{change maybe ... not only high Z but also low Z for HE~1327... Mainly on higher Z but not only}

The previous chapter focused on light elements, from C to Si. These elements are of particular interest for studying the origin of the most iron poor stars (with [Fe/H] $<-3$) that are mostly CEMP-no stars. 
%because they are often strongly enriched in such elements while, by definition, they are generally little enriched in heavy elements.
The reason is that these stars are often strongly enriched in such elements while, by definition, they are generally little enriched in heavier elements.
%The origin of such stars may consequently be inferred by considering only light elements.

Many CEMP stars show overabundances in s-elements, especially at higher metallicities ($ -3 \lesssim$ [Fe/H] $\lesssim -2$), but also sometimes at very low metallicity (e.g. HE~1327-2326 with [Fe/H] $<-5$ and [Sr/Fe] $=1.08$).
The weak s-process is expected to occur in massive stars (provided the metallicity is not zero). As mentioned in Sect.~\ref{sproctheo} (also Fig.~\ref{schemadiff}), the effect of rotation in massive stars can alter the production of s-elements. 
At very low/zero metallicity, the amount of seed (especially iron) is very low so that little s-elements can be produced. 
At higher metallicities, there are more seeds, hence possibly more s-elements. 

%The effect of rotation in massive stars can alter the production of s-elements (cf. Sect.~\ref{sproctheo}, also Fig.~\ref{schemadiff}). 
%In this chapter, we also consider the s-process in massive stars. The size of the nuclear network now comprises 737 species, from $^{1}$H to $^{212}$Po instead of 31 (cf. Sect~\ref{nucnetw}). 

The main purpose of this chapter is to extend the study to heavier elements that can be formed in the source stars through the s-process. 
The s-process is studied in a new grid of massive source stars at [Fe/H] $\sim -2$ together within lower metallicity source star models. %Lower metallicity models including s-process and in lower metallicity is also investigated
%I investigate what would be the nucleosynthetic signature of massive rotating stars including rotation and s-process. 
Then, I investigate whether the nucleosynthetic signatures of these massive stars can be recognized in observed metal-poor stars. % with [Fe/H] $\sim -2$.
A very recent paper (in Sect.~\ref{newgridsproc}) and a letter (in Sect.~\ref{origsinglecemp}) are directly included in this chapter. Additional results and discussions are included, particularly in Sect.~\ref{sproclower}.

\section{The weak s-process in a box}

Before investigating complete stellar models, the s-process in a single burning zone is discussed. 
Using a similar code as in Sect.~\ref{secbox} but that now mimics the He-burning core of massive stars (this is mostly where the s-process operates, cf. Sect.~\ref{sproctheo}), the s-process in massive stars can be studied qualitatively. 
The single burning zone considered burns at $T=300$ MK and $\rho = 500$ g~cm$^{-3}$, as characteristic values of the He-burning core of massive stars. The initial mass fractions in the zone are $\sim 0.4$ for $^{4}$He, $\sim 0.3$ for $^{16}$O, $\sim 0.3$ for $^{12}$C and $10^{-7}$ for $^{56}$Fe (in mass fraction). The initial mass fraction of $^{22}$Ne is varied from $10^{-4}$ to 0.05. The calculation is stopped when all the $^{4}$He is burnt. 
The final chemical patterns in the burning zone are shown in Fig.~\ref{boxsproc_Ne}. 
%With such initial conditions, the final composition in the burning box is shown by the solid pattern in Fig.~\ref{boxsproc_Fe}. The s-process has operated, starting from $^{56}$Fe, and has synthesized heavier elements. 
%When decreasing the initial $^{56}$Fe to $10^{-3}$~\% (dashed line), heavier s-elements are produced at the expense of lighter s-elements as a result of the higher initial source ($^{22}$Ne) over seed (${56}$Fe) ratio \textcolor{red}{REF}.

When increasing the initial $^{22}$Ne, heavier s-elements are produced at the expense of lighter s-elements. % as a result of the higher initial source ($^{22}$Ne) over seed (${56}$Fe) ratio \textcolor{red}{REF}.
%It can be understand by the fact that during the burning, heavy s-elements  isotopes heavier isotopes  seed 
%less seed is available and the final overall amount of s-elements decreases (Fig.~\ref{boxsproc_Fe}).
%However, If decreasing again the initial  $^{56}$Fe down to $10^{-5}$ and then $10^{-8}$~\%, although the source over seed ratio becomes very high, the entire s-process pattern is shifted down anyway.
%As discussed in \cite{prantzos90}, as the metallicity gets smaller, the abundances of source and seed, which are secondary, decrease. On the other hand, the abundances of primary neutron poisons like $^{16}$O remain the same. Consequently, the effect of primary neutron poisons becomes stronger at lower metallicities. This implies that the production of s-elements does not decrease linearly with metallicity.
%If setting $^{56}$Fe to $10^{-5}$~\% and changing $^{22}$Ne instead, the total amount of s-element is not increased but the pattern is shifted toward heavier elements (Fig.~\ref{boxsproc_Ne}). 
If little $^{22}$Ne is available, the neutron flux is small and only light s-elements are produced. When the initial $^{22}$Ne mass fraction is 0.05, the pattern becomes flatter. It shows the effect of changing the source ($^{22}$Ne) over seed ($^{56}$Fe) ratio. If this ratio increases (provided some reasonable amount of seed is present) the hs (heavy s-elements) over ls (light s-elements) ratio increases.

Fig.~\ref{boxsproc_O} illustrates the poisoning effect of $^{16}$O by changing the initial $^{16}$O mass fraction. For higher initial $^{16}$O mass fractions, the production of s-elements is reduced. This is because if $^{16}$O is more abundant, the $^{16}$O($n,\gamma$)$^{17}$O($\alpha,\gamma$)$^{21}$Ne chain is boosted, so that less neutrons are available for heavy elements.

 %  \begin{figure*}[t]
 %  \centering
  %    %\includegraphics[scale=0.55, trim = 2.5cm 12cm 2.5cm 0cm]{boxsproc_Fe.pdf}
  %    \includegraphics[scale=0.55, trim = 0cm 0cm 0cm 0cm]{boxsproc_Fe.pdf}
 %  \caption[a]{Final mass fraction $X$ (normalized to solar) of heavy elements in a single burning zone at $T=300$ MK and $\rho = 500$ g~cm$^{-3}$. The initial mass fraction of $^{56}$Fe is varied from $10^{-8}$~\% (dashed pattern) to 0.01~\% (solid line).}
%\label{boxsproc_Fe}
 %   \end{figure*}

   \begin{figure*}[t]
   \centering
      \includegraphics[scale=0.55, trim = 0cm 0cm 0cm 0cm]{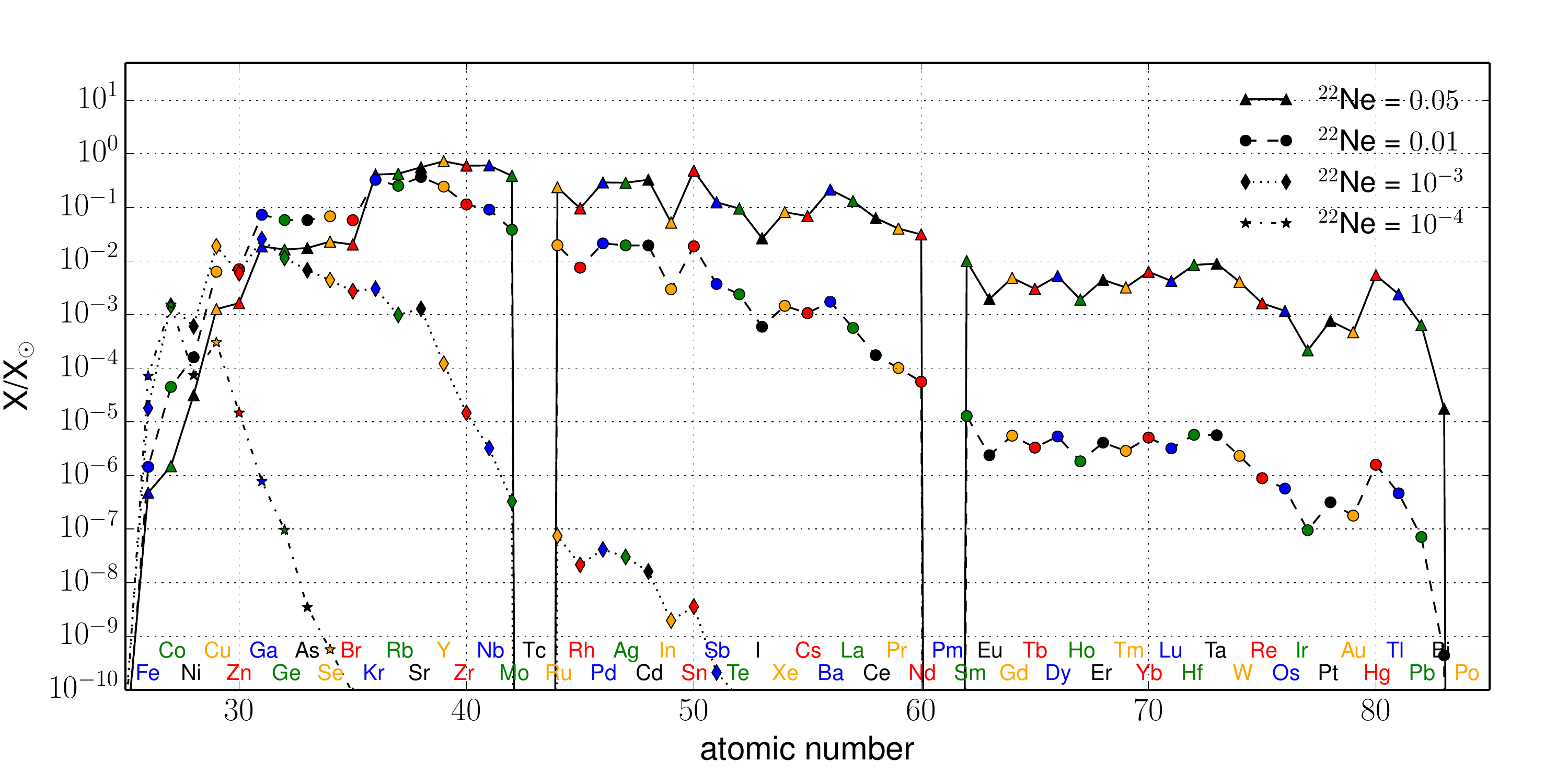}
   \caption[Composition of a one-zone model with various $^{22}$Ne content]{Final mass fraction $X$ (normalized to solar) of heavy elements in a single burning zone at $T=300$ MK and $\rho = 500$ g~cm$^{-3}$. The initial mass fraction of $^{56}$Fe is $10^{-7}$. The initial mass fraction of $^{22}$Ne is varied from $10^{-4}$ to 0.05.}
\label{boxsproc_Ne}
    \end{figure*}

   \begin{figure*}[t]
   \centering
      \includegraphics[scale=0.55, trim = 0cm 0cm 0cm 0cm]{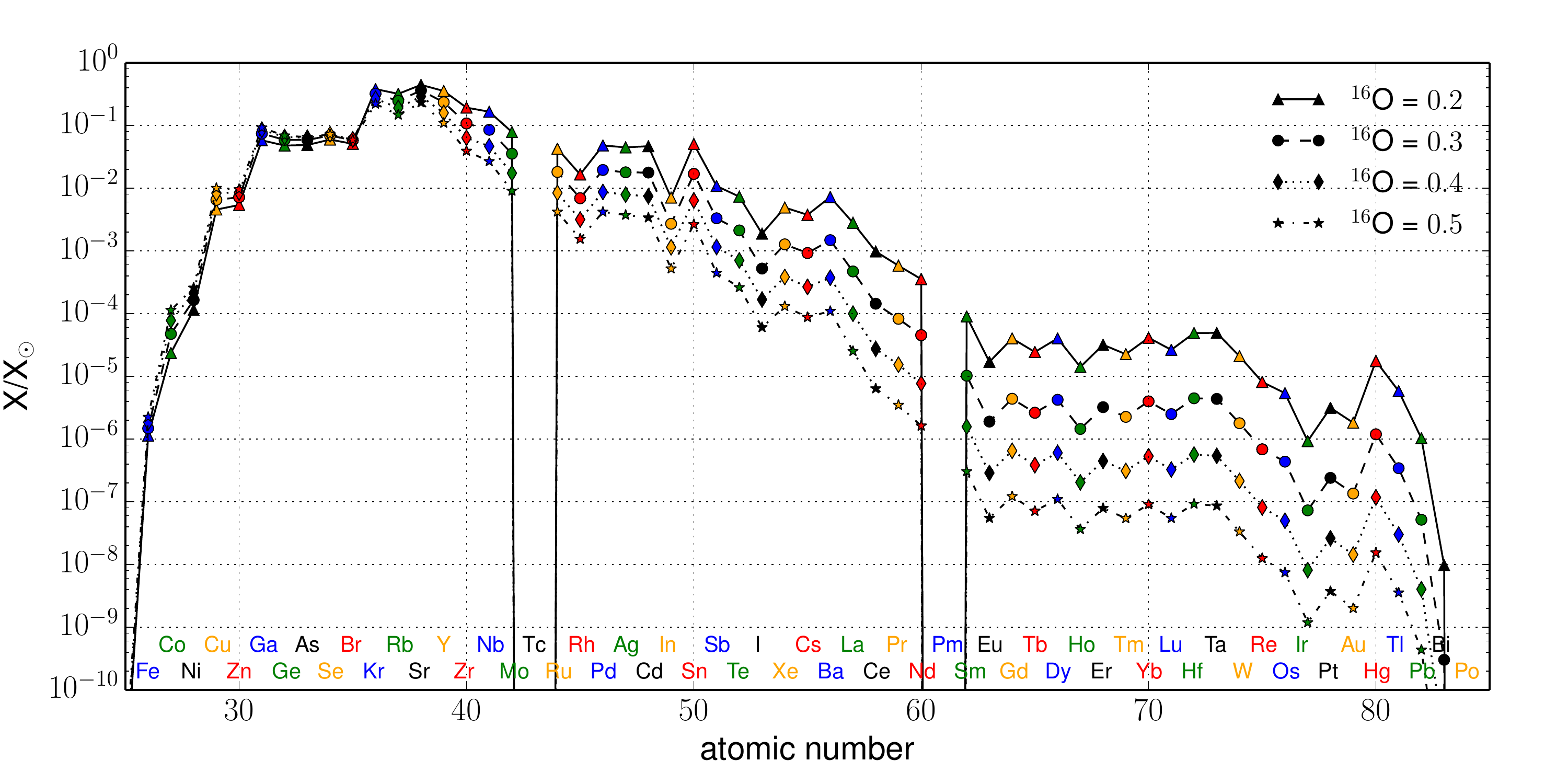}
   \caption[Composition of a one-zone model with various $^{16}$O content]{Same as Fig.~\ref{boxsproc_Ne} but when varying the initial mass fraction of $^{16}$O from 0.2 (solid pattern) to 0.5 (dot-dashed pattern).}
\label{boxsproc_O}
    \end{figure*}

\section{A new grid of massive stars with rotation and s-process}\label{newgridsproc}

%As mentioned in Sect.~\ref{sprocmassive}, the standard view of the weak s-process in massive stars is modified if including rotation.
%%%%Rotational mixing indeed produces more $^{22}$Ne during the core He-burning phase, hence provides more neutrons 
%(cf. Fig.~\ref{schemadiff} and discussion in Sect.~\ref{secback}). 
%At very low or zero metallicity, the amount of seed (iron) is very low so that little s-elements can be produced in any case. 
%At higher metallicities, there is more seed, hence possibly more s-elements. 
Grids of massive stellar models including rotation and full s-process network are needed in order to investigate the role of such stars in the chemical enrichment of the Universe. 
To date, only \cite{frischknecht16} and very recently \cite{limongi18} have provided such grids. An important part of my work was to compute a new grid of massive rotating star with full s-process, so as to extend the study of \cite{frischknecht16}. The final goal being to gain knowledge on the nature of such massive stars by comparing their yields to the abundances of observed low-mass metal-poor stars enriched in s-elements.

%In non-rotating models, the s-process efficiency peaks at $Z/Z_{\odot} = 10^{-1} - 10^{-2}$ (equivalent to $Z \sim 1.4~10^{-3} - 1.4~10^{-4}$)
In \cite{choplin18} we computed a new grid of massive stars with initial masses between 10 and 150~$M_{\odot}$, at a metallicity $Z=0.001$ in mass fraction (corresponding to [Fe/H] $=-1.8$). The models were computed either without rotation or with $\upsilon_{\rm ini}/\upsilon_{\rm crit} = 0.4$. We considered only one metallicity but significantly extended the range of mass compared to the study of \cite{frischknecht16}, that focused on $15-40$~$M_{\odot}$ models. We also investigated the impact of a faster initial rotation and a different $^{17}$O($\alpha,\gamma$)$^{21}$Ne reaction rate. %(cf. Sect.~\ref{sproctheo}).
Most of the physical ingredients of this grid are the same than for the models discussed previously. 
The main change is the size of the nuclear network that now comprises 737 species, from $^{1}$H to $^{212}$Po instead of 31 species (cf. Sect~\ref{nucnetw}).

The paper describing the models is directly included in the thesis (in the present section). It was very recently accepted in \aap~and I have therefore almost no discussion to add yet.
Before the paper itself, I discuss several aspects that were not published: several additional physical ingredients and some obstacles I had to overcome. In Appendix~\ref{sprocparam}, additional parameters, allowing to quantify the efficiency of the s-process, are defined and tabulated for the new grid of models. The table was not published in the paper but it may be useful to have these quantities for future comparisons with other models.
%with the aim to compare the yields with the abundances of observed low-mass metal-poor stars enriched in s-elements.

%\subsection{Physical ingredients}
%\subsection{Physics of the models}

%I highlight below two other differences.

%\subsubsection{Initial composition}

%The new version of the code includes slight changes regarding the initial abundances of light elements. The initial solar Li abundance is now $\log \epsilon_{\rm Li, \odot} = 3.25$ instead of 1.05. Both values are from \cite{asplund05} but 3.25 is derived from meteorites and 1.05 from the solar photosphere. The surface of the Sun has likely experienced Li depletion during its life while the meteoritic Li reflects more the initial value. Also, the primordial $^{6}$Li, $^{7}$Li, $^{9}$Be, $^{10}$B and $^{11}$B abundances \citep{asplund06, sbordone10, coc12} are taken as a threshold. It means that zero metallicity models have non-zero Li, Be and B initial abundances. 

\subsection{Nuclear network and reaction rates}\label{netandrates}

%The nucleosynthetic code BasNet \citep[Basel network, ][]{arnett85, thielemann85} is coupled to the Geneva stellar evolution code.

As shown in Table~2 of the paper below, 8 important nuclear reaction rates were updated compared to the study of \cite{frischknecht16}. 
Originally, the reaction rate format in stellar evolution codes is the following: one table for each reaction rate with two columns, one for the temperature, one for the associated rate. Then, interpolations are done in these tables to evaluate the rates at the desired temperatures.
With large networks, it is convenient to use analytical reaction rates such as in the \textsc{reaclib} format \citep{cyburt10} in order to save computational time. Such a format allows to express a reaction rate with $n$ sets of 7 parameters $a_0,..,a_6$. For 2-body reactions, it yields
\begin{equation}
\langle \sigma v \rangle N_A =  \sum_{k=1}^n e^{[a_0(k) + a_1(k) T_9^{-1} + a_2(k) T_9^{-1/3} + a_3(k) T_9^{1/3} + a_4(k) T_9 + a_5(k) T_9^{5/3} + a_6(k) \log (T_9)]}
\label{fitexp}
\end{equation}
with $T_9 = 10^{-9}~T$ and $N_A$ the Avogadro number. For 1- and 3-body reactions, the left side of Eq.~\ref{fitexp} is respectively equal to $\langle \sigma v \rangle N_A^2$ and $\lambda$, where $\lambda$ represents the photodisintegration or $\beta$-decay rate. If just the standard tabulated rate exists, one has to find the $a$ parameters that allow to fit correctly the rate.  A leastsquare method is used to estimate these parameters from a temperature-rate table. $n>1$ is often required to reach a reasonable accuracy. Finding the $a$ parameters can sometime be challenging, mainly because of the large range of temperature to be covered (several orders of magnitude, it depends on the reactions). Once the $a$ parameters are determined for the forward reaction, the $a_{rev}$ parameters for the reverse reaction can be easily calculated \citep{rauscher00}. 

\cite{best13} provided, in the tabular form, new rate measurements for the $^{17}$O($\alpha,n$)$^{20}$Ne and the $^{17}$O($\alpha,\gamma$)$^{21}$Ne reactions, that are of crucial importance for the s-process in massive stars. I used the fitting method described in \citet[][also nucastro.org]{rauscher00} to convert these tabulated rates into the \textsc{reaclib} format. For the two fits, I found that $n=3$ provides satisfactory results: the final deviation is less than 5~\% (Fig.~\ref{ne2021fit}, only one of the two fit is shown).

   \begin{figure}[t]
   \centering
      \includegraphics[scale=0.84, trim=0cm 0cm 1cm 0cm]{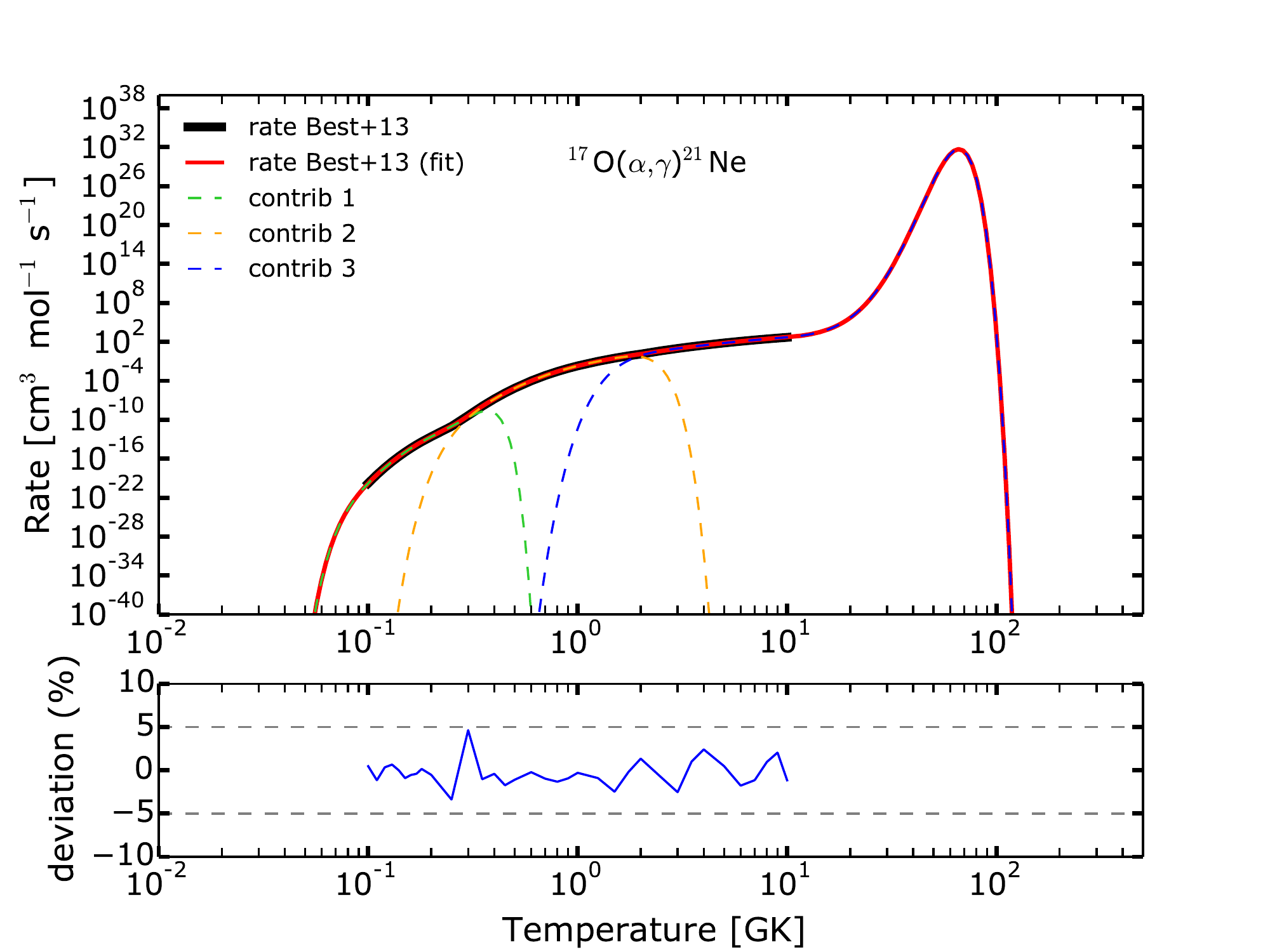}

\caption[Rate of the $^{17}$O($\alpha,\gamma$)$^{21}$Ne reaction]{Rate of the $^{17}$O($\alpha,\gamma$)$^{21}$Ne reaction. %(upper panel) and $^{17}$O($\alpha,n$)$^{20}$Ne (lower panel) reactions. 
The thick black line corresponds to the recommended values tabulated in \cite{best13}. Dashed lines show the different contributions, which were fitted according to the method described in \citet[][also nucastro.org]{rauscher00}. Three different contributions ($n=3$, see Eq.~\ref{fitexp}) were found to provide an acceptable fit. The thin red line shows the sum of the 3 contributions. The deviation from the original tabulated rate \citep{best13} is shown in the bottom panel.}
            
  \label{ne2021fit}
    \end{figure}

   \begin{figure}[t]
      \centering
   \begin{minipage}[c]{.49\linewidth}
      \includegraphics[scale=0.25, trim=2cm 0cm 0cm 0cm,clip]{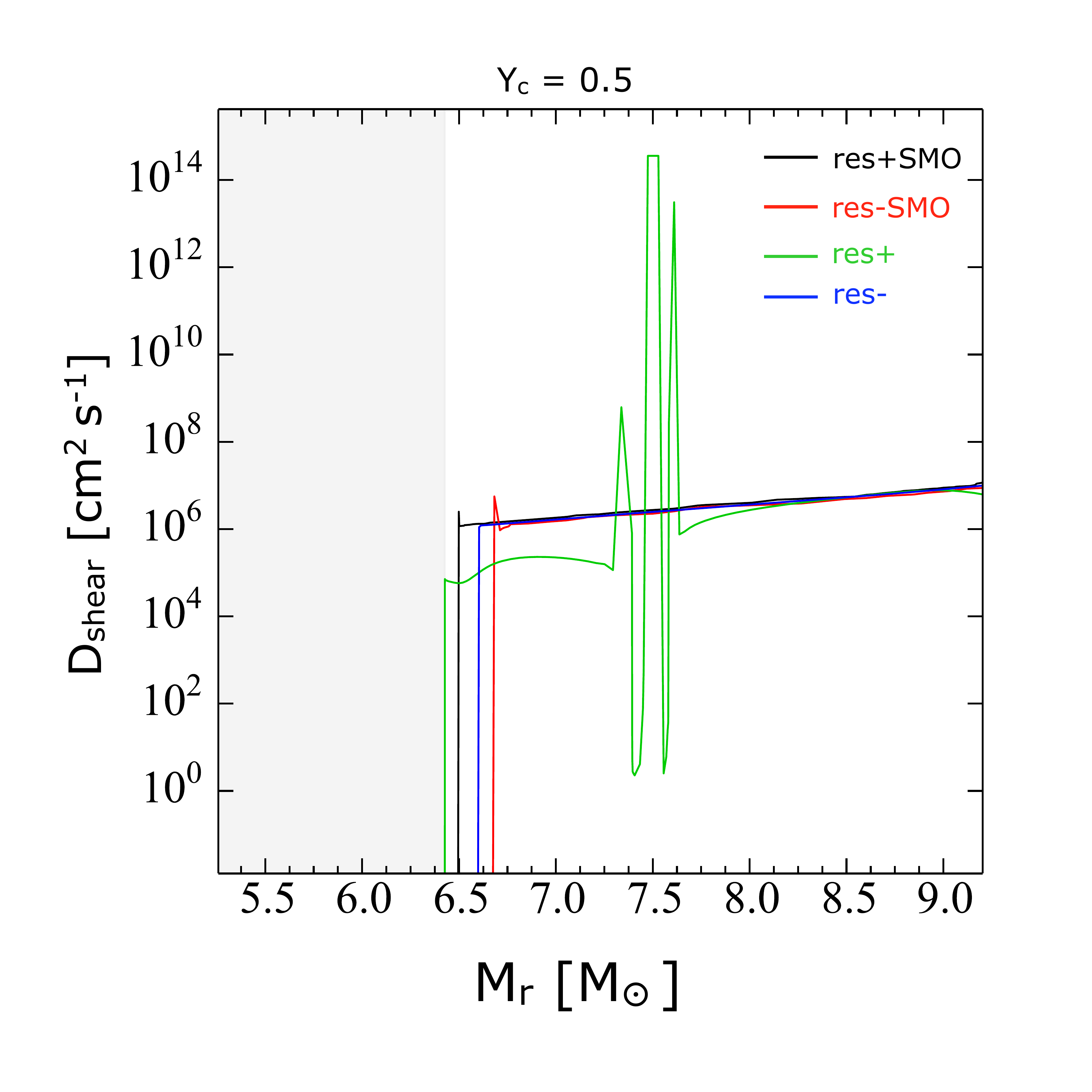}
   \end{minipage}
   \begin{minipage}[c]{.49\linewidth}
      \includegraphics[scale=0.25, trim=1cm 0cm 0cm 0cm,clip]{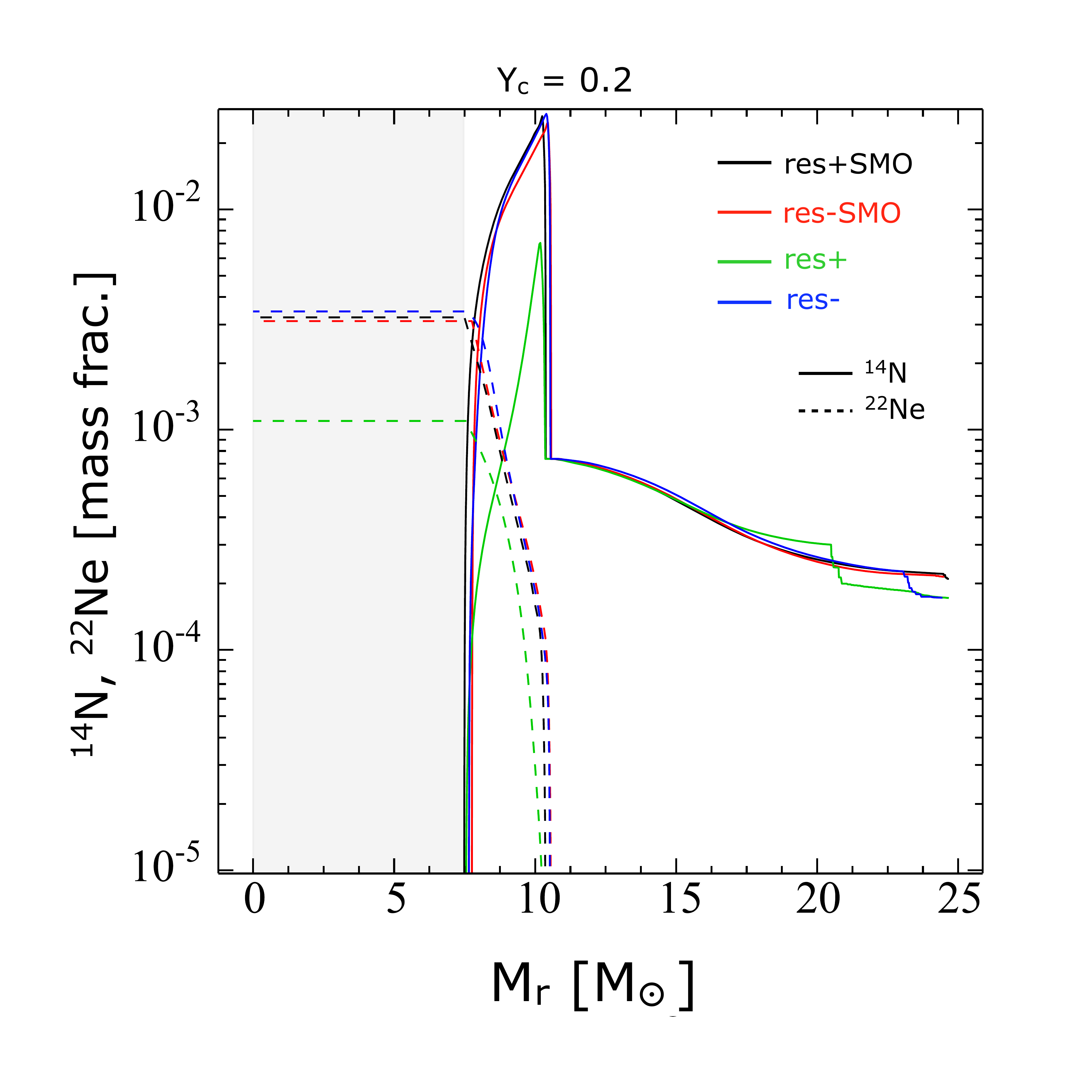}
   \end{minipage}

\caption[$D_{\rm shear}$, $^{14}$N and $^{22}$Ne profiles in 25~$M_{\odot}$ models]{\textit{Left panel:} $D_{\rm shear}$ coefficient between the He-core and H-shell. \textit{Right panel:} internal mass fraction of $^{14}$N and $^{22}$Ne. Models shown are rotating 25~$M_{\odot}$ at $Z = 10^{-3}$, during the core helium burning phase (the central mass fraction of $^{4}$He is 0.5 on the left panel and 0.2 on the right panel). Shaded area show convective zones. \textit{res+} models (black and green) are computed with high resolution (about twice more shells than \textit{res-} models). \textit{SMO} models (black and red) include the smoothing technique (see text and Appendix~\ref{oscillapp}).}
            
  \label{Dshlis}
    \end{figure}

   \begin{figure}[h!]
   \centering
      \includegraphics[scale=0.62, trim=0cm 0cm 0cm 0cm,clip]{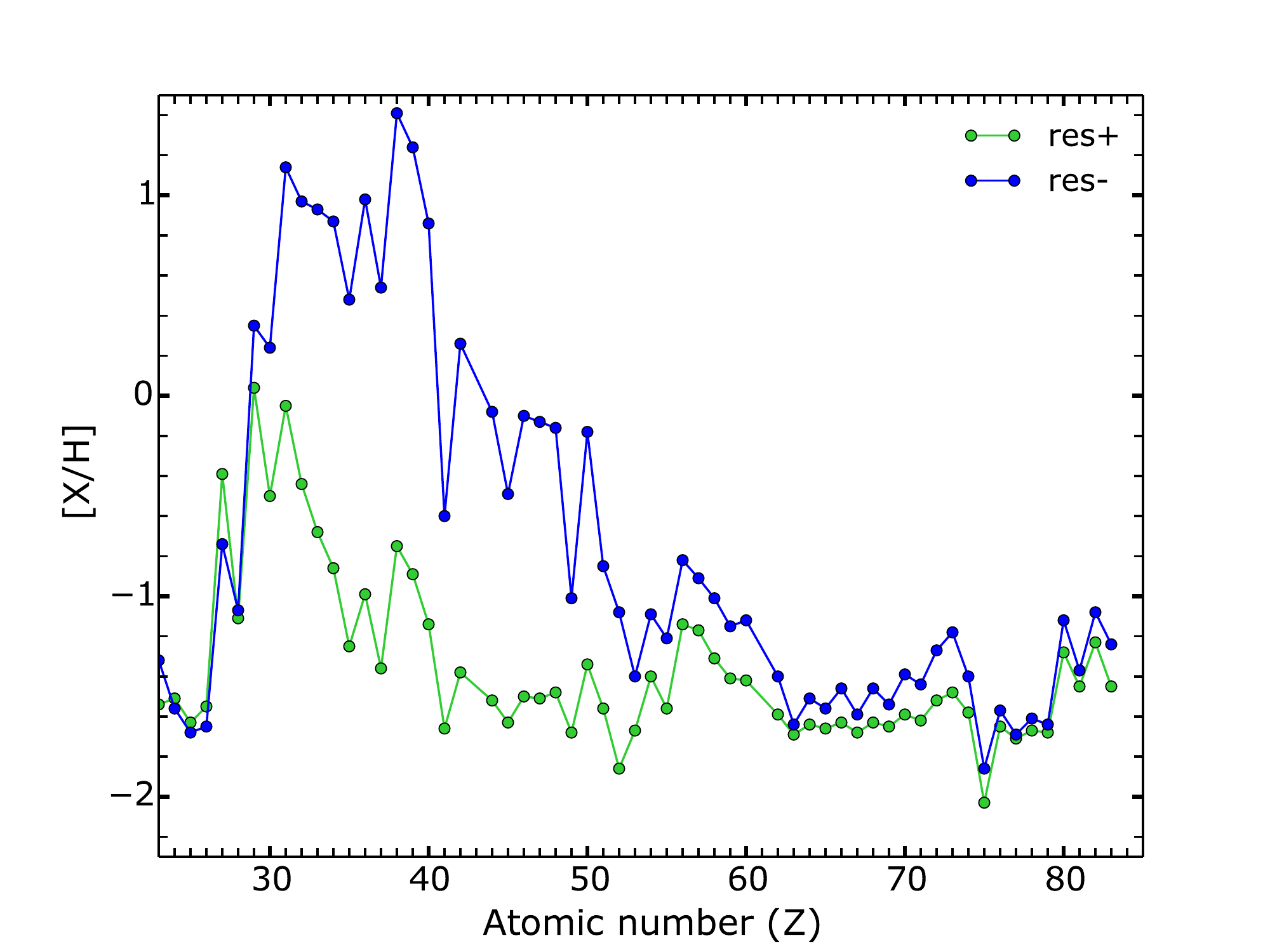}

\caption[Composition of the ejecta of models computed with low and high resolution]{[X/H] ratios in the ejecta of the green (high resolution) and blue (low resolution) models of Fig.~\ref{Dshlis}.}
            
  \label{sproclis}
    \end{figure}

%   \begin{figure*}
 %  \centering
 %  \begin{minipage}[c]{.49\linewidth}
 %     \includegraphics[scale=0.39]{/Users/Arthur/Obs/figs/figs_rate/ne2021.pdf}
 %  \end{minipage}
%   \begin{minipage}[c]{.49\linewidth}
%       \includegraphics[scale=0.39]{/Users/Arthur/Obs/figs/figs_rate/mg2526.pdf}
 %  \end{minipage}
 %  \caption[]{Ratios of the ($\alpha,n$) channel over the ($\alpha,\gamma$) channel. \textit{Left panel}: reactions $^{17}$O($\alpha,n$)$^{20}$Ne and $^{17}$O($\alpha,\gamma$)$^{21}$Ne. \textit{Right panel}: reactions $^{22}$Ne($\alpha,n$)$^{25}$Mg and $^{22}$Ne($\alpha,\gamma$)$^{26}$Ne. The black lines show the rates used in F16, the red lines show the rates used in this work.}
%\label{anag}
  %  \end{figure*}

\subsection{The $\mu$ and $\Omega$ profiles}

When first trying to reproduce the results of \cite{frischknecht16}, I found significantly different results: many s-elements were underproduced by a factor of 100 in the new models.
% This was due to the lower amount of extra $^{14}$N and then $^{22}$Ne (neutron source) synthesized by rotation during core He-burning. 
It was caused by strong oscillations of the $D_{\rm shear}$ coefficient in between the H- and He-burning zones, during the core He-burning stage. The green profile in the left panel of Fig.~\ref{Dshlis} shows such oscillations. The low values of $D_{\rm shear}$ at $M_r \simeq 7.5$ strongly reduces the exchanges of material between the He-core and the H-shell, leading to a smaller synthesis of primary $^{14}$N and $^{22}$Ne (Fig.~\ref{Dshlis}, right panel, green line). Such $D_{\rm shear}$ oscillations are likely not physical since their shape change or disappear when changing the model resolution. An example is shown by the blue profiles in Fig.~\ref{Dshlis}: the resolution of this model (\textit{res-} model in Fig.~\ref{Dshlis}) is reduced compared to the green model (\textit{res+} model) but it is still enough to see the eventual $D_{\rm shear}$ oscillations. In this case, the $D_{\rm shear}$ does not oscillate so that the mixing is not cut and more $^{14}$N and $^{22}$Ne are synthesized compared to the green model (Fig.~\ref{Dshlis}, right panel). While the changes induced by the presence or the absence of oscillations are not huge for $^{14}$N and $^{22}$Ne (factor of about 5), they are much more significant for s-elements (factor of about 100 at maximum, as shown in Fig.~\ref{sproclis}). Models of \cite{frischknecht16} are similar to the blue lines in Fig.~\ref{Dshlis} and \ref{sproclis} while new models to the green lines.

To improve the stability of the code and find similar results when changing the mesh number, I improved the way of calculating $D_{\rm shear}$ (Eq.~\ref{dshtz97}), by trying to better evaluate the derivative of $\mu$ and $\Omega$, that appeared to be responsible for the oscillation of the $D_{\rm shear}$ coefficient. The oscillation problem is solved if properly smoothing the $\mu$ and $\Omega$ profiles when calculating their derivatives (see Appendix~\ref{oscillapp} for more details). If applied, this technique gives similar results when changing the mesh number (see Fig.~\ref{Dshlis}, black and red profiles). It also gives similar s-process yields compared to the models of \cite{frischknecht16}.

%The results of \cite{frischknecht16} did not include smoothing and were computed at a lower resolution (blue line in Fig.~\ref{Dshlis} and \ref{sproclis}). Including the smoothing gives similar results, whatever the resolution.%In this case, oscillations are not present. , we adopted a similar resolution in order to save computational time. resolution case, as in \cite{frischknecht16},
%were reproduced are reproduced if the $D_{\rm shear}$ do not oscillate.

%\subsection{Grid of models}
%\subsection{Paper}\label{papersproc}

\newpage
\label{psproc}
%\includepdf[linkname= pdfsproc,scale=0.9,offset=0 -0.5cm]{aa33283-18_forth.pdf}
\includepdf[linkname= pdfsproc,scale=0.9,offset=0 -0.5cm]{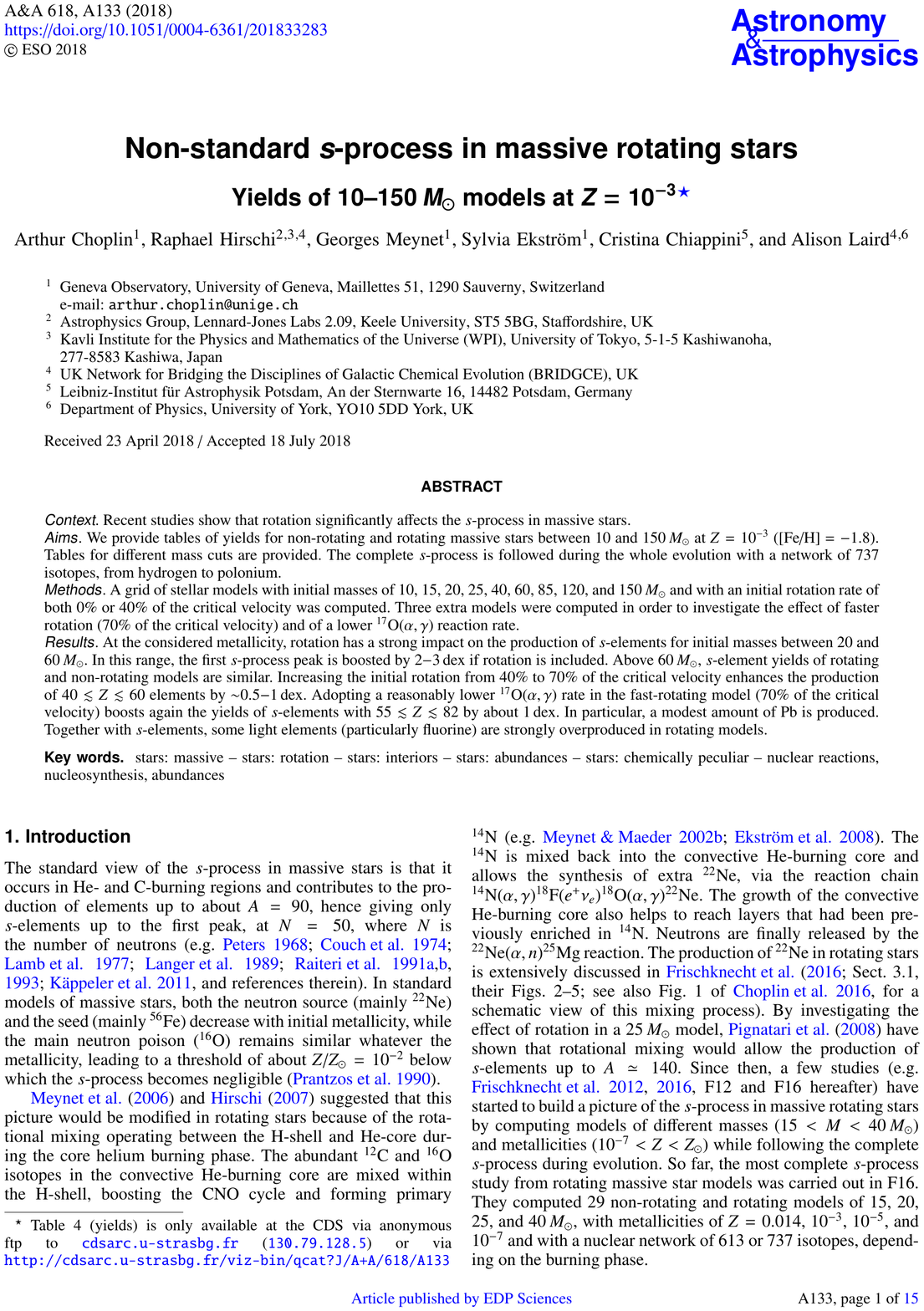}
\newpage

%== INTRO ======================================================================
\section{The origin of the single CEMP-s stars}\label{origsinglecemp}

As discussed in Sect.~\ref{origcemp}, CEMP-s stars are generally explained with the AGB binary scenario. This scenario predicts that the CEMP-s should have a white dwarf companion (cf. Fig.~\ref{cempsform}).
The CEMP-s stars that appear to be single \citep{hansen16a} may challenge the AGB scenario.

\subsection{Massive source stars}

In \citet[][see letter in the next section]{choplin17letter}, we have investigated the origin of the four apparently single CEMP-s stars in light of the new grid of massive stars including rotation and s-process presented in the previous paper. 
The abundances of the apparently single CEMP-s stars are reported in Table~\ref{tableCEMPs}. 
We found that 3 out of the 4 CEMP-s stars can be well reproduced by the yields of the fast rotating ($\upsilon_{\rm ini}/\upsilon_{\rm crit} = 0.7$) 25~$M_{\odot}$ model that was computed with the rate of the $^{17}$O($\alpha,\gamma$) reaction divided by 10. The dilution factors $D$ are between 4 and 20 and in all the cases, the mass cut $M_{\rm cut} = M_{\rm CO}$. In general, the other models of the grid do not provide enough s-elements. If deep layers are expelled some models may nevertheless provide enough s-elements but in this case, Na, Mg and Al are overestimated by several order of magnitudes compared to the observations. 
These results show that some s-rich metal-poor ([Fe/H] $\sim -2$) halo stars may have formed with a material that was enriched by the ejecta of a previous fast rotating massive source star. It gives support to the spinstar scenario at higher metallicities (compared to the lower metallicities considered in Chapter~\ref{cempno}). 

The late mixing process introduced in Sect.~\ref{seclate} is not considered in these new CEMP source star models. This process will mostly affect the C and N abundances. It may nevertheless also synthesize heavier elements through neutron captures \citep[][cf. also Sect.~\ref{secother}]{clarkson18}. Including the late mixing process in the models with an extended network is among the possible next steps of this work.
However, it is not excluded that the late mixing process occurs only at very low metallicity (maybe [Fe/H] $\lesssim -4$), and not in these new source star models, with [Fe/H]~$=-1.8$. The reason is that at very low metallicity, source stars are more compact so that the H- and He-burning shells are closer from each other. It might imply an easier interaction of these two shells. This is however speculative and deserves more investigation.

%As discussed in the previous chapter, this process will mostly affect the abundance of C and N elements. It can also affect 
%One qualitative 

%the nucleosynthetic signature of fast rotating massive star  Since only the 25~$M_{\odot}$ model was computed with fast rotation, 

%\textcolor{red}{schema agb massive star?}

\begin{table*}
\scriptsize{
\caption[Apparently single CEMP-s stars]{Apparently single CEMP-s stars \citep[abundances taken from the SAGA database,][]{suda08}. \label{tableCEMPs}} 
\begin{center}
\resizebox{13.5cm}{!} {
%\begin{tabular}{llllllllllllll}
\begin{tabular}{lrrrr}
\hline
\hline
 Star        & HE~0206-1916 & HE~1045+0226 & HE~2330-0555  & CS~30301-015	\\
\hline
$[$Fe/H] 			& -2.09 		&-2.20		& -2.78 		& -2.64		\\
$T_{\rm eff}$ 		& 5200 		& 5077 (100) 	& 4900 		& 4750		\\
$\log g$ 			& 2.7 		& 2.2 (0.25) 	& 1.7 		& 0.8			\\
\hline
%$^{12}$C/$^{13}$C 	& 15 (5) 		& -			& -			& 6 (2)		\\
$[$C/Fe] 			& 2.11 (0.19) 	& 1.18 		& 2.09 (0.19)	& 1.72 (0.28)	\\
$[$N/Fe] 			& 1.61 (0.33) 	& -			& 0.99 (0.42)	& 1.84		\\
$[$Na/Fe] 			& 0.72 (0.14) 	& 0.96 (0.05)	& 0.95 (0.12)	& -			\\
$[$Mg/Fe] 		& 0.50 (0.15) 	& 0.27 (0.09)	& 0.64 (0.17)	& 0.84 (0.22)	\\
$[$Al/Fe] 			& - 			& -			& -			& 0.17		\\
$[$Si/Fe] 			& -			& 0.45 (0.17)	& -			& -			\\
$[$K/Fe] 			& -			& 0.51 		& -			& -			\\
$[$Ca/Fe] 			& 0.13 (0.13) 	& 0.22 (0.16)	& 0.44 (0.19)	& 0.79 (0.16)	\\
$[$Sc/Fe] 			& -			& 0.09 (0.05)	& 0.20 (0.08)	& -			\\
$[$Ti I/Fe] 		& 0.41 (0.14) 	& 0.34 (0.1)	& 0.31 (0.11)	& 0.30 (0.13)	\\
$[$Ti II/Fe] 		& 0.59 (0.19) 	& 0.43 (0.13)	& 0.38 (0.23)	& 0.45 (0.24)	\\
$[$Cr/Fe] 			& 0.05 (0.13) 	& -0.05 (0.01)	& -0.05 (0.10)	& -			\\
$[$Mn/Fe] 		& -			& -0.23 (0.28)	& -			& -			\\
$[$Ni/Fe] 			& -			& 0.17 (0.16)	& -			& -			\\
$[$Zn/Fe] 			& -			& -			& -			& 0.37		\\
$[$Sr/Fe] 			& -			& -			& -			& 0.37 (0.23)	\\
$[$Y/Fe] 			& -			& 1.29 (0.25)	& -			& 0.31 (0.20)	\\
$[$Zr/Fe] 			& -			& 1.52 		& -			& -			\\
$[$Ba/Fe] 			& 2.01 (0.16) 	& 1.24 (0.1)	& 1.25 (0.25)	& 1.49 (0.16)	\\
$[$La/Fe] 			& -			& 0.92 (0.17)	& -			& 0.96 (0.25)	\\
$[$Ce/Fe] 			& -			& -			& -			& 1.21 (0.15)	\\
$[$Pr/Fe] 			& -			& 1.14	 	& -			& -			\\
$[$Nd/Fe] 			& -			& 0.85 (0.32)	& -			& 0.80 (0.17)	\\
$[$Sm/Fe] 		& -			& -			& -			& 0.87 (0.20)	\\
$[$Eu/Fe] 			& -			& 0.27		& -			& 0.22 (0.18)	\\
$[$Dy/Fe] 			& -			& -			& -			& 0.64 (0.20)	\\
$[$Pb/Fe] 			& -			& -			& -			& 1.70 (0.24)	\\
\hline
Source 			& 1			& 2			& 1			& 3, 4		\\
\hline
\end{tabular}
}
\end{center}

\textbf{References}. 1 - \cite{aoki07}; 2 - \cite{cohen13}; 3 - \cite{aoki02c}; 4 - \cite{aoki02b}
	
}
\end{table*}

\newpage
\label{pcemps}
\includepdf[linkname= pdfletter,scale=0.9,offset=0 -0.5cm]{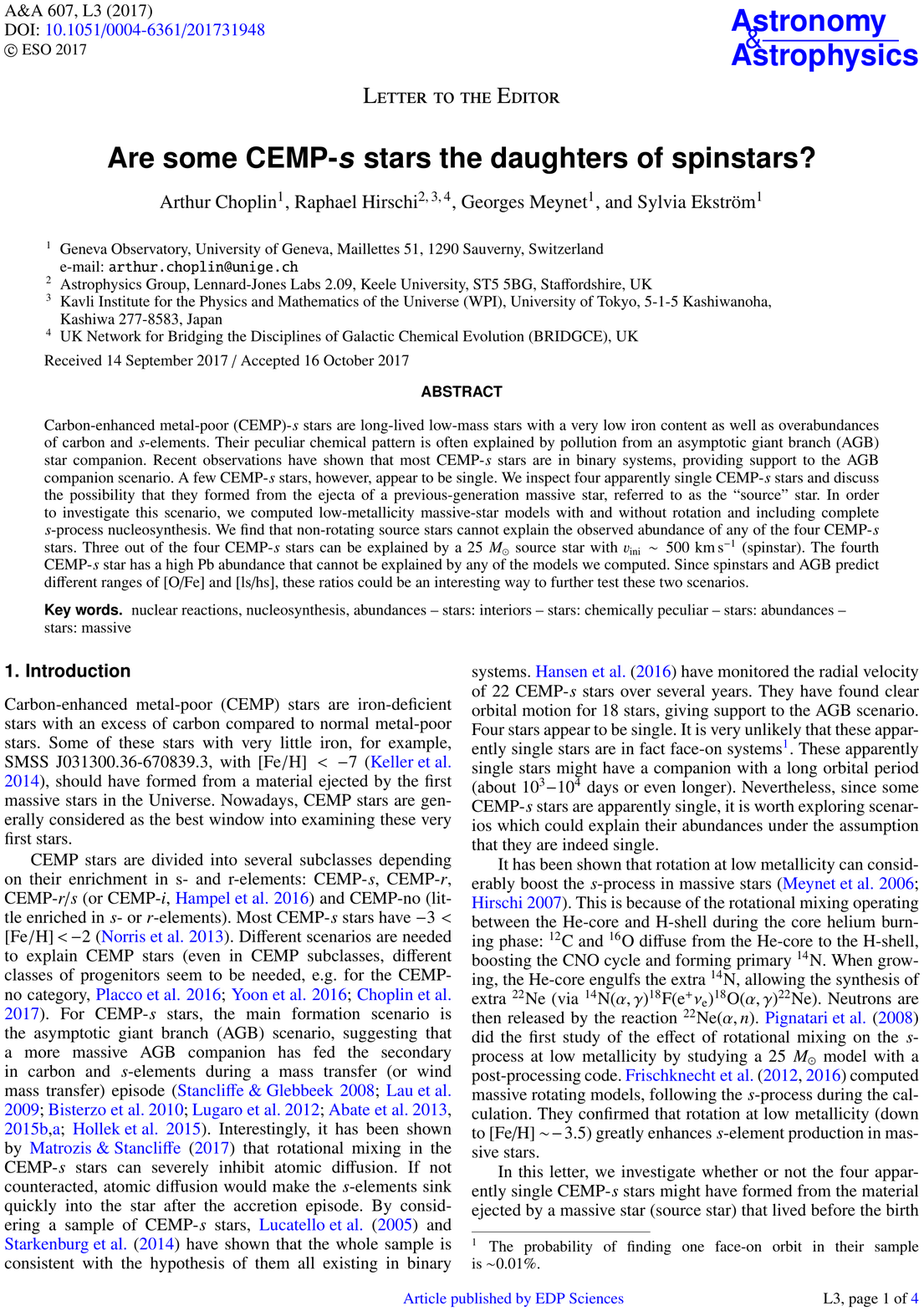}

\newpage

   \begin{figure}[h!]
   \begin{minipage}[c]{.99\linewidth}
      \centering
      \includegraphics[scale=0.35, trim=0cm 0cm 0cm 0cm,clip]{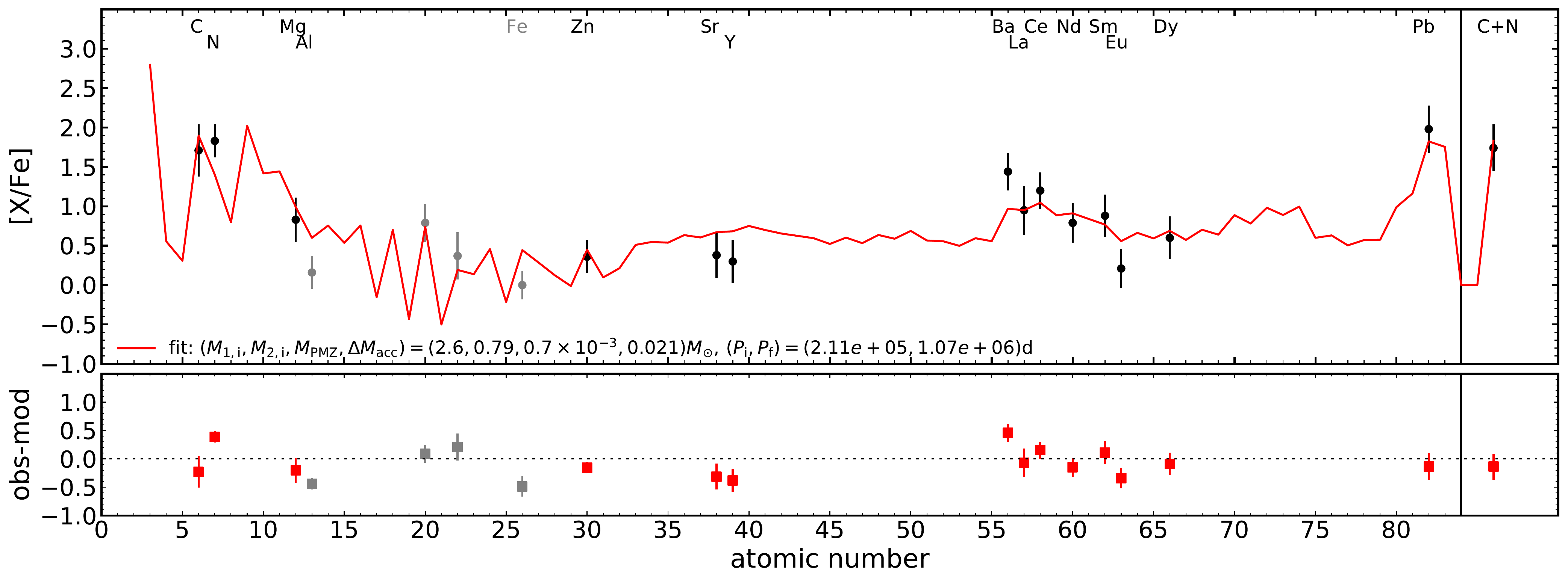}
   \end{minipage}
   \begin{minipage}[c]{.99\linewidth}
      \centering
      \includegraphics[scale=0.35, trim=0cm 0cm 0cm 0cm,clip]{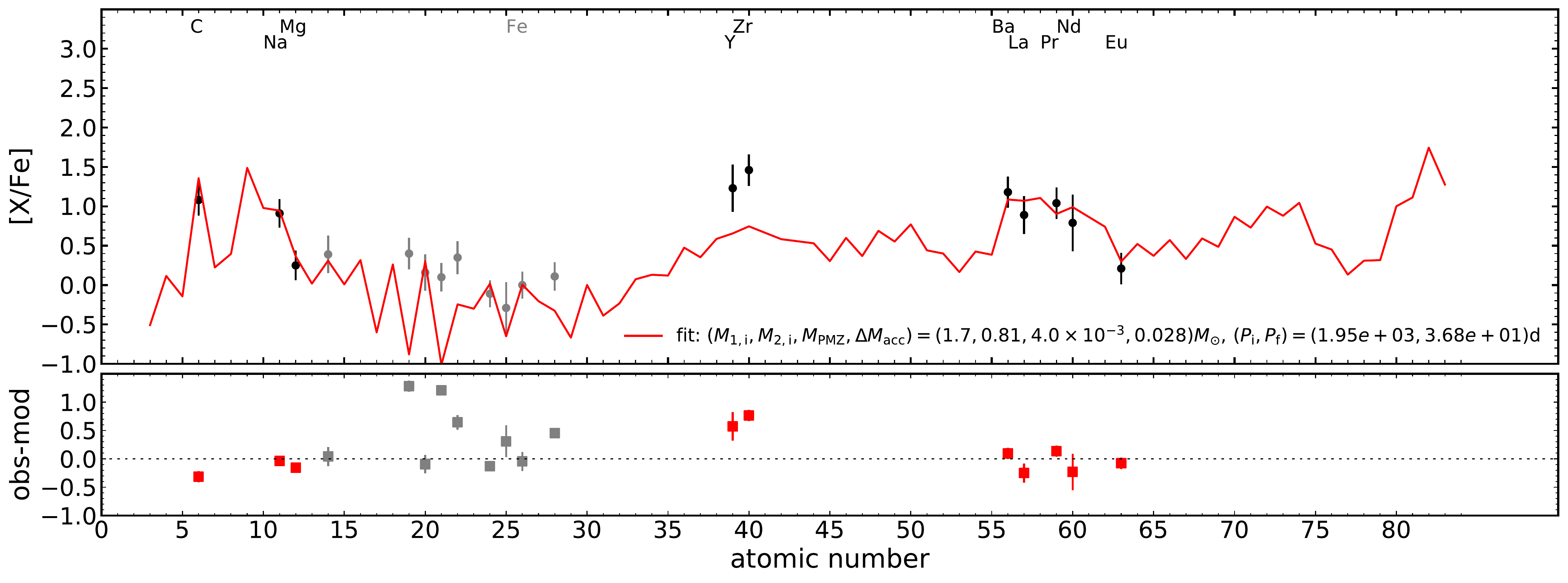}
   \end{minipage}
\caption[Composition of AGB models and apparently single CEMP-s stars]{Comparison of AGB models (red patterns) with the chemical composition of the apparently single CEMP-s stars CS~30301-015 (top panel) and HE~1045+0226 (bottom panel, Carlo Abate, priv. com.).}
            
  \label{Cabate}
\end{figure}

%\subsection{AGB star models}
\subsection{The AGB scenario}

%Two CEMP-s stars presented in the previous letter have only a barium abundance for elements heavier than iron. This is not a 
%As discussed in the previous letter, 
Below is discussed some aspects of the AGB binary scenario for the four stars of the previous letter. 
\cite{abate15b, abate15a} have used a binary evolution and population synthesis code to predict the orbital properties of the system (CEMP-s + AGB). The chemical composition in the AGB intershell region, from AGB models of \cite{lugaro12}, is used. The abundances are function of 3 quantities: the mass of the AGB star at the first thermal pulse, the thermal pulse number and the mass of the partial mixing zone (cf. Sect.~\ref{sprocagb}). The procedure is discussed in details in \cite{abate15b,abate15a}.

Because of its high Pb abundance, the apparently single CEMP-s star CS~30301-015 discussed in the previous letter likely cannot be reproduced if considering only the massive stars ejecta computed in this work. 
%Below is discussed  the possibility of explaining this CEMP-s star with the AGB mass transfer scenario. 
%The red pattern on the top panel of Fig.~\ref{Cabate} shows the predictions of a 2.6~$M_{\odot}$ AGB model. 
The best AGB fit for this CEMP-s star (shown in Fig.~\ref{Cabate}, top panel) is a 2.6~$M_{\odot}$ AGB star with a period (at the present day) of about $10^{6}$ days. Although not perfect, the fit is improved compared to what can be done with massive stars. In particular, Pb can be reproduced. Also, the long predicted period provides a simple explanation to why this CEMP-s star appears to be single: its orbits is so wide that no radial-velocity variations can be detected. 
%is consistent with the CEMP-s star being apparently single. 
It is consistent with the statement of \cite{hansen16a} that reported that any undiscovered binaries among their four apparently single CEMP-s stars would have periods of at least $10^3 - 10^4$ days.

For HE~1045+0226 (bottom panel in Fig.~\ref{Cabate}), AGB star models predict too low Y and Zr. Also, the predicted period for HE~1045+0226 is about 40 days, which is too short to explain the non-detection of radial velocity variations. For this star, a massive rotating source stars provides a better solution (cf. previous letter).
For the two other CEMP-s stars, only the Ba abundance is available among the heavy elements so that both massive stars and AGB star models can provide a reasonable solution.

Generally speaking, massive stars tend to produce larger light s-element (e.g. Sr) over heavy s-element (e.g. Ba, Pb) ratios than AGB stars. Another difference, as discussed in the previous letter, is the oxygen, that can be higher by $1-2$ dex in massive stars compared to AGB stars. 
Some CEMP-s stars are O-rich, a characteristic that AGB models may not be able to account for. 
For instance CS~31080-095 has [O/Fe] $=2.35 \pm 0.12$ \citep{sivarani06}, LP625-44 has [O/Fe] $=1.8$ \citep{aoki02c}, HE~2258-6358 has [O/Fe] $=1.8 \pm 0.1$ \citep{placco13}. 
Such high ratios do not exclude a contribution from an AGB companion but may indicate a contribution of one (or more) additional source (i.e. previous massive stars).
By detailed comparisons between AGB and massive star models, it would probably be interesting to try to spot the elements like oxygen that could help probing the nature of the additional source(s). %(in case an additional source is needed). \\

%All elements for 1 star

%Radial velocity of more stars

%AGB models starting with yields of rotating massive stars (already some s-elements)

%Maybe put figures in rapport doc\_latex slash rates slash
%\textcolor{red}{production factor at differnt stages,?}
%put fig. 18, 19, 20 of new\_sproc.tex. Effect of mcut and dilution for all zm3 models (just need to include the other models).

%Put fig 4 I think and also 1, 2, 3...

%\textcolor{red}{say a word about late mixing!! Why not here? natural explanation, higher Z...? Also, it almost only affects CC and C/N}

   \begin{figure*}[t]
   \centering
      \includegraphics[scale=0.6, trim = 0cm 0cm 0cm 0cm]{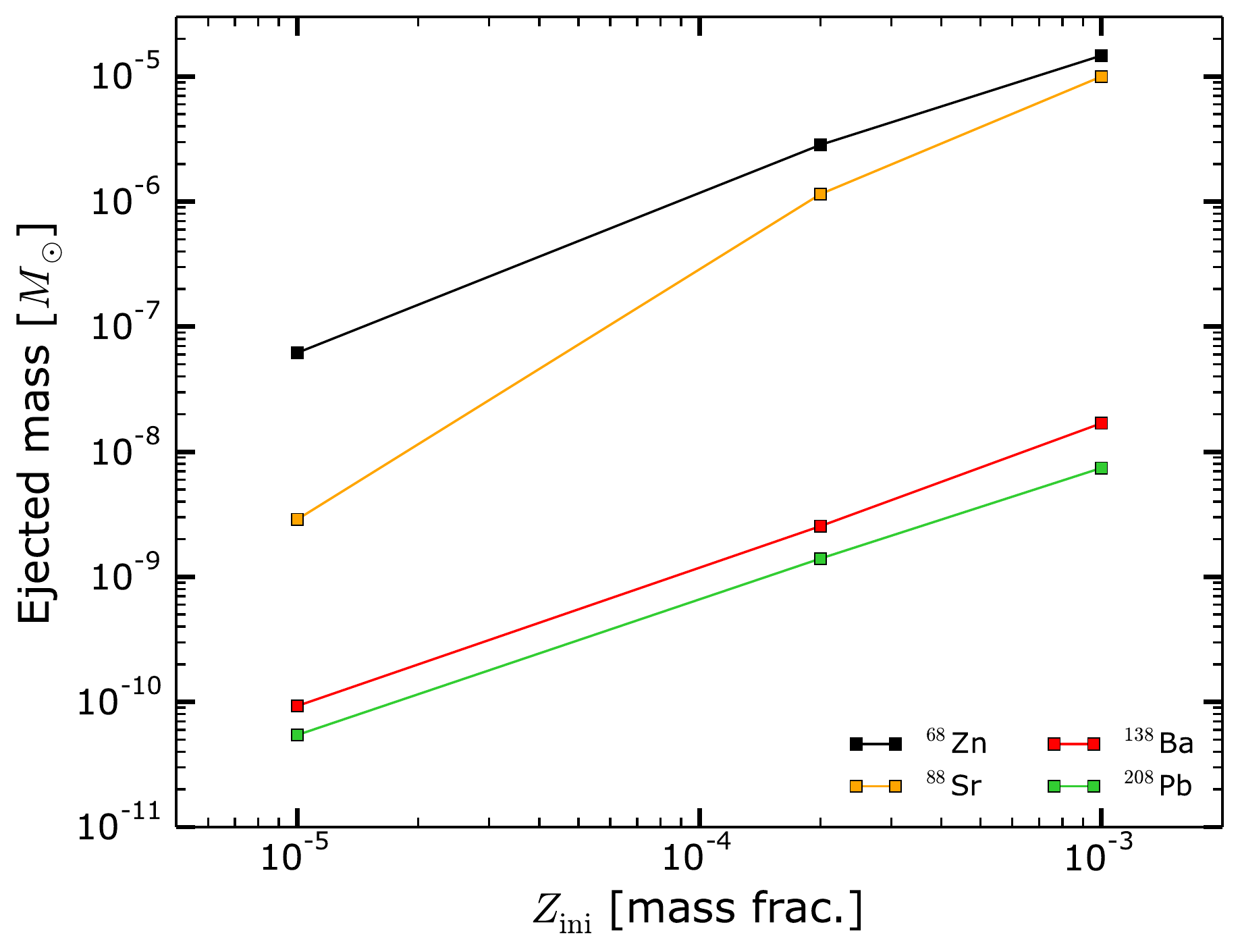}
   \caption[Ejected mass of 4 isotopes as a function of metallicity for rotating 25~$M_{\odot}$ models]{Ejected mass (Eq.~\ref{yie}) of 4 isotopes as a function of the initial metallicity for 25~$M_{\odot}$ models with $\upsilon_{\rm ini} / \upsilon_{\rm crit} = 0.4$. The mass cut is set according to \cite{maeder92}.}
\label{sprocz}
    \end{figure*}

%== INTRO ======================================================================
\section{The weak s-process at lower metallicities}\label{sproclower}

%While reduced, the s-process boost induced by rotation still exist at lower metallicity. 
%In this section, I discuss lower metallicity source star models including s-process and rotation. 

With decreasing metallicity, less seed is available so that less s-elements are produced. 
%In non-rotating models as the metallicity gets smaller, the abundances of source and seed, which are secondary, decrease. On the other hand, the abundances of primary neutron poisons like $^{16}$O remain the same. Consequently, the effect of primary neutron poisons becomes stronger at lower metallicities \citep{prantzos90}. This implies that the production of s-elements does not decrease linearly with metallicity. 
In non-rotating models, the production of s-nuclei in massive stars becomes negligible below about $Z$~$=$~$10^{-4}$ \citep{prantzos90}. 
In rotating models, this metallicity threshold may be around $Z=10^{-7}$: \cite{frischknecht16} have shown that at $Z=10^{-7}$, rotation in a 25~$M_{\odot}$ model has a modest effect on the abundances of s-elements. At very low metallicity, the lack of iron seed implies that even if the neutron source is boosted by rotation, there is no significant s-process boost.
Also, it is worth mentioning that above an given metallicity threshold, the s-process boost induced by rotation becomes very small: at $Z=Z_{\odot}$, \cite{frischknecht16} have shown that the abundance patterns of rotating and non-rotating models are almost identical. This is because rotation does not provide much additional $^{22}$Ne (neutron source) at this metallicity, as a result of the less efficient back-and-forth mixing process (Sect.~\ref{secback}). Indeed, at higher metallicity the distance between the He-core and the H-shell increases and the gradient of $\Omega$ is smaller so that the shear mixing between the He-core and the H-shell is less strong.
%Rotation provides more source but no more seed so that the total content in s-elements also decreases with metallicity in rotating models.

\subsection{A rotating 25~$M_{\odot}$ model with various initial metallicities}
%\textcolor{red}{maybe after the paper? And discuss the case of HE~1327?}

The computation of grids of massive star models at lower metallicities (including s-process) is a work in progress. Only 25~$M_{\odot}$ models are presented here. 
Fig.~\ref{sprocz} shows the ejected mass (Eq.~\ref{yie}) of four isotopes for rotating 25~$M_{\odot}$ models with $Z=10^{-3}$, $2~\times~10^{-4}$ and $10^{-5}$. As the metallicity decreases by 2 dex, the ejected mass of the considered isotopes decreases by $2-4$ dex, depending on the chemical specie. 
%However, with the additional (primary) $^{22}$Ne provided by rotation, the source over seed ratio is enhanced and first and second peak s-elements can still be produced at very low metallicity \citep{frischknecht16}.
%and, some s-elements up to the first and eventually second peak can still be produced at very low metallicity \citep{frischknecht16}. 
%Even with rotation, the \textit{absolute} amount of s-elements produced is strongly reduced with decreasing metallicity. 
%Nevertheless, the distribution of heavy elements can be heavily affected in very low rotation affects significantly the little amount of heavy elements that were initially present in the star. 
The green and red patterns in Fig.~\ref{prodzm5} show the production factors (Eq.~\ref{fact}) of a non-rotating and a rotating 25~$M_{\odot}$ model at $Z=10^{-5}$, respectively. Compared to their initial abundances, elements between Cu and Y are overproduced by $1-2$ dex in the rotating model. 
The black pattern shows an extreme case, with fast rotation and a lower $^{17}$O($\alpha,\gamma$) reaction rate (/10). In this case, the production of elements between Fe and Eu are enhanced by 1 up to about 3 dex compared to their initial abundances.
Overall, it shows that at very low metallicity, although the absolute yields are low (Fig.~\ref{sprocz}), rotation can strongly affect the distribution of heavy elements that are initially present in the star (Fig.~\ref{prodzm5}). If rotation is not considered, the initial distribution of heavy elements is barely modified at this metallicity.

%   \begin{figure*}[t]
%   \centering
%      \includegraphics[scale=0.6, trim = 0cm 0cm 0cm 0cm]{prodZ.pdf}
%   \caption[a]{Production factors \textcolor{red}{Eq.~XXX} for 25~$M_{\odot}$ models with $\upsilon_{\rm ini} / \upsilon_{\rm crit} = 0.4$ and with various metallicities. The mass cut is set according to \cite{maeder92}.}
%\label{prodz}
 %   \end{figure*}

   \begin{figure*}[t]
   \centering
      \includegraphics[scale=0.6, trim = 0cm 0cm 0cm 0cm]{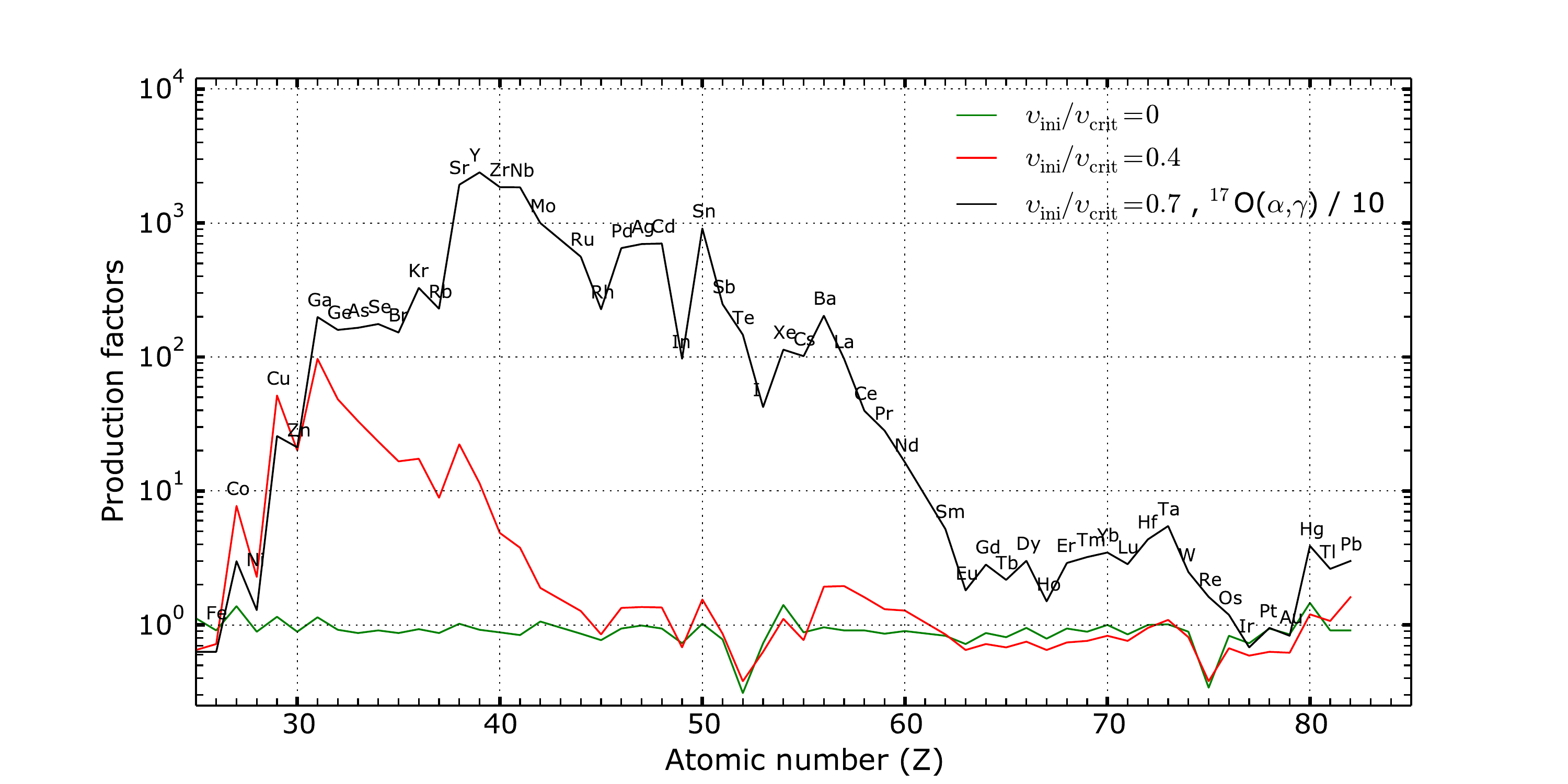}
   \caption[Production factors of 25~$M_{\odot}$ models with various initial rotation rates]{Production factors (Eq.~\ref{fact}) of 25~$M_{\odot}$ models with $Z=10^{-5}$ and with various initial rotation rates. For the fast rotating model, the rate of $^{17}$O($\alpha,\gamma$) was divided by 10. The mass cut is set according to \cite{maeder92}.}
\label{prodzm5}
    \end{figure*}

   \begin{figure*}[h!]
   \centering
      \includegraphics[scale=0.52, trim = 0cm 0cm 0cm 0cm]{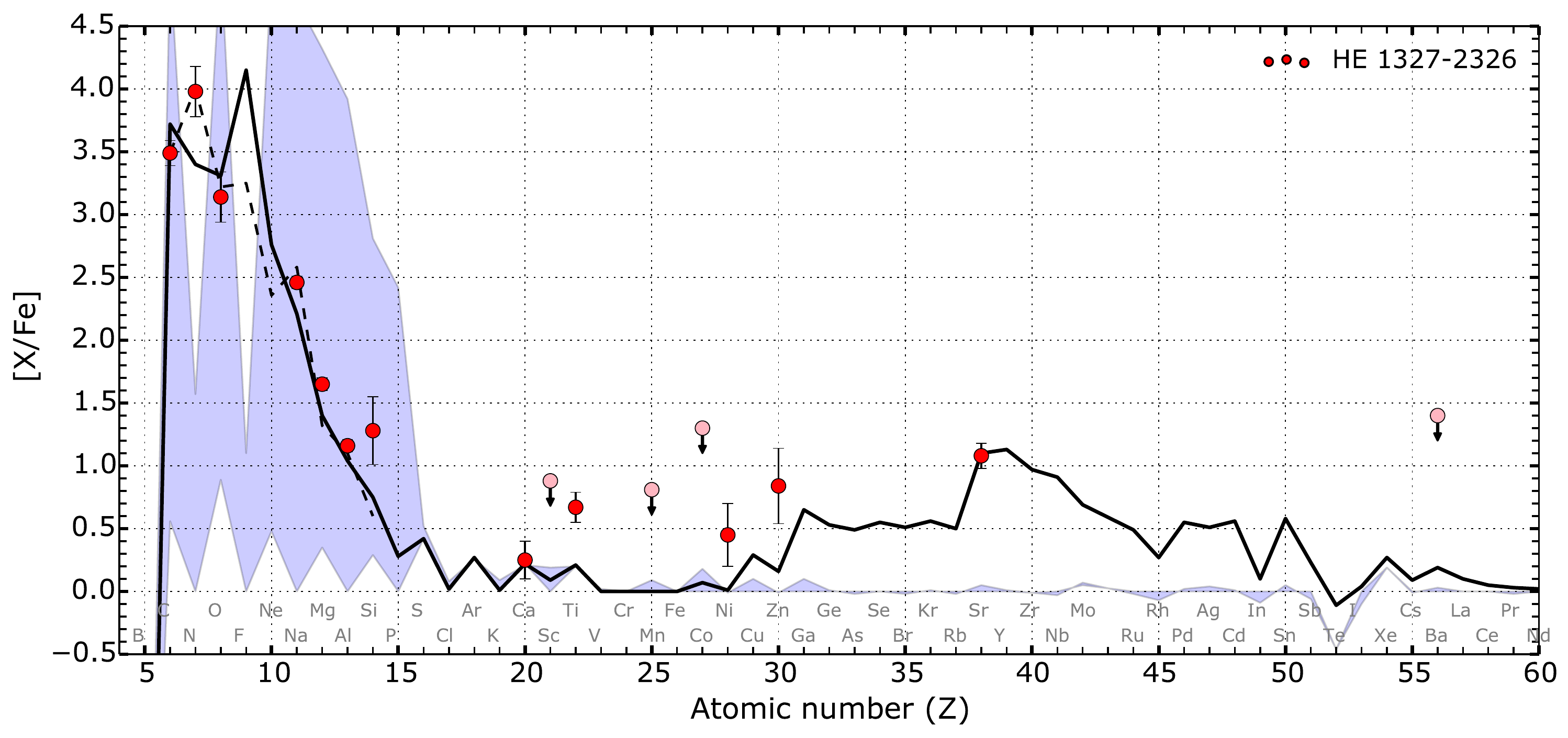}
   \caption[Composition of HE~1327-2326 and of a 25~$M_{\odot}$ source star models with s-process]{[X/Fe] ratios of the 25~$M_{\odot}$ model with $Z=10^{-5}$ and $\upsilon_{\rm ini}/\upsilon_{\rm crit}=0.7$ (thick black pattern). The mass cut is set to $M_{\rm cut} = 10.5$~$M_{\odot}$ (it corresponds to the bottom of the H-envelope). The purple shaded area stands for the same model without rotation. This area shows the range of yields one can obtain when varying the mass cut from the final mass of the model (only winds) to $M_{\rm rem}$ (see text for details). The dashed pattern shows the yields of the 20~$M_{\odot}$ model with $Z=10^{-5}$ and $\upsilon_{\rm ini}/\upsilon_{\rm crit}=0.7$ presented in Sect.~\ref{sec20} (computed with a smaller network). In this case, $M_{\rm cut} = 7.9$~$M_{\odot}$. $D=0$ in all the cases (i.e. no dilution with ISM). Symbols show the abundances of the star HE~1327-2326 (light red sumbols denote upper limits, as shown by arrows).}%\textcolor{red}{put in perspectives? or Previous chapter?}}
\label{HE1327sproc}
    \end{figure*}

\subsection{The origin of HE~1327-2326}\label{origHE1327}
%\subsection{The source of HE1327-2326: a fast rotating star experiencing a jet-supernova?}

The star HE~1327-2326 \citep{frebel05} is one of the most iron-poor star with [Fe/H] $-5.7$ and an excess of Sr compared to Fe. 
Red circles in Fig.~\ref{HE1327sproc} show its surface composition, according to a recent re-analysis (Ezzedine, priv. comm.). In particular, an overabundance of Zn was found ([Zn/Fe] $=0.84 \pm 30$).

All possible chemical compositions one can obtain with the non-rotating 25~$M_{\odot}$ model at $Z$~$=$~$10^{-5}$ are shown by the purple area. This area shows the range of [X/Fe] ratios one can obtain when varying the mass cut from the final mass of the model $M_{\rm fin} = 24.5$~$M_{\odot}$ (in this case only the material in the stellar winds is considered) to the remnant mass $M_{\rm rem} = 2.31$~$M_{\odot}$ \citep[according to the relation of][]{maeder92}. 
We see that whatever the mass cut, N is underproduced by more than 2 dex and Sr by more than 1 dex. 

The thick black pattern shows the [X/Fe] ratios of the fast rotating 25~$M_{\odot}$ model ($\upsilon_{\rm ini}/\upsilon_{\rm crit}$~$=$~$0.7$). The mass cut is set to 10.5~$M_{\odot}$, which is just below the H-rich envelope. No dilution with ISM was considered. In this case, the [Sr/Fe] ratio of HE~1327-2326 can be reproduced.  
The fit for light elements is also satisfactory. 
In particular, N is much closer to the observed value. It is however still underproduced by 0.5 dex. The dashed pattern, shows the yields of a model with same metallicity and rotation but with an initial mass of 20~$M_{\odot}$ and a small network (model of Sect.~\ref{sec20}). This model can produce enough N, suggesting that a source star with a bit different initial mass may provide a better fit (at least for the light elements).

In Sect.~\ref{secstat}, the source star model that best reproduced the abundance pattern (for light elements, C to Si) of HE~1327-2326 was a fast rotating 60~$M_{\odot}$ model without late mixing, without dilution with ISM and with a mass cut close to the bottom of the H-envelope (cf. Table~\ref{bestable}). 
All these characteristics are consistent with the new results obtained here, except for the initial mass of the source star ($20-25$~$M_{\odot}$ here vs. 60~$M_{\odot}$ before). 
%As discussed and shown in Fig.~\ref{HE1327sproc}, the $20 - 25$~$M_{\odot}$ models also provide a satisfactory fit for the light elements. 
The high [Sr/Fe] might favor rotating $\sim$ 25~$M_{\odot}$ instead of $\sim$~60~$M_{\odot}$ source stars because the s-process boost induced by rotation peaks around $20-40$~$M_{\odot}$ \citep{choplin18}. A rotating 60~$M_{\odot}$ source star may not produce enough Sr. In the future, the computation of additional source star models at this low metallicity but with different initial masses may help to further address this point.

The source star models presented here underproduce Ti, Ni and Zn by $\sim 0.5$ dex. 
Such elements can be affected by explosive nucleosynthesis. 
Interestingly, it was shown that Ti and Zn are overproduced in jet-induced SNe \citep{tominaga09}. 
This kind of asymmetric SN may be produced by fast rotation of the massive stellar core at the time of the explosion \citep{woosley93}.
Fast rotation of the core at the time of explosion may be due to fast rotation of the progenitor.
The fact that HE~1327-2326 is enriched in both the product of rotation (e.g. N, Sr) and of jet-induced SNe (e.g. Zn) may indicate that its source star was a massive rotating star that experienced a jet-induced SN.

In general, the determination of the abundances of both rotation products and jet-induced SNe products in very metal-poor stars might give clues on the impact of rotation on the explosion. An interesting perspective is to study these two families of elements in very metal-poor stars. However, for some interesting elements like Zn, almost only upper limits are available at the moment since they are very challenging to detect.
%At the moment, while N is available in many CEMP stars, almost only upper limits are available for Zn. 
%For instance, if a trend between N and Zn is found, one might link the initial velocity of the star 

%The [Zn/Fe] of 0.84 in HE~1327-2326 may suggest that its source experienced such a kind of explosion. It 

%A recent re-analysis of the composition of this star shows that it has [Fe/H] $=-5.84 \pm 0.25$ and [Sr/Fe] $=1.08 \pm 0.10$ (Ezzedine, priv. comm.). 

%Fig.~\ref{HE1327sproc} shows the 

\section{Summary}

By increasing the amount of available neutrons, rotation boosts the s-process in massive stars. A new grid of non-rotating and rotating ($\upsilon_{\rm ini}/\upsilon_{\rm crit} = 0.4$) stellar models at [Fe/H] $=-1.8$ with initial masses of 10, 15, 20, 25, 40, 60, 85, 120 and 150~$M_{\odot}$ was computed. Stellar yields are publicly available.
Rotation has a strong impact on the production of s-elements (especially the first peak, e.g. Sr) for initial masses between about 20 and 60~$M_{\odot}$. 
Two additional 25~$M_{\odot}$ models were computed: one with faster rotation ($\upsilon_{\rm ini}/\upsilon_{\rm crit} = 0.7$) and another with faster rotation ($\upsilon_{\rm ini}/\upsilon_{\rm crit} = 0.7$) and lower $^{17}$O($\alpha,\gamma$) rate (divided by 10, it reduces the poisoning effect of $^{16}$O). In these two models, the production of s-elements is boosted again. The fast rotating model with a lower $^{17}$O($\alpha,\gamma$) rate is the model where the s-process boost is the strongest. In particular, a modest amount of heavy s-elements (e.g. Pb) is synthesized.
In stellar models with initial masses $M\gtrsim 60$~$M_{\odot}$, the back-and-forth mixing is less efficient so that the production of s-elements is similar whether or not rotation is considered.
%Faster rotation ($\upsilon_{\rm ini}/\upsilon_{\rm crit} = 0.7$) enhances the production of $40 \lesssim?? Z\lesssim ?? 60$ elements by $0.5 - 1$ dex.

%The AGB mass transfer scenario faces difficultiesApparently single CEMP-s stars face difficulties 
Some CEMP-s stars appear to be single stars, which may challenge the AGB binary scenario (for these specific stars).
The yields of the fast rotating 25~$M_{\odot}$ model of the computed grid provide a material able to fit the abundance patterns of 3 out of the 4 apparently single CEMP-s stars.

The s-process boost induced by rotation still exists in lower metallicity massive stellar models (probably down to about $Z \sim 10^{-7}$).
HE~1327-2326, one of the most iron-poor star known, has [Sr/Fe] = 1.08. Its abundance pattern can be reproduced by a fast rotating very low metallicity $20-25$~$M_{\odot}$ source star. 
The fact that this CEMP star is also enriched in Zn, which is a product of jet-induced SNe, is consistent with a scenario proposing that its source star was a fast rotator that experienced a jet-induced SN. 
%the spinstar scenario: Zn is overproduced in energetic jet-like supernovae, that might be triggered because of fast rotation of the progenitor.

%This CEMP star is also Zn-rich, element that is overproduced in energetic jet-supernovae. Such peculiar supernovae

%\include{grid}
\chapter{Conclusions and perspectives}
\label{Cconclu}

%==============================================================================
\section{Conclusions}

%The nature of the early generation of massive stars is still largely unknown are crucial objects for that are long. 

This work mainly aimed at better understanding the nature of the early generations of massive stars. They are crucial objects to figure out the origin of the elements in the Universe. %entire chemical evolution of the Universe. 
%that are a crucial objects to figure out the origin of the elements in the Universe. 
Although long-dead, indirect clues on these stars can be obtained from the surface chemical composition of still alive and observable low-mass metal-poor stars, formed early in the Universe. %The surface chemical composition of these low-mass metal-poor stars likely has kept the nucleosynthetic imprint of the early generations of massive stars.
Some of these low-mass metal-poor stars, very little enriched in iron and showing highly non solar-like abundance patterns (CEMP stars, especially CEMP-no) are thought to have formed from the ejecta of only one or a few previous zero or very low-metallicity massive stars (the source stars).
To gain knowledge on the characteristics of the early massive stars, 
I computed models of massive source stars under various assumptions so as to determine what are the source star ingredients required to reproduce the abundances of CEMP stars.
%%%and compared the predicted yields of these models to the abundances of CEMP stars. 
%%%As suggested in different previous works, 
%Rotation may be an ingredient that has played a major role in the evolution and outputs (e.g. yields) of the early generations of massive stars. 
%Especially, the efficient rotational mixing operating at low metallicity triggers exchanges of material between the He-burning core and H-burning shell and lead to a very rich and varied nucleosynthesis.

Below are listed the main concrete works carried out during this PhD, followed by the main results.

\begin{itemize}
\item I developed a nucleosynthetic one-zone code mimicking the effect of rotation on nucleosynthesis in stellar interiors and made comparisons with abundances of CEMP stars \citep[Sect.~\ref{secbox} and][]{choplin16}.
\item I computed pre-supernovae massive source star models under various assumptions (initial mass, rotation, extra mixing...) and made comparisons with abundances of CEMP stars \citep[Sect.~\ref{sec20}, \ref{compacemp},  \ref{secstat} and][]{choplin17a}.
\item I developed a small module in \textsc{genec} allowing to generate extra mixing events in a specific zone of a stellar model \citep[Sect.~\ref{seclate} and][]{choplin17a}.
\item I computed a new grid of massive stars from 10 to 150~$M_{\odot}$, at $Z=10^{-3}$, with and without rotation and including full nucleosynthesis so as to follow the s-process \citep[Sect.~\ref{newgridsproc} and][]{choplin18}. Stellar yields are available online\footnote{See \href{https://www.unige.ch/sciences/astro/evolution/en/database/}{https://www.unige.ch/sciences/astro/evolution/en/database/} or the CDS database at \href{http://cdsarc.u-strasbg.fr}{cdsarc.u-strasbg.fr}.}. Then, comparisons with CEMP stars enriched in s-elements were done \citep[Sect.~\ref{origsinglecemp}, \ref{sproclower} and][]{choplin17letter}.
\item During a short term scientific mission with Prof. Jose Groh, I produced \textsc{CMFGEN} spectra for a population of main-sequence massive stars at $Z=10^{-5}$ and started to investigate the effect of fast rotation in the spectral features of such stars (see Sect.~\ref{distantold} for more details).
\item I included the effects of axion losses in \textsc{genec} and computed a grid of non rotating Pop~III stars with and without axion losses \citep[Sect.~\ref{secaxion}, Appendix~\ref{snumchem} and][]{choplin17ax}.
\item I included in \textsc{genec} the new diffusion coefficient of \cite{maeder13} that describes the interaction between various instabilities arising in rotating stars (Appendix~\ref{dshear}). I carried out several tests models to compare this prescription with other prescriptions (Sect.~\ref{transportrot}).
\item I included in \textsc{genec} the new mass loss recipe of \cite{moriya15} that considers the mass loss induced by pulsations in the late evolution of very massive metal-free stars (Appendix~\ref{pulsmassloss}).
\item I improved the way of treating the rotational mixing in \textsc{genec} by upgrading the calculation techniques of the $D_{\rm shear}$ coefficient (Appendix~\ref{oscillapp}).
\item I transformed the tabulated reaction rates of \citet[][$^{17}$O($\alpha,\gamma$) and $^{17}$O($\alpha,n$)]{best13} into the \textsc{reaclib} format (Sect.~\ref{netandrates}).
\end{itemize}

The main results of these works, presented in the chapters \ref{cempno} and \ref{cemps}, can be summarized as follow:

\begin{itemize}
\item Using the one-zone nucleosynthetic code, I found that a significant fraction of the CEMP stars with [Fe/H] $<-3$ can be overall well reproduced by a material processed by H-burning at a temperature and density characteristic of $20-60$~$M_{\odot}$ source stars. While burning, this material has to be enriched in $^{12}$C, $^{16}$O and occasionally in $^{22}$Ne. It suggests that the source stars experienced mid to strong rotational mixing. 
%\item Complete source star models with various initial parameters (especially rotation) were computed. While a global match can be found between source star yields and abundances of CEMP stars with [Fe/H] $<-3$, no material fro many model was found to account for the bulk of observation in the C/N
\item Complete source star models with various initial parameters (especially rotation) were computed. The efficient rotational mixing operating at low metallicity triggers exchanges of material between the He-burning core and the H-burning shell. It leads to a rich and varied nucleosynthesis, able to cover the ranges of abundances of light elements for the CEMP stars with [Fe/H]~$<-$~3.
\item Comparisons between source star models and observations in the [C/N] vs. $^{12}$C/$^{13}$C diagram revealed some discrepancies that can be alleviated if a late mixing process is included in the source stars. The mixing operates in between the H-burning shell and the He-burning shell, about 200 yr before the end of the source star evolution. The mixing may also operate before, provided the material in the H-burning shell (at least some of it) is saved from further processing shortly after the mixing event (e.g. it is expelled through stellar winds).
\item The abundance fitting (for light elements, C to Si) of 69 CEMP stars with [Fe/H] $<-3$ and not significantly enriched in s- and r-elements was performed with $\sim 35000$ ejecta composition of source star models. The best source stars are preferentially fast rotators (74~\% of the cases), have initial masses of 20~$M_{\odot}$ (70~\%) and experienced the late mixing process (72~\%). In most of the cases ($\sim 70$~\%), only the outer layers of the source stars (above or just below the interface between the H-rich and He-rich region) have to be expelled. Finally, the best abundance fits are mostly obtained with a modest dilution of the ejecta with the interstellar medium ($\sim 0-10$ times more interstellar material than source star ejecta).
\item A new grid of massive source stars at [Fe/H] $=-1.8$ was computed with an extended nuclear network, able to follow the full s-process during the evolution. Rotation was found to strongly boost the first s-process peak (e.g. Sr) for stars with initial masses between 20 and 60~$M_{\odot}$. If the initial rotation is high enough, models predict a significant synthesis of s-elements from the second peak (e.g. Ba). Elements from the third peak (e.g. Pb) were found to be little impacted by rotation.
\item CEMP-s stars, which often have $-3<$ [Fe/H] $<-2$, are generally associated with the AGB binary scenario. Four CEMP-s stars were recently found to be single stars, a characteristic that may challenge this scenario. The fast rotating 25~$M_{\odot}$ model of the new grid of model including s-process provides a material able to fit the abundance pattern of 3 out of the 4 apparently single CEMP-s stars.
\item The abundances of HE~1327-2326, one of the most iron-poor star, were examined in detail with the yields of new very low metallicity source star models including the s-process. Both the excess in light elements and Sr can be explained with the ejecta of a fast rotating $20-25$~$M_{\odot}$ source star model. Interestingly, the fact that HE~1327-2326 is enriched in both the products of rotation (e.g. N, Sr) and of jet-induced supernovae (e.g. Zn), is consistent with a scenario proposing that its source star was a fast rotator that experienced a jet-induced supernova.
%may be consistent with the yields of a fast-rotating source star that experienced a jet-like supernovae. 
\end{itemize}

%CEMP stars 
The results of this work suggest that rotation in the source star is a crucial ingredient, which is able to provide a solution for the abundances of light elements (C to Si) and s-elements of many CEMP stars. 
It supports the idea that rotation was a dominant effect in the early generations of massive stars.
The results propose that in most of the cases, a strong and late mixing event operated in between the H-burning and He-burning shell of early massive stars.
It was also shown that CEMP stars should be formed with only the relatively outer layers of the source stars (generally only the H-envelope or the H-envelope plus a small part of the He-rich region). 
%early massive stars should expel
This suggests that early massive stars expelled only their envelope and experienced strong fallback.
The present work does not offer a specific mechanism for the ejection of the source star envelope. Although stellar wind is a good candidate for the reasons mentioned in Sect.~\ref{secmassloss}, the wind material of the source star models presented here is generally not sufficiently enriched in light and heavy elements for reproducing the CEMP star abundances. An additional (generally modest) supernova contribution is required.
In any case, the fact that just the envelope of the source star should be expelled is consistent with the mixing \& fallback scenario that could finally co-exist with the fast rotating massive star scenario. %The mixing \& fallback scenario provides a solution to explain the abundances of the Fe-peak elements, which likely cannot be explained directly by rotating massive stars.

  \begin{figure*}[t]
  \centering
     \includegraphics[scale=0.42, trim = 4cm 8cm 4cm 0cm]{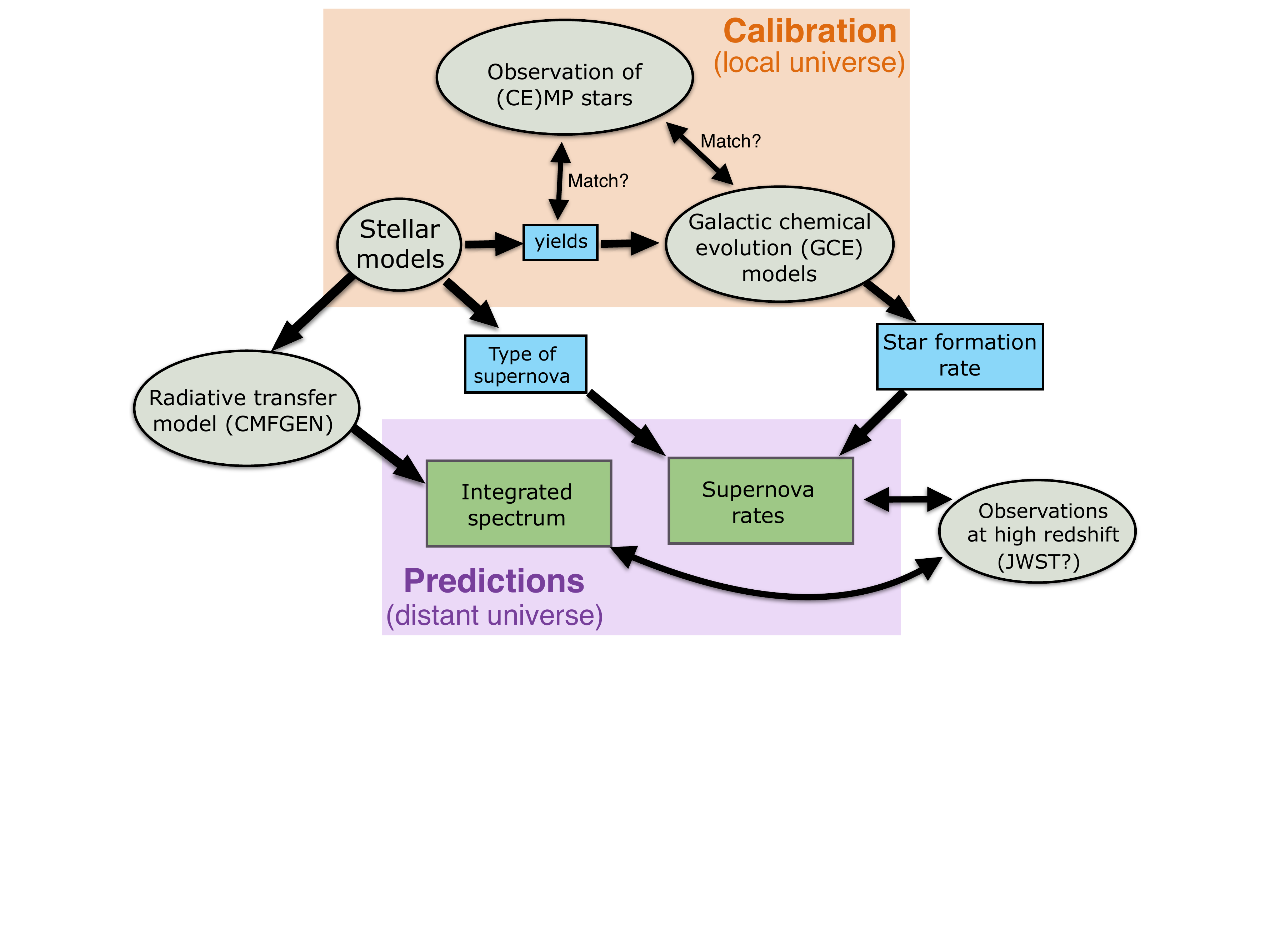}
   \caption[Link between the old and local Universe with the old and distant Universe]{The cartoon shows how the old and local Universe (i.e. metal-poor stars) could be used so as to make predictions in the very high redshift Universe. %high redshift predictions  of massive source stars using constraints from the old and local Universe (abundances of metal-poor stars). 
   %Green boxes represent the main predictions. 
   Individual stellar models and Galactic chemical evolution (GCE) models are used to determine the best massive source stars, able to reproduce the abundances of metal-poor stars. The appearance (integrated spectrum and supernova rates), in the very distant Universe, of this calibrated massive source star population may be predicted and confronted to (future) observations.
   Grey ovale shapes show the modeling tools and data that should be used. Blue boxes show, in this context, the important inputs/outputs from stellar and GCE models.}
\label{workflow}
    \end{figure*}

%==============================================================================
\section{Perspectives}

%The origin of the CEMP stars, which is tightly linked to the origin of the elements and 
The nature of the first stars is a topic where we are still scratching the surface and where to move on, modelers and observers should go hands in hands. 
Below are important possible future steps or projects (mainly on the modeling side) linked to the present work.

\begin{itemize}
\item The modeling of grids of source star models including s-process has to be continued, especially at very low metallicity. The aim is to get a more complete picture of the nucleosynthesis in rotating massive stars and perform extended comparisons with observations, with as much abundances as possible.
\item The new scenario for the origin of HE~1327-2326 raises the interest for studying the outputs of rotating massive stars experiencing a jet-induced supernova. A consistent prediction of the nucleosynthetic yields of a rotating massive star ending its life as a jet SN could be an interesting project. In the future, the comparison of such models with observations might give clues on how the explosion of early massive stars can be affected by rotation. %For instance, while N and Sr are overproduced by rotation, Zn is overproduced by jet SN. 
\item Although generally operating in very central regions, which may be finally locked into the remnant, the SN-shock may lead to further neutron capture nucleosynthesis in the He-shell. It could be included in source star models.
\item The late mixing process could be studied using multi dimension models. Also, when available, constraints from such multi dimension models (especially on the convective boundaries of the H- and He-burning shells) should be included in 1D stellar evolution codes so as to further test the hypothesis of the late mixing process in the source stars. %This process appeared to be a feature able to improve the fit between most CEMP stars and their source stars.
\item An important project would be to recover the initial surface composition of evolved CEMP stars by computing grids of CEMP star models including various mixing processes. To date, such a work on a large CEMP star sample has been done only for correcting effect of the first dredge-up on the C abundance \citep{placco14c}. Knowing well the initial surface composition of CEMP stars (especially for the most evolved CEMP stars) is crucial to connect them more consistently to their source stars. 
\end{itemize}

There are also different observational perspectives. One may be to try determining the abundances of both rotation and jet-induced SN products on the most metal poor stars (cf. point 2 just above). Isotopic ratios (especially $^{12}$C/$^{13}$C) are very constraining (especially on rather unevolved stars) and can provide strong clues on nucleosynthetic processes at work in source stars. Also, the oxygen is likely overproduced in massive stars compared to AGB stars \citep{choplin17letter} and might help to further assess the relative contribution of these sources (particularly in the case of the apparently single CEMP-s stars, cf. Sect.~\ref{origsinglecemp}). As first proposed in \cite{meynet10}, this work predicts that some CEMP stars are enriched in helium (Sect.~\ref{herich}). Detecting helium in CEMP stars is probably very challenging but it may be mentioned here as a possible observational perspective.
Generally, an observation campaign able to determine rather complete abundance patterns (as for HE~1045+0226 and CS~30301-015, cf. Sect.~\ref{origsinglecemp}) will help very much in choosing between the different existing models. Also, although probably tedious and challenging, radial velocity measurements of more CEMP stars over a long period of time would be very interesting \citep[][did such measurements for 22 CEMP-s and 24 CEMP-no stars]{hansen16b,hansen16a} and would help better understanding their origin.

\subsection{Linking the old and local with the old and distant Universe}\label{distantold}

To gain knowledge on the first generation of massive stars, the present work focused on their signatures in the local Universe. 
As mentioned in the introduction (Sect.~\ref{obssign}), an alternative way is to observe the distant Universe: either individual explosive/transient events or the integrated light of high redshift galaxies.
%This should be achievab the next generations of telescopes, e.g. JWST)
%An interesting project may be to use the observational constraints from the \textit{local} Universe (abundances of metal-poor stars) to predict what is happening at the edge of the observable Universe.
An interesting project would be to predict the specific signatures in the high redshift Universe of the CEMP massive source stars. %that best reproduce the abundances of observed metal-poor stars.
%In brief, the abundances of metal-poor stars are used to calibrate the characteristics (initial rotation, masses, ...) of a population of low/zero-metallicity massive stars.
It will establish a connection between the old and local Universe with the old and distant Universe. A schematic view of such a project is shown in Fig.~\ref{workflow}: the source star population that best account for observations of metal-poor stars is determined using stellar models (the aim of the present thesis, for CEMP stars) and also Galactic chemical evolution (GCE) models. GCE models are suitable to investigate the origin of \textit{normal} metal-poor stars, which probably formed from a reservoir well enriched by different previous sources and not just $\sim 1$ source star. %The present thesis was a . 
Type of supernovae, supernova rates, and emergent spectrum of this massive source star population may be estimated combining the outputs of stellar evolution modeling, GCE modeling and the predictions from a radiative transfer code. Such predictions could be compared to high redshift observations, especially observations from JWST, that may be able to catch the explosion of the first generations of massive stars \citep{whalen13c,whalen13a,whalen13b}.

  \begin{figure*}[t]
  \centering
     \includegraphics[scale=0.65, trim = 0.5cm 0cm 0cm 0cm]{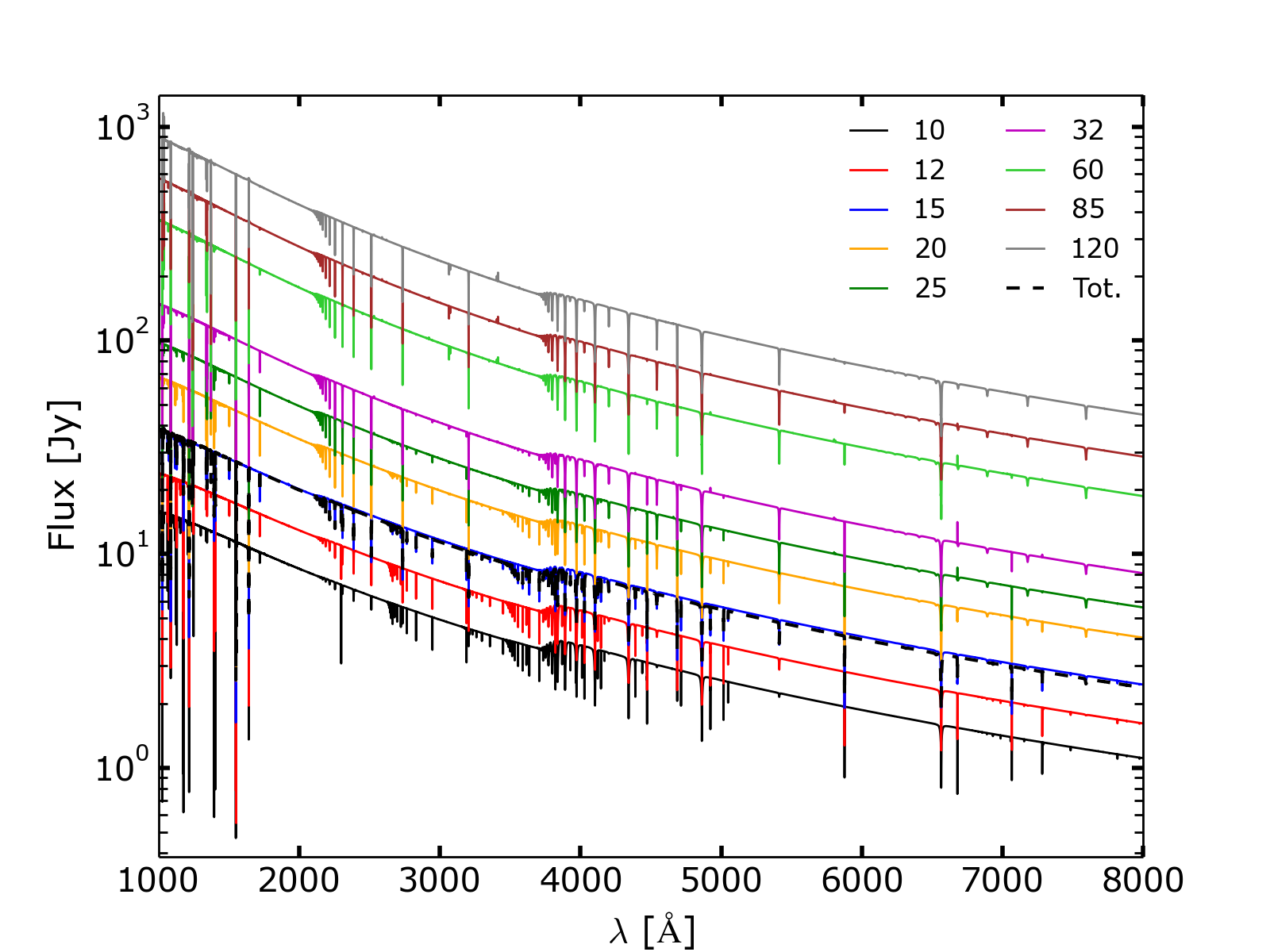}
   \caption[Emergent spectra of massive non-rotating models]{Emergent flux in Janskys (a distance of 1 kpc is assumed) as a function of the wavelength for non-rotating $10-120$~$M_{\odot}$ models at the middle of the main sequence (H in the core is half burnt).  The dashed line shows the integrated spectrum, where a constant star formation rate was assumed and where the mass function of \cite{chabrier03} was used (for a star of a mass $m$, $\phi(m) = Am^{-\alpha}$ with $A = 7.1~10^{-5}$ and $\alpha = 2.3$).}
\label{fluxs0}
    \end{figure*}

%In brief, knowing the population of low/zero metallicity massive stars that can best reproduce the old component of our Milky Way, what  The present work aimed at answering this question. 

During a Short Term Scientific Mission\footnote{STSM, in the frame of the COST action ChETEC (\href{http://www.chetec.eu/}{http://www.chetec.eu/}).} of two weeks with Pr. Jose Groh (Trinity college, Dublin), we started to work on one aspect of this project. %I computed a grid 
%As we have seen, fast rotation in the CEMP massive source stars is an interesting ingredient to explain the peculiar abundances of CEMP stars.
%Then a
%As a first step, 
As a very first step, I tried to see what would be the emergent spectrum of a population of main sequence fast rotating massive source stars compared to a population of main sequence non-rotating massive source stars.
I extended the small grid of massive stars presented in Sect.~\ref{seclate}. In addition to the 20, 32 and 60~$M_{\odot}$ models, I computed 10, 12, 15, 25, 85 and 120~$M_{\odot}$ models. The initial rotation is either 0 or 70~\% of the critical velocity. 
Only the main sequence was computed. %If a population of massive stars is observed, most of them should be in that stage since it is the longest one.
%Also, the late mixing process was not considered. This process only affects a small region in the star and very likely cannot alter significantly the emergent spectrum. 
To compute the output spectra of the models, the atmospheric radiative transfer code CMFGEN \citep{hillier98} is used. CMFGEN is a spherically-symmetric code that computes line and continuum formation in non-local thermodynamical equilibrium. Fig~\ref{fluxs0} shows the emergent spectra of the non-rotating models at the middle of the main sequence (H in the core is half burnt). Also the integrated spectrum is shown by the dashed line.
After normalizing the spectra with their continuum, comparisons can be done with the rotating population.
A detailed comparison of specific lines (e.g. HeI $\lambda$6680 or HeII $\lambda$4687) between the different models remains to be done.

%Of course here is a very first step. 
In the future, after the analysis of these first results, it would be worth computing a more realistic emergent integrated spectrum by calculating more CMFGEN spectra (at different stages of the main sequence but also after the main sequence), by considering a non-constant star formation rate (e.g. a burst of star formation), by maybe considering additional rotational velocities, by introducing some noise and convolving the spectra with a Gaussian (corresponding to a given spectral resolution) so as to predict which kind of observing facility may be able to detect a given spectral feature predicted by the models. % in the models.

}

{
\appendix

\chapter{Numerical aspects}
\label{Cnum}

%== ROTATION ==================================================================
%\section{Section \label{snumrot}}

%I will here recall the main developments of the effects of rotation for their inclusion in the Geneva code. The details of the derivations can be found in the series of papers published by the Geneva group under the generic name: ``Stellar evolution with rotation''.

%----------------------------------------------------------------------------------------------------------------------------------------
\section{Axion cooling\label{snumchem}}
The energy loss by axions can be written as
\begin{equation}
\epsilon_{\rm ax}=K g_{10}^2 T_8^7\rho_3^{-1} Z(\xi^2) C
\label{q:loss}
\end{equation}
where $K$ and $g_{10}$ are constants, $T_8 = 10^{-8}~T$, $\rho_3^{-1} = 10^{-3}~\rho$ and $Z(\xi^2) = A(\xi^2)\ln (B(\xi^2))$ with
\begin{equation}
A(\xi^2) = \frac{1.037 \xi^2}{1.01 + \xi^2 / 5.4} + \frac{1.037 \xi^2}{44 + 0.628 \xi^2} \hspace{0.5cm}	\text{and}  \hspace{0.5cm} B(\xi^2) = 3.85 + \frac{3.99}{\xi^2}
\end{equation}
and where
\begin{equation}
\xi\equiv\frac{\hbar\mathrm{c}\;k_S}{2\mathrm{k_B}T}.
\label{q:xi}
\end{equation}
$\hbar$ is the reduced planck constant, $\rm c$ the speed of light and $\rm  k_B$ the Boltzmann constant. The Debye--H\"uckel screening wavenumber $k_S$, is given by~\cite{raffelt08} as
\begin{equation}
k_S^2\equiv4\pi\alpha\left(\frac{\hbar\mathrm{c}}{\mathrm{k_B}T}\right)\sum_{i=e,{\rm ions}}{n_i}Z_i^2
\label{q:debye}
\end{equation}
where $\alpha$ is the fine structure constant, $Z_i$ the atomic number and $n_i$ the ions/electron number densities.
%\begin{equation}
%n_{\rm ions}=\rho\frac{X_i}{A_i}\mathcal{N}_A\;\mathrm{and}\;n_\mathrm{e}=\sum_{i={\rm ions}}{n_i}Z_i,
%\end{equation}
%with $\mathcal{N}_A$ the Avogadro number, 
The term $C$ in Eq.~\ref{q:loss} is equal to $C = \exp(-(\hbar \omega_0) / (k_B T))$ with $\omega_0$ the plasma frequency defined as
\begin{equation}
\omega_0 \equiv \left[4\pi\alpha\left(\frac{\hbar\mathrm{c}}{m_e c^2}\right)n_e\right]^\frac{1}{2}\mathrm{c}.
\label{q:damp}
\end{equation}

In \textsc{genec}, I have included the energy loss by axions $\epsilon_{\rm ax}$ by adding it to $\epsilon_{ \nu}$, the energy loss by neutrino. Axion losses are considered if the parameter \texttt{iaxions=1}. When solving the stellar structure equations with the Henyey method, the derivatives of $\epsilon_{ \nu}$ and $\epsilon_{\rm ax}$ as a function of $P$ and $T$ are required for convergence. Most of the time, $\epsilon_{\rm ax} \ll \epsilon_{\nu}$, so that $\epsilon_{\rm ax}$ can be neglected. At the end of He-burning however, axion losses become the dominant source of energy loss and must be included in the energy derivatives. In the case of \textsc{genec}, the following terms have to be evaluated:
\begin{equation}
\left(\frac{\partial \ln \epsilon}{\partial \ln P}\right)_{T} \hspace{0.5cm}	\text{and}  \hspace{0.5cm}  \left(\frac{\partial \ln \epsilon}{\partial \ln T}\right)_{P}
\label{enuept}
\end{equation}
where $\epsilon = \epsilon_{\nu} + \epsilon_{\rm ax}$. The term on the left in \ref{enuept} can be written as
\begin{equation}
\left(\frac{\partial \ln (\epsilon_{\nu} + \epsilon_{\rm ax} )}{\partial \ln P}\right)_{T} = \frac{\epsilon_{\nu}}{\epsilon_{\nu} + \epsilon_{\rm ax} } \left(\frac{\partial \ln \epsilon_{\nu}}{\partial  \ln P}\right)_{T} +  \frac{ \epsilon_{\rm ax}}{\epsilon_{\nu} + \epsilon_{\rm ax} } \left(\frac{\partial \ln \epsilon_{\rm ax}}{\partial  \ln P}\right)_{T}.
\end{equation}
The derivative of $\epsilon_{\rm ax}$ as a function of $P$ is equal to
\begin{equation}
\left(\frac{\partial \ln \epsilon_{\rm ax}}{\partial  \ln P}\right)_{T} = \left(\frac{\partial \ln \epsilon_{\rm ax}}{\partial  \ln \rho}\right)_{T} \left(\frac{\partial \ln \rho}{\partial  \ln P}\right)_{T}  = \left[ \frac{\rho}{Z(\xi^2)} \left( \frac{\partial Z(\xi^2)}{\partial \rho}\right)_{T} + \frac{\rho}{C}\left( \frac{\partial C}{\partial \rho} \right)_T -1 \right] \left(\frac{\partial \ln \rho}{\partial  \ln P}\right)_{T}
\label{eaxp}
\end{equation}
%\begin{equation}
%\left(\frac{\partial \ln \epsilon_{\rm ax}}{\partial \ln \rho}\right)_{T} \hspace{0.5cm}	\text{and}  \hspace{0.5cm}  \left(\frac{\partial \ln \epsilon_{\rm ax}}{\partial \ln T}\right)_{\rho}.
%\end{equation}
%In what follows we explicit the derivatives of $\epsilon_{\rm ax}$ as a function of $\rho$ and $T$. In the expression of $\epsilon_{\rm ax}$, 
%where the term in $\exp$ in the expression of $\epsilon_{\rm ax}$ (Eq.~\ref{q:loss}) was neglected for calculating the derivative since it always stays around 1 in our models. 
At this point, the derivatives of $Z(\xi^2)$ and $C$ have to be calculated:
\begin{equation}
\left( \frac{\partial Z(\xi^2)}{\partial \rho}\right)_{T} =  \left(\frac{\partial A(\xi^2)}{\partial \rho}\right)_{T} \ln (B(\xi^2)) + \left[ A(\xi^2) \frac{1}{B(\xi^2)} \left(\frac{\partial B(\xi^2)}{\partial \rho}\right)_{T} \right].
\label{zxi2}
\end{equation}
The derivatives of $A$ and $B$ as a function of $\rho$ are
\begin{equation}
\left(\frac{\partial A(\xi^2)}{\partial \rho}\right)_{T} = 1.037 \left(\frac{\partial \xi^2}{\partial \rho}\right)_{T} \left[ \frac{ 1.01 }{(1.01 + \xi^2 / 5.4)^2}    +      \frac{ 44}{(44 + 0.628\xi^2 )^2} \right]
\label{axi2}
\end{equation}
and
\begin{equation}
\left(\frac{\partial B(\xi^2)}{\partial \rho}\right)_{T} = - \frac{3.99}{\xi^4} \left(\frac{\partial \xi^2}{\partial \rho}\right)_{T} 
\label{bxi2}
\end{equation}
with
\begin{equation}
 \hspace{0.5cm} \left(\frac{\partial \xi^2}{\partial \rho}\right)_{T} = \frac{\xi^2}{\rho}
 \label{xirho2}
\end{equation}
since $\xi^2 \propto k_S^2 \propto n \propto \rho$. The derivative of $C$ in Eq.~\ref{eaxp} is non-zero because $C \propto \omega_0 \propto n_e \propto \rho$. It yields
\begin{equation}
\left( \frac{\partial C}{\partial \rho} \right)_T = -C\frac{\hbar \omega_0}{2 \rho k_B T}.
\end{equation}
A similar derivation has to be done for the second term of \ref{enuept}. Then, $\epsilon_{\rm ax}$ has to be derived as a function of $T$ (with $P$ constant but not $\rho$, so that $\rho$ has to be derived as a function of $T$). It yields
\begin{equation}
\left(\frac{\partial \ln \epsilon_{\rm ax}}{\partial  \ln T}\right)_{P} =  7 + \frac{T}{Z(\xi^2)} \left(\frac{\partial Z(\xi^2)}{\partial  T}\right)_{P} + \frac{T}{C}\left( \frac{\partial C}{\partial T} \right)_P  - \left(\frac{\partial \ln \rho}{\partial  \ln T}\right)_{P}.
\label{eaxt}
\end{equation}
The derivative of $Z(\xi^2)$ as a function of $T$ is evaluated similarly to Eq.~\ref{zxi2}, \ref{axi2} and \ref{bxi2}. Then, as in Eq.~\ref{xirho2}, one needs the derivative of $\xi$ but this time as a function of $T$:
\begin{equation}
 \left(\frac{\partial \xi^2}{\partial T}\right)_{P} = \frac{\xi^2}{\rho} \left( \frac{\partial \rho}{\partial T} \right)_{P} - \frac{3\xi^2}{T} = \frac{\xi^2}{T} \left[ \left( \frac{\partial \ln \rho}{\partial \ln T} \right)_{P}  - 3\right].
\end{equation}
Finally, the derivative of $C$ as a function of $T$ is
\begin{equation}
\left( \frac{\partial C}{\partial T} \right)_P = -C\frac{\hbar \omega_0}{2 \rho k_B T^2} \left[ \frac{1}{2}\left( \frac{\partial \ln \rho}{\partial \ln T} \right)_P -1 \right].
\end{equation}
From this point, all quantities are defined or available in the code so that Eq.~\ref{eaxp} and \ref{eaxt} can be calculated and the two quantities in \ref{enuept} evaluated.

\section{Secular shear mixing\label{dshear}}

%We recall here the other prescription for secular shear existing in the literature \citep[from][]{maeder97}. We discuss also the recent prescription of \cite{maeder13}, that includes various instabilities, and that I included in \textsc{genec}.

The $D_{\rm shear}$ used in this work is the one of \citet[][Eq.~\ref{dshtz97}]{talon97}. 
The other $D_{\rm shear}$ coefficient from \cite{maeder97} can be written as

\begin{equation}
%D_{\rm shear}^{\rm M97} = f_{energ} \frac{H_p}{g\delta} \frac{ K }{(\nabla_{ad} - \nabla) + \frac{\varphi}{\delta}\nabla_\mu } \left(\frac{9 \pi}{32} \Omega \frac{d \ln \Omega}{d \ln r}\right)^2.
D_{\rm shear}^{\rm M97} = f_{\rm energ} \frac{H_p}{g\delta} \frac{ K }{(\nabla_{ad} - \nabla) + \frac{\varphi}{\delta}\nabla_\mu } \left(\frac{9 \pi}{32} \Omega \frac{d \ln \Omega}{d \ln r}\right)^2.
\label{dshm97}
\end{equation}
%with $0\leq f_{\rm energ} \leq1 $, the fraction of the excess energy in the shear that contributes to mixing. It is generally calibrated on observations (e.g. set to reproduce the surface N/H of Galactic B-type at solar metallicity \textcolor{red}{REF}).
with $f_{\rm energ}$ the fraction of the excess energy in the shear that contributes to mixing. It is generally calibrated on observations (cf. Sect.~\ref{transportchem}).

It also exists other instabilities (e.g. thermohaline instability...) that, when included in 1D stellar evolution codes, are generally translated as individual diffusion coefficients. The global diffusion coefficient is then expressed as the sum of the coefficients of the various effects considered independently. However, it can exist physical interactions between the different instabilities, possibly leading to amplification or damping effects. \cite{maeder13} have proposed a global stability criterion, taking into account different instabilities together (thermohaline instability, shear instability, Rayleigh-Taylor instability, semiconvective instability, the instability characterized by the Solberg-H\text{\o}iland criterion and the baroclinic instability). From the stability criterium, I have derived the global diffusion coefficient and included it \textsc{genec}. 
The global stability criterium is written as

\begin{equation}
Ax^2 + Bx + C > 0
\label{poly}
\end{equation}

\noindent with\\

$A = N_{ad}^2 + N_{\mu}^2 + N_{\Omega - \delta v}^2$

$B = N_{ad}^2 D_h + N_{\mu}^2 (K+D_h) + N_{\Omega - \delta v}^2 (K+2D_h)$

$C = N_{\Omega - \delta v}^2 (D_h K+D_h^2)$\\

\noindent and where \\

$N_{ad}^2 = \displaystyle\frac{g \delta}{H_p}(\nabla_{ad} - \nabla)$  

$N_{\mu}^2 = \displaystyle\frac{g \varphi}{H_p}\nabla_{\mu}$

$N_{\Omega - \delta v}^2 = \displaystyle\frac{1}{\bar{\omega}^3} \frac{d(\Omega^2 \bar{\omega}^4)}{d\bar{\omega}} \sin \theta - Ri_c \left( \frac{dU}{dr} \right)^2 $\\

\noindent with $\bar{\omega}$ the distance to the rotational axis. It can be written $\bar{\omega} = r \sin \theta$ where $\theta$ is the colatitude. $U$ can be expressed as $U = r \Omega \sin \theta$. Let us express the first term of $N_{\Omega - \delta v}^2$:

\begin{equation}
\begin{split}
\frac{1}{\bar{\omega}^3} \frac{d(\Omega^2 \bar{\omega}^4)}{d\bar{\omega}} \sin \theta & =  \frac{1}{r^3 \sin^3 \theta}\frac{d(\Omega^2 r^4 \sin^4 \theta)}{d r}\frac{dr}{d \bar{\omega}}\sin \theta\\ 
&= 2\Omega \sin \theta \left( r\frac{ d\Omega}{dr} + 2\Omega \right).
\end{split}
\end{equation}

\noindent It yields

\begin{equation}
N_{\Omega - \delta v}^2 = 2\Omega \sin \theta \left( r\frac{ d\Omega}{dr} + 2\Omega \right) - Ri_c r^2 \sin^2 \theta\left(  \frac{d\Omega}{dr} \right)^2.
\end{equation}
In numerical calculations, one has to average $N_{\Omega - \delta v}^2$ over an isobar. We obtain

\begin{equation}
\begin{split}
\overline{N_{\Omega - \delta v}^2} & = \frac{\int_{0}^{\pi} N_{\Omega - \delta v}^2 \sin^3 \theta  d \theta}{\int_{0}^{\pi} \sin^3 \theta d \theta}  \\ &= 2\Omega \left( r\frac{ d\Omega}{dr} + 2\Omega \right)  \frac{\int_{0}^{\pi} \sin^4 \theta  d \theta}{\int_{0}^{\pi} \sin^3 \theta d \theta}  - Ri_c r^2  \left( \frac{d\Omega}{dr} \right)^2 \frac{\int_{0}^{\pi} \sin^5 \theta  d \theta}{\int_{0}^{\pi} \sin^3 \theta d \theta}\\
&= \frac{9 \pi}{16}\Omega^2 \left( \frac{d \ln \Omega}{d \ln r} + 2 \right) - \frac{4}{5}Ri_c \Omega^2 \left( \frac{d \ln \Omega}{d \ln r}\right)^2.
\end{split}
\end{equation}

The diffusion coefficient is $D_{\rm global}^{\rm M13} = (1/3) |v| l= 2|x|$ where $x$ are the two solutions of Eq. \ref{poly}. 
Test models were computed with both solutions. The first solution leads to convergence issues while the second does not. The second one also gives plausible results (Fig.~\ref{hrdico}). In \textsc{genec}, the second solution was implemented as a standard choice.

\section{Mass loss recipes\label{pulsmassloss}}

The pulsationally induced mass-loss rate derived by \cite{moriya15} are

\begin{equation}
\log_{10}\dot{M} = \left\{
    \begin{array}{ll}
        (10.38 \pm 1.58)\eta - (3.94 \pm 0.21) \text{ for } \epsilon = 0.1\\
        (15.51 \pm 2.92)\eta - (4.77 \pm 0.33) \text{ for } \epsilon = 0.3\\
        (12.55 \pm 2.88)\eta - (4.69 \pm 0.39) \text{ for } \epsilon = 0.5\\
        (5.08 \pm 2.87)\eta - (4.14 \pm 0.33) \text{ for } \epsilon = 0.8
    \end{array}
\right.
\end{equation}
where $\dot{M}$ is expressed in $M_{\odot}$ yr$^{-1}$ and with $\eta = (-8.30 \pm 0.59) \times 10^{-4}~T_{\rm eff} + (4.15 \pm 0.29)$. 
In \textsc{genec}, this mass loss is considered and added to other kinds of mass losses (mechanical, radiative) if the parameter \texttt{ipuls=1} and if $\log (T_{\rm eff}) < 3.698$ (it corresponds to $T_{\rm eff} = 4992$ K). This is the temperature below which the stellar envelope becomes pulsationally unstable \citep{moriya15}. The parameter $\eta$ is a free parameter that has to be chosen between the 4 possible values (0.1, 0.3, 0.5 or 0.8, parameter \texttt{epsil} in the code).

A small change was also done in the mass loss recipe of \citet[][\texttt{imloss=1}]{jager88} in order to correct a bug happening for very luminous stars. In the expression of $\dot{M}$, there is a term
\begin{equation}
T_j \left[ \frac{\log(L/L_{\odot})-4.6}{2.1} \right]
\label{tjjager}
\end{equation}
where $T_j$ is a function defined as $T_j(x) = \cos (j \arccos x)$. This definition implies that the term in brackets in Eq.~\ref{tjjager} must be $\leq 1$ or equivalently, $\log (L/L_{\odot}) \leq 6.7$. In extreme cases (e.g. very massive stars), it happens that this condition is not verified, leading to convergence issues. To solve this issue, I just took the minimum between 1 and the bracket value in Eq.~\ref{tjjager}.

   \begin{figure}[t]
   
      \centering
   \begin{minipage}[c]{.49\linewidth}
      \includegraphics[scale=0.59, trim= 0.4cm 0cm 0cm 0cm]{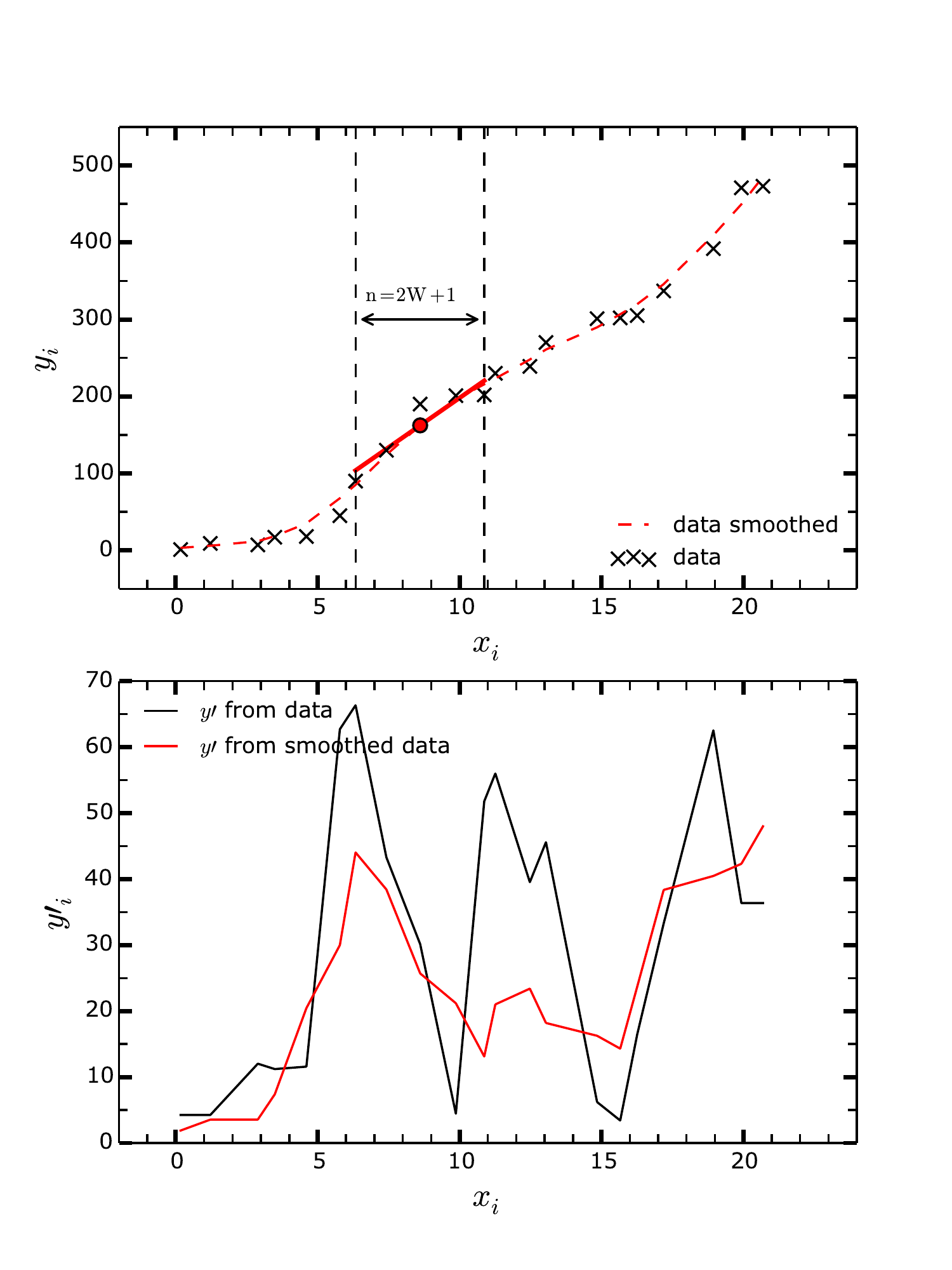}
   \end{minipage}
   \begin{minipage}[c]{.49\linewidth}
      \includegraphics[scale=0.59, trim= 0.4cm 0cm 0cm 0cm]{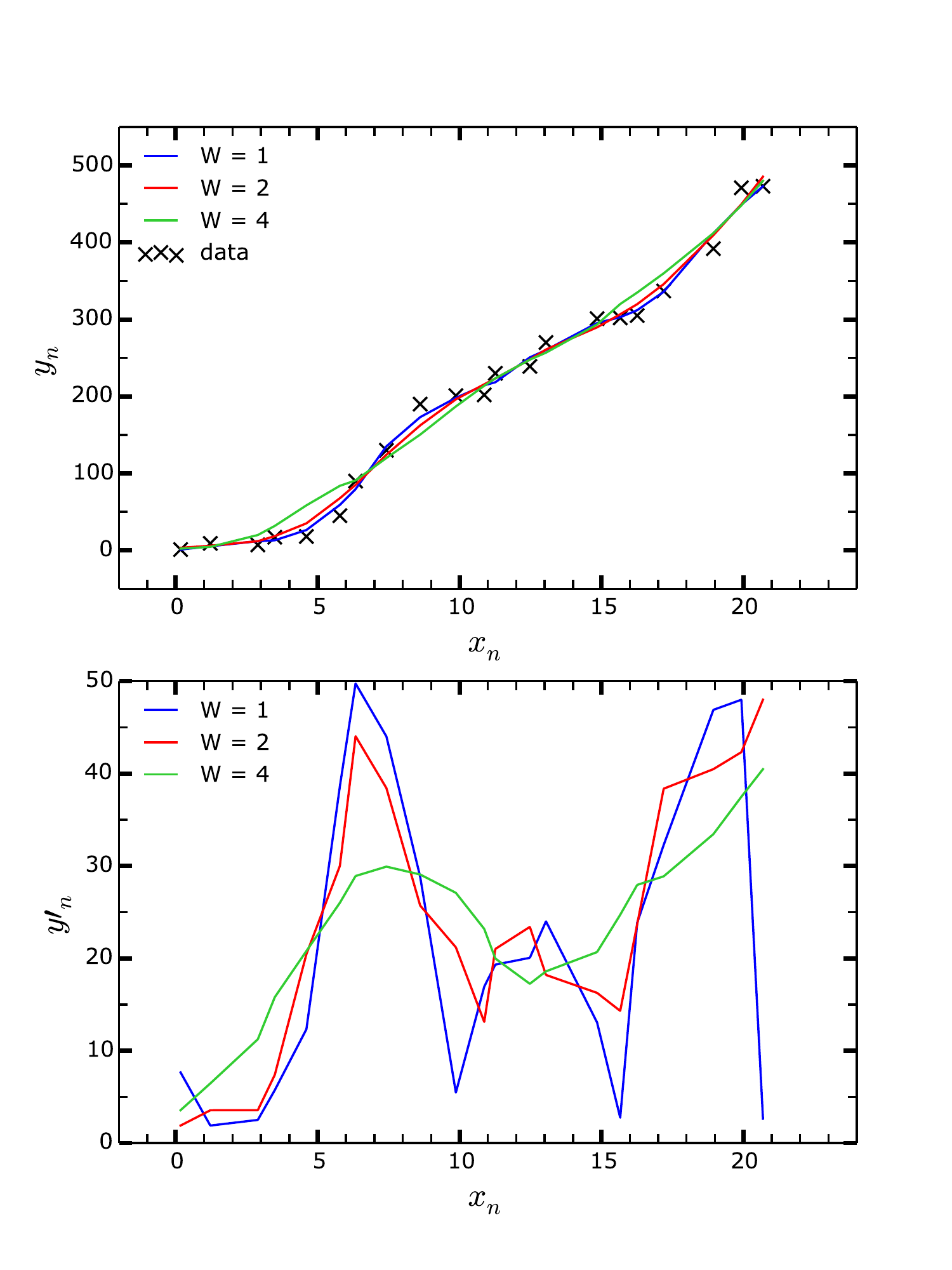}
   \end{minipage}
   \caption[Locally weighted scatterplot smoothing]{\textit{Upper left:} crosses represent the raw data. The red line shows the linear fit done around the $x_i \simeq 9$ point, considering the points between $i-W$ and $i+W$, where $W=2$.
   %$n$ is the number of points on which the fit is done. 
   The red dashed line show the resulting curve after the smoothing procedure done on each point. \textit{Lower left:} derivative of the raw data using the two neighborhood values (black) and of the smoothed data (red). \textit{Upper and lower right:}  effect of varying the size of the fitting window $W$.}
              \label{schemefit1}
    \end{figure}

  \begin{figure}[t]
   \centering
      \includegraphics[scale=0.68]{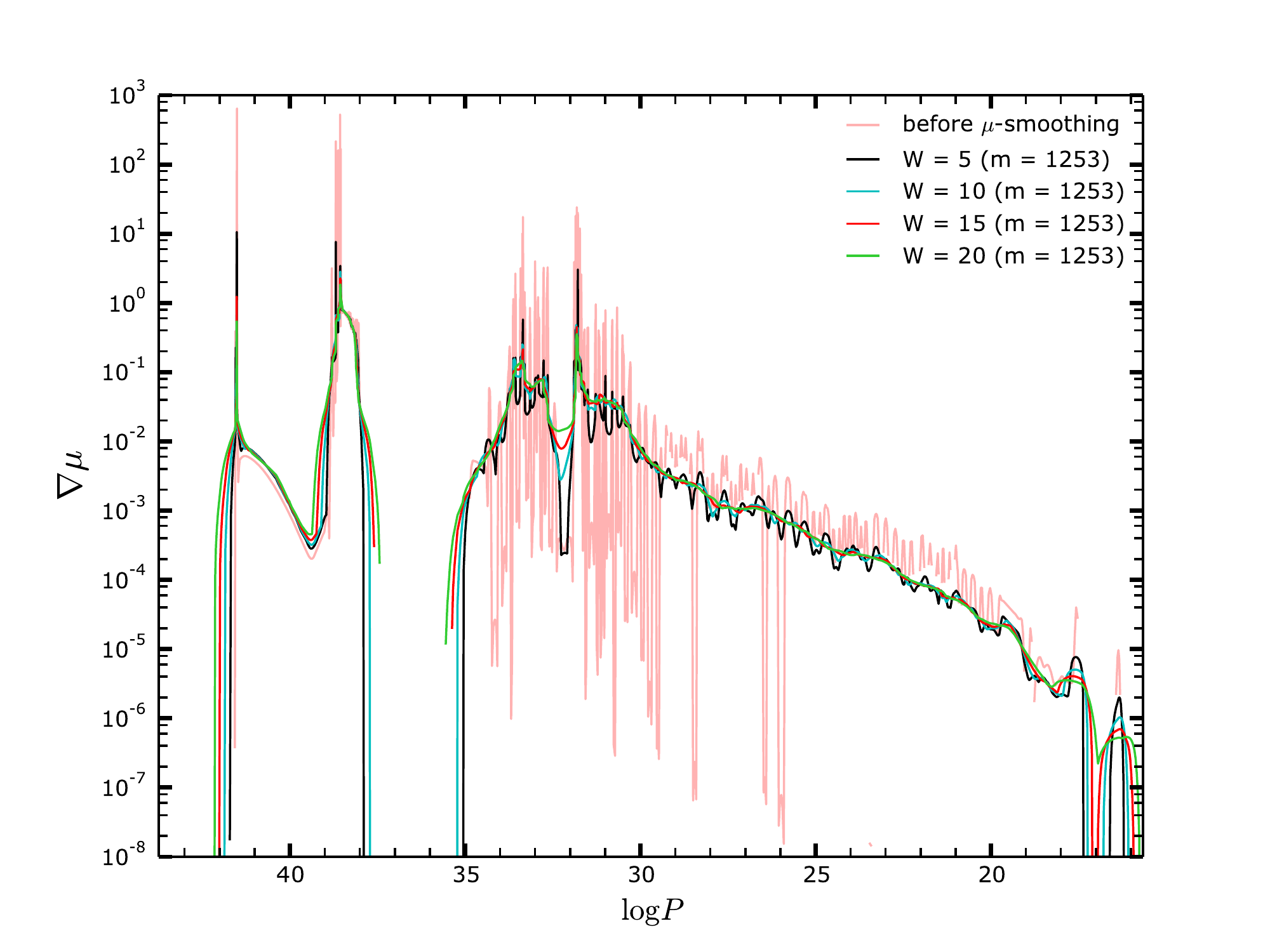}

   \caption[$\nabla \mu$ in a rotating 25 $\rm M_{\odot}$ model: the impact of smoothing]{$\nabla \mu$ as a function of $\log P$ in a rotating 25~$M_{\odot}$ model at $Z=0.001$ during the He-burning stage. The pink curve shows $\nabla \mu$ calculated with the raw $\mu$ profile. The other lines show $\nabla \mu$ calculated with the smoothing of $\mu$, with different $W$ values (see text for details). $m$ is the mesh number of the model.}
              \label{nabW}
    \end{figure}

%----------------------------------------------------------------------------------------------------------------------------------------
\section{Locally weighted scatterplot smoothing \label{oscillapp}}

%Talk about shell pb in phd, evoke smoothing to solve pb and then refer to appendix where smoothing is explained

In \textsc{genec}, a locally weighted scatterplot smoothing technique can be used for smoothing the profile of a quantity (e.g. $\mu$ or $\Omega$) when calculating its derivative. It is used at least during the main sequence of rotating models to smooth the $\Omega$, $\mu$ and $U$ (vertical component of the meridional circulation) profiles. Without it, the code may experience convergence issues. 
%The smoothing technique consists in doing a linear fit around one value (of e.g. $\mu$), (2) 
%by taking into account a given amount of neighborhood values. Once the linear fit is done, the considered value is replaced by the point on the linear fit, at the same abscissa. 
%The derivative at this point is taken equal to the slope of the linear fit. Once this step is done, we go to the next point and do the linear fit again, etc... 

Fig.~\ref{schemefit1} illustrates the technique. Let us consider $N$ data points with given coordinates $x_i$ and $y_i$ (black crosses in Fig.~\ref{schemefit1}, upper panels). Let us focus on the point at $x_i \simeq 9$ in the upper left panel of Fig.~\ref{schemefit1}. A window of $n = 2W+1$ points around that point is selected. $W$ is a parameter to chose. In this example, $W = 2$ so that $n = 5$. A linear fit with these $n = 5$ points is then performed and the derivative is taken equal to the slope of the fit. Fig.~\ref{schemefit1} (lower left) shows $y'$, the derivative of the data when considering just the two neighborhood values (black line) and the derivative if smoothing is done (red line). The smoothed data being more regular, the red profile is less sharp.

The size of the window $W$ used to smooth the profile plays an important role. The upper right panel of Fig.~\ref{schemefit1} shows the same data as well as the three smoothed profiles for $W = 1$ (blue), $W = 2$ and $W = 4$ (green). The larger the size of the window, the more damped the irregularity of the raw profile, because the fit is done using more points. It leads to smoother derivatives (Fig.~\ref{schemefit1}, lower right panel).
This smoothing technique can be seen as a filter that damps the irregularity of the $y_i$ profile. If $W$ is small, the smoothing is weak (i.e. close to the raw data), if $W$ is large, some important (physical) features in the profile can be lost.

The effect of varying $W$ in a complete stellar model is shown in Fig.~\ref{nabW}. The $\mu-$gradient calculated from just the two neighborhood values (pink line) presents strong oscillations which are likely not physical. The smoothing procedure described previously damps these features (other lines).

In \textsc{genec}, I introduced a mesh-adaptative $W$ parameter, defined as $W = m/a$ with $m$ the mesh number and $a=120$, a constant selected after carrying numerical tests with various values of $a$.

\section{S-process parameters\label{sprocparam}}

I remind here some quantities that can be used to quantify the efficiency of the s-process. These quantities are tabulated in Table~\ref{table:sprocparam} for the grid of models published in \cite{choplin18}. These quantities may be useful for future comparisons with other models. 

The central neutron exposure during core He-burning $\tau_{\rm c}$ is defined as

\begin{equation}
\tau_{\rm c} = \int_{t_{\rm ini-He}}^{t_{\rm end-He}} v_{\rm T} n_{\rm n} \text{d} t,
\label{exposure}
\end{equation}
where $t_{\rm ini-He}$ and $t_{\rm end-He}$ are the age of the model at the beginning and the end of the core He-burning phase, respectively, $n_{\rm n}$ the neutron density and $v_{\rm T} = \sqrt{2kT /m_{\rm n}}$ the thermal velocity (with $kT = 30$ keV as a typical value in the helium burning core). Because the neutron density $n_{\rm n}$ varies with time and mass coordinate in the model, it is interesting to define the neutron exposure averaged over the He-core and over the He-burning lifetime:

\begin{equation}
\langle \tau \rangle = \int_{t_{\rm ini-He}}^{t_{\rm end-He}} \langle n_{\rm n}(t) \rangle v_{\rm T} \text{d} t.
\label{meanexp}
\end{equation}
$\langle n_{\rm n}(t) \rangle$ is the neutron density at time $t$ averaged over the He-core. It is defined as
\begin{equation}
\langle n_{\rm n}(t) \rangle = \frac{1}{M_{\rm He} (t)} \int_{0}^{M_{\rm He}(t)} n_n (M_{\rm r}) \text{d}M_{\rm r}\label{meandens}
\end{equation}
where $M_{\rm He}(t)$ is the mass of the helium core at time $t$ and $n_n(M_{\rm r})$ the neutron density at coordinate $M_{\rm r}$.
%is averaged over the helium convective core. 
Finally, the average number of neutron captures per iron seed $n_{\rm c}$ is defined as \citep{kappeler90}

\begin{equation}
%n_{\rm c} = \frac{\sum\limits_{A=56}^{209} (A-56) (Y(A) - Y_{\rm 0} (A))}{\sum\limits_{Z=26} Y_{\rm 0} (A)}
n_{\rm c} = \frac{\sum\limits_{A=56}^{209} (A-56) (Y(A) - Y_{\rm 0} (A))}{ Y_{\rm 0} (^{56}\text{Fe})}
\label{capt}
\end{equation}
with $Y(A)$ and $Y_{\rm 0}(A)$ the final and initial number abundance of a nucleus with nuclear mass number A and $Y_0$($^{56}$Fe) the initial number abundance of $^{56}$Fe. %The quantities defined in Eq.~\ref{exposure}, \ref{meanexp}, and \ref{capt} are tabulated for our models in Table~\ref{table:sprocparam}.

\begin{table}
\scriptsize{
\caption[]{s-process parameters at the end of the core helium burning phase for the models in \cite{choplin18}: model label (column 1), central neutron exposure (column 2, Eq.~\ref{exposure}), neutron exposure averaged over He-core (column 3, Eq.~\ref{meanexp}), number of neutron capture per seed (Eq.~\ref{capt}) averaged over the He-core mass (column 4), maximum of the central neutron density (column 5), $^{22}$Ne burnt (column 6) and left (column 7). \label{table:sprocparam}}
\begin{center}
\resizebox{13.8cm}{!} {
%\begin{threeparttable}
\begin{tabular}{lcccccc} 
\hline % inserts double horizontal lines
\hline % inserts single horizontal line
Model  		& $\tau_c$ & $\langle{\tau}\rangle$ & $n_c$  & $n_{n,c,max}$  & $\Delta X$($^{22}$Ne) & $X_{\rm r}$($^{22}$Ne) \\ % table heading
 & [mb$^{-1}$]  &	[$10^{-1}$mb$^{-1}$]      &		&[cm${-3}$]  &	& 	\\
\hline % inserts single horizontal line
%\multicolumn{16}{c}{No rotation}\\
10s0		&0.23	&	 0.13	 & 0.13    &    8.01(5)  & 	6.73($-5$)		&	1.09($-3$)		\\
10s4		&0.68	&	 0.38 & 0.61    &      2.44(6) & 	1.69($-4$)		&	1.32($-3$)		\\
\hline
15s0		&0.93	&	0.45 	 & 0.62    &  5.07(6)    & 2.14($-4$)		& 9.46($-4$)		\\
15s4		&3.53	&	 1.82	 & 3.36    &  1.43(7)  	 & 7.66($-4$)			& 1.45($-3$)		\\
\hline
20s0		&2.10	&	0.88 	 & 1.45	    &  1.15(7)     & 	4.15($-4$)				& 7.58($-4$)	\\
20s4		&8.51	&	3.72 	 & 9.18	    &  2.66(7)     & 	2.08($-3$)				& 1.56($-3$)	\\
\hline
25s0		&3.29	&	1.29   & 2.32   &    1.71(7)  & 	5.98($-4$)		&	5.93($-4)$		\\
25s4		&12.47	& 	4.92  & 14.42  &     3.16(7) & 	3.32($-3$)		&	1.17($-3)$		\\
25s7		& 16.56	& 	6.52  & 22.54  &     7.11(7) & 	5.43($-3$)		&	1.87($-3)$		\\
25s7B$^{a}$	&  21.54	& 	8.57  & 33.19  &     4.21(7) & 	5.08($-3$)		&	1.83($-3)$		\\
\hline  
40s0		&6.16	&2.14	  &4.47	     &   2.23(7) 	& 	9.81($-4$)			&	2.64($-4$)		\\
40s4		&14.8	&5.11	  & 15.38	     &   2.89(7)	& 	3.37($-3$)			&	2.12($-4$)		\\
\hline  
60s0		&8.57	&	2.70   &6.26   	     &    2.04(7)   & 	1.24($-3$)		&	6.75($-5)$		\\
60s4		&11.06	&	3.43	  &8.56 	     &    2.15(7)	& 	1.68($-3$)		&	2.19($-5$)		\\
 \hline
85s0		&10.05	& 2.98   &  7.27    & 1.99(7)  & 	1.33($-3$)		&	1.53($-5)$		\\
85s4		&11.12	& 3.28  &   8.08   &  1.96(7) & 	1.44($-3$)		&	3.30($-6)$		\\
\hline
120s0		&11.08	& 3.19   & 8.08     &  1.94(7) & 	1.36($-3$)		&	2.99($-6)$		\\
120s0B$^{a}$&14.20	& 4.23   &  12.49    &  2.30(7) & 	1.30($-3$)		&	2.33($-6)$		\\
120s4		& 11.87	& 3.37  &  8.52    &   2.66(8)& 	1.39($-3$)		&	4.09($-7)$		\\
\hline
150s0		&11.95	& 3.38  & 8.66     &   1.99(7)& 	1.37($-3$)		&	4.49($-7)$		\\
150s4		& 12.68	& 3.52  &  9.04    &   1.94(8)& 	1.37($-3$)		&	7.07($-8)$		\\
\hline
\end{tabular}
%\begin{tablenotes}

%\end{tablenotes}
%\end{threeparttable}
}
\end{center}
}
$^{a}$ Models computed with the rate of $^{17}$O($\alpha,\gamma$) divided by 10.

\end{table}

%\chapter{Abundance fits of the most iron-poor stars}
\chapter{Abundance fits}
\label{fitstar}

%== ROTATION ==================================================================
%\section{Section \label{appstarp}}

This appendix contains the abundance fits of the 69 CEMP stars of Table~\ref{tabcemp} with the source star models presented in Sect.~\ref{seclate} and \ref{secstat}.

   \begin{figure*}[h!]
   \centering
   \begin{minipage}{.32\linewidth}
       \includegraphics[scale=0.3]{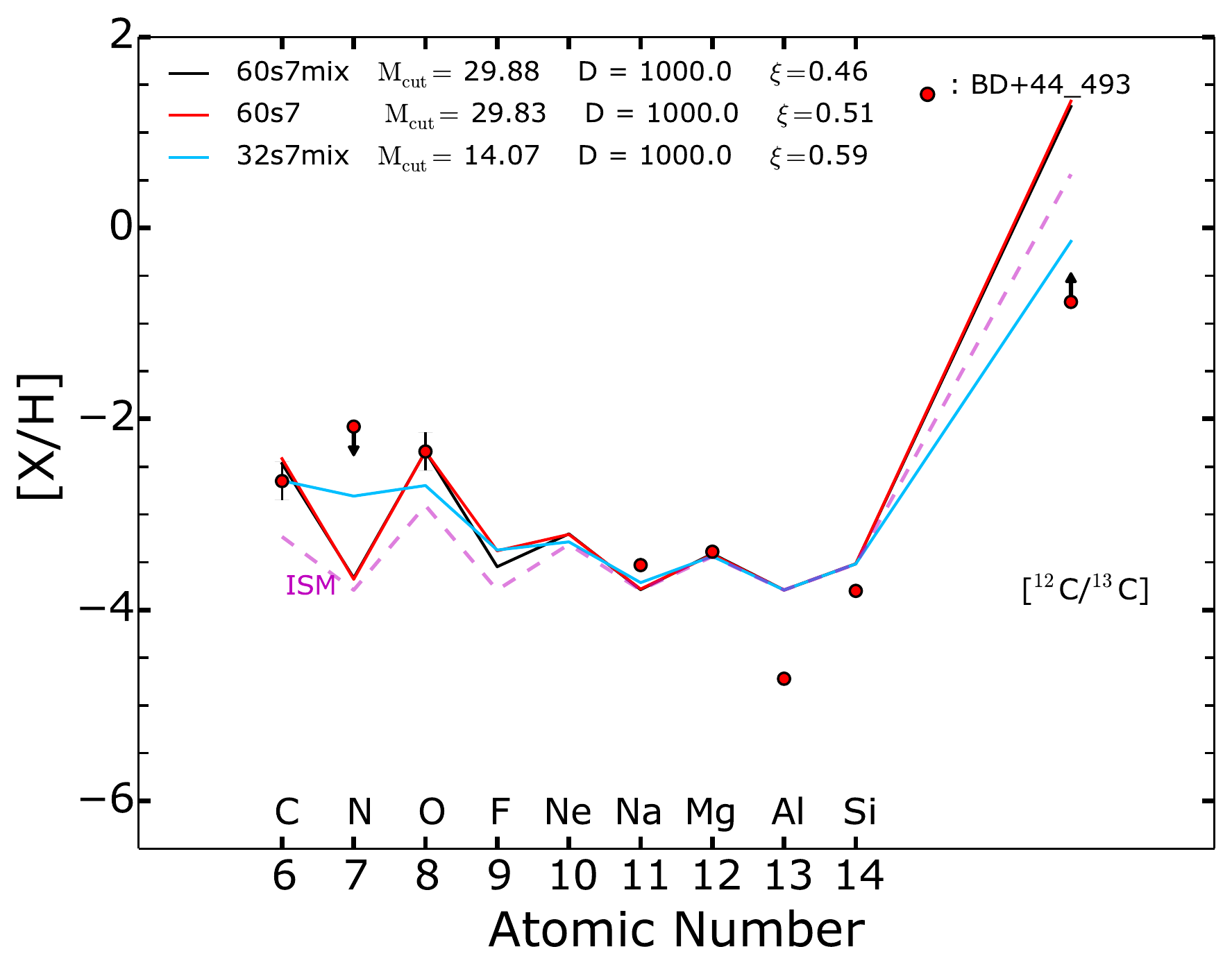}
   \end{minipage}
   \begin{minipage}{.32\linewidth}
       \includegraphics[scale=0.3]{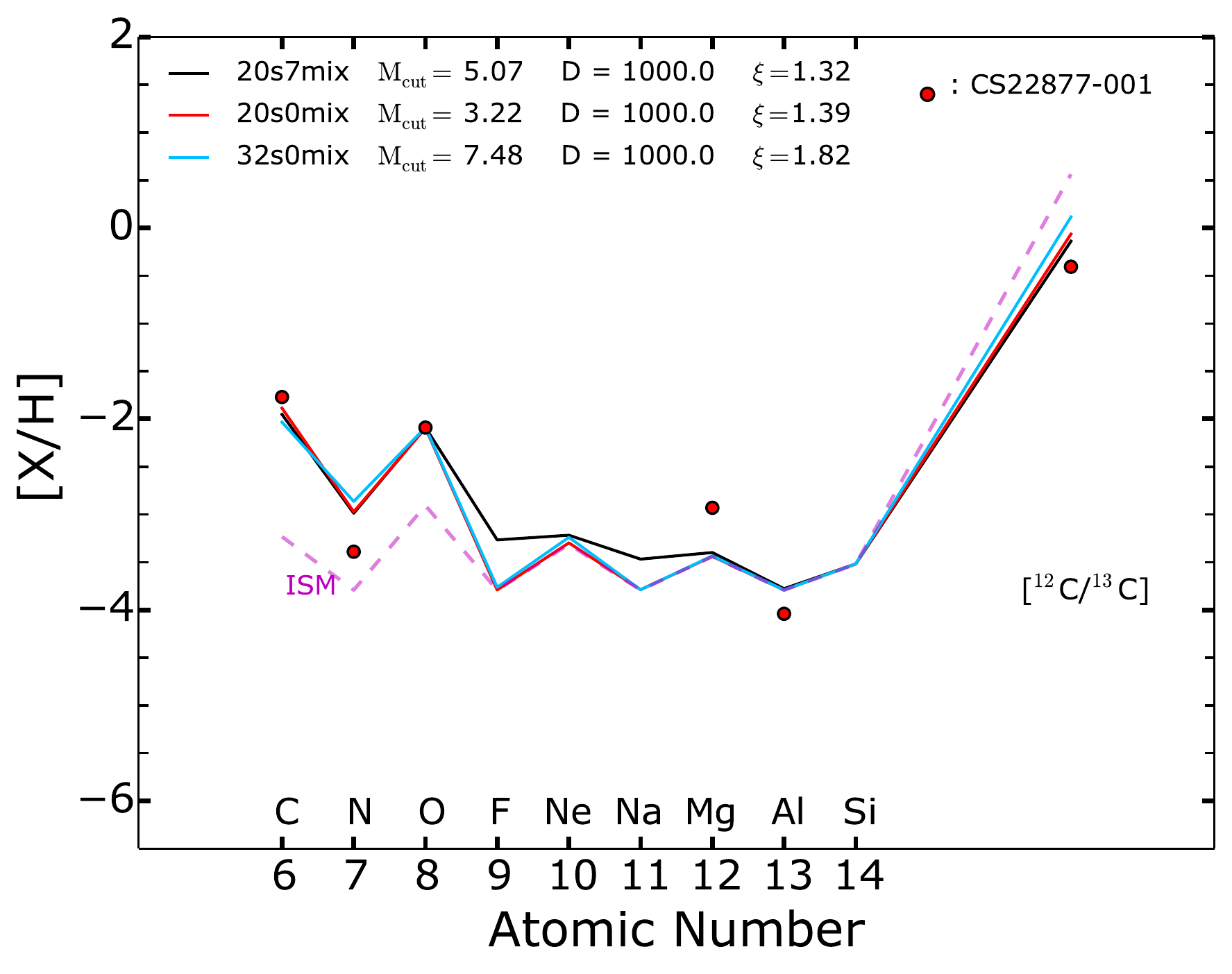}
   \end{minipage}
   \begin{minipage}{.32\linewidth}
       \includegraphics[scale=0.3]{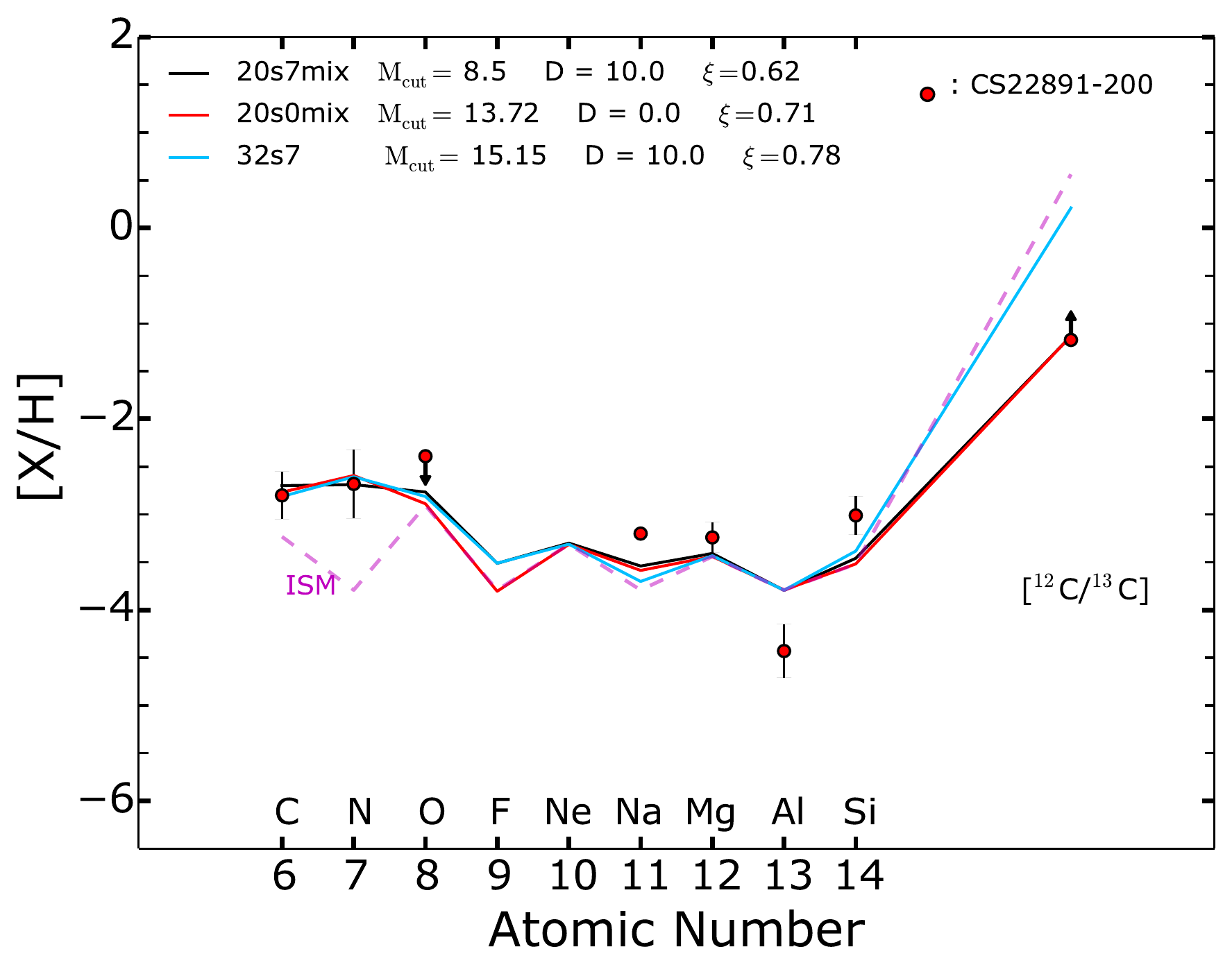}
   \end{minipage}
   \begin{minipage}{.32\linewidth}
       \includegraphics[scale=0.3]{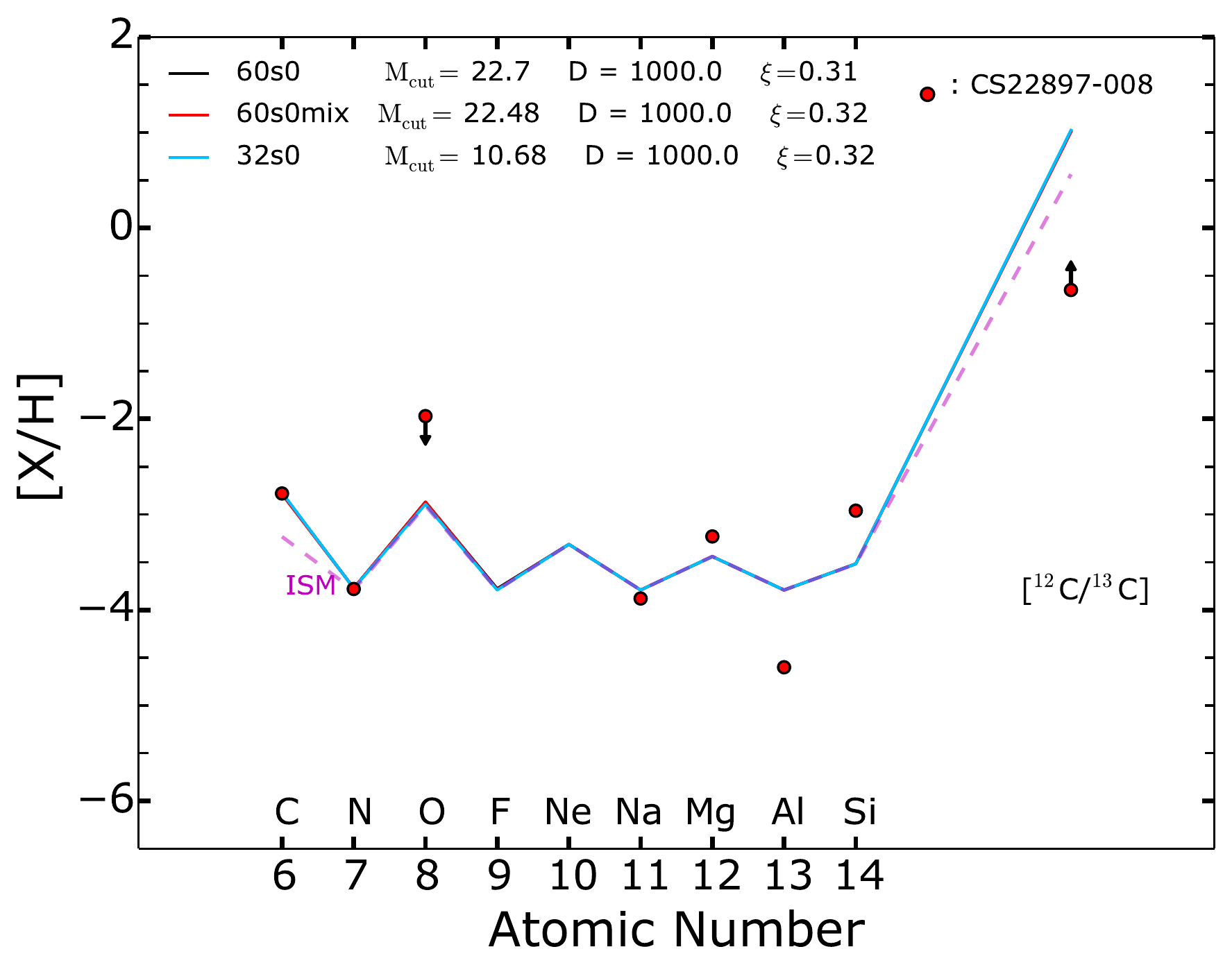}
   \end{minipage}
   \begin{minipage}{.32\linewidth}
       \includegraphics[scale=0.3]{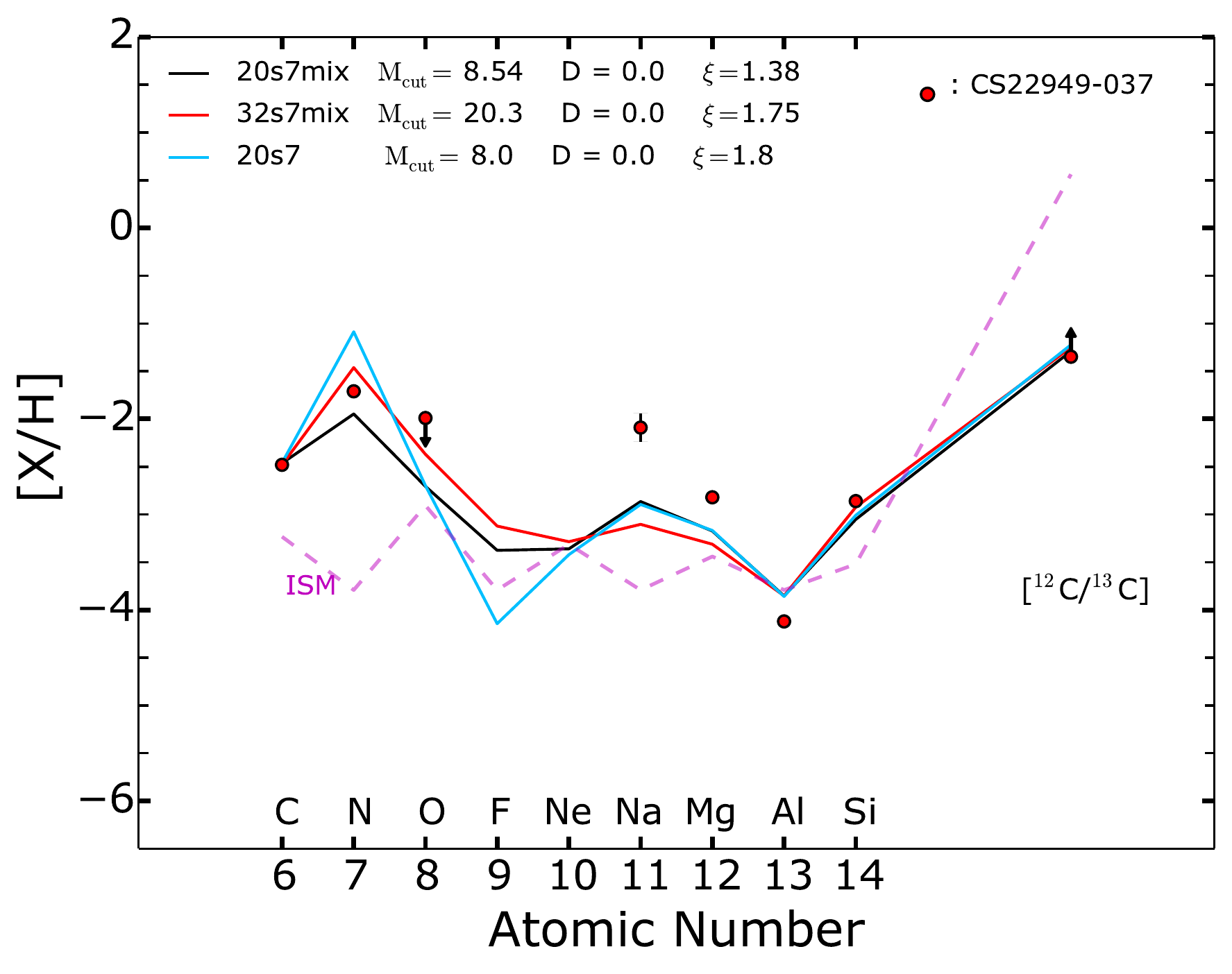}
   \end{minipage}
   \begin{minipage}{.32\linewidth}
       \includegraphics[scale=0.3]{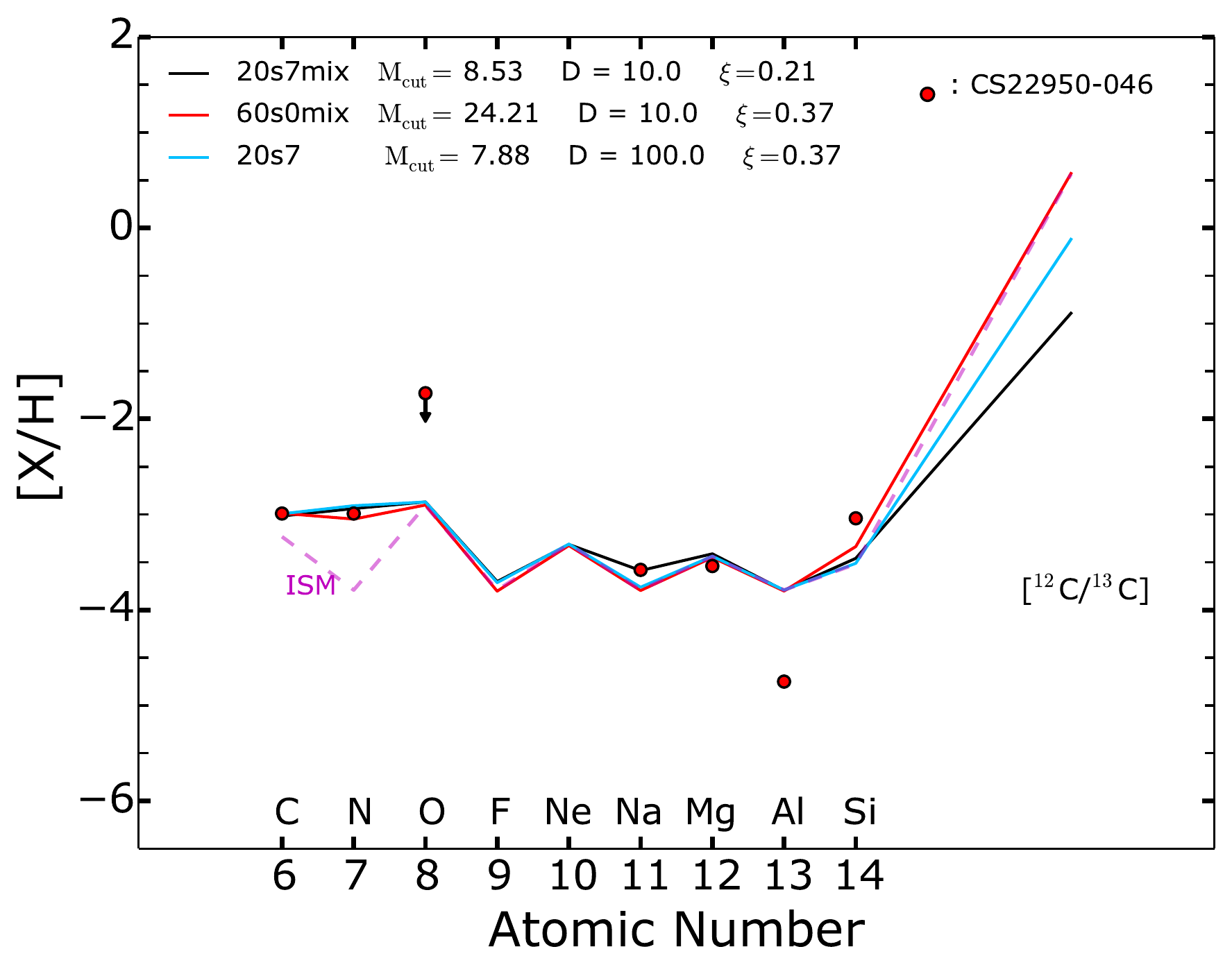}
   \end{minipage}
   \begin{minipage}{.32\linewidth}
       \includegraphics[scale=0.3]{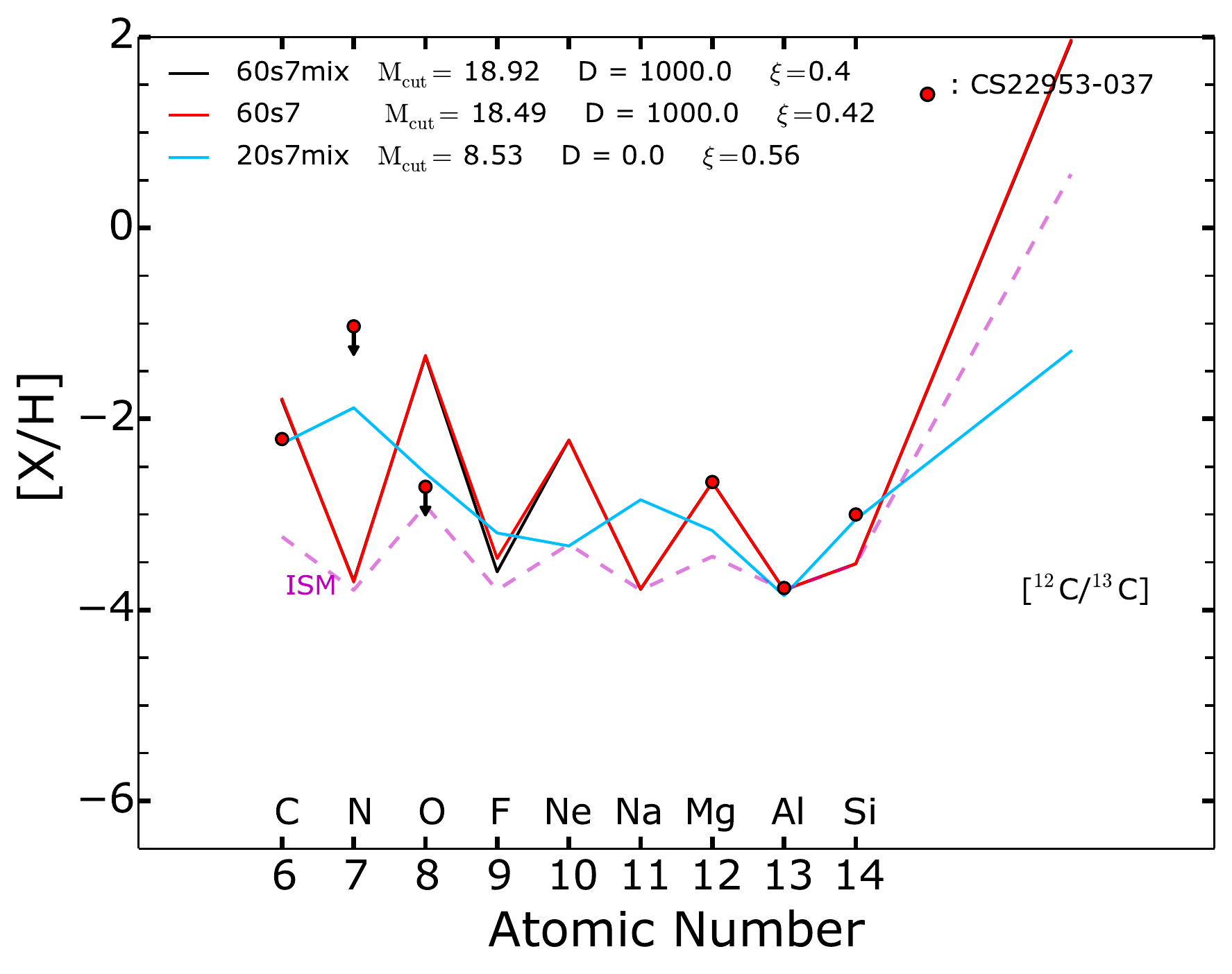}
   \end{minipage}
   \begin{minipage}{.32\linewidth}
       \includegraphics[scale=0.3]{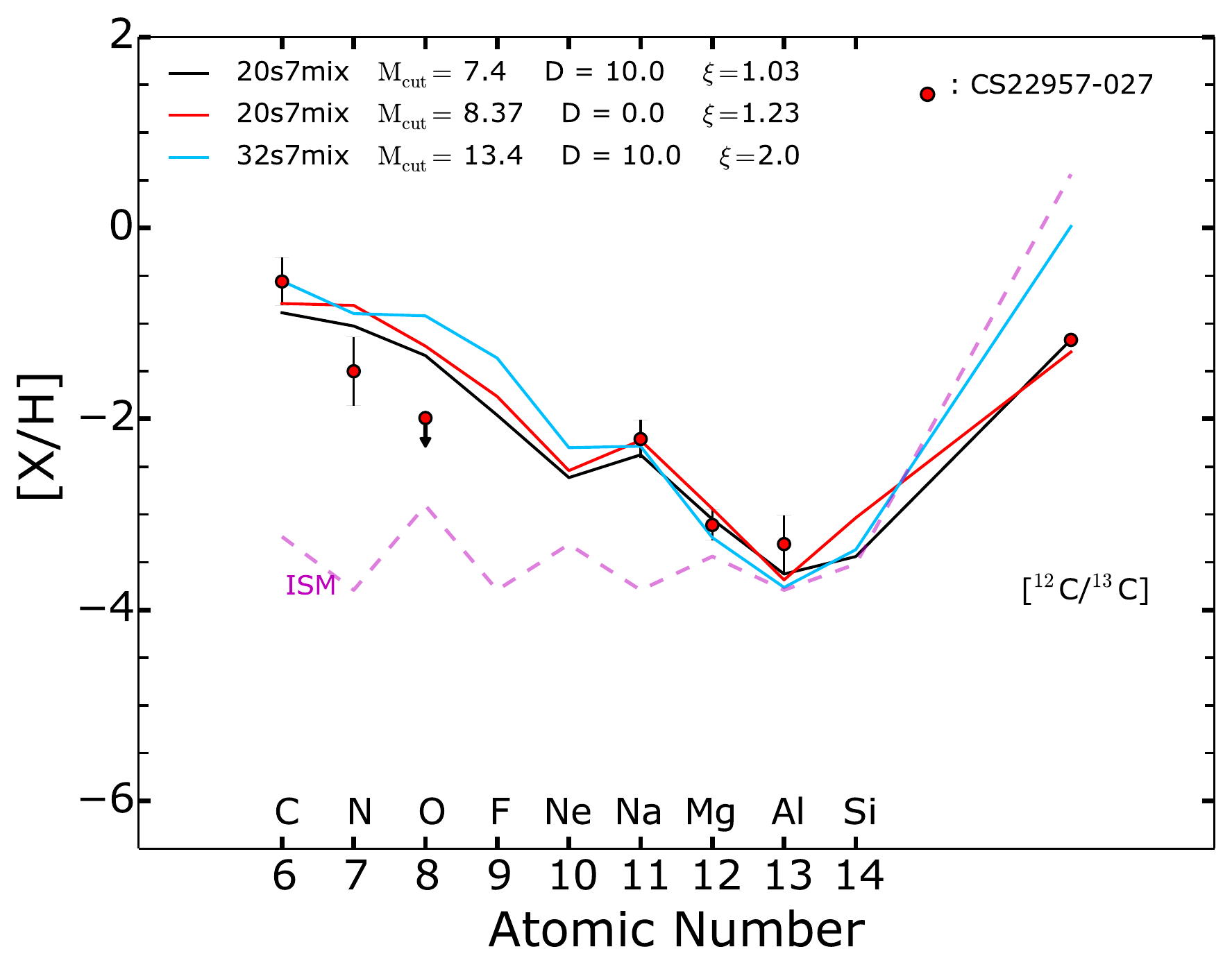}
   \end{minipage}
   \begin{minipage}{.32\linewidth}
       \includegraphics[scale=0.3]{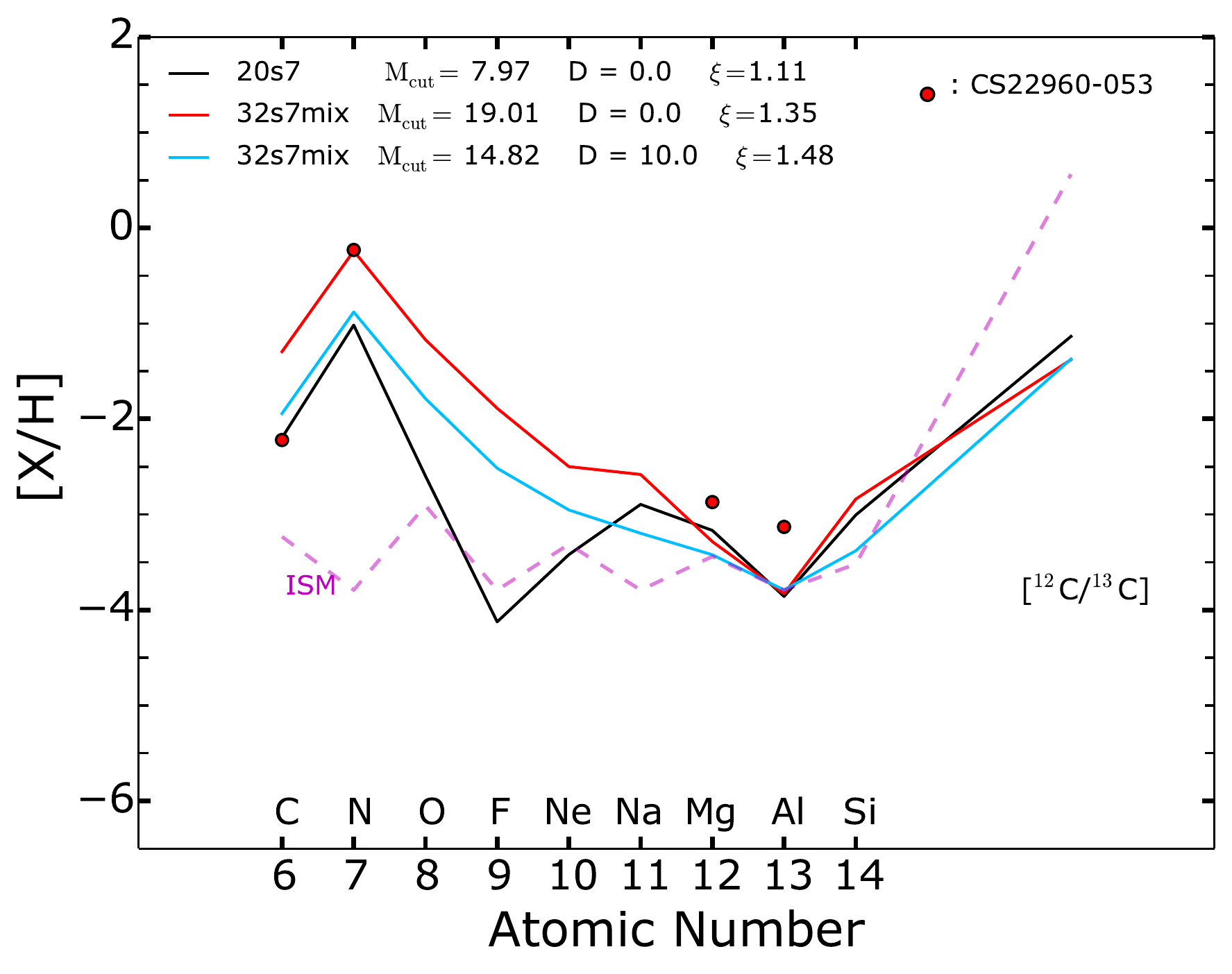}
   \end{minipage}
   \caption[Best fits of 69 CEMP stars with $20-60$ source star models]{Best fits of the 69 CEMP stars of Table~\ref{tabcemp} with the source star models of Table~\ref{modtab}, when considering the dilution factor and the mass cut as free parameters. The three best models, with the lowest $\xi$, are shown (black, red and green patterns sorted by increasing $\xi$). Uncertainties and limits on abundances are shown by vertical bars and arrows, respectively (cf. Sect.~\ref{secstat} for details).}
\label{allstar1}
    \end{figure*}

   \begin{figure*}
   \centering
      \begin{minipage}{.32\linewidth}
       \includegraphics[scale=0.3]{XH_3best_9.pdf}
   \end{minipage}
   \begin{minipage}{.32\linewidth}
       \includegraphics[scale=0.3]{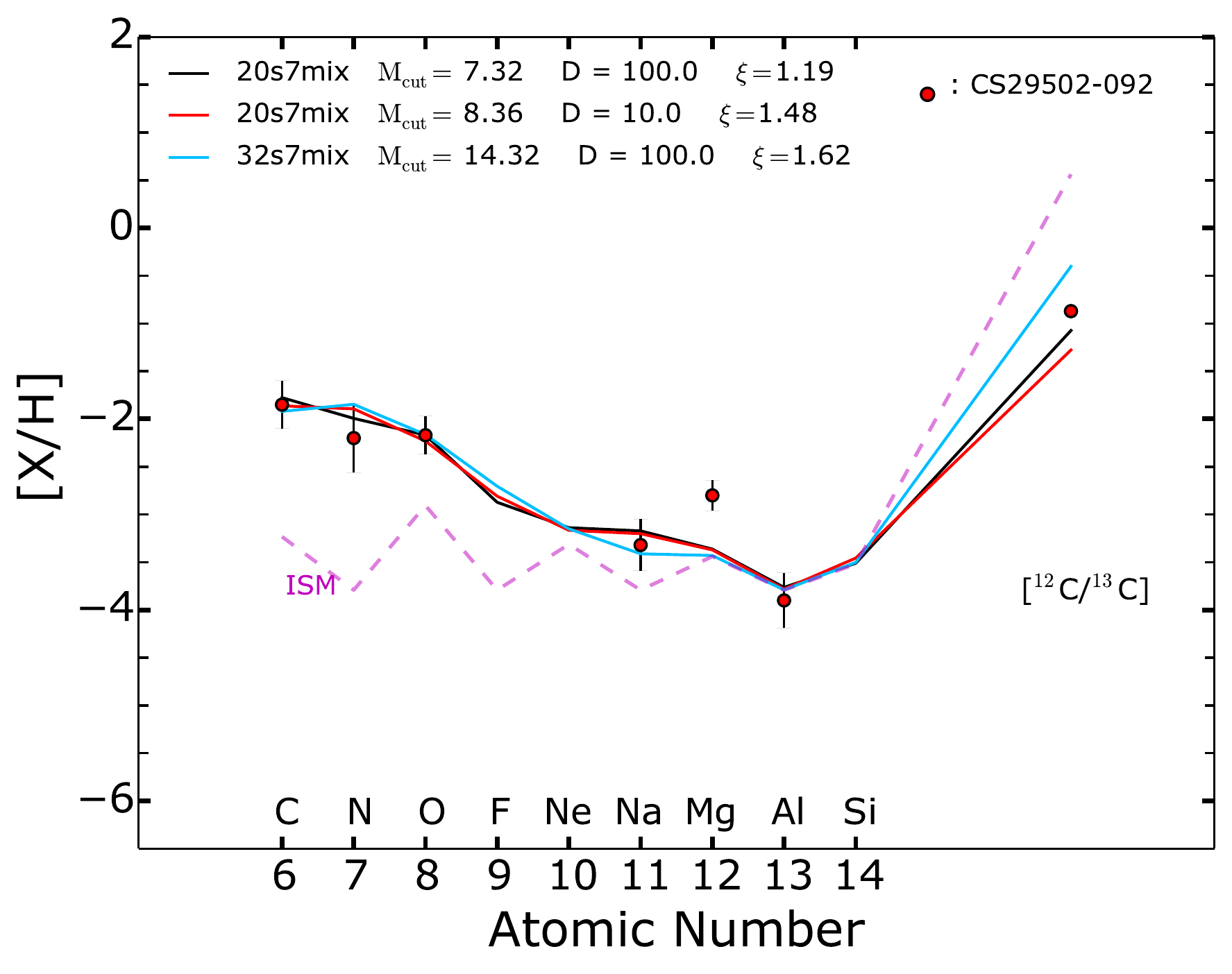}
   \end{minipage}
   \begin{minipage}{.32\linewidth}
       \includegraphics[scale=0.3]{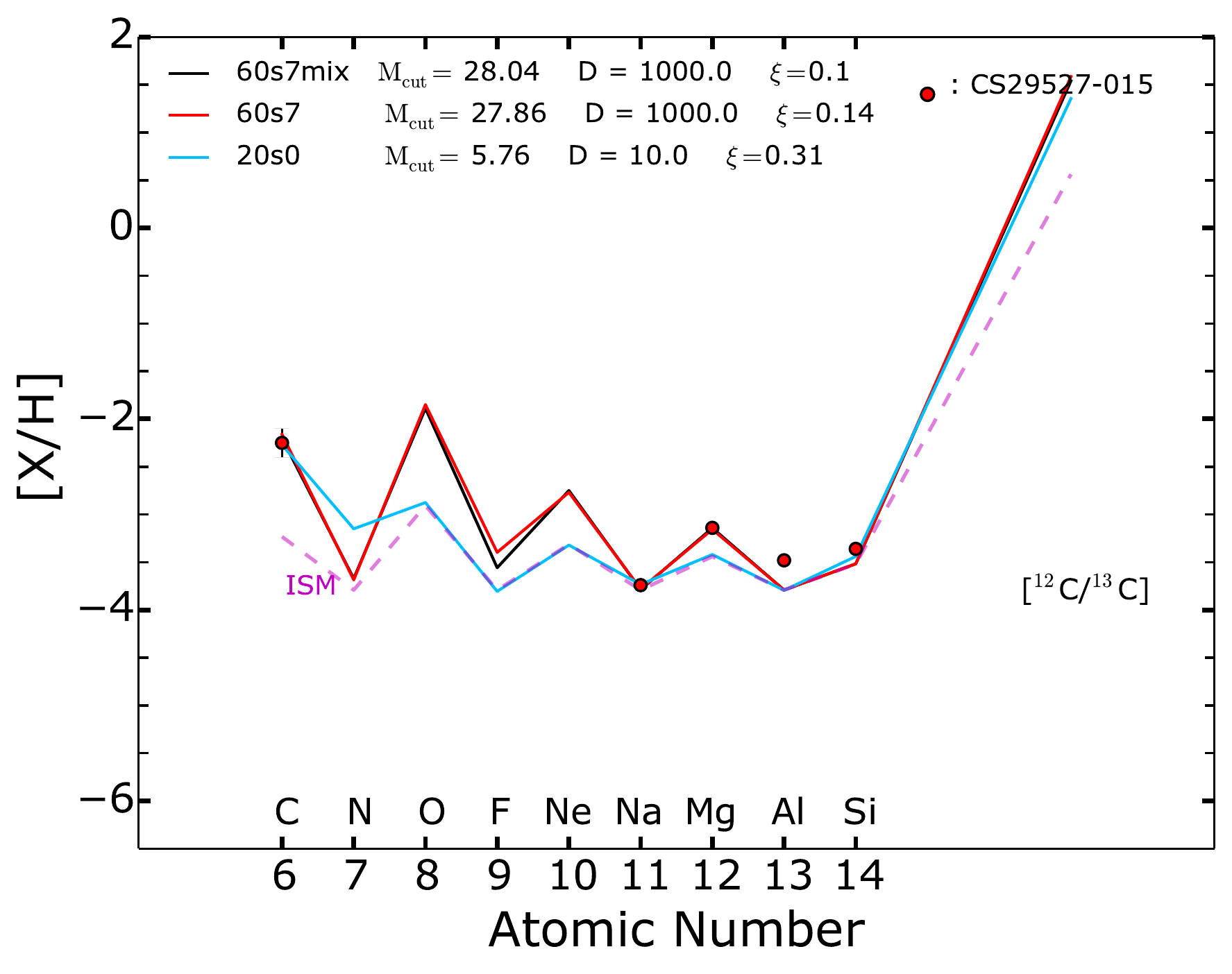}
   \end{minipage}
   \begin{minipage}{.32\linewidth}
       \includegraphics[scale=0.3]{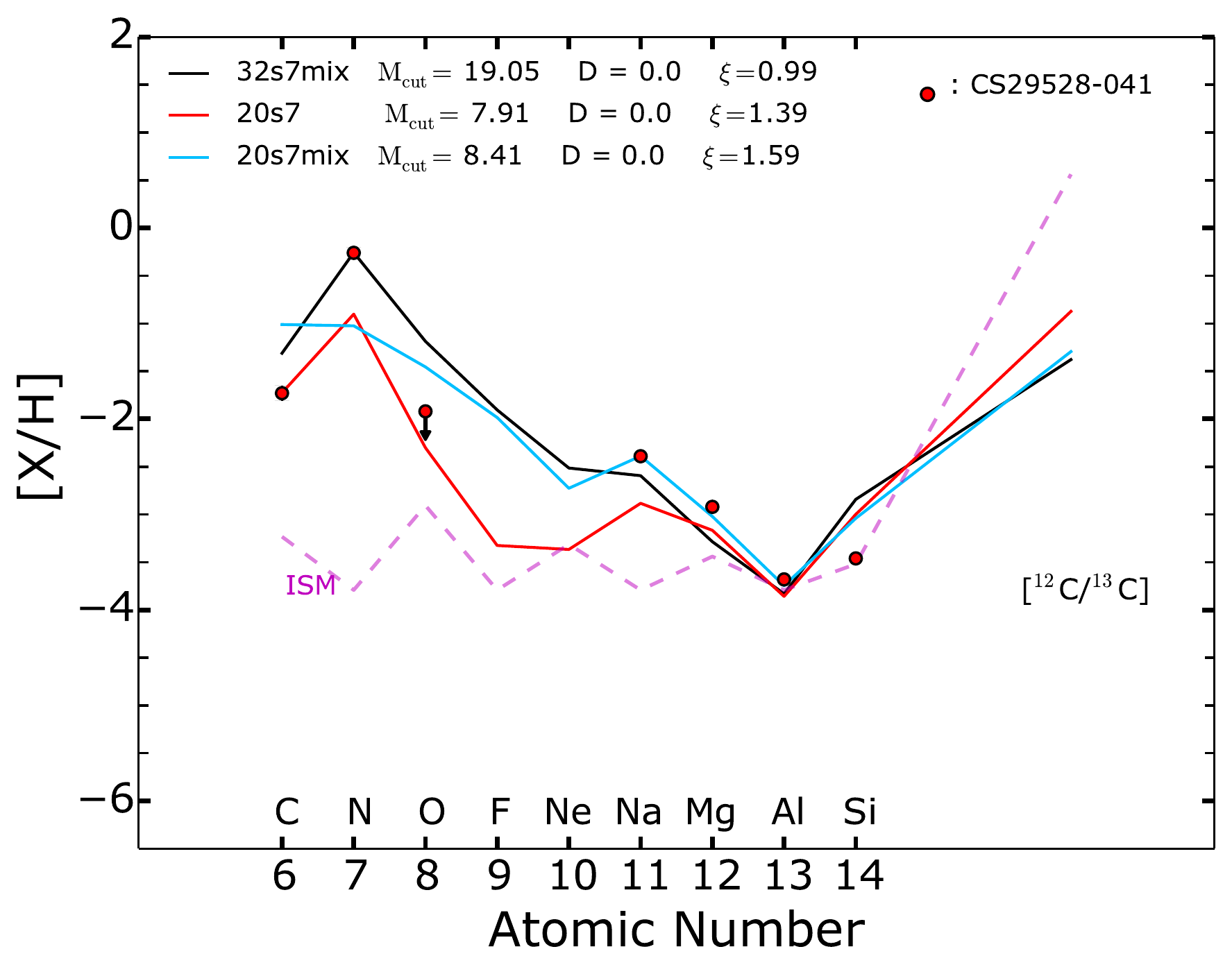}
   \end{minipage}
   \begin{minipage}{.32\linewidth}
       \includegraphics[scale=0.3]{XH_3best_13.pdf}
   \end{minipage}
   \begin{minipage}{.32\linewidth}
       \includegraphics[scale=0.3]{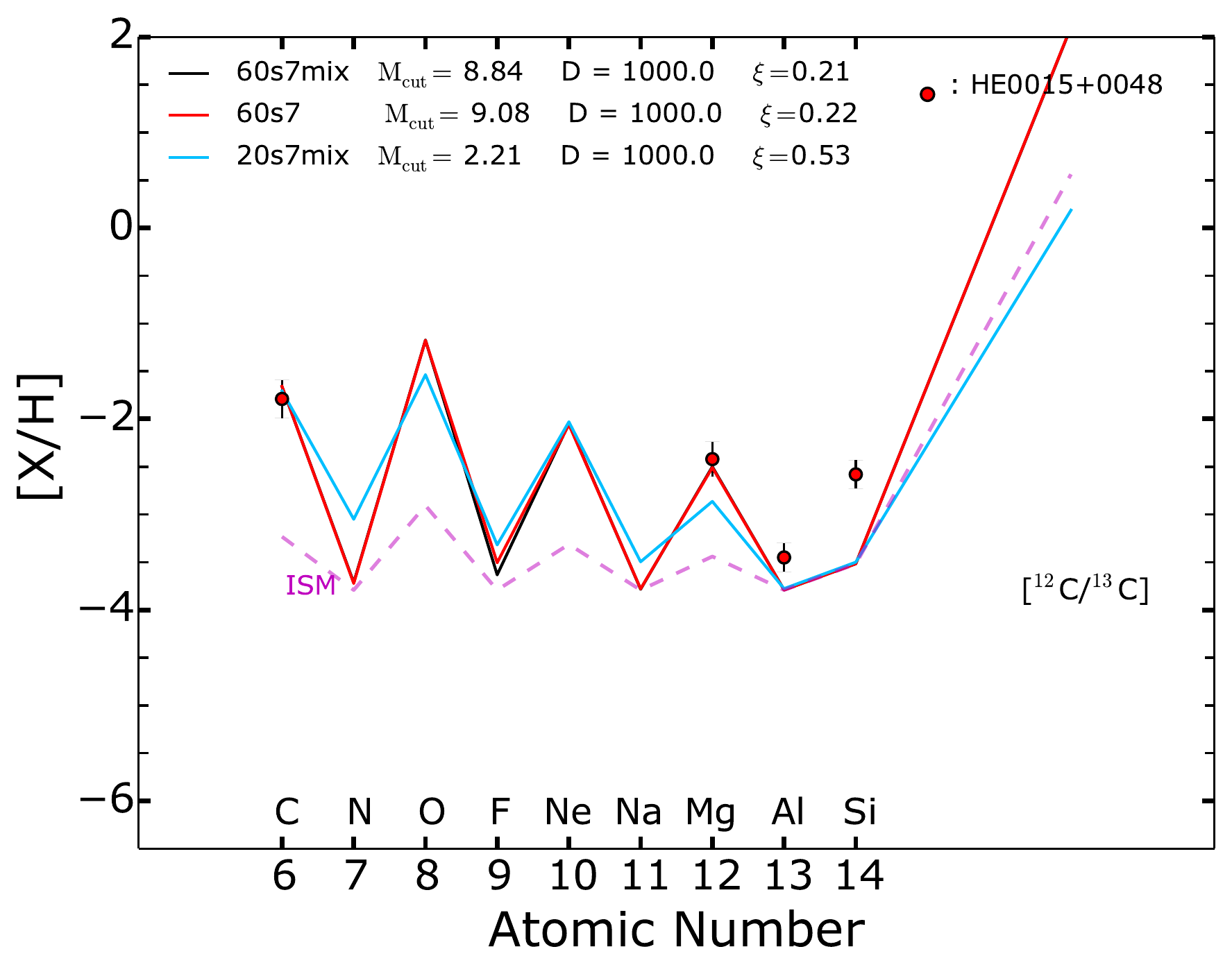}
   \end{minipage}
   \begin{minipage}{.32\linewidth}
       \includegraphics[scale=0.3]{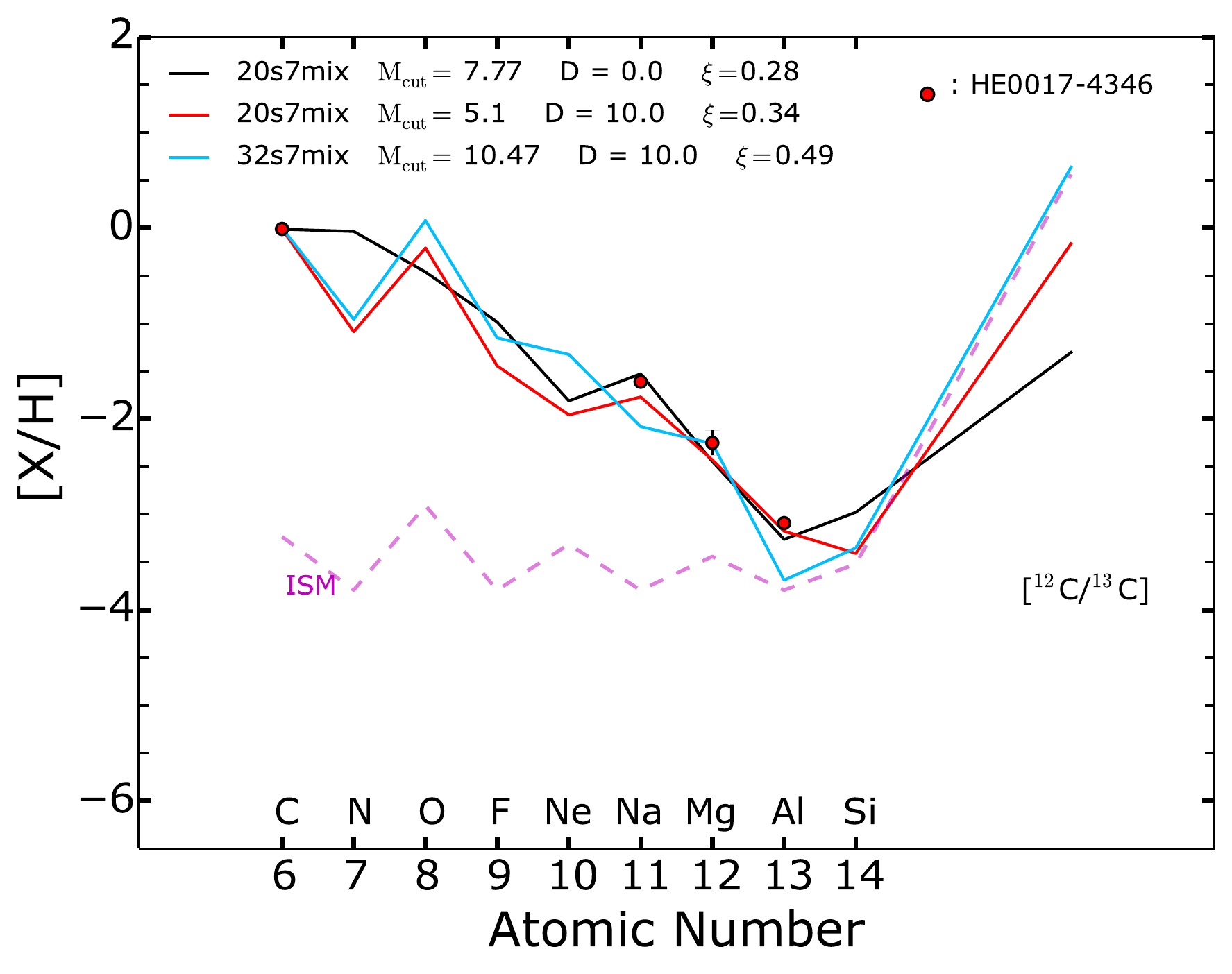}
   \end{minipage}
   \begin{minipage}{.32\linewidth}
       \includegraphics[scale=0.3]{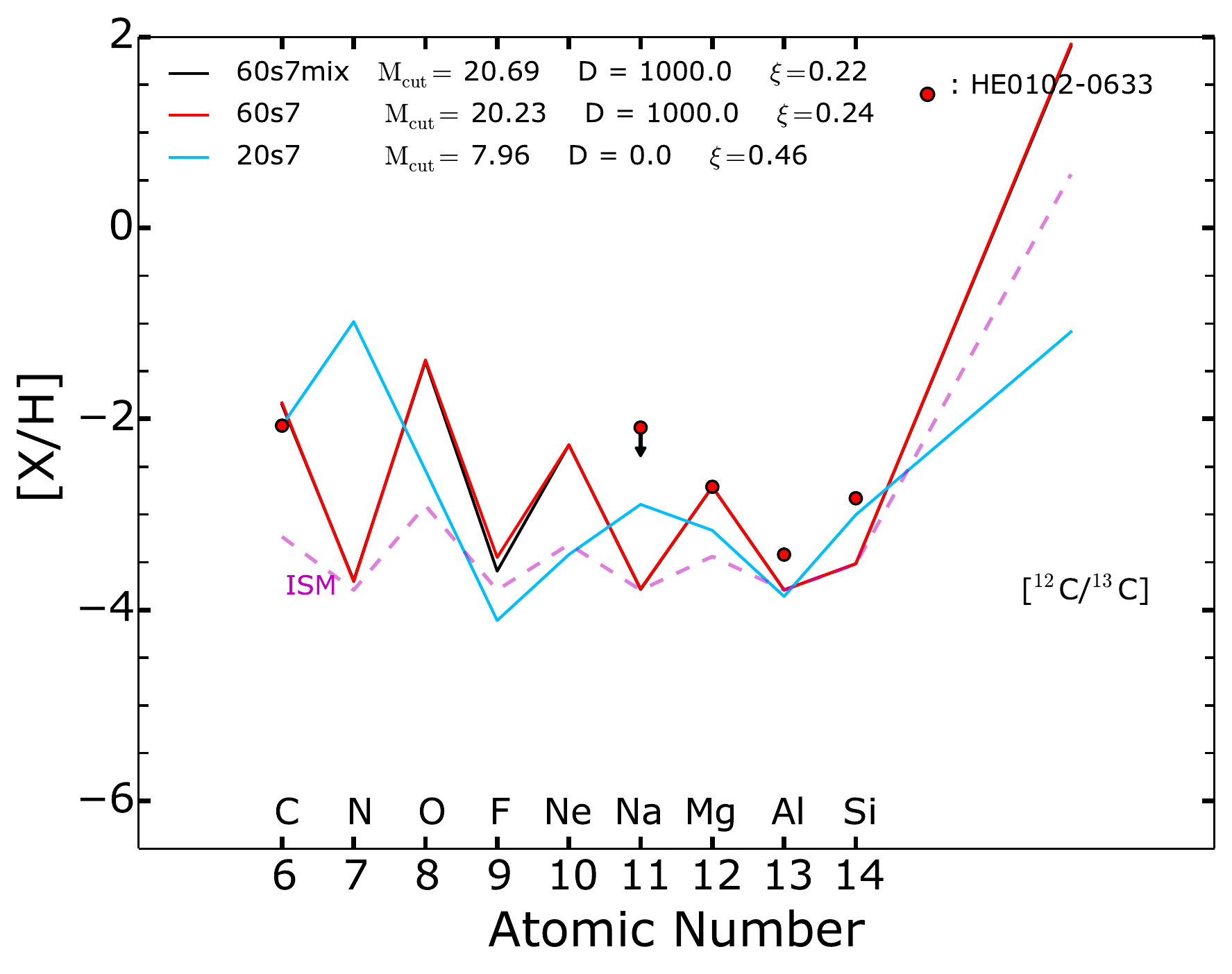}
   \end{minipage}
   \begin{minipage}{.32\linewidth}
       \includegraphics[scale=0.3]{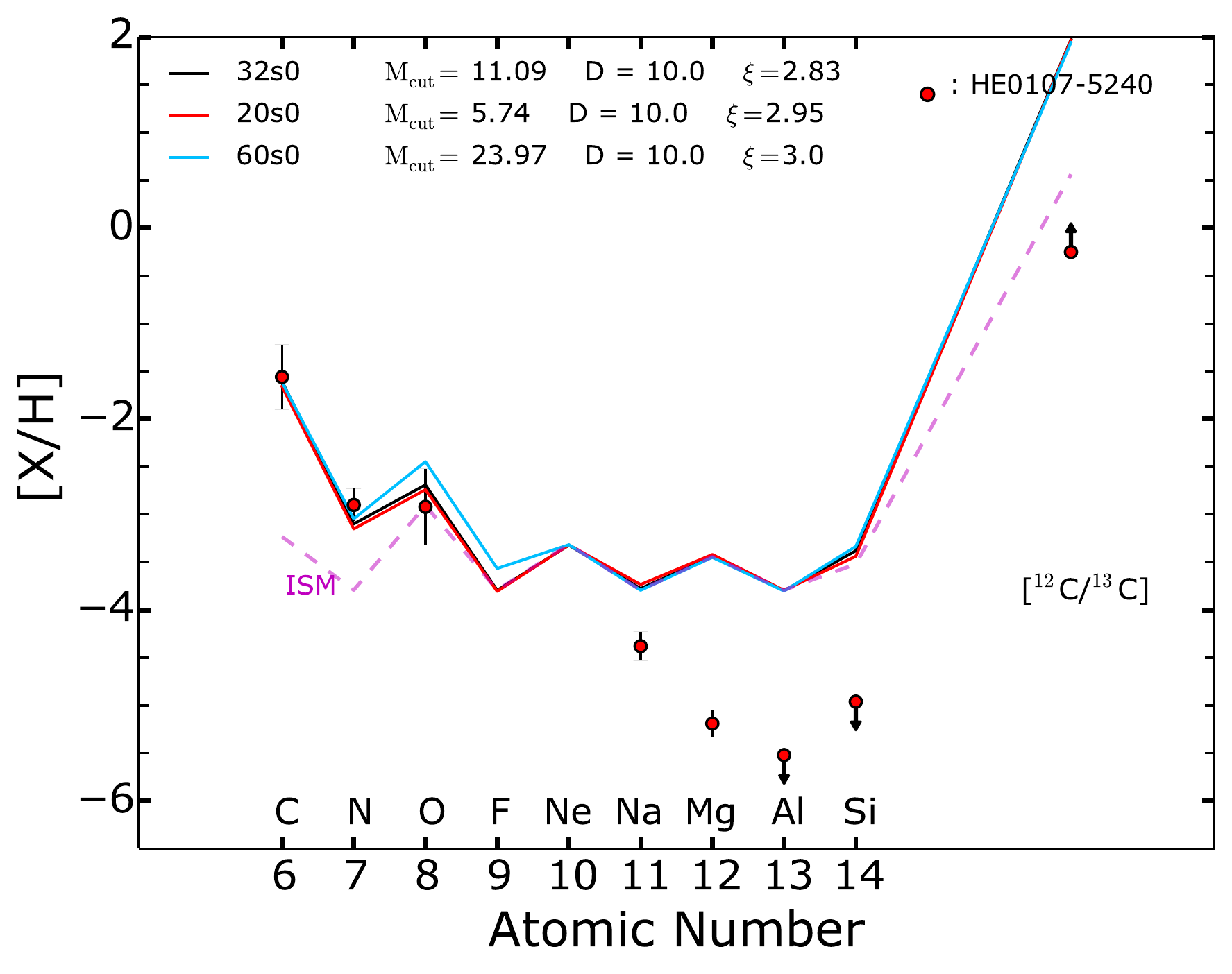}
   \end{minipage}
   \begin{minipage}{.32\linewidth}
       \includegraphics[scale=0.3]{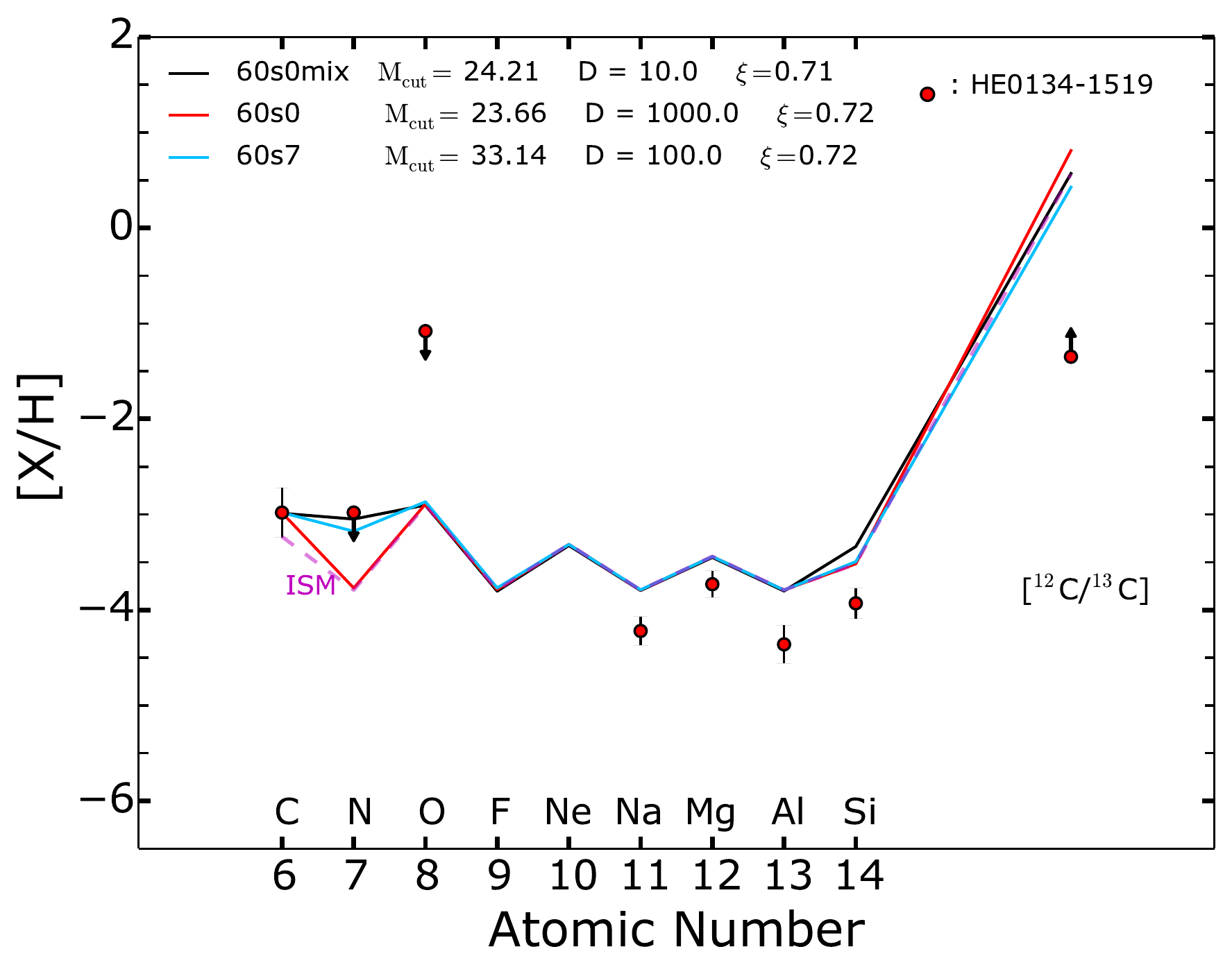}
   \end{minipage}
   \begin{minipage}{.32\linewidth}
       \includegraphics[scale=0.3]{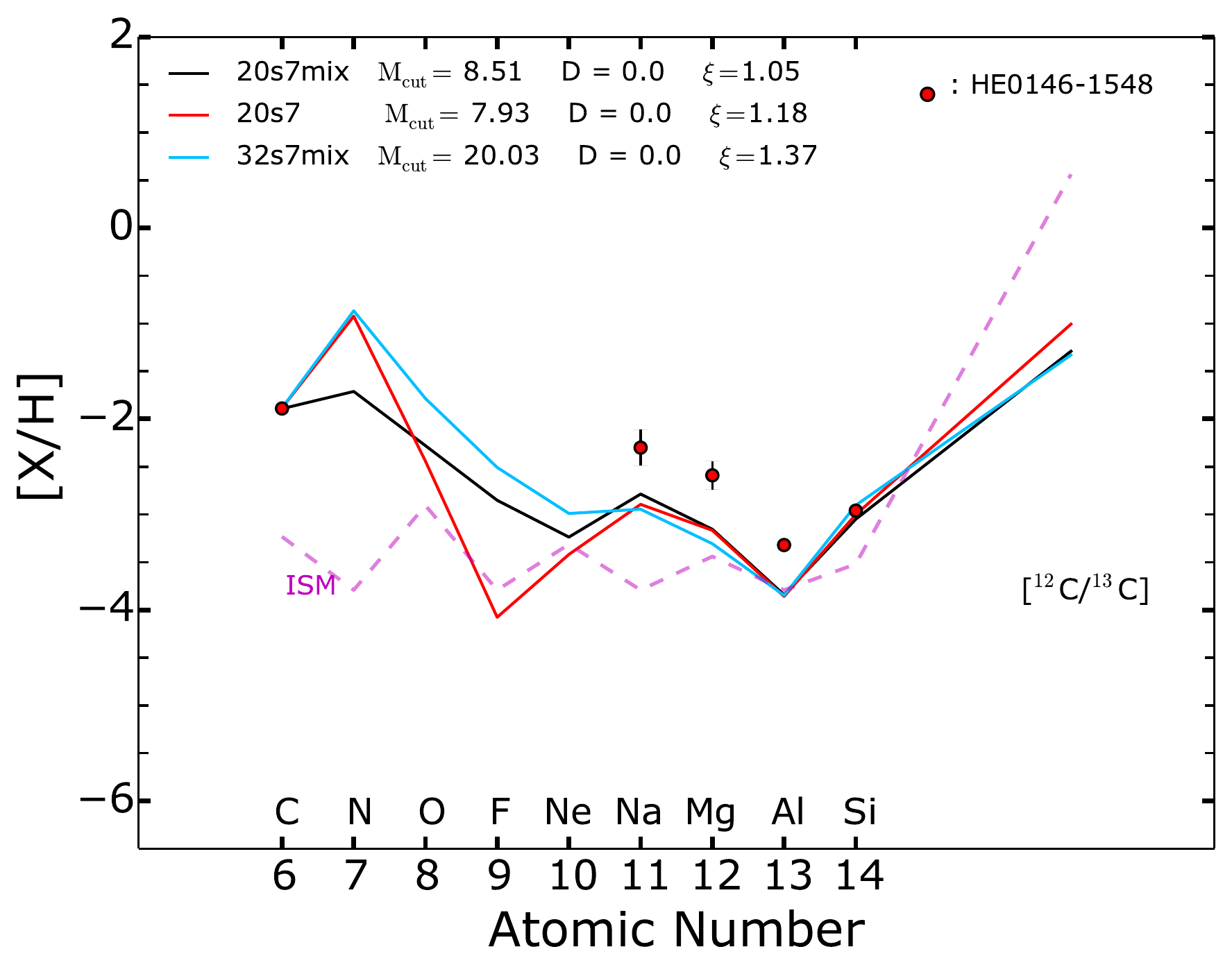}
   \end{minipage}
   \begin{minipage}{.32\linewidth}
       \includegraphics[scale=0.3]{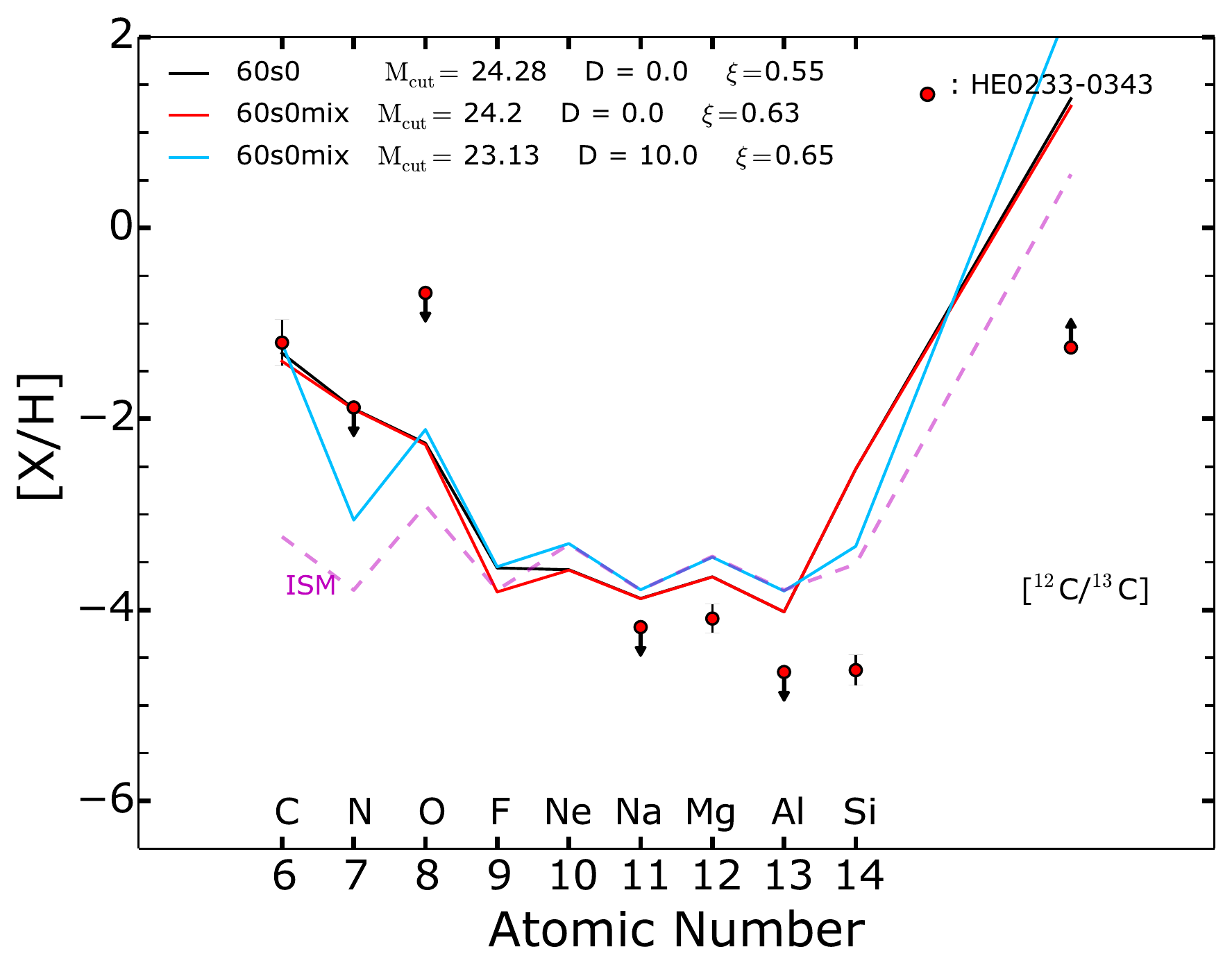}
   \end{minipage}
   \begin{minipage}{.32\linewidth}
       \includegraphics[scale=0.3]{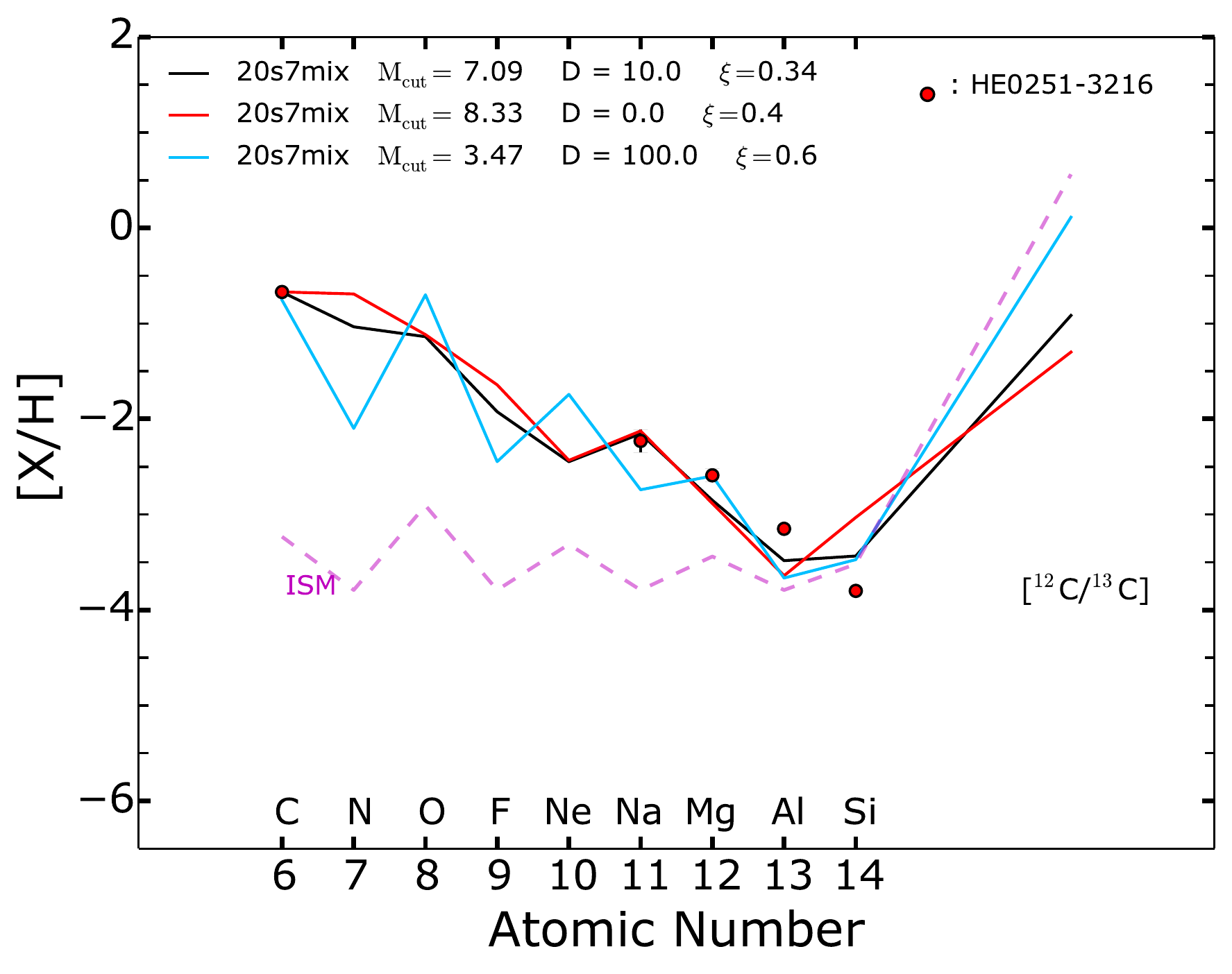}
   \end{minipage}
   \begin{minipage}{.32\linewidth}
       \includegraphics[scale=0.3]{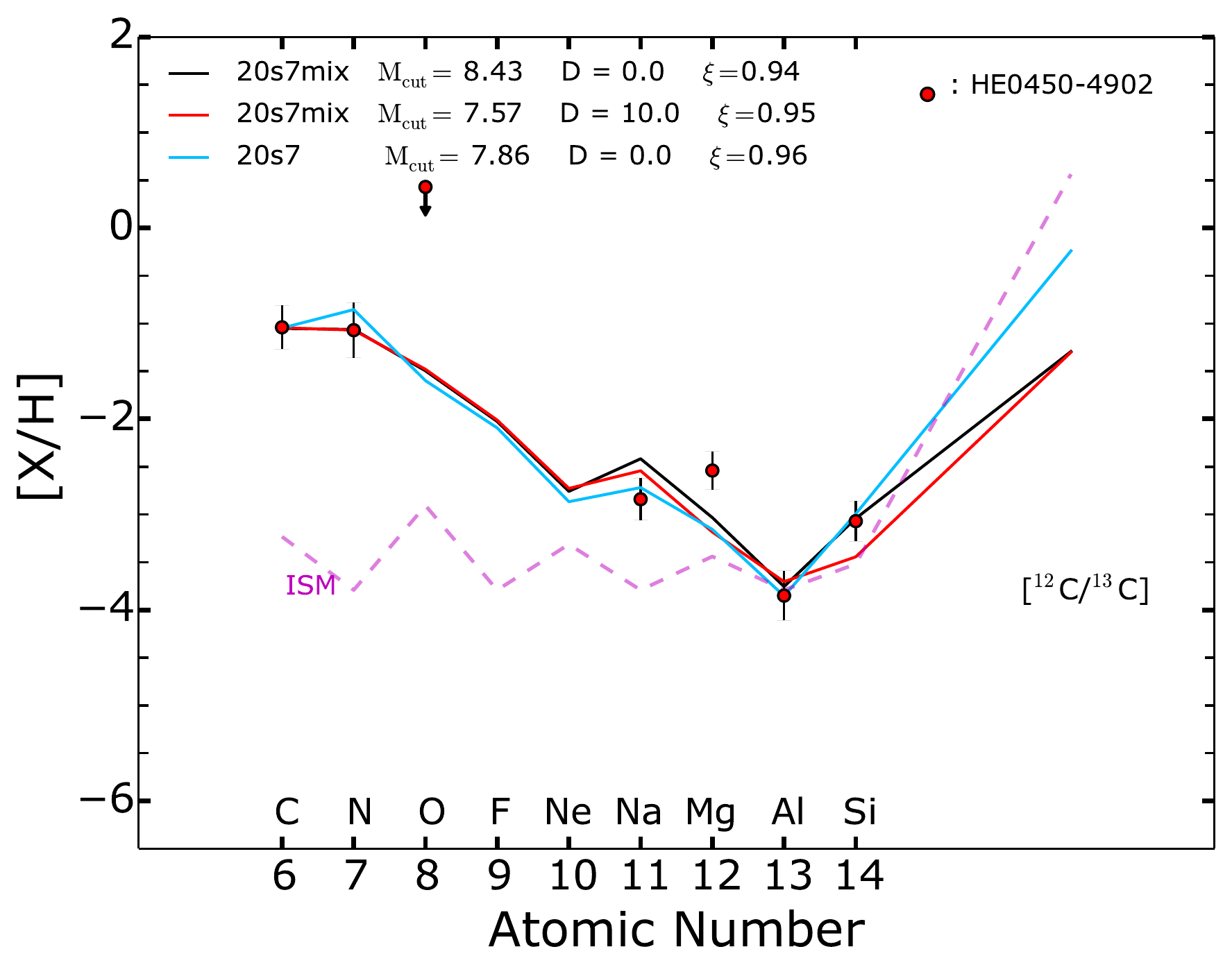}
   \end{minipage}
   \begin{minipage}{.32\linewidth}
       \includegraphics[scale=0.3]{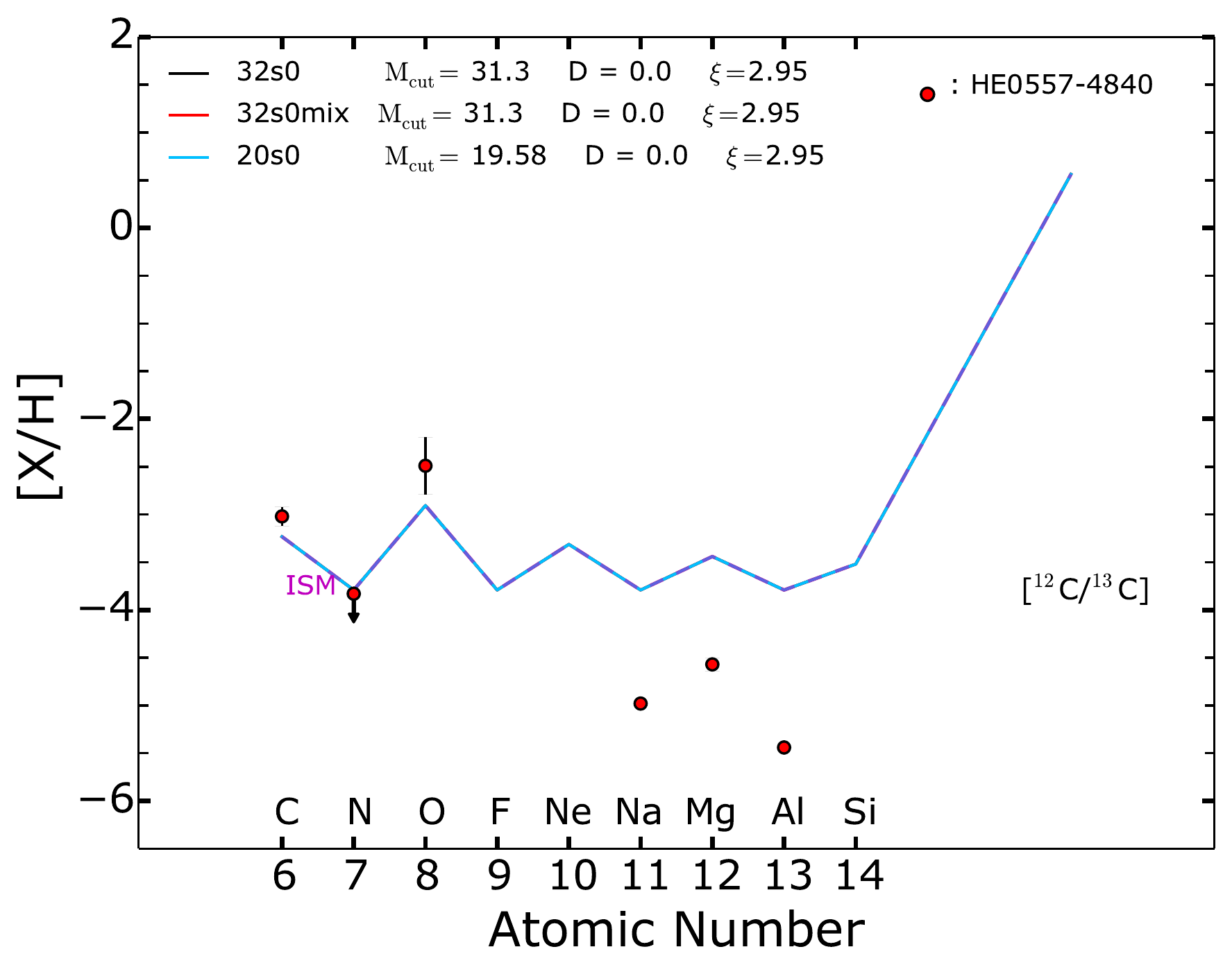}
   \end{minipage}
   \caption[Fig.~\ref{allstar1}, continued]{continued}
\label{allstar2}
    \end{figure*}

   \begin{figure*}
   \centering
      \begin{minipage}{.32\linewidth}
       \includegraphics[scale=0.3]{XH_3best_24.pdf}
   \end{minipage}
   \begin{minipage}{.32\linewidth}
       \includegraphics[scale=0.3]{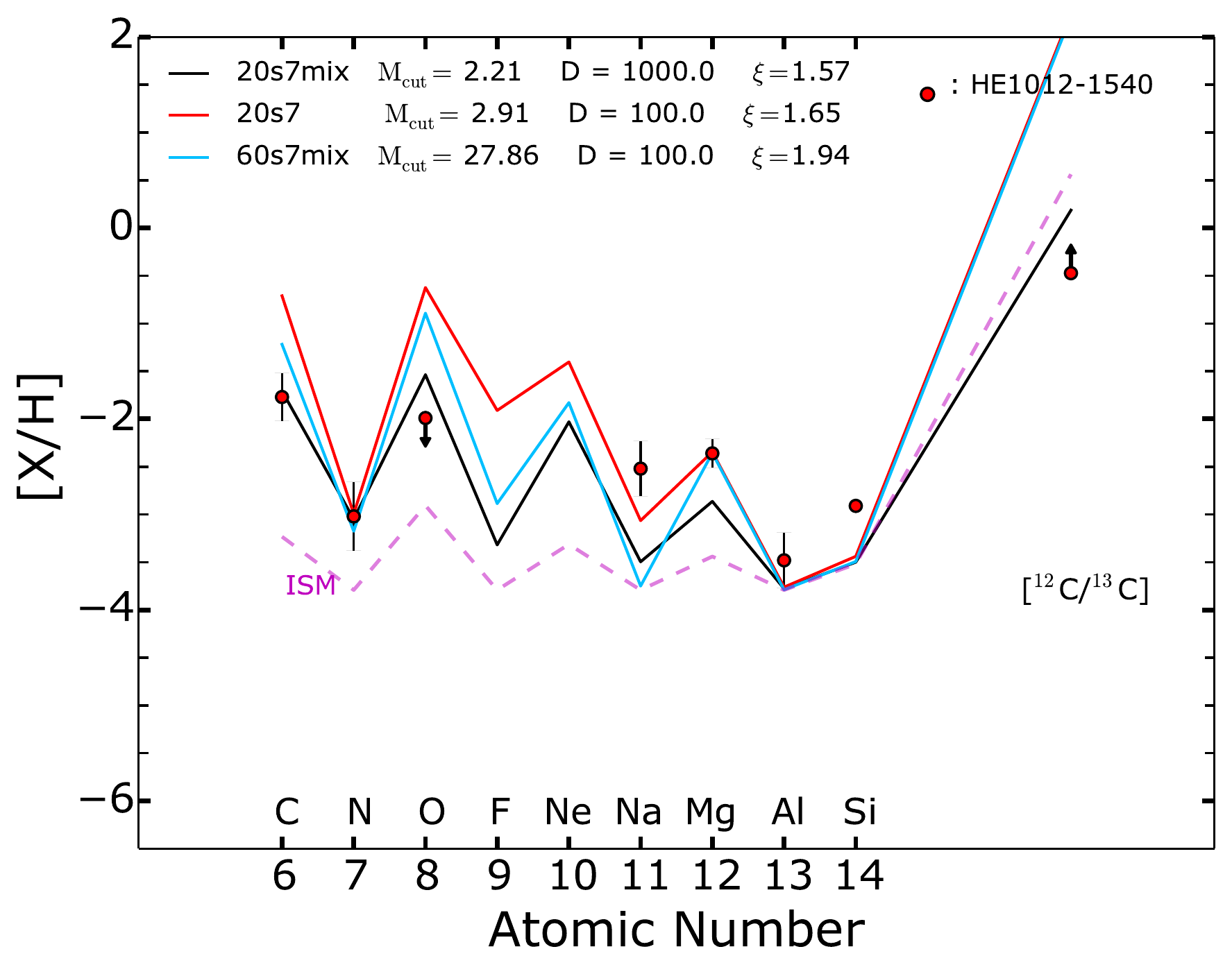}
   \end{minipage}
   \begin{minipage}{.32\linewidth}
       \includegraphics[scale=0.3]{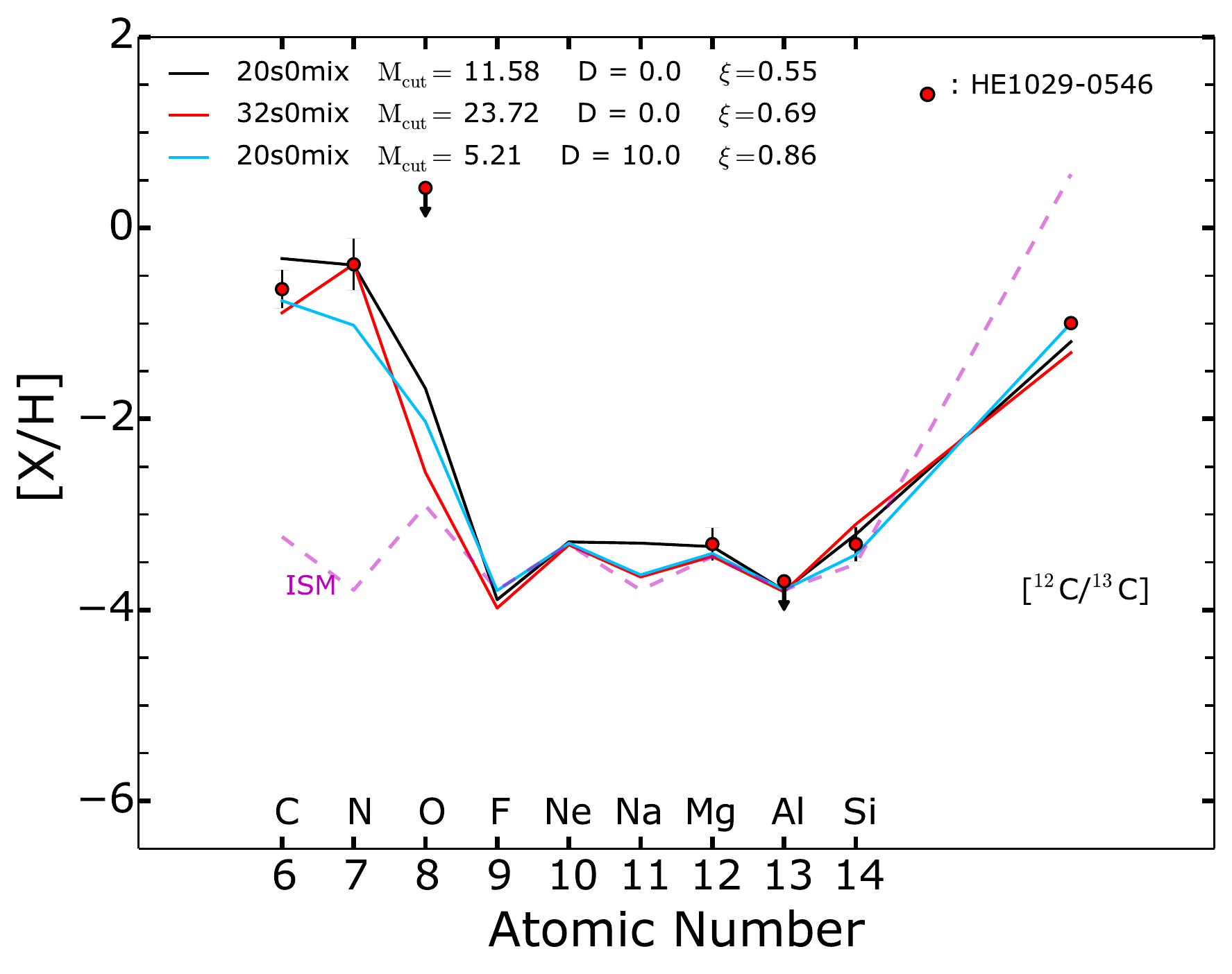}
   \end{minipage}
   \begin{minipage}{.32\linewidth}
       \includegraphics[scale=0.3]{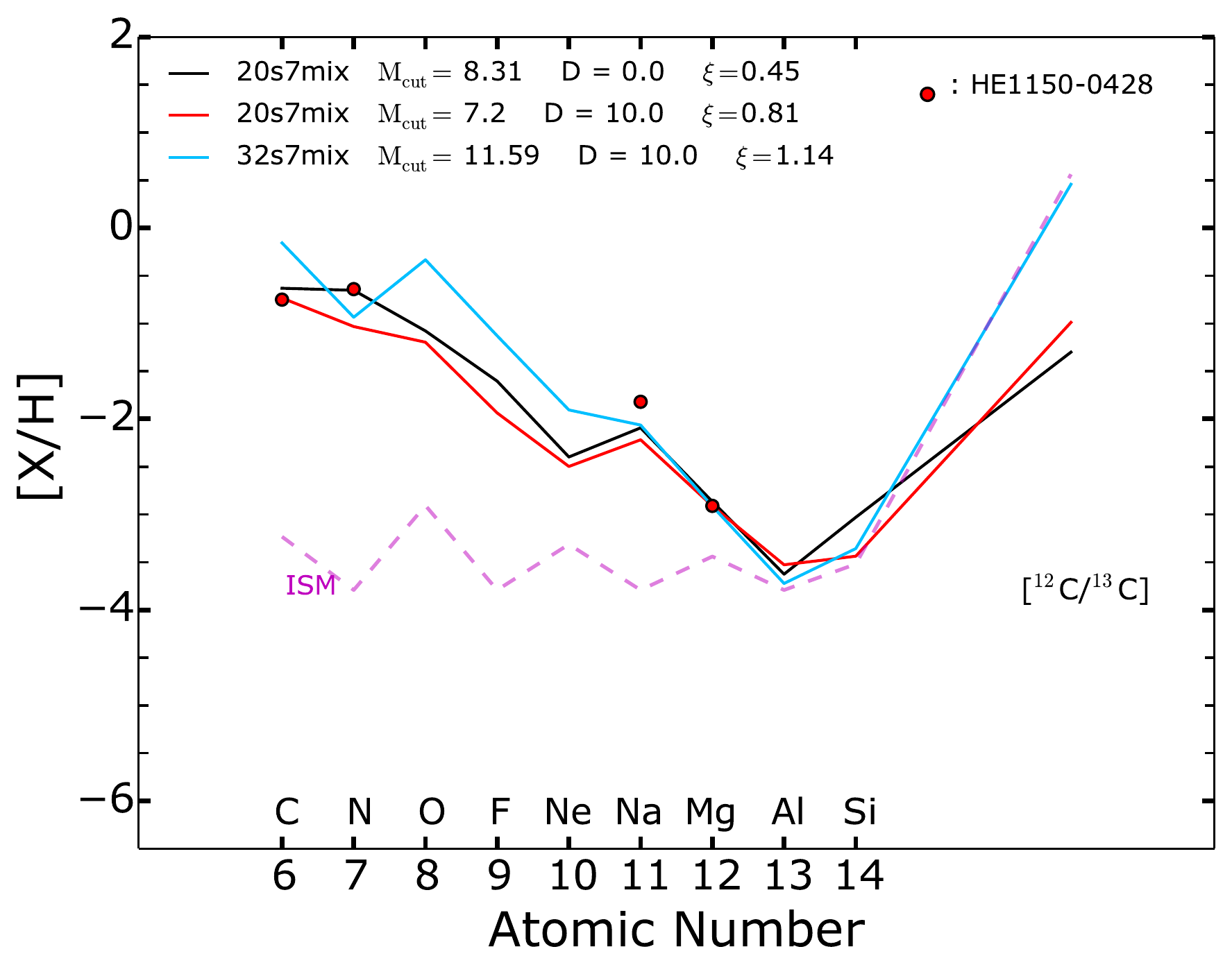}
   \end{minipage}
   \begin{minipage}{.32\linewidth}
       \includegraphics[scale=0.3]{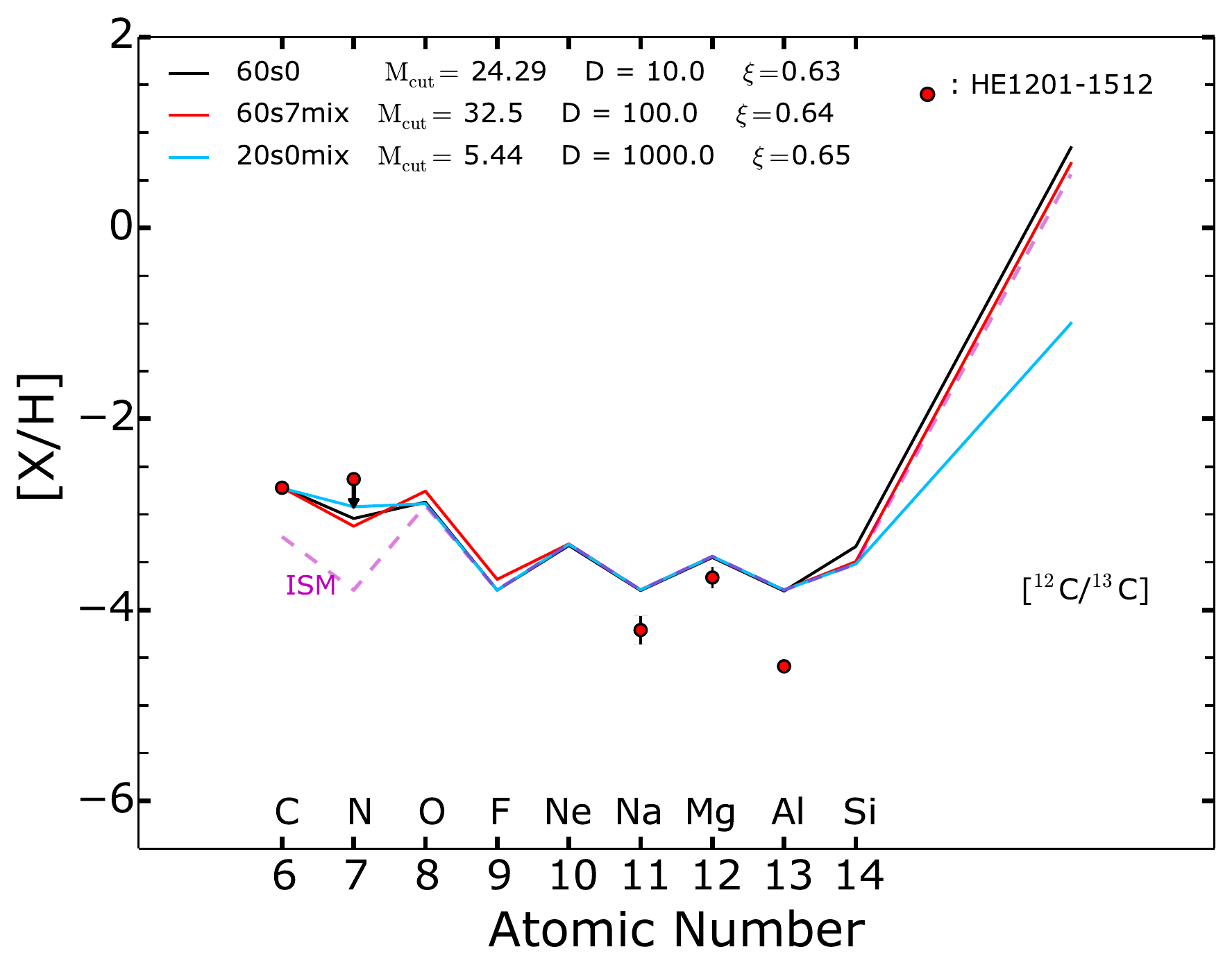}
   \end{minipage}
   \begin{minipage}{.32\linewidth}
       \includegraphics[scale=0.3]{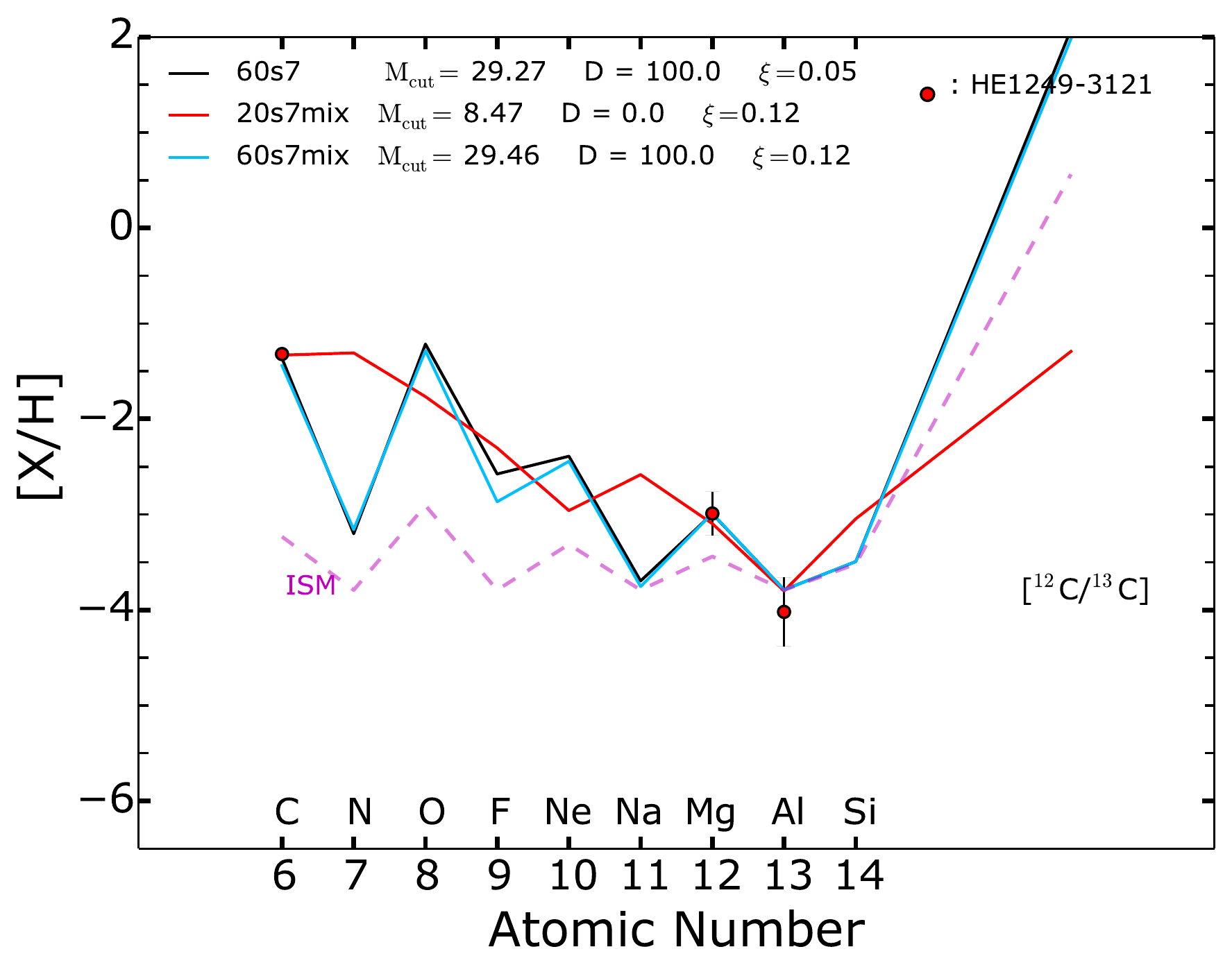}
   \end{minipage}
   \begin{minipage}{.32\linewidth}
       \includegraphics[scale=0.3]{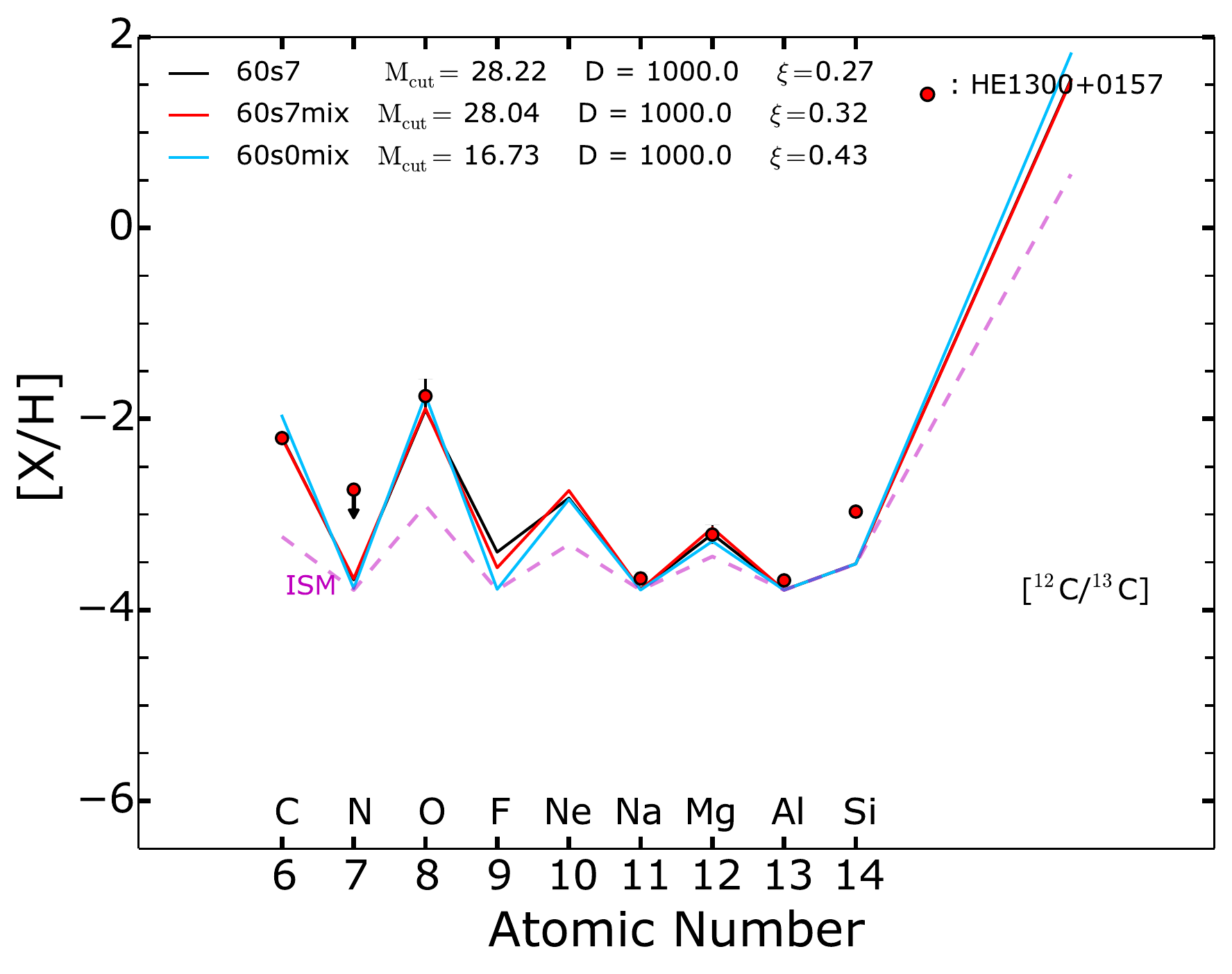}
   \end{minipage}
   \begin{minipage}{.32\linewidth}
       \includegraphics[scale=0.3]{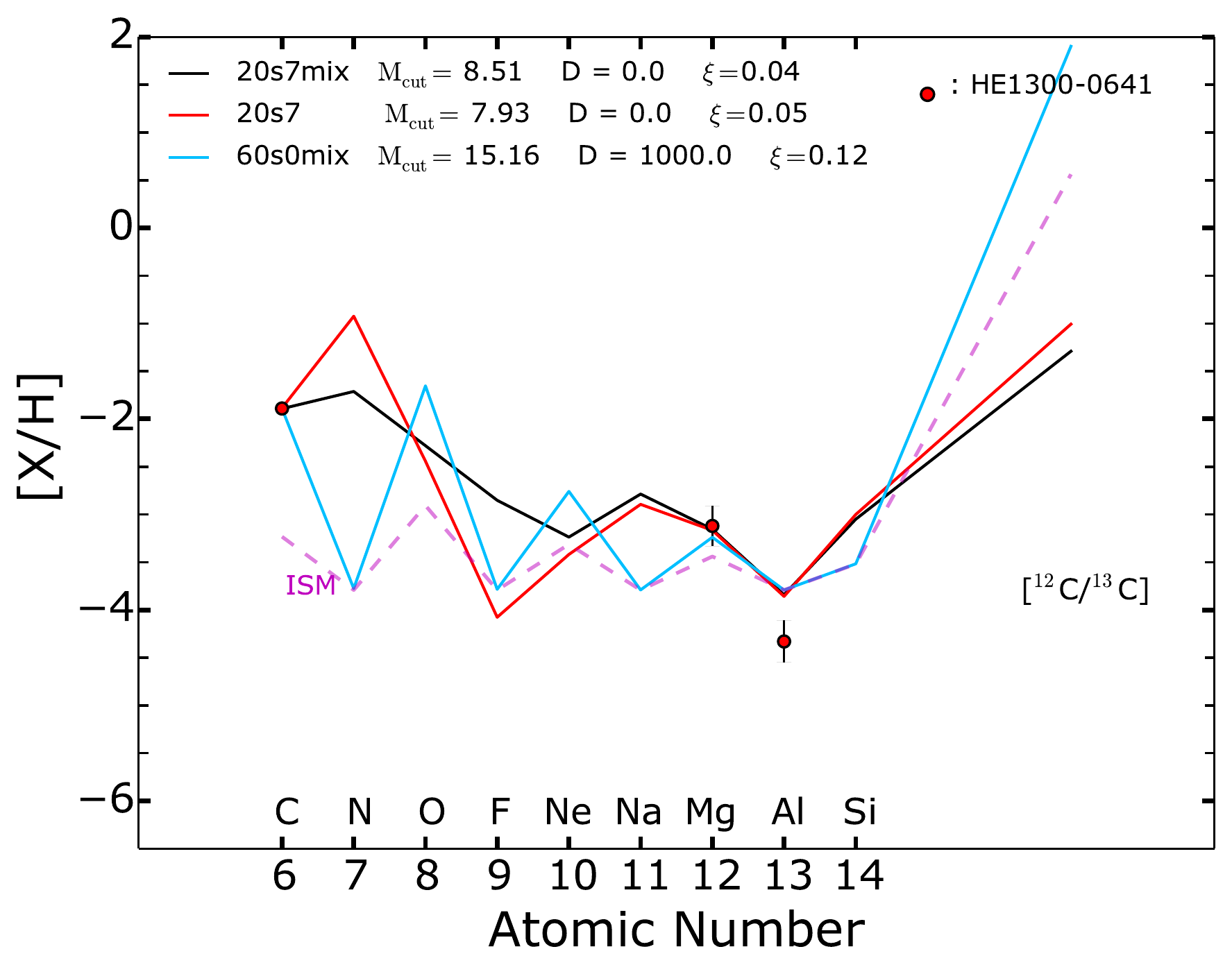}
   \end{minipage}
   \begin{minipage}{.32\linewidth}
       \includegraphics[scale=0.3]{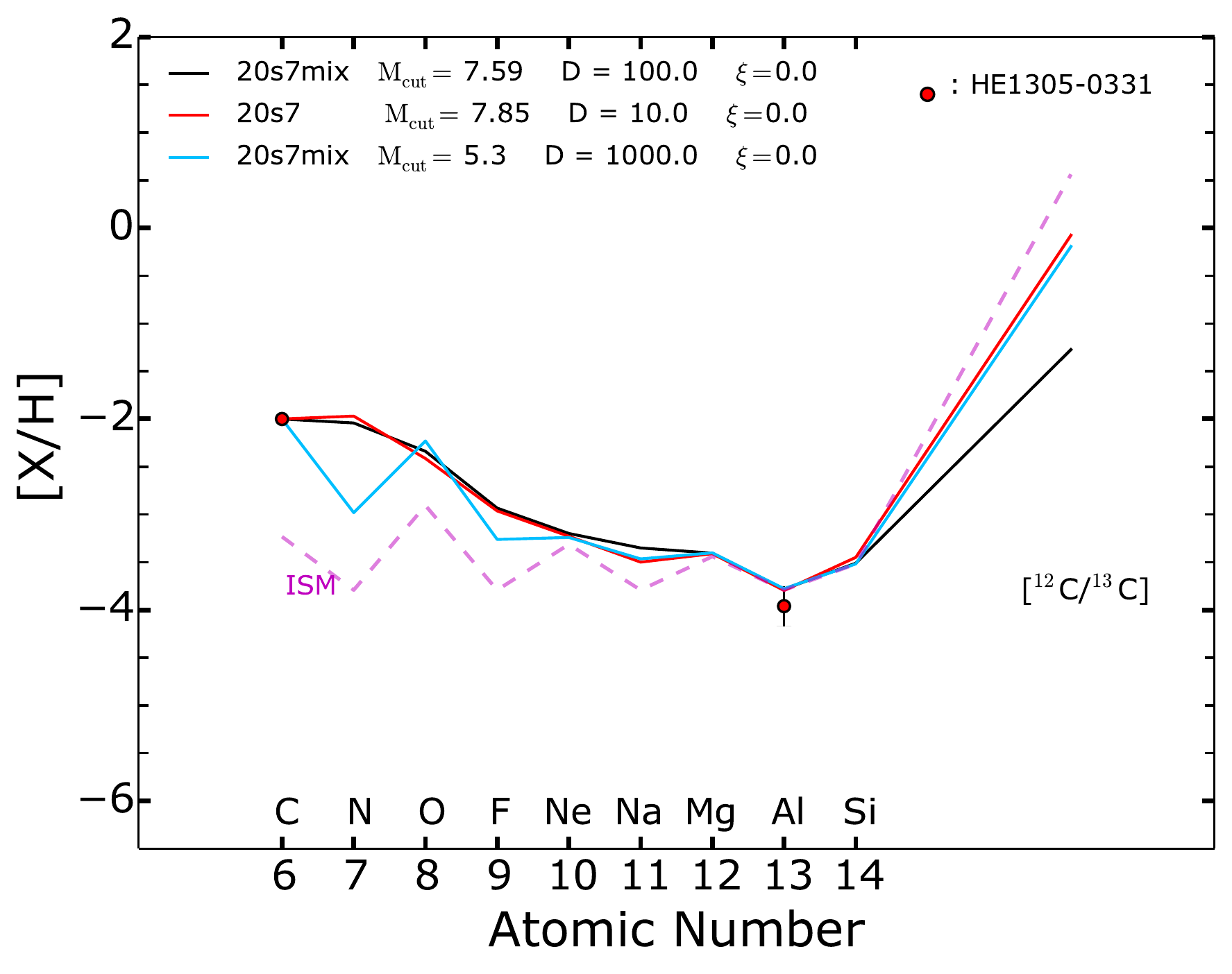}
   \end{minipage}
   \begin{minipage}{.32\linewidth}
       \includegraphics[scale=0.3]{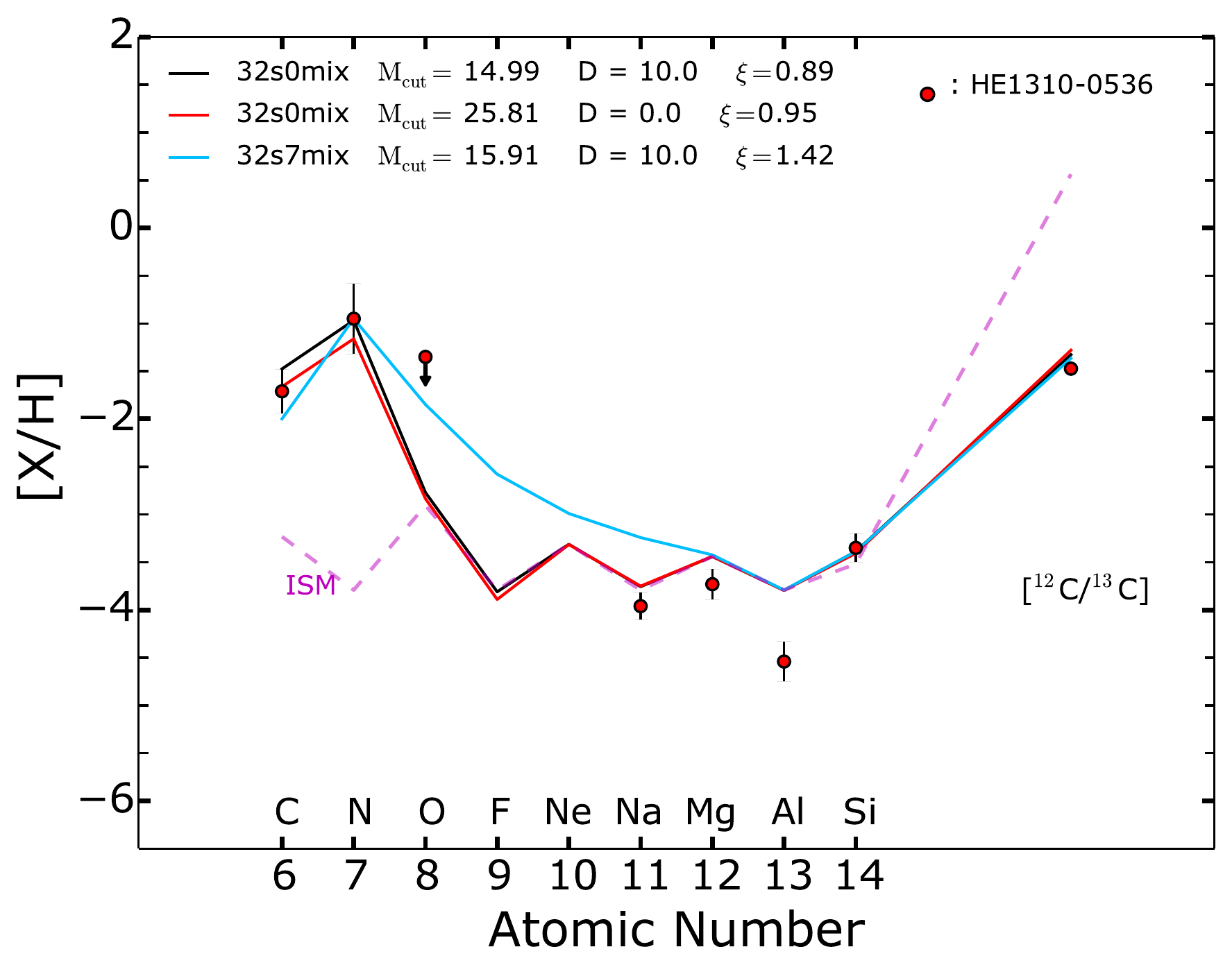}
   \end{minipage}
   \begin{minipage}{.32\linewidth}
       \includegraphics[scale=0.3]{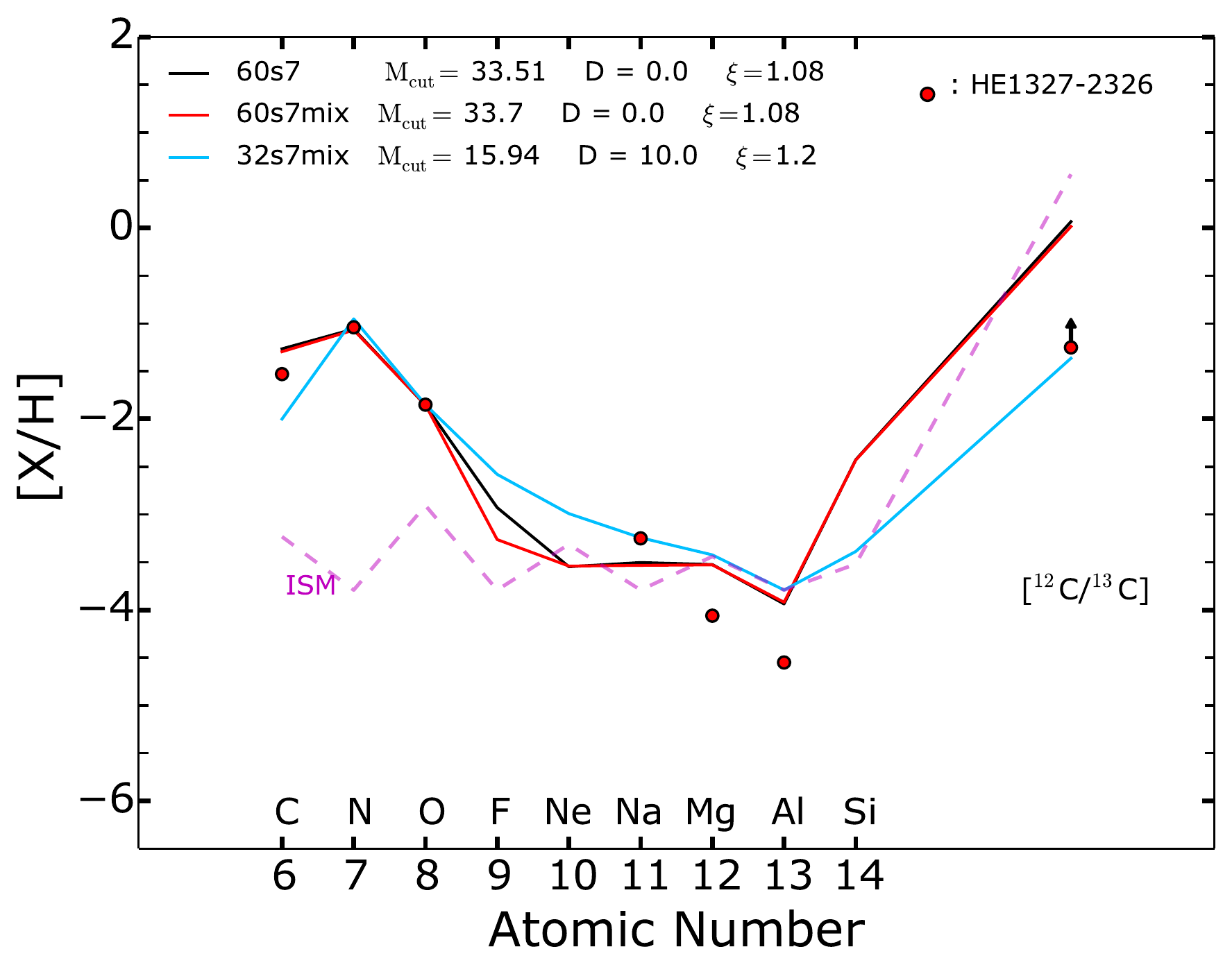}
   \end{minipage}
   \begin{minipage}{.32\linewidth}
       \includegraphics[scale=0.3]{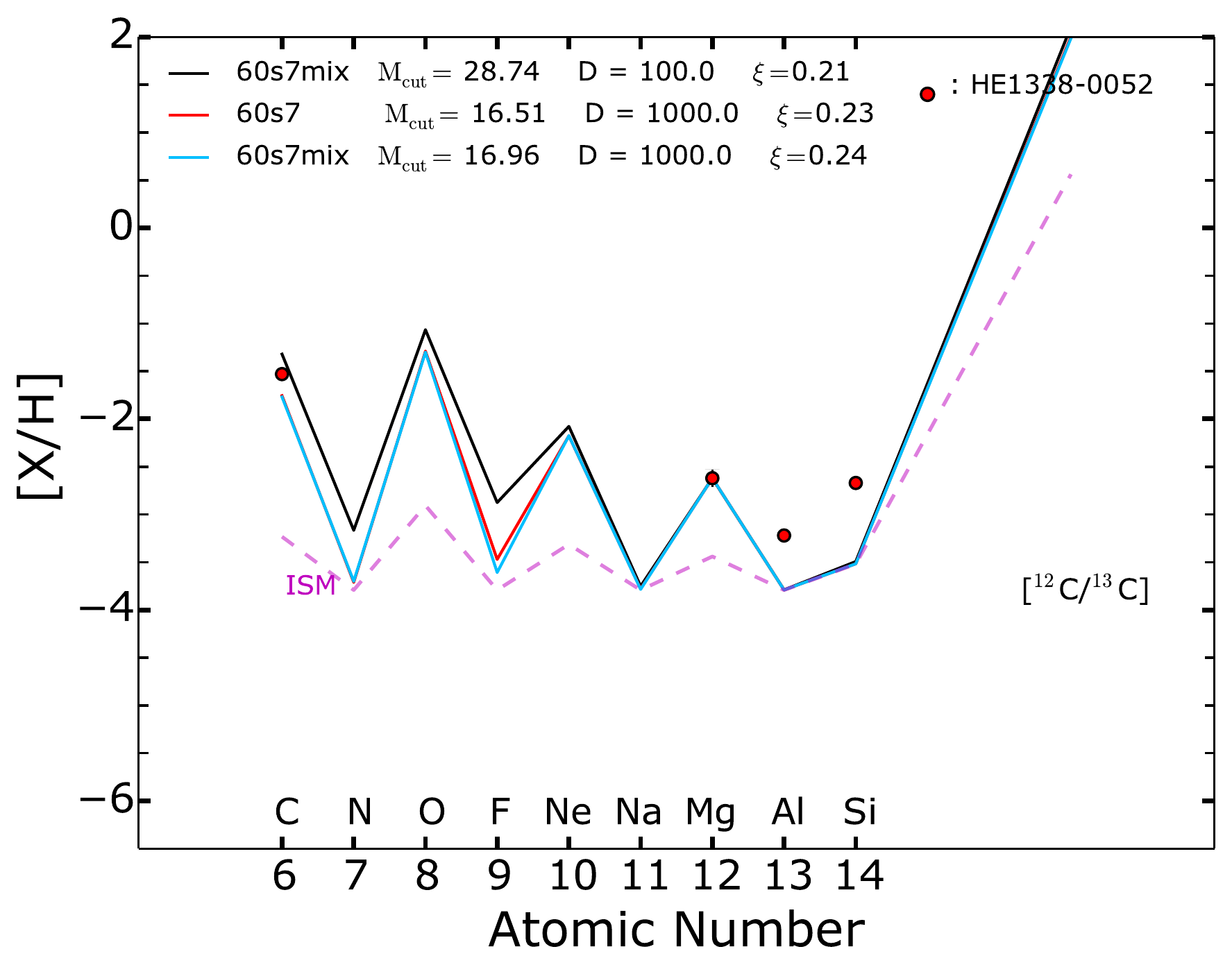}
   \end{minipage}
   \begin{minipage}{.32\linewidth}
       \includegraphics[scale=0.3]{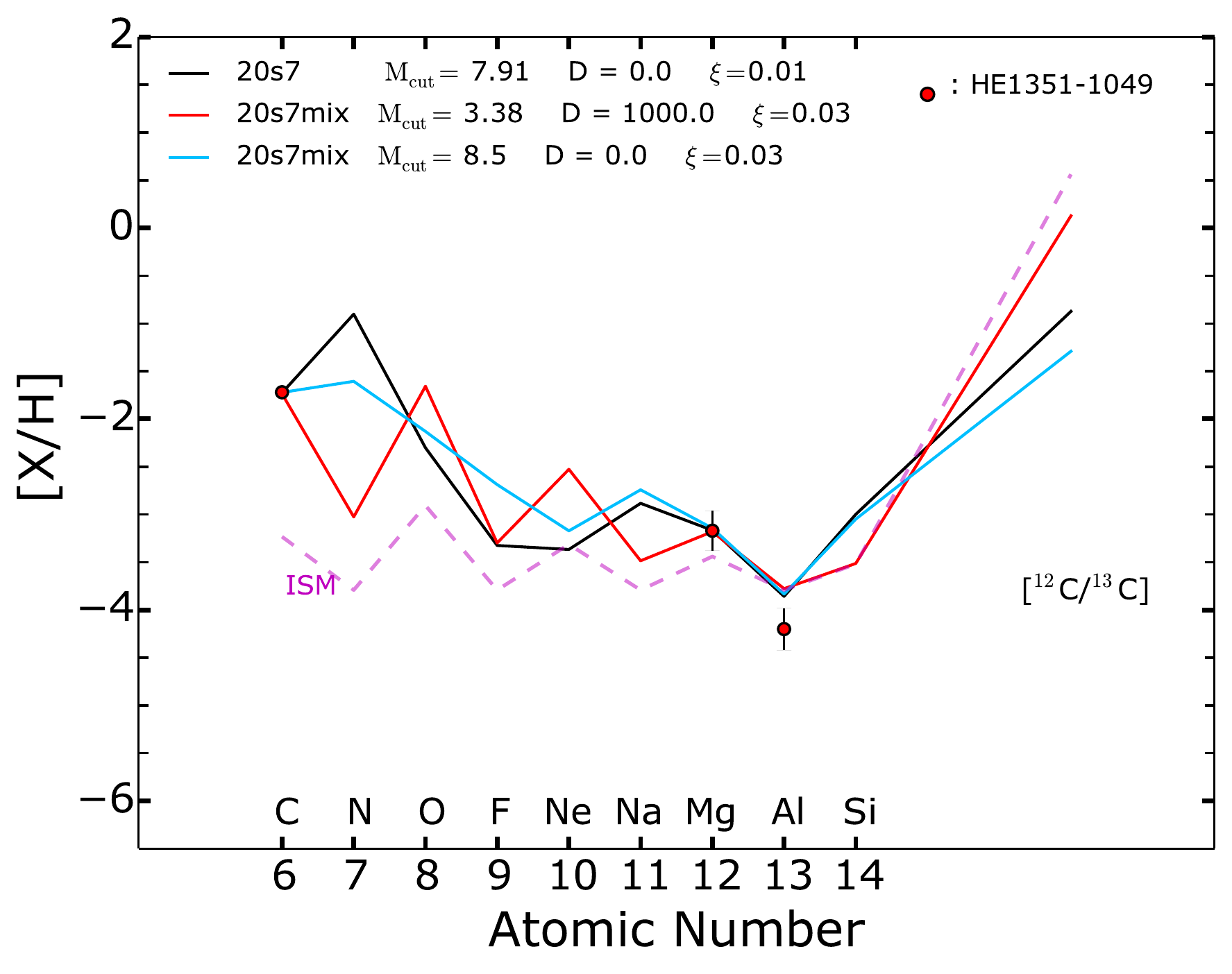}
   \end{minipage}
   \begin{minipage}{.32\linewidth}
       \includegraphics[scale=0.3]{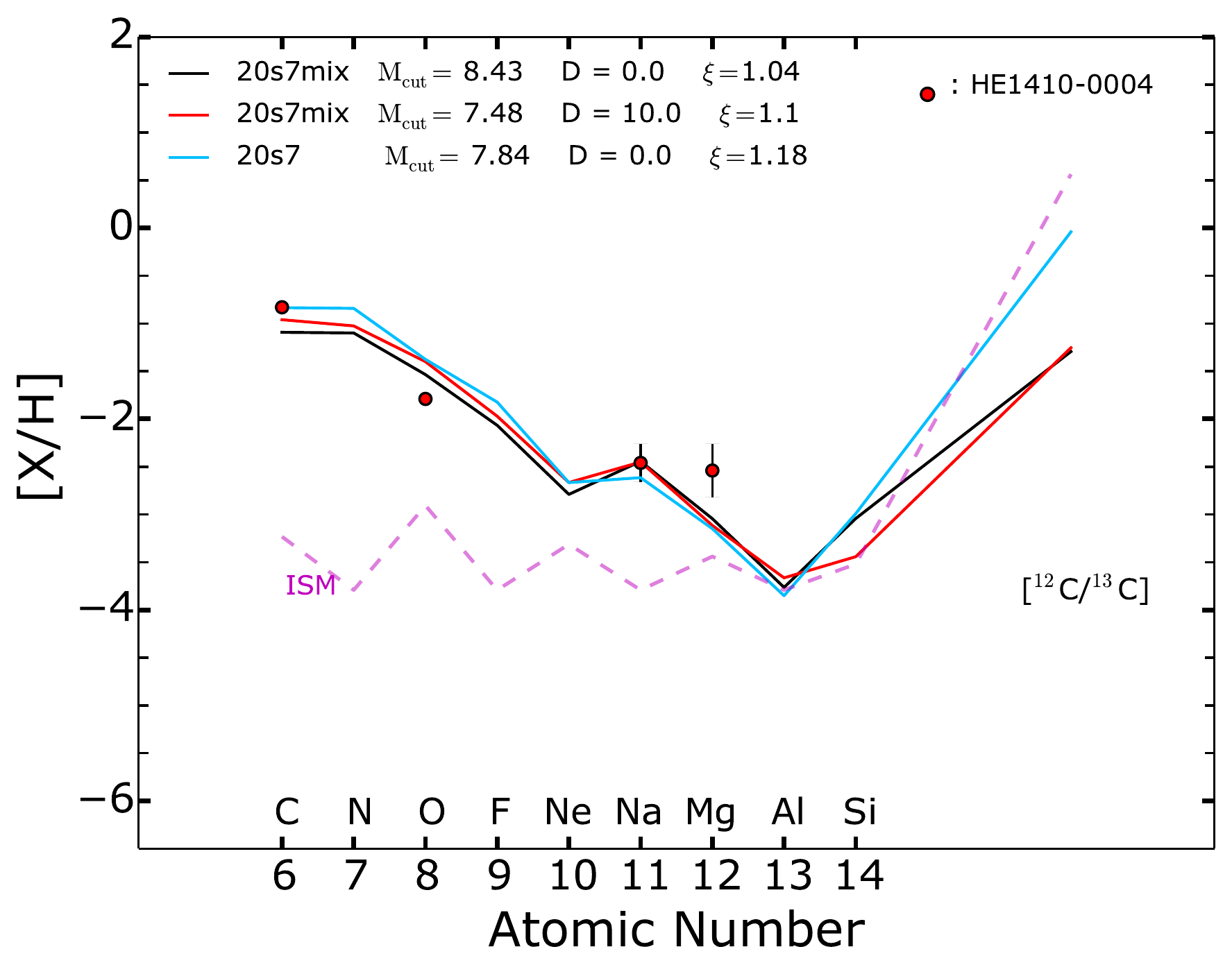}
   \end{minipage}
   \begin{minipage}{.32\linewidth}
       \includegraphics[scale=0.3]{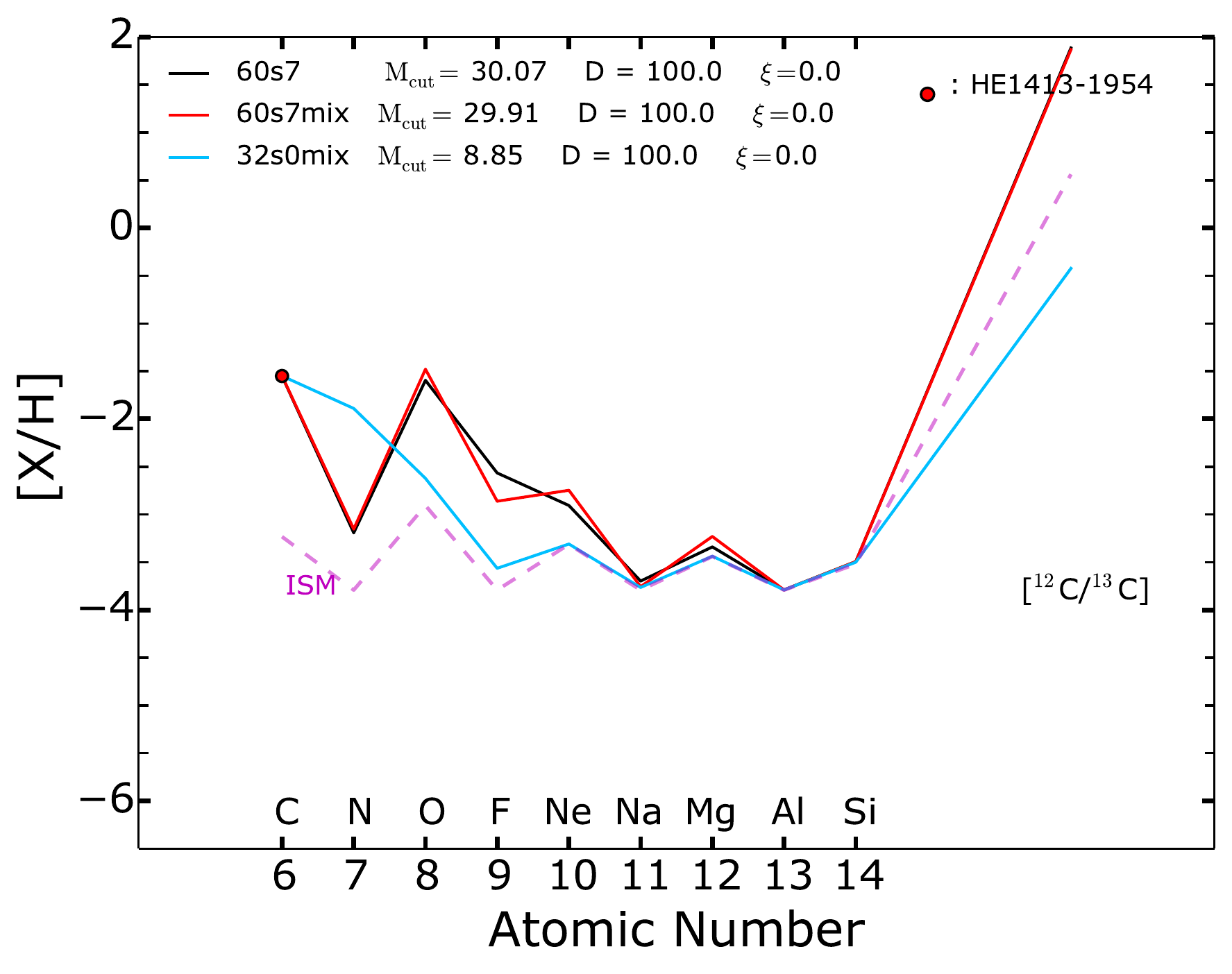}
   \end{minipage}
   \caption[Fig.~\ref{allstar1}, continued]{continued}
\label{allstar3}
    \end{figure*}

   \begin{figure*}
   \centering
      \begin{minipage}{.32\linewidth}
       \includegraphics[scale=0.3]{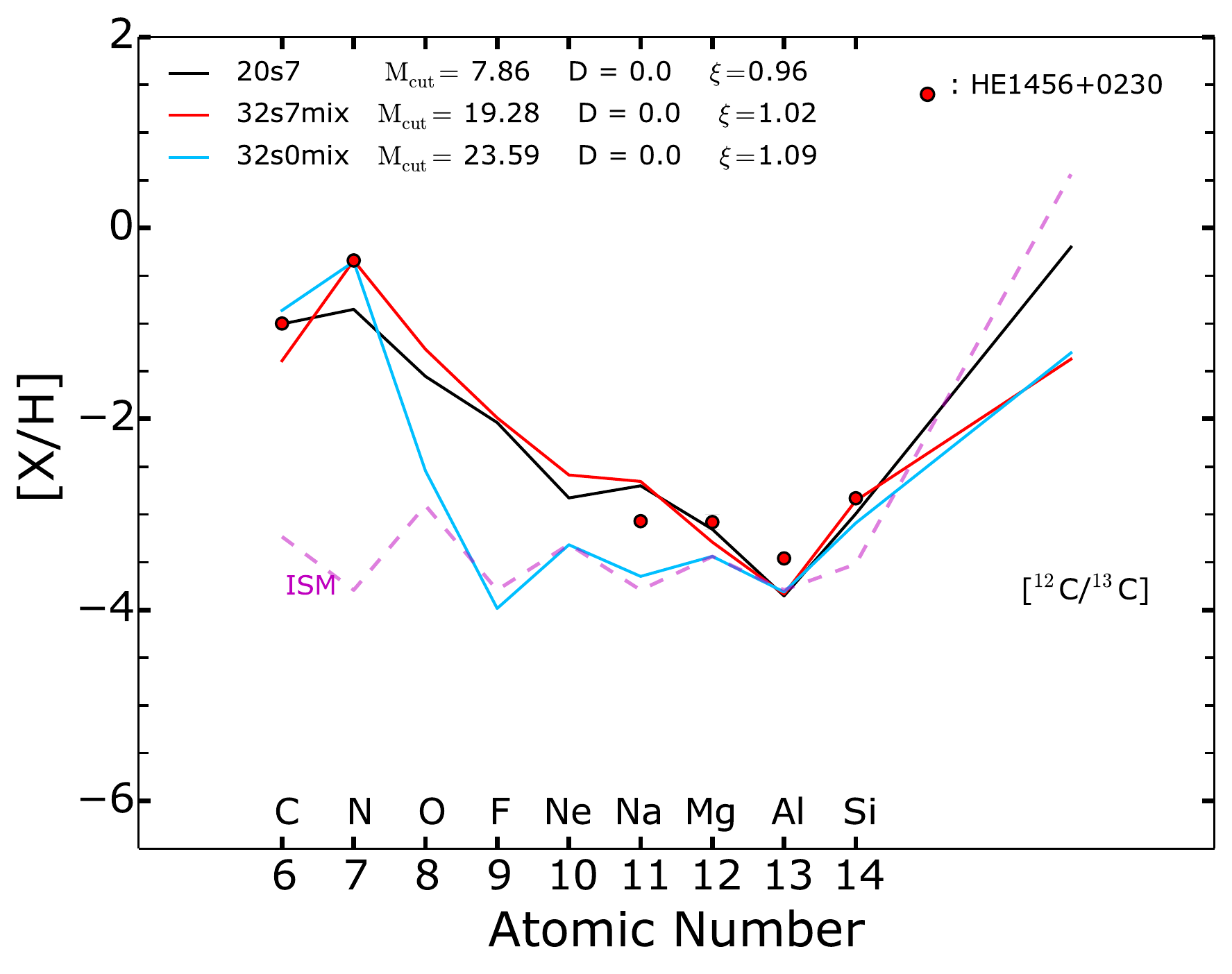}
   \end{minipage}
   \begin{minipage}{.32\linewidth}
       \includegraphics[scale=0.3]{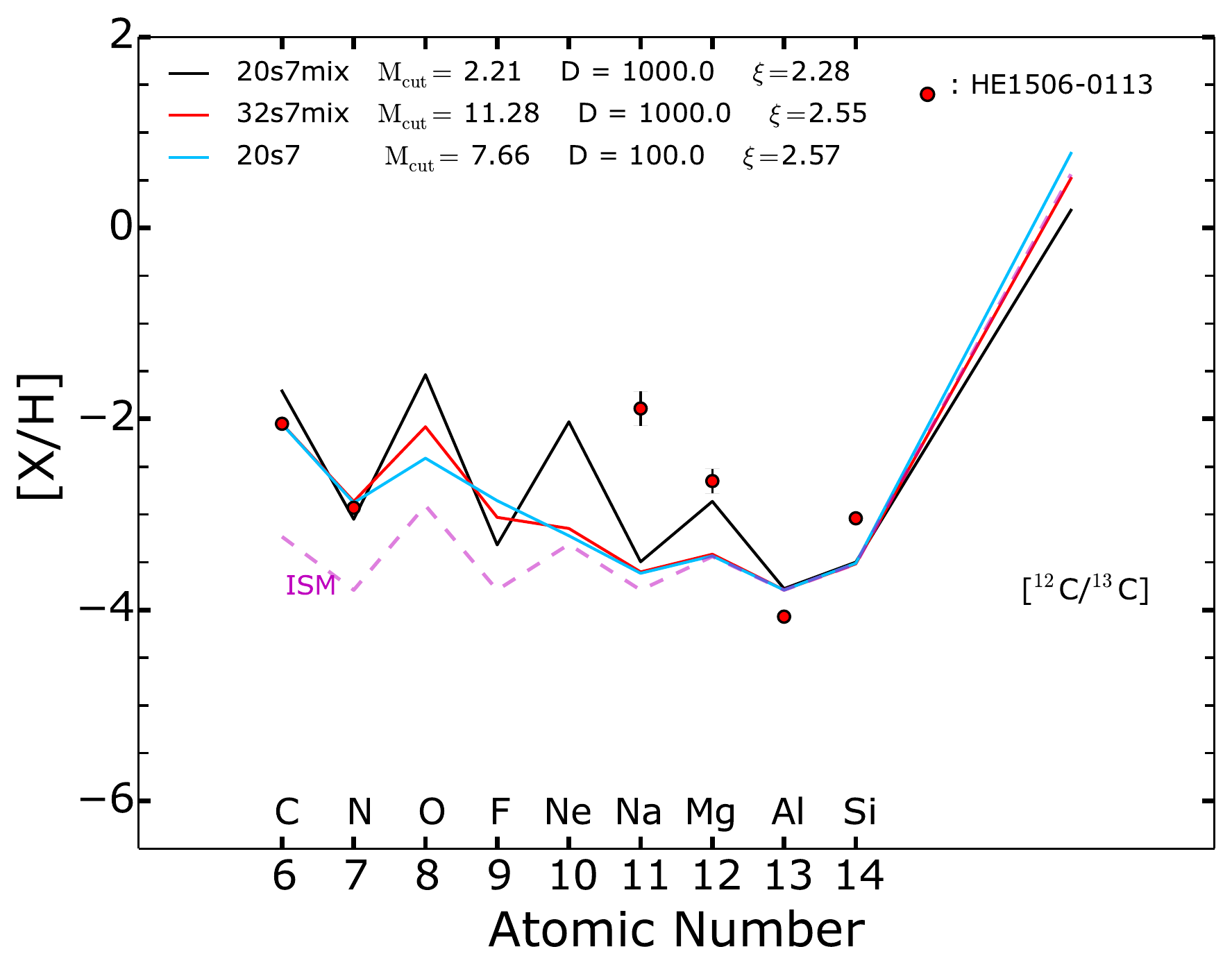}
   \end{minipage}
   \begin{minipage}{.32\linewidth}
       \includegraphics[scale=0.3]{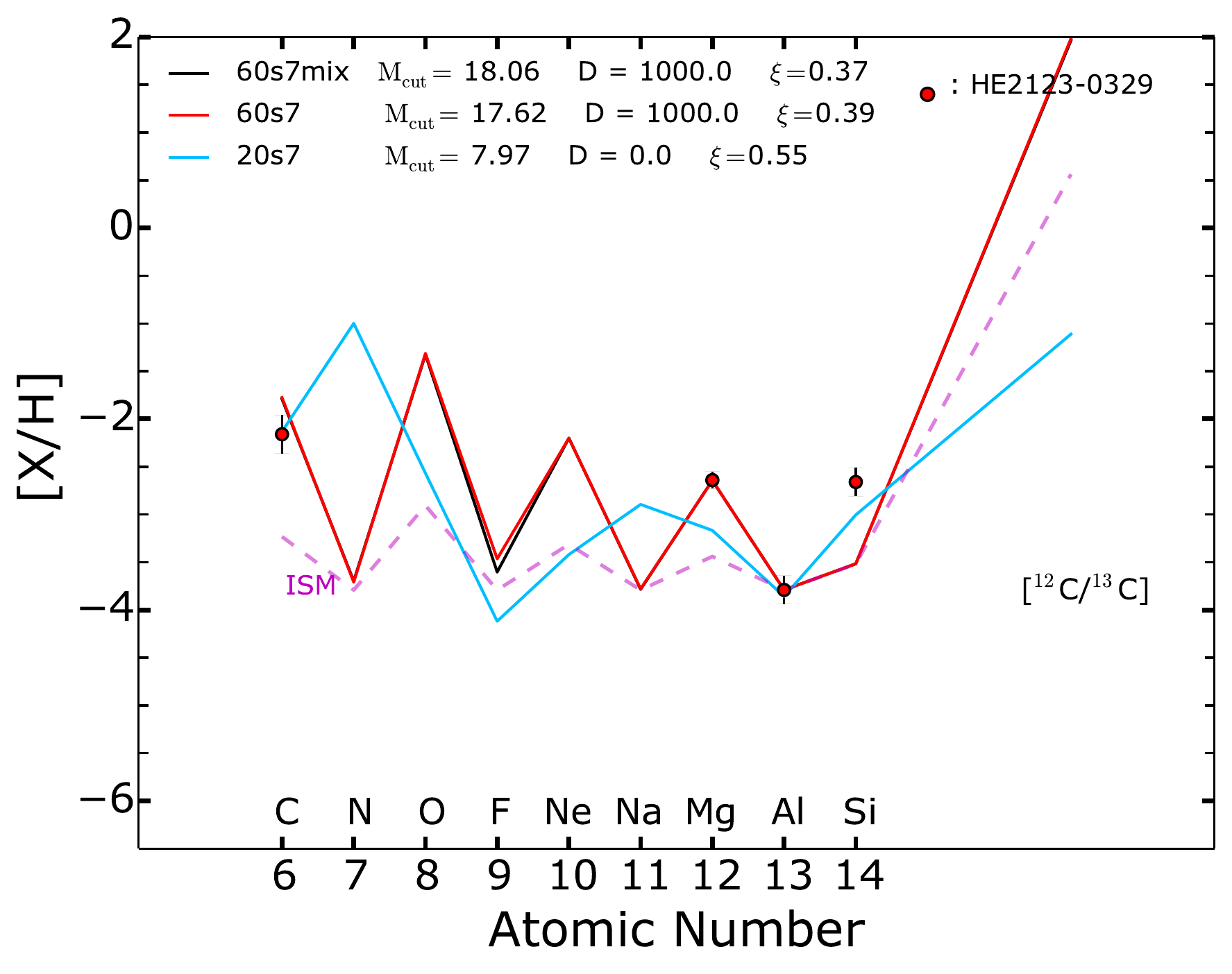}
   \end{minipage}
   \begin{minipage}{.32\linewidth}
       \includegraphics[scale=0.3]{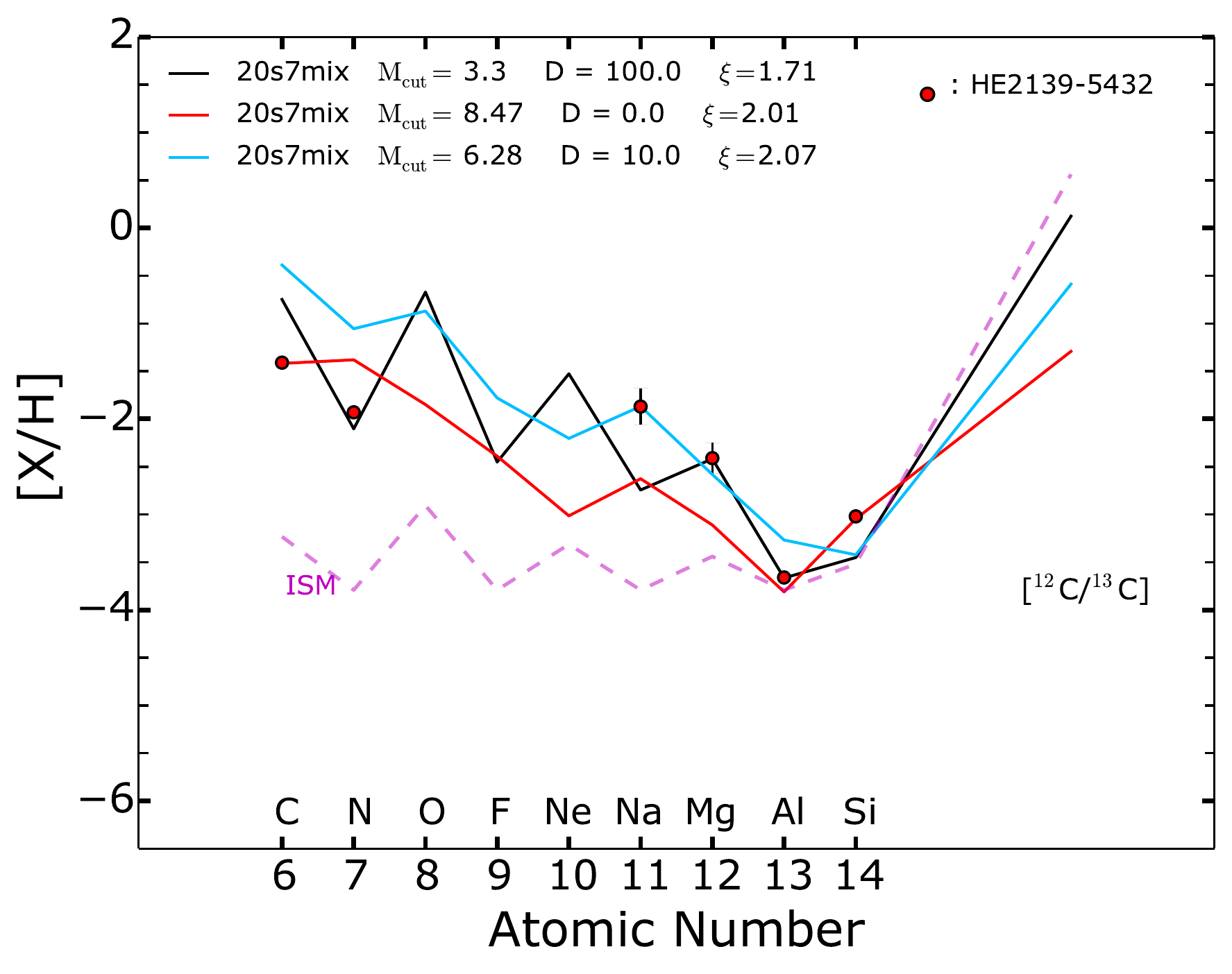}
   \end{minipage}
   \begin{minipage}{.32\linewidth}
       \includegraphics[scale=0.3]{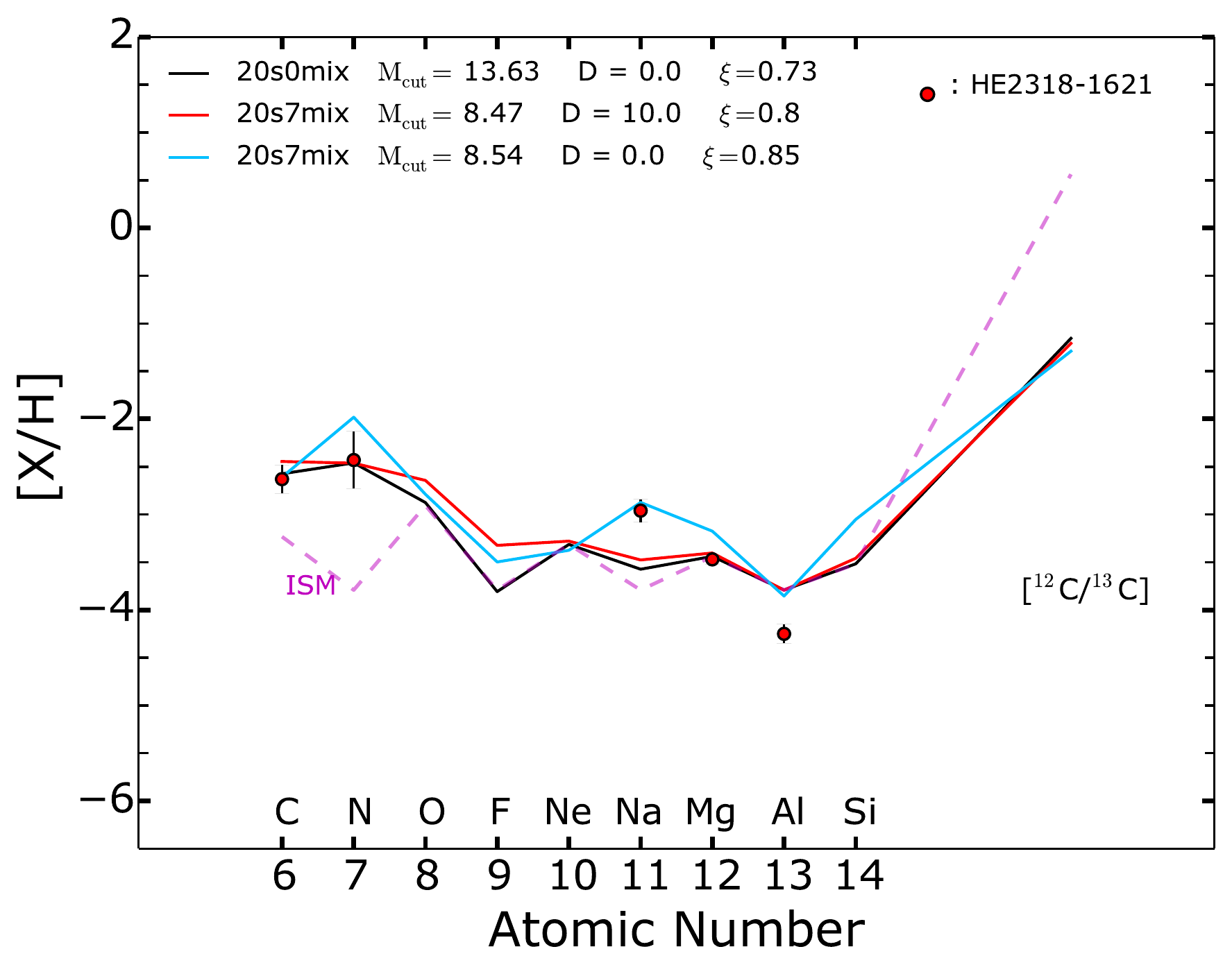}
   \end{minipage}
   \begin{minipage}{.32\linewidth}
       \includegraphics[scale=0.3]{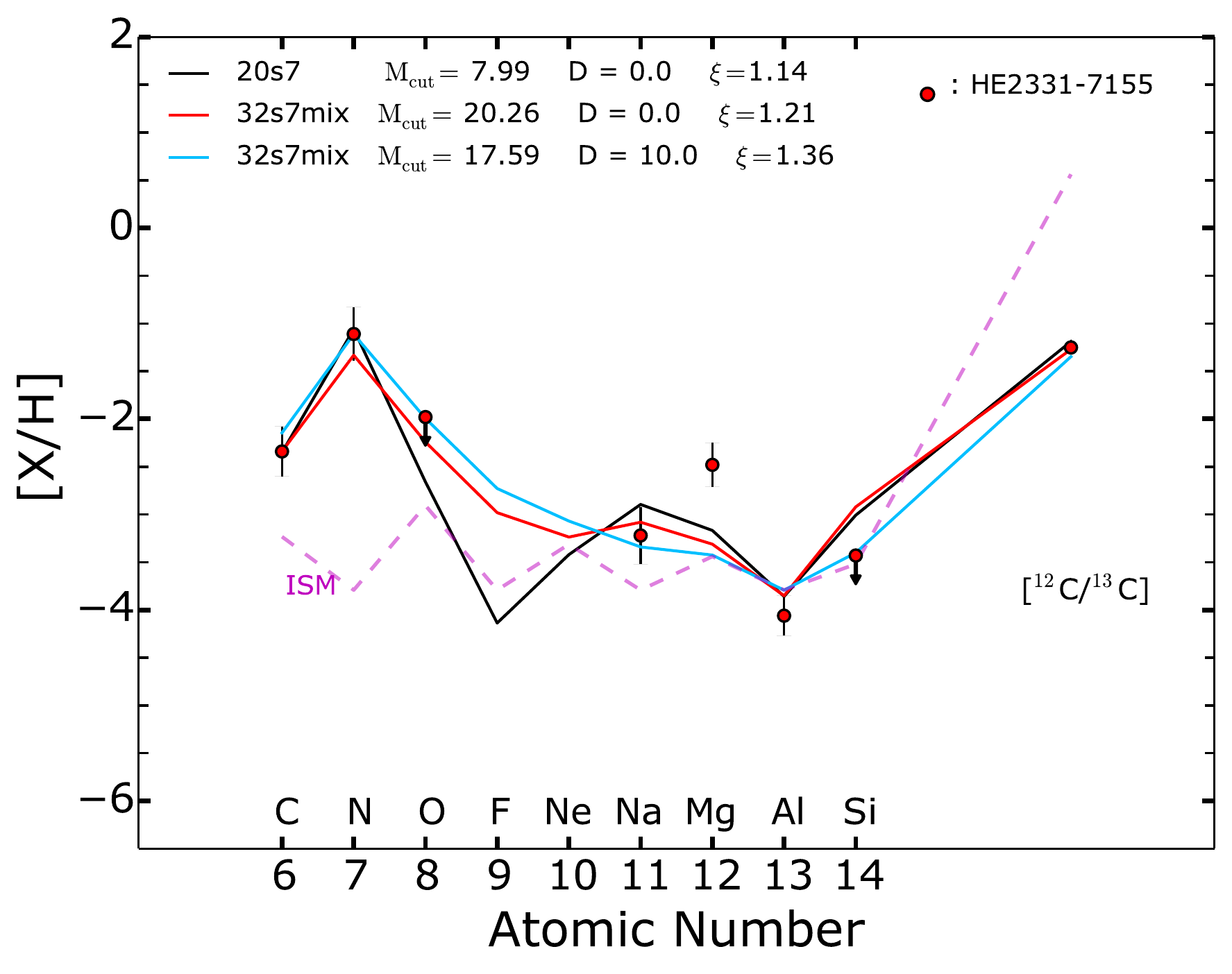}
   \end{minipage}
   \begin{minipage}{.32\linewidth}
       \includegraphics[scale=0.3]{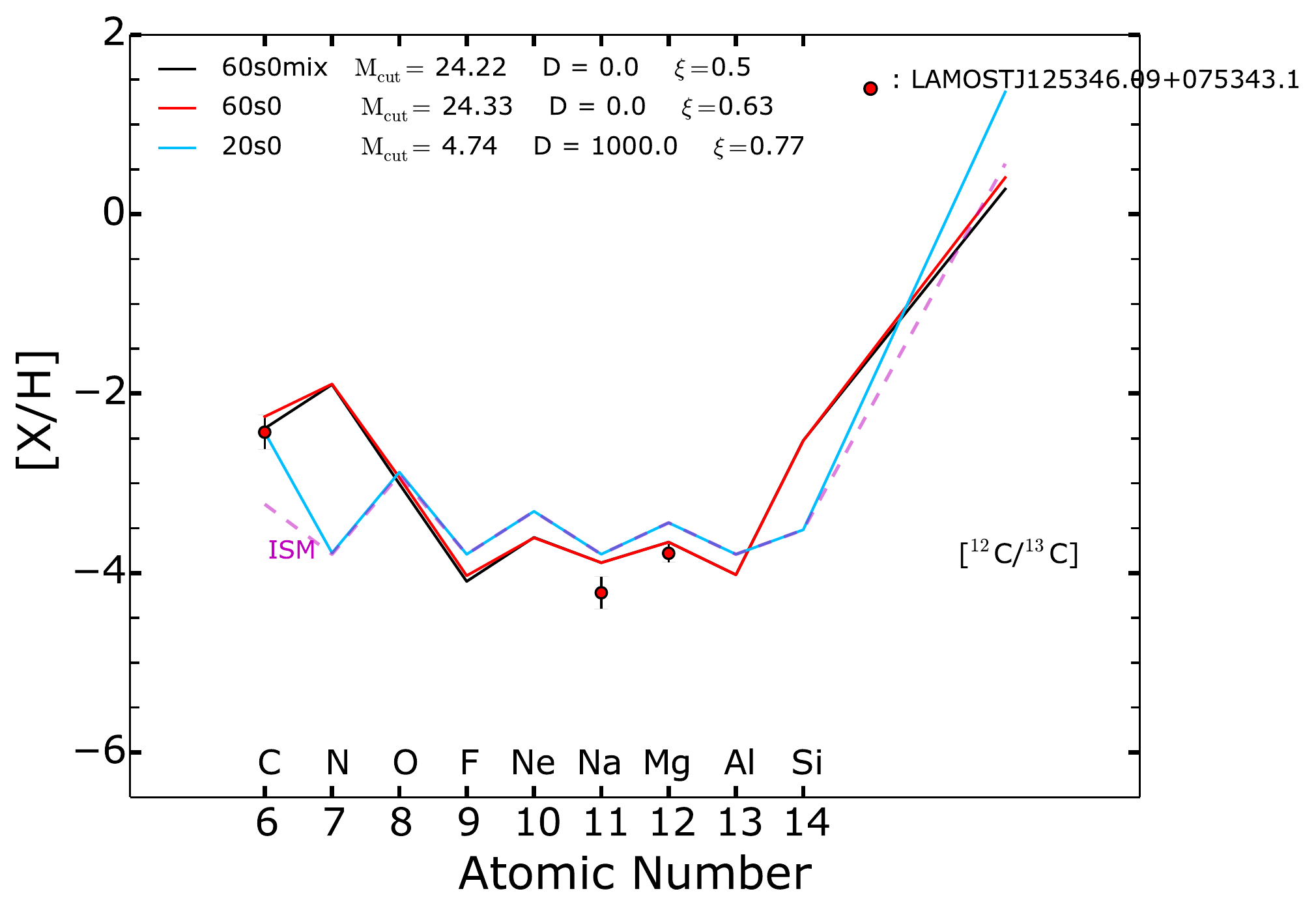}
   \end{minipage}
   \begin{minipage}{.32\linewidth}
       \includegraphics[scale=0.3]{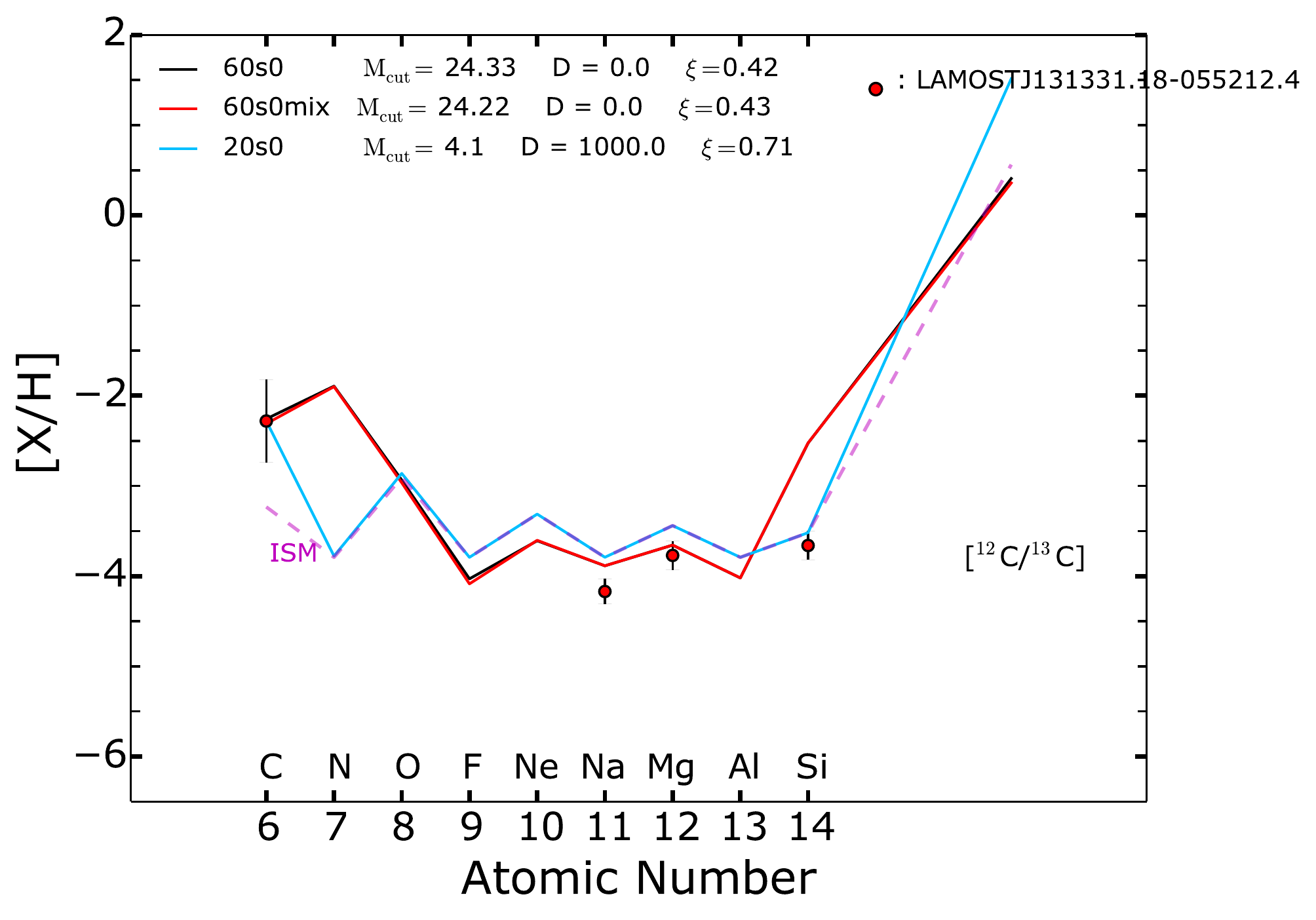}
   \end{minipage}
   \begin{minipage}{.32\linewidth}
       \includegraphics[scale=0.3]{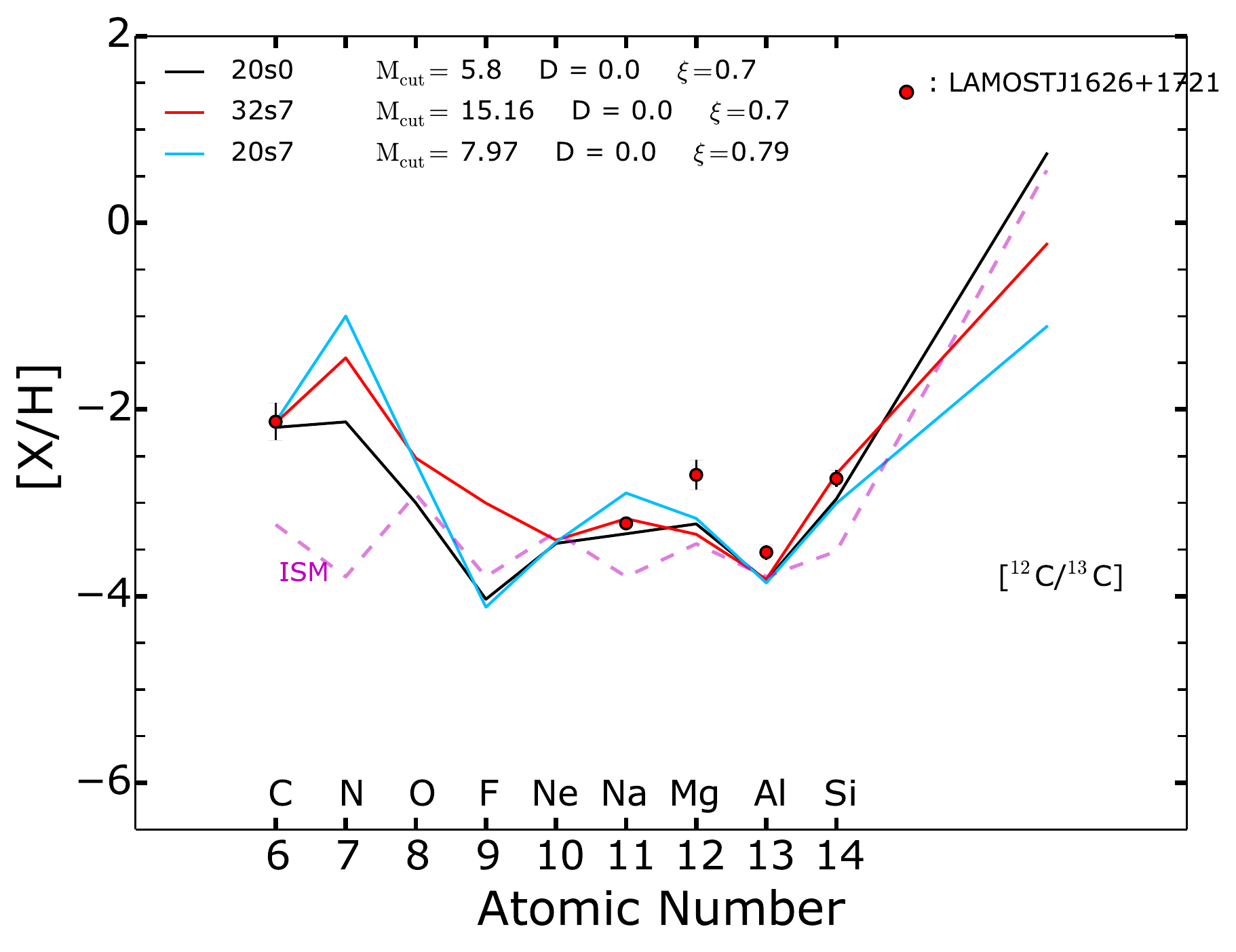}
   \end{minipage}
   \begin{minipage}{.32\linewidth}
       \includegraphics[scale=0.3]{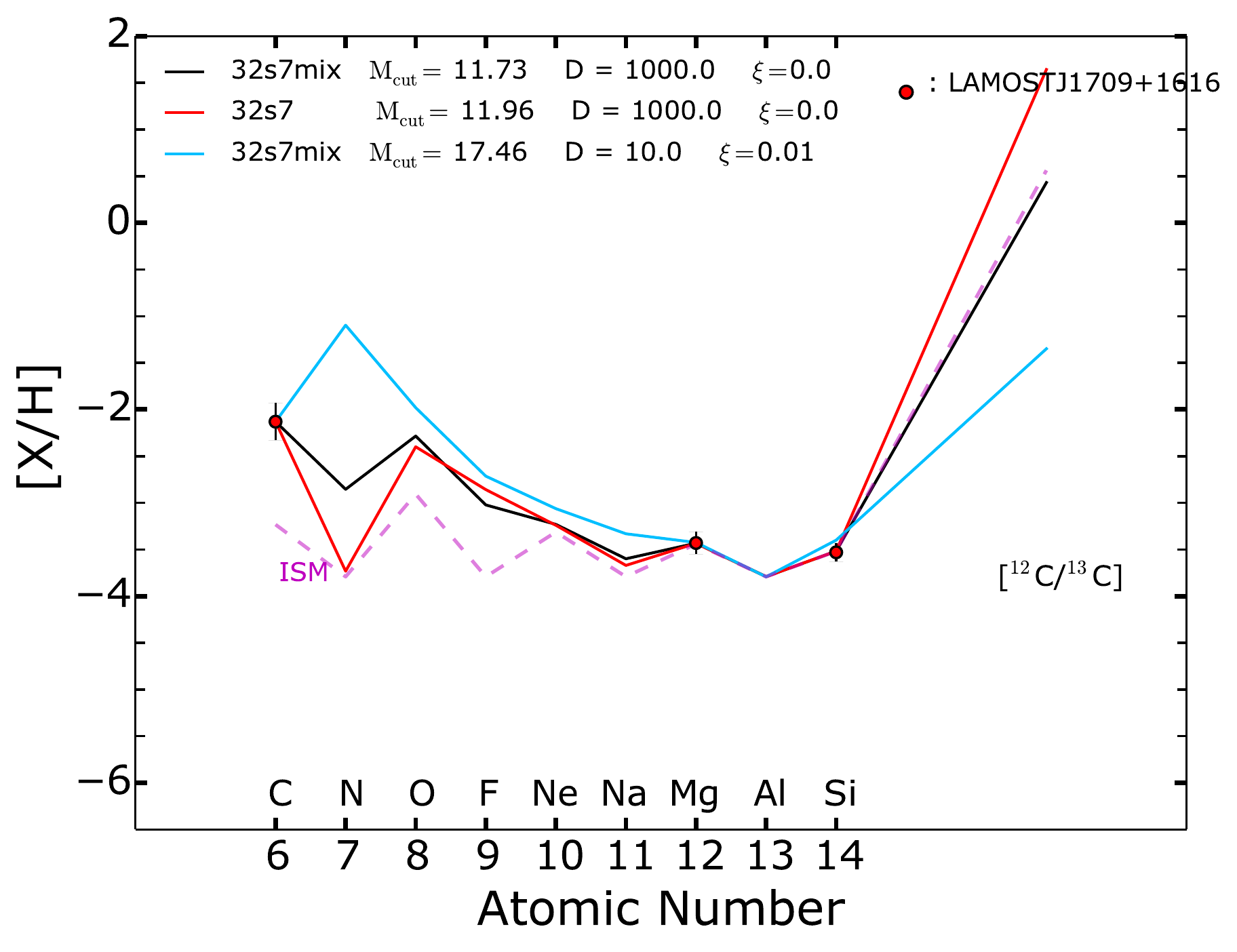}
   \end{minipage}
   \begin{minipage}{.32\linewidth}
       \includegraphics[scale=0.3]{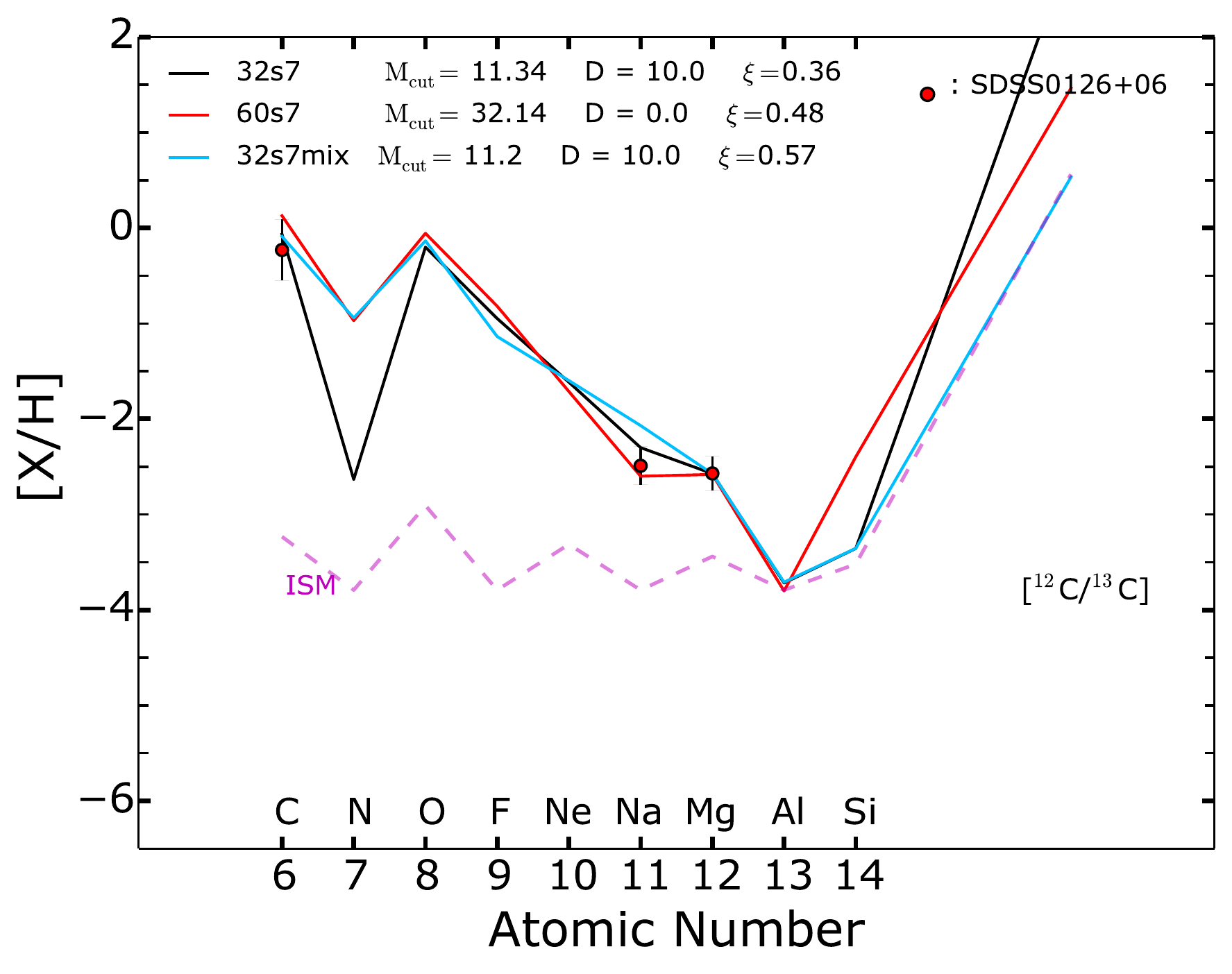}
   \end{minipage}
   \begin{minipage}{.32\linewidth}
       \includegraphics[scale=0.3]{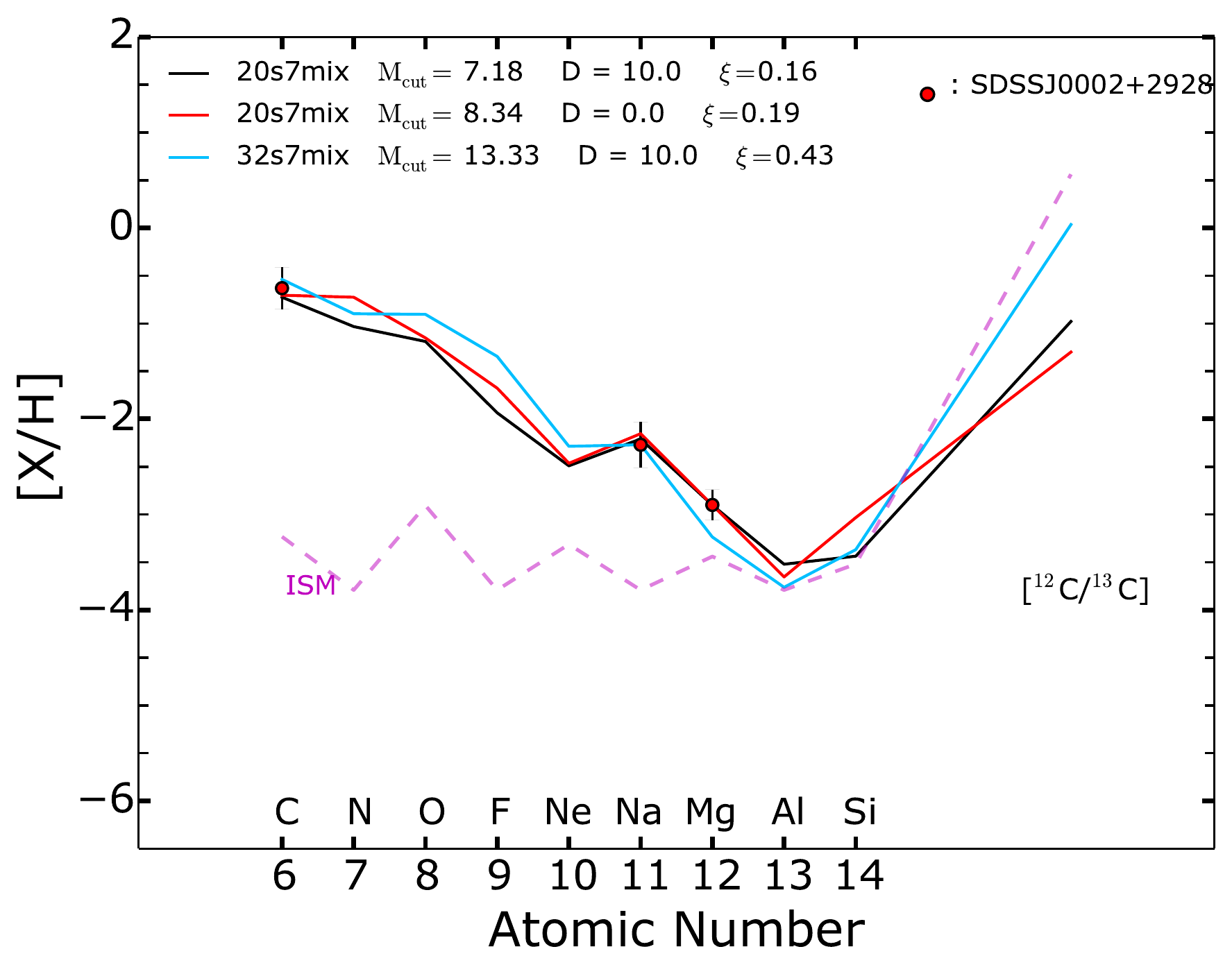}
   \end{minipage}
   \begin{minipage}{.32\linewidth}
       \includegraphics[scale=0.3]{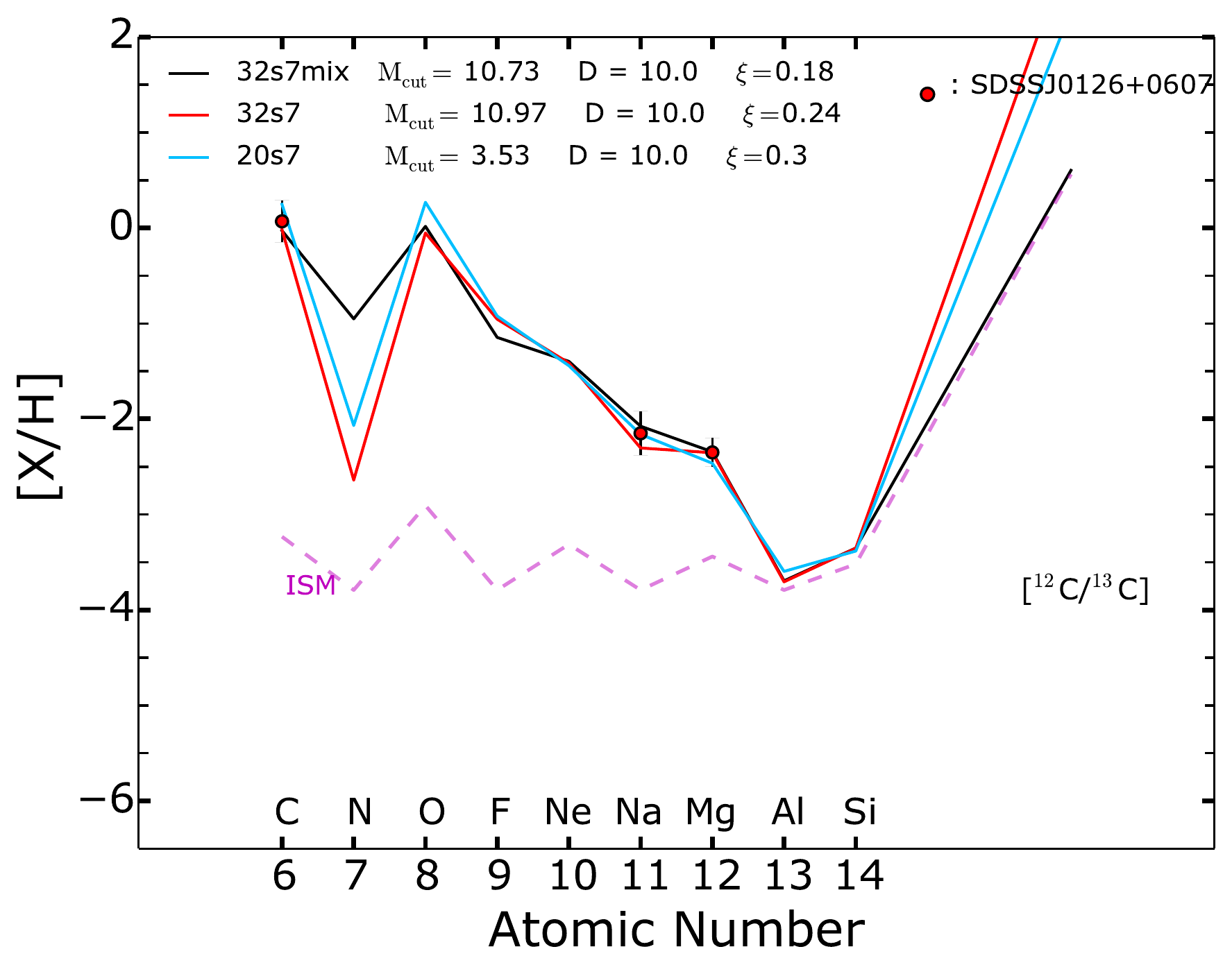}
   \end{minipage}
   \begin{minipage}{.32\linewidth}
       \includegraphics[scale=0.3]{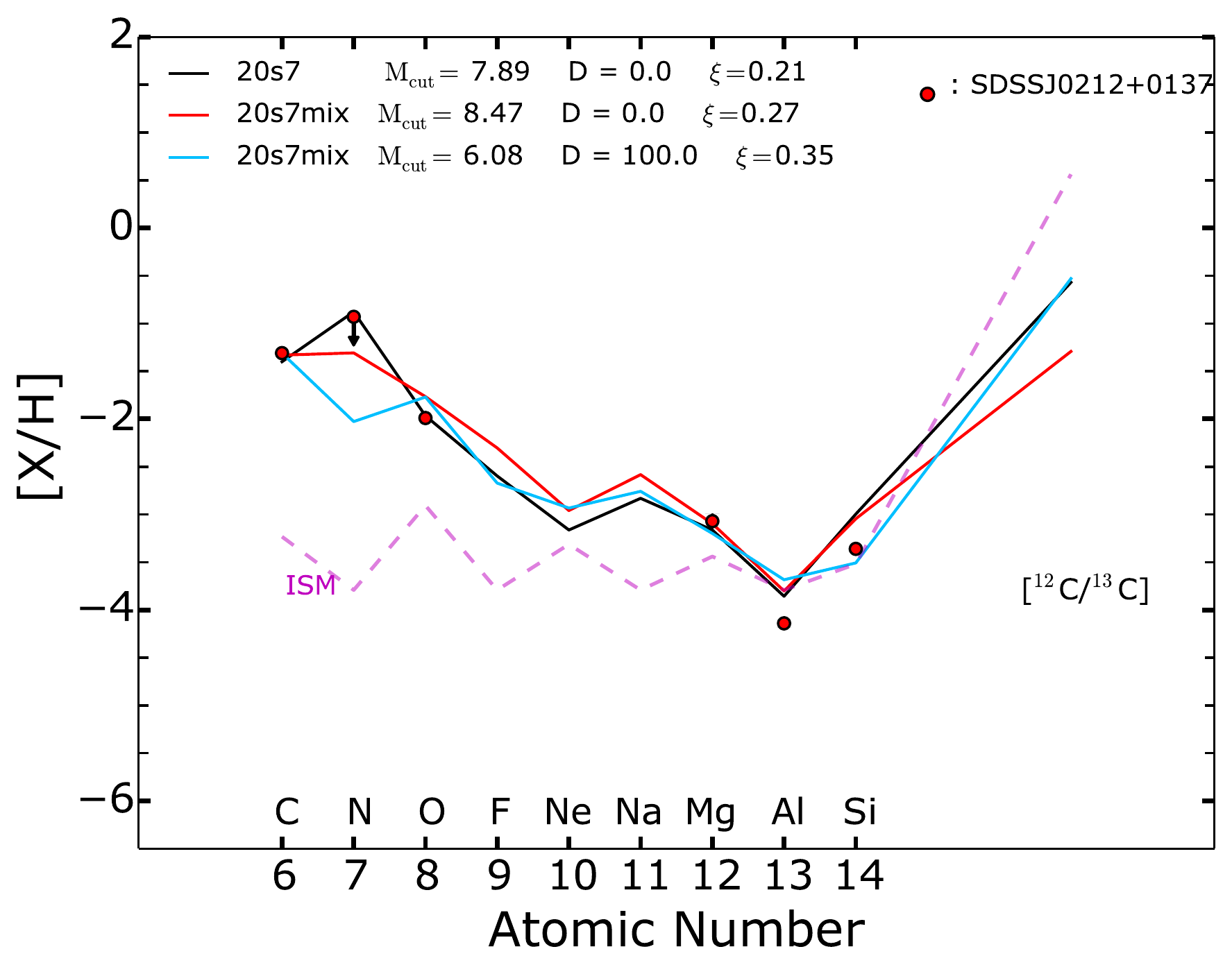}
   \end{minipage}
   \begin{minipage}{.32\linewidth}
       \includegraphics[scale=0.3]{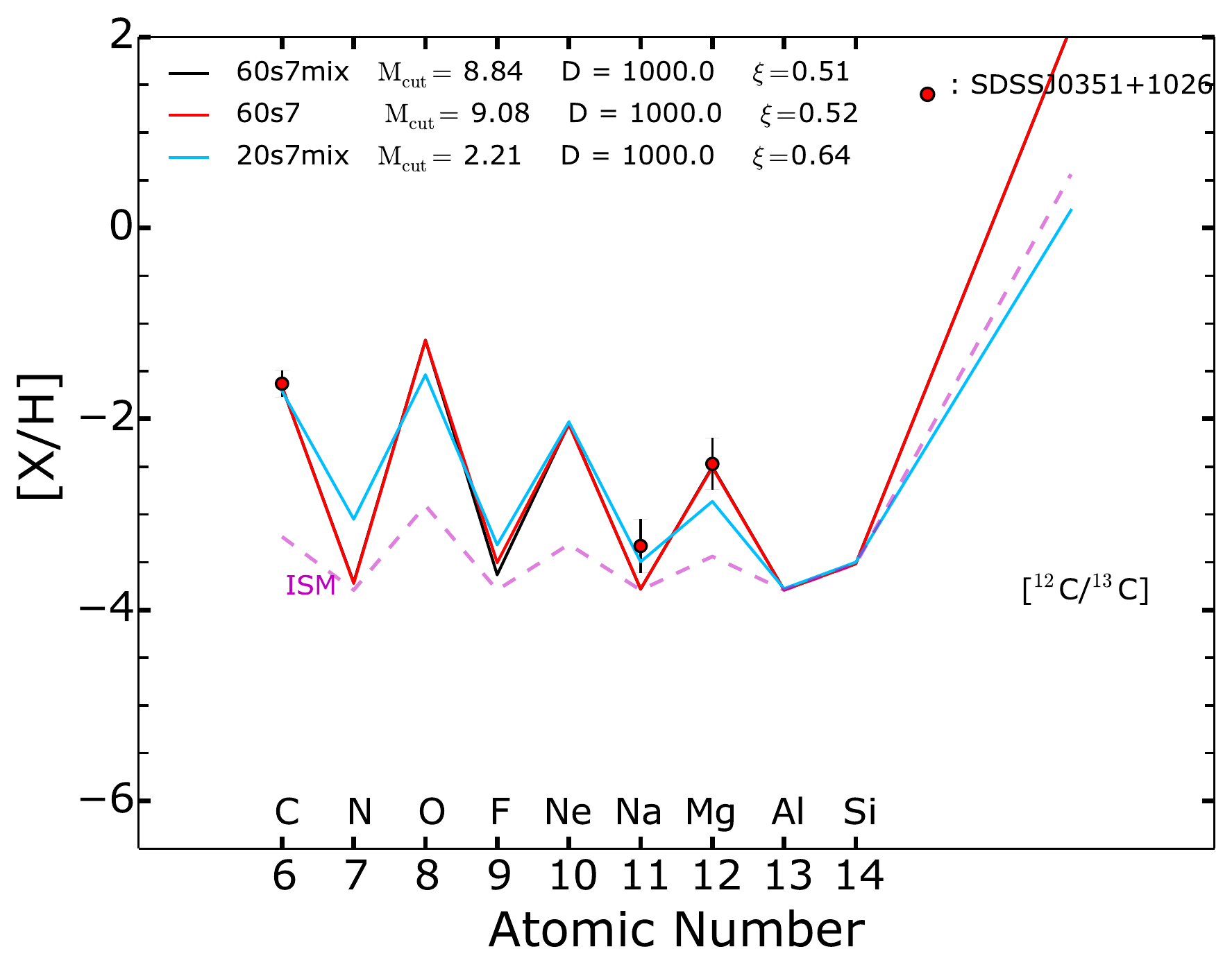}
   \end{minipage}
   \caption[Fig.~\ref{allstar1}, continued]{continued}
\label{allstar4}
    \end{figure*}

   \begin{figure*}
   \centering
   \begin{minipage}{.32\linewidth}
       \includegraphics[scale=0.3]{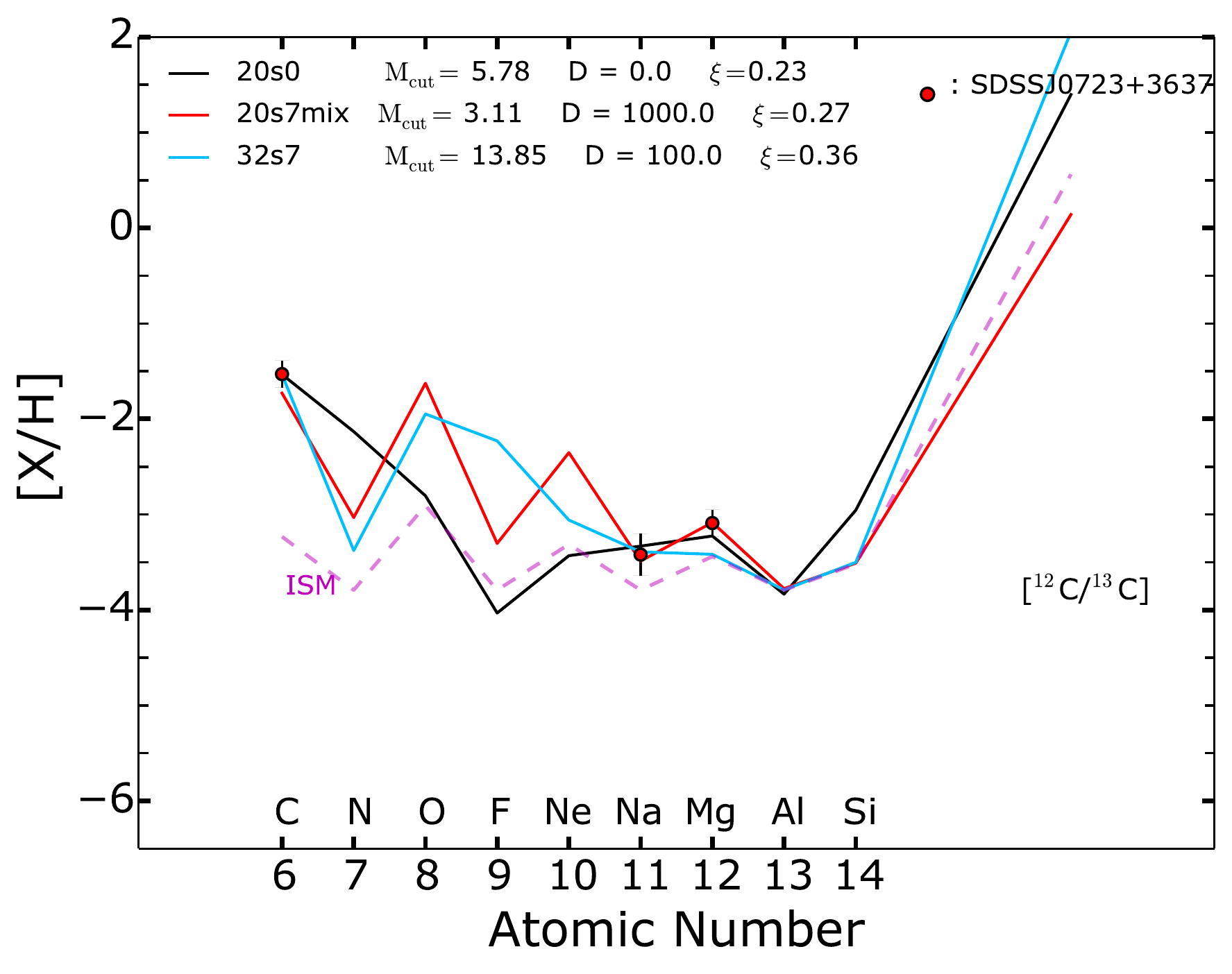}
   \end{minipage}
   \begin{minipage}{.32\linewidth}
       \includegraphics[scale=0.3]{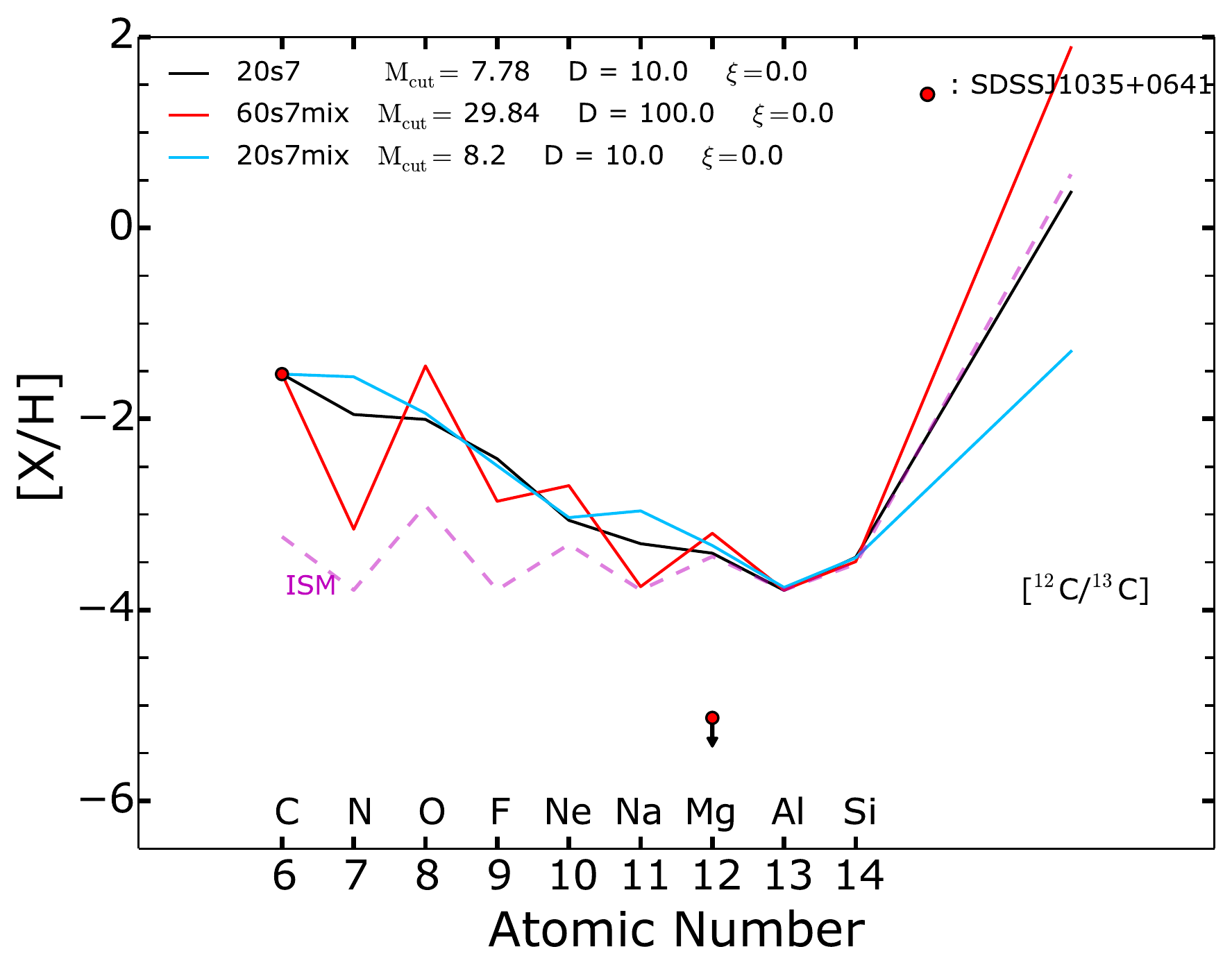}
   \end{minipage}
   \begin{minipage}{.32\linewidth}
       \includegraphics[scale=0.3]{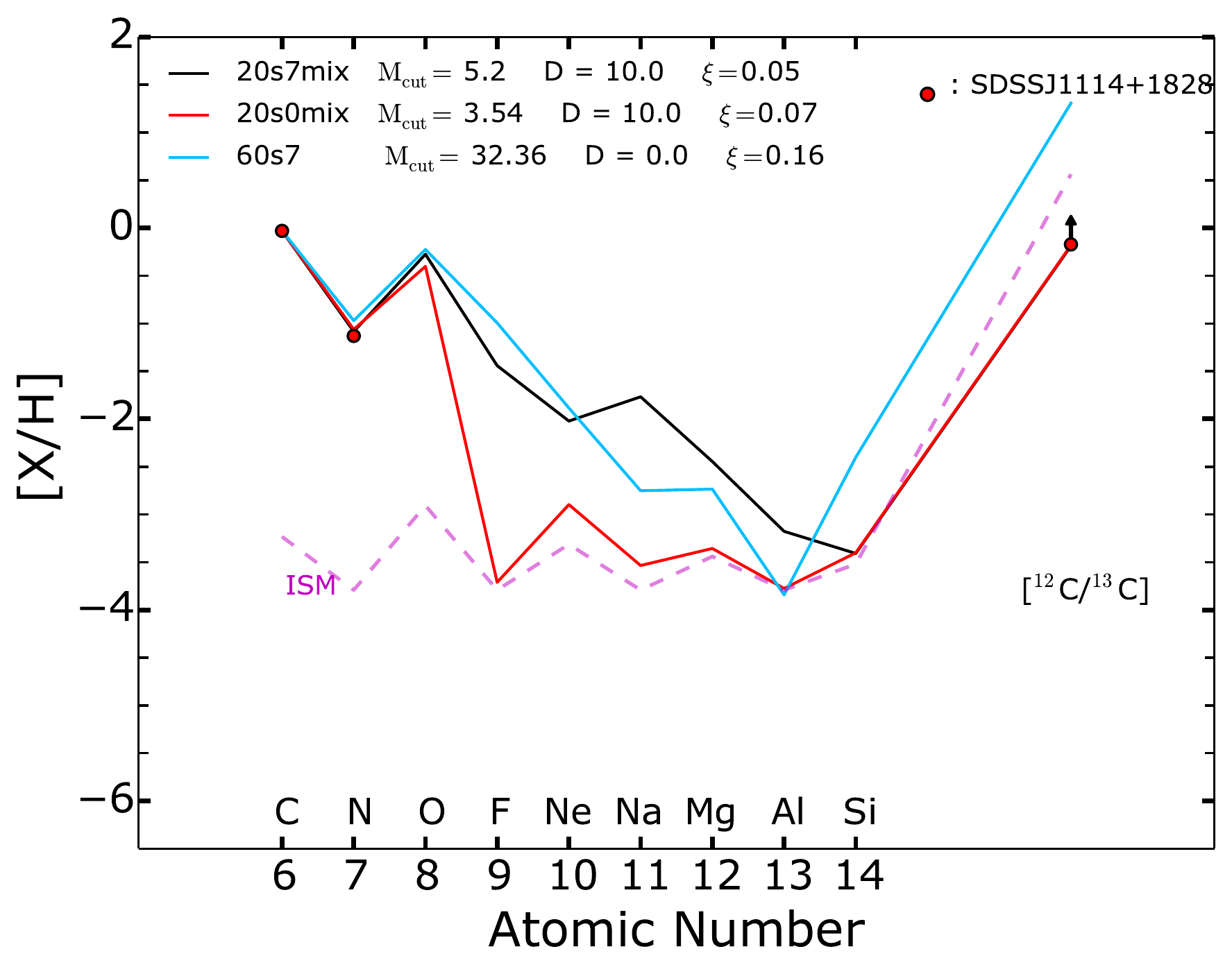}
   \end{minipage}
   \begin{minipage}{.32\linewidth}
       \includegraphics[scale=0.3]{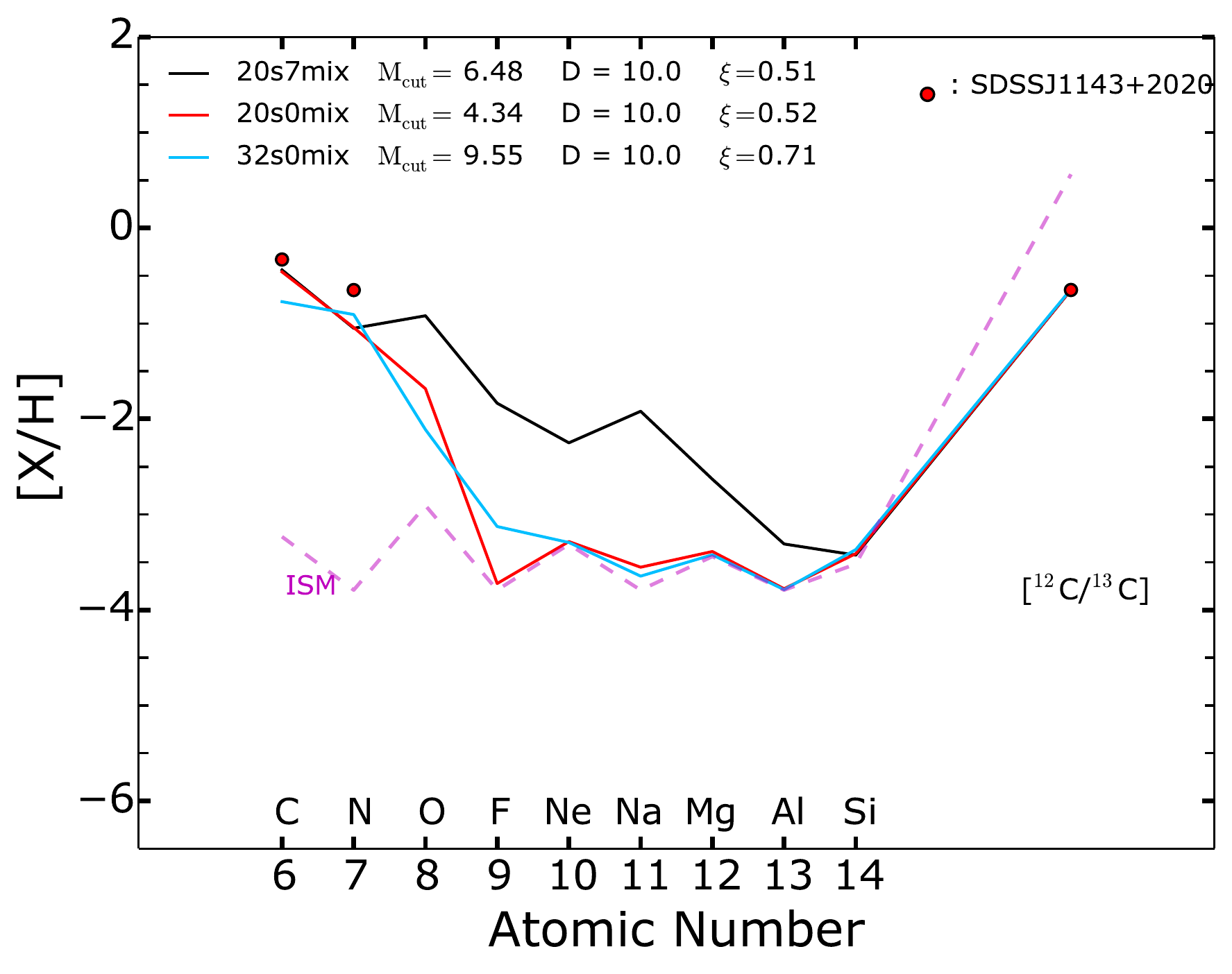}
   \end{minipage}
   \begin{minipage}{.32\linewidth}
       \includegraphics[scale=0.3]{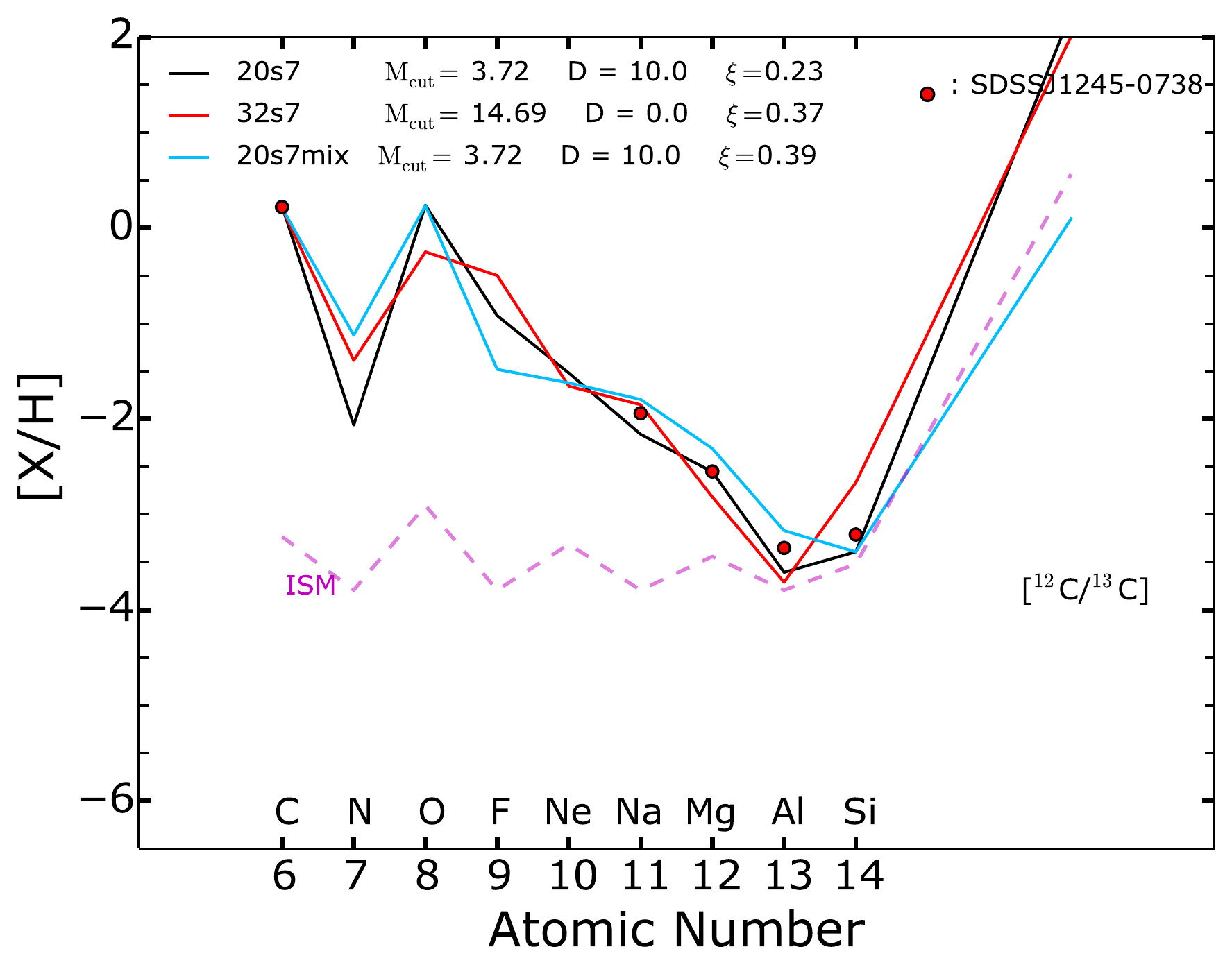}
   \end{minipage}
   \begin{minipage}{.32\linewidth}
       \includegraphics[scale=0.3]{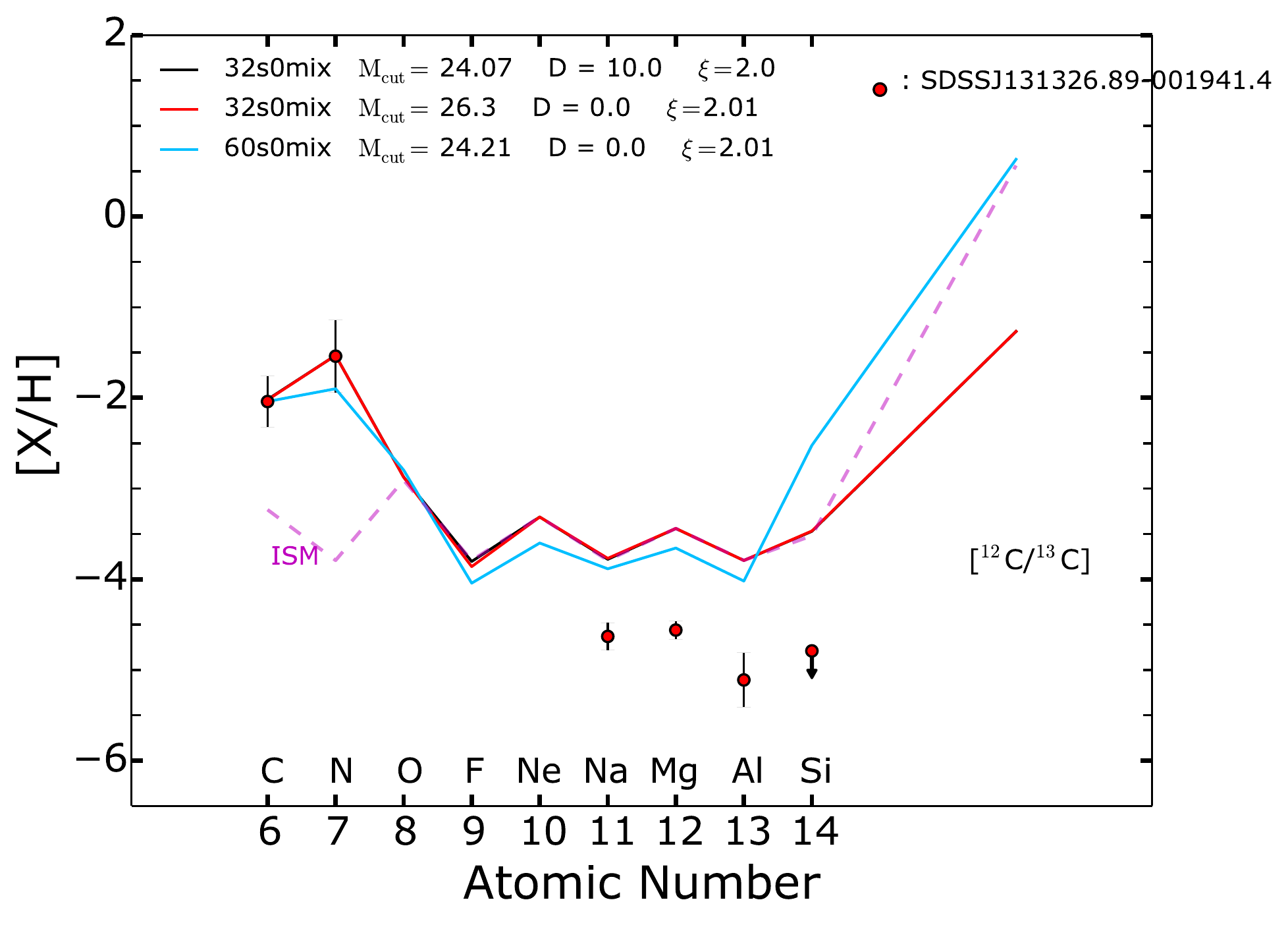}
   \end{minipage}
   \begin{minipage}{.32\linewidth}
       \includegraphics[scale=0.3]{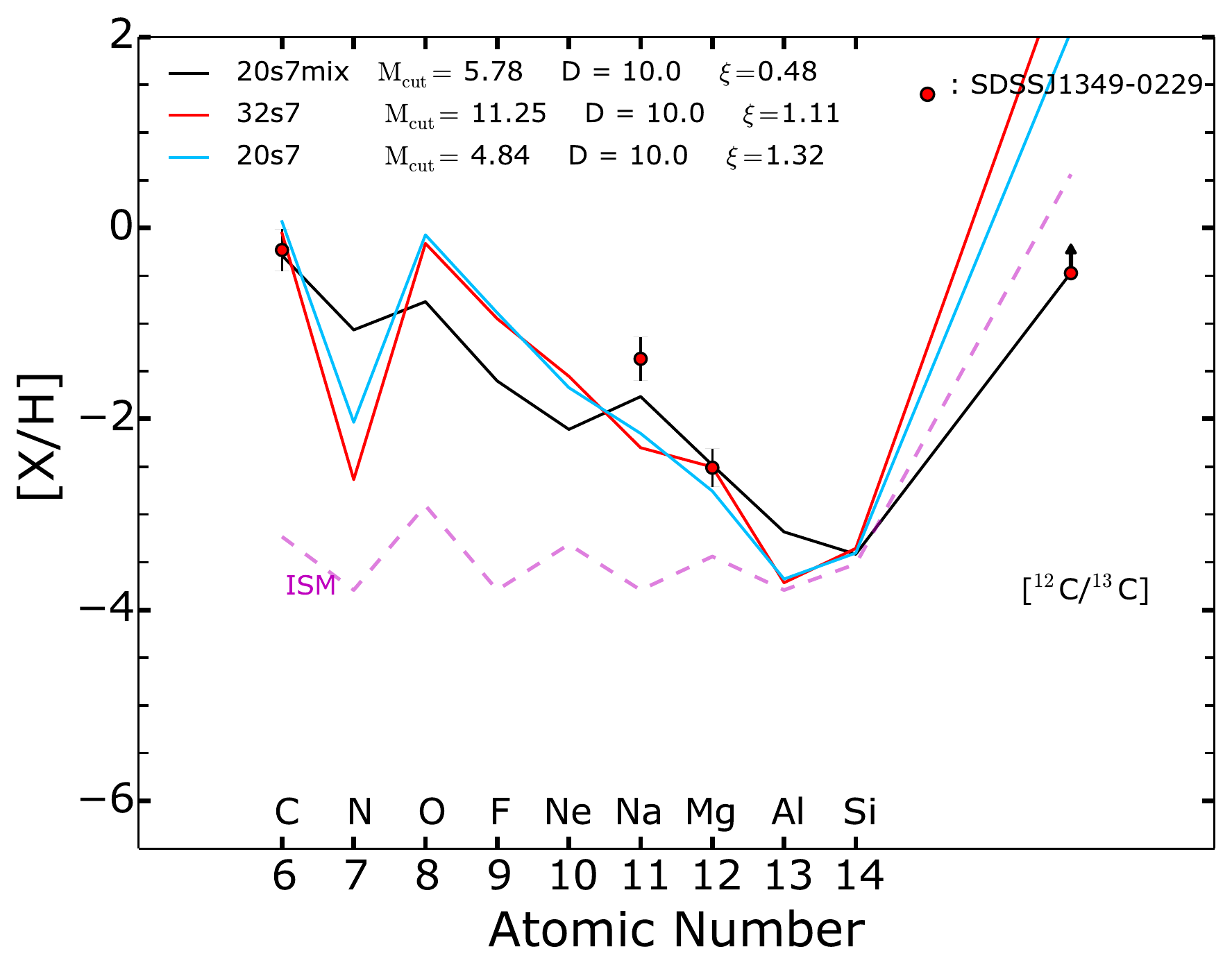}
   \end{minipage}
   \begin{minipage}{.32\linewidth}
       \includegraphics[scale=0.3]{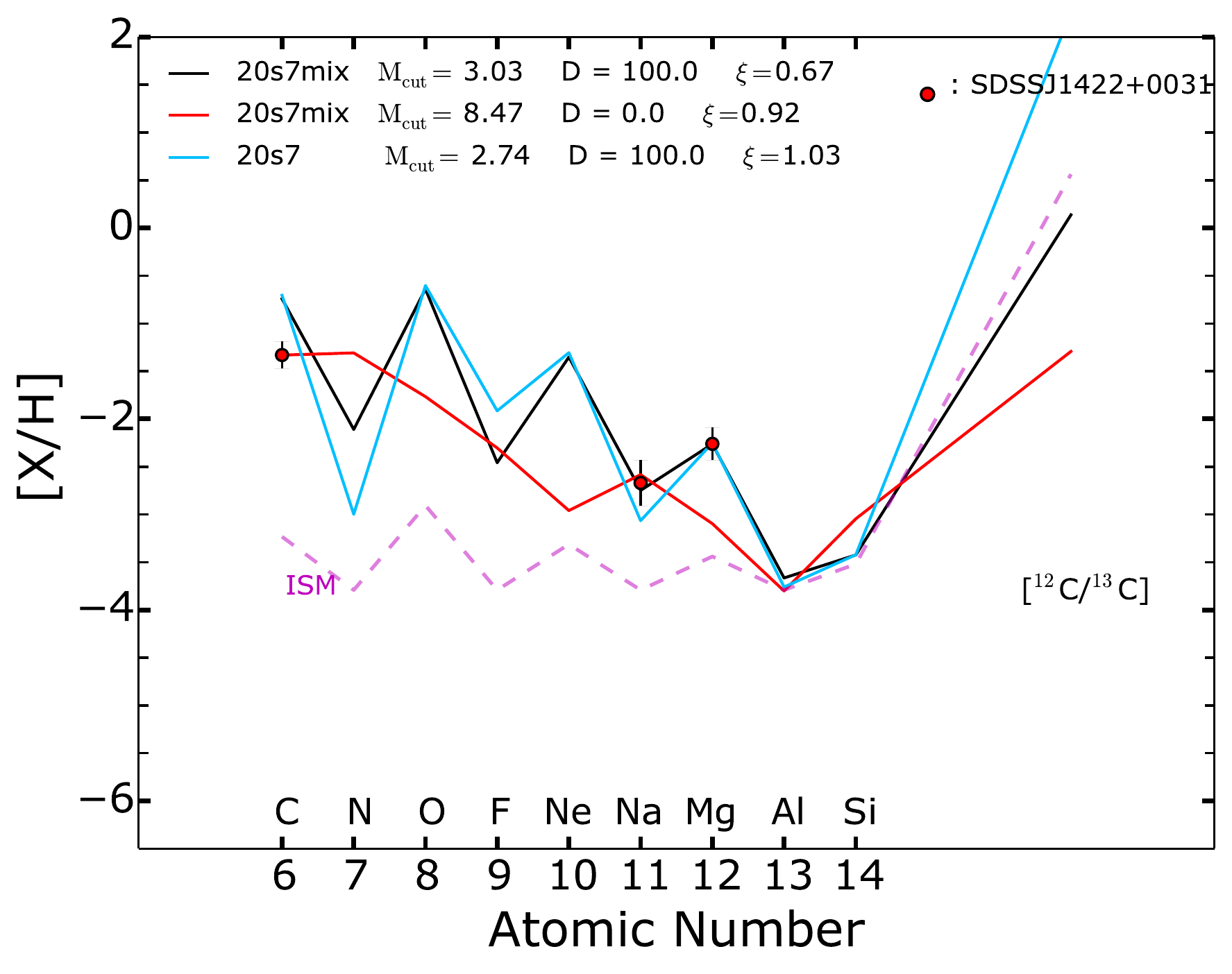}
   \end{minipage}
   \begin{minipage}{.32\linewidth}
       \includegraphics[scale=0.3]{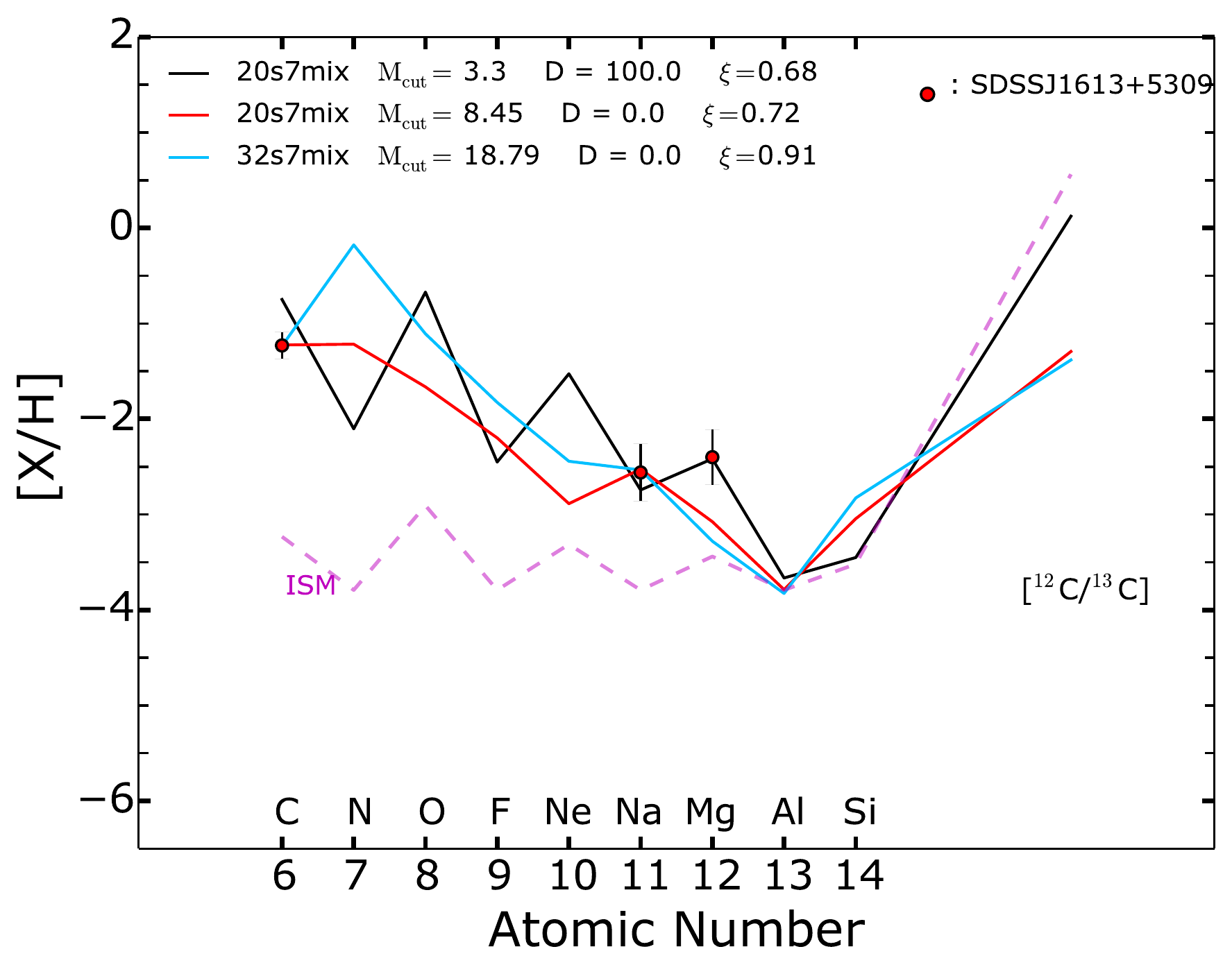}
   \end{minipage}
   \begin{minipage}{.32\linewidth}
       \includegraphics[scale=0.3]{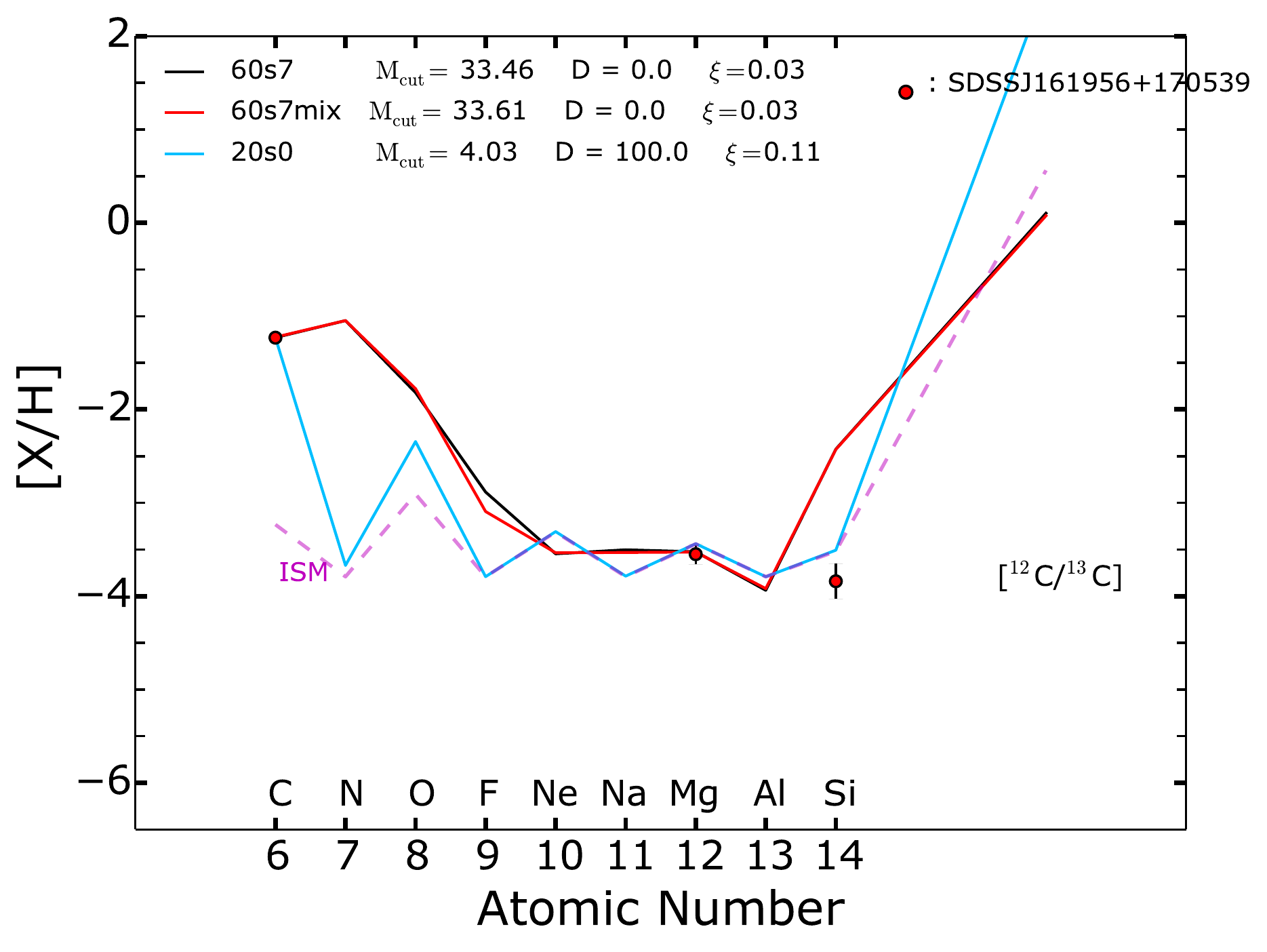}
   \end{minipage}
   \begin{minipage}{.32\linewidth}
       \includegraphics[scale=0.3]{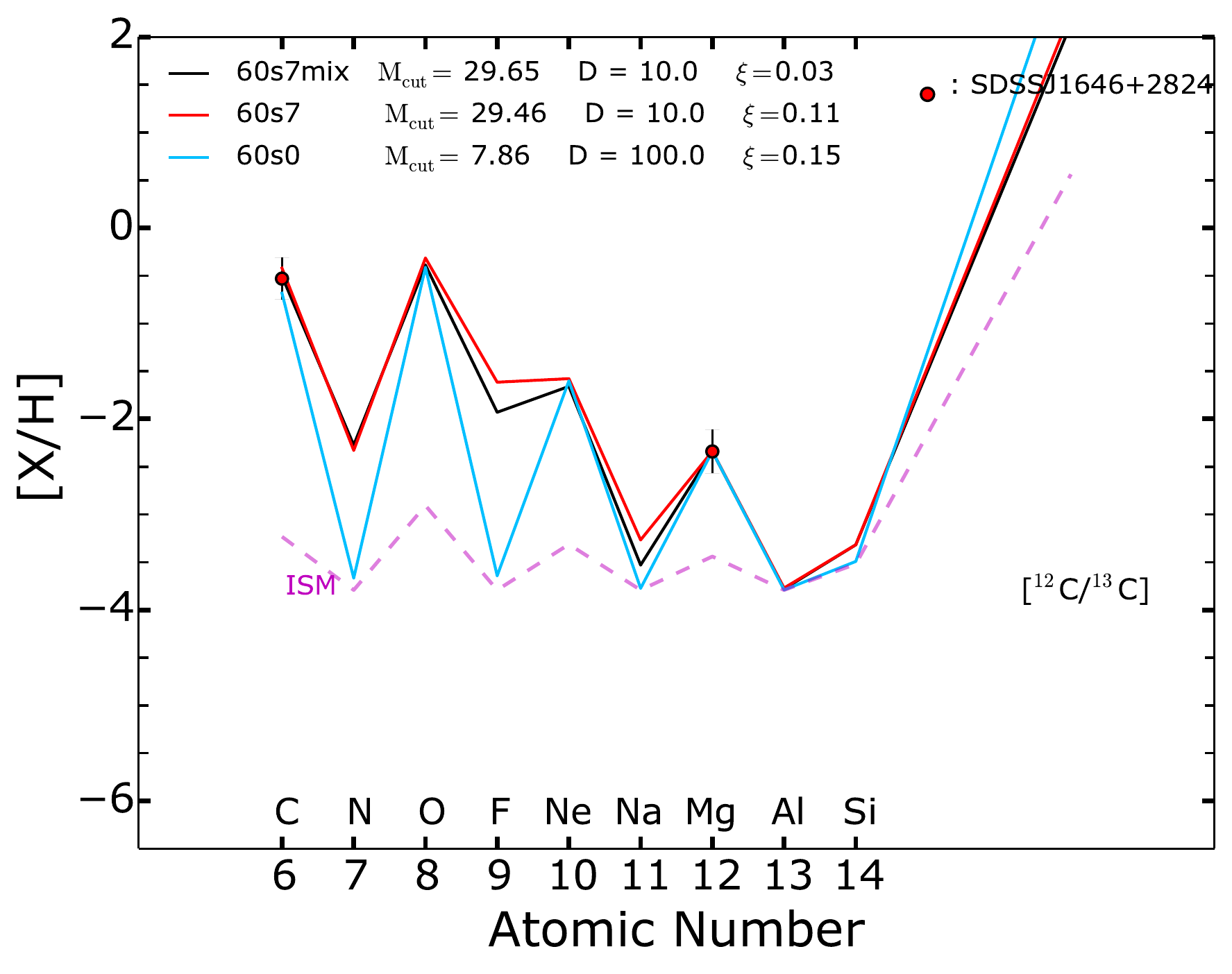}
   \end{minipage}
   \begin{minipage}{.32\linewidth}
       \includegraphics[scale=0.3]{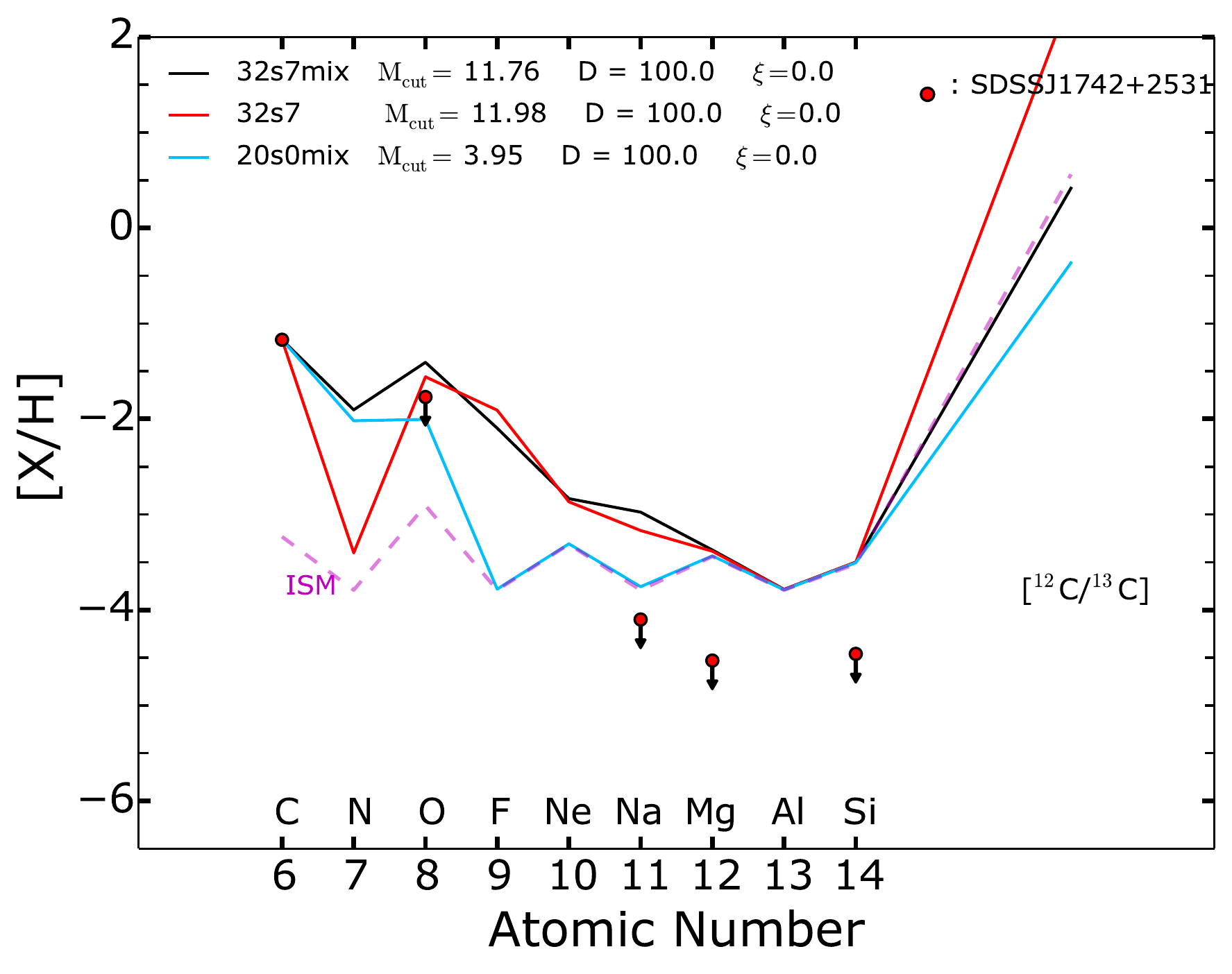}
   \end{minipage}
   \begin{minipage}{.32\linewidth}
       \includegraphics[scale=0.3]{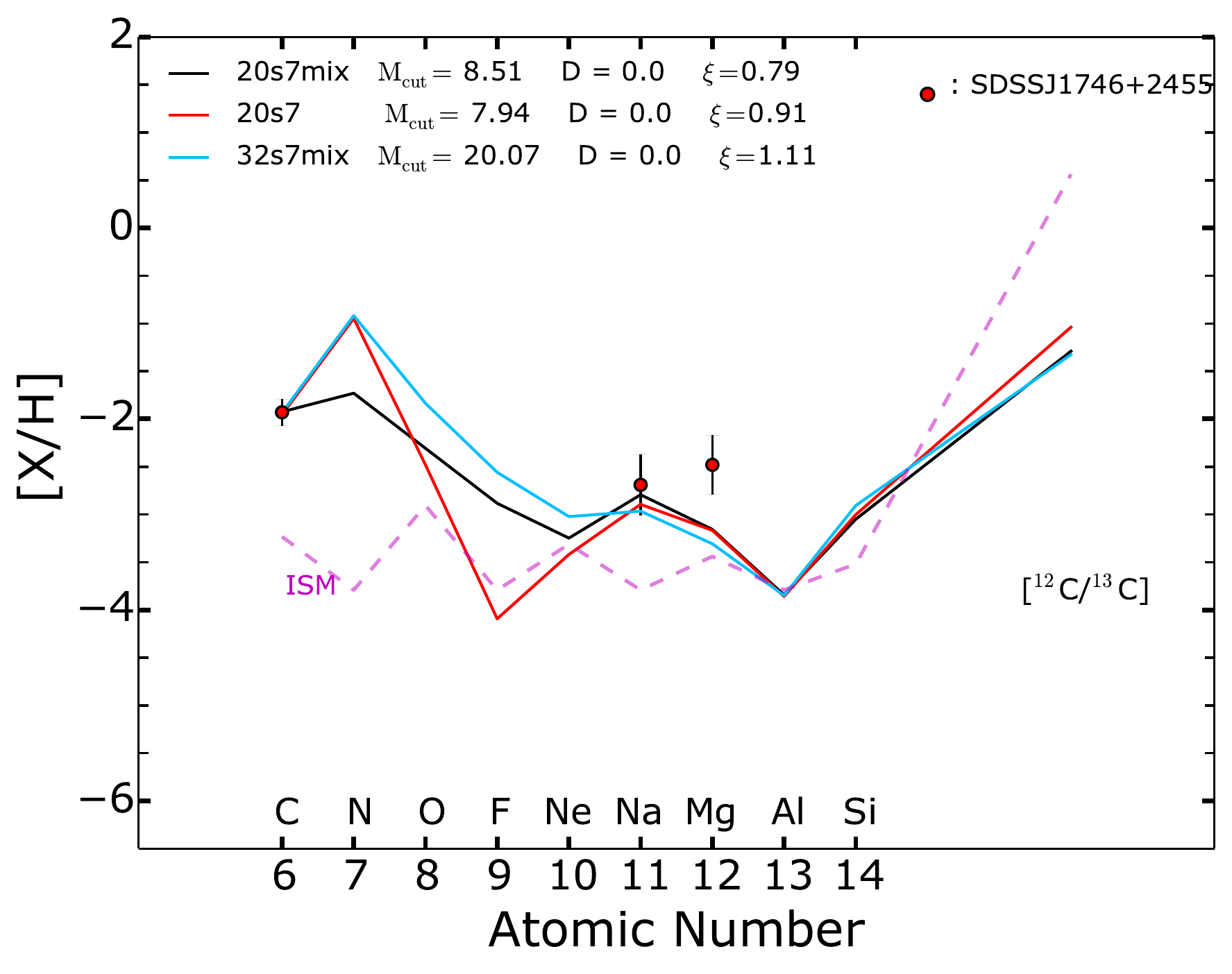}
   \end{minipage}
   \begin{minipage}{.32\linewidth}
       \includegraphics[scale=0.3]{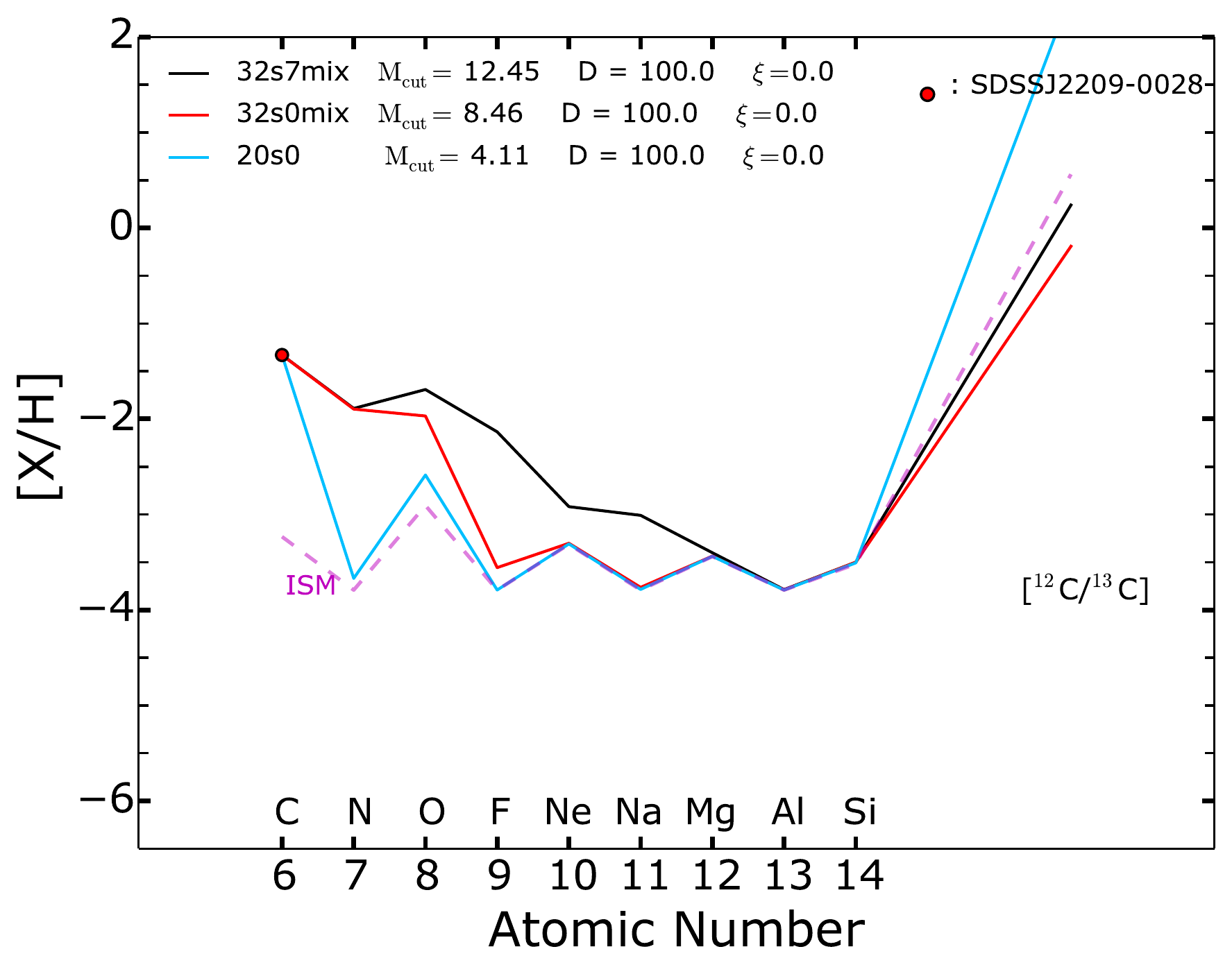}
   \end{minipage}
   \begin{minipage}{.32\linewidth}
       \includegraphics[scale=0.3]{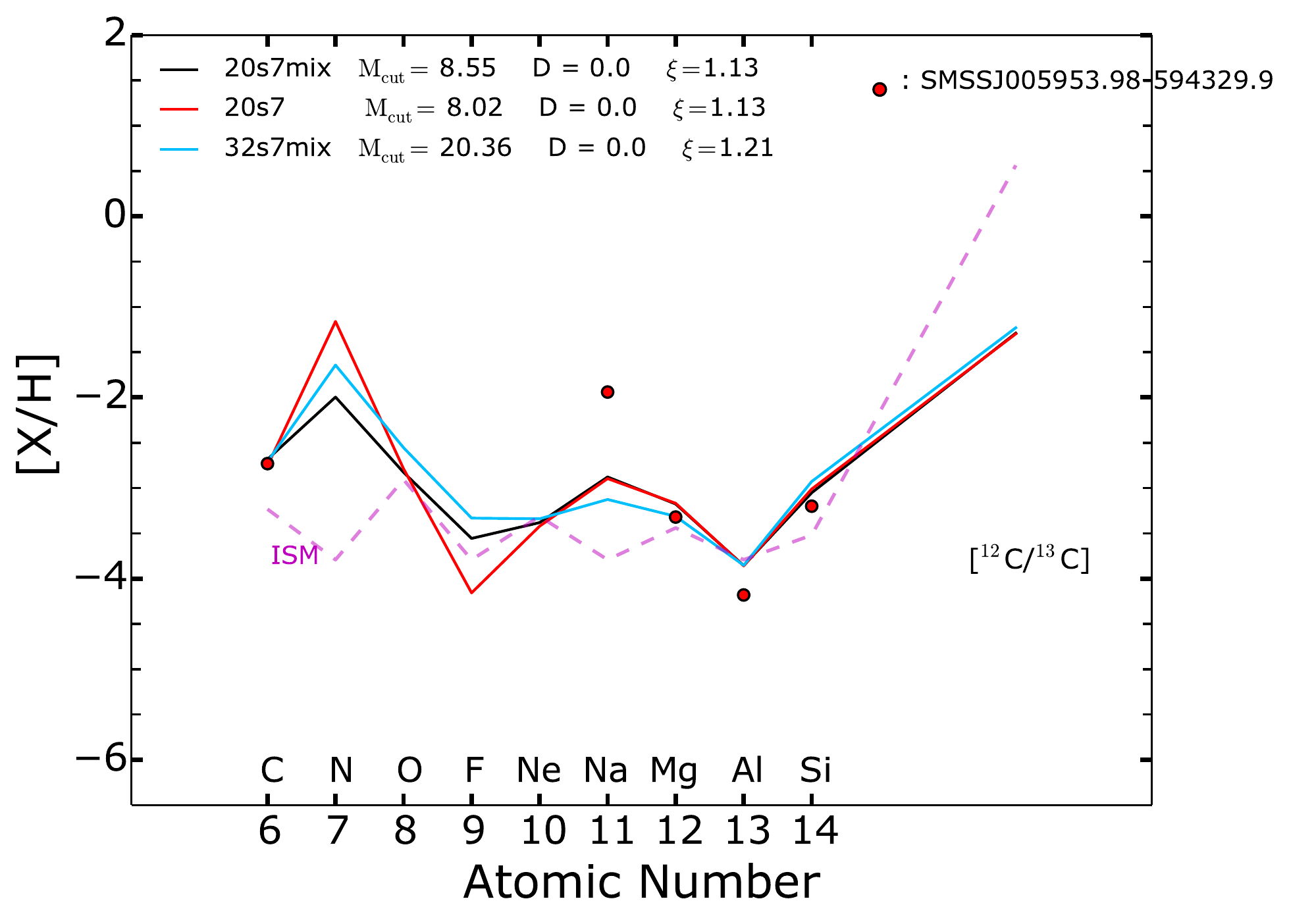}
   \end{minipage}
   \caption[Fig.~\ref{allstar1}, continued]{continued}
\label{allstar5}
    \end{figure*}  
    
%\chapter{Publications related to this thesis \label{cpublilist}}
%\chapter{Publications \label{cpublilist}}
\chapter{Publications \label{cpublilist}}
\markboth{PUBLICATIONS}{Publications}
%\section{List of papers \label{slistart}}

%This section contains the list of publication. After the list are the 3 refereed papers not directly included in the thesis. The 2 others papers can be found page~\hyperlink{pdfsproc}{\pageref{psproc}} and \ref{pdfcemps}{\pageref*{pcemps}} of this thesis.
This section contains the list of publication. After the list are the 3 refereed papers (page~\pageref{axions}, \pageref{pcempno} and \pageref{pbox}) that are not directly included in the thesis. The 2 others papers can be found page~\pageref{psproc} and \pageref{pcemps} of this thesis.\\

%Ce travail de th\`ese a donn\'e lieu \`a des publications dont la liste se trouve \`a la page \pageref{cpublilist}.

%-------------------------------------------------------------------------------
\noindent \uline{\large{Refereed articles}}

\begin{itemize}

    \item \textit{Non standard s-process in massive rotating stars. Yields of $10-150$ $M_{\odot}$ models at $Z=10^{-3}$}\\
    \textbf{Choplin, A}., Hirschi R., Meynet G., Ekstr\"om S., Chiappini C. \& Laird A.\\
    %\href{}{2018, accepted in \aap}
    \href{http://adsabs.harvard.edu/abs/2018A\%26A...618A.133C}{2018, \aap, 618, A133}
    
    \item \textit{Are some CEMP-s stars the daughters of spinstars?} \\
    \textbf{Choplin, A}., Hirschi R., Meynet G. \& Ekstr\"om S.\\    
    \href{http://cdsads.u-strasbg.fr/abs/2017A\%26A...607L...3C}{2017, \aap, 607, L3}

    \item \textit{Effects of axions on Population III stars}\\
    \textbf{Choplin, A}., Coc A., Meynet G., Olive K. A., Uzan J.-P. \& Vangioni E.\\
     \href{http://cdsads.u-strasbg.fr/abs/2017A\%26A...605A.106C}2017, \aap, 605, A106

    \item \textit{Pre-supernova mixing in CEMP-no source stars}\\
    \textbf{Choplin, A}., Ekstr\"om S., Meynet G., Maeder A., Georgy C. \& Hirschi R.\\
     \href{http://cdsads.u-strasbg.fr/abs/2017A\%26A...605A..63C}{2017, \aap, 605, A63}

    \item \textit{Constraints on CEMP-no progenitors from nuclear astrophysics}\\
    \textbf{Choplin, A}., Maeder A., Meynet G. \& Chiappini C.\\
    \href{http://cdsads.u-strasbg.fr/abs/2016A\%26A...593A..36C}{2016, \aap, 593, A36}

%\item \hyperlink{}{\textbf{\sc Non standard s-process in massive rotating stars. Yields of $10-150$ $M_{\odot}$ models at $Z=10^{-3}$}}\\
%A. Choplin, R. Hirschi, G. Meynet, S. Ekstr\"om, C. Chiappini \& A. Laird\\
%\href{}{2018 \aap, \textbf{\textcolor{red}{???}}, \textcolor{red}{???}}%, id. 123510}

%\item \hyperlink{}{\textbf{\sc Are some CEMP-s stars the daughters of spinstars?}}\\
%A. Choplin, R. Hirschi, G. Meynet \& S Ekstr\"om\\
%\href{http://adsabs.harvard.edu/abs/2017A\%26A...607L...3C}{2017 \aap, \textbf{607}, L3}%, id. 123510}

%\item \hyperlink{}{\textbf{\sc Effects of axions on Population III stars}}\\
%A. Choplin, A. Coc, G. Meynet, K. A. Olive, J.-P. Uzan \& E. Vangioni\\
%\href{http://adsabs.harvard.edu/abs/2017A\%26A...605A.106C}{2017 \aap, \textbf{605}, A106}%, id. 123510}

%\item \hyperlink{}{\textbf{\sc Pre-supernova mixing in CEMP-no source stars}}\\
%A. Choplin, S. Ekstr\"om, G. Meynet, A. Maeder, C. Georgy \& R. Hirschi\\
%\href{http://adsabs.harvard.edu/abs/2017A\%26A...605A..63C}{2017 \aap, \textbf{605}, A63}%, id. 123510}

%\item \hyperlink{}{\textbf{\sc Constraints on CEMP-no progenitors from nuclear astrophysics}}\\
%A. Choplin, A. Maeder, G. Meynet \& C. Chiappini\\
%\href{http://adsabs.harvard.edu/abs/2016A\%26A...593A..36C}{2016 \aap, \textbf{593}, A36}%, id. 123510}

%\item \hyperlink{pdfpop3.1}{\textbf{\sc Effects of rotation on the evolution of primordial stars}}\\
%\textbf{S. Ekstr\"om}, G. Meynet, C. Chiappini, R. Hirschi \& A. Maeder\\
%\href{http://cdsads.u-strasbg.fr/abs/2008A&A...489..685E}{2008 \aap\ \textbf{489}, 685}

\end{itemize}

\vspace{1em}

%------------------------------------------------------------------------------------------------------------------------------------------
\noindent \uline{\large{Publications in conference proceedings}}

\begin{itemize}

     \item \textit{Toward the first stars}\\
     		\textbf{Choplin, A}.\\
		Annual meeting of the French Society of Astronomy and Astrophysics, July 2017 %Paris, France, July 2017

     \item \textit{Massive stars: stellar models and stellar yields, impact on Galactic Archaeology}\\
     Meynet, G., \textbf{Choplin, A}., Ekstr\"{o}m, S. \& Georgy, C.\\
		\href{http://cdsads.u-strasbg.fr/abs/2017arXiv171104554M}{IAU Symposium 334, Rediscovering our Galaxy 2017}

     \item \textit{Evolution and Nucleosynthesis of Massive Stars}\\
     Meynet, G., Maeder, A., \textbf{Choplin, A}., Takahashi, K., Ekstr\"{o}m, S., Hirschi, R., Chiappini, C. \& Eggenberger, P.\\
        \href{http://cdsads.u-strasbg.fr/abs/2017nuco.confa0401M}{14th International Symposium on Nuclei in the Cosmos (NIC2016), id.010401, 8 pp, 2017}

     \item \textit{Impact of rotation on stellar models}\\
     Meynet, G., Maeder, A., Eggenberger, P., Ekstrom, S., Georgy, C., Chiappini, C., Privitera, G. \& \textbf{Choplin, A}.\\
	\href{http://cdsads.u-strasbg.fr/abs/2016AN....337..827M}{Astronomische Nachrichten, Vol.337, Issue 8-9, p.827 2016}

     \item \textit{Nucleosynthesis in the first massive stars}\\
     \textbf{Choplin, A}., Meynet, G., Maeder, A., Hirschi, R. \& Chiappini, C.\\
       \href{http://cdsads.u-strasbg.fr/abs/2016arXiv160204122C}{Nuclear in Astrophysics Conference, 2016}

     \item \textit{Clues about the first stars from CEMP-no stars}\\
     \textbf{Choplin, A}., Meynet, G. \& Maeder, A.\\
		 \href{http://cdsads.u-strasbg.fr/abs/2015sf2a.conf..355C}{Annual meeting of the French Society of Astronomy and Astrophysics, June 2015} %Toulouse, France, June 2015}

%\item \hyperlink{pdfiaus255.1}{\textbf{\sc Powerful explosions at \zzero\ ?}}\\
%S. Ekstr\"om, G. Meynet, R. Hirschi, A. Maeder\\
%in IAU Symposium 255: {\it Low-Metallicity Star Formation: From the First Stars to Dwarf Galaxies}\\ Eds L. K. Hunt, S Madden \& R. Schneider\\
%\href{http://adsabs.harvard.edu/abs/2008IAUS..255..194E}{2008 IAU Symp. 255, p. 194} (arXiv:\href{http://fr.arxiv.org/abs/0807.5050}{0807.5050})

%\item \hyperlink{pdfiaus250.1}{\textbf{\sc Can very massive stars avoid Pair-instability Supernovae?}}\\
%S. Ekstr\"om, G. Meynet, A. Maeder\\
%in IAU Symposium 250: {\it Massive Stars as Cosmic Engines}\\
%Eds F. Bresolin, P.A. Crowther \& J. Puls\\
%\href{http://cdsads.u-strasbg.fr/abs/2008IAUS..250..209E}{2008 IAU Symp. 250, p. 209} (arXiv:\href{http://fr.arxiv.org/abs/0801.3397}{0801.3397})\\\null\\
%and\\\null\\
%in {\it First Stars III}, Eds T. Abel, A. Heger \& B. O'Shea\\
%\href{http://cdsads.u-strasbg.fr/abs/2008AIPC..990..220E}{2007 AIP Conf. Ser. Vol. 990, p. 220} (arXiv:\href{http://fr.arxiv.org/abs/0801.3397}{0709.0202v1})

\end{itemize}

\vspace{1em}

%------------------------------------------------------------------------------------------------------------------------------------------
\noindent \uline{\large{Poster proceedings}}

\begin{itemize}

     \item \textit{Insights on the First Stars from CEMP-no Stars}\\
     \textbf{Choplin, A}., Meynet, G., Maeder, A., Hirschi, R., Ekstr\"{o}m, S. \& Chiappini, C.\\
        		\href{http://cdsads.u-strasbg.fr/abs/2017nuco.confb0202C}{International Symposium on Nuclei in the Cosmos, 2017}

     \item \textit{Clues on the first stars from CEMP-no stars}\\
     \textbf{Choplin, A}., Meynet, G., Maeder, A., Hirschi, R., Ekstr\"{o}m, S. \& Chiappini, C.\\
		\href{http://cdsads.u-strasbg.fr/abs/2016IAUS..317..282C}{IAU Symposium, Hawaii, USA, August 2016}

     \item \textit{Clues about the first stars from CEMP-no stars}\\
     Meynet, G., Maeder, A., \textbf{Choplin, A}., Hirschi, R., Ekstr\"{o}m, S. \& Chiappini, C.\\	
		\href{http://cdsads.u-strasbg.fr/abs/2015IAUGA..2255377M}{IAU General Assembly, 2015}

\end{itemize}

\vspace{1em}

%------------------------------------------------------------------------------------------------------------------------------------------
\noindent \uline{\large{Other oral contributions (no proceeding)}}

\begin{itemize}

     \item \textit{Nucleosynthesis in Rotating massive stars and Abundances of metal-poor stars.} (Invited talk)\\
     \textbf{Choplin, A}.\\
     %\NM{Invited talk}
      \href{https://indico.fnal.gov/event/15487/page/1}{the JINA-CEE Frontiers in Nuclear Astrophysics Meeting, South Bend IN, US, May 2018}

     \item \textit{Shedding light on the first stars with CEMP-no stars.} (Invited talk)\\
     \textbf{Choplin, A}., Meynet, G., Ekstr\"{o}m, S., Hirschi, R., Maeder, A., Chiappini, C. \& Georgy, C.\\ %\NM{Invited review}   
        \href{https://indico.fnal.gov/event/13478/}{A Celebration of CEMP and Gala of GALAH, Melbourne, Australia, November 2017}

\end{itemize}

\newpage

%== REFEREED ARTICLES ========================================================
%\section{\textcolor{red}{Refereed articles} \label{srefart}}
%\vp{.3cm}
%---------------------------------------------------------------------------------------------------------------------------------------

%: article axions ++
\label{axions}
%%%%\includepdf[pages=1,linkname= pdfwimps,offset=0 -2cm]{PhysRevD_78_123510.pdf}
\includepdf[linkname= pdfaxions,scale=0.9,offset=0 -0.5cm]{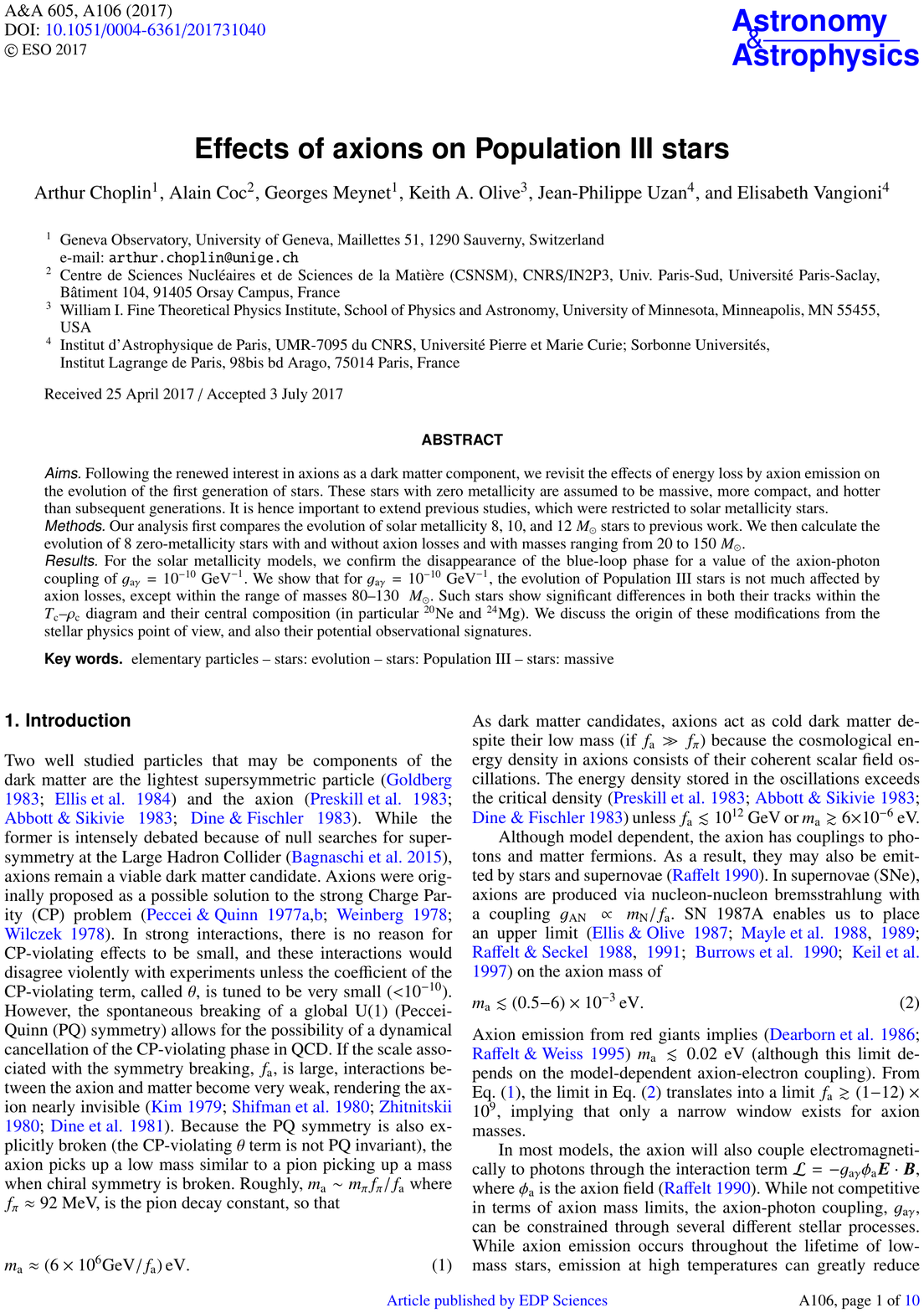}
\clearpage

%: article CEMPno modeles ++
\label{pcempno}
%%%%\includepdf[pages=1,linkname= pdfwimps,offset=0 -2cm]{PhysRevD_78_123510.pdf}
\includepdf[linkname= pdfcempno,scale=0.9,offset=0 -0.5cm]{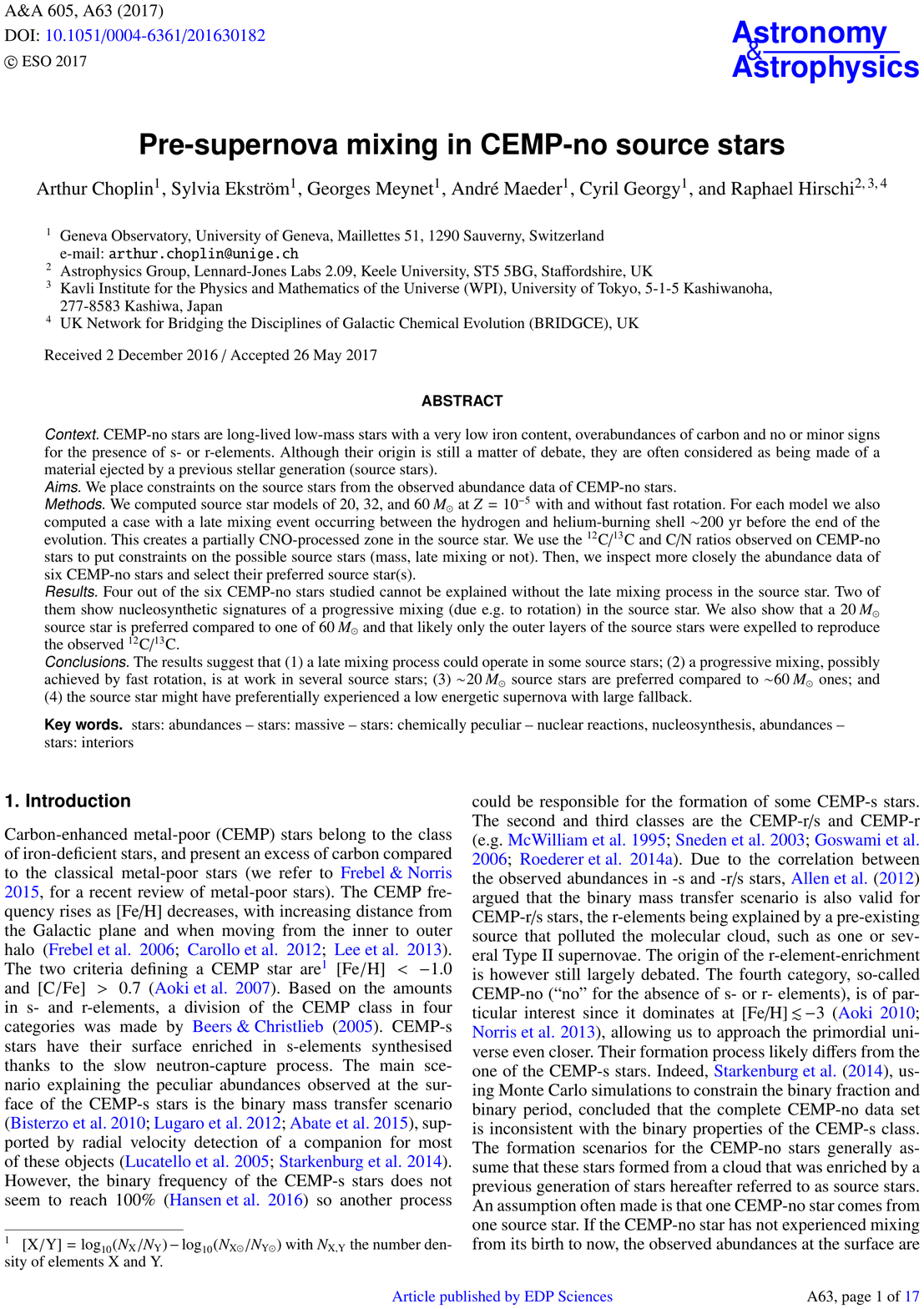}
\clearpage

%: article CEMPno box ++
\label{pbox}
%%%%\includepdf[pages=1,linkname= pdfwimps,offset=0 -2cm]{PhysRevD_78_123510.pdf}
\includepdf[linkname= pdfbox,scale=0.9,offset=0 -0.5cm]{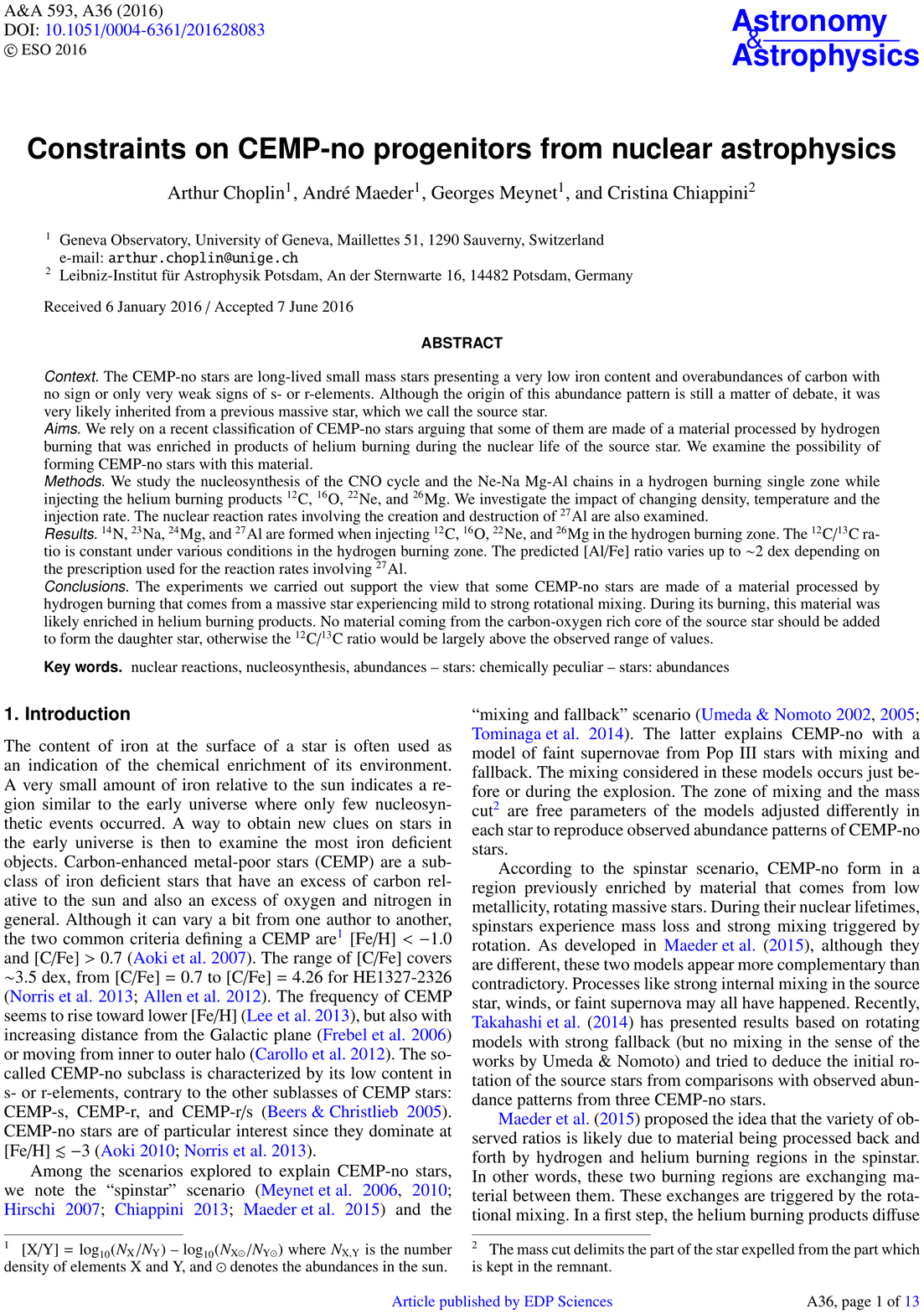}
\clearpage

%%%%%: article axions ++
%\label{letter}
%%%%\includepdf[pages=1,linkname= pdfwimps,offset=0 -2cm]{PhysRevD_78_123510.pdf}
%\includepdf[linkname= pdfletter,scale=0.9,offset=0 -0.5cm]{aa31948-17.pdf}
%\clearpage

%%%%\includepdf[pages=2-,link=false,offset=0 -2cm,scale=0.9]{PhysRevD_78_123510.pdf}
%\includepdf[pages=2-,link=false,scale=0.93]{PhysRevD_78_123510.pdf}
%\clearpage
%---------------------------------------------------------------------------------------------------------------------------------------
%: article POP3 ++
%\label{ppop3}
%\includepdf[linkname=pdfpop3,scale=0.93]{aa09633-08.pdf}
%\clearpage
%---------------------------------------------------------------------------------------------------------------------------------------

%%%%%%%%%%%%%%%%%%%%%%%%%%%%%%%%%%%%%%%%%%%%%%%%%%%
% BIBLIOGRAPHY %
%%%%%%%%%%%%%%%%%%%%%%%%%%%%%%%%%%%%%%%%%%%%%%%%%%%
\cleardoublepage
\markboth{REFERENCES}{References}
\phantomsection
\addcontentsline{toc}{chapter}{References}
\bibliographystyle{aa}
%\bibliography{ThesisBiblio}
%\bibliography{BibTexRefs}
\bibliography{biblio}

}

\end{document}